\documentclass[a4paper,11pt]{book}
\usepackage[top=2cm,bottom=2cm,left=2cm,right=2cm]{geometry}
\usepackage[english]{babel}
\usepackage{multirow}
\usepackage{graphicx}
\usepackage{mathtools}
\usepackage{lipsum}
\usepackage[T1]{fontenc}
\usepackage{amsmath, amssymb, multicol}
\usepackage{hyperref}
\usepackage{wrapfig}
\usepackage[export]{adjustbox}
\usepackage{amssymb}
\usepackage[font={small}]{caption}
\usepackage{float}
\usepackage{multirow}
\usepackage{newpxtext}
\usepackage{bbold}
\usepackage{feynmp}
\usepackage{pdfpages}
\usepackage{comment}
\usepackage{titling}
\usepackage[]{lipsum}
\usepackage{setspace}
\usepackage{breqn}
\usepackage{subcaption}
\usepackage{tikz-feynman}
\usepackage{csquotes}
\tikzfeynmanset{compat=1.1.0}
\usepackage{blindtext}
\usepackage{autobreak}
\usepackage{color}
\usepackage{xcolor}
\usepackage{url}
\usepackage{cite}
\allowdisplaybreaks
\usepackage{verbatim}
\usepackage{listings}

\usepackage{tikz}
\usetikzlibrary{shapes,arrows}
\usetikzlibrary{arrows.meta}
\usetikzlibrary{positioning,fadings}
\usetikzlibrary{decorations.pathmorphing}
\usetikzlibrary{decorations.pathreplacing}
\usetikzlibrary{decorations.markings}

\tikzfeynmanset{double_boson/.style={decorate,
/tikz/double,
/tikz/decoration={snake},
}
}

\newcommand{\mathsym}[1]{{}}
\newcommand{\unicode}[1]{{}}

\title{Master Thesis}
\author{Giacomo Brunello}
\date{March 2022}
\begin{document}
\frontmatter
\begin{titlepage}
\vspace{5mm}
\begin{figure}[hbtp]
\centering
\includegraphics[scale=.13]{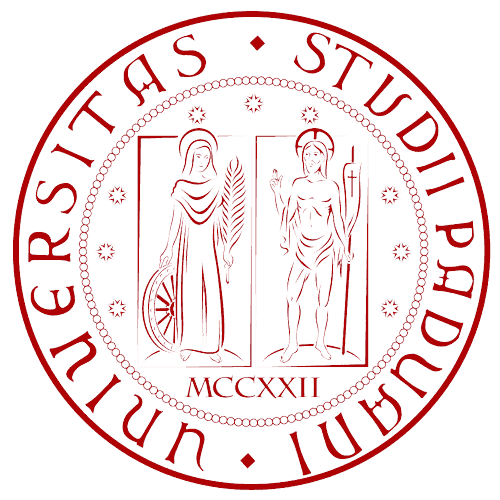}
\end{figure}
\vspace{5mm}
\begin{center}
{{\huge{\textsc{\bf UNIVERSIT\`A DEGLI STUDI DI PADOVA}}}\\}
\vspace{5mm}
{\Large{\bf Dipartimento di Fisica e Astronomia ``Galileo Galilei''}} \\
\vspace{5mm}
{\Large{\textsc{\bf Master Degree in Physics}}}\\
\vspace{20mm}
{\Large{\textsc{\bf Final Dissertation}}}\\
\vspace{30mm}
\begin{spacing}{3}
{\LARGE \textbf{Effective Field Theory Approach to General Relativity
and Feynman Diagrams for Coalescing Binary Systems}}\\
\end{spacing}
\vspace{8mm}
\end{center}

\vspace{20mm}
\begin{spacing}{2}
\begin{tabular}{ l  c  c c c  cc c c c c  l }
{\Large{\bf Thesis supervisor}} &&&&&&&&&&& {\Large{\bf Candidate}}\\
{\Large{\bf Prof. Pierpaolo Mastrolia}} &&&&&&&&&&& {\Large{\bf Giacomo Brunello}}\\
{\Large{\bf Thesis co-supervisor}}\\
{\Large{\bf Dr. Manoj K. Mandal}}\\
\end{tabular}
\end{spacing}
\vspace{15 mm}

\begin{center}
{\Large{\bf Academic Year 2021/2022}}
\end{center}
\end{titlepage}
\clearpage{\pagestyle{empty}\cleardoublepage}
\chapter*{Abstract}
The discovery of gravitational waves (GWs) emitted by a binary coalescing system has oriented the scientific community towards the development and the application of advanced techniques to reveal gravitational waves and to study waveforms, in order to capture the main characteristics of the astrophysical objects that generate them and to discover new features of the gravitational interaction. \\ 
A binary system composed of two coalescing objects such as black holes or neutron stars and the successive emission of gravitational waves has been studied in General Relativity (GR) following two different, but related, perturbative schemes, which are the Post-Newtonian (PN) and the Post-Minkowskian (PM) approaches, that allow to evaluate GR corrections to the Newtonian potential. \\ 
In modern gravitational waves physics such calculations are performed adopting an Effective Field Theory (EFT) approach and with the use of Feynman diagrams and modern Multi-loop techniques which are commonly used in Particle physics.
The master thesis focus on two different aspects of the EFT of a coalescing binary system in General Relativity, which are studied using Feynman diagrams.\\ 
During the first part of this project we are considering the issue of hereditary effects in the PN approach. The hereditary influence is the influence of the past evolution of a material system on its present gravitational internal dynamics, and it is due to GWs emitted by the system in the past and subsequently scattered-off the curvature of space-time back into the system. Such radiation terms give unavoidable contributions to the conservative dynamics of a binary system starting from 4 PN order, and it is mandatory to deal appropriately with this kind of processes in order to compute higher order PN corrections. \\
The current state-of-the-Art of the Post-Newtonian approach has reached the 5PN corrections to the Newtonian potential. Independent derivations of the conservative dynamics at such order present different predictions for physical observables, such as the scattering angle, meaning that we have not a complete understanding of the different contributions entering the conservative dynamics. Our aim is to support with alternative techniques the different derivations of the 5PN corrections, analyzing in detail the hereditary terms appearing using both Feynman and Schwinger-Keldysh formalism, paving the way to a derivation of PN corrections at higher order. The first original contribution of this thesis has been the development of a novel computational algorithm that generates and computes hereditary diagrams, in order to obtain the corresponding conservative Lagrangians in Feynman and in Schwinger-Keldysh formalism. It has been used to make an independent derivation of the 5PN hereditary terms, which involved the evaluation of several two-loop amplitudes, and it can be generalized at higher PN order. \\ 
Moreover, motivated by analogous studies done for other gauge theories such as QED and QCD, in the second part of the project we analyze the problem of the bending of light under the influence of a massive object in the PM approach, which can be seen as a scattering process between one massive and one massless particle in the low-energy effective theory of gravity. Our aim is to investigate additional insights of the gravitational force, by performing high precision predictions of physical observables in this framework. The second original contribution of this thesis has been the development of a computational setup fully automated in Mathematica which evaluates the 1-loop classical and quantum corrections to the bending process in dimensional regularization. By using a novel approach, new relevant quantum terms arise, that were not present in previous analyses, modifying predictions for physical observables. This algorithm has been also adapted to compute classical and quantum 1-loop corrections to the gravitational potential between two compact objects.
\tableofcontents
\mainmatter
\chapter*{Introduction}
\addcontentsline{toc}{chapter}{Introduction}
On September 15, 2015, LIGO \cite{LIGOScientific:2014pky} and Virgo \cite{VIRGO:2014yos} large interferometers made the first direct detection of Gravitational Waves (GWs) emitted by a binary coalescing black-hole system \cite{LIGOScientific:2016aoc}. This event was not only a direct proof of the existence of gravitational waves and of binary coalescing black-hole systems, but it also gave us a new instrument to probe our Universe and to test General Relativity (GR) in the strong field regime. Since then, more than $90$ events have been detected by the LIGO, Virgo, and Kagra\cite{KAGRA:2020agh} collaborations during the first three operative runs O1,O2,O3 \cite{LIGOScientific:2018mvr,LIGOScientific:2020ibl,LIGOScientific:2021qlt, LIGOScientific:2021djp}, in the upcoming years more detection runs have been planned, and exciting outcomes are expected. \\ 
In order to maximize the efficiency of GWs detection, the signals coming from the interferometers are processed via \textit{matched-filtering techniques} \cite{Usman:2015kfa}, which are particularly sensitive to the phase of the GWs signals. These techniques have the advantage to give us a unique probe of the quantitative complexities of the highly non-linear regime of General Relativity, but on the other side the accuracy of the parameters reconstruction relies on precise models of GWs signals. \\ High precision predictions are needed in the upcoming LVK runs, in space-based detectors such as LISA \cite{LISA:2017pwj}, and in future ground-based detectors such as LIGO-India \cite{Saleem:2021iwi}, Cosmic Explorer\cite{Reitze:2019iox} and Einstein Telescope\cite{Punturo:2010zz}. Precise studies of GWs waveforms are very cumbersome due to the multiple physical scales non-linearly coupled through General Relativity, and in order to obtain precision predictions for GWs observables one often finds approximate solutions of Einstein's field equations, expanded around the Newtonian results. The coalescence of a compact binary system can be divided in three distinctive phases: inspiral, merger, and ring-down, and GWs models are a mix of analytical techniques (see \cite{Blanchet:2013haa,Porto:2016pyg,Schafer:2018kuf,Barack:2018yvs} for reviews) used for the inspiral phase, of numerical-relativity methods used for the merger and the ring-down phases, which can be resummed within the \textit{effective-one-body} approach \cite{Buonanno:1998gg}.
During the \textit{inspiral phase}, the two binary constituents are very far away one from each other, they are slowly moving and weakly interacting, and they can be studied within the so-called \textit{Post-Newtonian} (PN) perturbative scheme.
The PN approximation is a low-velocity weak-field expansion, that allow us to study the dynamics of general-relativistic bound systems by adding perturbative correction in the dimensionless parameter $v/c$, where $v$ is the typical velocity of the system. Since for self-gravitating binary systems the virial theorem holds,
the potential energy of the binary is expected to be of the same order of magnitude as the kinetic energy, that is: 
\begin{equation}
v^2\approx \frac{G_N m }{r} \ , 
\end{equation}
where $G_N$ is the Newton constant, whereas $m$ and $r$ are the typical masses and separations of the system, respectively. By taking this relation into account, PN corrections guarantee that the system is bounded by its own gravity, and corrections are organized such that at $n$-PN order they scale as: $G_N^{n-l}v^{2\ell}$ where $0\leq\ell \leq n-1$. This method has been developed by Einstein in 1916 \cite{Einstein:1938yz}, and the 1PN corrections to the Newtonian potential where computed already in 1917 by Lorentz and Droste. The evaluation of higher order PN corrections is far from effortless\cite{Ohta:1973je,Jaranowski:1997ky,Damour:1999cr,Blanchet:2000nv,Damour:2001bu,Damour:2014jta,Jaranowski:2015lha,Damour:2016abl,Bini:2019nra,Bini:2020rzn,Bini:2020uiq}, and it took decades to obtain the 4PN corrections, evaluated in 2014 \cite{Damour:2014jta}. \\
The need for precision predictions, and the difficulties arising in this problem, have triggered the Particle Physics community, to apply the techniques that so far had been developed for collider experiments, to make precision predictions for this new field of Gravitational Waves Physics. \\  In particular, the two milestones that have been adapted to the study of the gravitational two-body problem are \textit{Effective Field Theories}\cite{Georgi:1990um} (EFT)(see \cite{Kaplan:2005es,Goldberger:2007hy,Manohar:2018aog} for  introductions) and modern \textit{Scattering Amplitudes Multi-loop techniques}. 
Effective Field Theories allow us to study a physical system at a given length scale by focusing only on the minimal set of degrees of freedom, which are relevant to describe a specific problem. 
They have been first applied in the case of a coalescing binary system in \cite{Damour:1995kt}, and later systematized in \cite{Goldberger:2004jt}, and in \cite{Porto:2005ac} in the case of binaries with spin (see \cite{Foffa:2013qca,Rothstein:2014sra,Porto:2016pyg,Levi:2018nxp,Goldberger:2022ebt,Blumlein:2022qjy} for modern reviews). They allow us to study the dynamics of a binary coalescing system using a tower of effective theories, each one focused on a specific length scale, and to compute Post-Newtonian corrections to General Relativity with a Feynman diagrammatic approach, characterized by a clear power counting in the expansion parameter $v/c$ \cite{Kol:2007bc,Kol:2007rx,Kol:2010ze,Kol:2010si}, and connections between physics at separate scales are made with renormalization group techniques. Thanks to the so-called \textit{method of regions} \cite{Beneke_1998,Smirnov:2002pj,Manohar:2006nz, Jantzen:2011nz}, one can organize the computation of PN corrections into two different EFTs, by splitting the carriers of the gravitational force, the \textit{gravitons}, into \textit{potential gravitons}, which are the short distance components that mediate the gravitational binding between the binary components, and \textit{radiation gravitons}, which are the long-range modes emitted by the system. In the so-called \textit{near zone} the system can be described by two point-like particles exchanging potential gravitons in a flat spacetime, whereas in the \textit{far zone} we can detail the problem with a single source expanded in multipole moments and emitting radiation gravitons \cite{RevModPhys.52.299,Goldberger:2009qd,Ross:2012fc}. The contributions from the two EFTs have to be summed in order to get consistently PN corrections.
The second milestone is the use of \textit{Quantum Field Theory techniques} for the evaluations of Feynman diagrams. PN contributions, which are computationally organized into \textit{near} and \textit{far} zone contributions, can be topologically mapped into massless multi-loop two points and one point functions, respectively, with a relation that has been first introduced in \cite{Foffa:2016rgu} in the case of potential gravitons. Exploiting this relation, one can perform computations using the machinery of modern amplitudes multi-loop techniques. Instead of directly evaluating all the Feynman diagrams appearing in the theory, we can divide the computations in three different stages: first one can make a \textit{tensor decomposition} of an amplitude containing different free tensorial indices, by writing it as a linear combination of Lorentz vectors and tensors, multiplied by coefficients called \textit{form factors}, where all indices are contracted. Then, one can use some special identities of dimensional regularization relating different integrals, known as \textit{Integration by parts identities} (IBPs) \cite{Chetyrkin:1981qh,Tkachov:1981wb}, Lorentz invariance identities, and other symmetries, to decompose all the scalar integrals appearing in the form factors in a minimal set of integrals, known as \textit{Master Integrals} (MIs), which form some sort of basis of the vector space of integrals. Eventually, one has to actually evaluate only the MIs, a procedure that can be done with direct evaluation or with the use of \textit{difference} \cite{Laporta:2000dsw,Laporta:2000dc} or \textit{differential equations} \cite{Kotikov:1990kg, Henn:2013pwa} (for reviews see \cite{Argeri:2007up,Argeri:2014qva}).
 These new methods allowed rapidly to obtain higher order Post-Newtonian corrections in this new approach: at 2PN in \cite{Gilmore:2008gq}, at 3PN in \cite{Foffa:2011ub}, at 4PN in \cite{Foffa:2012rn,Foffa:2016rgu}, and at 5PN in \cite{Foffa:2019hrb,Foffa:2019ahl,Foffa:2019rdf}, and up to NNNLO including Spin-Orbit interactions \cite{Mandal:2022nty,Levi:2022SO}. An alternative derivation up to 6PN can be found in \cite{Blumlein:2019zku,Blumlein:2020pog,Blumlein:2020pyo, Blumlein:2020znm,Blumlein:2021txe,Blumlein:2021txj}. Starting at 4PN order, there are some radiative terms contributing to the conservative dynamics appearing in the far zone known as \textit{Hereditary effects}\cite{Blanchet:1987wq,Blanchet:1992br,Blanchet:1993ec,Galley:2009px,Galley:2012qs,Foffa:2011np,Foffa:2013qca,Galley:2015kus,Foffa:2019eeb,Blumlein:2021txe,Henry:2021cek,Almeida:2021xwn,Foffa:2021pkg,Edison:2022cdu,Porto:2017dgs,Porto:2017shd,Foffa:2019yfl,Blanchet:2019rjs}, as first noticed in \cite{Blanchet:1987wq}. These terms are due to GWs emitted by the system and then scattered-back into the same GWs source. At the moment there are different derivations of these terms at 5PN, made by different communities and using different techniques \cite{Bini:2021gat,Foffa:2019eeb,Blumlein:2021txe,Bern:2021yeh}, which do not give the same predictions for physical observables, such as the scattering angle. Within the EFT approach that we have so far described, radiative terms have to be studied using the \textit{Schwinger-Keldysh} (or In-In or CTP) formalism \cite{Keldysh:1964ud}(\cite{Galley:2012hx,Galley:2014wla} for reviews), which was first developed by Schwinger in \cite{Schwinger:1960qe}, to compute expectation values in quantum mechanics from a path integral approach, and then adapted to NRGR to get the radiative terms at 2.5PN in \cite{Galley:2009px}, at 3.5PN in \cite{Galley:2012qs}, at 4PN in \cite{Foffa:2011np,Foffa:2013qca,Galley:2015kus}, at 5PN in \cite{Foffa:2019eeb,Blumlein:2021txe,Henry:2021cek,Almeida:2021xwn,Foffa:2021pkg}, and partially up to 7PN order in \cite{Edison:2022cdu}. These terms are not fully understood at the moment, and this puzzle represents one of the main obstacle for the computation of higher order PN corrections. \\ 
Another framework used to obtain PN corrections to the dynamics of a binary coalescing system was found in 1971 by Iwasaki \cite{Iwasaki:1971vb}, where he showed that one can obtain a 1PN gravitational potential by evaluating relativistic scattering amplitudes in a quantum field theory of gravity coupled to matter, described by an action made by the Einstein-Hilbert action plus the action of a scalar field minimally coupled to gravity. Later on, treating General Relativity as a low-energy Effective Field Theory of a Quantum theory of Gravity \cite{Donoghue:1993eb,Donoghue:1994dn, Donoghue:1995cz,Burgess:2003jk}, it was proved that the loop expansion used in this approach allow one to distinguish between classical and quantum terms appearing in the amplitude, because of a subtle cancellation of $\hbar$ factors \cite{Holstein:2004dn}, and this method allowed the evaluation of classical and quantum general relativistic corrections to the Newtonian potential \cite{Bjerrum-Bohr:2002gqz,Bjerrum-Bohr:2002fji,Holstein:2008sx, Holstein:2008sw, Khriplovich:2004cx,Holstein:2006pq, Bjerrum-Bohr:2013bxa}. In recent times this particular method has been receiving a renewed interest, since one can make computations using advanced amplitudes techniques originally developed for Yang-Mills theories computations, such as unitarity methods \cite{Bern:1994zx}. These new techniques extremely simplify the evaluation of gravity amplitudes and make such an approach a good candidate for further advance in the formulation of PN models. Moreover, scattering amplitudes are organized in the so-called \textit{Post-Minkowskian} (PM) perturbative scheme \cite{Bertotti:1956pxu,Bel:1981be,Damour:2016gwp}, which is a weak-field only expansion where at $n$-PM order one considers contributions that scale as $G_N^n$. The PM scheme has been lately proposed to compute the two-body potential at 2PM order, first in  \cite{Cheung:2018wkq} using an EFT matching procedure, and later in \cite{Cristofoli:2019neg} using instead a relativistic \textit{Lippmann-Schwinger} equation.
The evaluation has then been extended at 3PM order \cite{Bern:2019nnu,Bern:2019crd,Cheung:2020gyp,Kalin:2020fhe,Bjerrum-Bohr:2021din}, and at 4PM \cite{Bern:2021dqo,Bern:2021yeh,Dlapa:2021npj} using different techniques. This framework can be used to study other relevant physical problems, such as the bending of light under the influence of a heavy object like the sun \cite{Bjerrum-Bohr:2014zsa,Bjerrum-Bohr:2016hpa,Bjerrum-Bohr:2017dxw,Bai:2016ivl,Chi:2019owc}. In perturbative quantum gravity, one can study both classical and quantum corrections to this process and to different physical observables, like the Saphiro time delay, using Scattering Amplitudes multi-loop techniques. In particular, this kind of analysis can give us unique estimates of quantum gravity corrections, which can provide new insights on the UV completions of gravitational theories, helping in the development of EFTs of modified gravity \cite{AccettulliHuber:2020oou}. \\ Within this thesis, we will elaborate on both the approaches that we have described so far, and both analyses will present original contributions of this work. In the first part, we will work on the EFT approach to the binary two-body problem, and we will develop a computational algorithm to compute the Hereditary terms using Feynman diagrams up to 5PN order\footnote{This study was done in collaboration with Raj Patil.}, using the multi-loop technology to evaluate the complicated two-loop integrals appearing, and we will compare our independent results with the ones appearing in literature \cite{Foffa:2019eeb,Blumlein:2021txe,Almeida:2021xwn}. Then, after understanding that the correct approach to deal with radiation problems is the CTP formalism, we will repeat this same calculation using Schwinger-Keldysh variables, finding a new way to express In-In amplitudes in terms of Feynman diagrams, and we will recompute the 5PN conservative Lagrangians using this formalism. During the second part of the project instead, we will work in the PM perturbative scheme, and we will design an automatic procedure, fully implemented in \texttt{Mathematica}, to get the one-loop classical and quantum corrections to the scattering angle of a light-like scalar particle under the influence of a massive scalar. Working in d-dimensions within a novel approach, we will find new quantum contributions which were not appearing in previous analyses \cite{Bjerrum-Bohr:2014zsa,Bjerrum-Bohr:2016hpa,Bjerrum-Bohr:2017dxw}. The appearance of these terms will modify the predictions for the bending angle, and other studies obtained from these works have to be reviewed \cite{Bai:2016ivl,AccettulliHuber:2020oou,Chi:2019owc} according to our novel results. Eventually, we will adapt this computation to the case of two massive scalar fields, and we will evaluate the 1-loop classical and quantum corrections to the Newtonian potential, finding agreement with the results appearing in literature, proving the universality of the methods implemented. \\ 
The material is divided into seven Chapters, as follows: 
\begin{itemize}
    \item In the first Chapter, we will study Gravitational Waves in General Relativity: their propagation far from the source, and their generation from dynamical sources, introducing the concept of multipole expansion which is useful to study the radiation from a binary system. As an application, we will study gravitational waves emitted by a point-particle binary, deriving the predictions for physical observables like the phase or the amplitude of the GWs. Eventually we will show the limits of this approach, and we will introduce perturbative techniques which are necessary to study in a more systematic way deviations from the simple Newtonian results.
    \item In the second Chapter we will introduce the Effective Field Theory approach to the binary coalescing system, where PN corrections to the binary dynamics are given in terms of Feynman diagrams, organized in two different EFTs thanks to the method of regions. We will study a simplified model of scalar gravity to understand how the procedure works, and then we will describe how to obtain PN corrections in the Near and in the Far zone.
    \item In the third Chapter, since PN corrections to the Newtonian results are given by multi-loop Feynman diagrams, we will introduce modern scattering amplitudes multi-loop techniques which are necessary to compute them. We will describe the tensor decomposition procedure in which a generic integral can be expressed in terms of Lorentz tensors multiplied by form factors. We will classify Feynman diagrams, and we will outline how to reduce them in a basis of Master Integrals, via Integration-by-Parts reduction. Then we will directly evaluate some Master Integral which will be useful in the following. 
    \item In the fourth Chapter, we will use the multi-loop technology introduced to evaluate the Hereditary Effects up to 5PN order using Feynman diagram. To perform the computation we will develop a novel computational algorithm in \texttt{Mathematica}, which starting from the effective action of the theory: evaluates the Feynman rules, generate the diagrams and evaluates them, and which can be generalized at higher PN order. We will make also comparisons with the results appearing in literature.
    \item In the fifth Chapter we will introduce the so-called Schinger-Keldysh (or In-In, or CTP) formalism, useful to describe the dynamics of non-conservative problems such as radiation ones. Specializing this formalism to the case of NRGR, we will then evaluate again the Hereditary effects, this time in the In-In formalism, using a new approach which uses Feynman diagrams as building blocks for In-In computations, and we will compare the results with those appearing in literature.
    \item In the sixth Chapter, we will introduce another framework to study Binary Coalescing systems, based on the Post-Minkowskian perturbative scheme, which is based on treating the full theory of gravitons coupled to scalar fields as a relativistic effective field theory. We will describe a general method to compute corrections to the Newtonian potential between two massive objects using advanced scattering amplitudes techniques.
    \item Eventually, in the seventh Chapter, we will focus on the problem of the bending of a light-like scalar field under the influence of a heavy scalar, and we will compute classical and quantum corrections to the bending angle at one loop-order. For the evaluation we will develop a new computational method in \texttt{Mathematica} in dimensional regularization which can be generalized to higher order. Within this approach we will find new quantum terms arising, which were not present in previous analyses, which modify the predictions of the bending angle, and which require further studies. We will then adapt this computation to get the 1-loop classical and quantum corrections to the Newtonian potential between two massive objects, finding agreement with the results appearing in literature.
\end{itemize}
In conclusion, we will point out that the methods proposed can be used for planning new calculations that will be relevant for the future generation of gravitational waves detectors. On the PN side, the algorithmic methods developed can be generalized to compute Hereditary effects at higher order, which are essential to get a more precise knowledge of the gravitational waveforms. In the context of PM approach, the new quantum terms appearing in the estimation of the bending angle have opened new interesting and fascinating questions in the context of perturbative quantum gravity, and the analysis can be extended at higher loop order.
\chapter{Gravitational waves in General Relativity}
\label{Ch:grav_waves}
On 14 September 2015 the two detectors of the Laser Interferometer Gravitational-Wave Observatory (LIGO) made the first direct detection of Gravitational Waves generated by a binary coalescing system \cite{LIGOScientific:2016aoc}.  
The detection of GWs, aside from demonstrating the existence of binary stellar-mass coalescing black hole systems, has opened the possibility to test General Relativity (GR) in the strong field regime. 
Gravitational radiation is not only a prediction of Einstein's theory but also a crucial tool that allows us to study peculiar features of gravity. \\ 
Despite these remarkable successes of the theory, General relativity is a highly non-linear tensorial theory, where only few exact and physically meaningful solutions are known. Those that are known are highly symmetric and can be applied only to ideal cases. Two main ones are the Schwarzschild and Kerr geometries, extensively used to describe static and rotating astrophysical objects, respectively. Widely accepted is also the Friedman-Robertson-Walker solution, which is often used in cosmology.   \\ 
Clearly, any spacetime involving gravitational radiation is expected to be dynamical and highly non-symmetrical, so the known exact solutions of Einstein's equations are naturally not suitable in this case.
Therefore, our knowledge of the gravitational waves emission and propagation from a source mainly comes from approximations of the Einstein's field equations. \\ 
\subsubsection{Goals of the chapter:}
On this chapter we want to review gravitational waves in general relativity, and their application in the case of a compact binary system, and it is organized as follows:
\begin{itemize}
    \item First, we will study GWs in the context of linearized gravity, their propagation in the traceless tansverse gauge and its energy-momentum tensor;
    \item then we will focus of the generation of gravitational waves, showing how relativistic $v/c$ corrections arise by means of a so-called \textit{multipole expansion}, and we will apply this formalism to compact binaries;
    \item  finally, we will see that the GWs spectrum emitted by a coalescing binary system cannot be fully described within linearized general relativity, and we will introduce two methods used to parametrize General Relativity corrections as progressive deviations from Newtonian solutions.
\end{itemize}
\section{Gravitational waves in linearized gravity}
Let us consider the classical action in General Relativity (GR)(for detailed discussions on the topics see \cite{Carroll:2004st,Maggiore:2007ulw}) of matter coupled to gravity:
\begin{equation}
    S=S_{EH}+S_M \, 
\end{equation}
where: 
\begin{itemize}
    \item $S_{EH}$ is the Einstein-Hilbert action, and it is defined as:
    \begin{equation}
        S_{EH}= \frac{1}{16\pi G_N}\int d^4x\sqrt{-g}R \ , 
    \end{equation}
    where $g_{\mu\nu}$ is the metric tensor, $R_{\mu\nu\rho\sigma}$ is the Riemann curvature tensor, $R$ is the Ricci scalar, $G_N$ is the Newton constant, and $\sqrt{-g}=-\sqrt{-det(g_{\mu\nu})}$
    \item $S_M$ is a matter action that we will specify later.
\end{itemize}
From the least action principle, we can get the \textbf{Einstein's equations} by extremising the action with respect to the metric field $g_{\mu\nu}(x)$:
\begin{equation}
    R_{\mu\nu}-\frac{1}{2}g_{\mu\nu}R=8\pi G_N T_{\mu\nu}
\end{equation}
where $T_{\mu\nu}$ is the energy momentum tensor  of matter, defined as:
\begin{equation}
    T^{\mu\nu}=\frac{-2}{\sqrt{-g}}\frac{\delta S_M}{\delta g_{\mu\nu}} \ . 
\end{equation}
An essential property of GR is the diffeomorphism invariance, also known as gauge symmetry, that means that GR is invariant under the group of coordinate transformations: $x^\mu\to x^{'\mu}(x)$, with $x^{'\mu}$ differentiable with a differentiable inverse. \\
The metric transforms in a tensorial way under a local coordinate transformation, as:
\begin{equation}
    g_{\mu\nu}(x)\to g'_{\mu\nu}(x')=\frac{\partial x^{\rho}}{\partial x^{'\mu}}\frac{\partial x^{\sigma}}{\partial x^{'\nu}}g_{\rho\sigma}(x) \ . 
    \label{eq:diffeo_transformation}
\end{equation}
In order to study gravitational waves in GR, we need to move towards the so-called linearized GR, which can be obtained by expanding the metric tensor around the Minkowski solution $\eta_{\mu\nu}=diag(-1,1,1,1)$, in a \textit{weak field} regime, as:
\begin{equation}
    g_{\mu\nu}=\eta_{\mu\nu}+h_{\mu\nu}, \qquad |h_{\mu\nu}|\ll1 \ . 
    \label{eq:metric_expansion}
\end{equation}
By making this choice, we implicitly choose a reference frame in which the condition \eqref{eq:metric_expansion} is valid for a sufficient large region of spacetime, and we are breaking the diffeomorphism invariance. \\ 
Nevertheless, there is a residual gauge symmetry, meaning that the metric should be invariant under infinitesimal coordinate transformations defined as:
\begin{equation}
    x^{\mu}\to x^{\mu}+\xi^{\mu}\qquad \text{with} \qquad \xi\sim \mathcal{O}(h) \ . 
\end{equation}
Under such a coordinate transformation the field $h_{\mu\nu}(x)$, from Eq.\eqref{eq:diffeo_transformation}, at lowest order in $\xi$, transforms as:
\begin{equation}
    h_{\mu\nu}\to h_{\mu\nu}-(\partial_\mu\xi_\nu+\partial_\nu\xi_\mu) \ . 
\end{equation}
At linear order in $h_{\mu\nu}$ the Riemann curvature tensor $R_{\mu\nu\rho\sigma}$ reduces to: 
\begin{eqnarray}
R_{\mu\nu\rho\sigma}=\frac{1}{2}\left(\partial_\mu\partial_\sigma h_{\nu\rho}+\partial_\nu\partial_\rho h_{\mu\sigma}-\partial_\mu\partial_\rho h_{\nu\sigma}-\partial_\nu\partial_\sigma h_{\mu\rho}\right) \ , 
\end{eqnarray}
and the Einstein Equations can be linearized as:
\begin{equation}
    \Box h_{\mu\nu}+\partial_{\mu}\partial_{\nu}h-\partial_{\mu}\partial^{\rho}h_{\rho\nu}-\partial_{\nu}\partial^{\rho}h_{\rho\mu}=-16\pi G_NT_{\mu\nu} \ , \qquad h=\eta^{\mu\nu}h_{\mu\nu} \ . 
\end{equation}
We can then define the field $\bar{h}_{\mu\nu}=h_{\mu\nu}-\frac{1}{2}\eta_{\mu\nu}h$, and the linearized gravity equations become:
\begin{equation}
    \Box\bar{h}_{\mu\nu}+\eta_{\mu\nu}\partial^{\rho}\partial^{\sigma}\bar{h}_{\rho\sigma}-\partial^{\rho}\partial_{\nu}\bar{h}_{\mu\rho}-\partial^{\rho}\partial_{\mu}\bar{h}_{\nu\rho}=-16\pi G_NT_{\mu\nu} \ . 
      \label{eq:linearized_einstein}
\end{equation}
The field $\bar{h}_{\mu\nu}$ transforms under a gauge transformation, as:
\begin{equation}
    \bar{h}_{\mu\nu}\to \bar{h}_{\mu\nu}+\partial_\mu\xi_\nu+\partial_\nu\xi_\mu-\eta_{\mu\nu}\partial^\rho\xi_\rho \ , 
    \label{eq:gauge_hbar}
\end{equation}
At this stage the field $\bar{h}_{\mu\nu}$ has 10 degrees of freedom, since it is symmetric. We can partially remove the gauge redundancy by imposing the De-Dondler gauge fixing condition:
\begin{equation}
    \partial^{\nu}\bar{h}_{\mu\nu}=0 \ . 
\end{equation}
Such an imposition is always possible, as it is equivalent to require:
\begin{equation}
    \Box \xi_\mu = f_\mu \ ,
 \label{eq:de_dondler_xi}
\end{equation}
for a generic function $f_\mu $, while in turn \eqref{eq:de_dondler_xi} is guaranteed to admit solutions thanks to the invertibility of the d'Alambertian operator. 
By doing so, we are left with 6 independent components of $h_{\mu\nu}$ and the Linearized Einstein equations \eqref{eq:linearized_einstein} reduce to:
\begin{equation}
    \Box\bar{h}_{\mu\nu}=-16\pi G_N T_{\mu\nu}. 
    \label{eq:wave_equation}
\end{equation}
Eq.\eqref{eq:wave_equation} is a \textbf{wave equation} which is at the hearth of the study of the propagation and generation of Gravitational Waves within linearized GR, and will be much useful in the following sections. \\

\section{Propagation of gravitational waves in the traceless transverse gauge}

In order to understand the fundamental features of GWs propagation, we put our lens far away from the matter source, where $T_{\mu\nu}=0$. The vacuum Linearized Einstein's equations in the De-Dondler gauge become:
\begin{equation}
    \Box\bar{h}_{\mu\nu}=0 \ , \qquad \partial^{\nu}\bar{h}_{\mu\nu}=0 \ . 
    \label{eq:vacuum_ee} 
\end{equation}
Since $\Box= -(1/c^2)\partial_t^2+\nabla^2$, we immediately deduce that GWs travel at the speed of light. 
The gauge symmetry \eqref{eq:gauge_hbar} in terms of $\bar{h}_{\mu\nu}$ can be rewritten as: 
\begin{equation}
    \bar{h}_{\mu\nu}\to  \bar{h}_{\mu\nu}-\xi_{\mu\nu} \ ,
\end{equation}
where $\xi_{\mu\nu}=\partial_\mu \xi_\nu + \partial_\nu \xi_\mu +\eta_{\mu\nu}\partial^\rho \xi_\rho$. 
Hence, with a simple derivative, we find that:
\begin{equation}
    \partial^\nu\bar{h}_{\mu\nu} \to \partial^\nu \bar{h}_{\mu\nu}-\Box \xi_{\mu\nu} = \partial ^\nu \bar{h}_{\mu\nu}-\Box\xi_\mu \ , 
\end{equation}
and we see that a coordinate transformation $x^\mu \to x^\mu + x^{'\mu}$ does not spoil the De-Dondler condition, provided that $\Box\xi_{\mu\nu}=\Box\xi_\mu=0$. This gives us the opportunity to impose four other conditions on $\bar{h}_{\mu\nu}$. 
In particular, we  choose $\xi^0$ such that $\bar{h}=0$ and consequently:
\begin{eqnarray}
    \bar{h}_{\mu\nu}= h_{\mu\nu} \ ,
\end{eqnarray}
and we fix the remaining three components to set $h^{0i}=0$. \\ 
Combining this with the De-Dondler condition, we end up with the following set of conditions:
\begin{eqnarray}
    h^{0\mu}=0 \ , \qquad h^i_i =0 \ , \qquad \partial^j h_{ij}=0 \ ,
\end{eqnarray}
which constitute the so-called \textit{traceless-transverse gauge}, or TT gauge. In the TT gauge the metric tensor $h_{\mu\nu}$ has no more spurious degrees of freedom, and it is usually denoted by $h_{\mu\nu}^{TT}$.
In analogy to the Electromagnetic waves in the Lorentz gauge, we can write the general solution eq.\eqref{eq:vacuum_ee} in Fourier space, in terms of monochromatic plane waves, defined:
\begin{equation}
    \bar{h}_{\mu\nu}(k)=Re(H_{\mu\nu}(k)e^{ik_\rho x^{\rho}}) \ , 
\end{equation}
where $H_{\mu\nu}(k)$ is the polarization tensor and $k^{\mu}=(w/c,\mathbf{k})$, with $w/c=\vert\mathbf{k}\vert$, is the wave vector. \\
In momentum space we get the following conditions:
\begin{equation}
    k^{\rho}k_{\rho}H_{\mu\nu}(k)=0\qquad k^{\mu}H_{\mu\nu}(k)=0
\end{equation}
The polarization tensor in the TT gauge, satisfies:
\begin{equation}
    H_{0\mu}=0\ ,  \qquad H^{i}_{i}=0 \ , \qquad \hat{n}^jH_{ij}=0 \ ,
\end{equation}
with $\hat{\mathbf{n}}=\mathbf{k}/\vert \mathbf{k}\vert$ being the direction of propagation of the GWs. \\ 
A general solution of eq.\eqref{eq:vacuum_ee} is then given by a superposition of plane waves:
\begin{eqnarray}
    h_{ij}^{TT}=\int \frac{d^3\mathbf{k}}{(2\pi)^3}\left(H_{\mu\nu}^{TT}(\mathbf{k})e^{ikx}+c.c.\right)\bigg|_{k^0=\vert\mathbf{k}\vert} \ . 
\end{eqnarray}
Let us now restrict ourselves to a single plane wave propagating along the z axis, $k^{\mu}=w(1,0,0,1)^T$. 
In the TT gauge the gravitational field has only two independent degrees of freedom, and the polarization tensor can be written as:
\begin{equation}
    H_{\mu\nu}^{TT}=\begin{pmatrix}
0 & 0 & 0 & 0\\
0 & h_+ & h_{\times} & 0 \\
0 & h_{\times} & -h_{+} & 0 \\ 0& 0 & 0 & 0 \\
\end{pmatrix}
\end{equation}
where $h_+$ and $h_{\times}$ are known as "plus" and "cross" polarization. We can write the line element $ds^2$ in the TT gauge travelling along z direction as:
\begin{equation}
    ds^2=\eta_{\mu\nu}dx^{\mu}dx^{\nu}-h_+(t-z)(dx^2-dy^2)-2h_{\times}(t-z)dxdy \ .
\end{equation}
It is worth mentioning that the TT gauge we have worked in so far is not well defined inside the GW source, as in this case $\Box\bar{h}_{\mu\nu}\neq 0$. \\ There the De Dondler gauge could still be imposed and four other degrees of freedom could be gauged away, but we have no means of setting to zero any further component of $\bar{h}_{\mu\nu}$.

\section{Energy and momentum flux of GWs}

The aim of this section is to characterize GWs from the energetic point of view, computing the expressions for the energy and momentum flux within linearized theory. \\
In addition to its geometrical interpretation, one can think at this linearized theory as a classical theory of the field $h_{\mu\nu}$ living in the Flat Minkowski space-time with metric $\eta_{\mu \nu}$. 
In order to do that we need to find the action that governs the $h_{\mu\nu}$ dynamics, which is the so-called  \textit{Fierz-Pauli action}, and can be obtained  by expanding the Einstein-Hilbert action at second order in $h_{\mu\nu}$:
\begin{equation}
    S_{FP}=\frac{-1}{64\pi G_N}\int d^4x(\partial_{\rho}h_{\alpha\beta}\partial^{\rho}h^{\alpha\beta}-\partial_{\rho}h\partial^{\rho}h+2\partial_{\rho}h^{\rho\alpha}\partial_{\alpha}h-2\partial_{\rho}h_{\alpha\beta}\partial^{\beta}h^{\rho\alpha})\ . 
\end{equation}
From $S_{FP}$ we want to build an appropriate energy-momentum tensor for the GWs. \\ 
First, one can notice that $S_{FP}$ is invariant under global space-time translations $x^\mu\to x^{\mu}+a^{\mu}$. From Noether's theorem, exists an energy momentum tensor defined as the conserved current: 
\begin{equation}
    t_{\mu\nu}=-\frac{\partial\mathcal{L}}{\partial(\partial_{\mu}h_{\alpha\beta})}\partial_{\nu}h_{\alpha\beta}+\eta_{\mu\nu}\mathcal{L} \ , \qquad \partial^{\mu}t_{\mu\nu}=0 \ . 
    \label{eq:em_tensor}
\end{equation}
By computing the field derivatives, the final expression is given by:
\begin{equation}
    t_{\mu\nu}=\frac{c^4}{32\pi G_N}\partial_{\mu}h_{\alpha\beta}\partial_{\nu}h^{\alpha\beta}\ . 
    \label{eq:gw_em_tensor}
\end{equation}
A remarkable aspect of \eqref{eq:gw_em_tensor} is that, unlike electromagnetic waves, it is not possible to use it in order to define a local energy density which is gauge invariant. This is due to the fact that in GR the equivalence principle ensures the existence of a locally inertial frame in which any local quantity associated to the gravitational field can be set to zero, so that the effect of gravity is locally absent. \\
However, the lack of a gauge invariant local energy density it is not a problem for our purpose, since the only measurable quantity is the total energy emitted, which is not a local quantity. \\
In order to derive the energy flux of GWs, one needs first integrate on a fixed volume $V$ the equation on the right of \eqref{eq:em_tensor}, obtainining the following continuity equation:
\begin{equation}
    \int_Vd^3x\partial_\mu t^{\mu\nu}=0 \qquad \rightarrow \qquad \dot{P}^\nu=-\int_{\partial V}d\Sigma_it^{i\nu} \ , 
    \label{eq:continuity_equation}
\end{equation}
where  $P^{\nu}=\int_V d^3 x t^{0\nu}$ is the four momentum in a spatial volume $V$, in the region far from the source.
If we impose $\nu=0$ we get the conservation of the energy:
\begin{equation}
    \frac{dE}{dt}=-\int_{\partial V}d\Sigma_it^{i0}
\end{equation}
If we consider a spherical volume we have $d\Sigma_i=r^2d\Omega n_i$, by substituting it in Eq.\eqref{eq:gw_em_tensor} we obtain:
\begin{equation}
    \frac{dE}{dt}=\frac{c^4r^2}{32\pi G_N}\int d\Omega n_i\partial_0h_{\alpha\beta}\partial^ih^{\alpha\beta} \ , 
\end{equation}
where the overall sign is positive since we are interested in the received energy at infinity. \\
In particular for sufficiently large  $r$, it can be shown that in analogy with electromagnetic waves, the propagating GW can be assumed to have the general form:
\begin{equation}
    h_{ij}^{TT}(t,r)=\frac{1}{r}f_{ij}(t-r/c) 
\end{equation}
for an unspecified function $f_{ij}$ of the retarded time $t_{ret}=t-r/c$. \\
Choosing the direction $k$ parallel to the radial vector $\Vec{r}$ we have:
\begin{eqnarray}
    \partial_kh_{ij}^{TT}=-\partial_0h_{ij}^{TT}+\mathcal{O}(1/r^2) \ . 
\end{eqnarray}
Imposing the TT gauge outside the source, one obtains the following simple result for the energy flux:
\begin{equation}
    \frac{dE}{dt}=-\frac{c^3r^2}{32\pi G_N}\int d\Omega \dot{h}^{TT}_{ij}\dot{h}^{TT}_{ij} \ . 
    \label{eq:energy_flux}
    \end{equation}
    An analogous calculation can be done from eq.\eqref{eq:continuity_equation} to obtain the flux of momentum emitted by a gravitational wave in the TT gauge:
    \begin{equation}
        \frac{dP^k}{dt}=-\frac{c^3r^2}{32\pi G_N}\int d\Omega \dot{h}^{TT}_{ij}\partial^{k}\dot{h}^{TT}_{ij}
    \end{equation}

\section{Generation of gravitational waves}
In this section we are going to study how gravitational waves can be generated from the dynamics of physical sources, working in Linearized GR. \\
Let us recall the equations of motion in the De Dondler gauge, as given in Eq.\eqref{eq:wave_equation}:
\begin{equation}
    \Box \bar{h}_{\mu\nu}=-\frac{16\pi G_N}{c^4}T_{\mu\nu} \ , 
    \label{eq:gw_ee}
\end{equation}
where $T_{\mu\nu}$ is the energy-momentum tensor of the GW matter source. \\
A solution of \eqref{eq:gw_ee} can be found by introducing the Green's function $G(x-y)$ for the operator $\Box_x$, which is defined by identity:
\begin{equation}
    \Box G(x-y)=\delta^4(x-y) \ ,
    \label{eq:gw_green_function}
\end{equation}
and it is given by
\begin{equation}
    \bar{h}_{\mu\nu}(x)=-\frac{16\pi G_N}{c^4}\int d^4y G(x-y)T_{\mu\nu}(y) \ ,
    \label{eq:sol_green_function}
\end{equation}
where the integration in $y$ is over the region of spacetime containing the source.
For a radiation problem like this one, the correct boundary conditions are given by the retarded Green's function:
\begin{equation}
    G(x-y)=-\frac{1}{4\pi|\mathbf{x}-\mathbf{y}|}\delta(x^0_{ret}-y^0), \qquad x^0_{ret}=ct-|\mathbf{x}-\mathbf{y}| \ ,
\end{equation}
and by substituting it in Eq.\eqref{eq:sol_green_function} we get:
\begin{equation}
    \bar{h}_{\mu\nu}(x)=\frac{4G_N}{c^4}\int d^3\mathbf{y}\frac{1}{|\mathbf{x}-\mathbf{y}|}T_{\mu\nu}(ct-|\mathbf{x}-\mathbf{y}|,\mathbf{y})
\end{equation}
Outside the source we would like to impose the TT gauge. \\
To this end we introduce the transverse projector $P_{ij}(\hat{\mathbf{n}})=\delta_{ij}-n_in_j$ with $\mathbf{n}^2=1$,  and then the Lambda tensor:
\begin{equation}
    \Lambda_{ijkl}\equiv P_{ik}P_{jl}-\frac{1}{2}P_{ij}P_{kl} \ ,
\end{equation}
which satisfy the following properties:
\begin{equation}
    \Lambda_{ijkl}\Lambda_{klmn}=\Lambda_{ijmn}\ , \qquad n^i\Lambda_{ijkl}= n^j\Lambda_{ijkl}=\cdots = 0\ , \qquad \Lambda_{iikl}=\Lambda_{ijkk}=0\ .
\end{equation}
The main feature of this tensor is that it extracts the traceless-transverse part of any symmetric tensor $A_{ij}$:
\begin{equation}
    A_{ij}^{TT}=\Lambda_{ijkl}A_{kl}
\end{equation}
We can use it to write the gravitational waves in the TT gauge as:
\begin{equation}
    h_{ij}^{TT}=\frac{4G_N}{c^4}\Lambda_{ijkl}(\hat{\mathbf{x}})\int d^3y\frac{1}{|\mathbf{x}-\mathbf{y}|}T_{kl}(ct-|\mathbf{x}-\mathbf{y}|,\mathbf{y}) \ , 
    \label{eq:gw_tt}
\end{equation}
where $T_{00}$ and $T_{0k}$ can be omitted by recalling the conservation law $\partial_{\mu}T^{\mu\nu}=0$ that relates them to $T_{kl}$.

\subsection{Multipole expansion}   
\begin{figure}[H]
    \centering
    \includegraphics[width=0.5\textwidth]{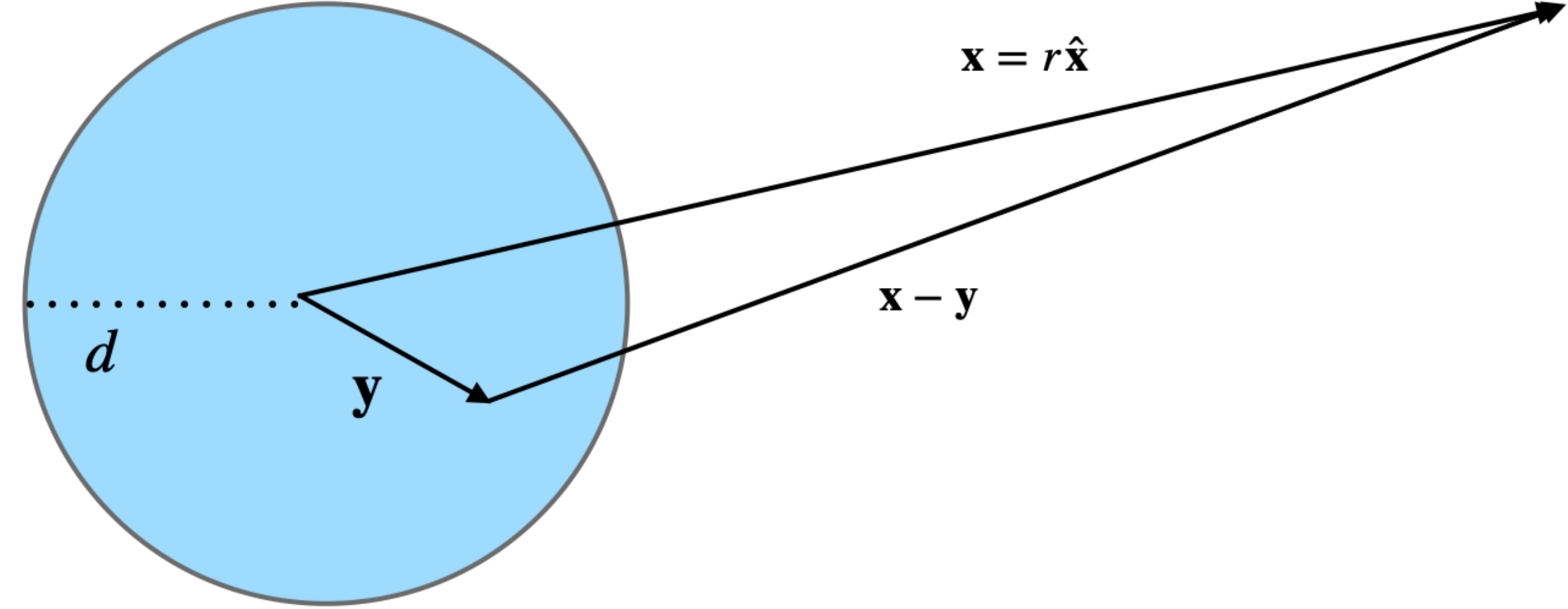}
    \caption{Sketch of the geometry of the multipole expansion.}
    \label{fig:multipole_exp}
\end{figure}
Let us consider a spherical source of radius $d$, and let us put ourselves far away from the source at a distance $r\gg d$. We can notice that the  integral in \eqref{eq:gw_tt} is restricted to $y\leq d$, since $T_{\mu\nu}$ vanishes outside the source.
In this situation, one can expand the retarded time for large $r$, as:
\begin{equation}
    ct- |\mathbf{x}-\mathbf{y}|=ct-r-\mathbf{y}\cdot \hat{\mathbf{x}} + \mathcal{O}\left(\frac{d^2}{r^2}\right), \qquad  \mathbf{{x}}=r\hat{\mathbf{x}} \ . 
\end{equation}
The expression for the GWs in the TT gauge of \eqref{eq:gw_tt} becomes:
\begin{equation}
    h_{ij}^{TT}(x)=\frac{4G_N}{rc^4}\Lambda_{ijkl}(\Vec{n})\int d^3\mathbf{y}T_{kl}\bigl(ct-r-\mathbf{y}\cdot \mathbf{n},\mathbf{y}\bigr) \ , 
    \label{eq:gw_multi}
\end{equation}
in which all terms $\mathcal{O}(1/r^2)$ have been neglected. \\ 
In general, it is not possible to evaluate this three-dimensional integral, and it is necessary to adopt other approximations beside the large distance one. We choose to work in a  perturbative approach called \textit{multipoles expansion}, which is valid only if the typical velocity of the source is non-relativistic $v \ll c$. \\
The magnitude of the typical velocity inside a source with size $d$ is $v\sim w_s d$, where $w_s$ is the typical frequency one could associate to its internal motion. \\ Furthermore, as we will show below in this chapter, the frequency $w_{gw}=2\pi c/\lambda$ of the radiated GWs presents a magnitude comparable to $w_s$, so that we have: 
\begin{equation}
    \lambda_{gw} \sim \frac{2\pi c}{w_{gw}} \sim \frac{2\pi c}{w_{s}} \frac{c}{v}d . 
\end{equation}
By imposing $v\ll c$, we end up with: 
\begin{equation}
    \lambda_{gw} \gg d \ .
\end{equation}
Such a condition states that, from the perspective of a GW observer, the motion inside a non-relativistic source is of little importance: the source is essentially probed as a whole. Therefore, we are naturally led to perform a multipole expansion of the source while neglecting, in first approximation, all the multipole moments beyond the lowest one since we already know they would provide corrections carrying along more and more details on the source internal motion. \\ 
With this in mind, let us Fourier transform the expression for the energy-momentum tensor in\eqref{eq:gw_multi} as:
\begin{equation}
    T_{kl}
(ct-r+\mathbf{y}\cdot \hat{\mathbf{x}}, \mathbf{y})=\int \frac{d^4 k}{(2\pi)^4}\Tilde{T}_{kl}(\omega,\mathbf{k})e^{-i\bigl(\omega t-\frac{r}{c}+\frac{\mathbf{y}\cdot \hat{\mathbf{x}}}{c}\bigr)+i\mathbf{k}\cdot \mathbf{y}}
\label{eq:gw_em_tensor_fourier}
\end{equation}
Being focused on a non-relativistic source, we should expect the energy momentum tensor $\Tilde{T}_{kl}(\omega,\mathbf{k})$ to be peaked around a typical frequency $\omega_s$, and the typical velocity of the source is $v\sim \omega_s d\ll c$. \\
Furthermore, mindful of the condition $\vert\mathbf{y}\vert \leq d$, we observe that :
\begin{equation}
    \frac{\omega_s}{c}\mathbf{y}\cdot \hat{\mathbf{x}} \leq \frac{\omega_s d}{c}\sim \frac{v}{c}<<1
    \label{eq:gw_multipole}
\end{equation}
Therefore we can use \eqref{eq:gw_multipole} to expand the exponential in \eqref{eq:gw_em_tensor_fourier}, using $\omega_s\sim\omega$, as:
\begin{equation}
    e^{i\omega(t-r/c)}\biggl(1-i\frac{\omega}{c}y^i\hat{x}^i-\frac{1}{2}\frac{\omega^2}{c^2}y^i y^j\hat{x}^i \hat{x}^j+\cdots\biggr)
    \label{eq:exponential_expansion}
\end{equation}
Using\eqref{eq:exponential_expansion}, we can expand the energy momentum tensor in\eqref{eq:gw_em_tensor_fourier} as:
\begin{equation}
    T_{kl}(ct-r+\mathbf{y}\cdot \hat{\mathbf{x}},\mathbf{y})\simeq \biggl[T_{kl}(ct-r,\mathbf{y})+\frac{y^i \hat{x}^i}{c}\partial_0T_{kl}+\frac{1}{2c^2}y^i y^j \hat{x}^i \hat{x}^j\partial_0^2T_{kl}+\cdots\biggr]_{(t-r/c,\mathbf{y})}
\end{equation}
where all the quantities on the RHS are evaluated at the point $(t-r/c,\mathbf{y})$. \\
This is the so-called multipole expansion, the first term is the \textit{monopole}, then there is the \textit{dipole},  the \textit{quadrupole} and so on.
Moreover, we introduce the momenta of $T^{ij}$ as:
\begin{eqnarray}
         S^{ij}(t)& \equiv &  \int d^3\mathbf{x} T^{ij}(t,\mathbf{x}) \qquad  \text{[stress monopole]}\ , \nonumber \\ 
     S^{ij,k}(t) &\equiv& \int d^3\mathbf{x}T^{ij}(t,\mathbf{x})x^k  \qquad \text{[stress dipole]} \ , \nonumber\\
         S^{ij,kl}(t) & \equiv&  \int d^3\mathbf{x}T^{ij}(t,\mathbf{x})x^k x^l \qquad \text{[stress quadrupole]} \ . \label{eq:momenta}
\end{eqnarray}
and similarly for higher order momenta. Note that the comma in $S^{ij,klm\cdots}$ stands between two completely symmetric groups of indices, while the exchange of two comma-separated indices is not a symmetry.
In terms of \eqref{eq:momenta}, we can rewrite \eqref{eq:gw_multi} as:
\begin{equation}
    h_{ij}^{TT}(t,\mathbf{x})=\frac{4G_N}{rc^4}\Lambda_{ij,kl}(\hat{\mathbf{x}})\biggl[S^{kl}(ct-r)+\frac{\hat{x}_m}{c}\dot{S}^{kl,m}(ct-r)+\frac{n_pn_m}{2c^2}\Ddot{S}^{kl,mp}(ct-r)+\cdots\biggr]_{(t-r/c)}
    \label{eq:gw_tt_multipole_2}
\end{equation}
Equation \eqref{eq:gw_tt_multipole_2} is the basis for the multipole expansion. \\
From \eqref{eq:momenta} we can notice that, with respect to $S^{ij}$, the term $S^{ij,k}$ has an additional factor $x^m\sim\mathcal{O}(d)$. Time derivatives bring a factor $\mathcal{O}(\omega_s)$, and so the term $\dot{S}^{ij,k}$ has an additional factor $\mathcal{O}(\omega_sd \sim v)$ with respect to $S^{ij}$. That means that $\frac{n_m}{c}\dot{S}^{ij,k}$ has an additional factor $\mathcal{O}(v/c)$ with respect to $S^{ij}$, and similarly the term $\Ddot{S}^{kl,mp}$ introduces corrections of order $\mathcal{O}(v^2/c^2)$ and so on. Therefore, the multipole expansion can be regarded as a non-relativistic expansion in powers of $v/c$. \\
Truncating \eqref{eq:gw_tt_multipole_2} to the lowest order we have: 
\begin{eqnarray}
    [h_{ij}^{TT}]_{leading}=\frac{1}{r}\frac{4G_N}{c^4}\Lambda_{ij,kl}(\hat{\mathbf{x}})S^{kl}(t-r/c)\ .
\end{eqnarray}
Now we want to rewrite the momentum $S^{kl}$ in terms of quantities with an easier physical interpretation. Using  the conservation of the associated energy momentum tensor we can express the various momenta $S^{ij},\dot{S}^{ij,k}, \ddot{S}^{ij,kl}$ and so on, in terms of the two set $M,M^{i},M^{ij},\cdots$ and $P^i,P^{ij},\cdot $.
As an example we will show that for the  first term in the multipole expansion we get:
\begin{equation}
    S^{ij}=\frac{1}{2}\ddot{M}^{ij}
\end{equation}
In order to do that is convenient to define the momenta of the other components of $T^{\mu\nu}$.
For $T^{00}/c^2$, the mass density in  the low-velocity weak-field limit, one has:
\begin{eqnarray}
M &\equiv& \frac{1}{c^2}\int d^3 x T^{00}(t,x) \qquad \text{[mass monopole]}\ , \nonumber  \\
M^i &\equiv &\frac{1}{c^2}\int d^3x T^{00}(t,x)x^i  \qquad \text{[mass dipole]}\ , \nonumber\\
M^{ij} &\equiv &\frac{1}{c^2}\int d^3xT^{00}(t,x)x^ix^j \nonumber \qquad \text{[mass quadrupole]} \ ,  \\
M^{ijk} &\equiv& \frac{1}{c^2}\int d^3xT^{00}(t,x)x^ix^jx^k \qquad \text{[mass octupole]}\ .\label{eq:em_momenta_m}
\end{eqnarray}
Correspondingly, the momenta associated to the momentum density $T^{0i}/c$ are:
\begin{eqnarray}
    P^i& \equiv & \frac{1}{c}\int d^3xT^{0i}(t,x) \qquad \text{[momentum monopole] }\ , \nonumber \\
    P^{i,j}& \equiv & \frac{1}{c}\int d^3xT^{0i}(t,x)x^j \qquad \text{[momentum dipole] }\ , \nonumber \\
    P^{i,jk}& \equiv & \frac{1}{c}\int d^3xT^{0i}(t,x)x^jx^k \qquad \text{[momentum quadrupole] }\ . 
\label{eq:em_momenta_p}
\end{eqnarray}
and so on. 
Let us rewrite the $S^{ij}$ definition in a convenient fashion: 
\begin{eqnarray}
    S^{ij}& =&  \int d^3 \mathbf{x} T^{ij} = \int d^3 \mathbf{x} T^{ik}\delta^j_k  \nonumber \\ 
    & = & \int d^3 \mathbf{x} T^{ik}\partial_k x^j = -\int d^3 \mathbf{x}\partial_k T^{ik}x^j \ . 
    \label{eq:S_in_T}
\end{eqnarray}
The last equality holds because $T^{\mu\nu}$ vanishes outside the source, so that we are free to extend the integral on a box with volume $V$ larger than the source, on whose boundaries $T^{\mu\nu}=0$, thus overall allowing integration by parts. \\ 
At the same time from the conservation law $\partial_\mu T^{\mu\nu}=0$ we have:
\begin{equation}
    \partial_k T^{k \nu}=- \frac{1}{c}\partial_t T^{0\nu} \ . 
    \label{eq:stress_cons}
\end{equation}
Inserting \eqref{eq:stress_cons} in \eqref{eq:S_in_T} and taking the symmetric part of the right term bring us to: 
\begin{equation}
    S^{ij}=\frac{1}{c}\int d^3\mathbf{x}\partial_tT^{0(k}x^{j)} =\frac{1}{2}\bigl(P^{i,j}+P^{j,i}\bigr)
\end{equation}
With the same trick we have:
\begin{eqnarray}
    P^{i,j} &=& \frac{1}{c}\int d^3\mathbf{x}\partial_t T^{0k}\delta^i_kx^j=\frac{1}{c}\int d^3\mathbf{x}T^{0k}\bigl(\partial_k x^i\bigr)x^j \nonumber \\ 
    & = & -\frac{1}{c}\int d^3\mathbf{x}\biggl(\partial_kT^{0k}x^ix^j+T^{0j}x^i\biggr)\nonumber \\ 
    & = & \partial_tM^{ij}-P^{j,i} \ . 
\end{eqnarray}
Altogether we obtain: 
\begin{equation}
    S^{ij}=\frac{1}{2}\ddot{M}^{ij}
    \label{eq:s_first_term}
\end{equation}

\subsection{The quadrupole radiation }
Using \eqref{eq:s_first_term} we can obtain the leading order of \eqref{eq:gw_tt_multipole_2} that is:
\begin{equation}
    [h_{ij}^{TT}(t,\mathbf{x})]_{leading}=\frac{2G_N}{rc^4}\Lambda_{ijkl}(\hat{\mathbf{x}})\ddot{M}_{kl}(ct-r)+\mathcal{O}(v/c)
\end{equation}
In other words the leading term of the expansion \eqref{eq:gw_tt_multipole_2} involves exclusively the mass quadrupole.  
This is the reason why truncating the multipole expansion to the lowest possible order is usually referred to as \textbf{quadrupole approximation}. It is possible to demonstrate that the leading quadrupole nature of GWs in linearized theory is actually preserved even in the full theory, and can be understood from a field-theoretic point of view of GR. Just like in electrodynamics, where photon states with angular momentum $j=0$ are not allowed, it can be shown that states  with total angular momentum $j=0$ or $j=1$ are forbidden for spin-2 massless particles with definite parity, such as gravity. If we imagine monopole and dipole gravitational waves as packets of gravitons in states with total angular momentum $j=0$ and $j=1$, respectively, it is clear that the multipole expansion cannot start but from $j=2$, because the lower terms would be forbidden.
We can define the quadrupole momentum as the traceless part of $M^{ij}$:
\begin{equation}
    Q^{ij}\equiv \biggl(M^{ij}-\frac{1}{3}\delta^{ij}M^k_k\biggr)
\end{equation}
and we can use it to arrive at the following equation:
\begin{equation}
    h_{ij}^{TT}=\frac{2G_N}{rc^4}\Lambda^{ijkl}(\hat{\mathbf{x}})\ddot{Q}_{kl}(ct-r)
    \label{eq:gw_quadrupole}
\end{equation}
We can use this last equation to derive an exact formula for the power radiated per unit solid angle in the quadrupole approximation. I can do that by inserting \eqref{eq:gw_quadrupole} in \eqref{eq:energy_flux}:
\begin{equation}
    \frac{dP}{d\Omega}=\frac{G_N}{8\pi c^5}\Lambda^{ijkl}(\hat{\mathbf{x}})\dddot{Q}_{ij}\dddot{Q}_{kl} \ . 
\end{equation}
We can then perform the integration over the solid angle, which can be easily done by noting that the angular dependence is in the projectors:
\begin{equation}
    \int d\Omega \Lambda^{ijkl}(\hat{\mathbf{x}})=\frac{2\pi}{15}(11\delta^{ik}\delta^{jl}-4\delta^{ij}\delta^{kl}+\delta^{il}\delta^{jk}) \ .
\end{equation}
By doing so we arrive at the total radiated power in quadrupole approximation:
\begin{equation}
    P_{quad}=\frac{G_N}{5c^5}\dddot{Q}^{ij}\dddot{Q}_{ij} \ .
    \label{eq:power_emitted_quad} 
\end{equation}
We will now use the formulas obtained for the waveform Eq.\eqref{eq:gw_quadrupole} and for the radiated power Eq.\eqref{eq:power_emitted_quad} in the quadrupole approximation to study systems emitting radiation. 

\section{Radiation from point particle sources}
Within linearized GR, the simplest source that one can study is represented by a closed system of point particles moving under their mutual gravitational attraction. \\ As we will see this is a good approximation for many physical situations, and it offers a first sight on the gravitational dynamics of a real binary system. \\
In particular, here we will describe a circular moving system of massive point particles, under the Newtonian regime. \\ 
The energy momentum tensor for point particles can be expressed in a non-covariant way as:
\begin{equation}
    T^{\mu\nu}_{tot}(t,\mathbf{x})=\sum_a\frac{p^\mu_ap^\nu_a}{\gamma_am_a}\delta^3(\mathbf{x}-\mathbf{x}_a(t))=\sum_a\gamma_a m_a\frac{dx^\mu_a}{dt}\frac{dx^{\nu}_a}{dt} \delta^{3}(\mathbf{x}-\mathbf{x}_a(t)) \ , 
\end{equation}
where $\gamma=(1-v^2/c^2)^{1/2}$ is the usual Lorentz factor and the index $a$ run over the number of constituents, $2$ for a binary.
We can expand the energy momentum tensors in powers of $v/c$, neglecting $\mathcal{O}(v^2/c^2)$ corrections, obtaining: 
\begin{equation}
    T^{\mu\nu}_{tot}(t,\mathbf{x})=\sum_am_a\frac{dx^\mu_a}{dt}\frac{dx^{\nu}_a}{dt} \delta^{3}(\mathbf{x}-\mathbf{x}_a(t)) +\sum_a\mathcal{O}\left(\frac{v_a^2}{c^2}\right) \ . 
    \label{eq:em_tensor_particles}
\end{equation}
By doing that we end up loosing some contributions which may seem necessary, but, in the quadrupole approximation, Eq.\eqref{eq:em_tensor_particles} can be proved to provide the same result of the full theory. \\ 
 We want to compute the total power radiated in the quadrupole approximation, in order to do that we need to compute the mass quadrupole momentum $M^{ij}$.
Before doing that we restrict to 2 point particles, and we move to the center of mass frame defined by:
\begin{equation}
    x^i_{cm}=\frac{m_Ax^i_A(t)+m_Bx^i_B(t)}{m_A+m_B} \ , \qquad x^i_{rel}(t)=x^i_A(t)-x^i_B(t) \ .
\end{equation}
We indicate with $m=m_A+m_B$ the total mass, and with $\mu=\frac{m_Am_B}{m_A+m_B}$ the reduced mass. \\
By doing so we have:
\begin{equation}
\begin{split}
    M^{ij}(t)=\frac{1}{c^2}\int d^3\mathbf{x}T^{00}x^ix^j=\sum_{k=A,B}m_kx^i_k(t)x^j_k(t)=mx^i_{cm}(t)x^{j}_{cm}(t)+\mu x^{i}_{rel}(t)x^j_{rel}(t) \ . 
\end{split}
\end{equation}
We can set the origin in the center of mass frame such that $x_{cm}=0$, yielding to the simpler expression:
\begin{equation}
    M^{ij}(t)=\mu x^i_{rel}x^{j}_{rel} \ . 
    \label{eq:quadrupole_mass_momentum}
\end{equation}
As for the relative motion, we consider for example a gravitational oscillator with mass m around a heavier mass M. The relative coordinate is moving along the z-axis with harmonic oscillations:
\begin{equation}
    z_{rel}(t)=a cos(w_st) \ . 
\end{equation}
The second mass momentum becomes:
\begin{equation}
    M^{ij}(t)=\delta^{i3}\delta^{j3}\frac{\mu a_1^2}{2}(1+2cos(2\omega_s t))\qquad \Rightarrow \qquad \ddot{M}^{ij}=-2\delta^{i3}\delta^{j3}\omega_s^2\mu a^2cos(2\omega_s t) \ . 
\end{equation}
Then the gravitational waves produced by the system are, in quadrupole approximation:
\begin{equation}
    h_{ij}^{TT}=\frac{2G_N}{rc^4}\Lambda^{ijkl}(\Vec{n})\ddot{M}_{kl}(ct-r)\qquad \Rightarrow \qquad h_{ij}^{TT}=C_{ij}cos(2\omega_s t) \ , 
    \label{eq:h_grav}
\end{equation}
with $C_{ij}$ time independent tensor. \\
It is important to notice that for a non-relativistic closed gravitational system, the gravitational waves, produced in quadrupole approximation, oscillate with a frequency $\omega_{gw}$ proportional to the typical frequency of the system with $\omega_{gw}=2\omega_s$. This proportionality is valid for more complex system, and retails its validity for higher multipole expansion with the only difference that the proportionality factor could change. \\
We can then compute the radiated power obtaining:
\begin{equation}
    \frac{dP}{d\Omega}=\frac{G_N\mu^2a^4\omega_s^6}{2\pi c^5}sin^4(\theta) \qquad \Rightarrow \qquad P_{quad}=\frac{16}{15}\frac{G_N\mu^2}{c^5}a^4\omega_s^6
    \label{eq:emitted_power}
\end{equation}
which is the final result of this section. \\
\section{Compact binaries as a radiating system}
The aim of this section is to apply what we have developed so far to study a more realistic system, as the inspiral of a compact binary system. We will assume a Minkowskian background, which means it is possible to apply Newton's gravitational law, in particular we will consider a compact binary made by two objects of masses $m_1$ and $m_2$ on a fixed Keplerian orbit, a circular one with constant frequency $w_s$. \\
Denoting the orbital distance with $R$ and the total mass as $m=m_1+m_2$, one has that the third Kepler's law holds: 
\begin{equation}
    w_s^2=\frac{G_N m}{R^3}
\end{equation}
We choose the frame so that the orbit lies on the x-y plane and the trajectories of the binaries are: 
\begin{equation}
    \mathbf{x}_{rel}(t) = \bigl(Rsin(w_s t),Rcos(w_s t ),0\bigr)
\end{equation}
Moving in the center of mass frame, one can proceed with the evaluation of the second mass momenta using \eqref{eq:quadrupole_mass_momentum}, which yields: 
\begin{eqnarray}
    M_{11} & = & \mu R^2 \biggl(\frac{1-cos(2w_s t)}{2}\biggr) \ , \nonumber \\ 
    M_{22} & = & \mu R^2 \biggl(\frac{1+cos(2w_s t )}{2}\biggr) \ , \nonumber \\ 
    M_{12} & = & -\frac{1}{2}\mu R^2 sin(2w_s t ) \ ,
\end{eqnarray}
while the other components are null. \\
Given these, one can derive from \eqref{eq:h_grav} the expression for the gravitational waves produced in the TT gauge. In terms of their two physical components: 
\begin{eqnarray}
    h_+(t,\theta,\phi)& =& \frac{1}{r}\frac{4 G_N \mu w_s^2 R^2}{c^4}\biggl(\frac{1+cos^2\theta}{2}\biggr)cos(2w_s t_{ret}) \ , \\ 
    h_\times (t,\theta \phi) & = & \frac{1}{r}\frac{4 G_N \mu w_s^2 R^2}{c^4}cos\theta cos (2 w_s t_{ret}) \label{eq:h_grav_circ}
\end{eqnarray}
being $\theta$ the angle between the line of sight and the normal to the orbit, and $\phi= 2 w_s t$ the angle lying on the $x-y$ plane. \\ 
Also in this case we recognize that the gravitational waves produced oscillates with a time frequency $w_{gw}$, related to the one of the motion inside the source, such that: 
\begin{equation}
    w_{gw}=2 w_s
\end{equation} 
At this point it is a straightforward calculation the evaluation of the total power radiated in quadrupole approximation. Introducing the so-called \textit{mass chirp} $M_c$ we have: 
\begin{equation}
    P_{quad} = \frac{32}{5}\frac{c^5}{G_N}\biggl(\frac{G_N M_C w_s}{c^3}\biggr)^{10/3} \ , \qquad M_c=\mu^{3/2}m^{2/5}
    \label{eq:emitted_energy}
\end{equation}

\subsection{The coalescence of compact binaries}
\label{subsec:coalescence_binaries}
The assumption that compact binaries are on a fixed Keplerian orbit is inconsistent with the conservation of energy, since the system constantly radiates energy into space as we found in \eqref{eq:emitted_energy}, and the release must take place at the expense of the internal energy of the system. \\
Therefore, we need to study time dependence of $R$ and $w_s$ and consequently understand the actual significance of our analysis in relation to the predicted orbital evolution. \\ 
First, let us recall that a Newtonian bound system must satisfy the \textit{virial theorem}: 
\begin{equation}
    E_{kin}=-\frac{1}{2}E_{pot} \qquad \Rightarrow \qquad \frac{1}{2}\mu v^2 = \frac{G_N \mu m}{2 R } \ , 
\end{equation}
namely, as long as $v=w R$: 
\begin{equation}
    \omega_s^2=\frac{G_N m}{R^3} \ , 
    \label{eq:freq_source}
\end{equation}
that is the well-known third Kepler's law. \\ 
In parallel, in linearized GR we can express the total energy of the system as the sum of the kinetic and potential energy:
\begin{equation} 
    E_{int}=E_{kin}+E_{pot}=-\frac{G_N \mu m }{2R} \ . 
    \label{eq:internal_energy}
\end{equation}
Due to the emission of gravitational waves, $E_{int}$ has to become more and more negative. That means that $R$ has to decrease with time, $\omega_s$ has to increase for \eqref{eq:freq_source}, and that the same for the emitted energy from \eqref{eq:emitted_energy}. The natural conclusion is that at the end there is the coalescence of the binary system. \\
In order to give an estimate of the process we can express the energy of the binary in terms of $\omega_s$, using \eqref{eq:freq_source}, as:
\begin{equation}
    E_{int}=-\biggl(\frac{G_N^2M_C^5w_s^2}{8}\biggr)^{1/3}
\end{equation}
Due to conservation of energy one must have that $P=\frac{-dE_{int}}{dt}$ which gives, unite with \eqref{eq:emitted_power}, the following differential equation in $w_s$:
\begin{equation}
    \dot{\omega}_s=\frac{192}{5}\biggl(\frac{G_NM_C}{c^3}\biggr)^{5/3}w_s^{11/3} \ , 
\end{equation}
and a similar differential equation for $R$. 
The integration exhibit a divergence for $\omega_s$ at a finite time, that we will call $t_{coal}$.
The generic solutions are: 
\begin{equation}
    R(t)=R_0\left(\frac{t_{coal}-t}{t_{coal}- t_0}\right)^{1/4} \ , \qquad w_s(t)= w_s(t_0) \left(\frac{t_{coal}-t}{t_{coal}-t_0}\right)^{-3/8} \ , 
    \label{eq:r_ws_generic}
\end{equation}
where $t_0$ is the initial time and $t_{coal}$ is the finite value of time we associate to the coalescence since: 
\begin{equation}
    \lim_{t\to t_{coal}}R(t)=0 \ , \qquad \lim_{t\to t_{coal}}w_s(t) = \infty \ . 
\end{equation}
Substituting $w(t_0)$ we obtain for $w_s$:
\begin{equation}
    \omega_s=\biggl(\frac{5}{256(t_{coal}-t)}\biggr)^{3/8}\biggl(\frac{G_NM_C}{c^3}\biggr)^{-5/8}=\frac{\omega_{gw}}{2}
    \label{eq:omega_final}
\end{equation}
From this last equation we can read the time behavior of the orbital frequency which, it must be remembered, is also proportional to the frequency of the emitted gravitational waves. With the passage of time the two binaries become closer and closer with an increasing orbital frequency, leading at the end to an artificial singularity which only reveals that our estimate is no longer valid when the two binary merges.
Several factors can explain the birth of this divergence: first having treated components of a binary system as point-like when in reality the binary ones have a physical extension. Nevertheless, this does not change the physical meaning of \eqref{eq:omega_final}, which can be used for phenomenological considerations. \\ 

\subsection{Estimates and numerical values for compact binaries }

We will now make some phenomenological considerations on \eqref{eq:omega_final} expressed in terms of the frequency $f_{gw}=\frac{w_{gw}}{2\pi}$. \\
If we express it in numerical values, including Newton's constant and velocity of light, in terms of the sun mass $M_{\odot}$ we get:
\begin{equation}
    f_{gw}=\frac{1}{\pi}\biggl(\frac{5}{256}\frac{1}{t_{coal}-t}\biggr)^{3/8}\biggl(\frac{G_NM_C}{c^3}\biggr)^{-5/8}\approx 143Hz\biggl(\frac{1.21M_{\odot}}{M_C}\biggr)^{5/8}\biggl(\frac{1s}{t_{coal}-t}\biggr)^{3/8}
    \label{eq:gw_frequency}
\end{equation}
We can invert this equation and get:
\begin{equation}
    t_{coal}-t\approx 2s\biggl(\frac{1.21M_{\odot}}{M_C}\biggr)^{5/3}\biggl(\frac{100Hz}{f_{gw}}\biggr)^{3/8}
\end{equation}
that gives an estimate of the time missing at the coalescence in terms of the frequency of the gravitational waves emitted. \\
Just by looking at \eqref{eq:gw_frequency} we can say that if a detector would receive a gravitational wave from a binary system with $M_C=1.21M_{\odot}$ this would mean that for frequencies detected around $f_{gw}=10Hz$ the system is at 17 minutes to coalescence, for frequencies at $f_{gw}=100Hz$ at few seconds, and so on, until the two merge and our model fail to be valid.\\
We can also estimate the number of cycles spent in the accessible detector bandwidth $[f_{min},f_{max}]$ which can be expressed as:
\begin{equation}
    N_{cycles}=\int_{t_i}^{t_f}dtf_{gw}=\int_{f_{min}}^{f_{max}}df_{gw}\frac{f_{gw}}{\dot{f_{gw}}}
\end{equation}
The result is:
\begin{equation}
    N_{cycles}\approx 1.6 \times 10^4\biggl(\frac{10Hz}{f_{min}}\biggr)^{5/3}
    \label{eq:number_cycles}
\end{equation}
where we have neglected the contribution coming from $f_{max}$ which for a typical detector is negligible.
From eq.\eqref{eq:number_cycles} we can see that for a detector sensitive to $mHz$ frequencies the number of cycles spent in the detector could be more than a million, which means that in principle the knowledge of the incoming gravitational wave is possible with great accuracy. This is the basis to test physics beyond linearized GR.

\subsection{The inspiral phase of a coalescing binary system}
For the purpose of this thesis we will be interested in the \textbf{inspiral phase} of the coalescing of a compact binary system, which corresponds to the time window $t \ll t_{coal}$, in which the binary components are moving at non-relativistic velocities, and the orbital separation is slowly decaying. This phase is characterized by a nearly smooth slope both in $R(t)$ and in $w_s(t)$. In this region we can in first approximation neglect time derivatives of $R$ and $w_s$, since $\dot{R}\ll R^2$ and $\dot{w_s}\ll w_s^2$, so roughly validating the results \eqref{eq:h_grav_circ}. We are in the so-called \textit{quasi-circular} motion regime. \\ 
In order to reach an acceptable first approximation we can partially reintroduce in \eqref{eq:h_grav_circ} the effects of the orbital evolution by replacing $w_s$ and $R$ with their expressions \eqref{eq:h_grav_circ}, together with the substitution: 
\begin{equation}
    w_{gw}t = 2w_s t \to \Phi(t)=2 \int_{t0}^{t}dt' w_{s}(t')
    \label{eq:gw_phase}
\end{equation}
where we defined the time-dependent phase $\Phi(t)$ of the emitted GW. \\
Therefore, after a shift of the time origin which let us replace $2w_s(t-r/c)$ with $2w_s t$, we get: 
\begin{eqnarray}
    h_{+}(\tau,\theta)& = & \frac{1}{r}\biggl(\frac{G_NM_c}{c^2}\biggr)^{5/4}\biggl(\frac{5}{c\tau}\biggr)^{1/4}\frac{(1+cos^2\theta)}{2}cos\bigl(\Phi(t)\bigr) \ , \label{eq:ampl_corrected_plus}\\
    h_{\times}(\tau,\theta)& = & \frac{1}{r}\biggl(\frac{G_NM_c}{c^2}\biggr)^{5/4}\biggl(\frac{5}{c\tau}\biggr)^{1/4}cos(\theta)sin\bigl(\Phi(t)\bigr) \ ,\label{eq:ampl_corrected_cross}
\end{eqnarray}
where $\tau= t_{coal} -t$ is the time to coalescence. \\ 
From \eqref{eq:r_ws_generic} and \eqref{eq:gw_phase} we obtain the time dependence of the GW phase $\Phi$: 
\begin{equation}
    \Phi(\tau)= -2\left(\frac{5G_NM_c}{c^3}\right)^{-5/8}\tau^{5/8}+\Phi_0
\end{equation}
in which we used $d\tau=-dt$ and the integration constant $\phi_0$ associated to the value of $\Phi$ at coalescence, namely $\Phi_0=\Phi(\tau=0)$. \\ 
Approaching the coalescence this quasi-circular orbit description eventually loses its validity, due to the visible growth of $\dot{R}$ and $\dot{w}_s$, however at this stage we would be definitely outside the inspiral phase, namely the one we are interested in. \\ 
Besides, even within the quasi-circular regime the currently achieved results represent only a first step in the modeling of binary-radiated GWs, suitable to support the detection experiments, since they are the ultimate offspring of the quadrupole approximation in the context of Newtonian dynamics. In the next section we will address the problems one encounters in trying to go beyond linearized General Relativity and introduce possible analytical methods to accomplish that.

\section{Beyond Linearized General relativity}
\label{sec:beyond_linearized_GR}
\begin{figure}[H]
    \centering
    \includegraphics[width=0.8\textwidth]{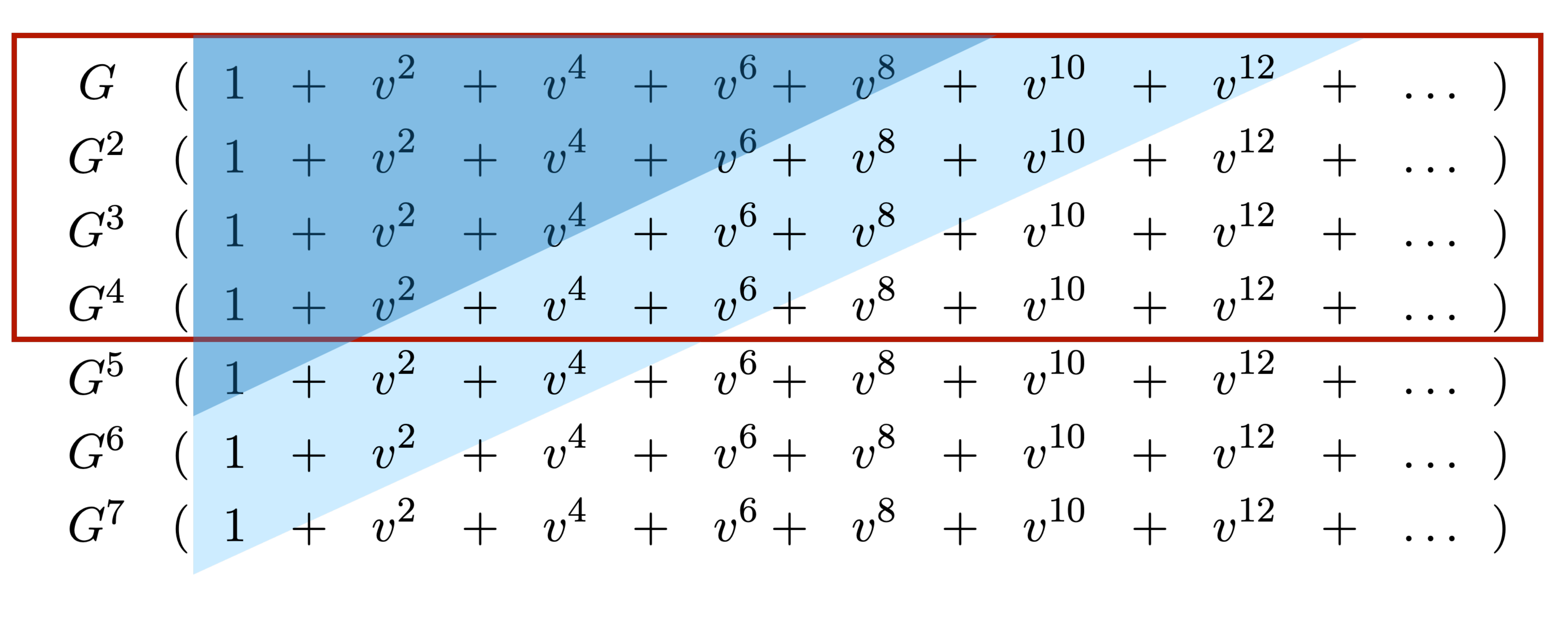}
    \caption{Contributions to PM and PN perturbative schemes at different orders. PM order scales with the number of horizontal lines, whereas PN order is increases along the diagonals}
    \label{fig:PN_PM}
\end{figure}
In order to describe the production of GWs in a multipole expansion in $v/c$, with $v$ the typical internal speed of the system, we assumed that the background space-time is the usual Minkowski one, and that the GWs sources do not contribute to the space-time curvature. However, this assumption is valid only if the background space-time curvature and the velocity of the source can be treated as independent variables. Unfortunately, this is not the case if the system is governed by gravitational forces. \\
Indeed, for a self-gravitational system the virial theorem holds:
\begin{equation}
    \frac{v^2}{c^2}\sim \frac{r_s}{r} \ , 
    \label{eq:virial_theorem}
\end{equation}
where $r_s=2Gm/c^2$ is the Schwarzschild radius, $m$ the total mass, and $r$ the typical size of the system. Since the ratio $r_s/r$ is a way to quantify the strength of the gravitational field around the corresponding system, if we want to increase multipoles we need to modify the background space-time. Therefore, we cannot proceed straightforwardly in the multipole expansion while remaining in the theoretical framework of the previous sections. We need to use more accurate models which give us general relativistic corrections as progressive deviations from the Minkowskian background metric. Two methods have been developed: the \textit{Post-Newtonian} and \textit{Post-Minkowskian} perturbative schemes. These two approaches allow us to built general relativistic two-body models by doing perturbative expansions around the simple Newtonian results. The main difference between the two schemes lies in how the respective expansions are performed: the PN one is a low-velocity and weak-field expansion, whose orders are organized following \eqref{eq:virial_theorem}, meaning that at order n, we have terms proportional to $G_N^{1+k}(v^2)^{n-k}$ with $k=1,..,n-1$. On the other side, PM is a weak field only expansion in which all orders in velocity, at fixed order in $G_N$, are included. \\ 
For a better understanding, the powers in $G_N$ and $v$ corresponding to each PM and PN order are summarized in Fig.\ref{fig:PN_PM}, where on horizontal lines there is the development of PM corrections whereas along the diagonals we have the PN corrections. The overlapping between the two expansion is evident and one has the possibility to evaluate the same corrections with both approaches, and to make crosschecks. \\ 
At the state-of-the-art the PN calculations have reached the 5PN order, whereas in PM approach the 4PM calculation have been accomplished \cite{Bern:2021dqo} (indicated with the red rectangle in figure) and it gives us contributions to the 6PN dynamics, beyond the component determined within PN framework. \\ 
Once the dynamics of the binary system is formalized (either in Hamiltonian, Lagrangian or Routhian language) up to the desired order in one of the two expansions, one can extract physical observables in the following way: 
\begin{itemize}
    \item compute the total energy $E_{int}$ and the radiated power $P$ of the system, up to the available expansion order.
    \item Obtain corrections relevant observables, principally the GW phase $\Phi$ and the radial separation between two bodies $R$, by solving the respective differential equations, which in turn are provided by the balance equation, similarly to what has been done in \ref{subsec:coalescence_binaries}. 
    \item Build the GWs waveforms by keeping higher order multipoles and thus obtain "side bands" with respect to the leading quadrupole waveforms, which oscillate at half-integer multiples of $\Phi$. \\ 
\end{itemize}
Carrying out this process is very cumbersome, especially when subtleties come into play: for instance the back-reaction effects of GWs on the dynamics of their sources, the secondary production of GWs from the gravitational field of other GWs (due to the non-linear character of GR), which we will explore in this thesis. \\ Moreover, in the current state of the art the PM and PN results describing the inspiral phase are combined with numerical relativity results within the so-called \textit{Effective one body approach} developed in \cite{Taracchini:2013rva,Bohe:2016gbl},  where the resulting description of the two-body dynamics manages to cover the entire inspiral-merger-ringdown sequence of the coalescence. \\ 
This provides templates which are used by the LIGO-Virgo-Kagra collaboration  to study gravitational waves. 
In the following chapters we will explore both the PN and PM approaches to General Relativity, evaluating corrections in a Lagrangian formalism, which can then be used to compute GWs observables. \\ 
Before proceeding further let us briefly analyze the consequences of the PN corrections on physical measurable quantities such as the phase and the amplitude of GWs. 
\subsection{PN corrections to the phase of a gravitational wave}
In this subsection we will analyze the consequences of the PN corrections on a physical measurable quantity as the phase of a gravitational wave:
\begin{equation}
    \Phi(t)=2\int_{t_i}^tdt'\omega_{gw}(t') \ , 
\end{equation}
where $\omega$ is the orbital angular frequency of the GWs. \\
Let us assume to have circular orbits, under PN corrections we should expect that the internal energy \eqref{eq:internal_energy} gets modified to an expression of the form:
\begin{equation}
    E=-\frac{\mu G_N}{2}v^2(1+e_{v^2}(\mu/m))v^2+e_{v^4}(\mu/m)v^4+...)
\end{equation}
where we denoted with $e_{^{2n}}(\mu/m)$ a generic correction.\\
Similarly, the radiated power \eqref{eq:emitted_power} should change as:
\begin{equation}
    P=\frac{32\mu^2}{5G_Nm^2}\frac{1}{G_Nc^5}v^{10}(1+f_{v^2}(\mu/m)v^2+f_{v^3}(\mu/m)v^3+\ldots) \ , 
\end{equation}
where we have expressed all formulas using $v^2$ instead of $\omega$ by means of: $v^3=\omega mG_N$ that follows from the third Kepler's law and from the virial theorem. \\
We can use these quantities to estimate the phase of a gravitational wave since:
\begin{equation}
    \Phi=2\int_{t_i}^tdt'\omega(t')=2\int_{v_i}^{v_f}dv\omega(v)\frac{dE}{dv}\frac{dt}{dE}=\frac{2}{G_Nm}\int_{v_i}^{v_f}dv\frac{v^3}{P}\frac{dE}{dv}
\end{equation}
Plugging the expanded expressions for $E$ and $P$ we get:
\begin{equation}
    \Phi=\frac{5m}{16\mu}\int_{v_i}^{v_f}dv\frac{1}{v^6}(1+p_{v^2}v^2+p_{v^3}v^3+...)
    \label{eq:phase_pn}
\end{equation}
In this last expression the first term gives the Newtonian phase previously computed while the other terms gives the post-Newtonian corrections parametrized in terms of $p_{v^n}$. \\
The presence of these extra terms modify the prediction on the number of cycles spent by a gravitational wave in the frequency band, which is a physical quantity that a detector can measure with high accuracy.
Analogously, following Eq.\eqref{eq:ampl_corrected_plus}\eqref{eq:ampl_corrected_cross}, also the amplitude of GWs gets modified in PN scheme.

\chapter{EFT of a Coalescing Binary System in General Relativity}
Given the impossibility to compute higher order predictions for GWs observables following a simple approach based on multipole expansion, as shown in Sec.\ref{sec:beyond_linearized_GR}, on this chapter we want now to develop a framework that allow us to describe the slow inspiral phase of a Binary Coalescing System in classical General Relativity, and the successive emission of Gravitational Waves, using an Effective Field Theory approach (see \cite{Goldberger:2022ebt,Foffa:2013qca,Rothstein:2014sra,Porto:2016pyg,Levi:2018nxp} for modern reviews ). Effective Field Theory methods (see \cite{Kaplan:2005es,Goldberger:2007hy,Manohar:2018aog} for  introductions) were first introduced in particle physics for the study of heavy quark field theory (see \cite{Georgi:1990um}). They were then adapted to gravitational physics by Damour and Farése in \cite{Damour:1995kt}, and later systematized by Goldberger and Rothstein in \cite{Goldberger:2004jt}. Within this approach the action for a binary system can be evaluated using Feynman diagrams, with the use of modern scattering amplitudes multi-loop techniques for high precision calculations, as first seen in \cite{Foffa:2016rgu}. During the slow inspiral phase the velocity $v\sim r_s/r \ll 1$ is a small parameter, and so the dynamics can be treated perturbatively. Hence, computations can be organized in the Post-Newtonian perturbation scheme, where General Relativity corrections are organized in powers of velocity $v$, and of Newton constant $G_N$, such that at $n$PN order we consider corrections that scale as: $G_N^{n-l}l^{2(l)}$ with $0\leq l\leq n-1$. In the treatment of the EFT approach of a binary coalescing system we will adopt the "mostly plus" convention: $\eta_{\mu\nu}=diag(-,+,+,+)$.
\subsubsection{Goals of the Chapter:}
The chapter will be organized as follows: 
\begin{itemize}
    \item we introduce the concept of EFT, showing how to obtain a low-energy effective action from a complete one, and we encounter the so-called method of regions, which allow us to study different regions of momenta using different physical theories. 
    \item We apply it to the case of a Binary Coalescing System, using the hierarchy of length scales appearing in this problem to build a tower of EFTs, where the computation is divided into internal, near and far zone;
    \item We deal with a simple model of scalar gravity in order to understand how to get the effective action of a binary from a Feynman diagrammatic approach;
    \item We focus on the near zone, and we show an algorithmic procedure to obtain Near zone contributions to the conservative dynamics at any PN order.
    \item Eventually we develop an EFT for the far zone, and we set the bases to compute far zone contributions to the conservative dynamics, that will be dealt in the following chapters.
\end{itemize}
\section{Effective Field Theory approach}
Effective field theory (EFT) methods are tailor-suited for dealing with problems characterized by multiple length scales.
They allow us to focus only on the relevant degrees of freedom, which are necessary to study a physical problem at a given length scale.
With an effective field theory approach we can make predictions for physical observables, at a given precision, with a finite number of measurements. \\  
Let us consider for example the computation of the effects of a short range physics $r_s$, characterized by a high energy scale $\Lambda$, on the dynamics at a low energy scale $w\ll\Lambda$. In the EFT description we can keep track of the effects of $\Lambda$ by doing a systematic expansion in the ratio $w/\Lambda \ll 1$. \\
Let us consider a quantum field theory characterized by a light degree of freedom $\phi$ with mass $m$, and a heavy one $\chi$ with mass near the high energy scale $\Lambda$, and let $S[\phi,\chi]$ be the action of the full theory. \\ 
Suppose that we are interested in describing experimental observables at low energy, involving only the field $\phi$. We can obtain an effective action $S_{eff}[\phi]$ by integrating out the modes $\chi$:
\begin{equation}
    e^{iS_{eff}[\phi]}=\int D\chi(x)\ e^{iS[\phi,\chi]} \ , 
\end{equation}
this procedure can be carried out diagrammatically, order by order, with the use of Feynman diagrams. \\ 
The resulting effective action can be expressed in terms of local functional of the fields $\phi(x)$:
\begin{equation}
    S_{eff}[\phi]=\sum_iC_i\int d^4xO_i(x)
\end{equation}
for local operators $O_i(x)$ and with $C_i$ known as \textit{Wilson coefficients}. \\
If $O_i(x)$ has mass dimension $\Delta_i$, then the Wilson coefficients, evaluated at a renormalization scale $\mu$ of order $\Lambda$, scale as:
\begin{equation}
    c_i(\mu=\Lambda)=\frac{\alpha_i}{\Lambda^{\Delta_i-4}}
\end{equation}
We can conclude that the short distance physics can have two types of effects on the dynamics at energies $w<<\Lambda$:
\begin{itemize}
    \item Renormalization of the coefficients of operators with mass dimension $\Delta\leq 4$,
    \item Generation of an infinite tower of irrelevant operators with mass dimension $\Delta>4$.
\end{itemize}
Doing that, all the ultraviolet (UV) dependence appears directly in the coefficients of the effective Lagrangian, and the $\Lambda$ dependence of an observable follows from power counting.
The number of operators is infinite but since:
\begin{itemize}
    \item An operator $O(x)$ with mass dimension $\Delta>4$ contribute to an observable at relative order: $\left(\frac{w}{\Lambda}\right)^{\Delta-4}<<1$,
    \item A given observable can only be determined up to a finite experimental resolution $\epsilon<<1$,
\end{itemize}
we can truncate the series at a certain order N, so up to operators with mass dimension $\Delta < N+4$, where:
\begin{equation}
    \epsilon\sim \biggl(\frac{w}{\Lambda}\biggr)^N
\end{equation}
We can obtain the explicit predictions for the Wilson Coefficients with a matching procedure. To do that we need to calculate some observables in the full theory, expanding the result in powers of $w/\Lambda$ and in the effective field theory, adjusting the EFT parameters in order to reproduce the full theory result. 
If we did not know the fundamental theory underlying a physical process, we could have built an effective action by specifying: the field content of the theory, the symmetries of the system, and an appropriate expansion parameter, and by constructing the most general Lagrangian which is invariant under the symmetry group of the theory.
\subsection{The method of regions}
\label{sec:method_of_regions}
In order to perform computations we will use a technique known as method of regions\footnote{See \cite{Becher:2014oda} for a comprehensive review in the context of SCET.} which allows us to carry out asymptotic expansions of loop integrals \cite{Beneke_1998} in dimensional regularization around various limits \cite{Smirnov:2002pj}. The expansion is obtained by splitting the integration in different regions and by expanding the integrand in each case. In the effective field theory approach, the different regions will be represented by different effective theory fields. The expanded integrals obtained by means of the strategy of regions technique are in one-to-one correspondence to Feynman diagrams of effective field theories regularized in dimensional regularization.  If one is simply interested to expand some perturbative result in a small parameter, one can use the strategy of regions without the need of an effective Lagrangian. However, the use of an effective field theory offers some important advantages when one is interested in deriving all-order statements, like factorization theorems or to resum logarithmic contributions at all orders in the coupling constants using Renormalization Group (RG) techniques.
In addition, in the EFT gauge invariant is manifest at the Lagrangian level, while this is not true for individual diagrams. The effective Lagrangian also provides a systematic way to organize higher power corrections, by including subleading terms in the effective Lagrangian. 

\subsubsection{A simple example}
In order to understand how the strategy of regions works, let us consider the following integral:
\begin{equation}
    I = \scalebox{0.6}{\begin{tikzpicture}[baseline=(a1)] 
        \begin{feynman}
        \vertex (a1) ;
        \vertex[left=0.3cm of a1] (a0);
        \vertex[right=3cm of a1] (a2); 
        \vertex[right=0.3cm of a2] (a7); 
        \vertex[below=1.4cm of a1] (a3); 
        \vertex[right=0.5cm of a3] (a4); 
        \vertex[right=2.5cm of a3] (a5); 
        \diagram* {
        (a0) -- (a1),
        (a1) -- [ half left] (a2),
        (a1) -- [ ultra thick, half right] (a2),
        (a2) -- (a7),
        };
        \end{feynman} 
        \end{tikzpicture}} = \int_0^\infty \ dk \frac{k}{(k^2+m^2)(k^2+M^2)}=\frac{1}{M^2-m^2}\log\biggl(\frac{M}{m}\biggr) \ ,
    \label{eq:1loop_reg} 
\end{equation}
which is a self-energy one-loop integral with two different masses at zero external momentum, evaluated in $d=2$. We will expand it using two different methods, first using a cutoff to separate two different regions and then with dimensional regularization. 
Let us assume that $m^2\ll M^2$, we can expand the integral in $m/M$, just by expanding the denominator on the r.h.s of \eqref{eq:1loop_reg} as: 
\begin{equation}
    I=\frac{\log\biggl(\frac{M}{m}\biggr)}{M^2}\biggl(1+\frac{m^2}{M^2}+\frac{m^4}{M^4}+\cdots\biggr) \ . \label{eq:1loop_exp}
\end{equation}
Note that the integral is non-analytic in the expansion parameter $m/M$ because of the presence of the logarithm. Expansions of functions around points where they have essential singularities are also called asymptotic expansions. We want now to get the expansion in Eq.\eqref{eq:1loop_exp} by performing an expansion at the level of the integrand in Eq.\eqref{eq:1loop_reg}. This is important for cases where the full result is not available, and it will also tell us what kind of degrees of freedom the effective theory will contain. \\
 If one tries naively to expand the integrand, IR divergences arise, and 
 \begin{equation}
    \frac{k}{(k^2+m^2)(k^2+M^2)}=\frac{k}{k^2(k^2+M^2)}\biggl(1-\frac{m^2}{k^2}+\frac{m^2}{k^4}+\cdots\biggr)
    \label{eq:1loop_taylor}
 \end{equation}
cannot be used in the integrand of Eq.\eqref{eq:1loop_reg}:
\begin{equation}
    I\neq \int_0^{\infty} \ dk \frac{k}{k^2(k^2+M^2)}\biggl(1-\frac{m^2}{k^2}+\frac{m^4}{k^4}+\cdots\biggr) \ . 
    \label{eq:1loop_exp_naive}
\end{equation}
This was to be expected since the result is non-analytic in  $m/M$, and so we cannot obtain it by Taylor expanding the integrand. 
So just from the form of the result in Eq.\eqref{eq:1loop_exp}, it is clear that expansion and integration do not commute:  the series expansion in Eq.\eqref{eq:1loop_taylor} is valid only for $k^2\gg m^2$, while the integration domain in Eq.\eqref{eq:1loop_reg} includes a region in which $k^2\sim m^2$, which contributes to the integral. We can try to split the integration in two regions, by introducing a new cutoff scale $\Lambda$ such that $m\ll \Lambda \ll M$, obtaining: 
\begin{equation}
    I \ = \ \int_0^\Lambda \ dk \frac{k}{(k^2+m^2)(k^2+M^2)}+\int_\Lambda^{\infty} \ dk \ \frac{k}{(k^2+m^2)(k^2+M^2)} = I_{(I)}+I_{(II)}\ .
\end{equation}
We call the region $[0,\Lambda]$ the \textit{low-energy} region. In this region $k\sim m \ll M$, and therefore we can expand the integrand in the integral $I_{(I)}$ as follows: 
\begin{equation}
    I_{(I)}=\int_0^\Lambda \ dk \frac{k}{(k^2+m^2)(k^2+M^2)}= \int_0^\Lambda \ dk \frac{k}{(k^2+m^2)M^2}\biggl(1-\frac{k^2}{M^2}+\frac{k^4}{M^4}+\ldots \biggr) \ . 
    \label{eq:1loop_lambda_1}
\end{equation}
The scale $\Lambda$ acts as an ultraviolet cutoff for the integrals on the r.h.s. of the Eq. \eqref{eq:1loop_lambda_1}. \\ 
The region $[\Lambda,\infty] $ is referred to as the \textit{high-energy} region; in that region $m\ll k \sim M $, and one can expand the integrand according to: 
\begin{equation}
    I_{(II)}\ = \ \int_\Lambda^{\infty} \ dk \ \frac{k}{(k^2+m^2)(k^2+M^2)} = \int_\Lambda^{\infty} \ dk \ \frac{k}{k^2(k^2+M^2)}\biggl(1-\frac{m^2}{k^2}+\frac{m^4}{k^4}+\ldots\biggr) \ ,
    \label{eq:1loop_lambda_2}
\end{equation}
where in this case $\Lambda$ acts as an infrared cutoff. \\ 
By integrating the first two terms on the r.h.s. of \eqref{eq:1loop_lambda_1} one gets: 
\begin{eqnarray}
    I_{(I)} &\approx & \frac{M^2+m^2}{2 M^4}\log\biggl(1+\frac{\Lambda^2}{m^2}\biggr)-\frac{\Lambda^2}{2M^4} \nonumber \\ 
    & = & -\frac{1}{M^2}\log\biggl(\frac{m}{\Lambda}\biggr)-\frac{\Lambda^2}{2M^4}+\mathcal{O}\biggl(\frac{\Lambda^4}{M^6},\frac{m^2}{M^4}\log\biggl(\frac{\Lambda}{m}\biggr)\biggr) \ , 
\end{eqnarray}
Similarly, by integrating the first term on the r.h.s of \eqref{eq:1loop_lambda_2} one finds: 
\begin{equation}
    I_{(II)} \approx \frac{1}{2M^2}\log\biggl(1+\frac{M^2}{\Lambda^2}\biggr) = -\frac{1}{M^2}\log\biggl(\frac{\Lambda}{M}\biggr)+\frac{\Lambda^2}{2M^4}+\mathcal{O}\biggl(\frac{\Lambda^4}{M^6}\log\biggl(\frac{M}{\Lambda}\biggr)\biggr)\ . 
\end{equation}
Adding up the two pieces one obtains: 
\begin{equation}
    I=I_{(I)}+I_{(II)}=-\frac{1}{M^2}\log\biggl(\frac{m}{M}\biggr)+\mathcal{O}\biggl(\frac{m^2}{M^4}\log\biggl(\frac{M}{m}\biggr)\biggr) \ , 
    \label{eq:1loop_lambda_result}
\end{equation}
which is exactly what we got before in \eqref{eq:1loop_reg}. When summing the results for the low-energy and high-energy regions, the terms which depend on the cutoff $\Lambda$ cancel out as expected. \\ 
It is possible to separate the low- and high-energy regions without introducing this additional scale by using dimensional regularization, rewriting the integral as: 
\begin{equation}
    I = \int_0^\infty \ dk \ k^{-\epsilon} \frac{k}{(k^2+m^2)(k^2+M^2)} \ , 
\end{equation}
where we eventually send $\epsilon\to 0$ at the end of the calculation. \\ 
The integral in the low-energy region $k\sim m \ll M$ will be: 
\begin{equation}
    I_{(I)} = \int_0^\infty \ dk \ k^{-\epsilon} \frac{k}{(k^2+m^2)M^2}\biggl(1-\frac{k^2}{M^2}+\frac{k^4}{M^4}+\cdots \biggr)
    \label{eq:1loop_dimreg_1} \ ,
\end{equation}
which is infrared safe in the region in which $k\to 0 $. The dimensional regulator $\epsilon$ can be chosen positive, so that the integrand is also ultraviolet finite. \\ 
The integral in the high-energy region will be: 
\begin{equation}
  I_{(II)} = \int_0^\infty \ dk \ k^{-\epsilon}\frac{k}{k^2(k^2+M^2)}\biggl(1-\frac{m^2}{k^2}+\frac{m^4}{k^4}+\ldots\biggr) \ .
  \label{eq:1loop_dimreg_2} 
\end{equation}
The integral is ultraviolet safe, and we consider $\epsilon$ negative, so that the integrand does not give rise to an infrared singularity in the region where $k\to 0 $. By integrating the first term on the r.h.s. of Eq.\eqref{eq:1loop_dimreg_1} one finds, at leading power in the expansion around $m/M$: 
\begin{equation}
    I_{(I)}=\frac{m^{-\epsilon}}{2M^2}\Gamma\biggl(1-\frac{\epsilon}{2}\biggr)\Gamma\biggl(\frac{\epsilon}{2}\biggr)=\frac{1}{M^2}\biggl(\frac{1}{\epsilon}-\log m +\mathcal{O}(\epsilon)\biggr) \ . 
    \label{eq:1loop_dimreg_1_result}
\end{equation}
The integral of the first term on the r.h.s of Eq.\eqref{eq:1loop_dimreg_2} is: 
\begin{equation}
    I_{(II)}= -\frac{M^{-\epsilon}}{2M^2}\Gamma\biggl(1-\frac{\epsilon}{2}\biggr)\Gamma\biggl(\frac{\epsilon}{2}\biggr)=\frac{1}{M^2}\biggl(-\frac{1}{\epsilon}+\log M +\mathcal{O}(\epsilon)\biggr) \ . 
    \label{eq:1loop_dimreg_2_result}
\end{equation}
The poles in $\epsilon$ cancel in the sum of Eqs.(\eqref{eq:1loop_dimreg_1_result},\eqref{eq:1loop_dimreg_2_result}), and the final result is the same one as  Eq.\eqref{eq:1loop_lambda_result}. 
Let us notice that in both Eq.\eqref{eq:1loop_dimreg_1} and Eq.\eqref{eq:1loop_dimreg_2} the integration domain coincides with the full integration domain of the original integral.  One could fear that this leads to additional contributions and double counting. This does not happen because the two integrals scale differently:  $I_{(I)}$ factors out $m^{-\epsilon}$, while  $I_{(II)}$ factors out $M^{-\epsilon}$. This statement remains true even if we are considering subleading terms. When keeping the complete dependence on $m$ and $M$ the result is: 
\begin{equation}
    I=\frac{1}{2}\Gamma\biggl(1-\frac{\epsilon}{2}\biggr)\Gamma\biggl(\frac{\epsilon}{2}\biggr)\frac{m^{-\epsilon}-M^{-\epsilon}}{M^2-m^2} \ . 
\end{equation}
The result clearly displays the low-energy and the high-energy part. Expanding in one region, one loses the other part and the full integral is recovered after adding the two contributions.  
To demonstrate directly from the integral that there is indeed no double counting, let us now see what happens if we insist in restricting the integration domain of the low- and high- energy region integrals when using dimensional regularization. The integral in the low-energy region would become in this case: 
\begin{eqnarray}
    I_{(I)}^{\Lambda}
    & = & 
    \int_0^{\Lambda} \ dk \ k^{-\epsilon}\frac{k}{(k^2+m^2)M^2}\biggl(1-\frac{k^2}{M^2}+\frac{k^4}{M^4}+\ldots \biggr) \nonumber \\ 
    & = & 
    \biggl[\int_0^{\infty} \ dk -\int_\Lambda^\infty\ dk \biggr]k^{-\epsilon}\frac{k}{(k^2+m^2)M^2}\biggl(1-\frac{k^2}{M^2}+\frac{k^4}{M^4}+\ldots \biggr) \nonumber \\ 
    & = & 
    I_{(I)}-R_{(I)}
\end{eqnarray}
The first integral in the second line of the equation above is the same at the one in \eqref{eq:1loop_dimreg_1}. In the integrand of $R_{(I)}$, which depends on the cutoff $\Lambda$, one can use the fact that $k\geq \Lambda \gg m^2$ to expand in the small $m$ limit: 
\begin{eqnarray}
    R_{(I)} 
    & = & 
    \int_{\Lambda}^{\infty}\ dk k^{-\epsilon}\frac{k}{(k^2+m^2)M^2}\biggl(1-\frac{k^2}{M^2}+\ldots\biggr) \nonumber \\ 
    & = & \int_\lambda^\infty \ dk k^{-\epsilon} \frac{k}{k^2 M^2}\biggl(1-\frac{m^2}{k^2}-\frac{k^2}{M^2}+\ldots\biggr) \ . 
\end{eqnarray}
For the remainder part $R_{(I)}$, we thus have performed two expansions. First the low energy expansion, which is equivalent to expanding the integrand in the $M\to \infty$ limit. Then we have expanded the result around $m\to 0 $, which is equivalent to the high-energy expansion. At this point it is sufficient to observe that for dimensional reasons the integrals in the equation above must behave as follows: 
\begin{equation}
    \int_\Lambda^{\infty} \ dk k^{n-\epsilon}\sim \Lambda^{n+1-\epsilon} \ . 
\end{equation}
So the cutoff pieces scale as fractional powers of the cutoff. Since the $\Lambda$ dependent terms must cancel out completely in the calculation of $I$, one can as well drop the $\Lambda$ dependent integrals from the start. 
Therefore, when regulating divergences by means of dimensional regularization one can integrate over the complete integration domain, in this case $k\in [0,\infty]$. We can explicitly verify that the cutoff pieces vanish if we also consider the high energy integral $I_{(II)}$ in Eq.\eqref{eq:1loop_dimreg_2} with a lower cutoff $\Lambda$ on the integration. Proceeding in the same way as before, we can rewrite the high-energy integral as the expanded integral without a cutoff and a remainder which depends on the cutoff: 
\begin{eqnarray}
    R_{(II)}
    & = & 
    \int_0^\Lambda \ dk k^{-\epsilon}\frac{k}{k^2(k^2+M^2)}\biggl(1-\frac{m^2}{k^2}+\ldots\biggr) \nonumber \\ 
    & = & 
    \int_0^\Lambda \ dk k^{-\epsilon}\frac{k}{k^2M^2}\biggl(1-\frac{m^2}{k^2}-\frac{k^2}{M^2}+\ldots\biggr) \ , 
\end{eqnarray}
where we have expanded the integrand in both the limit of small $m$ and also in the limit of large $M$, but in opposite order as in $R_{(I)}$. However, the two expansions commute so that the integrands of $R_{(I)}$ and $R_{(II)}$ are identical. 
Adding up the two pieces, we find that: 
\begin{equation}
    R=R_{(I)}+R_{(II)}=\int_0^{\infty}\ dk k^{-\epsilon}\frac{k}{k^2M^2}\biggl(1-\frac{m^2}{k^2}-\frac{k^2}{M^2}+\ldots\biggr) \ . 
\end{equation}
This is manifestly independent on the cutoff. It is also manifestly zero, because it is given by a series of scaleless integrals. 
In the context of effective field theories where the method of region is used, such as NRGR of SCET, the overlap contribution $R$ is usually referred as the "zero-bin contribution" \cite{Manohar:2006nz}. One can obtain the full overlap either by expanding the high-energy integral $I_{(II)}$ around the low-energy limit, or the integrand of the low-energy integral $I_{(I)}$ around the high-energy limit. Since the overlap is obtained by expanding single-scale integrals, $I_{(I)}$ or $I_{(II)}$ it is given by scaleless integrals which vanish in dimensional regularization. 
We will now apply these techniques in the context of NRGR.

\section{EFT for a Binary Coalescing System in General Relativity}
\begin{figure}[H]
    \centering
    \includegraphics[width=0.8\textwidth]{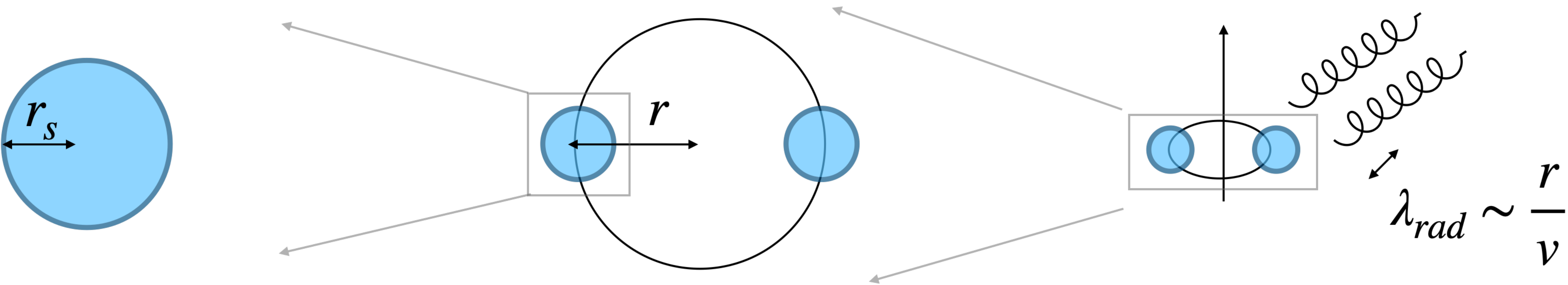}
    \caption{Hierarchy of length scales present in the slow inspiral phase of a coalescing binary system}
\end{figure}
Let us focus on the slow inspiral phase of the dynamics of a binary coalescing system in General Relativity, that corresponds to the period in the evolution of the binary in which the system is moving at non-relativistic velocities $v \ll 1 $, and the bound orbit is slowly decaying due to the emission of gravitational radiation. \\
This problem is characterizes by multiple length scales:
\begin{itemize}
    \item the size of the compact objects $r_s$,
    \item the orbital distance $r$, 
    \item the wavelength $\lambda_{rad}\approx \frac{r}{v}$ of the emitted radiation.
\end{itemize}
The orbital distance is much greater than the size of the compact object $r_s\ll r$, 
Moreover, from a multipole expansion of the radiation field coupled to non-relativistic sources it follows that $r\ll \lambda_{rad}$.  \\ 
Hence, there is a precise \textbf{hierarchy} of widely separated length scales: 
\begin{equation}
  r_s \ll r \ll \lambda_{rad} \ ,
\end{equation}
all controlled by the same expansion parameter $v$:
\begin{eqnarray}
    \frac{r_s}{r}\sim v^2  \qquad \frac{r}{\lambda}\sim v \ . 
    \end{eqnarray}
Then, at any order in the PN expansion, qualitatively different effects, due to physics at different length scales, have to be taken into account. \\ 
As explained in the previous section, a convenient way to deal with multiscale processes is by using an effective field theory approach \cite{Goldberger:2009qd}.
Hence, we can use this hierarchy of length scales to construct a tower of Effective Field Theories, that captures only the relevant physics at each scale, separately: 
\begin{enumerate}
    \item \textbf{Internal zone} $r_s$: can be described with an EFT of a point particle action, plus higher order terms with unknown coefficients which encode the internal structure of the compact object, 
    \item \textbf{Near zone} $r_s$: can be described by two point particles EFTs coupled to gravitons in a flat spacetime, 
    \item \textbf{Far zone} $\lambda_{rad}$: can be described by a multipole-expanded source, emitting radiation gravitons.
\end{enumerate}

\subsection{Internal zone $r_s$}
Let us consider an isolated compact object, such as a black hole or a neutron star, coupled to gravity.  \\ 
The microscopic full theory that describe this problem is given by classical general relativity, coupled to some source $T^{\mu\nu}$ of conserved energy-momentum that describes the internal structure of each compact object.
In the EFT approach, since the modes of interest in a non-relativistic binary dynamics have wavelength much larger than the scale that characterizes the internal structure of the object $\lambda_{rad}\gg r_s$, the compact object can be approximated as a point-like particle, that carries internal degrees of freedom coupled to gravity. We can use the principles of Effective Field Theories (EFT) to build an effective action in a bottom-up approach. This action will contain Wilson coefficients, which encode the internal structure of the object. \\ 
In order to do that we need to identify the relevant degrees of freedom at the scale of interest and the corresponding symmetries, and then we need to construct the most general Lagrangian for these degrees of freedom consistent with the symmetries.
The relevant degrees of freedom of the system are:
\begin{enumerate}
    \item the gravitational field $g_{\mu\nu}(x)$, 
    \item the black hole's worldline coordinate $x^{\mu}(\lambda)$, which is a function of an arbitrary affine parameter $\lambda$, and in principle a spin degree of freedom $S^{\mu\nu}(\lambda)=-S^{\nu\mu}(\lambda)$ localized on each worldline, 
    \item an orthonormal frame $e^{\mu}_a(\lambda)$ localized on the particle worldline that describes the orientation of the object relative to local inertial frames.
\end{enumerate}
Ignoring spin effects, the symmetries of the problem are:
\begin{enumerate}
    \item coordinate invariance $x^{\mu}\to x^{'\mu}(x)$,
    \item worldline reparametrization invariance $\lambda \to \lambda'(\lambda)$,
    \item $SO(3)$ invariance to guarantee that the compact object is spherical.
\end{enumerate}
Due to the omission of the spin degrees of freedom and to this last assumption it is clear that the EFT that we are constructing is appropriate to describe Schwarzschild black holes interacting with gravitational fields. \\
We can then construct a Lagrangian that is generically a sum over an infinite number of invariants containing $g_{\mu\nu},dx^\mu/d\lambda$, and their derivatives. \\ 
The simplest term that we can construct, consistent with these criteria is a point-particle action: 
\begin{eqnarray}
    S_{pp}[x^\mu_{pp},g^{\mu\nu}]
    & = & 
    -m\int d\tau \nonumber \\ 
    & = & 
    -m \int d\lambda d^4 x \delta^4(x^\mu-x^\mu_{pp}(\lambda))\sqrt{-g_{\mu\nu}\frac{dx^\mu}{d\lambda}\frac{dx^\nu}{d\lambda}} \label{eq:point_particle}
\end{eqnarray}
where $x^\mu_{pp}(\lambda)$ is the trajectory of the particle, and $\tau$ is the proper time. 
At this order, by Equivalence Principle, compact object dynamics is universal, with worldlines that follow timelike geodesics of $g_{\mu\nu}$.
Higher order terms can be constructed by making use of the Riemann tensor, by contracting it with the metric or with the four-velocity $u^\alpha(\lambda)$. The internal structure of the object is then encoded in Wilson coefficients appearing in front of the compact object.
By adding the first higher order terms we obtain an effective action of the form: 
\begin{equation}
    S_{eff}[x_{pp},g^{\mu\nu}]=-m\int d\tau+C_E\int d\tau E_{\mu\nu}E^{\mu\nu}+c_B\int d\tau B_{\mu\nu}B^{\mu\nu}+\cdots
\end{equation}
The tensors $E_{\mu\nu}$, $B_{\mu\nu}$ denote the decomposition of the Riemann tensor $R_{\mu\nu\alpha\beta}$ into components of electric and magnetic type parity, respectively. \\
Explicitly, they are given by:
       \begin{equation}
        E_{\mu\nu}= R_{\mu\nu\alpha\beta}\dot{x}^{\alpha}\dot{x}^{\beta} \qquad B_{\mu\nu}=\epsilon_{\mu\alpha\beta\rho}\dot{x}^{\rho}R^{\alpha\beta}_{\sigma\nu}\dot{x}^{\sigma} \ . 
       \end{equation}
The coefficient $C_{E,B}$ measure the leading (quadrupolar) tidal response of the compact object to an external gravitational field.
There is a principle known as \textit{effacement principle} for which we can neglect the internal structure of the compact object up to the 5PN corrections. For our future interests we can restrict our attention to the simple point particle action \eqref{eq:point_particle}. \\ 
The gravitational dynamics instead can be described by the simple Einstein Hilbert action: 
\begin{equation}
    S_{EH}=2\Lambda^2\int d^4x\sqrt{-g}R(g_{\mu\nu}) \ , \qquad \Lambda^{-2}=32\pi G_N \ . 
\end{equation}
Moreover, since Einstein's theory enjoys a coordinate invariance, we can impose a gauge fixing condition by adding an appropriate gauge-fixing term. \\
We choose the harmonic gauge, where $\Gamma_\mu=0$ with 
\begin{equation}
    \Gamma^\mu=\Gamma^{\mu}_{\alpha\beta}g^{\alpha\beta} \ , 
\end{equation}
with $\Gamma^\mu=\Gamma^\mu_{\nu\rho}g^{\mu\nu}$. 
We impose it at the level of the action by adding the gauge-fixing term:
\begin{equation}
    S_{GF}=-\Lambda^2\int d^4x \sqrt{-g} \ \Gamma_\mu\Gamma^\mu
\end{equation}
The introduction of this term can be justified within the Faddev-Popov procedure, which makes possible to physically define a consistent functional integral by first fixing a proper gauge. 
Let us notice that in this case there are no ghosts, since we are only interested in the evaluation of classical contributions to the effective action (proportional to $\hbar^0$). \\ 
The effective action consistent with these criteria is then given by:
\begin{equation}
    S[x^{\mu}_{pp},g_{\mu\nu}]=S_{EH}[g_{\mu\nu}]+S_{GF}[g_{\mu\nu}]+S_{pp}[x^{\mu}_{pp},g_{\mu\nu}]
\end{equation}

\subsection{Near zone $r$}
The system at the scale $r$ can be described by two point particles moving on a flat spacetime and weakly interacting with the gravitational fields. 
Hence, the effective action for this theory is given by:
\begin{equation}
    S_{tot}[x_a^\mu,g_{\mu\nu}]=S_{EH}[g_{\mu\nu}]+S_{GF}[g_{\mu\nu}]+\sum_aS^a_{pp}[x_a^\mu,g_{\mu\nu}] \ , 
    \label{eq:binary_near}
\end{equation}
where $a=1,2$.
\subsection{Far zone $\lambda_{rad}$}
As we move far away from the source we are not able to distinguish anymore the binary constituents, and the system can effectively be described by an effective source, expanded in multipole moments $\{Q_i\}$, and emitting gravitational waves.
An effective action for this theory is given by: 
\begin{equation}
    S_{rad}[g_{\mu\nu},\{Q_i\}]=S_{EH}[g_{\mu\nu}]+S_{GF}[g_{\mu\nu}]+S_{mult}[g_{\mu\nu},\{Q_i\}]
\end{equation}
where $S_{mult}[g_{\mu\nu},\{Q_i\}]$ contains couplings between gravitational fields and the source multipoles, and it will be derived later in this chapter. \\ 

\subsection{Calculating observables from the EFT of a binary}
Before proceeding further, let us understand how to calculate observables from the EFT of a binary. Let us remark also that the inclusion of spin degrees of freedom, which we will not consider, is of crucial importance in phenomenology (\cite{Porto:2005ac}).
In order to understand how to evaluate physical observables from the EFT approach, one should remember that the phase $\phi$ of a gravitational wave emitted from a binary during the slow inspiral phase can be expressed in terms of its internal energy $E(v)$ and its emitted power $P(v)$ as:
\begin{eqnarray}
\phi =2\int dv \frac{v^3}{P}\frac{dE}{dv} \ , 
\end{eqnarray}
 where $E(v)$ and $P(v)$ can be derived from the action of a binary once the gravitational field has been integrated out. \\
In the weak-field regime we can expand the metric tensor as:
\begin{eqnarray}
    g_{\mu\nu}=\eta_{\mu\nu}+\frac{h_{\mu\nu}}{\Lambda} \ . 
\end{eqnarray}
Given the effective action of a binary system in the near zone \eqref{eq:binary_near}, one can start integrating out gravitons in the presence of fixed worldlines $x^\mu_{a=1,2}(\lambda)$, in a saddle point approximation, in order to define a new effective action $S_{eff}$ involving only the binary constituents $x^\mu_a$:
\begin{eqnarray}
    e^{iS_{eff}[x_a]}=\int Dh e^{iS_{EH}[h]+iS_{GF}[h]+iS_{PP} [h,x_a]}
    \label{eq:integrating_out}
\end{eqnarray}
Given the non-linearity of the gravitational theory, this procedure must be done perturbatively, order by order, and we will show how to perform it with the use of Feynman diagrams.
Since the extreme with respect of $h_{\mu\nu}$ of the exponent in \eqref{eq:integrating_out} are simply gauge-fixed classical solutions $h_{\mu\nu}(x_a)$ of Einstein's field equations at given worldlines, one can immediately see that:
\begin{eqnarray}
S_{eff}[x_a]=S_{eff}[x_a,h_{\mu\nu}(x_a)]+\mathcal{O}(\hbar) \ , 
\label{eq:Seff}
\end{eqnarray}
where quantum corrections can be neglected for the phenomenology we are aiming. \\
Once the action has been derived, one can use the real part of \eqref{eq:Seff} to define the classical equations of motion for worldlines:
\begin{eqnarray}
\delta_{x_a}[Re(S_{eff}[x_a])]=0 \ , 
\end{eqnarray}
which can be further used to define the internal energy of the binary $E(v)$ as a constant of motion, using Noether's theorem. \\
In the same manner, the imaginary part of \eqref{eq:Seff} can be used to define the total radiated power as:
\begin{eqnarray}
P=\lim_{T->\infty}\frac{2}{T}[Im(S_{eff}(x))] \ , 
\end{eqnarray}
the demonstration of which will be given in the following section using a simplified model of a binary system with scalar gravity. 
In the following we will be interested in studying the conservative dynamics of the binary system, which will arise  from the real part of $S_{eff}[x_a]$.
 
\section{The EFT of a binary system in scalar gravity}
\label{sec:eft_scalar_gravity}
Let us apply the Effective field theory approach to a simplified model, where a binary system interacts linearly with a massless scalar field $\phi(x)$ representing gravity. To be more familiar to what we will do in the following, let us work in dimensional regularization, with $d=3+\epsilon$ the continuous dimension of space. The action for the system is given by:
\begin{eqnarray}
S[\phi,x^{\mu}_a]=\frac{1}{2}\int d^{d+1}x \partial_\mu\phi \partial^\mu\phi -m_a\int d\tau_a\biggl(1+\frac{\phi(x_a)}{2\sqrt{2}\Lambda}\biggr) \ ,
\end{eqnarray}
where $\Lambda=\bigl(\sqrt{32\pi G_d}\bigr)^{-1}$ is the d-dimensional coupling constant, with $G_d=G_N\mu^{d-3}$.
The action can be rewritten as: 
\begin{eqnarray}
    S[\phi,x^{\mu}_a] & = & -m_a\int d\tau_a+\int d^{d+1}x \biggl(\frac{1}{2}\partial_\mu\phi \partial^\mu\phi +J(x)\phi(x)\biggr)  \ , \\ 
    J(x) & = & \sum_{a=1}^{2}-\frac{m_a}{2\sqrt{2}\Lambda}\int d \tau_a\delta^{d+1}(x-x_a) \ . 
\end{eqnarray}
We can obtain an effective action containing only the worldlines degrees of freedom by integrating out the scalar field modes as:
\begin{eqnarray}
e^{iS_{eff}[x^\mu_a]}=\int D\phi e^{iS[\phi,x^\mu_a]} = e^{iS_{kin}[x_a^\mu]}Z[J]\ , \qquad \qquad Z[J]=\int D\phi e^{i\int d^{d+1} x(\partial_\mu \phi\partial^\mu \phi+iJ(x)\phi(x))}
\label{eq:functional_scalar}\ ,
\end{eqnarray}
where $S_{kin}[x^\mu_a]$ is the elapsed proper time on the black hole worldlines, and $Z[J]$ is the so-called \textit{functional generator of disconnected n-points functions}, typical of quantum field theories. \\
This procedure can be done exactly, since in eq.\eqref{eq:functional_scalar} the functional integral is Gaussian, and one gets:
\begin{eqnarray}
Z[J]&=& e^{-\frac{1}{2}\int d^{d+1}x d^{d+1} y J(x)J(y)G_2(x-y)} \ , \\
G_2(x-y)&=&\int \frac{d^{d+1}k}{(2\pi)^4}e^{ik\cdot (x-y)}\frac{i}{k^2+i\epsilon} \ , 
\end{eqnarray}
where $G_2(x-y)$ is the so-called \textit{2-point function of a free scalar field}. \\
If one considers self-interactions between scalar fields then it is necessary to introduce a diagrammatic approach typical of QFT to get the effective action. \\
As a warm-up for future calculations, let's define the Feynman rules for this theory in momentum space, for the interaction between $\phi(x)$ and $J(x)$ and for the scalar propagator:

\begin{eqnarray}
\begin{tikzpicture}[baseline=(current bounding box.center)]
\begin{feynman}
\vertex (a);
\vertex[right=2cm of a] (b);
\vertex[below =1cm of a] (c);
\vertex[right=1 of a] (d);
\vertex[below=1cm of d] (e);
\diagram* {
(a) -- [] (b),
(d) --  [scalar](e), 
};
\end{feynman}
\end{tikzpicture}&=& 
i\int d^{d+1}xJ(x)e^{-ik\cdot x}=\sum_{a=1}^2-i\frac{m_a}{2\sqrt{2}\Lambda}\int d\tau_ae^{-ik\cdot x} \ , \\
\begin{tikzpicture}[baseline=(current bounding box.center)]
\begin{feynman}
\vertex (a);
\vertex[right=2cm of a] (b);
\diagram* {
(a) -- [scalar] (b),
};
\end{feynman}
\end{tikzpicture}&=& 
\frac{i}{k^2+i\epsilon} \ , 
\end{eqnarray}

\noindent where the black horizontal line denotes the worldline $x_a$, whereas dotted lines denote the scalar gravitons. \\
We can now use these rules to build amplitudes in momentum space, whose Fourier transform gives the contribution to the desired functional generator.\\
Z[J] has a diagrammatic expansion in terms of "ladder diagrams", which are made by wordlines exchanging gravitons, as:

\begin{eqnarray}
Z[J] &=& \begin{tikzpicture}[baseline=(current bounding box.center)]
    \begin{feynman}
    \vertex (a);
    \vertex[right=0.5cm of a] (a2);
    \vertex[right=0.5cm of a2] (a3);
    \vertex[right=0.25cm of a3] (b1);
    \vertex[right=0.5cm of a3] (a4);
    \vertex[right=0.5cm of a4] (a5);
    \vertex[right=0.5cm of a5] (a6);
    \vertex[below=1cm of a] (e);
    \vertex[right=0.5cm of e] (e2);
    \vertex[right=0.5cm of e2] (e3);
    \vertex[right=0.5cm of e3] (e4);
    \vertex[right=0.25cm of e3] (b2);
    \vertex[right=0.5cm of e4] (e5);
    \vertex[right=0.5cm of e5] (e6);
    \diagram* {
    (a2) -- [] (a5),
    (e2)-- [] (e5),
    (b1) --  [scalar] (b2), 
  }; 
\end{feynman}
\end{tikzpicture}
    + \begin{tikzpicture}[baseline=(current bounding box.center)]
        \begin{feynman}
        \vertex (a);
        \vertex[right=0.5cm of a] (a2);
        \vertex[right=0.5cm of a2] (a3);
        \vertex[right=0.25cm of a3] (b1);
        \vertex[right=0.5cm of a3] (a4);
        \vertex[right=0.5cm of a4] (a5);
        \vertex[right=0.5cm of a5] (a6);
        \vertex[below=1cm of a] (e);
        \vertex[right=0.5cm of e] (e2);
        \vertex[right=0.5cm of e2] (e3);
        \vertex[right=0.5cm of e3] (e4);
        \vertex[right=0.25cm of e3] (b2);
        \vertex[right=0.5cm of e4] (e5);
        \vertex[right=0.5cm of e5] (e6);
        \diagram* {
        (a2) -- [] (a5),
        (e2)-- [] (e5),
        (a3) --  [scalar] (e3), 
        (a4) --  [scalar] (e4), 
      }; 
    \end{feynman}
\end{tikzpicture}
        + \begin{tikzpicture}[baseline=(current bounding box.center)]
            \begin{feynman}
            \vertex (a);
            \vertex[right=0.2cm of a] (a3);
            \vertex[right=0.5cm of a3] (a4);
            \vertex[right=0.5cm of a4] (a5);
            \vertex[right=0.2cm of a5] (a6);
            \vertex[below=1cm of a] (e);
            \vertex[right=0.2cm of e] (e3);
            \vertex[right=0.5cm of e3] (e4);
            \vertex[right=0.5cm of e4] (e5);
            \vertex[right=0.2cm of e5] (e6);
            \diagram* {
            (a) -- [] (a6),
            (e)-- [] (e6),
            (a3) -- [scalar] (e3),
            (a4) --  [scalar] (e4), 
            (a5) -- [scalar] (e5),};
            \end{feynman}
            \end{tikzpicture} 
            + \begin{tikzpicture}[baseline=(current bounding box.center)]
                \begin{feynman}
                \vertex (a);
                \vertex[right=0.2cm of a] (a2);
                \vertex[right=0.5cm of a2] (a3);
                \vertex[right=0.5cm of a3] (a4);
                \vertex[right=0.5cm of a4] (a5);
                \vertex[right=0.2cm of a5] (a6);
                \vertex[below=1cm of a] (e);
                \vertex[right=0.2cm of e] (e2);
                \vertex[right=0.5cm of e2] (e3);
                \vertex[right=0.5cm of e3] (e4);
                \vertex[right=0.5cm of e4] (e5);
                \vertex[right=0.2cm of e5] (e6);
                \diagram* {
                (a) -- [] (a6),
                (e)-- [] (e6),
                (a2) --  [scalar] (e2), 
                (a3) -- [scalar] (e3),
                (a4) --  [scalar] (e4), 
                (a5) -- [scalar] (e5),};
                \end{feynman}
                \end{tikzpicture} + \ldots\  = \  exp\left(\ \begin{tikzpicture}[baseline=(current bounding box.center)]
                        \begin{feynman}
                        \vertex (a);
                        \vertex[right=0.2cm of a] (a2);
                        \vertex[right=0.5cm of a2] (a3);
                        \vertex[right=0.25cm of a3] (b1);
                        \vertex[right=0.5cm of a3] (a4);
                        \vertex[right=0.5cm of a4] (a5);
                        \vertex[right=0.2cm of a5] (a6);
                        \vertex[below=1cm of a] (e);
                        \vertex[right=0.2cm of e] (e2);
                        \vertex[right=0.5cm of e2] (e3);
                        \vertex[right=0.5cm of e3] (e4);
                        \vertex[right=0.25cm of e3] (b2);
                        \vertex[right=0.5cm of e4] (e5);
                        \vertex[right=0.2cm of e5] (e6);
                        \diagram* {
                        (a2) -- [] (a5),
                        (e2)-- [] (e5),
                        (b1) --  [scalar] (b2), 
                      }; 
                    \end{feynman}
                \end{tikzpicture}
               \  \right)
\end{eqnarray}
Notice that intermediate particle lines do not propagate, so diagrams with multiple graviton exchanges are simply products of the diagram with a single scalar exchange. 
For example, one can consider a generic amplitude describing the emission and absorption of $n$ particles from the two black holes:
\begin{eqnarray}
Z_n(ki)&=& \begin{tikzpicture}[baseline=(current bounding box.center)]
\begin{feynman}
\vertex (a);
\vertex[right=0.5cm of a] (a2);
\vertex[right=0.5cm of a2] (a3);
\vertex[right=0.5cm of a3] (a4);
\vertex[right=0.5cm of a4] (a5);
\vertex[right=0.5cm of a5] (a6);
\vertex[below=1.5cm of a] (e);
\vertex[right=0.5cm of e] (e2);
\vertex[right=0.5cm of e2] (e3);
\vertex[right=0.5cm of e3] (e4);
\vertex[right=0.5cm of e4] (e5);
\vertex[right=0.5cm of e5] (e6);
\diagram* {
(a) -- [] (a6),
(e)-- [] (e6),
(a2) --  [scalar] (e2), 
(a3) -- [scalar] (e3),
(a4) --  [scalar] (e4), 
(a5) -- [scalar] (e5),};
\end{feynman}
\end{tikzpicture}
\end{eqnarray}
Substituting the Feynman rules we get:
\begin{eqnarray}
Z_n(k_i)& =& \frac{1}{2^nn!}\prod_{i=1}^n\biggl[i\int d^{d+1}x_iJ(x_i)e^{-ik_i\cdot x_i}\biggr]\biggl[\frac{1}{k_i^2+i\epsilon}\biggr]\biggl[i\int d^{d+1}y_iJ(y_i)e^{ik_i\cdot y_i}\biggr] \nonumber \\
& = & \frac{1}{n!}\prod_{i=1}^n\biggl[-\frac{1}{2}\int d^{d+1}x_id^{d+1}y_iJ(x_i)J(y_i)e^{-ik_i\cdot(x_i-y_i)}\frac{1}{k_i^2+i\epsilon}\biggr]
\end{eqnarray}
where the index $i$ runs over the intermediate lines, while $2^{n}n!$ is a symmetry factor associated to the topology of the diagram.
After a Fourier transform of each $k_i$ momentum we get:
\begin{eqnarray}
Z_n[J]&=& \frac{1}{n!}\prod_{i=1}^n\biggl[-\frac{1}{2}\int d^{d+1}x_id^{d+1}y_iJ(x_i)J(y_i)G_2(x_i-y_i)\biggr] \nonumber \\
& = & \frac{1}{n!}\biggl[-\frac{1}{2}\int d^{d+1}xd^{d+1}yJ(x)J(y)G_2(x-y)\biggr]^n \ , 
\label{eq:zn}
\end{eqnarray}
where we can recognize in \eqref{eq:zn} the $n-$th term of an exponential expansion. \\
By summing over all $n$, we recover the expression for the functional generator:
\begin{eqnarray}
Z[J]& =& \sum_{n=1}^{\infty}Z_n[J]\ =\ e^{iW[J]}  \ , \\
W[J]&=& \frac{i}{2}\int d^{d+1}x\ d^{d+1}y\ J(x)G_2(x-y)J(y) \ , 
\end{eqnarray}
where we introduced the \textit{functional generator of connected n-th points functions} $W(J)$, which as the name states depends only on connected diagrams. From \eqref{eq:functional_scalar} the effective action for the binary system is:
\begin{eqnarray}
e^{iS_{eff}[x_a^\mu]}=e^{iS_{kin}[x_a^\mu]}e^{iW[J]}\qquad \Leftrightarrow \qquad S_{eff}(x)=S_{kin}(x)+W(J) \ , 
\end{eqnarray}
and so we get:
\begin{eqnarray}
S_{eff}(x_a)=-m_a\int dt+\frac{i}{2}\sum_{a,b}^{a\neq b}\frac{m_am_b}{8\Lambda^2}\int d\tau_ad\tau_bG_2(x_a-x_b)
\label{eq:effective_non_local} \ ,
\end{eqnarray}
where divergent terms due to $a=b$ have been removed since they cannot affect physical observables. In principle, they can be reabsorbed in renormalization of the particle masses. \\
Notice that the exponentiation of the diagrams contributing to $Z[J]$ implies that $S_{eff}(x_a)$ receives contribution only from diagrams that remain connected after the particle worldlines are stripped off. \\ 
This fact remain true in real gravity, where the graphs can have graviton self-interaction vertices 
To make the model more realistic one can add self-interactions for the scalar field as a potential of the type $V(\phi)=\lambda\phi^n$ for some coupling $\lambda$. \\ This would have generated new Feynman rules, and therefore additional contributions, in form of coupling powers, to the effective action.
\subsection{Method of regions and local EFT via a non-relativistic limit}
The effective action derived in the previous section is non-local in proper times of the two black holes. \\
From a relativistic point of view this is natural: the dynamics of a worldline at a certain time, depends on the motion of the other at a retarded one, which in turn depends on the dynamics of the previous one at another delayed time and so on. However, since we will be interested in the slow inspiral phase of a binary, we can apply a proper non-relativistic limit to eq.\eqref{eq:effective_non_local} in order to neglect these time non-localities, at least up to a given PN order. \\
The procedure relies on the so-called method of regions introduced in Sec.\ref{sec:method_of_regions}. For slowly moving objects, $v\ll 1$, the wavelength of the radiation is much longer than their separation, $\lambda_{rad}\sim r/v\gg r$. In such scenario we can split in two the scalar field, each part representing a graviton in a specific region of momenta: 
\begin{equation}
    \phi(x)= \Phi+\Bar{\phi}
\end{equation}
where:
\begin{enumerate}
    \item $\Phi$ denotes a short distance component, or potential graviton, with scaling rules $(k_0,\mathbf{k})_{pot}\approx (v/r,1/r)$, 
    \item $\Bar{\phi}$ denotes a radiation graviton with scaling rules $(k_0,\mathbf{k})_{rad}\approx (v/r,v/r)$,
\end{enumerate}
with $r$ the orbital separation and $v\ll 1$ the typical three velocity. \\
Radiation gravitons represent the on-shell propagating degrees of freedom ($k^2=0$) whereas potential ones are the off-shell ($k^2\neq 0$) modes, spacelike, which mediate the binding forces between the constituents of the binary. \\
For a static source we only find potential modes, which do not have a temporal component, and simply $\vert\mathbf{k}\vert\sim 1/r$.
Departures from instantaneity are then parametrized in powers of $k_0$, which satisfy $k_0\sim v/r\ll 1/r$ for slowly moving sources. \\
The presence of non-localities is due to the integration on the radiation region in the two point function \eqref{eq:effective_non_local}, and in order to avoid this region we can restrict ourselves to the potential one where $k^0\ll \vert\Vec{k}\vert$.\\
If we restrict to this region, we can power expand the 2-point function  in the non-relativistic limit as:
\begin{eqnarray}
G_2^{NR}(x_a-x_b)& =& i\int\frac{d^{d+1}k}{(2\pi)^{d+1}}\frac{e^{-ik\cdot(x_a-x_b)}}{\mathbf{k}^2-k_0^2 +i\epsilon} \nonumber\\
& = & i\int \frac{d^{d+1}k}{(2\pi)^{d+1}}\frac{e^{-ik\cdot(x_a-x_b)}}{\mathbf{k}^2}\sum_{n=0}^{\infty}\biggl(\frac{k_0^2}{\mathbf{k}^2}\biggr)^n
\end{eqnarray}
Defining $w=k_0$ we have:
\begin{eqnarray}
G_2^{NR}(x_a-x_b)=i\sum_{n=0}^{\infty}\int \frac{d^dk}{(2\pi)^d}\frac{e^{-i\mathbf{k}\cdot(\mathbf{x}_a-\mathbf{x}_b)}}{(\mathbf{k}^2)^{n+1}}\int \frac{dw}{2\pi}e^{iw(t_a-t_b)}w^{2n}
\end{eqnarray}
The $w$ integration is straightforward:
\begin{eqnarray}
\int \frac{dw}{2\pi}e^{iw(t_a-t_b)}w^{2n}=(-1)^n\frac{\partial^{2n}}{\partial^{2n}(t_a-t_b)}\int \frac{dw}{2\pi}e^{iw(t_a-t_b)}=(-1)^n\frac{\partial^{2n}\delta(\tau_a-\tau_b)}{\partial^{2n}(\tau_a-\tau_b)}
\end{eqnarray}
As for the $\mathbf{k}$ integral, we use the following formula that we will prove in the next chapter:
\begin{eqnarray}
    I_F(d,a)=\int \frac{d^d\mathbf{k}}{(2\pi)^d}\frac{e^{-i\mathbf{k}\cdot(\mathbf{x}_a-\mathbf{x}_b)}}{(\mathbf{k})^{2a}}= \frac{\Gamma[d/2-a]}{(4\pi)^{d/2}\Gamma[a]}\left(\frac{r}{2}\right)^{2a-d} \ ,
    \label{eq:d_dim_scalar_integral}
\end{eqnarray}
with $r=\vert \mathbf{x}_a-\mathbf{x}_b\vert$ the Euclidean distance between $x_a$ and $x_b$. \\
By combining the two results we have that the non-relativistic expansion of the two point function is:
\begin{eqnarray}
G_2^{NR}(x_a-x_b)=i\sum_{a=1}^{\infty}(-1)^{a}\frac{\partial^{2(n-1)}}{\partial^{2(n-1)}(\tau_a-\tau_b)}\biggl[\delta(\tau_a-\tau_b)\biggr]\frac{\Gamma[d/2-a]}{(4\pi)^{d/2}\Gamma[a]}\left(\frac{r}{2}\right)^{2a-d}
\label{eq:g2_nr}
\end{eqnarray}
As an example let us evaluate the effects of the first term of \eqref{eq:g2_nr} in \eqref{eq:effective_non_local}. \\
This is given by:
\begin{eqnarray}
G_2^{NR}(x_a-x_b)=-i\delta(\tau_a-\tau_b) \frac{\Gamma[d/2-1]}{(4\pi)^{d/2}}\left(\frac{r}{2}\right)^{2-d} \ . 
\end{eqnarray}
The corresponding contribution to the effective action is, writing $d=3+\epsilon$, and taking the limit $\epsilon\to 0$:
\begin{eqnarray}
S_{eff}[x_a^\mu]& =& 
\lim_{d\to3}\frac{1}{2}\sum_{a,b}^{a\neq b}\frac{m_am_b}{8\Lambda^2}\int d\tau_ad\tau_b\delta(\tau_a-\tau_b)\frac{\Gamma[d/2-1]}{(4\pi)^{d/2}}\left(\frac{r}{2}\right)^{2-d}\nonumber \\ 
&=& 
+\frac{1}{2}\sum_{a,b}^{a\neq b}\frac{m_am_b}{8\Lambda^2}\int d\tau_ad\tau_b\frac{\delta(\tau_a-\tau_b)}{4\pi r}
\end{eqnarray}
Then evaluating the non-relativistic expression for the elapsed proper time we arrive at:
\begin{eqnarray}
S_{eff}[x_a^\mu]=-m_a\int d\tau_a+\frac{1}{2}\sum_{a,b}^{a\neq b}\frac{m_am_b}{8\Lambda^2}\int d\tau_ad\tau_b\frac{\delta(\tau_a-\tau_b)}{4\pi r}
\end{eqnarray}
We can now define in the non-relativistic limit a unique time $\tau_a=t$ such that:
\begin{eqnarray}
S_{eff}[x_a^\mu]=\int dt \biggl(\frac{m_a v_a^2}{2}+\frac{m_1m_2G_N}{r}\biggr) \ , 
\end{eqnarray}
which is the well known Lagrangian for the classical two body problem. \\
In addition to this term, one should also consider terms coming from higher order term in the non-relativistic expansion of the Feynman propagators.\\
These terms are proportional to time derivatives of Dirac's delta, and one can eliminate non-localities after an integration by parts, with the effect of adding time derivatives to the worldlines. \\
We can obtain a local Lagrangian for a binary, which can be easily used to define the energy $E$ of the system as a conserved quantity via Noether's theorem. \\ At a certain point also radiative contributions need to be considered, and they will contribute at higher order in the conservative dynamics. 
As for the evaluation of $P$, we should evaluate the imaginary part of $S_{eff}[x_a]$ which is possible by restricting to the radiative region with $k^2=0$. \\
\begin{eqnarray}
G_2^{Rad}(x_a-x_b)=i\int \frac{d^{d+1}k}{(2\pi)^{d+1}}\frac{e^{-ik\cdot(x_a-x_b)}}{k^2+i\epsilon}=\pi \int \frac{d^{d+1}k}{(2\pi)^{d+1}}e^{-ik\cdot (x_a-x_b)}\delta(k^2)
\end{eqnarray}
We can then use the following property of $\delta$ function:
\begin{eqnarray}
\delta(k^2)=\frac{\delta(w-\vert\mathbf{k}\vert)}{2\vert\mathbf{k}\vert}+\frac{\delta(w+\vert\mathbf{k}\vert)}{2\vert\mathbf{k}\vert} \ , 
\end{eqnarray}

and perform an integration with respect to $w$ to arrive at:
\begin{eqnarray}
Im(S_{eff}[x_a])=\frac{1}{16\Lambda^2}\int \frac{d^dk}{(2\pi)^d}\frac{1}{2\vert\mathbf{k}\vert}\bigg|\sum_a m_a \int d\tau_ae^{-ik\cdot x_a}\bigg|^2_{w=\vert\mathbf{k}\vert} \ . 
\end{eqnarray}
From this last equation we can get the differential power emitted as:
\begin{eqnarray}
\frac{dP}{d\Omega d\vert\mathbf{k}\vert}=\frac{1}{T}\frac{G}{4\pi^2}\vert\mathbf{k}\vert^2\bigg|\sum_a m_a\int d\tau_a e^{-ik\cdot x_a}\bigg|^2_{w=\vert\mathbf{k}\vert} \ . 
\end{eqnarray}  
Perturbative calculations within General Relativity can be done by generalizing this scalar gravity model. \\

\section{Boundary conditions for Feynman diagrams in NRGR}
A coalescing binary system is a radiation problem, and it is important to respect causality since there cannot be incoming radiation from past infinity. In dealing with this diagrammatic approach, causality conditions can be imposed by appropriately choosing boundary conditions for the propagators appearing in the diagrams. \textit{Feynman propagators} for example, ensure a pure in-going wave at past infinity and out-going at future infinity, leading to a non-causal evolution of the system. One has then to use appropriately retarded or advanced propagators in dealing with radiation problems. Actually, it has been shown that an action for non-conservative systems should be formulated in terms of the so-called \textit{in-in formalism}, and it will be the main topic of Chapter\ref{chapter:inin}. \\ However, for reasons that will soon be clear, we can avoid the use of this formalism for the evaluation of most diagrams, choosing appropriately the Green's functions appearing in the diagram computation
\subsection{Boundary conditions for Feynman diagrams in field theory}
Let us consider again the toy model introduced in \ref{sec:eft_scalar_gravity}, supposing now to have two different sources $J_1$ and $J_2$:
\begin{eqnarray}
    S_{eff} &=& \frac{i}{2}\int d^{d+1}xd^{d+1}yJ_1(x)G_2(x-y)J_2(y) \ . 
\end{eqnarray}
Once a choice for the boundary conditions of the Green's function is made, one can derive the equations of motion. Notice that this choice is independent on the fundamental underlying theory, but depends on the specific physical setup under investigation. \\
In particular, the potential generated at position $2$ by particle $1$ will be obtained using the \textit{retarded} Green's function with $J_2$.\\
In particle physics, the appropriate Green's function is the Feynman one, $G_F$, where we have symmetric boundary conditions with particles ingoing for $t\to -\infty$ and outgoing for $t\to \infty$:
\begin{eqnarray}
G_F(t,\mathbf{x})=\int \frac{d^d\mathbf{k}}{(2\pi)^d}\frac{dk_0}{(2\pi)}\frac{e^{-ik_0t+i\mathbf{k}\cdot \mathbf{x}}}{\mathbf{k^2}-k_0^2+ia}
\end{eqnarray}
where $a$ is an arbitrary small positive number, which is usually denoted with $\epsilon$ in literature. \\
Note that $G_F$ is even under $t\leftrightarrow -t, \mathbf{x}\leftrightarrow -\mathbf{x}$ exchange, and it is complex in direct space as its Fourier transform, shows that:
\begin{equation}
    \tilde{G}_F(w,\mathbf{k})\neq \tilde{G}^*_F(-w,-\mathbf{k}) \qquad 
    \tilde{G}_F(w,\mathbf{k})\neq -\tilde{G}^*_F(-w,-\mathbf{k})
\end{equation}
Advanced and retarded Green's functions $G_{A,R}$ are defined by:
\begin{eqnarray}
    G_{A,R}(t,\mathbf{x})=\int \frac{d^d\mathbf{k}}{(2\pi)^d}\frac{dk_0}{2\pi}\frac{e^{-ik_0t+i\mathbf{k}\cdot \mathbf{x}}}{\mathbf{k}^2-(k_0\mp ia)^2} \overset{d=3}{=} -\frac{1}{4\pi}\frac{\delta(t\pm \vert \mathbf{x}\vert)}{\vert\mathbf{x}\vert}
\end{eqnarray}
where $d\to 3$ limit has been written explicitly, showing that $G_{A,R}$ are real functions in any dimensions as $\tilde{G}_{A}(w,\mathbf{k})=\tilde{G}_R(-w,\mathbf{k})$, and they fulfil the relationship: $G_A(t,\mathbf{x})=G_R(-t,\mathbf{x})$. \\
Introducing the Wightman functions $\Delta_{\pm}(t,\mathbf{x})$ defined by:
\begin{eqnarray}
    \Delta_{\pm}(t,\mathbf{x})=\int \frac{d^d\mathbf{k}}{(2\pi)^d}\frac{dk_0}{2\pi}\frac{e^{\mp ik_0t+i\mathbf{k}\cdot \mathbf{x}}}{2k}=\int \frac{d^d\mathbf{k}}{(2\pi)^d}\frac{dk_0}{2\pi}\theta(\pm k_0)\delta(k_0^2-\mathbf{k}^2)e^{-ik_0t+i\mathbf{k}\cdot\mathbf{x}}
\end{eqnarray}
one can find the relations:
\begin{eqnarray}
    G_F(t,\mathbf{x})&=& 
    i[\theta(t)\Delta_+(t,\mathbf{x})+\theta(-t)\Delta_-(t,\mathbf{x})] \ , \\ 
    G_R(t,\mathbf{x})&=& 
   - i\theta(t)[\Delta_+(t,\mathbf{x})-\Delta_-(t,\mathbf{x})]
    \ , \\ 
    G_A(t,\mathbf{x})&=& 
    i\theta(-t)[\Delta_+(t,\mathbf{x})-\Delta_-(t,\mathbf{x})]
     \ ,
\end{eqnarray}
which enable to rewrite the Feynman green function as:
\begin{eqnarray}
    G_F(t,\mathbf{x})=\frac{1}{2}\left( G_A(t,\mathbf{x})+ G_R(t,\mathbf{x})\right)-\frac{i}{2}\left(\Delta_+(t,\mathbf{x})+\Delta_-(t,\mathbf{x})\right) \ .
\end{eqnarray}
The real part of Feynman Green's function is given by the time-symmetric combination $G_A+G_R$, and its imaginary part is given by the time-symmetric combination of Wightman functions known as \textit{Hadamard} function.\\
\subsection{Boundary conditions involving only potential gravitons}
When in \ref{sec:eft_scalar_gravity} we restrict ourselves to potential gravitons we did not ask ourselves anything regarding the choice of boundary conditions for propagators. \\
This is due to the fact that for potential gravitons we Taylor expanded the Green's functions in powers of $(k_0/\vert\mathbf{k}\vert)^2$ as:
\begin{eqnarray}
    \frac{1}{\mathbf{k^2}-k_0^2}\approx N(k)= \frac{1}{\mathbf{k}^2}\sum_{n\geq 0}\left(\frac{k_0^2}{\mathbf{k}^2}\right)^n,
\end{eqnarray}
making immaterial the $ia$ part in the Green's function denominator. \\ 
In this approximation then $G_F,G_A,G_R,(G_A+G_R)/2$ collapse to the same quantity. The Fourier space version of the Wightman functions have support only on-shell, which is for $k_0=\pm \vert\mathbf{k}\vert$, hence they vanish in the Taylor expansion of the near zone, which assumes $k_0\ll \vert\mathbf{k}\vert$. 
\subsection{Boundary conditions with radiation gravitons}
\label{sec:boundary_conditions_far}
The situation is different when also radiation gravitons are involved since modes can go on-shell, and so the choice of Green's functions with Feynman or causal boundary conditions leads to in-equivalent results.
We did not consider radiation gravitons in sec.\ref{sec:eft_scalar_gravity} as they were not present in that simple model, however they will be present in the general theory.
Following the analysis done in \cite{Foffa:2021pkg}, one can try to classify the diagrams appearing according to the number of radiative modes involved. In Chapter\ref{chapter:hereditary} we will consider hereditary diagrams, where radiation gravitons are emitted by the gravitational waves source, scattered and then reabsorbed by the same system. The process of emission and absorption involve a time direction, requiring the use of advanced and retarded, or causal, Green's functions.      \\ 
In case of processes with \textit{only one} radiative mode involved, since internal line momentum integration is unrestricted over all real values $w$ of the time component of the momentum of the radiative mode, the result of the amplitude is proportional to the time-symmetric combination $G_A+G_R$, coinciding with the real part of the Feynman Green's function. Note that while in Fourier space all Green's functions have a non-vanishing imaginary part, in position space causal Green's functions are real, as $\tilde{G}_{R/A}(w)=\tilde{G}_{R/A}^*(-w)$. \\ 
As a consequence no time-asymmetric effects are found, and what is obtained is the contribution of the exchange of radiative modes to the conservative dynamics of the system. \\ 
Using the Feynman Green's function we can get the correct real, time-symmetric contribution to the effective action, and we will get also an imaginary part which is related via optical theorem to the power emission. \\ 
When \textit{two} Green's functions of radiative modes are involved, by using momentum conservation the process involve products of the type $\tilde{G}_R(w)\tilde{G}_A(-w)=\tilde{G}^2_R(w)$ with $w$ integrated over the entire real axis.
If we take the difference between two causal Green's functions and two Feynman ones we get:
\begin{equation}
    \tilde{G}_F(w)^2-\frac{1}{2}\left(\tilde{G}_R(w)^2+\tilde{G}_A(w)^2\right)\sim -\tilde{\Delta}_+(w)\tilde{\Delta}_-(w)-i(\tilde{G}_A(w)+\tilde{G}_R(w))\tilde{G}_H(w)
\end{equation}
with $\tilde{G}_H(w)$ the Fourier transform of the Hadamard Green's function. \\ 
The real part of the difference vanishes because $\tilde{\Delta}_+(w)$ and $\tilde{\Delta}_-(w)$ have no common support, while the imaginary part is again related via the optical theorem to the probability loss. \\
By using Feynman boundary conditions then, one obtains the correct conservative dynamics, with the addition of the imaginary part related to the energy flux.\\
For processes where three radiative modes are involved, products of propagators cannot be expressed purely in terms of Feynman Green's functions. Here we cannot use Feynman propagators, but we will have to make use of the in-in formalism.   

\section{Perturbative calculations in the effective field theory of a Binary System}

Let us now go back to the EFT describing the slow inspiral phase of a coalescing binary system. \\  
As shown in \ref{sec:eft_scalar_gravity}, we want to integrate out the gravitational fields from \eqref{eq:binary_near} in order to obtain an effective action which depends only on the two worldlines: 
\begin{equation}
    e^{iS_{eff}[x_a^\mu]}\ = \ \int \ Dh_{\mu\nu}e^{iS_{tot}[x_a^\mu,h_{\mu\nu}]}  
\end{equation}
In order to perform this computation, let us work in dimensional regularization, moving to $d+1$-dimensions, where $\epsilon=d-3$ and $d$ denotes the continuous space dimensions. Consider again the effective action, neglecting finite size effects:
\begin{equation}
    S_{tot}[x^{\mu}_a,g_{\mu\nu}]=2\Lambda^2\int d^dx dt\sqrt{-g}\biggl(R[g]-\frac{1}{2}\Gamma_\mu\Gamma^\mu\biggr)-m_a\int d\tau_a\sqrt{-g_{\mu\nu}(x_a)\dot{x}^{\mu}_a\dot{x}^{\nu}_a} \ , 
    \label{eq:action_near}
\end{equation}
where $\Lambda$ is the d-dimensional coupling constant defined as: $\Lambda=\left(\sqrt{32\pi G_N\mu^{d-3}}\right)^{-1}$. \\ 
We can now rewrite the metric $g_{\mu\nu}$ in a weak field approximation as:
\begin{equation}
    g_{\mu\nu}=\eta_{\mu\nu}+\frac{h_{\mu\nu}}{\Lambda}\ , 
\end{equation} 
and substitute it in the action. We can then expand the action in powers of $\Lambda^{-1}$ to get the Feynman rules for the theory. 
From the Einstein-Hilbert action one gets the graviton propagator, and the self interactions among $h_{\mu\nu}$ fields, that scale as $\Lambda^{2-n}$, with $n$ the number of $h_{\mu\nu}$ fields involved. \\
\begin{equation}
\begin{split}
    S_{bulk}=2\Lambda^2\int d^dxdt\sqrt{-detg_{\mu\nu}}\biggl(R-\frac{1}{2}\Gamma^\mu\Gamma_\mu\biggr)\quad\Rightarrow \quad & 2\int d^dxdt\biggl[(\partial h)^2+\frac{h(\partial h)^2}{\Lambda}+\frac{h^2(\partial h)^2}{\Lambda^2}+...\biggr] \\
    &  \feynmandiagram [baseline=(b.base), small, horizontal=a to b] {
a -- [gluon] b, }; +
\feynmandiagram [baseline=(b.base), small, horizontal=d to b] {
a -- [gluon] b -- [gluon] c,
b --  [gluon] d, };+
\begin{tikzpicture}[baseline=(current bounding box.center)]
\begin{feynman}
\vertex (a);
\vertex[right=2cm of a] (b);
\vertex[below =1cm of a] (c);
\vertex[right=1 of c] (d);
\vertex[below=2cm of a] (e);
\vertex[right=2cm of e] (f);
\diagram* {
(a) -- [gluon] (d) -- [gluon] (b),
(e) --  [gluon](d) -- [gluon](f), 
};
\end{feynman}
\end{tikzpicture}+\cdots\ , 
\end{split}
\end{equation}
where curly lines denote gravitons. 
On the other side, from the point-particle actions we can get source-gravity interactions proportional to $\Lambda^{-n}$, with $n$ the number of $h_{\mu\nu}$ fields involved:
\begin{equation}
\begin{split}
    S_{pp}=-m_a\int dt \sqrt{-g_{\mu\nu}(x_a)\dot{x}^{\mu}_a\dot{x}^{\nu}_a}\quad  \Rightarrow \quad & -m_a\int dt\sqrt{-\dot{x}^2_a}-\frac{-m_a}{2\Lambda}\int dt h_{\mu\nu}\frac{\dot{x}^\mu_a\dot{x}^\nu_a}{\sqrt{-\dot{x}^2_a}}+\cdots\\
    & \begin{tikzpicture}[baseline=(current bounding box.center)]
\begin{feynman}
\vertex (a);
\vertex[right=2cm of a] (b);
\vertex[below =1cm of a] (c);
\vertex[right=1 of a] (d);
\vertex[below=1cm of d] (e);
\diagram* {
(a) -- [] (b),
(d) --  [draw=none](e)
};
\end{feynman}
\end{tikzpicture}+
\begin{tikzpicture}[baseline=(current bounding box.center)]
\begin{feynman}
\vertex (a);
\vertex[right=2cm of a] (b);
\vertex[below =1cm of a] (c);
\vertex[right=1 of a] (d);
\vertex[below=1cm of d] (e);
\diagram* {
(a) -- [] (b),
(d) --  [gluon](e), 
};
\end{feynman}
\end{tikzpicture}+
\begin{tikzpicture}[baseline=(current bounding box.center)]
\begin{feynman}
\vertex (a);
\vertex[right=2cm of a] (b);
\vertex[below =1cm of a] (c);
\vertex[right=1 of a] (d);
\vertex[below=1cm of b] (e);
\diagram* {
(a) -- [] (b),
(d) --  [gluon](c), 
(d) -- [gluon](e),
};
\end{feynman}
\end{tikzpicture}+... \ , 
\end{split}
\end{equation}
where the black horizontal line denote the particle worldline.
We can use these rules to build effective diagrams containing two worldlines and an arbitrary number of gravitons exchanged between them, that contribute to the diagrammatic expansion of $S_{eff}(x_a)$ and scale with a definite power of $\Lambda^{-1}$ only, not velocities:
\begin{eqnarray}
    \begin{tikzpicture}[baseline=(current bounding box.center)]
        \begin{feynman}
        \vertex (a);
        \vertex[right=0.5cm of a] (a2);
        \vertex[right=0.5cm of a2] (a3);
        \vertex[right=0.25cm of a3] (b1);
        \vertex[right=0.5cm of a3] (a4);
        \vertex[right=0.5cm of a4] (a5);
        \vertex[right=0.5cm of a5] (a6);
        \vertex[below=1cm of a] (e);
        \vertex[right=0.5cm of e] (e2);
        \vertex[right=0.5cm of e2] (e3);
        \vertex[right=0.5cm of e3] (e4);
        \vertex[right=0.25cm of e3] (b2);
        \vertex[right=0.5cm of e4] (e5);
        \vertex[right=0.5cm of e5] (e6);
        \diagram* {
        (a2) -- [] (a5),
        (e2)-- [] (e5),
        (b1) --  [gluon] (b2), 
      }; 
    \end{feynman}
\end{tikzpicture} \qquad 
\begin{tikzpicture}[baseline=(current bounding box.center)]
    \begin{feynman}
    \vertex (a);
    \vertex[right=0.5cm of a] (a2);
    \vertex[right=0.5cm of a2] (a3);
    \vertex[right=0.25cm of a3] (b1);
    \vertex[right=0.5cm of a3] (a4);
    \vertex[right=0.5cm of a4] (a5);
    \vertex[right=0.5cm of a5] (a6);
    \vertex[below=1cm of a] (e);
    \vertex[right=0.5cm of e] (e2);
    \vertex[right=0.5cm of e2] (e3);
    \vertex[right=0.5cm of e3] (e4);
    \vertex[right=0.25cm of e3] (b2);
    \vertex[right=0.5cm of e4] (e5);
    \vertex[right=0.5cm of e5] (e6);
    \diagram* {
    (a2) -- [] (a5),
    (e2)-- [] (e5),
    (b1) --  [gluon] (e3), 
    (b1) --  [gluon] (e4), 
  }; 
\end{feynman}
\end{tikzpicture} 
\qquad 
\begin{tikzpicture}[baseline=(current bounding box.center)]
    \begin{feynman}
    \vertex (a);
    \vertex[right=0.5cm of a] (a2);
    \vertex[right=0.5cm of a2] (a3);
    \vertex[right=0.25cm of a3] (b1);
    \vertex[right=0.5cm of a3] (a4);
    \vertex[right=0.5cm of a4] (a5);
    \vertex[right=0.5cm of a5] (a6);
    \vertex[below=1cm of a] (e);
    \vertex[below=0.5cm of b1] (c);
    \vertex[right=0.5cm of e] (e2);
    \vertex[right=0.5cm of e2] (e3);
    \vertex[right=0.5cm of e3] (e4);
    \vertex[right=0.25cm of e3] (b2);
    \vertex[right=0.5cm of e4] (e5);
    \vertex[right=0.5cm of e5] (e6);
    \diagram* {
    (a2) -- [] (a5),
    (e2)-- [] (e5),
    (b1) --  [gluon] (c), 
    (c) --  [gluon] (e3), 
    (c) --  [gluon] (e4), 
  }; 
\end{feynman} 
\end{tikzpicture}  \ . 
\end{eqnarray}
This covariant formulation is the post-Minkowskian (PM) approach, and it is useful when no $v/c \l 1$ assumption is made. In order to proceed further we need to be able to identify a velocity \textit{power counting} scheme for the diagram appearing. Since in the end we are interested in computing gravitational wave signals to a fixed order in $v$, it is important to develop a set of rules that assigns a unique power of $v$ to each diagram in the theory, and again the method of region comes to play. 
\subsection{Non relativistic limit: method of regions}
As done for the scalar gravity case in Sec. \ref{sec:eft_scalar_gravity}, in order to disentangle physics at scale $r$ from physics at scale $\lambda_{rad}\gg r$ we can apply the method of regions and split the metric perturbations into two classes:
\begin{equation}
    h_{\mu\nu}= H_{\mu\nu}+\Bar{h}_{\mu\nu} \ , 
\end{equation}
where:
\begin{enumerate}
    \item $H_{\mu\nu}$ denotes a \textbf{potential graviton} with scaling rules $(k_0,\mathbf{k})_{pot}\approx (\frac{v}{r},\frac{1}{r})$,
    \item $\Bar{h}_{\mu\nu}$ denotes a \textbf{radiation graviton} with scaling rules $(k_0,\mathbf{k})_{rad}\approx (\frac{v}{r},\frac{v}{r})$ \ , 
\end{enumerate}
with $r$ the orbital separation and $v\ll 1$ the typical three velocity. \\
Radiation gravitons represent the on-shell propagating degrees of freedom ($k^2=0$), the long-wavelength modes, that satisfy:
\begin{eqnarray}
    \partial_\alpha \bar{h}_{\mu\nu}\sim \frac{v}{r}\bar{h}_{\mu\nu}
\end{eqnarray}  
meaning that the field $\bar{h}_{\mu\nu}$ varies slowly over spacetime, with a typical length scale $r/v$. \\
It can then been regarded as a slow background field in which potential fields $H_{\mu\nu}$, which are off-shell modes ($k^2\neq 0$) that mediate the binding forces between the constituents of the binary, propagate, with:
\begin{eqnarray}
    \partial_{0}H_{\mu\nu}\sim\frac{v}{r}\ , \qquad \partial_i H_{\mu\nu} \sim \frac{1}{r}H_{\mu\nu} \ .
\end{eqnarray}  
If we move to momentum space, we are able to disentangle hard momenta $\mathbf{k}\sim 1/r $ from the long wavelength scale $\lambda_{rad}\sim \frac{v}{r}$:
\begin{itemize}
    \item for potential modes the propagators can be expanded in non-relativistic limit as:
    \begin{equation}
        \frac{1}{\mathbf{k}^2-k_0^2}\approx \frac{1}{\mathbf{k^2}}\sum_{n=0}^{\infty}\bigl(k_0^2\bigr)^n \ , 
    \end{equation}
    \item for radiative modes we cannot expand the gravitons propagators:
    \begin{equation}
        \frac{1}{\mathbf{k}^2-k_0^2} \ . 
    \end{equation}
\end{itemize}
Now that we have made this splitting, we can integrate out separately potential and radiation gravitons in order to get contributions to the conservative dynamics: 
\begin{equation}
    e^{iS_{eff}[x_a]}=\int  \ D\Bar{h} \int D H e^{iS_{tot}[x_a,\Bar{h},H]} \ . 
\end{equation}
In the following sections we will describe a systematic procedure to do that.
\section{Near zone contributions to the conservative dynamics}
If we focus on the near zone, in order to evaluate contributions to the conservative dynamics, we can start from $S_{eff}(x_a,H_{\mu\nu},\Bar{h}_{\mu\nu})$ and integrate out the potential modes to get $S_{NR}(x_a,\Bar{h})$:
    \begin{eqnarray}
        e^{iS_{NR}(x_a,\bar{h})} & =&   \int D H_{\mu\nu}e^{i(S_{bulk}(\Bar{h}+H)+S_{pp}(x_a,\Bar{h}+H))}\nonumber \\
         &=&  exp\biggl\{\ \scalebox{0.7}{\begin{tikzpicture} 
            \begin{feynman}
            \vertex (a1) ;
            \vertex[right=2cm of a1] (a2); 
            \diagram* { 
            (a1) -- [double, thick] (a2),
            };
            \end{feynman} 
            \end{tikzpicture}}\quad +\quad \scalebox{0.7}{\begin{tikzpicture} 
                \begin{feynman}
                \vertex (a1) ;
                \vertex[right=1cm of a1,square dot,red ] (a2) {}; 
                \vertex[right=1cm of a2] (a3); 
                \vertex[above=1cm of a2] (b1); 
                \diagram* { 
                (a1) -- [double, thick] (a3),
                (a2) -- [gluon] (b1)
                };
                \end{feynman} 
                \end{tikzpicture}\ } \ , 
                \quad +\quad \scalebox{0.7}{\begin{tikzpicture} 
                    \begin{feynman}
                    \vertex (a1) ;
                    \vertex[right=1cm of a1,square dot,red ] (a2) {}; 
                    \vertex[right=1cm of a2] (a3); 
                    \vertex[above=1cm of a2] (b1); 
                    \vertex[left=0.5cm of b1] (b2);
                    \vertex[right=0.5cm of b1] (b3);
                    \diagram* { 
                    (a1) -- [double, thick] (a3),
                    (a2) -- [gluon] (b2),
                    (a2) -- [gluon] (b3)
                    };
                    \end{feynman} 
                    \end{tikzpicture}}
                \biggr\} \ , 
    \end{eqnarray} 
where curly lines denotes radiation gravitons whereas the double solid line represents the compact binary.
In that way  $S_{NR}(x_a,\Bar{h})$ contains two-body forces between the point particles, written as an explicit expansion in powers of $v$, and the couplings of the worldlines to radiation. \\
Diagrammatically, $S_{eff}(x_a,\Bar{h})$ can be obtained by summing diagrams that have the following topological properties:
\begin{enumerate}
    \item diagrams must remain connected if the particles worldlines are stripped off,
    \item diagrams may only contain internal lines corresponding to propagators for the potential modes $H_{\mu\nu}$. Diagrams \textit{cannot} contain external potential graviton lines,
    \item diagrams can only contain external $\Bar{h}_{\mu\nu}$. Diagrams cannot contain propagators corresponding to internal radiation graviton lines. 
\end{enumerate}
The point of splitting the original graviton $h_{\mu\nu}$ into the new modes $H_{\mu\nu},\Bar{h}_{\mu\nu}$ is that the diagrams written in terms of these new variables have definite powers of the expansion parameter $v$.
The power counting rules for determining how many powers of $v$ to assign to a given diagram follow simply from the fact that the three momentum of a potential graviton scales as $\mathbf{k}\sim 1/r$, since this is the range of the force it mediates, and that the spacetime variation of a radiation graviton is $\lambda_{rad}\sim r/v$. With these two observations we can assign powers of $v$ to any term in the action, and by extension to the Feynman rules. \\
Diagrammatically, we can denote the two black hole worldlines with horizontal straight lines, potential modes with dotted lines and radiative modes with wiggled ones. \\
The conservative dynamic is given by the sum of diagrams containing no radiation gravitons, as: 
\begin{equation}
\scalebox{0.7}{\begin{tikzpicture} 
    \begin{feynman}
    \vertex (a1) ;
    \vertex[right=2cm of a1] (a2); 
    \diagram* { 
    (a1) -- [double, thick] (a2),
    };
    \end{feynman} 
    \end{tikzpicture}} \quad = \quad  \begin{tikzpicture}[baseline=(current bounding box.center)]
        \begin{feynman}
        \vertex (a);
        \vertex[right=0.5cm of a] (a2);
        \vertex[right=0.5cm of a2] (a3);
        \vertex[right=0.25cm of a3] (b1);
        \vertex[right=0.5cm of a3] (a4);
        \vertex[right=0.5cm of a4] (a5);
        \vertex[right=0.5cm of a5] (a6);
        \vertex[below=1cm of a] (e);
        \vertex[right=0.5cm of e] (e2);
        \vertex[right=0.5cm of e2] (e3);
        \vertex[right=0.5cm of e3] (e4);
        \vertex[right=0.25cm of e3] (b2);
        \vertex[right=0.5cm of e4] (e5);
        \vertex[right=0.5cm of e5] (e6);
        \diagram* {
        (a) -- [thick] (a6),
        (e)-- [thick] (e6),
      };
    \end{feynman}
    \fill[gray] (a2) rectangle (e5);
    \end{tikzpicture} \quad = \quad 
        \begin{tikzpicture}[baseline=(current bounding box.center)]
        \begin{feynman}
        \vertex (a);
        \vertex[right=1cm of a] (b);
        \vertex[right=1cm of b] (c);
        \vertex[below=1cm of a] (d);
        \vertex[right=1cm of d] (e);
        \vertex[right=1cm of e] (f);
        \diagram* {
        (a) --  (b) --  (c),
        (b) --  [scalar] (e),
        (d) -- (e) -- (f)};
        \end{feynman}
        \end{tikzpicture} \quad +  \quad 
        \begin{tikzpicture}[baseline=(current bounding box.center)]
            \begin{feynman}
                \vertex (a);
                \vertex[right=1cm of a] (b);
                \vertex[right=1cm of b] (c);
                \vertex[below=1cm of a] (d);
                \vertex[right=1cm of d] (e);
                \vertex[left= 0.5cm of e] (g);
                \vertex[right=0.5cm of e] (h);
            \diagram* {
            (a) --  (b) --  (c),
            (b) --  [scalar] (g), 
            (b) -- [scalar] (h),
            (d) -- (e) -- (f), };
            \end{feynman}
            \end{tikzpicture} 
            \quad +  \quad 
            \begin{tikzpicture}[baseline=(current bounding box.center)]
                \begin{feynman}
                    \vertex (a);
                    \vertex[right=1cm of a] (b);
                    \vertex[right=1cm of b] (c);
                    \vertex[below=1cm of a] (d);
                    \vertex[below=0.5cm of b] (i);
                    \vertex[right=1cm of d] (e);
                    \vertex[left= 0.5cm of e] (g);
                    \vertex[right=0.5cm of e] (h);
                \diagram* {
                (a) --  (b) --  (c),
                (b) -- [scalar] (i),
                (i) --  [scalar] (g), 
                (i) -- [scalar] (h),
                (d) -- (e) -- (f), };
                \end{feynman}
                \end{tikzpicture} \quad +\cdots
\end{equation}
Being interested in studying processes with one emitted radiation graviton, one should consider all diagrams containing one radiation field: 
\begin{equation}
    \scalebox{0.7}{\begin{tikzpicture} 
        \begin{feynman}
        \vertex (a1) ;
        \vertex[right=1cm of a1,square dot,red ] (a2) {}; 
        \vertex[right=1cm of a2] (a3); 
        \vertex[above=1cm of a2] (b1); 
        \diagram* { 
        (a1) -- [double, thick] (a3),
        (a2) -- [gluon] (b1)
        };
        \end{feynman} 
        \end{tikzpicture}\ } \quad = \quad   \begin{tikzpicture}[baseline=(f)]
            \begin{feynman}
            \vertex (a);
            \vertex[right=0.5cm of a] (a2);
            \vertex[right=0.5cm of a2] (a3);
            \vertex[right=0.25cm of a3] (b1);
            \vertex[above=0.75cm of b1] (c1);
            \vertex[right=0.5cm of a3] (a4);
            \vertex[right=0.5cm of a4] (a5);
            \vertex[right=0.5cm of a5] (a6);
            \vertex[below=0.4cm of a] (f);
            \vertex[below=0.8cm of a] (e);
            \vertex[right=0.5cm of e] (e2);
            \vertex[right=0.5cm of e2] (e3);
            \vertex[right=0.5cm of e3] (e4);
            \vertex[right=0.25cm of e3] (b2);
            \vertex[right=0.5cm of e4] (e5);
            \vertex[right=0.5cm of e5] (e6);
            \diagram* {
            (a) -- [thick] (a6),
            (e)-- [thick] (e6),
            (b1)-- [gluon] (c1),
          };
        \end{feynman}
        \fill[gray] (a2) rectangle (e5);
        \end{tikzpicture} \quad = \quad 
        \begin{tikzpicture}[baseline=(g)]
            \begin{feynman}
            \vertex (a);
            \vertex[right=1cm of a] (b);
            \vertex[right=1cm of b] (c);
            \vertex[below=1cm of a] (e);
            \vertex[right=1cm of e] (d);
            \vertex[right=1cm of d] (f);
            \vertex[below=0.5cm of b] (g);
            \vertex[below=0.5cm of g] (l);
            \vertex[right=0.5 cm of d] (h);
            \vertex[left=0.5 cm of d] (i);
            \vertex[above=0.5cm of c] (m);
            \diagram* {
            (b) -- [gluon] (m),
            (a) --  (b) --  (c),
            (e) -- (d) -- (f), };
            \end{feynman}
            \end{tikzpicture}+
            \begin{tikzpicture}[baseline=(g)]
            \begin{feynman}
            \vertex (a);
            \vertex[right=1cm of a] (b);
            \vertex[right=1cm of b] (c);
            \vertex[below=1cm of a] (e);
            \vertex[right=1cm of e] (d);
            \vertex[right=1cm of d] (f);
            \vertex[below=0.5cm of b] (g);
            \vertex[below=0.5cm of g] (l);
            \vertex[right=0.5 cm of d] (h);
            \vertex[left=0.5 cm of d] (i);
            \vertex[above=0.5cm of c] (m);
            \diagram* {
            (b) -- [gluon] (m),
            (a) --  (b) --  (c),
            (b) --  [scalar] (g), 
            (g) -- [scalar] (l),
            (e) -- (d) -- (f), };
            \end{feynman}
            \end{tikzpicture}+\begin{tikzpicture}[baseline=(g)]
                \begin{feynman}
                \vertex (a);
                \vertex[right=1cm of a] (b);
                \vertex[right=1cm of b] (c);
                \vertex[below=1cm of a] (e);
                \vertex[right=1cm of e] (d);
                \vertex[right=1cm of d] (f);
                \vertex[below=0.5cm of b] (g);
                \vertex[right=1cm of g] (k);
                \vertex[below=0.5cm of g] (l);
                \vertex[right=0.5 cm of d] (h);
                \vertex[left=0.5 cm of d] (i);
                \vertex[above=0.5cm of c] (m);
                \diagram* {
                (a) --  (b) --  (c),
                (b) --  [scalar] (g), 
                (g) -- [scalar] (l),
                (g) -- [gluon] (k),
                (e) -- (d) -- (f), };
                \end{feynman}
                \end{tikzpicture}+\ldots \ , 
        \label{eq:radiation_emitted}
    \end{equation}
    where it should be noticed that radiation gravitons can be coupled either to worldlines or to potential gravitons.
\subsection{Kol-Smolkin Variables}
To simplify the diagram computation we can take advantage of the diffeomorphism invariance of General Relativity to impose a Kaluza-Klein parametrization \cite{Kol:2007bc,Kol:2007rx, Kol:2010ze,Kol:2010si} for the metric tensor, which is based on the use of the \textit{Kol-Smolkin} variables. \\
We can decompose the symmetric tensor $g_{\mu\nu}$ in terms of a scalar field $\phi$, a d-dimensional vector field $A_i$ and a $d\times d$ symmetric tensor field $\sigma_{ij}$ where the indices $i,j$ run from $1$ to $d$:
\begin{equation}
    g_{\mu\nu}=e^{2\phi/\Lambda}\Biggl(\begin{matrix}
 -1& A_j/\Lambda \\
A_i/\Lambda & e^{-c_d\frac{\phi}{\lambda}}\gamma_{ij}-A_iA_j/\Lambda^2
\end{matrix} \Biggr)\ , \qquad  \qquad \gamma_{ij}=\left(\delta_{ij}+\frac{\sigma_{ij}}{\Lambda}\right) \ , 
\end{equation}
where $c_d=2\bigl(\frac{d-1}{d-2}\bigr)$.
In 4 dimensions the 10 degrees of freedom of $g_{\mu\nu}$ are decomposed as: 1 for the scalar field $\phi$, 3 for the  three-vector field $A_i$ and 6 for the  $3\times 3$ symmetric tensor $\sigma_{ij}$. \\
We can rewrite the point particle action $S_{pp}$ in the Kol-Smolkin variables, first by parametrizing the BH worldline with $t$, time of an external static observer:
\begin{equation}
    S_{pp}=-m_a\int dt\sqrt{-g_{\mu\nu}(x_a)\dot{x}^\mu_a\dot{x}^\nu_a} \qquad with \qquad \dot{x}^{\mu}=(1,v^{i}/c) \ ,
\end{equation}
then by substituting the variables, obtaining:
\begin{equation}
    S_{pp}=-m_a\int dt e^{\frac{\phi}{\Lambda}}\sqrt{1-\frac{2v_iA^i}{\Lambda}-\gamma_{ij}v^{i}v^{j}e^{-c_d\frac{\phi}{\Lambda}}+\frac{(A_iv^i)^2}{\Lambda^2}} \ .
    \label{eq:pp_action_kk}
\end{equation}
Similarly, we can express $S_{bulk}$ in terms of Kol-Smolkin variables, and reporting only the part needed for calculations that will be performed within this thesis we get: 
\begin{eqnarray}
    S_{bulk} 
    & \supset &
    \int d^{d+1}x\sqrt{-\gamma}\Biggl\{ \frac{1}{4}\biggl[\bigl(\Vec{\nabla}\sigma\bigr)^2-2\bigl(\Vec{\nabla}\sigma_{ij}\bigr)^2-\bigl(\dot{\sigma}^2-2(\dot{\sigma}_{ij})^2\bigr)^{-\frac{c_d\phi}{\Lambda}}\biggr]\nonumber \\ 
    & & 
    -c_d\biggl[\bigl(\Vec{\nabla}\phi\bigr)^2-\dot{\phi}^2-\dot{\phi}^2e^{-\frac{c_d\phi}{\Lambda}}\biggr] +\biggl[\frac{F_{ij}^2}{2}+\bigl(\Vec{\nabla}\cdot\Vec{A}\bigr)^2-\dot{\Vec{A}}^2e^{-\frac{c_d\phi}{\Lambda}}\biggr]e^{\frac{c_d\phi}{\Lambda}}\nonumber \\ 
    & & 
    +\frac{2}{\Lambda}\biggl[\bigl(F_{ij}A^{i}\dot{A}^j+\Vec{A}\cdot\dot{\Vec{A}}(\Vec{\nabla}\cdot\Vec{A})\bigr)-c_d\dot{\phi}\Vec{A}\cdot\Vec{\nabla}\phi\biggr]+2c_d\biggl(\dot{\phi}\Vec{\nabla}\cdot\Vec{A}-\dot{\Vec{A}}\cdot\Vec{\nabla}\phi\biggr)\nonumber \\ 
    & & 
    +\frac{\dot{\sigma}_{ij}}{\Lambda}\biggl(-\delta^{ij}A_l\hat{\Gamma}^l_{k k}+2A_k\hat{\Gamma}^k_{ij}-2A^i\hat{\Gamma}^j_{k k }\biggr)\nonumber \\ 
    & & 
    -\frac{1}{\Lambda}\biggl(\frac{\sigma}{2}\delta^{ij}-\sigma^{ij}\biggr)\biggl(\sigma_{ik}^{,l}\sigma_{jl}^{,k}-\sigma_{ik}^{,k}\sigma_{jl}^{,l}+\sigma_{, i }\sigma_{jk}^{,k}-\sigma_{ik,j}\sigma^{,k}\biggr)\Biggr\}
 \end{eqnarray}
 where: $\Lambda=\sqrt{32\pi G_d}^-1$, $G_d=G_N^{d-3} $ is the $d$-dimensional coupling constant, $F_{ij}=A_{j,i}-A_{i,j}$ and $\hat{\Gamma}^i_{jk}$ is the connection of the purely spatial $d-$dimensional metric $\gamma_{ij}=\delta_{ij}+\sigma_{ij}/\Lambda$, which is also used above to raise and contract spatial indices. All spatial derivatives are understood as simple (not covariant) derivatives and when ambiguities might rise gradients are always meant to act on controvariant fields, e.g. $\Vec{\nabla}\cdot\Vec{A}=\gamma^{ij}A_{i,j}$, $F_{ij}^2=\gamma^{ik}\gamma^{jl}F_{ij}F_{kl}$, and $\sigma=\sigma^i_i=\gamma^{ij}\sigma_{ij}$. 
From the total action we can extract the Feynman rules, which will be reported in Appendix\ref{chapter:feynman_rules_pn}, which are needed to build effective Feynman diagrams that scale with fixed powers of $v^2$ and $G_N$. 
Diagrammatically we will denote the $\phi\ ,\  A\ , \ \sigma$ fields in the following way: 
\begin{equation}
    \textcolor{blue}{\phi} \ 
     \begin{tikzpicture}[baseline=(current bounding box.center)]
        \begin{feynman}
        \vertex (a);
        \vertex[right=1cm of a] (b);
        \diagram* {
        (a) --  [blue,scalar] (b)};
        \end{feynman}
        \end{tikzpicture} \qquad \textcolor{red}{A} \ \begin{tikzpicture}[baseline=(current bounding box.center)]
            \begin{feynman}
            \vertex (a);
            \vertex[right=1cm of a] (b);
            \diagram* {
            (a) --  [red,boson] (b)};
            \end{feynman}
            \end{tikzpicture} \qquad \textcolor{black!60!green}{\sigma} \ \begin{tikzpicture}[baseline=(current bounding box.center)]
                \begin{feynman}
                \vertex (a);
                \vertex[right=1cm of a] (b);
                \diagram* {
                (a) --  [double_boson,black!60!green] (b)};
                \end{feynman}
                \end{tikzpicture}
\end{equation}
We introduced these variables because they provide important calculations advantages over the $h_{\mu\nu}$ parametrization.
Working in the $h_{\mu\nu}$ parametrization, one often consider the components of the metric tensor $h_{00},h_{0i},h_{ij}$ separately, since their leading order coupling to the worldline scale as $\mathcal{O}(v^0),\mathcal{O}(v^1)\mathcal{O}(v^2)$, respectively.\\
However, when one consider propagators and interaction vertices, different modes mix together, and it is difficult to keep track of the velocity scaling. In the Kol-Smolkin parametrization instead, mixing between different field modes are removed, since 2-point functions between different fields are zero: $\langle T(\phi A_i)\rangle= =\langle T(\phi \sigma_{ij})\rangle= =\langle T(A_i \sigma_{ij})\rangle=0$. 
That reduces drastically the calculation effort. \\
\subsection{Systematic procedure to get near-zone contributions}
In these new variables, considering only potential modes ($\Bar{h}=0$), we can get conservative contributions to the effective action by integrating out the fields $\phi, \ A_i, \ \sigma_{jk}$:
\begin{equation}
    e^{iS_{eff}(x_a)}=\int D\phi DA_iD\sigma_{jk}e^{iS_{EH}[\phi,A_i,\sigma_{jk}]+iS_{pp}[x_a,\phi,A_i,\sigma_{jk}]+iS_{GF}[\phi,A_i,\sigma_{jk}]} \ , 
\end{equation}
and has already discussed the procedure needs to be done perturbatively, with the use of Feynman diagrams. \\
From the total action one can extract the Feynman rules in the Kol-Smolkin variables, which are reported in Chapter.\ref{chapter:feynman_rules_pn}. They are characterized by a definite scaling of velocities $v$ and Newton constant $G_N$.
We can evaluate corrections to the conservative dynamics, order by order, in the following way:
\begin{enumerate}
    \item Generate all the diagrams at a given PN order,
    \item Evaluate all the diagrams,
    \begin{eqnarray}
    \mathcal{M}=\sum_i\mathcal{M}_i
    \end{eqnarray}
    \item Take the Fourier transform of the evaluated amplitude to get the contributions to the effective action:
    \begin{eqnarray}
    \mathcal{L}=-i \lim_{d\to 3}\int \frac{d^d p}{(2\pi)^d}e^{ip\cdot r}\mathcal{M}
    \end{eqnarray}
\end{enumerate}
\subsection{Diagram Generation}
In order to determine the diagram generation let us define an \textit{effective diagram}, as an amplitude in momentum space that can be built from the Feynman rules,  which scales with definite powers of the Newton constant $G_N$ and $v^2$, with $v$ the typical velocity of the system. \\
It gives a $n$ post-Newtonian correction, if it scales as $G_N^{n-l}v^{2l}$ with $0\leq l \leq n-1$. \\
At this point, given an effective diagram, one should be aware to follow the subsequent rules:
\begin{itemize}
    \item Only connected diagrams contribute to the effective action.
    \item Given a connected diagram, one is required to draw all the possible arrangements which are topological inequivalent after having identified all the external worldlines.
    \item Once a proper effective diagram has been build, one has to integrate over momentum variables by taking care that external momenta (those coming from worldlines) has to be Fourier transformed, while internal one has to be integrated. 
\end{itemize}
There can be two types of Loop diagrams, and we do not want to consider quantum corrections: 
\begin{itemize}
     \item  The ones containing at least an external leg of a worldline, as an example:
    \begin{eqnarray}
\begin{tikzpicture}[baseline=(current bounding box.center)]
\begin{feynman}
\vertex (a);
\vertex[right=1cm of a] (b);
\vertex[right=1cm of b] (c);
\vertex[below=1.5cm of a] (e);
\vertex[right=1cm of e] (d);
\vertex[right=1cm of d] (f);
\vertex[below=0.75cm of b] (g);
\vertex[below=1.5cm of b] (l);
\vertex[right=0.5 cm of l] (h);
\vertex[left=0.5 cm of l] (i);
\vertex[above=0.5cm of c] (m);
\diagram* {
(a) --  (b) --  (c),
(b) --  [scalar] (g), 
(g) -- [scalar] (h),
(g) -- [scalar] (i),
(e) -- (d) -- (f), };
\end{feynman}
\end{tikzpicture}
\end{eqnarray}

    Since there is no propagator associated to the worldlines, this integration cannot give quantum contributions, therefore, besides they formally resemble loop integral, these are just classical contributions. \item The one containing closed graviton loops, these corrections bring non null powers of $\hbar$, and they will not be considered in the following:
     \begin{eqnarray}
\begin{tikzpicture}[baseline=(current bounding box.center)]
\begin{feynman}
\vertex (a);
\vertex[right=1cm of a] (b);
\vertex[right=1cm of b] (c);
\vertex[below=1.5cm of a] (e);
\vertex[right=1cm of e] (d);
\vertex[right=1cm of d] (f);
\vertex[below=0.5cm of b] (g);
\vertex[below=1cm of b] (l);
\vertex[below=0.5cm of l] (h);
\diagram* {
(a) --  (b) --  (c),
(b) --  [scalar] (g), 
(g) -- [scalar,half left] (l),
(g) -- [scalar,half right] (l),
(l) -- [scalar] (h),
(e) -- (d) -- (f), };
\end{feynman}
\end{tikzpicture}
\end{eqnarray}
\end{itemize}

\subsection{Diagram Evaluation: from effective Feynman diagrams to QFT amplitudes}
In order to properly evaluate the effective diagrams, we have to convert them in a QFT amplitudes language.
In general, within the EFT approach, since the sources (black lines) are static and do not propagate, any gravity amplitude of order $G_N^l$ can be mapped into a $(l-1)$-loop 2-point function (as first noticed in \cite{Foffa:2016rgu}) with massless internal lines and external momentum $p$, where $p^2=s\neq0$:
\begin{eqnarray}
    \begin{tikzpicture}[baseline=(current bounding box.center)]
        \begin{feynman}
        \vertex (a);
        \vertex[right=0.5cm of a] (a2);
        \vertex[right=0.5cm of a2] (a3);
        \vertex[right=0.25cm of a3] (b1);
        \vertex[right=0.5cm of a3] (a4);
        \vertex[right=0.5cm of a4] (a5);
        \vertex[right=0.5cm of a5] (a6);
        \vertex[below=1.2cm of a] (e);
        \vertex[right=0.5cm of e] (e2);
        \vertex[right=0.5cm of e2] (e3);
        \vertex[right=0.5cm of e3] (e4);
        \vertex[right=0.25cm of e3] (b2);
        \vertex[right=0.5cm of e4] (e5);
        \vertex[right=0.5cm of e5] (e6);
        \diagram* {
        (a) -- [thick] (a6),
        (e)-- [thick] (e6),
      };
    \end{feynman}
    \fill[gray] (a2) rectangle (e5);
    \end{tikzpicture} \qquad \Leftrightarrow \qquad 
     \begin{tikzpicture}[baseline=(current bounding box.center)]
        \begin{feynman}
        \vertex (a);
        \vertex[below=0.4cm of a] (a2);
         \vertex[below=0.5cm of a2] (c);
        \vertex[below=1cm of a2] (a3);
        \vertex[below=0.4cm of a3](a4);
        \diagram* {
        (a) -- [thick] (a2),
        (a3)-- [thick] (a4),
      };
    \end{feynman}
    \draw[gray,fill=gray] (c) circle (.6cm);
    \end{tikzpicture}
\end{eqnarray}
Once this identification has been made, the amplitude can be evaluated with usual QFT multi-loop techniques that will be explained in the next chapter.\\
\subsection{The Newtonian potential from a 0PN calculation}
In order to have a taste of what we are going to compute and give an example of Post-Newtonian calculations with diagrammatic approach, let us show that from a simple 0PN calculations we recover the Newtonian potential. \\
Following the prescription given in the previous section, we look for diagrams $\mathcal{M}\sim G_N$. There is only one diagram contributing that is given by: 
\begin{eqnarray}
    \mathcal{M}_{0PN}\ = \  \begin{tikzpicture}[baseline=(current bounding box.center)]
        \begin{feynman}
        \vertex (a);
        \vertex[right=0.5cm of a] (a2);
        \vertex[right=0.5cm of a2] (a3);
        \vertex[right=0.25cm of a3] (b1);
        \vertex[right=0.5cm of a3] (a4);
        \vertex[right=0.5cm of a4] (a5);
        \vertex[right=0.5cm of a5] (a6);
        \vertex[below=1cm of a] (e);
        \vertex[right=0.5cm of e] (e2);
        \vertex[right=0.5cm of e2] (e3);
        \vertex[right=0.5cm of e3] (e4);
        \vertex[right=0.25cm of e3] (b2);
        \vertex[right=0.5cm of e4] (e5);
        \vertex[right=0.5cm of e5] (e6);
        \diagram* {
        (a2) -- [] (a5),
        (e2)-- [] (e5),
        (b1) --  [scalar,blue] (b2), 
      }; 
    \end{feynman}
    \end{tikzpicture}
\end{eqnarray}
Substituting the Feynman rules we get:
\begin{eqnarray}
    \mathcal{M}_{0PN}=\frac{im_1m_2}{2c_d\Lambda^2}\frac{1}{\mathbf{p}^2}
\end{eqnarray}
The corresponding contribution to the effective action is given by the Fourier transform of such amplitude: 
\begin{eqnarray}
    \mathcal{L}_{0PN}& =& -i\lim_{d\to 3}\int \frac{d^d\mathbf{p}}{(2\pi)^d}e^{i\mathbf{p}(\mathbf{x}_1-\mathbf{x}_2)}\left(\ \begin{tikzpicture}[baseline=(current bounding box.center)]
        \begin{feynman}
        \vertex (a);
        \vertex[right=0.5cm of a] (a2);
        \vertex[right=0.5cm of a2] (a3);
        \vertex[right=0.25cm of a3] (b1);
        \vertex[right=0.5cm of a3] (a4);
        \vertex[right=0.5cm of a4] (a5);
        \vertex[right=0.5cm of a5] (a6);
        \vertex[below=1cm of a] (e);
        \vertex[right=0.5cm of e] (e2);
        \vertex[right=0.5cm of e2] (e3);
        \vertex[right=0.5cm of e3] (e4);
        \vertex[right=0.25cm of e3] (b2);
        \vertex[right=0.5cm of e4] (e5);
        \vertex[right=0.5cm of e5] (e6);
        \diagram* {
        (a2) -- [] (a5),
        (e2)-- [] (e5),
        (b1) --  [scalar,blue] (b2), 
      }; 
    \end{feynman}
    \end{tikzpicture}\ \right) \nonumber\\ 
        & = & -i\lim_{d\to 3}\int \frac{d^d\mathbf{p}}{(2\pi)^d}e^{i\mathbf{p}(\mathbf{x}_1-\mathbf{x}_2)}\frac{im_1m_2}{2c_d\Lambda^2}\frac{1}{\mathbf{p}^2} \nonumber\\
        & = & -i\lim_{d\to 3}\frac{im_1m_2}{2c_d\Lambda^2}I_F(d,1) \nonumber \\
        & = & -i\lim_{d\to 3}\frac{im_1m_2}{2c_d\Lambda^2}\frac{\Gamma(d/2-1)}{(4\pi)^{d/2}}\left(\frac{r}{2}\right)^{2-d} \nonumber\\ 
        & = & \frac{G_N m_1 m_2}{r} 
        \label{eq:0PN}
\end{eqnarray}
where $r=\vert\mathbf{x}_1-\mathbf{x}_2\vert$, and we used the formula for the d-dimensional scalar integral  Eq.\eqref{eq:d_dim_scalar_integral}.
As expected from a 0PN calculation we recover the Lagrangian of two point particles attracted by the Newtonian potential:
\begin{eqnarray}
    \mathcal{L}_{0PN}=\frac{1}{2}m_1v_1^2+\frac{1}{2}m_2v_2^2+\frac{G_Nm_1m_2}{r} \ . 
\end{eqnarray} 
\subsection{The Einstein-Infeld-Hoffmann Lagrangian from a 1PN calculation}
At 1PN order we look for diagrams: $\mathcal{M}\sim G_N^2,G_Nv^2$. \\ 
There are $3$ diagrams contributing, $2$ from the $G_N$ topology and one from the $G_N^2$ ones, given by:
\begin{equation}
    \mathcal{M}_{1PN}\quad = \quad \begin{tikzpicture}[baseline=(current bounding box.center)]
\begin{feynman}
\vertex (a);
\vertex[right=1cm of a] (b);
\vertex[right=1cm of b] (c);
\vertex[below=1cm of a] (e);
\vertex[right=1cm of e] (d);
\vertex[right=1cm of d] (f);
\diagram* {
(a) --  (b) --  (c),
(b) -- [blue, scalar] (d), 
(e) -- (d) -- (f), };
\end{feynman}
\end{tikzpicture}\  + \  \begin{tikzpicture}[baseline=(current bounding box.center)]
\begin{feynman}
\vertex (a);
\vertex[right=1cm of a] (b);
\vertex[right=1cm of b] (c);
\vertex[below=1cm of a] (e);
\vertex[right=1cm of e] (d);
\vertex[right=1cm of d] (f);
\diagram* {
(a) --  (b) --  (c),
(b) -- [red , photon] (d), 
(e) -- (d) -- (f), };
\end{feynman}
\end{tikzpicture}
 \ + \ \begin{tikzpicture}[baseline=(current bounding box.center)]
\begin{feynman}
\vertex (a);
\vertex[right=1cm of a] (b);
\vertex[right=1cm of b] (c);
\vertex[below=1cm of a] (d);
\vertex[right=0.5cm of d] (e);
\vertex[right=1cm of e] (f);
\vertex[right=0.5cm of f] (g);
\diagram* {
(a) --  (b) --  (c),
(b) -- [blue, scalar] (e), 
(b) -- [blue, scalar] (f),
(d)-- (e) -- (f) -- (g), };
\end{feynman}
\end{tikzpicture}
\end{equation}
The corresponding contribution to the effective action is given by the Fourier transform of such diagrams, and by summing up this result with the 0PN one,  we get the Einstein-Infeld-Hoffmann Lagrangian (\cite{Einstein:1938yz}):
\begin{equation}
    \mathcal{L}_{EIH}= \frac{m_av_a^2}{2}+\frac{m_av_a^4}{8}+\frac{G_Nm_1m_2}{2r}\biggl[2+3(v_1^2+v_2^2)-7(v_1\cdot v_2)-(v_1\cdot\hat{r})(v_2\cdot\hat{r})-\frac{G_N(m_1+m_2)}{r}\biggr] \ . 
\end{equation}
This calculation gives us the opportunity to show explicitly the relation between effective diagrams and 2-point functions by considering the third diagram:
\begin{equation}
     \mathcal{M}_{\phi\phi} \ = \ \begin{tikzpicture}[baseline=(current bounding box.center)]
\begin{feynman}
\vertex (a);
\vertex[right=1cm of a] (b);
\vertex[right=1cm of b] (c);
\vertex[below=1cm of a] (d);
\vertex[right=0.5cm of d] (e);
\vertex[right=1cm of e] (f);
\vertex[right=0.5cm of f] (g);
\diagram* {
(a) --  (b) --  (c),
(b) -- [blue, scalar] (e), 
(b) -- [blue, scalar] (f),
(d)-- (e) -- (f) -- (g), };
\end{feynman}
\end{tikzpicture}
\end{equation}
If we did not know about this relation we would have proceeded by assigning a different momentum to each internal graviton line, and we would have taken a Fourier transform for each coupling to the source obtaining: 
\begin{eqnarray}
  S_{\phi\phi} =   -i \lim_{d\to 3 }\frac{m_1m_2^2}{8\Lambda^4c_d^2}\int d t \biggl[\int \frac{d^d\mathbf{k}_1}{(2\pi)^d}\frac{d^d\mathbf{k}_2}{(2\pi)^d}e^{i(\mathbf{k}_1+\mathbf{k}_2)\cdot (\mathbf{x}_1(t)-\mathbf{x}_2(t))}\frac{1}{\mathbf{k}_1^2}\frac{1}{\mathbf{k}_2^2}\biggr]
\end{eqnarray}
At this point we can define $\mathbf{r}=\mathbf{x}_1-\mathbf{x}_2$, and we can make the following shift of momentum variables: 
\begin{eqnarray}
    \mathbf{k}_1+\mathbf{k}_2= \mathbf{p} \ , \qquad \mathbf{k}_1= \mathbf{k} \ , 
\end{eqnarray}
obtaining: 
\begin{eqnarray}
    S_{\phi\phi} =   -i \lim_{d\to 3 }\frac{m_1m_2^2}{8\Lambda^4c_d^2}\int dt\Biggl\{ \int \frac{d^d\mathbf{p}}{(2\pi)^d}e^{i\mathbf{p}\cdot\mathbf{x}(t)}\biggl[\int \frac{d^d\mathbf{k}}{(2\pi)^d}\frac{1}{\mathbf{k}^2(\mathbf{k}-\mathbf{p})^2}\biggr]\Biggr\} \ , 
\end{eqnarray}
which is exactly the Fourier transform of a 1-loop two-point function. 
\subsection{Higher order PN corrections}
This procedure is very powerful and can be generalized at any PN order (see \cite{Cristofoli:2018bex} for other pedagogical examples of PN calculations in the near zone). 
The binary dynamics in the near zone at 2PN order in the EFT framework has been computed in \cite{Gilmore:2008gq}. It has been extended to 3PN order in \cite{Foffa:2011ub}, and at 4PN in \cite{Foffa:2012rn,Foffa:2016rgu}, and at 5PN in \cite{Foffa:2019hrb,Foffa:2019ahl}. An alternative derivation up to 6PN can be found in \cite{Blumlein:2019zku,Blumlein:2020pog,Blumlein:2020pyo, Blumlein:2020znm,Blumlein:2021txe,Blumlein:2021txj}. \\ 
However, as noticed by Damour and Blanchet in. \cite{Blanchet:1987wq}, the evaluation of the conservative dynamics, by focusing only on the near zone contribution, breaks down at 4PN order due to the appearance of infrared divergences. Indeed, at this level of accuracy, is crucial to take into account that the gravitational propagator in the curved space-time, generated by the binary system, contains a significant tail contribution whose support is not limited to light-like intervals, but extends to strongly time non-local intervals. \\
These so-called hereditary effects, can be included in the computation by studying contributions to the conservative dynamics coming from the far zone. 
By adding these contributions the IR divergence of the near-zone exactly cancels out with UV divergences coming from the far zone theory with a procedure known as \textit{zero bin subtraction}. Let us now introduce an EFT that appropriately describes the far zone physics.
\section{Long-wavelength EFT}
In order to consider contributions to the conservative dynamics coming from the far zone, we need to consider physical effects at distances from the source compatible with the wavelength of the emitted gravitational waves $\lambda_{rad}$, which is much larger than the size of the orbital radius $r\ll \lambda_{rad}$. 
At such distances, one can effectively see the binary system, with the exchange of potential gravitons between the binary components, as a single linear source $T_{\mu\nu}$ of size $r$, emitting radiation graviton $\Bar{h}_{\mu\nu}$, that can be described using the following action: 
\begin{eqnarray}
    S_{source}=-\frac{1}{2}\int d^4x T^{\mu\nu}\bar{h}_{\mu\nu} \ , 
    \label{eq:Source_Action}
\end{eqnarray}
where $\partial_{\mu}T^{\mu\nu}=0$. We can diagrammatically represent this coupling as:
\begin{equation}
    \begin{tikzpicture} \begin{feynman}
\vertex (a1) ; 
\vertex[right=1cm of a1, square dot, red] (a2) {}; 
\vertex[below= 0.3cm of a2] (c1){\(T_{\mu\nu}\)};
\vertex[right=1cm of a2] (a3); 
\vertex[above=1cm of a2] (b2);
\diagram* { 
(a1)-- [double, thick] (a3),
(b2) -- [gluon] (a2)
};
\end{feynman} \end{tikzpicture} 
\end{equation}
where the double line denotes the non-propagating source while the curly one represents the radiation graviton. We know that radiation gravitons scale as:
\begin{eqnarray}
    \bar{h}_{\mu\nu}\sim\frac{v}{r} \ . 
\end{eqnarray}
However, to obtain an EFT for radiation that has manifest velocity power counting, the decomposition of the graviton into potential and radiation modes is not sufficient.
It is necessary to Taylor expand the action in the radiation gravitons around a point, which lies within the source, with a \textit{multipole expansion} procedure, that we will soon describe in detail. The Wilson coefficients of this action, the so-called \textit{multipole moments}, are organized in irreducible representations of the rotation group $SO(3)$, in particular we will adopt symmetric trace free tensors (STF). 
This has several important advantages:
 \begin{itemize}
    \item it makes the calculation simpler and more transparent,
    \item it ensures the absence of mixing of multipole moments. 
 \end{itemize}
\noindent The multipoles are not determined by the symmetries of the system and need to be fixed through a matching calculation. Before proceeding further, let us describe how one can perform the procedure of multipole expansion at the level of the action for the case of a simple scalar field, and then we will turn to the one for the graviton field.
\subsection{Multipole expansion of a scalar field}
Let us consider a scalar field $\phi$ coupled linearly to a source $J(x)$, described by an action of the type: 
\begin{eqnarray}
    S=\int d^4x \biggl(\frac{1}{2}\partial_{\mu}\phi\partial^\mu\phi+J\phi\biggr) \ . 
\end{eqnarray}
The corresponding equation of motion is:
\begin{eqnarray}
    \Box\phi=J \ .
\end{eqnarray}
Let us focus only on field configurations where the spatial variation of the field outside the source is much larger than the size of the source $J$. Meaning, if the size of the source is $r$ and the spatial derivatives scale as $\partial_i\phi \approx \frac{1}{\lambda}\phi$, we have $r \ll \lambda$. \\
In this case we can see $\lambda$ as a wavelength, and we can say that we are considering radiation scalar fields, meaning that we are working in a long wavelength approximation. \\
We can Taylor expand the field $\phi$ in the source term in the action $S$, around a point within the source. \\
Let us choose our coordinates such that the point we expand around is the origin $\mathbf{x}=0$.
Then we get the Taylor expansion:
\begin{eqnarray}
    \phi(t,\mathbf{x})=\sum_{n=0}^{\infty}\frac{1}{n!}\mathbf{x}^{k_1}...\mathbf{x}^{k_n}(\partial_{k_1}...\partial_{k_n})\phi(t,0)=\sum_{n=0}^{\infty}\frac{1}{n!}x^N(\partial_N\phi)(t,0) \ ,
    \end{eqnarray}
    We can plug it into the source term in the action:
\begin{eqnarray}
    S_{source}& = & \int dt\int d^3\mathbf{x} J(t,\mathbf{x})=\sum_{n=0}^{\infty}\frac{1}{n!}x^N(\partial_N\phi)(t,0)=\int dt \sum_{n=0}^{\infty}Q^N\partial_N\phi \ , 
\end{eqnarray}
where we defined the moments:
\begin{eqnarray}
    Q^N(t)=\int d^3\mathbf{x} \ J(t,\mathbf{x})x^N \ .
\end{eqnarray}
The moments $Q^N$ are already symmetric in $k_1,...,k_n$. \\
In order to bring the action into the form of a multipole expansion we need to decompose the moments $Q^N$ in irreducible representation of the rotation group $SO(3)$, for which we use symmetric trace free (STF) tensors.
The tensors $Q^N$ can be expressed in terms of STF tensors starting form the formula for arbitrarily symmetric tensors $S^N$:
\begin{eqnarray}
    S^N=S^N_{STF}+\sum_{p=1}^{\left[\frac{n}{2}\right]}\frac{(-1)^{p+1}n!(2n-2p-1)!!}{(n-2p)!(2n-1)!!(2p)!!}\delta^{(k_1 k_2}...\delta^{k_{2p-1}k_{2p}}S^{k_{2p+1}...k_n)a_1a_1...a_pa_p} \ , 
    \label{eq:STF_formula}
\end{eqnarray}
where $S^N_{STF}$ is a symmetric trace free tensor and $\left[\frac{n}{2}\right]$ denotes the largest integer $\leq n/2$.\\
In Eq. \eqref{eq:STF_formula} the tensors of lower rank, $S^{k_{2p+1}...k_n)a_1a_1...a_pa_p}$, are still not trace free in their free indices $k_{1p+1}...k_n$. Therefore, we need to use recursively Eq.\eqref{eq:STF_formula} to make all tensors trace free, where we mean that all free indices in $\{k_1,...,k_n\}$ which are not on a $\delta^{k_1k_j}$ should become free. Then we obtain: 
\begin{eqnarray}
    S^N=\sum_{p=0}^{\left[\frac{n}{2}\right]}c_p^{(n)}\delta^{(k_1k_2}...\delta^{k_{2p-1}k_{2p}}S_{STF}^{k_{2p+1}...k_n)a_1a_1...a_pa_p}
\end{eqnarray}
where the STF prescription only applies to the uncontracted indices $k_{2p+1}...k_n$ and where the coefficients are:
\begin{eqnarray}
    c_p^{(n)}=\frac{n!(2n-4p+1)!!}{(2p)!!(n-2p)!(2n-2p+1)!!}
\end{eqnarray}
The source action becomes:
\begin{eqnarray}
    S_{source} &=& 
    \int dt \sum_{n=0}^{\infty}\sum_{p=0}^{\left[\frac{n}{2}\right]}\frac{c_p^{(n)}}{n!}\int d^3\mathbf{x}J r^{2p}x^{N-2P}_{STF}(\nabla^2)^p\partial_{N-2P}\phi \nonumber\\
    &=& \int dt\sum_{l=0}^{\infty}\frac{1}{l!}\sum_{j=0}^{\infty}\frac{(2l+1)!!}{(2j)!!(2l+2j+1)!!}\int d^3\mathbf{x} J r^{2j}x^L_{STF}(\nabla^2)^j\partial_L\phi
\end{eqnarray}
where we define $r=\vert\mathbf{x}\vert$. We can now use the equation of motion outside the source $\Box\phi=0$ to convert contracted spatial derivatives to time derivatives, and in turn integrate by parts to let them act on the moments rather than on the field. \\
This yields the source action in multipole expanded form:
\begin{eqnarray}
    S_{source}=\int dt \sum_{l=0}^{\infty}\frac{1}{l!}\mathcal{I}^L\partial_L\phi
\end{eqnarray}
where the multipole moments are given by:
\begin{eqnarray}
    \mathcal{I}^L=\sum_{p=0}^{\infty}\frac{(2l+1)!!}{(2p)!!(2l+2p+1)!!}\int d^3\mathbf{x}\partial_t^{2p}Jr^{2p}x^L_{STF}
\end{eqnarray}
The normalization of the multipole moments $\mathcal{I}_L$ is chosen such that for $p=0$ their expression reads:
\begin{eqnarray}
    \mathcal{I}_0^
    L=\int d^3\mathbf{x}Jx^L_{STF}
\end{eqnarray}
where we note that the static case of a time independent source is given only by the $p=0$ components of the multipoles. \\
A similar procedure can be done in NRGR to get the multipole action.
\subsection{Multipole expansion in NRGR}
Let us now go back to the action of the gravitational field coupled to a linear source defined in Eq.\eqref{eq:Source_Action}.
As done in the previous case we can Taylor expand the gravitational field $h_{\mu\nu}$ in the source action around a point within the source choosing our coordinates such that the point we expand around is the origin $\mathbf{x}=0$.
The source term in the action then is:
\begin{eqnarray}
    S_{source}=-\frac{1}{2}\int dt \sum_{n=0}^{\infty}\int d^3\mathbf{x}T^{\mu\nu}(t,\mathbf{x})x^{N}(\partial_N \bar{h}_{\mu\nu})(t,0) \ . 
\end{eqnarray}
After few technical steps that can be found in \cite{Ross:2012fc}, we can obtain the multipole expanded source action, first introduced in \cite{RevModPhys.52.299}, and then in  \cite{Ross:2012fc}:
\begin{eqnarray}
    S_{source}&=& 
    \frac{1}{2}\int dt[\ E\bar{h}_{00}+\mathbf{P}^i\bar{h}_{0i}-E\mathbf{Z}^i\partial_i\bar{h}_{00}-\mathbf{L}^i\epsilon_{ijk}\partial_j\bar{h}_{0k}] \nonumber\\
    & & -\int dt \sum_{l=2}^{\infty}\frac{1}{l!}I^L\partial_{L-2}\mathcal{E}_{k_{l-1}k_l}-\int dt \sum_{l=2}^{\infty}\frac{2l}{(l+1)!}J^L\partial_{L-2}B_{k_{l-1}k_l}
\end{eqnarray}
where $E,\mathbf{P}^i,M\mathbf{X}^i,\mathbf{L}^i,I^L,J^L$ are the multipole moments, 
$\mathcal{E}$ and $B$ are the electric and magnetic components of the metric tensors and at lowest order in $\bar{h}$ read:
    \begin{eqnarray}
        \mathcal{E}_{ij}&=& R_{0i0j}\approx \frac{-1}{2}\biggl(\bar{h}_{00,ij}+\ddot{\Bar{h}}_{ij}-\dot{\Bar{h}}_{0i,j}-\dot{\Bar{h}}_{0j,i}\biggr)+\mathcal{O}(h^2)\label{eq:multipole_E} \\
        B_{ij}&=&\frac{1}{2}\epsilon_{ikl}R_{0jkl}\approx\frac{1}{4}\epsilon_{ikl}\biggl(\dot{\bar{h}}_{jk,l}-\dot{\bar{h}}_{jl,k}+\bar{h}_{0l,jk}-\bar{h}_{0k,jl}\biggr)+\mathcal{O}(h^2) \ . \label{eq:multipole_B}
    \end{eqnarray}
Notice that this expression is in a system with four-velocity $v^\mu=(1,\mathbf{0})$.
The exact expressions for the multipole moments are given by:
\begin{eqnarray}
    E & = & \int d^3\mathbf{x}T^{00} \ , \\ 
    \mathbf{P}^i & = & \int d^3\mathbf{x}T^{0i} \ , \\ 
    E\mathbf{Z}^i & = & \int d^3\mathbf{x}T^{0i}\mathbf{x}^i \ , \\ 
    \mathbf{L}^i & = & -\int d^3\mathbf{x}\epsilon^{ijk}T_{0j}\mathbf{x}_k \ , \\ 
    I^L & = & 
    \sum_{p=0}^{\infty}c(l,p)\left(1+\frac{8p(l+p+1)}{(l+1)(l+2)}\right)\left[\int d^3\mathbf{x}\partial_t^{2p}T^{00}r^{2p}x^L\right]^{STF} \nonumber\\
    & & 
    +\sum_{p=0}^{\infty}c(l,p)\left(1+\frac{4p}{(l+1)(l+2)}\right)\left[\int d^3\mathbf{x}\partial_t^{2p}T^{aa}r^{2p}x^L\right]^{STF} \nonumber\\
     & & 
     -\sum_{p=0}^{\infty}c(l,p)\frac{4}{l+1}\left(1+\frac{2p}{l+2}\right)\left[\int d^3\mathbf{x}\partial_t^{2p+1}T^{0a}r^{2p}x^{aL}\right]^{STF}  \nonumber\\
      & & 
      +\sum_{p=0}^{\infty}c(l,p)\left(1+\frac{4p}{(l+1)(l+2)}\right)\left[\int d^3\mathbf{x}\partial_t^{2p}T^{aa}r^{2p}x^L\right]^{STF} \nonumber\ , \\ 
     & &
      -\sum_{p=0}^{\infty}c(l,p)\frac{2}{(l+1)(l+2)}\left[\int d^3\mathbf{x}\partial_t^{2p+2}T^{ab}r^{2p}x^{abL}\right]^{STF} \\
     J^L & = &  +\sum_{p=0}^{\infty}c(l,p)\left(1+\frac{2p}{l+2}\right)\left[\int d^3\mathbf{x}\epsilon^{k_lba}\partial_t^{2p}T^{0a}r^{2p}x^{bL-1}\right]^{STF} \nonumber\\
     & &
      -\sum_{p=0}^{\infty}c(l,p)\frac{1}{l+2}\left[\int d^3\mathbf{x}\epsilon^{k_lba}\partial_t^{2p+1}T^{ac}r^{2p}x^{bcL-1}\right]^{STF}   \ , 
\end{eqnarray}
where: 
\begin{equation}
    c(l,p)= \frac{(2l+1)!!}{(2p)!!(2l+2p+1)!!} \ . 
\end{equation}
We will refer to $I^L$ and $J^L$ as electric and magnetic multipoles, respectively. \\ Notice that in the center of mass frame: $\mathbf{Z}=\mathbf{P}=0$.
One could have instead built this action from a bottom-up approach based on symmetry argument, obtaining \cite{Goldberger:2009qd}: 
\begin{eqnarray}
    S_{source}&=& 
    -E\int d\tau -\frac{1}{2}\int dx^\mu L^i\epsilon_{ijk}w_\mu^{jk}-\int d\tau \sum_{l=2}^{\infty}\frac{1}{l!}I^L\nabla_{L-2}\mathcal{E}_{k_{l-1}k_l}\nonumber \\ 
    & & 
    -\int d\tau \sum_{l=2}^{\infty}\frac{2l}{(l+1)!}J^L\nabla_{L-2}B_{k_{l-1}k_l} \ , 
\end{eqnarray}
where we can replace $d\tau=dt\sqrt{g_{00}}$ and $dx^\mu=dt\delta^{\mu 0}$,  when we parametrize the world-line in terms of coordinate time t.

\subsection{Long-Wavelength Action}

For the purpose of our calculations, we will only need the first few terms of the multipole action $S_{mult}$:
\begin{eqnarray}
    S_{mult}[\Bar{h},\{Q_a\}]=\int dt \biggl[\frac{1}{2}E \bar{h}_{00}-\frac{1}{2}\epsilon_{ijk}L^i\bar{h}_{0j,k}-\frac{1}{2}Q^{ij}\mathcal{E}_{ij}-\frac{1}{6}O^{ijk}\mathcal{E}_{ij,k}-\frac{2}{3}J^{ij}B_{ij}+\ldots\biggr]
    \label{eq:multipole_action}
\end{eqnarray}
where:
\begin{itemize}
    \item $E,\Vec{L},Q^{ij},O^{ijk}$ are respectively the energy, spin, mass quadrupole and octupole moments of the system, 
    \item $J^{ij}$ is the current quadrupole moment.
\end{itemize}
The total radiative action is then given by the sum of $S_{mult}$ with the "bulk" gravitational action for radiation gravitons:
\begin{eqnarray}
    S_{rad}[\bar{h},\{Q_a\}]=S_{mult}[\bar{h},\{Q_a\}]+S_{EH}[\bar{h}]+S_{GF}[\bar{h}]
    \label{eq:radiative_action}
\end{eqnarray}

\subsection{Matching procedure}
If we want to fix the \textit{multipole moments} appearing in eq.\eqref{eq:multipole_action}, we need to do a matching procedure with the "full theory", the one containing both potential and radiation degrees of freedom, which is described by the action $S_{eff}(x_a,H_{\mu\nu},\bar{h}_{\mu\nu})$ introduced before. This procedure is done perturbatively, evaluating order by order Feynman diagrams with one external radiation mode and arbitrarily potential modes being exchanged between the two particles, as shown in the previous section in eq.\eqref{eq:radiation_emitted}:
\begin{equation}
    \scalebox{0.7}{\begin{tikzpicture} 
        \begin{feynman}
        \vertex (a1) ;
        \vertex[right=1cm of a1,square dot,red ] (a2) {}; 
        \vertex[right=1cm of a2] (a3); 
        \vertex[above=1cm of a2] (b1); 
        \diagram* { 
        (a1) -- [double, thick] (a3),
        (a2) -- [gluon] (b1)
        };
        \end{feynman} 
        \end{tikzpicture}\ } \quad = \quad   \begin{tikzpicture}[baseline=(f)]
            \begin{feynman}
            \vertex (a);
            \vertex[right=0.5cm of a] (a2);
            \vertex[right=0.5cm of a2] (a3);
            \vertex[right=0.25cm of a3] (b1);
            \vertex[above=0.75cm of b1] (c1);
            \vertex[right=0.5cm of a3] (a4);
            \vertex[right=0.5cm of a4] (a5);
            \vertex[right=0.5cm of a5] (a6);
            \vertex[below=0.4cm of a] (f);
            \vertex[below=0.8cm of a] (e);
            \vertex[right=0.5cm of e] (e2);
            \vertex[right=0.5cm of e2] (e3);
            \vertex[right=0.5cm of e3] (e4);
            \vertex[right=0.25cm of e3] (b2);
            \vertex[right=0.5cm of e4] (e5);
            \vertex[right=0.5cm of e5] (e6);
            \diagram* {
            (a) -- [thick] (a6),
            (e)-- [thick] (e6),
            (b1)-- [gluon] (c1),
          };
        \end{feynman}
        \fill[gray] (a2) rectangle (e5);
        \end{tikzpicture} 
\end{equation}
where we treat the coupling of the effective action $T^{\mu\nu}$ to the long wavelength modes linearly, while the interactions of short-distances modes can be included at arbitrary order.

\subsection{Integrating out the radiation graviton}
Now that we have built an effective theory for extended objects in terms of the multipole moments, and we have shown how to match those moments with the full theory, we can use this action to integrate out the gravitational radiation to obtain an effective action for the multipoles alone:
\begin{eqnarray}
    e^{iS_{eff}(\{Q_i\})}=\int D\bar{h}e^{iS_{rad}(\bar{h},\{Q_i\})} \ , 
\end{eqnarray}
where $\{Q_i\}$ denotes the collection of multipole moments.
As done for the near zone contributions, the procedure should be done diagrammatically, with the use of Kol-Smolkin variables to simplify the computations. \\
Integrating out radiation gravitons will give rise to contributions both to the conservative and dissipative dynamics, but in the following we will be interested in the computation of contributions to the conservative dynamics only, in particular to the relevant diagrams at 4PN and 5PN. \\
Corrections to the conservative dynamics can be computed in the following way:
\begin{itemize}
    \item Generate all the diagrams at a given PN order,
    \item Evaluate all the diagrams: \begin{equation}
        \mathcal{M}=\sum_i\mathcal{M}_i
    \end{equation}
    \item The corresponding contributions to the effective action is given by: \begin{equation}
        S_{eff}=-i\lim_{d\to 3}\mathcal{M }
    \end{equation}
\end{itemize}

\subsection{Diagram generation}
\label{sec:diagram_generation_far_zone}
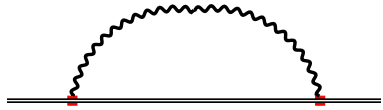
\begin{figure}[H]
\centering
    \scalebox{0.8}{\begin{tikzpicture} 
        \begin{feynman}
        \vertex (a1) ;
        \vertex[right=1cm of a1,square dot, red ] (a2) {}; 
        \vertex[right=2cm of a2] (a3); 
        \vertex[above=1.5cm of a3] (b3);
        \vertex[right=2cm of a3,square dot, red ] (a4) {};
        \vertex[right=1cm of a4] (a5); 
        \diagram* { 
        (a1) -- [double,thick] (a5),
        (a2) -- [boson, ultra thick, quarter left] (b3),
        (b3) -- [boson, ultra thick, quarter left] (a4)
        };
        \end{feynman} 
        \end{tikzpicture}}
    \caption{Example of effective diagram in Long-wavelength EFT. The horizontal double line denotes the source term while curly lines denote radiation gravitons. This is a 1-loop diagram giving classical contributions}
\end{figure}
As for the case of near-zone contributions, we want to generate effective diagrams, which scale with definite powers of $G_N$ and $v^2$ in order to know exactly at which PN order they are contributing.\\
Here the counting in powers of velocity is much simpler than in the case of potential gravitons and can be easily made without the use of Kol-Smolkin variables.
Notice that in this case we do not expand propagators in the non-relativistic limit, since for radiated gravitons $k_0\sim\mathbf{k}\sim v/r$. 
The diagrams are built in the following way: 
\begin{itemize}
    \item The non-propagating source term is depicted with a single horizontal double line, while radiated gravitons are denoted by curly lines.
    \item Coupling of radiation gravitons to the source term are identified by a specific multipole moments, each one with a specific scaling in velocity powers. 
    \item Only connected diagrams contribute to the effective action.
    \item We do not want to consider quantum contributions, loop diagrams need to contain at least an external leg of the source term, which is not propagating.
    \item Once a proper effective diagram has been build, one has to integrate over momentum variables.
\end{itemize}
In order to understand at which order a given diagram is contributing we use the following velocity counting rules (\cite{Blumlein:2021txe}):
\begin{itemize}
    \item each graviton propagator scales with $v^{-2}$,
    \item each momentum integral yields a factor of $v^3$,
    \item each post-Newtonian correction more adds a factor of $v^2$,
    \item A triple graviton vertex is $\sim v^3$,
    \item Graviton coupling to: $E \sim v,\ L_k \sim v^2,\ Q_{ij},J_{ij}\sim v^3$ and $O_{ijk}\sim v^4$
    \item multipole moments $L_k,J_{ij}\sim v$
    \item double graviton vertex to $Q_{ij}\sim v^4$
\end{itemize}

\subsection{Diagram Evaluation: from effective Feynman diagrams to QFT amplitudes}
In order to properly evaluate these effective diagrams we have to convert them in a QFT amplitudes language.
In general, within the EFT approach, since the source is static and do not propagate, any amplitude can be mapped into a multi-loop 1-point functions with massive internal lines, so with no external momenta.
This case, since we are dealing with radiation gravitons, we cannot expand the propagators in the non-relativistic limit, and we treat energy components as mass terms:
\begin{equation}
    \scalebox{0.7}{\begin{tikzpicture} [baseline=(a1)]
\begin{feynman}
\vertex (a1) ;
\vertex[right=1cm of a1] (a2); 
\vertex[right=2cm of a2] (a3); 
\vertex[above=1.5cm of a3] (b3);
\vertex[right=2cm of a3] (a4);
\vertex[right=1cm of a4] (a5); 
\diagram* { 
(a1) -- [double, thick] (a5)
};
\end{feynman} 
\draw[gray,fill=gray] (a2) -- (a4) arc(0:180:2) --cycle;
\end{tikzpicture}} \qquad \Leftrightarrow \qquad
\begin{tikzpicture}[baseline=(b)]
\begin{feynman}
        \vertex (a);
        \vertex[below =1 cm of a] (b);
        \vertex[left =1 cm of b] (c);
        \vertex[right =1  cm of b] (d);
        \diagram* {
        (c) -- [double, thick] (d),
      };
    \end{feynman}
   \draw[gray,fill=gray] (a) circle (1cm); 
\end{tikzpicture}
\label{eq:radiation_multi_loop}
\end{equation}
Once this identification has been made, we can evaluate amplitudes using multi-loop techniques, that will be described in the next chapter.

\chapter{Multi-loop techniques in Quantum Field Theory}
\label{Ch:multiloop}

In the previous chapter we saw that within the Effective Field Theory of General Relativity, near-zone computations involve multi-loop two-point functions whereas far-zone computations require the evaluation of multi-loop one-point functions. 
These Feynman integrals are difficult to compute in general, and we need to develop a systematic procedure to make efficient computations.
The evaluation of each Feynman Integral in general can be very time-consuming, therefore each identity which relate different integrals, can greatly simplify the problem of their determination and consequently the calculation of loop corrections as a whole. \\
The general procedure that one adopts is to use the several relations among the different integrals to decompose them in a minimal set of integrals, known as Master Integrals (MI), which constitute a basis. To do that we will follow a three-step procedure, which will be described in the following sections, which are:
\begin{enumerate}
    \item Tensor decomposition: in which we decompose the tensorial structure of the amplitude in terms of a linear combination of independent Lorentz vectors and Tensor, multiplied by some coefficients known as \textit{form factors}. 
    \item Reduction to Master Integrals (MI): in which we will use the powerful identities relating the different Feynman integrals, such as the \textit{integration-by-parts} (IBP) identities \cite{Chetyrkin:1981qh,Tkachov:1981wb}, to decompose the form factors in a basis of integrals, which are the Master Integrals.
    \item Master Integral evaluation: in which we will evaluate the Master Integrals appearing. This procedure can be done directly or using other methods such as \textit{difference} \cite{Laporta:2000dsw,Laporta:2000dc} or \textit{differential equations} (DEs) \cite{Kotikov:1990kg, Henn:2013pwa} (for reviews see \cite{Argeri:2007up,Argeri:2014qva}). 
\end{enumerate}
\subsubsection{Chapter organization}
The chapter will be divided as follows: 
\begin{itemize}
    \item We describe how to tensor decompose a generic amplitude in terms of Lorentz tensor multiplied by form factors, which can be obtained by means of projectors. We will then show how to compute the projectors that will be necessary for future calculations. 
    \item Then, we describe a general classification of multi-loop Feynman integrals, and how they are organized in topologies. 
    \item Furthermore, we will describe the symmetry relations among the various Feynman integrals, and how they can be used to decompose them in a basis of Master Integrals. 
    \item Eventually, we will directly compute some master integrals and scalar integrals that will be used in the following chapters. 
\end{itemize}
\section{Tensor Decomposition}
\label{sec:tensor_dec}
The first step to take when one is dealing with a generic Feynman diagrams is to separate Lorentz and Dirac structures from the integrals, and an efficient way to do that is by using the so-called \textit{tensor decomposition}. Let us consider a generic amplitude $\mathcal{M}$, where we expose all external polarization vectors as:
\begin{equation}
    \mathcal{M}=\epsilon_1^{\mu_1}\cdots\epsilon_k^{\mu_k}\mathcal{M}_{\mu_1 \cdots \mu_k} \ .
\end{equation}
We can write an ansatz for the tensor $\mathcal{M}_{\mu_1\cdots\mu_k}$ in terms of independent Lorentz vectors and tensors, and of the possible Dirac structures:
\begin{eqnarray}
    \mathcal{M}_{\mu_1 \cdots \mu_k}=\sum_i f_i T_{\mu_1 \cdots \mu_k, i}
\end{eqnarray}
where $f_i$ are called \textit{form factors}. 
The form factors $f_i$ can be computed by \textit{projections}, namely they can be extracted from the diagrams by the action of suitably chosen projectors $P_i$,
\begin{equation}
    f_i= P_i \mathcal{M} \ . 
\end{equation}
Through the tensor decomposition, the numerators of the Feynman integrals appearing in the form factors $f_i$ may include only scalar products build from the external momenta and the loop momenta. \\
An efficient way to find the projectors $P_i$ is to use an ansatz of the same form of the one used for the tensor $\mathcal{M}_{\mu_1\cdots \mu_k}$:
\begin{equation}
    P_i= \sum_jC_{ij}T_{j} \ , 
\end{equation}
and to fix the coefficients by solving the following system of equations:
\begin{eqnarray}
    \begin{cases}
        P_i \cdot T_j = 0 \qquad if i\neq j \\
        P_i T_i = 1 
      \end{cases}\,.
\end{eqnarray}
\subsection{Example: One-loop vacuum diagrams}
Let us discuss an example which will be useful in the evaluations of PN corrections, which is the tensor decomposition of one-loop vacuum diagrams of the form:
\begin{equation}
    \int \frac{d^d\mathbf{k}}{(2\pi)^d}\frac{k^ik^j;k^i k^j k^k k^l;k^i k^j k^k k^l k^m k^n}{\mathbf{k}^2-k_0^2}
\end{equation}
Notice that in this case we are integrating over the $d$ spatial dimensions, whereas the energy component is treated as an effective mass term.
Since there are no external momenta involved in this process, these integrals can be tensor decomposed only as linear combinations of products of metric tensors. In the next chapter we  will deal with tensor integrals of this type with at most 6 free indices, and so we will describe the tensor decomposition needed for these cases.
\subsubsection{2-indices tensor decomposition}
Let us consider a 1-loop integral with 2 free indices:
\begin{eqnarray}
I_2^{ij}=\int \frac{d^d\mathbf{k}}{(2\pi)^d}\frac{k^ik^j}{\mathbf{k}^2-k_0^2}
\end{eqnarray}
It has only two free indices, and so it must be proportional to the spatial metric $\delta^{ij}$ as:
\begin{eqnarray}
I_2^{ij}=A \delta^{ij}
\label{eq:I2}
\end{eqnarray}
In order to get the form factor $A$ we can contract the integral $I_2^{ij}$ with an appropriate projector $P_A^{ij}$ such that:
\begin{eqnarray}
A=I_2^{ij}P_{A\ ij}
\end{eqnarray}
We can write an ansatz for $P_A$ and write it in terms of the Lorentz tensor appearing on RHS of eq.\eqref{eq:I2} as:
\begin{eqnarray}
P_A^{ij}=C_A\delta^{ij}
\end{eqnarray}
and we can fix the coefficient $C_A$ by solving the following equation:
\begin{eqnarray}
P_{A}^{ij}\delta_{ij}=1 , 
\end{eqnarray}
which is solved for $C_A=\frac{1}{d}$, and so the projector is:
\begin{eqnarray}
P_A^{ij}=\frac{1}{d}\delta^{ij} , 
\end{eqnarray}
and the form factor can be obtained as:
\begin{eqnarray}
A=I_2^{ij}\frac{\delta_{ij}}{d} \ . 
\end{eqnarray}
This simple prescription can be generalized for tensors involving more indices. 
\subsubsection{4-indices tensor decomposition}
Consider the following 1-loop integral with 4 free Lorenz indices:
\begin{eqnarray}
I_4^{ijkl}=\int \frac{d^d\mathbf{k}}{(2\pi)^d}\frac{k^i k^j k^k k^l}{\mathbf{k}^2-k_0^2}
\end{eqnarray}
It can be written in terms of form factors as:
\begin{eqnarray}
I_4^{ijkl}=A\delta^{ij}\delta^{kl}+B\delta^{ik}\delta^{jl}+C\delta^{il}\delta^{jk}
\end{eqnarray}
As in the previous case, we want to obtain form factors by contracting $I_4^{ijkl}$ with appropriate projectors that have the following ansatz:
\begin{eqnarray}
P^{ijkl}_{a}=C_{1\ a }\delta^{ij}\delta^{kl}+C_{2\ a }\delta^{ik}\delta^{jl}+C_{3\ a }\delta^{il}\delta^{jk}\qquad with \ a=A,B,C
\end{eqnarray}
where the coefficients can be fixed by solving the following system of equations:
\begin{eqnarray}
P_a^{ijkl}g_{V\ ijkl}= V_a 
\end{eqnarray}
where $g_{V\ ijkl}=(\delta^{ij}\delta^{kl},\delta^{ik}\delta^{jl},\delta^{il}\delta^{jk} )^T$ and $V_a$ is a unit vector with values 1 in the $a$ position and $0$ in the rest. 
\subsubsection{6-indices tensor decomposition}
Consider the following 1-loop integral with 6 free Lorenz indices:
\begin{eqnarray}
I_6^{ijklmn}=\int \frac{d^d\mathbf{k}}{(2\pi)^d}\frac{k^ik^jk^kk^lk^mk^n}{\mathbf{k}^2-k_0^2}
\end{eqnarray}
It can be written in terms of 15 form factors as:
\begin{eqnarray}
I_6^{ijklmn}=\sum_a F_a g_{F a}^{ijklmn}
\end{eqnarray}
where $g_F^{ijklmn}$ is a 15 components vector defined as:
\begin{eqnarray}
g_F^{ijklmn}=&(& 
\delta^{ij}\delta^{kl}\delta^{mn},
\delta^{ij}\delta^{km}\delta^{ln},
\delta^{ij}\delta^{kn}\delta^{ml},
\delta^{ik}\delta^{jl}\delta^{mn},
\delta^{ik}\delta^{jm}\delta^{ln},
\delta^{ik}\delta^{jn}\delta^{ml},
\delta^{il}\delta^{kj}\delta^{mn},
\delta^{il}\delta^{mj}\delta^{kn},\nonumber\\
& &
\delta^{il}\delta^{nj}\delta^{mk},
\delta^{im}\delta^{kl}\delta^{jn},
\delta^{im}\delta^{kn}\delta^{jl},
\delta^{im}\delta^{nl}\delta^{jm},
\delta^{in}\delta^{kl}\delta^{mj},
\delta^{in}\delta^{km}\delta^{lj},
\delta^{in}\delta^{ml}\delta^{kj},
)^T
\end{eqnarray}
As in the previous case we can obtain form factors by contracting $I_6^{ijklmn}$ with appropriate projectors that have the following ansatz:
\begin{eqnarray}
P^{ijklmn}_{a}=\sum_b C_{a b}g_{F b\ ijklmn}\qquad with \ a=1\ldots15
\end{eqnarray}
where the coefficients can be fixed by solving the following system of equations:
\begin{eqnarray}
P_a^{ijklmn}g_{F\ ijklmn}= V_a 
\end{eqnarray}
where  and $V_a$ is a unit vector with 15 components with values 1 in the $a$ position and $0$ in the rest. 
After performing appropriately the tensor decomposition, we can express our integrals as a linear combination of form factors multiplying products of metric tensors. \\ 

\section{Classification of Feynman integrals and reduction to scalar integrals}
Now that we dealt with tensor structure, let us classify a generic Feynman integrals. \\ 
A typical $L$-loop Feynman integral, in $d$ space-time dimensions, which appear in a form factor, is given by:
\begin{eqnarray}
    \mathcal{M}=\int \prod_{i=1}^L\biggl[\frac{d^dk_i}{(2\pi)^d}\biggr]\frac{\mathcal{N}(k_i,p_j)}{D_1^{\alpha_1}\cdots D_m^{\alpha_m}} \ , 
\end{eqnarray}
where $k_i$ are the loop momenta, $p_i$ are external momenta and $m$ is the number of denominators corresponding to internal lines of the diagrams. \\ 
Any diagram with $n$ external legs, because of momentum conservation depends only on $n-1$ momenta.  The number $N$ of possible scalar products between loop momenta, $k_i\cdot k_j$, and loop momenta and external momenta, $k_i \cdot p_j$, can be computed as:
\begin{eqnarray}
    N=L(n-1)+\frac{L(L+1)}{2} = L\left(n+\frac{L+1}{2}\right) \ , 
\end{eqnarray}
where $L(n-1)$ is the number of scalar products between the $L$ loop momenta and the $n-1$ independent external momenta, whereas $L(L+1)/2$is the number of scalar products involving loop momenta only.
For $L> 1$ the number of scalar products can exceed the number of denominators of the diagram. Therefore, there are some scalar products which can be written in terms of denominators, the so-called \textit{reducible scalar products} (RSPs), and other which cannot be written in terms of denominators, dubbed \textit{irreducible scalar products} (ISPs).\\ 
We can easily deal with ISPs by artificially enlarge the set of denominators appearing, introducing \textit{auxiliary denominators}, such that every scalar product can be written in terms of denominators. \\ 
Consequently, each scalar product in the numerator can be completely written in terms of the enlarged set of propagators, such that a generic Feynman integral can be written as a sum of scalar integrals:
\begin{eqnarray}
    \mathcal{M}& = & \sum_i I_i  \ , \\
    I_i &=& \int \prod_{i=1}^L\biggl[\frac{d^d k_i}{(2\pi)^d}\biggr]\mathcal{I}_i= \int \prod_{i=1}^L\biggl[\frac{d^d k_i}{(2\pi)^d}\biggr]\frac{1}{D_1^{\alpha_{i1}}\cdots D_m^{\alpha_{in}}} \ .
\end{eqnarray}
Before proceeding further let us fix some definitions that will be useful in the following:
\begin{itemize}
\item Integral family: the set of denominators spanning the complete space of scalar products,
\item topology: a subset of the integral family, where all powers of the denominators are positive, and which identifies a graph with momentum conservation at each vertex,
\item sub-topology: a subset of the topology. 
\end{itemize}

\section{Decomposition of a Feynman diagram in Master Integrals}
Now that we have written a generic form factor in terms of scalar integrals, we want to express all integrals appearing in terms of a smaller set of independent integrals, known as \textit{master integrals}. The master integrals $\mathbf{I}^{MI}$ are a basis of integrals in dimensional regularization.\\ 
 An amplitude can be decomposed in terms of master Integrals as:
\begin{eqnarray}
    \mathcal{M}=\sum_i\alpha_i I_i = \sum_i c_i \mathbf{I}_i^{MI}
\end{eqnarray}
The evaluation of a generic multi-loop amplitude then is achieved in two steps:
\begin{itemize}
    \item \textit{reduction}: decomposition in terms of MIs. This amount to determine the coefficients $c_i$ of each MI $\mathbf{I}_i^{MI}$. 
    \item MIs \textit{evaluation}: the actual calculation of the MI $\mathbf{I}_i^{MI}$
\end{itemize}
\subsection{Reduction to Master Integrals}

Most of the Feynman integrals in a topology are related one to each other by relations that are due to \textit{symmetry relations}, \textit{Lorentz invariance identities} and \textit{integration-by-parts} identities. \\ 
These relations are due to general properties of Feynman integrals, such as graph symmetries, Lorentz invariance of scalar functions, and to invariance under loop momenta redefinition of the integrals.
By using these relations we are able to express all integrals within any topology in terms of a basis of master integrals. 

\subsubsection{Symmetry Relations}
The first set of relations between different integrals can be derived from trivial relabeling of the loop momenta (shifts), which leave the value of the integral unchanged since they have a trivial Jacobian. \\ 
Two classes of transformations are particularly useful:
\begin{itemize}
    \item the set of transformations that maps different topologies into each other, which allow us to decrease the number of independent topologies,
    \item the set of transformations that map each topology onto itself, that allow us to derive identities between integrals within the same topology.
\end{itemize}
\subsubsection{Lorentz Invariance Identities (LI)}
Our scalar integrals $I_i$ are Lorentz scalars, therefore they are invariant under all Lorentz transformations. \\ 
In particular, under an infinitesimal shift of the external momenta:
\begin{eqnarray}
p_i^\mu \to p_i^\mu + w^{\mu\nu}p_{i,\nu},
\end{eqnarray}
where $w^{\mu\nu}$ is a totally antisymmetric tensor, our Feynman integral transforms as: 
\begin{eqnarray}
I_i \to \biggl( 1+w^{\mu\nu}\sum_{i=1}^{n-1}p_{i,\mu} \frac{\partial}{\partial_\mu}\biggr)I_i
\end{eqnarray}
From Lorentz invariance of the latter expression one can obtain relations of the form:
\begin{eqnarray}
    \sum_{i=1}^{n-1}\biggl(p_{i,\mu}\frac{\partial}{\partial_i^\nu}-p_{i,\nu}\frac{\partial}{\partial_i^\mu}\biggr)I_i=0,
\end{eqnarray}
which can be contracted with all possible symmetric tensors built from the external momenta, like $p_j^\mu p_k^\nu$.
The differentiation acting at the integrand level will generate integrals with different powers of denominators, which can be used as non-trivial relations among them called \textit{Lorentz invariance identities} (LIs).
\subsubsection{Integration by parts identities (IBP)} 
The most important set of identities that we will use are the integration-by-parts identities. \\ 
A generic Feynman integral is invariant under the redefinition of the loop momenta $k_i$, with $i=1,\ldots, L$, which may also involve any combination of loop and external momenta $p_j$, with $j=1,\cdots ,n-1$: 
\begin{equation}
    k_i^\mu \to A_{ij}k_j^\mu + B_{ij}p_j^\mu.
\end{equation}
By considering the action of infinitesimal transformations:
\begin{eqnarray}
    k_i^\mu \to k_i^\mu + \beta_{ij}q_j^\mu \ , \qquad q_j=(k_1,\cdots, k_L,p_1,\cdots p_{n-1}),
\end{eqnarray}
the integrand transforms as:
\begin{eqnarray}
    \mathcal{I}_i\to \biggl(1+\beta_{ij} q_j \cdot \frac{\partial}{\partial_{q_i}}\biggr)\mathcal{I}_i.
\end{eqnarray}
If $j=i$, then also the integration measure changes as: 
\begin{eqnarray}
    d^dk_i\to (1+\beta_{ii}d)d^dk_i\ ,
\end{eqnarray}
because $\partial k_i^\mu/\partial k_j^\mu = d\delta_{ij}$. We can combine these results in a single equation: 
\begin{eqnarray}
    I_i=\int \prod_{i=1}^L d^d k_i \mathcal{I}_ \qquad \to \qquad \int \prod_{i=1}^L d^d k_i \biggl( 1+ \beta_{ij}\frac{\partial }{\partial k_i^\mu}q_j^\mu\biggr)\mathcal{I}_i \ . 
\end{eqnarray}
The invariance of the integral implies:
\begin{eqnarray}
    \int \prod_{i=1}^L d^d k_i \frac{\partial}{\partial k_i^\mu}\bigl( q_j^\mu \mathcal{I}_i \bigr) = 0 \ , 
\end{eqnarray}
where $q_j^\mu$ is any linear combination of vectors chosen between the external legs or the integration momenta. \\ 
The differentiation acting on the integrand will generate integrals with different powers of denominators, which can be used as non-trivial relations among them, called integration-by-parts identities (IBPs). 
These relations can be solved to write the initial scalar integrals in terms of more simple ones, that are the so-called \textit{Master Integrals}.  
The procedure has been implemented by Laporta \cite{Laporta:2000dsw} developing a method for high precision calculations of multi-loop, and it has been used in several works. 

\subsection{Master Integral evaluation}
The third step of this procedure consists in the evaluation of the Master Integrals. \\ There are different ways to compute the Master Integrals: we can use direct integration (Feynman parameters, Schwinger parameters,$\ldots$), but when it becomes prohibitive one can use \textit{differential equations} or \textit{difference equations}.
\subsubsection{Evaluation of massive tadpole via direct integration}
We can compute the massive tadpole integral via direct integration. 
Consider the following MI in $d$-dimensions:
\begin{eqnarray}
    I_1(d)=  \begin{tikzpicture}[baseline=(current bounding box.center)]             \begin{feynman}
        \vertex (a) ;
        \vertex[right=0.2cm of a] (b);
        \vertex[right=1cm of b] (c); 
        \vertex[right=0.2cm of c] (d);
        \diagram* {
        (b) -- [half left, thick] (c),
        (b) -- [half right, thick] (c),
        };
        \end{feynman} \end{tikzpicture} = \int \frac{d^d p}{(2\pi)^{d}}\frac{1}{p^2+m^2} \ . 
\end{eqnarray}
We can rewrite it, with the following change of variables:  
\begin{eqnarray}
    k^2=\frac{p^2}{m^2} \ , \qquad d^d k = ( m^2)^{\frac{d-2}{2}}m^2 d^dk \ , 
\end{eqnarray}
obtaining
\begin{eqnarray}
    I_1(d)&=& (m^2)^{\frac{d-2}{2}}\int d^d k \frac{1}{k^2+1}= (m^2)^{\frac{d-2}{2}}X(d) \ . 
\end{eqnarray}
Let us rewrite the integration measure appearing in $X(d)$ in spherical coordinates, integrating over the $d-1$ solid angle we obtain:
\begin{eqnarray}
    d^d k=\Omega_{d-1} k^{d-1}dk \ , \qquad \Omega_{d-1}=\frac{2\pi^{d/2}}{\Gamma\left(\frac{d}{2}\right)} \ . 
\end{eqnarray}
Then we perform the change of variables:
\begin{eqnarray}
    t=k^2 \ , \qquad dt=2kdk \ , \qquad \to \qquad k^{d-1}dk = \frac{1}{2}t^{\frac{d-2}{2}}dt \ ,
\end{eqnarray}
obtaining:
\begin{eqnarray}
    X(d) & = & \frac{\Omega_{d-1}}{(2\pi)^d}\frac{1}{2}\int_0^\infty dt \frac{t^{\frac{d-2}{2}}}{t+1} \nonumber \\ 
    & = & \frac{\Omega_{d-1}}{(2\pi)^d}\frac{1}{2} \beta\left(\frac{d}{2},1-\frac{d}{2}\right) \ , 
\end{eqnarray}
where we introduced the Euler's $\beta$ function, defined as:
\begin{eqnarray}
    \beta(x,y)=\int_0^\infty dt \frac{t^{x-1}}{(1+t)^{x+y}} = \frac{\Gamma[x]\Gamma[y]}{\Gamma[x+y]} \ . 
    \label{eq:beta_function}
\end{eqnarray}

Then the integral becomes:
\begin{eqnarray}
    X(d)= \frac{2\pi^{\frac{d}{2}}}{\Gamma(d/2)}\frac{1}{2(2\pi)^d}\frac{\Gamma(d/2)\Gamma(1-d/2)}{\Gamma(1)}=\frac{1}{(4\pi)^{d/2}}\Gamma\left(1-\frac{d}{2}\right) \ . 
\end{eqnarray}

Eventually we find:
\begin{eqnarray}
    I_1(d)=  \begin{tikzpicture}[baseline=(current bounding box.center)]             \begin{feynman}
        \vertex (a) ;
        \vertex[right=0.2cm of a] (b);
        \vertex[right=1cm of b] (c); 
        \vertex[right=0.2cm of c] (d);
        \diagram* {
        (b) -- [half left, thick] (c),
        (b) -- [half right, thick] (c),
        };
        \end{feynman} \end{tikzpicture} = \frac{1}{(4\pi)^{d/2}} (m^2)^{\frac{d-2}{2}}\Gamma\left(1-\frac{d}{2}\right) \ . 
\end{eqnarray}

\subsubsection{Vanishing of the massless tadpole}
Using the calculation done for the massive case we can show in a simple way that the \textit{massless tadpole} is vanishing:
\begin{eqnarray}
    \int\frac{d^d k}{k^2}=0 
\end{eqnarray}
This is somehow obvious in the framework of dimensional regularization, since it is a scaleless integral.  \\
Let us compute the following integral:
\begin{eqnarray}
    \int \frac{d^d k}{k^2(k+1)^2 }& = & \frac{1}{2}\Omega_{d-1}\int_{-\infty}^\infty dt \frac{t^{\frac{d}{2}-2}}{t+1} = \frac{1}{2}\Gamma_{d-1}\beta \left( \frac{d}{2}-1,2-\frac{d}{2}\right) \nonumber \\ 
    & = & \frac{1}{2}\Omega_{d-1}\Gamma\left(\frac{d}{2}-1\right)\Gamma\left(2-\frac{d}{2}\right) = -\frac{1}{2}\Omega_{d-1}\Gamma\left(\frac{d}{2}\right)\Gamma\left(1-\frac{d}{2}\right) \nonumber \\ 
     &=&  - \int \frac{d^d k}{k^2+1}
\end{eqnarray}
but we have also:
\begin{eqnarray}
    \int \frac{d^d k}{k^2(1+k^2)}=\int \frac{d^d k}{k^2}-\int \frac{d^d k}{k^2+1}
\end{eqnarray}
hence we should have:
\begin{eqnarray}
    \int \frac{d^d k}{k^2}= 0 
\end{eqnarray}

\subsection{Direct evaluation of sunrise integral}
We are interested in evaluating the following scalar integral in $d-$dimensions:
\begin{equation}
    I_S(d,a,b)=\int \frac{d^d k }{(2\pi)^d}\frac{1}{k^{2a}(k^2+\Delta)^b}
\end{equation}
Here $a,b$ and $\Delta$ are arbitrary real numbers, for which the integral could converge or not, while the $k$ terms in the integrand stands for the Euclidean norm of $\mathbf{k}$. 
In order to perform the evaluation we split the integration $\mathbb{R}^d$ using spherical coordinates in $d$ dimension: 
\begin{eqnarray}
    I_S(d,a,b) 
    & = & 
    \int \frac{d\Omega_d}{(2\pi)^d}\int_0^{\infty} dk \frac{k^{d-1}}{k^{2a}(k^2+\Delta)^d} \nonumber \\ 
    & = & 
    \frac{2^{1-d}\pi^{-d/2}}{\Gamma[d/2]}\int_0^{+\infty}\ dk \frac{k^{d-1-2a}}{(k^2+\Delta)^b} \label{eq:sunrise} \\ 
    & = & 
    \frac{2^{1-d}\pi^{-d/2}}{\Gamma[d/2]}I_{|k|}(d,a,b) \\ 
    I_{|k|}(d,a,b)
     & = & 
     \int_0^{+\infty}\ dk \frac{k^{d-1-2a}}{(k^2+\Delta)^b} \ . \label{eq:sunrise_1}
\end{eqnarray}
As done for the tadpole case, we can collect the $\Delta$ term in \eqref{eq:sunrise_1}, and perform a rescaling in the variable of integration in order to obtain: 
\begin{equation}
    I_{|k|}(d,a,b)= \int_0^{+\infty}dk\frac{k^{d-1-2a}}{(k^2+\Delta)^b}=\Delta^{d/2-a-b}\int_0^{+\infty}dk\frac{k^{d-1-2a}}{(k^2+1)^b}
    \label{eq:sunrise_k}
\end{equation}
Introducing the change of coordinates $k=\sqrt{x}$, the following identities hold: 
\begin{equation}
    \int_0^{+\infty}dk\frac{k^{d-1-2a}}{(k^2+1)^b}=\frac{1}{2}\int_0^{+\infty}dx \frac{x^{d/2-a-1}}{(x+1)^b}=\frac{1}{2}\beta\biggl(\frac{d}{2}-a,a+b-\frac{d}{2}\biggr)
    \label{eq:sunrise_identity}
\end{equation}
where in the last equivalence we have introduced the $\beta-$function, as defined in Eq.\eqref{eq:beta_function}. Inserting \eqref{eq:sunrise_identity} into \eqref{eq:sunrise_k}, we arrive at the final expression for \eqref{eq:sunrise_1}:
\begin{equation}
    I_{|k|}(d,a,b)=\Delta^{\frac{d}{2}-a-b}\frac{\Gamma[d/2-a]\Gamma[a+b-d/2]}{2\Gamma[b]} \ , 
\end{equation}
and so \eqref{eq:sunrise} becomes: 
\begin{equation}
    I_s(d,a,b)=\frac{\Delta^{d/2-a-b}}{(4\pi)^{d/2}}\frac{\Gamma[d/2-a]\Gamma[a+b-d/2]}{\Gamma[b]\Gamma[d/2]} \ . 
    \label{eq:sunrise_final}
\end{equation}

\section{The scalar integral $I_F(d,a)$}
All our calculations will be made in Fourier space, and in order to give back a contribution to the effective action, we will have to perform a Fourier transform. \\ 
Is then necessary to evaluate a class of scalar integral regarded as a Fourier transform. The simplest, is the following Euclidean $d-$dimensional integral:
\begin{equation}
    I_F(d,a) = \int \ \frac{d^d q}{(2\pi)^d}\frac{e^{iq\cdot r}}{q^{2 a}} \ . 
\end{equation}
Let us assume that $\Vec{r}$ represents a $d-$dimensional $z-$axis, we then decompose the measure of integration as: 
\begin{equation}
    d^d q=dq_zd^{d-1}Q_\perp = dq_z dQ Q^{d-2}d\Omega_{d-1} \ , 
    \label{eq:measure_d_dim}
\end{equation}
where $Q_\perp$ is orthogonal to $q_z$, which is parallel to $\Vec{r}$, while $Q$ stand for its Euclidean norm. \\ 
Using  \eqref{eq:measure_d_dim}, the previous integral becomes: 
\begin{equation}
    I_F(d,a)=\frac{\Omega_{d-1}}{(2\pi)^d}\int_{-\infty}^{+\infty}dq_z\int_0^{+\infty} dQ Q^{d-2} \frac{e^{iq_z r}}{(q_z^2+Q^2)^a} \ . 
\end{equation}
At this point it is useful to introduce the following change of coordinates on the $(q_z,Q)$ upper-half plane: 
\begin{equation}
    \begin{cases}
      q_z = x Q \\ 
      Q=\frac{y}{x r }
      \end{cases}
      \qquad \rightarrow \qquad
      \begin{cases}
        q_z = \frac{y}{r}\\ 
        Q=\frac{y}{x r }
        \end{cases} \ , 
\end{equation}
which is a map form $\mathcal{R}_{q_z}\times \mathcal{R}^+_{Q}$ to $R^2_{x,y}$ with determinant $\frac{y}{(xr)^2}$\ . 
Applying this change to our integral gives: 
\begin{equation}
    I_F(d,a)=\frac{\Omega_{d-1}}{(2\pi)^d}r^{2a-d}\int_{-\infty}^{+\infty} dx \int_{-\infty}^{+\infty}dy \frac{e^{iy}}{(y^2)^{1/2-d/2+a}}\frac{x^{2a-d}}{(1+x^2)^a} \ , 
\end{equation}
which can be reduced to the product of two scalar integrals as: 
\begin{equation}
    I_F(d,a)=\frac{\Omega_{d-1}}{(2\pi)^d}r^{2a-d}\int_{-\infty}^{+\infty}dx \frac{x^{2a-d}}{(1+x^2)^a}\int_{-\infty}^{+\infty}dy \frac{e^{iy}}{(y^2)^{1/2-d/2+a}} \ . 
    \label{eq:if_xy}
\end{equation}
The integral in the $x$ variable can be easily performed by noticing that it is equivalent to a 1-loop sunrise integral (Eq.\eqref{eq:sunrise_final})with $d=\Delta=1$: 
\begin{equation}
    \int_{-\infty}^{+\infty}dx\frac{x^{2a-d}}{(1+x^2)^a}= \frac{\Gamma[a-d/2+1/2]\Gamma[(d-1)/2]}{2\Gamma[a]} \ .
    \label{eq:L_int_x}
\end{equation}
We are left with the integral on the second bracket which comes from the more general one: 
\begin{equation}
    L(\alpha)=\int_{-\infty}^{+\infty}dy\frac{e^{iy}}{y^{\alpha}} \ ,
    \label{eq:L_int}
\end{equation}
for a specific choice of $\alpha$. \\ 
The integral Eq.\eqref{eq:L_int} can be evaluated using integration by parts, assuming $\alpha\in\mathbb{N}$, in order to derive a useful recurrence relation:
\begin{equation}
    L(\alpha+1)=\frac{iL(\alpha)}{\alpha}\qquad \Rightarrow \qquad L(\alpha)=\frac{i^{\alpha-1}}{\Gamma[\alpha]}L(1) \ ,
\end{equation}
where $L(1)$ is given by the principal value of a contour integral in the complex plane. \\ 
To show this, let us first define the line of integration by means of a counterclockwise closed curve avoiding $z=0$ given by $\Omega=\{\Omega_R\cup(\mathbb{R}/[-r,+r])\cup\Omega_{r}\}$ where $\Omega_r=\{z\in \mathbb{C}| z=re^{i\phi},0\le \phi \le \pi\}$, the same for $\Omega_R$. \\ 
Since the integrand of $L(1)$ is a meromorphic function inside $\Omega$, one has:
\begin{equation}
    \oint_\Omega \frac{e^{iz}}{z}=\int_{\Omega_R}\frac{e^{iz}}{z}dz + \int_{\mathbb{R}/[-r,r]}\frac{e^{iz}}{z}dz+\int_{\Omega_r}\frac{e^{iz}}{z}dz =  0 \ . 
\end{equation}
Now, sending $R\to \infty$ the first integral is zero by means of Jordan's lemma, while the second is equal to $L(1)$ given that $r\to 0 $. This means that: 
\begin{equation}
    L(1)= -\lim_{r\to 0 }\int_{\Omega_r}\frac{e^{iz}}{z}dz
    \label{eq:L_int_1}
\end{equation}
Integral \eqref{eq:L_int_1} is easily evaluated using the fact that in the integration region $r$ is fixed, $0\le \phi \le \pi$, and $dz=izd\phi$:
\begin{equation}
    L(1)=-i\lim_{r\to 0 }\int_\pi^0e^{ir(cos(\phi)+isin(\phi)}d\phi =-\int_{\pi}^{0}d\phi = i\pi \ . 
\end{equation}
The result is: 
\begin{equation}
    \int_{-\infty}^{+\infty}dy\frac{e^{iy}}{y^\alpha}=\frac{\pi i^\alpha}{\Gamma[\alpha]} \ , \qquad \alpha\in \mathbb{N} \ . 
    \label{eq:L_int_alpha}
\end{equation}
In order to relate \eqref{eq:L_int_alpha} to 
the integral in Eq.\eqref{eq:if_xy}, we restrict ourselves to even $\alpha$ since the integrand we are looking for depend on $y^2$, not $y$ only. 
With this choice, we can set $\alpha=2n $ with $n=a+\frac{1}{2}-\frac{d}{2}$ into \eqref{eq:L_int_alpha}, which gives via analytic continuation into $a$ and $d$ the following result: 
\begin{equation}
    \int_{-\infty}^{+\infty}dy \frac{e^{iy}}{(y^2)^{a+1/2-d/2}}= \frac{\pi (-1)^{a+1/2-d/2}}{\Gamma[2a+1-d]} \ . 
    \label{eq:L_int_y}
\end{equation}
Inserting  \eqref{eq:L_int_y},\eqref{eq:L_int_x} into \eqref{eq:if_xy}, after few manipulations with the Gamma function involved, one can prove that our scalar integral is equal to: 
\begin{equation}
    I_F(d,a)=\frac{\Gamma[d/2-a]}{(4\pi)^{d/2}\Gamma[a]}\left(\frac{r}{2}\right)^{2a-d} \ . 
    \label{eq:scalar_integral_fourier}
\end{equation}

\chapter{Hereditary Effects in the EFT approach}
\label{chapter:hereditary}
Now that we have shown how to systematically compute multi-loop Feynman integrals, on this chapter we will evaluate far-zone contributions to the conservative dynamics, known as \textbf{Hereditary effects}, up to 5PN order. The hereditary effects, whose importance has been first noticed in \cite{Blanchet:1987wq}, are the influences of the past evolution of a material system on its present gravitational internal dynamics, and with such terms the dynamic of the binary system depend on the full past history of the source. \\ They are due to gravitational waves emitted by the system in the past and subsequently scattered-off the curvature of space-time back into the system. Such radiative terms give unavoidable contributions to the conservative binary dynamics, and it is mandatory to deal appropriately with this kind of processes in order to compute higher order PN corrections. 
They can be classified in 3 different types of terms known as \textit{Back-scattering}, \textit{Tail-Effects}, \textit{Memory Effects}:
\begin{figure}[H]
\centering
\scalebox{0.7}{\begin{tikzpicture}[baseline=(a1)] 
    \begin{feynman}
    \vertex (a1) ;
    \vertex[right=1cm of a1,square dot, red ] (a2) {}; 
    \vertex[right=2cm of a2] (a3); 
    \vertex[above=1.5cm of a3] (b3);
    \vertex[right=2cm of a3,square dot, red ] (a4) {};
    \vertex[right=1cm of a4] (a5); 
    \vertex[below=0.2cm of a3] (e3) {\((a)\)};
    \diagram* { 
    (a1) -- [double, thick] (a5),
    (a2) -- [boson, black!60!green,  ultra thick, quarter left] (b3),
    (b3) -- [boson,black!60!green, ultra thick, quarter left] (a4),
    };
    \end{feynman} 
    \end{tikzpicture}}\qquad
    \scalebox{0.7}{\begin{tikzpicture}[baseline=(a1)] 
\begin{feynman}
\vertex (a1) ; 
\vertex[right=1cm of a1,square dot, red ] (a2) {}; 
\vertex[right=2cm of a2,square dot, red ] (a3) {}; 
\vertex[above=1.5cm of a3] (b3);
\vertex[right=2cm of a3,square dot, red ] (a4) {};
\vertex[right=1cm of a4] (a5); 
\vertex[below=0.4cm of a3] (e3) {\((b)\)};
\diagram* { 
(a1) -- [double, thick] (a5),
(a2) -- [boson, black!60!green, ultra thick, quarter left] (b3),
(b3) -- [boson, black!60!green, ultra thick, quarter left] (a4),
(b3) -- [scalar, blue, ultra thick] (a3)
};
\end{feynman} 
\end{tikzpicture}} \qquad  \scalebox{0.7}{\begin{tikzpicture}[baseline=(a1)] 
\begin{feynman}
\vertex (a1) ;
\vertex[right=1cm of a1,square dot, red ] (a2) {}; 
\vertex[right=2cm of a2,square dot, red ] (a3) {}; 
\vertex[above=1.5cm of a3] (b3);
\vertex[right=2cm of a3,square dot, red ] (a4) {};
\vertex[right=1cm of a4] (a5); 
\vertex[below=0.4cm of a3] (e3) {\((c)\)};
\diagram* { 
(a1) -- [double, thick] (a5),
(a2) -- [boson, black!60!green, ultra thick, quarter left] (b3),
(b3) -- [boson, black!60!green, ultra thick, quarter left] (a4),
(b3) -- [boson, black!60!green, ultra thick] (a3)
};
\end{feynman}  
\end{tikzpicture}}\qquad 
 \scalebox{0.7}{\begin{tikzpicture} 
    \begin{feynman}
    \vertex (a1) ;
    \vertex[right=1cm of a1, square dot, red] (a2) {}; 
    \vertex[right=2cm of a2, square dot, red] (a3) {}; 
    \vertex[above=1.5cm of a3] (b3);
    \vertex[right=2cm of a3,square dot, red] (a4) {};
    \vertex[right=1cm of a4] (a5); 
    \vertex[below=0.4cm of a3] (e3) {\((d)\)};
    \vertex[below=0.2cm of a2] (f3) {};
    \vertex[below=0.2cm of a4] (g3) {};
    \diagram* { 
    (a1) -- [double,thick] (a5),
    (a2) -- [boson, black!60!green, ultra thick,half left] (a3),
    (a3) -- [boson, black!60!green, ultra thick,half left] (a4),
    };
    \end{feynman} 
    \end{tikzpicture}}
\caption{The diagrams appear in the computation of Hereditary Effects. Diagram $(a)$ is known as back-reaction, diagram $(b)$ is a tail term whereas diagrams $(c)$ and $(d)$ are memory effects. In order to distinguish between $(c)$ and $(d)$ we will refer to diagram $(d)$ as double emission. The double thick line denotes the multipole source, whereas curly and dashed lined correspond to gravitons coupled to time-dependent or conserved multipoles, respectively}
\label{fig:hereditary_effects}
\end{figure}
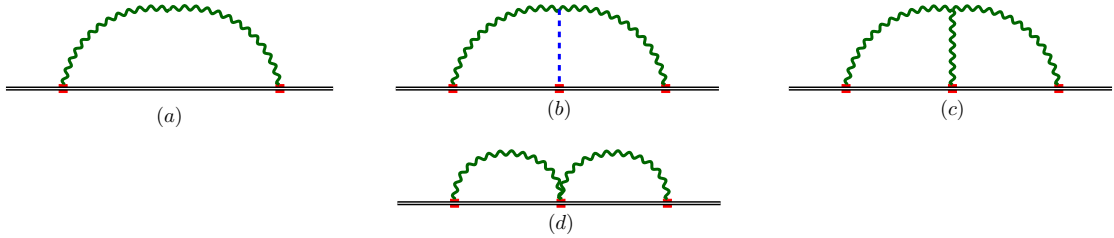
\begin{itemize}
    \item \textbf{Back-scattering} processes: diagram $(a)$ in Fig.\ref{fig:hereditary_effects}, are processes where a gravitational wave is emitted by the binary system and then reabsorbed into the same GW source. They are the lowest-order contributions to the effective action, but as we will show they do not affect the conservative dynamics, but they only contribute to the dissipative one.
    \item \textbf{Tail terms}: diagram $(b)$ in Fig.\ref{fig:hereditary_effects}, are processes where a Gravitational Wave is emitted by the binary system and scattered off the quasi-static curvature into the same GW source. The dotted line denotes a graviton attached to a conserved charge, which imply that the corresponding Green's function has \textit{exactly zero} time component momentum. They contribute to the conservative dynamics starting at 4PN as first noticed in \cite{Blanchet:1992br,Blanchet:1993ec}, and in the context of EFT in \cite{Goldberger:2009qd}.
    \item \textbf{Memory terms}: diagram $(c)$ and $(d)$ in Fig.\ref{fig:hereditary_effects}, are processes where a GW is emitted by the binary system and scattered off the curvature induced by GW themselves. In that case all the gravitons are attached to dynamical multipoles, and we must pay attention in addressing their computation. They contribute to the conservative dynamics starting at 5PN as first noticed in \cite{Zeldovich:1974gvh} in the context of linearized gravity, and then in \cite{Christodoulou:1991cr}. In order to distinguish between the two diagrams, we will refer to diagram $(d)$ as double emission.
\end{itemize}

\subsubsection{State of the art}
The computation of Hereditary Effects in the EFT approach was started in \cite{Galley:2009px}, where it has been computed the LO radiation reaction effect at 2.5PN, and then at 3.5PN in \cite{Galley:2012qs}. Tail effects were computed at 4PN in \cite{Foffa:2011np,Foffa:2013qca,Galley:2015kus}, at 5PN in \cite{Foffa:2019eeb,Blumlein:2021txe}, and then extended to $d-$dimensional multipoles following \cite{Henry:2021cek}, and for all multipoles, in \cite{Almeida:2021xwn}. Tail of tail effects have been computed up to 7PN order in \cite{Edison:2022cdu}. Memory effects instead have been computed at 5PN order in  \cite{Foffa:2019eeb,Blumlein:2021txe,Foffa:2021pkg}. Conservative contributions coming from hereditary effects at 4PN order, and in particular the cancellation of IR divergences arising in intermediate stages have been studied in \cite{Galley:2015kus,Porto:2017dgs,Porto:2017shd,Foffa:2019rdf,Foffa:2019yfl}.  \\ 
The EFT approach is not the only way of computing hereditary effects. An analogous derivation has been performed from the "Tutti Frutti" approach \cite{Bini:2019nra,Bini:2020hmy}, and presented in \cite{Bini:2021gat}, and using PM techniques \cite{Bern:2021yeh,Kalin:2022hph,Jakobsen:2022psy}. 
However, at 5PN order, the different results proposed \cite{Foffa:2021pkg,Blumlein:2021txe,Bini:2021gat,Bern:2021yeh} do not agree in the prediction of physical observables like the two-body scattering angle. The 5PN dynamics is not fully understood at the moment, and there are several subtle puzzles that needs to be studied and understood before proceeding in the evaluation of higher order PN corrections. 

\subsubsection{Goals of the chapter}
The goal of this chapter is to perform an independent derivation of the hereditary processes up to 5PN order, to verify the results proposed in \cite{Foffa:2019eeb}, and later explained in \cite{Foffa:2021pkg} and \cite{Almeida:2021xwn}, and the ones computed in \cite{Blumlein:2021txe}. We want to shed new light in the computation of these processes, by constructing a new computational algorithm in \texttt{Mathematica} which starting from the effective action in the far zone introduced in Eq.\eqref{eq:multipole_action}, computes the Feynman rules, generate the diagrams and evaluates them using modern multi-loop techniques, and which can be generalized at arbitrary PN order. 
From the multipole action \eqref{eq:multipole_action} we can generate a wide variety of triple multipole diagrams following the power counting rules described in the previous chapter \ref{sec:diagram_generation_far_zone}. \\ 
The lowest contributing terms are those containing two electric quadrupole moments, because $E$ and $L_k$ are conserved in the stationary case we are considering.
We will consider the following diagrams:
\begin{itemize}
    \item \textbf{QQ}, which is of order $\mathcal{O}(v^{7})$, and contributes at lowest order at 2.5 PN,
    \item \textbf{QEQ}, which is of order $\mathcal{O}(v^{10})$ and contributes at lowest order at 4PN,
    \item \textbf{QLQ},\textbf{JEJ}, \textbf{OEO} and  \textbf{QQQ} which are of order  $\mathcal{O}(v^{12})$ and contributes at lowest order at 5PN. 
\end{itemize}
\section{The method}
The complete evaluation of the Hereditary terms reported in Fig.\ref{fig:hereditary_effects} has been performed using \texttt{Mathematica}, following a systematic procedure that we will now present.
The generic contribution to the effective action is of the form: 
\begin{equation}
    S_{eff} = -i \lim_{d\to 3}A_{her}\ , 
\end{equation}
\noindent where $A_{her}$ is the generic amplitude for a hereditary process: 
\begin{eqnarray}
 A_{her}&=& \scalebox{0.4}{\begin{tikzpicture} [baseline=(a1)]
\begin{feynman}
\vertex (a1) ;
\vertex[right=1cm of a1] (a2); 
\vertex[right=2cm of a2] (a3); 
\vertex[above=1.5cm of a3] (b3);
\vertex[right=2cm of a3] (a4);
\vertex[right=1cm of a4] (a5); 
\diagram* { 
(a1) -- [double, thick] (a5)
};
\end{feynman} 
\draw[gray,fill=gray] (a2) -- (a4) arc(0:180:2) --cycle;
\end{tikzpicture}} = \int \prod_{i=1}^{n-1}\biggl[\frac{dk_{i0}}{(2\pi)}\biggr]\Biggl( \scalebox{0.4}{\begin{tikzpicture}[baseline=(b)]
    \begin{feynman}
            \vertex (a);
            \vertex[below =1 cm of a] (b);
            \vertex[left =1 cm of b] (c);
            \vertex[right =1  cm of b] (d);
            \diagram* {
            (c) -- [double, thick] (d),
          };
        \end{feynman}
       \draw[gray,fill=gray] (a) circle (1cm); 
    \end{tikzpicture}}\Biggr)\ . 
\end{eqnarray}
 In the following we will focus only on the amplitude in momentum space: 
 \begin{eqnarray}
\mathcal{M}_{her}^{Q^n}\ = \ \scalebox{0.4}{\begin{tikzpicture}[baseline=(b)]
    \begin{feynman}
            \vertex (a);
            \vertex[below =1 cm of a] (b);
            \vertex[left =1 cm of b] (c);
            \vertex[right =1  cm of b] (d);
            \diagram* {
            (c) -- [double, thick] (d),
          };
        \end{feynman}
       \draw[gray,fill=gray] (a) circle (1cm); 
    \end{tikzpicture}} \ ,
\end{eqnarray}
which can be computed using the  Feynman diagrammatic techniques introduced in the previous chapter. 
The procedure can be divided in different steps which are summarized in figure Fig.\ref{fig:flowalgoritm_hereditary}:
\begin{figure}[!ht]
\centering
\tikzstyle{decision} = [diamond, draw, fill=blue!20,
   , text badly centered, inner sep=0pt]
\tikzstyle{block} = [rectangle, draw, fill=blue!20,
    text centered, rounded corners, minimum height=4em]
\tikzstyle{block2} = [rectangle, draw, fill=blue!20,
    text centered, rounded corners, minimum height=4em]
\tikzstyle{inoroutput} = [trapezium, draw, fill=blue!20,
    text centered, minimum height=4em,trapezium left angle=60, trapezium right angle=120]
\tikzstyle{inoroutput} = [trapezium, draw, fill=blue!20,
   text centered, minimum height=4em,trapezium left angle=60, trapezium right angle=120]
\tikzstyle{line} = [draw, -latex']
\tikzstyle{cloud} = [draw, ellipse,fill=red!20, node distance=3cm,
    minimum height=2em]
\begin{tikzpicture}[auto]
\node at (0,0) [block, text width=20em] (start) {Integrand generation};
\node [below=0.8cm of start,block, text width=20em] (tcontr) {Tensor Contractions via \texttt{xAct}};
\node [below=0.8cm of tcontr,block2, text width=20em] (tdec) {Tensor Decomposition};
\node [below=0.8cm of tdec, block, text width=20em] (intred) {Integrand reduction
};

\node [below=0.8cm of intred, block, text width=20em](ibpdec){IBP decomposition \texttt{Litered}
\begin{equation*}
\mathcal{M}_i=\sum_j c_{ij} I_{j}^{M.I.}
\end{equation*}};
\node [below=0.8cm of ibpdec, block, text width=20em](mi){Master Integrals Evaluation};
\node [below =0.8cm of mi, block,text width=20em](results){Results
\begin{equation*}
\mathcal{M}= \sum_i \mathcal{M}_i = \sum_j c_j I^{M.I.}_j
\end{equation*}
};
\path [line] (start) -- (tcontr);
\path [line] (tcontr) -- (tdec);
\path [line] (tdec) -- (intred);
\path [line] (intred) -- (ibpdec);
\path [line] (ibpdec) -- (mi);
\path [line] (mi) -- (results);
\end{tikzpicture}
\captionof{figure}{Flow chart of the algorithm for the evaluation of Hereditary Effects}\label{fig:flowalgoritm_hereditary}
\end{figure}
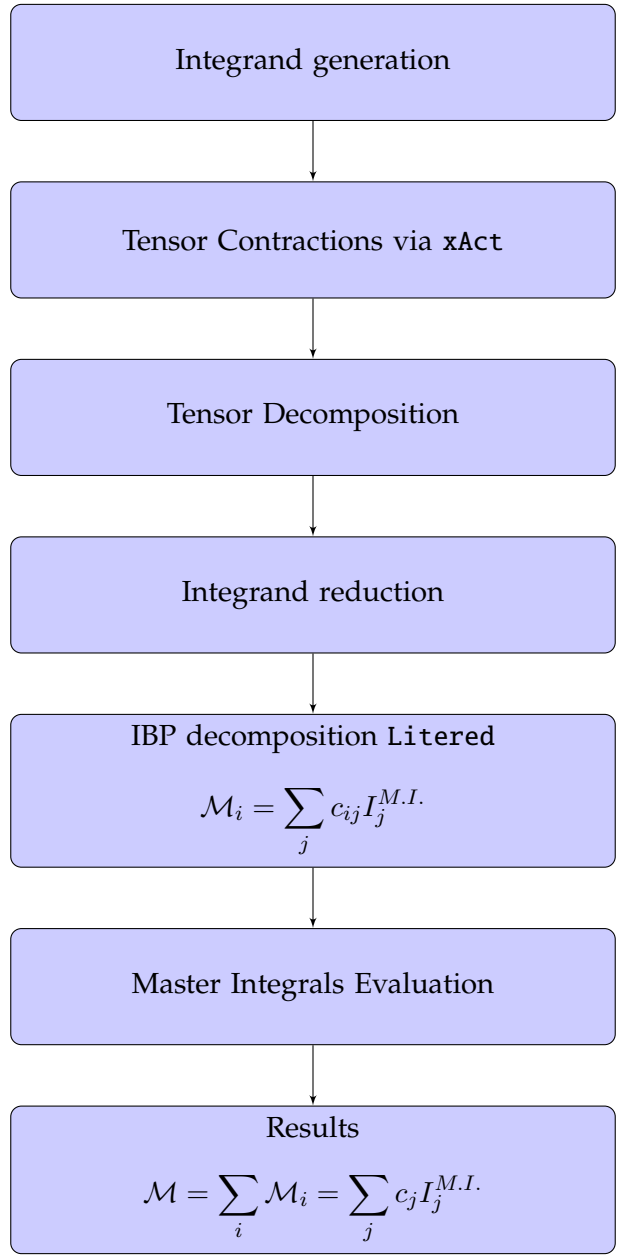
\subsection{Integrand generation}
The diagrams appearing in figure \ref{fig:hereditary_effects} are only \textbf{skeleton diagrams}, and propagating lines denote generic gravitational fields without specifying the polarization $\phi,A,\sigma$. In particular, green curly lines denote couplings to time-dependent multipoles whereas blue dashed lines denote gravitons coupled to conserved quantities such as the energy and the angular momentum. If we want to compute them we need to \textbf{specify the polarization} of the fields, and by doing so we obtain a linear combination of several diagrams with appropriate symmetry factors.
We need then to substitute the explicit expressions of the \textbf{Feynman Rules}, which are reported in Appendix.\ref{chapter:feynman_rules_pn}, for propagators, triple graviton vertices, and multipoles couplings. We can then perform tensor contractions using the package \texttt{xAct}\cite{xactref}. Following the key observation made in the previous chapter in eq.\eqref{eq:radiation_multi_loop}, since the source is static and do not propagate, any amplitude can be mapped into a \textbf{multi-loop 1 point function} with massive internal lines, where the masses are the energy components of the propagators: $D(k)=\frac{1}{\mathbf{k}^2-k_0^2}$. Once this identification has been made, we can evaluate amplitudes using multi-loop techniques. \\ 
$\mathcal{M}_{her}^{Q_n}$ turns out to have the following structure: 
\begin{eqnarray}
\mathcal{M}_{her}^{Q_n}& = &  
\int \prod_{i=1}^{n-1}\biggl[\frac{d^d\mathbf{k_i}}{(2\pi)^d}\biggr]Q^{(1)}_{J_1}(k_{10})\cdots Q^{(n)}_{J_n}(k_{n0})\frac{N^{J_1\cdots J_n }(\mathbf{k}_1,\cdots\mathbf{k}_n,k_{10},\cdots k_{n0})}{D_1D_2\ldots D_n}
\label{eq:hereditary}
\end{eqnarray}
where: 
\begin{itemize}
    \item $J_i=j_{i1},...j_{il_i}$
    \item $Q^{(i)}_{J_1}(k_{i0})$ denotes the Fourier-transform of a generic multipole moment,
    \item $\mathbf{k_i}$ are the loop momenta, $k_{i0}$ are the energy components.
    \item $N^{J_1,\cdots J_N}$ is a generic tensorial numerator,
    \item $D_i= (\mathbf{k_i}^2-k_{i0}^2)$, where $k_n=-\sum_{i=1}^{n-1}k_i$, are the denominators coming from the propagators.
\end{itemize}
Since we are dealing with radiative gravitons and not with potential ones, the scalings of momentum components are: $\left( k_0,\vert \mathbf{k}\vert  \right)\approx \left(\frac{v}{r},\frac{v}{r}\right)$. We cannot take anymore the non-relativistic expansion of the propagators, and so we treat the energy components of loop momenta as effective masses, and we integrate over the spatial components of the momenta.\\
\subsection{Tensor decomposition}
Since we are dealing with tensorial quantities we need to appropriately decompose tensor integrals in terms of form factors as explained in Sec. \ref{sec:tensor_dec} . Moreover, since multipoles are symmetric trace free tensors and there are no external momenta, the tensor decomposition will be hugely simplified and in the end all multipoles will be contracted among each others, obtaining an expression of the form: 
\begin{equation}
    \mathcal{M}_{her}^{Q^n}= Tr[Q_{1}\cdots Q_{n}] \tilde{\mathcal{M}}_{her}^{Q^n}
\end{equation}
\subsection{Integrand reduction}
Once we obtain a form factor of the form  $\tilde{\mathcal{M}}_{her}^{Q^n}$, we need to rewrite it in terms of scalar integrals. In order to do that, we need to rewrite the scalar products appearing in the numerator of our integrand in terms of denominators, and then expand the expression to get a sum of scalar integrals: 
\begin{equation}
    \tilde{\mathcal{M}}_{her}^{Q^n}=\sum_i b_i I_i
\end{equation}
\subsection{IBP Decomposition}
In order to get the result we need to decompose the form factor written in scalar integrals $ \tilde{\mathcal{M}}_{her}^{Q^n}$ in terms of Master Integrals, using the package \texttt{Litered}\cite{Lee:2013mka}, as:
\begin{eqnarray}
    \tilde{\mathcal{M}}_{her}^{Q^n} \ =\ \sum_i C_iJ_i^{MI}
\end{eqnarray}
\subsection{Results}
We can then substitute the explicit expression for the MIs, and expand around $d=3$ to obtain a correction to the effective action: 
\begin{eqnarray}
    S_{eff}=-i\lim_{d\to 3}A_{her}
\end{eqnarray}
\section{Back-scattering at leading order}
\label{Sec:back_scattering}
In this section we will evaluate the 1-loop back-scattering process at leading order. \\ 
As we will see this diagram will not contribute to the conservative dynamics, but only to the dissipative one. This diagram contains a single radiation graviton. Hence, as discussed in Sec.\ref{sec:boundary_conditions_far} and in \cite{Foffa:2021pkg}, using Feynman propagator we will get the correct answer for the contribution to the conservative dynamics, and we will proceed by doing so. It contains two electric quadrupole moments $Q$:
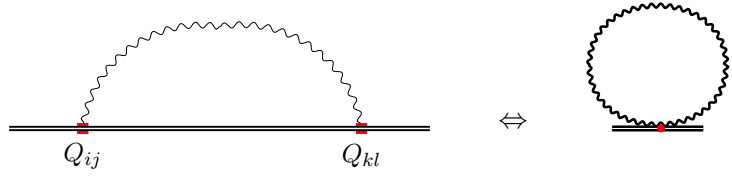
\begin{figure}[H]
    \centering
     \scalebox{0.9}{\begin{tikzpicture} [baseline=(a1)] 
    \begin{feynman}
    \vertex (a1) ;
    \vertex[right=1cm of a1,square dot, red ] (a2) {}; 
    \vertex[right=2cm of a2] (a3); 
    \vertex[above=1.5cm of a3] (b3);
    \vertex[right=2cm of a3,square dot, red ] (a4) {};
    \vertex[right=1cm of a4] (a5); 
    \vertex[below=0.2cm of a3] (e3);
    \vertex[below=0.4cm of a2] (f3) {\(Q_{ij}\)};
    \vertex[below=0.4cm of a4] (g3) {\(Q_{kl}\)};
    \diagram* { 
    (a1) -- [double,thick] (a5),
    (a2) -- [boson,quarter left] (b3),
    (b3) -- [boson,quarter left] (a4),
    };
    \end{feynman} 
    \end{tikzpicture}} $\qquad \Leftrightarrow \qquad $\scalebox{0.6}{\begin{tikzpicture}[baseline=(a4)] 
        \begin{feynman}
        \vertex (a1) ;
        \vertex[right=3cm of a1] (a2); 
        \vertex[below=1.4cm of a1] (a3); 
        \vertex[right=0.5cm of a3] (a4); 
        \vertex[right=2.5cm of a3] (a5); 
        \diagram* {
        (a1) -- [boson, ultra thick, half left] (a2),
        (a1) -- [boson, ultra thick, half right] (a2),
        (a4) -- [double, ultra thick] (a5)
        };
        \vertex[right=1.5cm of a3, dot, red] (a6) {};
        \end{feynman} 
        \end{tikzpicture}}
    \caption{\textit{Radiation reaction}}
    \end{figure}
\noindent
The generic 1-loop amplitude is of the form:
\begin{eqnarray}
A^{Q^2}&=&\int \frac{dk_0}{(2\pi)}\Biggl( \scalebox{0.4}{\begin{tikzpicture}[baseline=(a1)] 
    \begin{feynman}
    \vertex (a1) ;
    \vertex[right=3cm of a1] (a2); 
    \vertex[below=1.4cm of a1] (a3); 
    \vertex[right=0.5cm of a3] (a4); 
    \vertex[right=2.5cm of a3] (a5); 
    \diagram* {
    (a1) -- [boson, ultra thick, half left] (a2),
    (a1) -- [boson, ultra thick, half right] (a2),
    (a4) -- [double, ultra thick] (a5)
    };
    \vertex[right=1.5cm of a3, dot, red] (a6) {};
    \end{feynman} 
    \end{tikzpicture}}\Biggr) \\ 
   \mathcal{M}^{Q^2} & = & \scalebox{0.4}{\begin{tikzpicture}[baseline=(a1)] 
    \begin{feynman}
    \vertex (a1) ;
    \vertex[right=3cm of a1] (a2); 
    \vertex[below=1.4cm of a1] (a3); 
    \vertex[right=0.5cm of a3] (a4); 
    \vertex[right=2.5cm of a3] (a5); 
    \diagram* {
    (a1) -- [boson, ultra thick, half left] (a2),
    (a1) -- [boson, ultra thick, half right] (a2),
    (a4) -- [double, ultra thick] (a5)
    };
    \vertex[right=1.5cm of a3, dot, red] (a6) {};
    \end{feynman} 
    \end{tikzpicture}} =  \int \frac{d^d\mathbf{k}}{(2\pi)^d}Q_{ij}^{(1)}(k_0)Q^{(2)}_{kl}(-k_0)\frac{N^{ij kl }(\mathbf{k},k_0)}{D_1}
\end{eqnarray}
where: 
\begin{equation}
    D_1= \mathbf{k}^2-k_0^2+ia 
\end{equation}
\subsubsection{Contributing diagrams}
In order to compute the effective diagrams we need to specify the polarisations of the graviton fields and sum up all the non-vanishing combinations.\\ 
There are 3 diagrams contributing, where we substitute, respectively, the fields $\phi,A,\sigma$:
\begin{eqnarray}
 \scalebox{0.6}{\begin{tikzpicture} [baseline=(a1)] 
\begin{feynman}
\vertex (a1) ;
\vertex[right=1cm of a1,square dot, red ] (a2) {}; 
\vertex[right=2cm of a2] (a3); 
\vertex[above=1.5cm of a3] (b3);
\vertex[right=2cm of a3,square dot, red ] (a4) {};
\vertex[right=1cm of a4] (a5) ; 
\vertex[below=0.2cm of a3] (e3);
\vertex[below=0.4cm of a2] (f3) {\(Q_{ij}\)};
\vertex[below=0.4cm of a4] (g3) {\(Q_{kl}\)};
\diagram* { 
(a1) -- [double,thick] (a5),
(a2) -- [boson,quarter left] (b3),
(b3) -- [boson,quarter left] (a4)
};
\end{feynman} 
\end{tikzpicture}}& =&  \scalebox{0.5}{\begin{tikzpicture}[baseline=(a1)]  
\begin{feynman}
\vertex (a1) ;
\vertex[right=1cm of a1,square dot, red ] (a2) {}; 
\vertex[right=2cm of a2] (a3); 
\vertex[above=1.5cm of a3] (b3);
\vertex[right=2cm of a3,square dot, red ] (a4) {};
\vertex[right=1cm of a4] (a5) ; 
\vertex[below=0.2cm of a3] (e3);
\vertex[below=0.4cm of a2] (f3) {\(Q_{ij}\)};
\vertex[below=0.4cm of a4] (g3) {\(Q_{kl}\)};
\diagram* { 
(a1) -- [double,thick] (a5),
(a2) -- [scalar,blue,,quarter left] (b3),
(b3) -- [scalar,blue,quarter left] (a4),
};
\end{feynman} 
\end{tikzpicture}}+\scalebox{0.5}{\begin{tikzpicture}[baseline=(a1)]  
\begin{feynman}
\vertex (a1) ;
\vertex[right=1cm of a1,square dot, red ] (a2) {}; 
\vertex[right=2cm of a2] (a3); 
\vertex[above=1.5cm of a3] (b3);
\vertex[right=2cm of a3,square dot, red ] (a4) {};
\vertex[right=1cm of a4] (a5) ; 
\vertex[below=0.2cm of a3] (e3);
\vertex[below=0.4cm of a2] (f3) {\(Q_{ij}\)};
\vertex[below=0.4cm of a4] (g3) {\(Q_{kl}\)};
\diagram* { 
(a1) -- [double,thick] (a5),
(a2) -- [boson,red,quarter left] (b3),
(b3) -- [boson, red, quarter left] (a4),
};
\end{feynman} 
\end{tikzpicture}}+\scalebox{0.5}{\begin{tikzpicture}[baseline=(a1)] 
\begin{feynman}
\vertex (a1) ;
\vertex[right=1cm of a1,square dot, red ] (a2) {}; 
\vertex[right=2cm of a2] (a3); 
\vertex[above=1.5cm of a3] (b3);
\vertex[right=2cm of a3,square dot, red ] (a4) {};
\vertex[right=1cm of a4] (a5) ; 
\vertex[below=0.2cm of a3] (e3);
\vertex[below=0.4cm of a2] (f3) {\(Q_{ij}\)};
\vertex[below=0.4cm of a4] (g3) {\(Q_{kl}\)};
\diagram* { 
(a1) -- [double,thick] (a5),
(a2) -- [double_boson,black!60!green,quarter left] (b3),
(b3) -- [double_boson,black!60!green,quarter left] (a4),
};
\end{feynman} 
\end{tikzpicture}}
\end{eqnarray}
The total amplitude is given by:
\begin{eqnarray}
 \mathcal{M}^{Q^2}= \mathcal{M}_{\phi}^{Q^2}+\mathcal{M}_{A}^{Q^2}+\mathcal{M}_{\sigma}^{Q^2}
\end{eqnarray}
where:
\begin{eqnarray}
 \mathcal{M}_{\phi}^{Q^2}&=& \frac{1}{2}\int \frac{d^d\mathbf{k}}{(2\pi)^d}\left(T_{Q\phi}(-\mathbf{k},-k_0)T_{Q\phi}(\mathbf{k},k_0)P_{\phi}(\mathbf{k},k_0)\right) \ , \\
  \mathcal{M}_{A}^{Q^2}&=& \frac{1}{2}\int \frac{d^d\mathbf{k}}{(2\pi)^d}\left(T_{QA}^i(-\mathbf{k},-k_0)T_{Q A}^j(\mathbf{k},k_0)P_{Aij}(\mathbf{k},k_0)\right) \ , \\
 \mathcal{M}_{\sigma}^{Q^2}&=& \frac{1}{2}\int \frac{d^d\mathbf{k}}{(2\pi)^d}\left(T_{Q\sigma}^{ij}(-\mathbf{k},-k_0)T_{Q\sigma}^{kl}(\mathbf{k},k_0)^{kl}P_{\phi ijkl}(\mathbf{k},k_0)\right) \ .
 \end{eqnarray}
 Substituting the appropriate Feynman rules and applying tensor contractions using \texttt{xTensor} we get:
 \begin{eqnarray}
    \mathcal{M}_{\phi}^{Q^2}&=& \frac{i (d-2)}{16 (d-1) \Lambda ^2}  \int \frac{d^d\mathbf{k}}{(2\pi)^d}\frac{k_a k_b k_c k_d}{D_1} Q^{ab}(k_0)Q^{cd}(-k_0)\ , \\
    \mathcal{M}_{A}^{Q^2}&=& -\frac{i }{8 \Lambda ^2}k_0^2\int \frac{d^d\mathbf{k}}{(2\pi)^d} \frac{k_a k_b}{D_1} Q^{ac}(k_0) Q_c^b(-k_0)\ , \\
    \mathcal{M}_{\sigma}^{Q^2}&=& \frac{i}{16 \Lambda ^2} \int \frac{d^d\mathbf{k}}{(2\pi)^d}\frac{1}{D_1}k_0^4 Q_{ab}(k_0)Q^{ab}(-k_0) \ .
    \end{eqnarray}
    After applying tensor decomposition we obtain an expression of the form: 
    \begin{equation}
    \mathcal{M}^{Q^2}=Q_{ab}(k_0)Q^{ab}(-k_0)\tilde{\mathcal{M}}^{Q^2}
    \end{equation}
where $\tilde{\mathcal{M}}^{Q^2}$ is of the type:
\begin{eqnarray}
    \tilde{\mathcal{M}}^{Q^2}\ = \  \frac{i}{32 (d-1) d (d+2) \Lambda ^2}\int \frac{d^d\mathbf{k}}{(2\pi)^d}\frac{\left(d^2+d-2\right) k_0^2 \left(d \ k_0^2-2 \mathbf{k}^2\right)+2 (d-2)
   \mathbf{k}^4}{D_1}
\end{eqnarray}
    \subsubsection{Integrand reduction}
In order to get an expression in terms of scalar integrals, we need to rewrite the scalar products appearing in terms of denominators, which in this case can just be performed by substituting: 
\begin{equation} 
    \mathbf{k}^2 \ = \ D_1 + k_0^2 \ , 
\end{equation}
and expanding the expression in order to obtain a linear combination of integrals. 
    \subsubsection{IBP decomposition}

 We can now perform an IBP decomposition of the integrals appearing using \texttt{Litered}, to get an expression in terms of a single 1-loop Master Integral:
 \begin{equation}
 \tilde{\mathcal{M}}^{Q^2}= C_1^{Q^2}  \scalebox{0.8}{\begin{tikzpicture}[baseline=(current bounding box.center)]             \begin{feynman}
    \vertex (a) ;
    \vertex[right=0.2cm of a] (b);
    \vertex[right=1cm of b] (c); 
    \vertex[right=1cm of c] (d);
    \diagram* { 
    (b) -- [half left,very thick ] (c),
    (b) -- [half right,very thick ] (c)
    };
    \end{feynman} \end{tikzpicture}}
 \end{equation}
 \subsubsection{Coefficient}
 The coefficient is given by: 
 \begin{equation}
    C_1^{Q^2}= \frac{(d-2) (d+1) }{32 (d-1) (d+2) \Lambda ^2}
 \end{equation}
 \subsubsection{MI evaluation}
 The master integral appearing $j_1(k_0)$ is the 1-loop massive tadpole given by:
 \begin{eqnarray}
    j_1(k_0)=\scalebox{0.8}{\begin{tikzpicture}[baseline=(current bounding box.center)]             \begin{feynman}
        \vertex (a) ;
        \vertex[right=0.2cm of a] (b);
        \vertex[right=1cm of b] (c); 
        \vertex[right=1cm of c] (d);
        \diagram* { 
        (b) -- [half left,very thick ] (c),
        (b) -- [half right,very thick ] (c)
        };
        \end{feynman} \end{tikzpicture}}= \int \frac{d^d\mathbf{k}}{(2\pi)^d}\frac{1}{\mathbf{k^2}-k_0^2+i\epsilon}=  \frac{\Gamma[1-d/2]}{(4\pi)^{d/2-1}}(-k_0^2)^{d/2-1}
 \end{eqnarray}
 \subsubsection{Result}
 In order to get the contribution to the conservative dynamics we need to substitute the explicit expression for the MI, to write $d=3+\epsilon$, and to take the limit $\epsilon\to 0$ as:
\begin{eqnarray}
 S^{Q^2}_{eff\ 2.5PN}&=& 
 -i \lim_{\epsilon\to 0 }\int_{-\infty}^{\infty}\frac{dk_0}{(2\pi)}Q_{ij}(k_0)Q^{ij}(-k_0)\tilde{\mathcal{M}}^{Q^2} \nonumber \\ 
 & = & 
i\frac{G_N}{10}\int_{-\infty}^{\infty}\frac{dk_0}{(2\pi)}\vert k_0\vert k_0^4Q_{ij}(k_0)Q^{ij}(-k_0)
\label{eq:INOUT_RR}
\end{eqnarray}
From \eqref{eq:INOUT_RR}, one can obtain the corresponding contribution in position space: 
\begin{eqnarray}
 S_{eff \ 2.5PN}^{Q^2}
  & = & 
 \frac{G_N}{10} \int \ dt \ Q_{ij}(t)\frac{d^5}{dt^5}Q^{ij}(t)   \nonumber \\ 
 & = & 
 \frac{G_N}{20} \int \ dt \ \frac{d}{dt}\biggl(\ddot{Q}^{ij}(t)\ddot{Q}_{ij}(t) \biggr)  \nonumber \\ 
\end{eqnarray}
This is a total time derivative, which gives zero assuming that the multipoles moments and their derivatives vanish at infinity, and hence it does not contribute to the equations of motion for the conservative dynamics.
\section{Tail Effects}
This section will be dedicated to the evaluation of the tail diagrams. \\
Tail effects are due to gravitational waves emitted by the binary system and scattered off the quasi-static curvature onto the same gravitational-wave source. 
The leading contribution to the conservative dynamics is at 4PN order, whereas the NLO is at 5PN. They are characterized by 3 multipoles, hence they can be seen as 2-loops one-point functions.
\begin{eqnarray}
    \centering
     \scalebox{0.9}{\begin{tikzpicture}[baseline=(a1)]
    \begin{feynman}
    \vertex (a1);
    \vertex[right=1cm of a1, square dot, red] (a2) {}; 
    \vertex[right=2cm of a2, square dot, red] (a3) {}; 
    \vertex[above=1.5cm of a3] (b3);
    \vertex[right=2cm of a3, square dot, red] (a4) {};
    \vertex[right=1cm of a4] (a5); 
    \diagram* { 
    (a1) -- [double,thick] (a5),
    (a2) -- [boson, thick, quarter left] (b3),
    (b3) -- [boson, thick, quarter left] (a4),
    (b3) -- [scalar, thick] (a3)
    };
    \end{feynman} 
    \end{tikzpicture}} \qquad \Leftrightarrow \qquad
    \scalebox{0.6}{\begin{tikzpicture}[baseline=(a4)] 
        \begin{feynman}
        \vertex (a1) ;
        \vertex[right=3cm of a1] (a2); 
        \vertex[below=1.4cm of a1] (a3); 
        \vertex[right=0.5cm of a3] (a4); 
        \vertex[right=2.5cm of a3] (a5); 
        \diagram* {
        (a1) -- [boson, ultra thick, half left] (a2),
        (a1) -- [boson, ultra thick, half right] (a2),
        (a4) -- [double, ultra thick] (a5)
        };
        \vertex[right=1.5cm of a3, dot, red] (a6) {};
         \vertex[above=1.4cm of a1] (a7);
        \vertex[right=1.5cm of a7] (a8);
        \diagram* {
        (a6) -- [scalar, ultra thick] (a8)
        }; 
        \end{feynman} 
        \end{tikzpicture}}
    \end{eqnarray}
The main feature about tail effects is that one of the graviton is coupled to a conserved source, which is either the energy $E$ of the system or the angular momentum $L_i$. 
That means that we can send:
\begin{eqnarray}
    Q^{(1)}_{i_1\cdots i_n}(q_0)\to  (2\pi)\delta(q_0)\tilde{Q}_{i_1\cdots i_n}^{(1)}
    \end{eqnarray}
    and the propagator of the corresponding line becomes instantaneous.
In this case the generic amplitude given in eq.\eqref{eq:hereditary} simplifies in:
\begin{eqnarray}
    A_{her}^{tail}
    &=& 
    \int \frac{dk_0}{(2\pi)}\frac{dq_0}{(2\pi)}\Biggl(  \scalebox{0.4}{\begin{tikzpicture}[baseline = (a1)]
        \begin{feynman}
        \vertex (a1) ;
        \vertex[right=3cm of a1] (a2); 
        \vertex[below=1.4cm of a1] (a3); 
        \vertex[right=0.5cm of a3] (a4); 
        \vertex[right=2.5cm of a3] (a5); 
        \diagram* {
        (a1) -- [boson, ultra thick, half left] (a2),
        (a1) -- [boson, ultra thick, half right] (a2),
        (a4) -- [double, ultra thick] (a5)
        };
        \vertex[right=1.5cm of a3, dot, red] (a6) {};
         \vertex[above=1.4cm of a1] (a7);
        \vertex[right=1.5cm of a7] (a8);
        \diagram* {
        (a6) -- [scalar, ultra thick] (a8)
        }; 
        \end{feynman} 
        \end{tikzpicture}}\Biggr) \nonumber \\ 
   & = & 
   \int \frac{dk_0}{(2\pi)}\frac{dq_0}{(2\pi)}\frac{d^d\mathbf{k}}{(2\pi)^d}\frac{d^d\mathbf{q}}{(2\pi)^d}Q^{(1)}_{i_1...i_l}(q_0)Q^{(2)}_{j1,...,j_m}(k_0)Q^{(3)}_{k_1...k_n}(p_{0})\nonumber \\
   & & 
   \frac{N^{i_1...i_l\ j_1...j_m \ k_1...k_n}(\mathbf{k},\mathbf{q},k_0,q_0,p_0)}{D_1D_2D_3} \nonumber \\ 
   & = &  
   \int \frac{dk_0}{(2\pi)}\frac{d^d\mathbf{k}}{(2\pi)^d}\frac{d^d\mathbf{q}}{(2\pi)^d}\tilde{Q}_{i_1...i_l}Q^{(2)}_{j1,...,j_m}(k_0)Q^{(3)}_{k_1...k_n}(-k_{0})\frac{N^{i_1...i_l\ j_1...j_m \ k_1...k_n}(\mathbf{k},\mathbf{q},k_0)}{(D_1D_2D_3)\big|_{q_0=0}} \\
   \mathcal{M}_{her}^{tail} & = & 
   \scalebox{0.4}{\begin{tikzpicture}[baseline=(a1)] 
    \begin{feynman}
    \vertex (a1) ;
    \vertex[right=3cm of a1] (a2); 
    \vertex[below=1.4cm of a1] (a3); 
    \vertex[right=0.5cm of a3] (a4); 
    \vertex[right=2.5cm of a3] (a5); 
    \diagram* {
    (a1) -- [boson, ultra thick, half left] (a2),
    (a1) -- [boson, ultra thick, half right] (a2),
    (a4) -- [double, ultra thick] (a5)
    };
    \vertex[right=1.5cm of a3, dot, red] (a6) {};
     \vertex[above=1.4cm of a1] (a7);
    \vertex[right=1.5cm of a7] (a8);
    \diagram* {
    (a6) -- [scalar, ultra thick] (a8)
    }; 
    \end{feynman} 
    \end{tikzpicture}} \nonumber \\ 
    & = & 
    \int \frac{d^d\mathbf{k}}{(2\pi)^d}\frac{d^d\mathbf{q}}{(2\pi)^d}\tilde{Q}_{i_1...i_l}Q^{(2)}_{j1,...,j_m}(k_0)Q^{(3)}_{k_1...k_n}(-k_{0})\frac{N^{i_1...i_l\ j_1...j_m \ k_1...k_n}(\mathbf{k},\mathbf{q},k_0)}{(D_1D_2D_3)\big|_{q_0=0}}
\label{eq:tail} \ ,
\end{eqnarray}
where the denominators appearing are defined as: 
\begin{eqnarray}
    D_{1}= [\mathbf{k}^2-(k_0)^2] \ ,  \qquad D_2=   [(\mathbf{k}+\mathbf{q})^2-(p_0)^2]\ , \qquad D_3 =  [\mathbf{q}^2-(q_0)^2]  \ ,
\end{eqnarray}
and once we take the limit $q_0\to 0$ they collapse in: 
\begin{eqnarray}
    D_{T1} = (\mathbf{k}^2-k_0^2)\ , \qquad D_{T2}= [(\mathbf{p}+\mathbf{q})^2-k_0^2]\ , \qquad D_{T3}=  \mathbf{q^2}\ . 
\end{eqnarray}
In the computation of these diagrams we encounter two radiation propagators. Hence, as discussed in Sec.\ref{sec:boundary_conditions_far}, using Feynman propagators we will obtain the correct answer for the contribution to the conservative dynamics. 

\subsection{Tail Effect at LO:  $EQ^2$}
The LO Tail Effect is contributing at 4PN to the conservative dynamics, and it contains the multipoles $EQ^2$:
\begin{figure}[H]
\centering
 \scalebox{0.9}{\begin{tikzpicture} 
\begin{feynman}
\vertex (a1) ;
\vertex[right=1cm of a1, square dot, red] (a2) {}; 
\vertex[right=2cm of a2, square dot, red] (a3) {}; 
\vertex[above=1.5cm of a3] (b3);
\vertex[right=2cm of a3, square dot, red] (a4) {};
\vertex[right=1cm of a4] (a5); 
\vertex[below=0.4cm of a3] (e3) {\(E\)};
\vertex[below=0.4cm of a2] (f3) {\(Q_{ij}\)};
\vertex[below=0.4cm of a4] (g3) {\(Q_{kl}\)};
\diagram* { 
(a1) -- [double,thick] (a5),
(a2) -- [boson, black!60!green,quarter left] (b3),
(b3) -- [boson, black!60!green,quarter left] (a4),
(b3) -- [scalar, blue] (a3)
};
\end{feynman} 
\end{tikzpicture}}
\caption{\textit{Quadrupole tail}}
\end{figure}
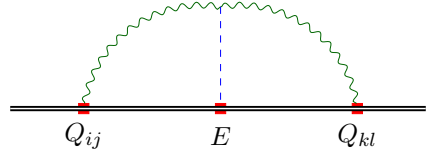
\noindent The multipole $E$ is conserved and so the computation is simplified as explained in the previous subsection. 

\subsubsection{Contributing diagrams}
In order to compute the effective diagrams we need to specify the polarisations of the graviton fields and sum up all the non-vanishing combinations.\\ In the case of a coupling to the effective mass $E$ only the field $\phi$ couples to $E$. \\
The symmetry factors can be fixed in the following way: we assign symmetry factor $1/2$ to diagrams having identical polarizations coupled to $Q's$.
There are 6 diagrams contributing:
\begin{eqnarray}
 \scalebox{0.6}{\begin{tikzpicture} [baseline=(a1)] 
\begin{feynman}
\vertex (a1) ;
\vertex[right=1cm of a1, square dot, red] (a2) {}; 
\vertex[right=2cm of a2, square dot, red] (a3) {}; 
\vertex[above=1.5cm of a3] (b3);
\vertex[right=2cm of a3, square dot, red] (a4) {};
\vertex[right=1cm of a4] (a5); 
\vertex[below=0.4cm of a3] (e3) {\(E\)};
\vertex[below=0.4cm of a2] (f3) {\(Q_{ij}\)};
\vertex[below=0.4cm of a4] (g3) {\(Q_{kl}\)};
\diagram* { 
(a1) -- [double,thick] (a5),
(a2) -- [boson,quarter left] (b3),
(b3) -- [boson,quarter left] (a4),
(b3) -- [scalar] (a3)
};
\end{feynman} 
\end{tikzpicture}}& =&  \scalebox{0.5}{\begin{tikzpicture}[baseline=(a1)]  
\begin{feynman}
\vertex (a1) ;
\vertex[right=1cm of a1, square dot, red] (a2) {}; 
\vertex[right=2cm of a2, square dot, red] (a3) {}; 
\vertex[above=1.5cm of a3] (b3);
\vertex[right=2cm of a3, square dot, red] (a4) {};
\vertex[right=1cm of a4] (a5); 
\vertex[below=0.4cm of a3] (e3) {\(E\)};
\vertex[below=0.4cm of a2] (f3) {\(Q_{ij}\)};
\vertex[below=0.4cm of a4] (g3) {\(Q_{kl}\)};
\diagram* { 
(a1) -- [double,thick] (a5),
(a2) -- [double_boson,black!60!green,quarter left] (b3),
(b3) -- [double_boson,black!60!green,quarter left] (a4),
(b3) -- [scalar,blue] (a3)
};
\end{feynman} 
\end{tikzpicture}}+\scalebox{0.5}{\begin{tikzpicture}[baseline=(a1)]  
\begin{feynman}
\vertex (a1) ;
\vertex[right=1cm of a1, square dot, red] (a2) {}; 
\vertex[right=2cm of a2, square dot, red] (a3) {}; 
\vertex[above=1.5cm of a3] (b3);
\vertex[right=2cm of a3, square dot, red] (a4) {};
\vertex[right=1cm of a4] (a5); 
\vertex[below=0.4cm of a3] (e3) {\(E\)};
\vertex[below=0.4cm of a2] (f3) {\(Q_{ij}\)};
\vertex[below=0.4cm of a4] (g3) {\(Q_{kl}\)};
\diagram* { 
(a1) -- [double,thick] (a5),
(a2) -- [boson,red,quarter left] (b3),
(b3) -- [double_boson,black!60!green,quarter left] (a4),
(b3) -- [scalar,blue] (a3)
};
\end{feynman} 
\end{tikzpicture}}+\scalebox{0.5}{\begin{tikzpicture}[baseline=(a1)] 
\begin{feynman}
\vertex (a1) ;
\vertex[right=1cm of a1, square dot, red] (a2) {}; 
\vertex[right=2cm of a2, square dot, red] (a3) {}; 
\vertex[above=1.5cm of a3] (b3);
\vertex[right=2cm of a3, square dot, red] (a4) {};
\vertex[right=1cm of a4] (a5); 
\vertex[below=0.4cm of a3] (e3) {\(E\)};
\vertex[below=0.4cm of a2] (f3) {\(Q_{ij}\)};
\vertex[below=0.4cm of a4] (g3) {\(Q_{kl}\)};
\diagram* { 
(a1) -- [double,thick] (a5),
(a2) -- [double_boson,black!60!green,quarter left] (b3),
(b3) -- [scalar,blue,quarter left] (a4),
(b3) -- [scalar,blue] (a3)
};
\end{feynman} 
\end{tikzpicture}}\nonumber\\
& & +\scalebox{0.5}{\begin{tikzpicture}[baseline=(a1)]  
\begin{feynman}
\vertex (a1) ;
\vertex[right=1cm of a1, square dot, red] (a2) {}; 
\vertex[right=2cm of a2, square dot, red] (a3) {}; 
\vertex[above=1.5cm of a3] (b3);
\vertex[right=2cm of a3, square dot, red] (a4) {};
\vertex[right=1cm of a4] (a5); 
\vertex[below=0.4cm of a3] (e3) {\(E\)};
\vertex[below=0.4cm of a2] (f3) {\(Q_{ij}\)};
\vertex[below=0.4cm of a4] (g3) {\(Q_{kl}\)};
\diagram* { 
(a1) -- [double,thick] (a5),
(a2) -- [boson,red,quarter left] (b3),
(b3) -- [boson,red,quarter left] (a4),
(b3) -- [scalar,blue] (a3)
};
\end{feynman} 
\end{tikzpicture}}+\scalebox{0.5}{\begin{tikzpicture}[baseline=(a1)]  
\begin{feynman}
\vertex (a1) ;
\vertex[right=1cm of a1, square dot, red] (a2) {}; 
\vertex[right=2cm of a2, square dot, red] (a3) {}; 
\vertex[above=1.5cm of a3] (b3);
\vertex[right=2cm of a3, square dot, red] (a4) {};
\vertex[right=1cm of a4] (a5); 
\vertex[below=0.4cm of a3] (e3) {\(E\)};
\vertex[below=0.4cm of a2] (f3) {\(Q_{ij}\)};
\vertex[below=0.4cm of a4] (g3) {\(Q_{kl}\)};
\diagram* { 
(a1) -- [double,thick] (a5),
(a2) -- [boson,red,quarter left] (b3),
(b3) -- [scalar,blue,quarter left] (a4),
(b3) -- [scalar,blue] (a3)
};
\end{feynman} 
\end{tikzpicture}}+\scalebox{0.5}{\begin{tikzpicture}[baseline=(a1)]  
\begin{feynman}
\vertex (a1) ;
\vertex[right=1cm of a1, square dot, red] (a2) {}; 
\vertex[right=2cm of a2, square dot, red] (a3) {}; 
\vertex[above=1.5cm of a3] (b3);
\vertex[right=2cm of a3, square dot, red] (a4) {};
\vertex[right=1cm of a4] (a5); 
\vertex[below=0.4cm of a3] (e3) {\(E\)};
\vertex[below=0.4cm of a2] (f3) {\(Q_{ij}\)};
\vertex[below=0.4cm of a4] (g3) {\(Q_{kl}\)};
\diagram* { 
(a1) -- [double,thick] (a5),
(a2) -- [scalar,blue,quarter left] (b3),
(b3) -- [scalar,blue,quarter left] (a4),
(b3) -- [scalar,blue] (a3)
};
\end{feynman} 
\end{tikzpicture}}
\end{eqnarray}
The total amplitude is given by:
\begin{eqnarray}
    \mathcal{M}^{EQ^2}= \left(\mathcal{M}_{\phi\sigma^2}^{EQ^2}+\mathcal{M}_{\phi A\sigma}^{EQ^2}+\mathcal{M}_{\phi^2\sigma}^{EQ^2}+\mathcal{M}_{\phi A^2}^{EQ^2}+\mathcal{M}_{\phi^2 A}^{EQ^2}+\mathcal{M}_{\phi^3}^{EQ^2} \right)
\end{eqnarray}
and the single components are given by: 
\begin{eqnarray}
\mathcal{M}_{\phi\sigma^2}^{EQ^2}&=& 
\frac{1}{2}\int \frac{d^d\mathbf{k}}{(2\pi)^d}\frac{d^d\mathbf{q}}{(2\pi)^d}T_{Q\sigma}^{ij}(-\mathbf{k},-k_0)T_{Q\sigma}^{kl}(\mathbf{k}+\mathbf{q},p_0)T_{E\phi}(-\mathbf{q},-q_0)\nonumber\\ 
& & P_{\sigma ijmn}(\mathbf{k},k_0)P_{\sigma kl op }(\mathbf{k}+\mathbf{q},p_0)P_\phi V_{\phi\sigma^2}^{mnop}(\mathbf{k},k_0,\mathbf{q},q_0) \ , \\
 \mathcal{M}_{\phi A \sigma}^{EQ^2}&=& 
 \frac{1}{2}\int \frac{d^d\mathbf{k}}{(2\pi)^d}\frac{d^d\mathbf{q}}{(2\pi)^d}T_{Q A}^{i}(-\mathbf{k},-k_0)T_{Q\sigma}^{kl}(\mathbf{k}+\mathbf{q},p_0)T_{E\phi}(-\mathbf{q},-q_0)\nonumber\\
 & & P_{A  ij}(\mathbf{k},k_0)P_{\sigma kl op }(\mathbf{k}+\mathbf{q},p_0)P_\phi V_{\phi A \sigma}^{jop}(\mathbf{k},k_0,\mathbf{q},q_0) \ , \\
 \mathcal{M}_{\phi^2\sigma}^{EQ^2}&=& 
 \frac{1}{2}\int \frac{d^d\mathbf{k}}{(2\pi)^d}\frac{d^d\mathbf{q}}{(2\pi)^d}T_{Q\sigma}^{ij}(-\mathbf{k},-k_0)T_{Q \phi}(\mathbf{k}+\mathbf{q},p_0)T_{E\phi}(-\mathbf{q},-q_0)\nonumber\\
 & & P_{\sigma ijmn}(\mathbf{k},k_0)P_{\phi}(\mathbf{k}+\mathbf{q},p_0)P_\phi V_{\phi^2\sigma}^{mn}(\mathbf{k},k_0,\mathbf{q},q_0) \ , \\
 \mathcal{M}_{\phi A^2}^{EQ^2}&=& 
 \frac{1}{2}\int \frac{d^d\mathbf{k}}{(2\pi)^d}\frac{d^d\mathbf{q}}{(2\pi)^d}T_{Q A}^{i}(-\mathbf{k},-k_0)T_{Q A}^{k}(\mathbf{k}+\mathbf{q},p_0)T_{E\phi}(-\mathbf{q},-q_0)\nonumber\\
 & & P_{A i m}(\mathbf{k},k_0)P_{A k o }(\mathbf{k}+\mathbf{q},p_0)P_\phi V_{\phi\sigma^2}^{mo}(\mathbf{k},k_0,\mathbf{q},q_0) \ , \\
 \mathcal{M}_{\phi^2 A}^{EQ^2}&=& 
 \frac{1}{2}\int \frac{d^d\mathbf{k}}{(2\pi)^d}\frac{d^d\mathbf{q}}{(2\pi)^d}T_{Q A }^{i}(-\mathbf{k},-k_0)T_{Q\phi}(\mathbf{k}+\mathbf{q},p_0)T_{E\phi}(-\mathbf{q},-q_0)\nonumber\\
 & & P_{A i m}(\mathbf{k},k_0)P_{\phi}(\mathbf{k}+\mathbf{q},p_0)P_\phi V_{\phi^2 A}^{m}(\mathbf{k},k_0,\mathbf{q}) \ , \\
 \mathcal{M}_{\phi^3}^{EQ^2}&=& 
 \frac{1}{2}\int \frac{d^d\mathbf{k}}{(2\pi)^d}\frac{d^d\mathbf{q}}{(2\pi)^d}T_{Q\phi}(-\mathbf{k},-k_0)T_{Q\phi}(\mathbf{k}+\mathbf{q},p_0)T_{E\phi}(-\mathbf{q},-q_0)\nonumber\\
 & & P_{\phi}(\mathbf{k},k_0)P_{\phi}(\mathbf{k}+\mathbf{q},p_0)P_{\phi}(\mathbf{q},q_0) V_{\phi^3}(\mathbf{k},k_0,\mathbf{q},q_0) \ .
 \end{eqnarray}
After applying tensor contractions and the tensor decomposition procedure we obtain an expression of the form:
\begin{equation}
    \mathcal{M}^{EQ^2}=E Q_{ij}(k_0)Q^{ij}(-k_0)\tilde{\mathcal{M}}^{EQ^2}
\end{equation}

\subsubsection{IBP Decomposition}

Applying IBP decomposition on $\tilde{\mathcal{M}}^{EQ^2}$ we get an expression in terms of a single Master Integral:
\begin{eqnarray}
    \tilde{\mathcal{M}}^{EQ^2}= C^{EQ^2}j_{T, 1,1,0}
\label{eq:tail_decomposed}
\end{eqnarray}
\subsubsection{Coefficient}
The coefficient is given by:
\begin{eqnarray}
    C^{EQ^2} & = &  \frac{k_0^4}{128 \Lambda ^4 \epsilon  (\epsilon +2)^2 (\epsilon +3)
    (\epsilon +5) } \left(\epsilon ^5+9 \epsilon ^4+31 \epsilon ^3+51 \epsilon ^2+52 \epsilon +24\right)\ . \\
\end{eqnarray}
\subsubsection{Master Integrals}
$j_{T 1,1,0}$ is the only MI of the following topology:
\begin{eqnarray}
    j_{T n_1,n_2,n_3}=\  \scalebox{0.8}{\begin{tikzpicture}[baseline=(b)]             \begin{feynman}
        \vertex (a);
        \vertex[right=0.2cm of a] (b);
        \vertex[right=1.5 cm of b] (c); 
        \vertex[right=1.5 cm of c] (d);
        \diagram* { 
        (b) -- [half left,very thick ] (c),
        (b) -- [half right,very thick] (c),
        (b) -- [] (c),
        };
        \end{feynman} \end{tikzpicture}}\ = \int \frac{d^d\mathbf{k}}{(2\pi)^d}\frac{d^d\mathbf{q}}{(2\pi)^d}\frac{1}{D_{T1}^{n_1}D_{T2}^{n_2}D_{T3}^{n_3}} \ , 
\end{eqnarray} 
where:
\begin{eqnarray}
    D_{T1} = \mathbf{k}^2-k_0^2+i a \ , \qquad D_{T2}= (\mathbf{p}+\mathbf{q})^2-k_0^2+ia \ , \qquad D_{T3}=  \mathbf{q^2} +i a \ ,
\end{eqnarray}
and the thin line in the diagram denote the massless propagator $D_{T3}$.
The explicit solution of $j_{T, 1,1,0}$ is immediate to find, since the integral factorizes in the product of two massive tadpoles of mass $k_0$ as:
\begin{eqnarray}
j_{T, 1,1,0}= \int \frac{d^d\mathbf{k}}{(2\pi)^d}\frac{d^d\mathbf{q}}{(2\pi)^d}\frac{1}{D_{T1}D_{T2}}= \scalebox{0.8}{\begin{tikzpicture}[baseline=(current bounding box.center)]             \begin{feynman}
    \vertex (a) ;
    \vertex[right=0.2cm of a] (b);
    \vertex[right=1cm of b] (c); 
    \vertex[right=1cm of c] (d);
    \diagram* { 
    (b) -- [half left,very thick ] (c),
    (b) -- [half right,very thick ] (c),
    (c) -- [half left,very thick ] (d),
    (c) -- [half right,very thick ] (d),
    };
    \end{feynman} \end{tikzpicture}}= \left(\scalebox{0.8}{\begin{tikzpicture}[baseline=(current bounding box.center)]              \begin{feynman}
        \vertex (a) ;
        \vertex[right=0.2cm of a] (b);
        \vertex[right=1cm of b] (c); 
        \vertex[right=1cm of c] (d);
        \diagram* { 
        (b) -- [half left,very thick ] (c),
        (b) -- [half right,very thick ] (c)
        };
        \end{feynman} \end{tikzpicture}}\right)^2=  \frac{\Gamma[1-d/2]^2}{(4\pi)^{d-2}}(-k_0^2)^{d-2}
\end{eqnarray}
 \subsubsection{Result}
 After applying the reduction procedure outlined in the previous section we get:
\begin{eqnarray}
 S^{EQ^2}_{eff\ 4PN}= -\frac{G_N^2 E}{5}\int_{-\infty}^{\infty}\frac{dk_0}{(2\pi)}k_0^6\left[\frac{1}{\epsilon}-\frac{41}{30}-i \pi + \log\left(\frac{k_0^2}{\bar{\mu}^2}\right)+\mathcal{O}(\epsilon)\right]Q_{ij}(k_0)Q^{ij}(-k_0)
\end{eqnarray}
where $\bar{\mu}^2=\mu^2\pi/e^{\gamma_E}$.
This result agrees with Eq.10 of \cite{Foffa:2019eeb}
 \subsubsection{Divergent part}
The divergent part of the tail amplitude receives contributions only from processes involving the same graviton polarizations attaching to the two radiative sources, as they are the ones diverging when $\mathbf{q}\to 0$. \\
The double integration over space momenta $\mathbf{k},\mathbf{q}$ on purely heuristic arguments, leads to an amplitude result $\sim G_d^3(-k_0^2)^{d-3}$, from which we can infer that the presence of an imaginary part is invariably linked to a divergence:
\begin{eqnarray}
 \frac{G_d^2}{\epsilon}(-k_0^2)^\epsilon=G_N^2\left(\frac{1}{\epsilon}+\log\left(\frac{k_0^2}{\mu^2}\right)-i\pi+\mathcal{O}(\epsilon)\right)
 \end{eqnarray}
 implying that pole residual not only fixes the logarithmic term but also the imaginary one.
 The divergent part can be regularized by adding a similarly divergent term coming from the near zone dynamics as shown in \cite{Foffa:2019yfl}, with a procedure known as \textit{zero-bin subtraction scheme}  (see \cite{Manohar:2006nz} for a generic treatment of cancellation of divergences using the method of regions)\\
\subsection{Octupole Tail: $EO^2$}
The octupole Tail is contributing at 5PN to the conservative dynamics and it contains the multipoles $EO^2$:
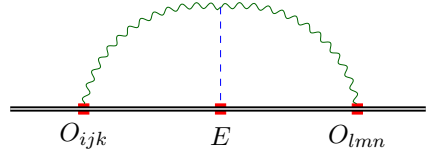
\begin{figure}[H]
\centering
 \scalebox{0.9}{\begin{tikzpicture} 
\begin{feynman}
\vertex (a1) ;
\vertex[right=1cm of a1, square dot, red] (a2) {}; 
\vertex[right=2cm of a2, square dot, red] (a3) {}; 
\vertex[above=1.5cm of a3] (b3);
\vertex[right=2cm of a3, square dot, red] (a4) {};
\vertex[right=1cm of a4] (a5); 
\vertex[below=0.4cm of a3] (e3) {\(E\)};
\vertex[below=0.4cm of a2] (f3) {\(O_{ijk}\)};
\vertex[below=0.4cm of a4] (g3) {\(O_{lmn}\)};
\diagram* { 
(a1) -- [double,thick] (a5),
(a2) -- [boson, black!60!green,quarter left] (b3),
(b3) -- [boson, black!60!green,quarter left] (a4),
(b3) -- [scalar, blue] (a3)
};
\end{feynman} 
\end{tikzpicture}}
\caption{\textit{Octupole tail}}
\end{figure}
As in the previous case the multipole $E$ is conserved. 
\subsubsection{Contributing diagrams}
We need then to specify the polarizations and there are 6 diagrams contributing:
\begin{eqnarray}
 \scalebox{0.6}{\begin{tikzpicture} [baseline=(a1)] 
\begin{feynman}
\vertex (a1) ;
\vertex[right=1cm of a1, square dot, red] (a2) {}; 
\vertex[right=2cm of a2, square dot, red] (a3) {}; 
\vertex[above=1.5cm of a3] (b3);
\vertex[right=2cm of a3, square dot, red] (a4) {};
\vertex[right=1cm of a4] (a5); 
\vertex[below=0.4cm of a3] (e3) {\(E\)};
\vertex[below=0.4cm of a2] (f3) {\(O_{ijk}\)};
\vertex[below=0.4cm of a4] (g3) {\(O_{lmn}\)};
\diagram* { 
(a1) -- [double,thick] (a5),
(a2) -- [boson,quarter left] (b3),
(b3) -- [boson,quarter left] (a4),
(b3) -- [scalar] (a3)
};
\end{feynman} 
\end{tikzpicture}}& =&  \scalebox{0.5}{\begin{tikzpicture}[baseline=(a1)]  
\begin{feynman}
\vertex (a1) ;
\vertex[right=1cm of a1, square dot, red] (a2) {}; 
\vertex[right=2cm of a2, square dot, red] (a3) {}; 
\vertex[above=1.5cm of a3] (b3);
\vertex[right=2cm of a3, square dot, red] (a4) {};
\vertex[right=1cm of a4] (a5); 
\vertex[below=0.4cm of a3] (e3) {\(E\)};
\vertex[below=0.4cm of a2] (f3) {\(O_{ijk}\)};
\vertex[below=0.4cm of a4] (g3) {\(O_{lmn}\)};
\diagram* { 
(a1) -- [double,thick] (a5),
(a2) -- [double_boson,black!60!green,quarter left] (b3),
(b3) -- [double_boson,black!60!green,quarter left] (a4),
(b3) -- [scalar,blue] (a3)
};
\end{feynman} 
\end{tikzpicture}}+\scalebox{0.5}{\begin{tikzpicture}[baseline=(a1)]  
\begin{feynman}
\vertex (a1) ;    
\vertex[right=1cm of a1, square dot, red] (a2) {}; 
\vertex[right=2cm of a2, square dot, red] (a3) {}; 
\vertex[above=1.5cm of a3] (b3);
\vertex[right=2cm of a3, square dot, red] (a4) {};
\vertex[right=1cm of a4] (a5); 
\vertex[below=0.4cm of a3] (e3) {\(E\)};
\vertex[below=0.4cm of a2] (f3) {\(O_{ijk}\)};
\vertex[below=0.4cm of a4] (g3) {\(O_{lmn}\)};
\diagram* { 
(a1) -- [double,thick] (a5),
(a2) -- [boson,red,quarter left] (b3),
(b3) -- [double_boson,black!60!green,quarter left] (a4),
(b3) -- [scalar,blue] (a3)
};
\end{feynman} 
\end{tikzpicture}}+\scalebox{0.5}{\begin{tikzpicture}[baseline=(a1)] 
\begin{feynman}
\vertex (a1) ;
\vertex[right=1cm of a1, square dot, red] (a2) {}; 
\vertex[right=2cm of a2, square dot, red] (a3) {}; 
\vertex[above=1.5cm of a3] (b3);
\vertex[right=2cm of a3, square dot, red] (a4) {};
\vertex[right=1cm of a4] (a5); 
\vertex[below=0.4cm of a3] (e3) {\(E\)};
\vertex[below=0.4cm of a2] (f3) {\(O_{ijk}\)};
\vertex[below=0.4cm of a4] (g3) {\(O_{lmn}\)};
\diagram* { 
(a1) -- [double,thick] (a5),
(a2) -- [double_boson,black!60!green,quarter left] (b3),
(b3) -- [scalar,blue,quarter left] (a4),
(b3) -- [scalar,blue] (a3)
};
\end{feynman} 
\end{tikzpicture}}\nonumber\\
& & +\scalebox{0.5}{\begin{tikzpicture}[baseline=(a1)]  
\begin{feynman}
\vertex (a1) ;
\vertex[right=1cm of a1, square dot, red] (a2) {}; 
\vertex[right=2cm of a2, square dot, red] (a3) {}; 
\vertex[above=1.5cm of a3] (b3);
\vertex[right=2cm of a3, square dot, red] (a4) {};
\vertex[right=1cm of a4] (a5); 
\vertex[below=0.4cm of a3] (e3) {\(E\)};
\vertex[below=0.4cm of a2] (f3) {\(O_{ijk}\)};
\vertex[below=0.4cm of a4] (g3) {\(O_{lmn}\)};
\diagram* { 
(a1) -- [double,thick] (a5),
(a2) -- [boson,red,quarter left] (b3),
(b3) -- [boson,red,quarter left] (a4),
(b3) -- [scalar,blue] (a3)
};
\end{feynman} 
\end{tikzpicture}}+\scalebox{0.5}{\begin{tikzpicture}[baseline=(a1)]  
\begin{feynman}
\vertex (a1) ;
\vertex[right=1cm of a1, square dot, red] (a2) {}; 
\vertex[right=2cm of a2, square dot, red] (a3) {}; 
\vertex[above=1.5cm of a3] (b3);
\vertex[right=2cm of a3, square dot, red] (a4) {};
\vertex[right=1cm of a4] (a5); 
\vertex[below=0.4cm of a3] (e3) {\(E\)};
\vertex[below=0.4cm of a2] (f3) {\(O_{ijk}\)};
\vertex[below=0.4cm of a4] (g3) {\(O_{lmn}\)};
\diagram* { 
(a1) -- [double,thick] (a5),
(a2) -- [boson,red,quarter left] (b3),
(b3) -- [scalar,blue,quarter left] (a4),
(b3) -- [scalar,blue] (a3)
};
\end{feynman} 
\end{tikzpicture}}+\scalebox{0.5}{\begin{tikzpicture}[baseline=(a1)]  
\begin{feynman}
\vertex (a1) ;
\vertex[right=1cm of a1, square dot, red] (a2) {}; 
\vertex[right=2cm of a2, square dot, red] (a3) {}; 
\vertex[above=1.5cm of a3] (b3);
\vertex[right=2cm of a3, square dot, red] (a4) {};
\vertex[right=1cm of a4] (a5); 
\vertex[below=0.4cm of a3] (e3) {\(E\)};
\vertex[below=0.4cm of a2] (f3) {\(O_{ijk}\)};
\vertex[below=0.4cm of a4] (g3) {\(O_{lmn}\)};
\diagram* { 
(a1) -- [double,thick] (a5),
(a2) -- [scalar,blue,quarter left] (b3),
(b3) -- [scalar,blue,quarter left] (a4),
(b3) -- [scalar,blue] (a3)
};
\end{feynman} 
\end{tikzpicture}}
\end{eqnarray}
The total amplitude is given by:
\begin{eqnarray}
 \mathcal{M}^{EO^2}\ = \ \mathcal{M}^{EO^2}_{\phi\sigma^2}+\mathcal{M}^{EO^2}_{\phi A\sigma}+\mathcal{M}^{EO^2}_{\phi^2\sigma}+\mathcal{M}^{EO^2}_{\phi A^2}+\mathcal{M}^{EO^2}_{\phi^2 A}+\mathcal{M}^{EO^2}_{\phi^3} 
\end{eqnarray}
 \subsubsection{Result}
 After applying the reduction procedure outlined in the previous section we get:
\begin{eqnarray}
 S^{EO^2}_{eff\ 5PN}= -\frac{G_N^2 E}{189}\int_{-\infty}^{\infty}\frac{dk_0}{(2\pi)}k_0^8\left[\frac{1}{\epsilon}-\frac{82}{35}-i \pi + \log\left(\frac{k_0^2}{\bar{\mu}^2}\right)+\mathcal{O}(\epsilon)\right]O_{ijk}(k_0)O^{ijk}(-k_0)
\end{eqnarray}
This result agrees with Eq.18 of \cite{Foffa:2019eeb}
\subsection{Magnetic quadrupole tail: $EJ^2$}
The Magnetic quadrupole Tail is contributing at 5PN to the conservative dynamics and it contains the multipoles $EJ^2$:
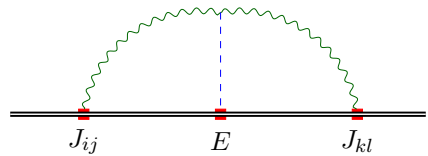
\begin{figure}[H]
\centering
 \scalebox{0.9}{\begin{tikzpicture} 
\begin{feynman}
\vertex (a1) ;
\vertex[right=1cm of a1, square dot, red] (a2) {}; 
\vertex[right=2cm of a2, square dot, red] (a3) {}; 
\vertex[above=1.5cm of a3] (b3);
\vertex[right=2cm of a3, square dot, red] (a4) {};
\vertex[right=1cm of a4] (a5); 
\vertex[below=0.4cm of a3] (e3) {\(E\)};
\vertex[below=0.4cm of a2] (f3) {\(J_{ij}\)};
\vertex[below=0.4cm of a4] (g3) {\(J_{kl}\)};
\diagram* { 
(a1) -- [double,thick] (a5),
(a2) -- [boson,black!60!green,quarter left] (b3),
(b3) -- [boson, black!60!green,quarter left] (a4),
(b3) -- [scalar, blue] (a3)
};
\end{feynman} 
\end{tikzpicture}}
\caption{\textit{Magnetic quadrupole tail}}
\end{figure}
\subsubsection{Contributing diagrams}
As in the previous case we need to specify the polarizations, in this case there are only 3 diagrams contributing because $\phi$ does not couple with the magnetic quadrupole moment $J^{ij}$:
\begin{eqnarray}
 \scalebox{0.6}{\begin{tikzpicture} [baseline=(a1)] 
\begin{feynman}
\vertex (a1) ;
\vertex[right=1cm of a1, square dot, red] (a2) {}; 
\vertex[right=2cm of a2, square dot, red] (a3) {}; 
\vertex[above=1.5cm of a3] (b3);
\vertex[right=2cm of a3, square dot, red] (a4) {};
\vertex[right=1cm of a4] (a5); 
\vertex[below=0.4cm of a3] (e3) {\(E\)};
\vertex[below=0.4cm of a2] (f3) {\(J_{ij}\)};
\vertex[below=0.4cm of a4] (g3) {\(J_{kl}\)};
\diagram* { 
(a1) -- [double,thick] (a5),
(a2) -- [boson,quarter left] (b3),
(b3) -- [boson,quarter left] (a4),
(b3) -- [scalar] (a3)
};
\end{feynman} 
\end{tikzpicture}}& =&  \scalebox{0.5}{\begin{tikzpicture}[baseline=(a1)]  
\begin{feynman}
\vertex (a1) ;
\vertex[right=1cm of a1, square dot, red] (a2) {}; 
\vertex[right=2cm of a2, square dot, red] (a3) {}; 
\vertex[above=1.5cm of a3] (b3);
\vertex[right=2cm of a3, square dot, red] (a4) {};
\vertex[right=1cm of a4] (a5); 
\vertex[below=0.4cm of a3] (e3) {\(E\)};
\vertex[below=0.4cm of a2] (f3) {\(J_{ij}\)};
\vertex[below=0.4cm of a4] (g3) {\(J_{kl}\)};
\diagram* { 
(a1) -- [double,thick] (a5),
(a2) -- [double_boson,black!60!green,quarter left] (b3),
(b3) -- [double_boson,black!60!green,quarter left] (a4),
(b3) -- [scalar,blue] (a3)
};
\end{feynman} 
\end{tikzpicture}}+\scalebox{0.5}{\begin{tikzpicture}[baseline=(a1)]  
\begin{feynman}
\vertex (a1) ;
\vertex[right=1cm of a1, square dot, red] (a2) {}; 
\vertex[right=2cm of a2, square dot, red] (a3) {}; 
\vertex[above=1.5cm of a3] (b3);
\vertex[right=2cm of a3, square dot, red] (a4) {};
\vertex[right=1cm of a4] (a5); 
\vertex[below=0.4cm of a3] (e3) {\(E\)};
\vertex[below=0.4cm of a2] (f3) {\(J_{ij}\)};
\vertex[below=0.4cm of a4] (g3) {\(J_{kl}\)};
\diagram* { 
(a1) -- [double,thick] (a5),
(a2) -- [boson,red,quarter left] (b3),
(b3) -- [double_boson,black!60!green,quarter left] (a4),
(b3) -- [scalar,blue] (a3)
};
\end{feynman} 
\end{tikzpicture}} +\scalebox{0.5}{\begin{tikzpicture}[baseline=(a1)]  
\begin{feynman}
\vertex (a1) ;
\vertex[right=1cm of a1, square dot, red] (a2) {}; 
\vertex[right=2cm of a2, square dot, red] (a3) {}; 
\vertex[above=1.5cm of a3] (b3);
\vertex[right=2cm of a3, square dot, red] (a4) {};
\vertex[right=1cm of a4] (a5); 
\vertex[below=0.4cm of a3] (e3) {\(E\)};
\vertex[below=0.4cm of a2] (f3) {\(J_{ij}\)};
\vertex[below=0.4cm of a4] (g3) {\(J_{kl}\)};
\diagram* { 
(a1) -- [double,thick] (a5),
(a2) -- [boson,red,quarter left] (b3),
(b3) -- [boson,red,quarter left] (a4),
(b3) -- [scalar,blue] (a3)
};
\end{feynman} 
\end{tikzpicture}}
\end{eqnarray}
The total amplitude is given by:
\begin{eqnarray}
 \mathcal{M}^{EJ^2}= \mathcal{M}^{EJ^2}_{\phi\sigma^2}+\mathcal{M}^{Ej^2}_{\phi A\sigma}+\mathcal{M}^{EJ^2}_{\phi A^2}
\end{eqnarray}
 \subsubsection{Result}
 After applying the reduction procedure outlined in the previous section we get:
\begin{eqnarray}
 S^{EJ^2}_{eff\ 5PN}= -\frac{16}{45}G_N^2 E\int_{-\infty}^{\infty}\frac{dk_0}{(2\pi)}k_0^6\left[\frac{1}{\epsilon}-\frac{127}{60}-i \pi + \log\left(\frac{k_0^2}{\bar{\mu}^2}\right)+\mathcal{O}(\epsilon)\right]J_{ij}(k_0)J^{ij}(-k_0)
\end{eqnarray}
This result agrees with Eq.21 of \cite{Foffa:2019eeb}. 
A comment should be made before proceeding further. In the Feynman rules for the coupling of $\phi,A,\sigma$ with the magnetic quadrupole tail $J_{ij}$ ,appears a Levi-Civita tensor $\epsilon_{ijk}$. Following the prescriptions given in \cite{Foffa:2021pkg} we assume that this $\epsilon_{ijk}$ is 3-dimensional and no d-dimensional, hence we should be careful when contracting two Levi-Civita tensors in using the appropriate 3-dimensional rules. A $d$-dimensional generalization of this multipole has been proposed in \cite{Henry:2021cek,Almeida:2021xwn}. 
\subsubsection{d-dimensional generalization}
Following \cite{Henry:2021cek,Almeida:2021xwn}, one can rewrite the magnetic quadrupole coupling appearing in Eq.\eqref{eq:multipole_action} in d-dimensions, by specifying the  magnetic component of the Riemann tensor defined in Eq.\eqref{eq:multipole_B}, and by making the following substitution: 
\begin{equation}
    S_{mult}\supset -\frac{2}{3}\int dt J^{ij}B_{ij} =  -\frac{1}{3}\int dt J^{ij}\epsilon_{ikl}R_{0jkl}\ \ \to \ \   -\frac{1}{3}\int \ dt J^{k|jl}R_{0jkl}  \ , 
\end{equation}
where $J^{k|jl}$ is the d-dimensional generalization of the magnetic quadrupole,  it is symmetric trace free and antisymmetric under $k\leftrightarrow l $, and in the limit $d\to 3$ it reproduces the known coupling: 
\begin{equation}
    J^{k|jl}  \ \overset{\mathrm{d=3}}{=} \ J^{ij}\epsilon_{ikl} \ . 
\end{equation} 
By using this new coupling, one can recompute the magnetic quadrupole tail obtaining: 
\begin{eqnarray}
    S^{EJ^2}_{eff\ 5PN}& = &-\frac{16}{135}G_N^2 E\int_{-\infty}^{\infty}\frac{dk_0}{(2\pi)}k_0^6\left[\frac{1}{\epsilon}-\frac{49}{20}-i \pi + \log\left(\frac{k_0^2}{\bar{\mu}^2}\right)+\mathcal{O}(\epsilon)\right]\times \nonumber \\ 
    & & \times \biggl(J_{i|jk}(k_0)J^{i|jk}(-k_0)+J_{i|jk}(k_0)J^{j|ik} (-k_0)\biggr) \ , 
   \end{eqnarray}
which agrees with Eq.(9) of \cite{Almeida:2021xwn}.
\subsection{Angular momentum "failed" tail: $LQ^2$}
The Angular momentum "failed" tail is contributing at 5PN to the conservative dynamics, and it contains the multipoles $LQ^2$:
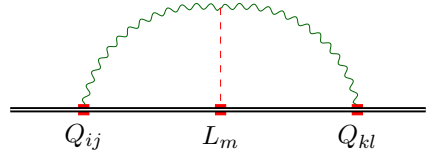
\begin{figure}[H]
\centering
 \scalebox{0.9}{\begin{tikzpicture} 
\begin{feynman}
\vertex (a1) ;
\vertex[right=1cm of a1, square dot, red] (a2) {}; 
\vertex[right=2cm of a2, square dot, red] (a3) {}; 
\vertex[above=1.5cm of a3] (b3);
\vertex[right=2cm of a3, square dot, red] (a4) {};
\vertex[right=1cm of a4] (a5); 
\vertex[below=0.4cm of a3] (e3) {\(L_m\)};
\vertex[below=0.4cm of a2] (f3) {\(Q_{ij}\)};
\vertex[below=0.4cm of a4] (g3) {\(Q_{kl}\)};
\diagram* { 
(a1) -- [double,thick] (a5),
(a2) -- [boson, black!60!green,quarter left] (b3),
(b3) -- [boson, black!60!green,quarter left] (a4),
(b3) -- [scalar, red] (a3)
};
\end{feynman} 
\end{tikzpicture}}
\caption{\textit{Angular Momentum "failed" tail}}
\end{figure}
In this case $L_i$ is conserved and we can send:
\begin{eqnarray}
L_i(q_0)\to L_i (2\pi)\delta(q_0)
\end{eqnarray}
and the propagator of the corresponding line becomes instantaneous.
\subsubsection{Contributing diagrams}
We need now to specify the polarizations, there are 6 diagrams contributing as only the $A$ field couples with the angular momentum $L_i$:
\begin{eqnarray}
 \scalebox{0.6}{\begin{tikzpicture} [baseline=(a1)] 
\begin{feynman}
\vertex (a1) ;
\vertex[right=1cm of a1, square dot, red] (a2) {}; 
\vertex[right=2cm of a2, square dot, red] (a3) {}; 
\vertex[above=1.5cm of a3] (b3);
\vertex[right=2cm of a3, square dot, red] (a4) {};
\vertex[right=1cm of a4] (a5); 
\vertex[below=0.4cm of a3] (e3) {\(L_m\)};
\vertex[below=0.4cm of a2] (f3) {\(Q_{ij}\)};
\vertex[below=0.4cm of a4] (g3) {\(Q_{kl}\)};
\diagram* { 
(a1) -- [double,thick] (a5),
(a2) -- [boson,quarter left] (b3),
(b3) -- [boson,quarter left] (a4),
(b3) -- [scalar] (a3)
};
\end{feynman} 
\end{tikzpicture}}& =&  \scalebox{0.5}{\begin{tikzpicture}[baseline=(a1)]  
\begin{feynman}
\vertex (a1) ;
\vertex[right=1cm of a1, square dot, red] (a2) {}; 
\vertex[right=2cm of a2, square dot, red] (a3) {}; 
\vertex[above=1.5cm of a3] (b3);
\vertex[right=2cm of a3, square dot, red] (a4) {};
\vertex[right=1cm of a4] (a5); 
\vertex[below=0.4cm of a3] (e3) {\(L_m\)};
\vertex[below=0.4cm of a2] (f3) {\(Q_{ij}\)};
\vertex[below=0.4cm of a4] (g3) {\(Q_{kl}\)};
\diagram* { 
(a1) -- [double,thick] (a5),
(a2) -- [double_boson,black!60!green,quarter left] (b3),
(b3) -- [double_boson,black!60!green,quarter left] (a4),
(b3) -- [boson, red] (a3)
};
\end{feynman} 
\end{tikzpicture}}+\scalebox{0.5}{\begin{tikzpicture}[baseline=(a1)]  
\begin{feynman}
\vertex (a1) ;
\vertex[right=1cm of a1, square dot, red] (a2) {}; 
\vertex[right=2cm of a2, square dot, red] (a3) {}; 
\vertex[above=1.5cm of a3] (b3);
\vertex[right=2cm of a3, square dot, red] (a4) {};
\vertex[right=1cm of a4] (a5); 
\vertex[below=0.4cm of a3] (e3) {\(L_m\)};
\vertex[below=0.4cm of a2] (f3) {\(Q_{ij}\)};
\vertex[below=0.4cm of a4] (g3) {\(Q_{kl}\)};
\diagram* { 
(a1) -- [double,thick] (a5),
(a2) -- [boson,red,quarter left] (b3),
(b3) -- [double_boson,black!60!green,quarter left] (a4),
(b3) -- [boson,red] (a3)
};
\end{feynman} 
\end{tikzpicture}}+\scalebox{0.5}{\begin{tikzpicture}[baseline=(a1)] 
\begin{feynman}
\vertex (a1) ;
\vertex[right=1cm of a1, square dot, red] (a2) {}; 
\vertex[right=2cm of a2, square dot, red] (a3) {}; 
\vertex[above=1.5cm of a3] (b3);
\vertex[right=2cm of a3, square dot, red] (a4) {};
\vertex[right=1cm of a4] (a5); 
\vertex[below=0.4cm of a3] (e3) {\(L_m\)};
\vertex[below=0.4cm of a2] (f3) {\(Q_{ij}\)};
\vertex[below=0.4cm of a4] (g3) {\(Q_{kl}\)};
\diagram* { 
(a1) -- [double,thick] (a5),
(a2) -- [double_boson,black!60!green,quarter left] (b3),
(b3) -- [scalar,blue,quarter left] (a4),
(b3) -- [boson,red] (a3)
};
\end{feynman} 
\end{tikzpicture}}\nonumber\\
& & +\scalebox{0.5}{\begin{tikzpicture}[baseline=(a1)]  
\begin{feynman}
\vertex (a1) ;
\vertex[right=1cm of a1, square dot, red] (a2) {}; 
\vertex[right=2cm of a2, square dot, red] (a3) {}; 
\vertex[above=1.5cm of a3] (b3);
\vertex[right=2cm of a3, square dot, red] (a4) {};
\vertex[right=1cm of a4] (a5); 
\vertex[below=0.4cm of a3] (e3) {\(L_m\)};
\vertex[below=0.4cm of a2] (f3) {\(Q_{ij}\)};
\vertex[below=0.4cm of a4] (g3) {\(Q_{kl}\)};
\diagram* { 
(a1) -- [double,thick] (a5),
(a2) -- [boson,red,quarter left] (b3),
(b3) -- [boson,red,quarter left] (a4),
(b3) -- [boson,red] (a3)
};
\end{feynman} 
\end{tikzpicture}}+\scalebox{0.5}{\begin{tikzpicture}[baseline=(a1)]  
\begin{feynman}
\vertex (a1) ;
\vertex[right=1cm of a1, square dot, red] (a2) {}; 
\vertex[right=2cm of a2, square dot, red] (a3) {}; 
\vertex[above=1.5cm of a3] (b3);
\vertex[right=2cm of a3, square dot, red] (a4) {};
\vertex[right=1cm of a4] (a5); 
\vertex[below=0.4cm of a3] (e3) {\(L_m\)};
\vertex[below=0.4cm of a2] (f3) {\(Q_{ij}\)};
\vertex[below=0.4cm of a4] (g3) {\(Q_{kl}\)};
\diagram* { 
(a1) -- [double,thick] (a5),
(a2) -- [boson,red,quarter left] (b3),
(b3) -- [scalar,blue,quarter left] (a4),
(b3) -- [boson,red] (a3)
};
\end{feynman} 
\end{tikzpicture}}+\scalebox{0.5}{\begin{tikzpicture}[baseline=(a1)]  
\begin{feynman}
\vertex (a1) ;
\vertex[right=1cm of a1, square dot, red] (a2) {}; 
\vertex[right=2cm of a2, square dot, red] (a3) {}; 
\vertex[above=1.5cm of a3] (b3);
\vertex[right=2cm of a3, square dot, red] (a4) {};
\vertex[right=1cm of a4] (a5); 
\vertex[below=0.4cm of a3] (e3) {\(L_m\)};
\vertex[below=0.4cm of a2] (f3) {\(Q_{ij}\)};
\vertex[below=0.4cm of a4] (g3) {\(Q_{kl}\)}; 
\diagram* { 
(a1) -- [double,thick] (a5),
(a2) -- [scalar,blue,quarter left] (b3),
(b3) -- [scalar,blue,quarter left] (a4),
(b3) -- [boson,red] (a3)
};
\end{feynman} 
\end{tikzpicture}}
\end{eqnarray}
The total amplitude is given by:
\begin{eqnarray}
 \mathcal{M}^{LQ^2}\ = \ \mathcal{M}_{A\sigma^2}^{LQ^2}+\mathcal{M}_{\phi A\sigma}^{LQ^2}+\mathcal{M}_{A^2\sigma}^{LQ^2}+\mathcal{M}_{ A^3}^{LQ^2}+\mathcal{M}_{\phi^2 A}^{LQ^2}+\mathcal{M}_{A^2\phi}^{LQ^2} 
\end{eqnarray}
 \subsubsection{Result}
 After applying the reduction procedure outlined in the previous section we get:
\begin{eqnarray}
 S^{LQ^2}_{eff\ 5PN}= -\frac{i G_N^2}{15}\int_{-\infty}^{\infty}\frac{dk_0}{(2\pi)}k_0^7L_m\epsilon_{mik}Q_{ij}(k_0)Q_{kj}(-k_0)
\end{eqnarray}
and if we move in time domain the effective action becomes:
\begin{eqnarray}
 S^{LQ^2}_{eff\ 5PN}= -\frac{8}{15}G_N^2\int dt \ddddot{Q}_{ij}\dddot{Q}_{jl}\epsilon_{ijk}L_k
\end{eqnarray}
This result agrees with Eq.(24) of \cite{Foffa:2019eeb}, where we notice that we are using a different definition for the angular momentum $\Vec{L}$, in particular: $\Vec{L}=-\Vec{L}^{Ref.[81]}$. This diagram involving the conserved angular momentum $\Vec{L}$ can be dubbed as "failed" tail because it gives just an instantaneous contribution to the conservative dynamics.\\
Indeed, for this diagram at least one of the graviton polarizations must be an $A$, and it presents a gradient coupling proportional to momentum $\mathbf{q}$ that kills any divergence of the amplitude, which in the case of $q_0=0$ occurs for $\mathbf{q}\to 0$. \\
Then all the diagrams involving the conserved quantity $\Vec{L}$ and any higher multipole are qualitatively different from the ones involving $E$ in that the former are real, finite and local. 
\section{Memory Effects}
Let us now compute the memory term diagram contributing at 5PN, which is the $QQQ$ diagram. \\
In this case there are 3 mass quadrupoles, which are not conserved quantities, hence no simplification occurs.
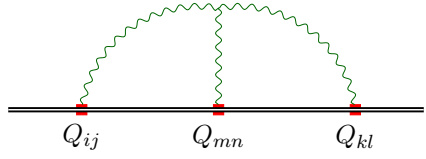
\begin{figure}[H]
\centering
 \scalebox{0.9}{\begin{tikzpicture} 
\begin{feynman}
\vertex (a1) ;
\vertex[right=1cm of a1,square dot, red] (a2) {}; 
\vertex[right=2cm of a2,square dot, red] (a3) {}; 
\vertex[above=1.5cm of a3] (b3);
\vertex[right=2cm of a3,square dot, red] (a4) {};
\vertex[right=1cm of a4] (a5); 
\vertex[below=0.4cm of a3] (e3) {\(Q_{mn}\)};
\vertex[below=0.4cm of a2] (f3) {\(Q_{ij}\)};
\vertex[below=0.4cm of a4] (g3) {\(Q_{kl}\)};
\diagram* { 
(a1) -- [double,thick] (a5),
(a2) -- [boson, black!60!green, quarter left] (b3),
(b3) -- [boson, black!60!green, quarter left] (a4),
(b3) -- [boson, black!60!green] (a3)
};
\end{feynman} 
\end{tikzpicture}}
\caption{\textit{Memory term}}
\end{figure}
In this case the right description for the problem is the use of the in-in formalism since we are dealing with radiation. However, as discussed in Sec.\ref{sec:boundary_conditions_far}, we can deal with this problem by using retarded BC for two propagators and advanced for the third one.
Following \cite{Foffa:2019eeb}, we use retarded boundary conditions for $k,q$ and advanced ones for $p$, and at the of the computation we will take $p_0=-k_0-q_0$.
The generic form of the amplitude in this case is: 
\begin{eqnarray}
 A_{her}^{Q^3} &=& \int \frac{dk_0}{(2\pi)}\frac{dq_0}{(2\pi)}\Biggl(\scalebox{0.4}{\begin{tikzpicture}[baseline=(a1)] 
    \begin{feynman}
    \vertex (a1) ;
    \vertex[right=3cm of a1] (a2); 
    \vertex[below=1.4cm of a1] (a3); 
    \vertex[right=0.5cm of a3] (a4); 
    \vertex[right=2.5cm of a3] (a5); 
    \vertex[above=1.4cm of a1] (a7);
    \vertex[right=1.5cm of a7] (a8);
    
    \diagram* {
    (a1) -- [black!60!green, boson, ultra thick, half left] (a2),
    (a1) -- [black!60!green,boson, ultra thick, half right] (a2),
    
    (a4) -- [double, ultra thick] (a5)
    }; 
    \vertex[right=1.5cm of a3, dot, red] (a6) {};
    \diagram* {
    (a6) -- [black!60!green, boson, ultra thick] (a8)
    }; 
    \end{feynman} 
    \end{tikzpicture}}\Biggr)  \\ 
 \mathcal{M}_{her}^{Q^3} &= & 
 \scalebox{0.4}{\begin{tikzpicture}[baseline=(a1)] 
    \begin{feynman}
    \vertex (a1) ;
    \vertex[right=3cm of a1] (a2); 
    \vertex[below=1.4cm of a1] (a3); 
    \vertex[right=0.5cm of a3] (a4); 
    \vertex[right=2.5cm of a3] (a5); 
    \vertex[above=1.4cm of a1] (a7);
    \vertex[right=1.5cm of a7] (a8);
    
    \diagram* {
    (a1) -- [black!60!green, boson, ultra thick, half left] (a2),
    (a1) -- [black!60!green,boson, ultra thick, half right] (a2),
    
    (a4) -- [double, ultra thick] (a5)
    }; 
    \vertex[right=1.5cm of a3, dot, red] (a6) {};
    \diagram* {
    (a6) -- [black!60!green, boson, ultra thick] (a8)
    }; 
    \end{feynman} 
    \end{tikzpicture}} \nonumber \\ 
    &=& 
    \int\frac{d^d\mathbf{k}}{(2\pi)^d}\frac{d^d\mathbf{q}}{(2\pi)^d} Q^{(1)}_{i_1...i_l}(q_0)Q^{(2)}_{j1,...,j_m}(k_0)Q^{(3)}_{k_1...k_n}(p_{0})\nonumber  \frac{N^{i_1...i_l\ j_1...j_m \ k_1...k_n}(\mathbf{k},\mathbf{q},k_0,q_0,p_0)}{D_1D_2D_3} 
\end{eqnarray}
where the denominators are defined as:
\begin{eqnarray}
    D_1= [\mathbf{k}^2-(k_0+ia )^2] \ ,  \qquad D_2=   [(\mathbf{k}+\mathbf{q})^2-(p_0-ia)^2]\ , \qquad D_3 =  [\mathbf{q}^2-(q_0+ia)^2]  \ .
\end{eqnarray}
\subsection{Contributing diagrams}
There are 27 different diagrams contributing, but we can compute only 10 of them and symmetrize over the possible choices of momenta. \\
We will assign symmetry factor $1/2$ for diagrams with 2 identical internal legs and $1/6$ for diagrams with 3 identical graviton polarizations.
\begin{eqnarray}
 \scalebox{0.6}{\begin{tikzpicture} [baseline=(a1)] 
\begin{feynman}
\vertex (a1) ;
\vertex[right=1cm of a1, square dot, red] (a2) {}; 
\vertex[right=2cm of a2, square dot, red] (a3) {}; 
\vertex[above=1.5cm of a3] (b3);
\vertex[right=2cm of a3, square dot, red] (a4) {};
\vertex[right=1cm of a4] (a5); 
\vertex[below=0.4cm of a3] (e3) {\(Q_{mn}\)};
\vertex[below=0.4cm of a2] (f3) {\(Q_{ij}\)};
\vertex[below=0.4cm of a4] (g3) {\(Q_{kl}\)};
\diagram* { 
(a1) -- [double,thick] (a5),
(a2) -- [boson,quarter left] (b3),
(b3) -- [boson,quarter left] (a4),
(b3) -- [boson] (a3)
};
\end{feynman} 
\end{tikzpicture}}& =&  
\scalebox{0.5}{\begin{tikzpicture}[baseline=(a1)]  
\begin{feynman}
\vertex (a1) ;
\vertex[right=1cm of a1, square dot, red] (a2) {}; 
\vertex[right=2cm of a2, square dot, red] (a3) {}; 
\vertex[above=1.5cm of a3] (b3);
\vertex[right=2cm of a3, square dot, red] (a4) {};
\vertex[right=1cm of a4] (a5); 
\vertex[below=0.4cm of a3] (e3) {\(Q_{mn}\)};
\vertex[below=0.4cm of a2] (f3) {\(Q_{ij}\)};
\vertex[below=0.4cm of a4] (g3) {\(Q_{kl}\)};
\diagram* { 
(a1) -- [double,thick] (a5),
(a2) -- [double_boson,black!60!green,quarter left] (b3),
(b3) -- [double_boson,black!60!green,quarter left] (a4),
(b3) -- [double_boson,black!60!green] (a3)
};
\end{feynman} 
\end{tikzpicture}}+
\scalebox{0.5}{\begin{tikzpicture}[baseline=(a1)]  
\begin{feynman}
\vertex (a1) ;
\vertex[right=1cm of a1, square dot, red] (a2) {}; 
\vertex[right=2cm of a2, square dot, red] (a3) {}; 
\vertex[above=1.5cm of a3] (b3);
\vertex[right=2cm of a3, square dot, red] (a4) {};
\vertex[right=1cm of a4] (a5); 
\vertex[below=0.4cm of a3] (e3) {\(Q_{mn}\)};
\vertex[below=0.4cm of a2] (f3) {\(Q_{ij}\)};
\vertex[below=0.4cm of a4] (g3) {\(Q_{kl}\)};
\diagram* { 
(a1) -- [double,thick] (a5),
(a2) -- [double_boson,black!60!green,quarter left] (b3),
(b3) -- [double_boson,black!60!green,quarter left] (a4),
(b3) -- [boson,red] (a3)
};
\end{feynman} 
\end{tikzpicture}}+
\scalebox{0.5}{\begin{tikzpicture}[baseline=(a1)] 
\begin{feynman}
\vertex (a1) ;
\vertex[right=1cm of a1, square dot, red] (a2) {}; 
\vertex[right=2cm of a2, square dot, red] (a3) {}; 
\vertex[above=1.5cm of a3] (b3);
\vertex[right=2cm of a3, square dot, red] (a4) {};
\vertex[right=1cm of a4] (a5); 
\vertex[below=0.4cm of a3] (e3) {\(Q_{mn}\)};
\vertex[below=0.4cm of a2] (f3) {\(Q_{ij}\)};
\vertex[below=0.4cm of a4] (g3) {\(Q_{kl}\)};
\diagram* { 
(a1) -- [double,thick] (a5),
(a2) -- [double_boson,black!60!green,quarter left] (b3),
(b3) -- [double_boson,black!60!green,quarter left] (a4),
(b3) -- [scalar,blue] (a3)
};
\end{feynman} 
\end{tikzpicture}}\nonumber\\
& & +\scalebox{0.5}{\begin{tikzpicture}[baseline=(a1)]  
\begin{feynman}
\vertex (a1) ;
\vertex[right=1cm of a1, square dot, red] (a2) {}; 
\vertex[right=2cm of a2, square dot, red] (a3) {}; 
\vertex[above=1.5cm of a3] (b3);
\vertex[right=2cm of a3, square dot, red] (a4) {};
\vertex[right=1cm of a4] (a5); 
\vertex[below=0.4cm of a3] (e3) {\(Q_{mn}\)};
\vertex[below=0.4cm of a2] (f3) {\(Q_{ij}\)};
\vertex[below=0.4cm of a4] (g3) {\(Q_{kl}\)};
\diagram* { 
(a1) -- [double,thick] (a5),
(a2) -- [double_boson,black!60!green,quarter left] (b3),
(b3) -- [boson,red,quarter left] (a4),
(b3) -- [scalar,blue] (a3)
};
\end{feynman} 
\end{tikzpicture}}+
\scalebox{0.5}{\begin{tikzpicture}[baseline=(a1)]  
\begin{feynman}
\vertex (a1) ;
\vertex[right=1cm of a1, square dot, red] (a2) {}; 
\vertex[right=2cm of a2, square dot, red] (a3) {}; 
\vertex[above=1.5cm of a3] (b3);
\vertex[right=2cm of a3, square dot, red] (a4) {};
\vertex[right=1cm of a4] (a5); 
\vertex[below=0.4cm of a3] (e3) {\(Q_{mn}\)};
\vertex[below=0.4cm of a2] (f3) {\(Q_{ij}\)};
\vertex[below=0.4cm of a4] (g3) {\(Q_{kl}\)};
\diagram* { 
(a1) -- [double,thick] (a5),
(a2) -- [double_boson,black!60!green,quarter left] (b3),
(b3) -- [boson,red,quarter left] (a4),
(b3) -- [boson,red] (a3)
};
\end{feynman} 
\end{tikzpicture}}+
\scalebox{0.5}{\begin{tikzpicture}[baseline=(a1)]  
\begin{feynman}
\vertex (a1) ;
\vertex[right=1cm of a1, square dot, red] (a2) {}; 
\vertex[right=2cm of a2, square dot, red] (a3) {}; 
\vertex[above=1.5cm of a3] (b3);
\vertex[right=2cm of a3, square dot, red] (a4) {};
\vertex[right=1cm of a4] (a5); 
\vertex[below=0.4cm of a3] (e3) {\(Q_{mn}\)};
\vertex[below=0.4cm of a2] (f3) {\(Q_{ij}\)};
\vertex[below=0.4cm of a4] (g3) {\(Q_{kl}\)};
\diagram* { 
(a1) -- [double,thick] (a5),
(a2) -- [double_boson,black!60!green,quarter left] (b3),
(b3) -- [scalar,blue,quarter left] (a4),
(b3) -- [scalar,blue] (a3)
};
\end{feynman} 
\end{tikzpicture}}\nonumber\\
& & +\scalebox{0.5}{\begin{tikzpicture}[baseline=(a1)]  
\begin{feynman}
\vertex (a1) ;
\vertex[right=1cm of a1, square dot, red] (a2) {}; 
\vertex[right=2cm of a2, square dot, red] (a3) {}; 
\vertex[above=1.5cm of a3] (b3);
\vertex[right=2cm of a3, square dot, red] (a4) {};
\vertex[right=1cm of a4] (a5); 
\vertex[below=0.4cm of a3] (e3) {\(Q_{mn}\)};
\vertex[below=0.4cm of a2] (f3) {\(Q_{ij}\)};
\vertex[below=0.4cm of a4] (g3) {\(Q_{kl}\)};
\diagram* { 
(a1) -- [double,thick] (a5),
(a2) -- [boson,red,quarter left] (b3),
(b3) -- [boson,red,quarter left] (a4),
(b3) -- [boson,red] (a3)
};
\end{feynman} 
\end{tikzpicture}}+\scalebox{0.5}{\begin{tikzpicture}[baseline=(a1)]  
\begin{feynman}
\vertex (a1) ;
\vertex[right=1cm of a1, square dot, red] (a2) {}; 
\vertex[right=2cm of a2, square dot, red] (a3) {}; 
\vertex[above=1.5cm of a3] (b3);
\vertex[right=2cm of a3, square dot, red] (a4) {};
\vertex[right=1cm of a4] (a5); 
\vertex[below=0.4cm of a3] (e3) {\(Q_{mn}\)};
\vertex[below=0.4cm of a2] (f3) {\(Q_{ij}\)};
\vertex[below=0.4cm of a4] (g3) {\(Q_{kl}\)};
\diagram* { 
(a1) -- [double,thick] (a5),
(a2) -- [boson,red,quarter left] (b3),
(b3) -- [boson,red,quarter left] (a4),
(b3) -- [scalar,blue] (a3)
};
\end{feynman} 
\end{tikzpicture}}+\scalebox{0.5}{\begin{tikzpicture}[baseline=(a1)]  
\begin{feynman}
\vertex (a1) ;
\vertex[right=1cm of a1, square dot, red] (a2) {}; 
\vertex[right=2cm of a2, square dot, red] (a3) {}; 
\vertex[above=1.5cm of a3] (b3);
\vertex[right=2cm of a3, square dot, red] (a4) {};
\vertex[right=1cm of a4] (a5); 
\vertex[below=0.4cm of a3] (e3) {\(Q_{mn}\)};
\vertex[below=0.4cm of a2] (f3) {\(Q_{ij}\)};
\vertex[below=0.4cm of a4] (g3) {\(Q_{kl}\)};
\diagram* { 
(a1) -- [double,thick] (a5),
(a2) -- [boson,red,quarter left] (b3),
(b3) -- [scalar,blue,quarter left] (a4),
(b3) -- [scalar,blue] (a3)
};
\end{feynman} 
\end{tikzpicture}}\nonumber\\
& & +\scalebox{0.5}{\begin{tikzpicture}[baseline=(a1)]  
\begin{feynman}
\vertex (a1) ;
\vertex[right=1cm of a1, square dot, red] (a2) {}; 
\vertex[right=2cm of a2, square dot, red] (a3) {}; 
\vertex[above=1.5cm of a3] (b3);
\vertex[right=2cm of a3, square dot, red] (a4) {};
\vertex[right=1cm of a4] (a5); 
\vertex[below=0.4cm of a3] (e3) {\(Q_{mn}\)};
\vertex[below=0.4cm of a2] (f3) {\(Q_{ij}\)};
\vertex[below=0.4cm of a4] (g3) {\(Q_{kl}\)};
\diagram* { 
(a1) -- [double,thick] (a5),
(a2) -- [scalar,blue,quarter left] (b3),
(b3) -- [scalar,blue,quarter left] (a4),
(b3) -- [scalar,blue] (a3)
};
\end{feynman} 
\end{tikzpicture}}
\end{eqnarray}

The corresponding contribution to the effective action is given by:

\begin{eqnarray}
 S_{EFF\ 5PN}^{Q^3}= -i \int_{-\infty}^{\infty}\frac{dk^0}{(2\pi)}\int_{-\infty}^{\infty}\frac{dq^0}{(2\pi)}\left(\mathcal{M}^{Q^3}\right)
\end{eqnarray}
where:

\begin{eqnarray}
\mathcal{M}^{Q^3} & = &\biggl(V(k,p,q)+V(p,q,k)+V(q,k,p)\biggr)  \\ 
 V(k,p,q)& =& \biggl(\mathcal{M}_{\sigma^3}^{Q^3}+\mathcal{M}_{\sigma^2 A}^{Q^3}+\mathcal{M}_{\sigma^2\phi}^{Q^3}+\mathcal{M}_{ \phi A\sigma}^{Q^3}+\mathcal{M}_{\sigma A^2}^{Q^3} \nonumber \\
 & & +\mathcal{M}_{\sigma \phi^2}^{Q^3}+\mathcal{M}_{A^3}^{Q^3}+\mathcal{M}_{A^2\phi}^{Q^3}+\mathcal{M}_{A\phi^2}^{Q^3}+\mathcal{M}_{\phi^3}^{Q^3} \biggr)
\end{eqnarray}
We can substitute the Feynman rules, apply tensor contractions and tensor decomposition to get:
\begin{eqnarray} 
    \mathcal{M}^{Q^3}=Q_{ij}(k_0)Q_{jk}(q_0)Q_{ki}(p_0)\tilde{\mathcal{M}}^{Q^3} \ . 
\end{eqnarray}
\subsubsection{IBP decomposition}
$\tilde{\mathcal{M}}^{Q^3}$ is an expression in terms of scalar integrals, belonging to the same topology $j_{n_1,n_2,n_3}$ defined as:
\begin{eqnarray}
 j_{n_1,n_2,n_3} \ =  \scalebox{0.8}{\begin{tikzpicture}[baseline=(current bounding box.center)]\begin{feynman}
    \vertex (a);
    \vertex[right=0.2cm of a] (b);
    \vertex[right=1.5 cm of b] (c); 
    \vertex[right=1.5 cm of c] (d);
    \diagram* { 
    (b) -- [half left,very thick ] (c),
    (b) -- [half right,very thick] (c),
    (b) -- [] (c),
    };
    \end{feynman} \end{tikzpicture}} \ = \ \int \frac{d^d\mathbf{k}}{(2\pi)^d}\int \frac{d^d\mathbf{q}}{(2\pi)^d}\frac{1}{(D_1)^{n_1}(D_2)^{n_2}(D_3)^{n_3}} \ , 
\label{eq:topology_memory}
\end{eqnarray}
where, given the prescriptions for the propagators, the denominators are defined as:
\begin{eqnarray}
    D_1= [\mathbf{k}^2-(k_0+ia )^2] \ ,  \qquad D_2=   [(\mathbf{k}+\mathbf{q})^2-(p_0-ia)^2]\ , \qquad D_3 =  [\mathbf{q}^2-(q_0+ia)^2]  \ .
\end{eqnarray}
At this point we can IBP decompose the scalar integrals appearing, since this procedure is not affected by the choice of boundary conditions.
The total amplitude can then be written as a linear combination of 4 Master Integrals:
\begin{eqnarray}
    \tilde{\mathcal{M}}^{Q^3}=C_{1}^{Q^3}\ \scalebox{0.9}{\begin{tikzpicture}[baseline=(current bounding box.center)]             \begin{feynman}
        \vertex (a) ;
        \vertex[right=0.2cm of a] (b);
        \vertex[right=1cm of b] (c); 
        \vertex[right=1cm of c] (d);
        \diagram* { 
        (b) -- [half left,very thick, edge label'=\(1\) ] (c),
        (b) -- [half right,very thick ] (c),
        (c) -- [half left,very thick,edge label'=\(2\) ] (d),
        (c) -- [half right,very thick ] (d),
        };
        \end{feynman} \end{tikzpicture}}+C_{2}^{Q^3}\ \scalebox{0.9}{\begin{tikzpicture}[baseline=(current bounding box.center)]             \begin{feynman}
            \vertex (a) ;
            \vertex[right=0.2cm of a] (b);
            \vertex[right=1cm of b] (c); 
            \vertex[right=1cm of c] (d);
            \diagram* { 
            (b) -- [half left,very thick ,edge label'=\(1\)] (c),
            (b) -- [half right,very thick ] (c),
            (c) -- [half left,very thick ,edge label'=\(3\)] (d),
            (c) -- [half right,very thick ] (d),
            };
            \end{feynman} \end{tikzpicture}}+C_{3}^{Q^3}\ \scalebox{0.9}{\begin{tikzpicture}[baseline=(current bounding box.center)]             \begin{feynman}
                \vertex (a) ;
                \vertex[right=0.2cm of a] (b);
                \vertex[right=1cm of b] (c); 
                \vertex[right=1cm of c] (d);
                \diagram* { 
                (b) -- [half left,very thick ,edge label'=\(2\)] (c),
                (b) -- [half right,very thick ] (c),
                (c) -- [half left,very thick ,edge label'=\(3\)] (d),
                (c) -- [half right,very thick ] (d),
                };
                \end{feynman} \end{tikzpicture}}+C_{4}^{Q^3}\ \scalebox{0.9}{\begin{tikzpicture}[baseline=(current bounding box.center)]             \begin{feynman}
                    \vertex (a);
                    \vertex[right=0.2cm of a] (b);
                    \vertex[right=1cm of b] (c); 
                    \vertex[right=1cm of c] (d);
                    \diagram* { 
                    (b) -- [half left,very thick ] (c),
                    (b) -- [half right,very thick ] (c),
                    (b) -- [very thick] (c),
                    };
                    \end{feynman} \end{tikzpicture}} \ ,
\label{eq:memory_decomposed}
\end{eqnarray}
\subsubsection{Coefficients}
The coefficients, writing $d=3+\epsilon$ are given by: 
\begin{eqnarray}
    C_1^{Q^3} & = & 
    \frac{k_0}{384 \Lambda ^4 (\epsilon +2)^2 (\epsilon +3)
       (\epsilon +5) (\epsilon +7)} \biggl[k_0^3 q_0^2 \left(\epsilon ^5+18 \epsilon ^4+107 \epsilon ^3+308 \epsilon ^2+406 \epsilon +200\right)\nonumber \\
       & & 
       +2 k_0^2 q_0^3
       \left(\epsilon ^5+18 \epsilon ^4+115 \epsilon ^3+351 \epsilon ^2+507 \epsilon +284\right) \nonumber\\
       & & 
       -2 k_0^4 q_0 \left(8 \epsilon ^3+61 \epsilon ^2+163 \epsilon
       +140\right)  -2 q_0^5 \left(\epsilon ^3+8 \epsilon ^2+19 \epsilon +12\right)\nonumber \\ 
       & &
       -2 k_0^5 \left(4 \epsilon ^3+35 \epsilon ^2+97 \epsilon +84\right)+k_0 q_0^4 \left(\epsilon ^5+18 \epsilon ^4+113 \epsilon ^3+326
       \epsilon ^2+438 \epsilon +232\right)\biggr]\\
    C_2^{Q^3}  & = &  -\frac{k_0 q_0 }{384 \Lambda ^4 (\epsilon +2)^2 (\epsilon +3) (\epsilon +5) (\epsilon +7)}\biggl[2 k_0^2 q_0^2 \bigl(\epsilon ^5+18 \epsilon ^4+121 \epsilon ^3+381 \epsilon ^2+559 \epsilon \nonumber \\
    & & +300\bigr)+k_0^3
       q_0 \left(\epsilon ^5+18 \epsilon ^4+123 \epsilon ^3+406 \epsilon ^2+628 \epsilon +352\right)\nonumber \\
       & &+2 k_0^4 \left(\epsilon ^3+8 \epsilon ^2+19 \epsilon
       +12\right) +k_0 q_0^3 \left(\epsilon ^5+18 \epsilon ^4+115 \epsilon ^3+318 \epsilon ^2+372 \epsilon +128\right)\nonumber \\
       & &+2 q_0^4 \left(\epsilon ^3+8 \epsilon
       ^2+19 \epsilon +12\right)\biggr] \\
    C_3^{Q^3} & = & \frac{q_0}{384 \Lambda ^4 (\epsilon +2)^2 (\epsilon +3) (\epsilon
       +5) (\epsilon +7)} \biggl[-2 k_0^3 q_0^2 \left(\epsilon ^5+18 \epsilon ^4+99
       \epsilon ^3+175 \epsilon ^2-5 \epsilon -164\right)\\
       & & -k_0^2 q_0^3
       \left(\epsilon ^5+18 \epsilon ^4+67 \epsilon ^3-132 \epsilon ^2-874 \epsilon
       -920\right)-k_0^4 q_0 \bigl(\epsilon ^5+18 \epsilon ^4\nonumber \\ 
      & &   +105 \epsilon
       ^3+238 \epsilon ^2+182 \epsilon +8\bigr) +2 k_0^5 \left(\epsilon ^3+8 \epsilon
       ^2+19 \epsilon +12\right)\nonumber \\
       & & +2 k_0 q_0^4 \left(16 \epsilon ^3+149 \epsilon
       ^2+419 \epsilon +364\right)+2 q_0^5 \left(4 \epsilon ^3+35 \epsilon ^2+97
       \epsilon +84\right)\biggr]   \\
      C_4^{Q^3} &=& 0
    \end{eqnarray}
Notice that, since the coefficients are polynomials in $k_0,p_0,q_0$, we could easily replace: $p_0= -k_0-q_0$, since this part is independent on the choice of boundary conditions in the propagators.
We need to be much more careful instead in keeping tracks of $k_0,p_0,q_0$ in the Master Integrals. \\
It is remarkable that the coefficient $C_4^{Q^3}$ is exactly $0$ in $d$-dimensions. 

\subsubsection{Master Integrals}

The Master Integrals are defined as: 
\begin{eqnarray}
    J_1^{MI} & = &  \scalebox{0.8}{\begin{tikzpicture}[baseline=(current bounding box.center)]             \begin{feynman}
        \vertex (a) ;
        \vertex[right=0.2cm of a] (b);
        \vertex[right=1cm of b] (c); 
        \vertex[right=1cm of c] (d);
        \diagram* { 
        (b) -- [half left,very thick, edge label'=\(1\) ] (c),
        (b) -- [half right,very thick ] (c),
        (c) -- [half left,very thick,edge label'=\(2\) ] (d),
        (c) -- [half right,very thick ] (d),
        };
        \end{feynman} \end{tikzpicture}}=   j_{1,1,0}\ = \ \int \frac{d^d\mathbf{k}}{(2\pi)^d}\int \frac{d^d\mathbf{q}}{(2\pi)^d}\frac{1}{D_1D_2}\ ,  \\
        J_2^{MI} & = & \scalebox{0.8}{\begin{tikzpicture}[baseline=(current bounding box.center)]             \begin{feynman}
            \vertex (a) ;
            \vertex[right=0.2cm of a] (b);
            \vertex[right=1cm of b] (c); 
            \vertex[right=1cm of c] (d);
            \diagram* { 
            (b) -- [half left,very thick ,edge label'=\(1\)] (c),
            (b) -- [half right,very thick ] (c),
            (c) -- [half left,very thick ,edge label'=\(3\)] (d),
            (c) -- [half right,very thick ] (d),
            };
            \end{feynman} \end{tikzpicture}}=    j_{1,0,1} \ =\ \int \frac{d^d\mathbf{k}}{(2\pi)^d}\int \frac{d^d\mathbf{q}}{(2\pi)^d}\frac{1}{D_1D_3} \ , \\
            J_3^{MI} & = &  \scalebox{0.8}{\begin{tikzpicture}[baseline=(current bounding box.center)]             \begin{feynman}
                \vertex (a) ;
                \vertex[right=0.2cm of a] (b);
                \vertex[right=1cm of b] (c); 
                \vertex[right=1cm of c] (d);
                \diagram* { 
                (b) -- [half left,very thick ,edge label'=\(2\)] (c),
                (b) -- [half right,very thick ] (c),
                (c) -- [half left,very thick ,edge label'=\(3\)] (d),
                (c) -- [half right,very thick ] (d),
                };
                \end{feynman} \end{tikzpicture}}=  j_{0,1,1} \ =\ \int \frac{d^d\mathbf{k}}{(2\pi)^d}\int \frac{d^d\mathbf{q}}{(2\pi)^d}\frac{1}{D_2D_3} \ , \\
                J_4^{MI} & = &   \scalebox{0.8}{\begin{tikzpicture}[baseline=(current bounding box.center)]             \begin{feynman}
                    \vertex (a);
                    \vertex[right=0.2cm of a] (b);
                    \vertex[right=1cm of b] (c); 
                    \vertex[right=1cm of c] (d);
                    \diagram* { 
                    (b) -- [half left,very thick ] (c),
                    (b) -- [half right,very thick ] (c),
                    (b) -- [very thick] (c),
                    };
                    \end{feynman} \end{tikzpicture}}=   j_{1,1,1} \ =\ \int \frac{d^d\mathbf{k}}{(2\pi)^d}\int \frac{d^d\mathbf{q}}{(2\pi)^d}\frac{1}{D_1D_2D_3} \ ,
                    \label{eq:MI_memory}
    \end{eqnarray}
where $J_1^{MI},J_2^{MI},J_3^{MI}$ are factorizable in a product of two massive tadpoles, whereas $J_4$ is irreducible. \\
The coefficient $C_4^{Q^3}$ is zero, and that hugely simplifies the computation since we just need the explicit expression for the other 3 MI appearing, which are just products of massive tadpoles, with masses $k_0,q_0,p_0$ and are given by: 
\begin{eqnarray}
    J_1^{MI} & = &  \frac{\Gamma \left(1-\frac{d}{2}\right)^2}{(4\pi)^d}(-k_0^2)^{\frac{d-2}{2}} (-p_0^2)^{\frac{d-2}{2}}  \ , \\
    J_2^{MI} & = &  \frac{\Gamma \left(1-\frac{d}{2}\right)^2}{(4\pi)^d}(-k_0^2)^{\frac{d-2}{2}} (-q_0^2)^{\frac{d-2}{2}} \ , \\
    J_3^{MI} & = & \frac{\Gamma \left(1-\frac{d}{2}\right)^2}{(4\pi)^d}(-p_0^2)^{\frac{d-2}{2}} (-q_0^2)^{\frac{d-2}{2}}  \ .
\end{eqnarray}
If 
When we substitute the explicit expression for the MI, and we take the limit $\epsilon\to 0$, we encounter some square root terms of the type $\sqrt{-k_0^2}$, where we must be careful in taking the correct prescriptions, which are given by: 
\begin{eqnarray}
    \sqrt{-k_0^2}  = i k_0 \ , \qquad \sqrt{-q_0^2} =  i q_0 \ ,  \qquad 
    \sqrt{-p_0^2} = - i p_0 \ .  
\end{eqnarray}

 \subsubsection{Result}
We can obtain the final expression by substituting the explicit expression for the MI and taking the limit $\epsilon\to 0$:

\begin{eqnarray}
 S^{Q^3}_{eff\ 5PN}& =& 
 -\frac{G_N^2}{315} \int_{-\infty}^{\infty}\frac{dk_0}{(2\pi)}\frac{dq_0}{(2\pi)} \biggl(k_0^4 \left(-32 p_0^2 q_0^2+28 p_0^3 q_0+21 p_0^4+74 p_0 q_0^3-21 q_0^4\right)\nonumber \\
 & & 
 +32
   k_0^2 p_0^3 q_0^3+2 k_0^3 p_0 q_0^3 (2 p_0+23 q_0)+28 k_0 p_0^4 q_0^3+21 p_0^4
   q_0^4\biggr)Q_{ij}Q_{jk}Q_{ki} 
   \label{eq:memory_momentum}
\end{eqnarray}
If we move to time domain we get:
\begin{eqnarray}
 S^{Q^3}_{eff\ 5PN}= -\frac{G_N^2}{15}\int dt\biggl[ \ddddot{Q}_{ij}\ddddot{Q}_{jk}Q_{ki}+\frac{4}{7}\dddot{Q}_{ij}\dddot{Q}_{jk}\ddot{Q}_{ki}\biggr]
 \label{eq:memory_position}
\end{eqnarray}
This result agrees with Eq.(27) of \cite{Foffa:2019eeb}.

\section{Double Emission diagram}

This section will be dedicated to the evaluation of the double emission diagram:
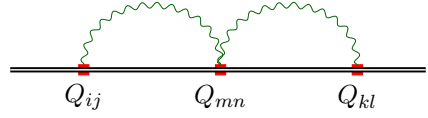
\begin{figure}[H]
    \centering
     \scalebox{0.9}{\begin{tikzpicture} 
    \begin{feynman}
    \vertex (a1) ;
    \vertex[right=1cm of a1, square dot, red] (a2) {}; 
    \vertex[right=2cm of a2, square dot, red] (a3) {}; 
    \vertex[above=1.5cm of a3] (b3);
    \vertex[right=2cm of a3, square dot, red] (a4) {};
    \vertex[right=1cm of a4] (a5); 
    \vertex[below=0.4cm of a3] (e3) {\(Q_{mn}\)};
    \vertex[below=0.4cm of a2] (f3) {\(Q_{ij}\)};
    \vertex[below=0.4cm of a4] (g3) {\(Q_{kl}\)};
    \diagram* { 
    (a1) -- [double,thick] (a5),
    (a2) -- [boson, black!60!green,half left] (a3),
    (a3) -- [boson, black!60!green,half left] (a4),
    };
    \end{feynman} 
    \end{tikzpicture}}
    \caption{\textit{Double emission}}
    \end{figure}
\subsection{Contributing diagrams}
There are six different diagrams contributing to this process, moreover, we should add a symmetry factor 1/2 in dealing with identical sources. 
\begin{eqnarray}
    \scalebox{0.6}{\begin{tikzpicture}[baseline=(a1)] 
        \begin{feynman}
        \vertex (a1) ;
    \vertex[right=1cm of a1, square dot, red] (a2) {}; 
    \vertex[right=2cm of a2, square dot, red] (a3) {}; 
    \vertex[above=1.5cm of a3] (b3);
    \vertex[right=2cm of a3, square dot, red] (a4) {};
    \vertex[right=1cm of a4] (a5); 
    \vertex[below=0.4cm of a3] (e3) {\(Q_{mn}\)};
    \vertex[below=0.4cm of a2] (f3) {\(Q_{ij}\)};
    \vertex[below=0.4cm of a4] (g3) {\(Q_{kl}\)};
        \diagram* { 
        (a1) -- [double,thick] (a5),
        (a2) -- [boson, half left] (a3),
        (a3) -- [boson, half left] (a4),
        };
        \end{feynman} 
        \end{tikzpicture}} & = &  \scalebox{0.5}{\begin{tikzpicture}[baseline=(a1)]  
            \begin{feynman}
            \vertex (a1) ;
    \vertex[right=1cm of a1, square dot, red] (a2) {}; 
    \vertex[right=2cm of a2, square dot, red] (a3) {}; 
    \vertex[above=1.5cm of a3] (b3);
    \vertex[right=2cm of a3, square dot, red] (a4) {};
    \vertex[right=1cm of a4] (a5); 
    \vertex[below=0.4cm of a3] (e3) {\(Q_{mn}\)};
    \vertex[below=0.4cm of a2] (f3) {\(Q_{ij}\)};
    \vertex[below=0.4cm of a4] (g3) {\(Q_{kl}\)};
            \diagram* { 
            (a1) -- [double,thick] (a5),
            (a2) -- [double_boson, black!60!green,half left] (a3),
            (a3) -- [double_boson, black!60!green,half left] (a4),
            };
            \end{feynman} 
            \end{tikzpicture}} + 
            \scalebox{0.5}{\begin{tikzpicture}[baseline=(a1)]
                \begin{feynman}
                \vertex (a1) ;
    \vertex[right=1cm of a1, square dot, red] (a2) {}; 
    \vertex[right=2cm of a2, square dot, red] (a3) {}; 
    \vertex[above=1.5cm of a3] (b3);
    \vertex[right=2cm of a3, square dot, red] (a4) {};
    \vertex[right=1cm of a4] (a5); 
    \vertex[below=0.4cm of a3] (e3) {\(Q_{mn}\)};
    \vertex[below=0.4cm of a2] (f3) {\(Q_{ij}\)};
    \vertex[below=0.4cm of a4] (g3) {\(Q_{kl}\)};
                \diagram* { 
                (a1) -- [double,thick] (a5),
                (a2) -- [boson, red,half left] (a3),
                (a3) -- [boson, red,half left] (a4),
                };
                \end{feynman} 
                \end{tikzpicture}} + \scalebox{0.5}{\begin{tikzpicture}[baseline=(a1)]  
                    \begin{feynman}
                    \vertex (a1) ;
    \vertex[right=1cm of a1, square dot, red] (a2) {}; 
    \vertex[right=2cm of a2, square dot, red] (a3) {}; 
    \vertex[above=1.5cm of a3] (b3);
    \vertex[right=2cm of a3, square dot, red] (a4) {};
    \vertex[right=1cm of a4] (a5); 
    \vertex[below=0.4cm of a3] (e3) {\(Q_{mn}\)};
    \vertex[below=0.4cm of a2] (f3) {\(Q_{ij}\)};
    \vertex[below=0.4cm of a4] (g3) {\(Q_{kl}\)};
                    \diagram* { 
                    (a1) -- [double,thick] (a5),
                    (a2) -- [scalar, blue,half left] (a3),
                    (a3) -- [scalar, blue,half left] (a4),
                    };
                    \end{feynman} 
                    \end{tikzpicture}} \nonumber \\ 
                    & & 
                    +\scalebox{0.5}{\begin{tikzpicture}[baseline=(a1)]  
                        \begin{feynman}
                        \vertex (a1) ;
                    \vertex[right=1cm of a1, square dot, red] (a2) {}; 
                    \vertex[right=2cm of a2, square dot, red] (a3) {}; 
                    \vertex[above=1.5cm of a3] (b3);
                    \vertex[right=2cm of a3, square dot, red] (a4) {};
                    \vertex[right=1cm of a4] (a5); 
                    \vertex[below=0.4cm of a3] (e3) {\(Q_{mn}\)};
                    \vertex[below=0.4cm of a2] (f3) {\(Q_{ij}\)};
                    \vertex[below=0.4cm of a4] (g3) {\(Q_{kl}\)};
                        \diagram* { 
                        (a1) -- [double,thick] (a5),
                        (a2) -- [double_boson, black!60!green,half left] (a3),
                        (a3) -- [boson, red,half left] (a4),
                        };
                        \end{feynman} 
                        \end{tikzpicture}}
                        +\scalebox{0.5}{\begin{tikzpicture}[baseline=(a1)]  
                            \begin{feynman}
                            \vertex (a1) ;
    \vertex[right=1cm of a1, square dot, red] (a2) {}; 
    \vertex[right=2cm of a2, square dot, red] (a3) {}; 
    \vertex[above=1.5cm of a3] (b3);
    \vertex[right=2cm of a3, square dot, red] (a4) {};
    \vertex[right=1cm of a4] (a5); 
    \vertex[below=0.4cm of a3] (e3) {\(Q_{mn}\)};
    \vertex[below=0.4cm of a2] (f3) {\(Q_{ij}\)};
    \vertex[below=0.4cm of a4] (g3) {\(Q_{kl}\)};
                            \diagram* { 
                            (a1) -- [double,thick] (a5),
                            (a2) -- [double_boson, black!60!green,half left] (a3),
                            (a3) -- [scalar, blue,half left] (a4),
                            };
                            \end{feynman} 
                            \end{tikzpicture}}
                            +\scalebox{0.5}{\begin{tikzpicture}[baseline=(a1)]  
                                \begin{feynman}
                                \vertex (a1) ;
    \vertex[right=1cm of a1, square dot, red] (a2) {}; 
    \vertex[right=2cm of a2, square dot, red] (a3) {}; 
    \vertex[above=1.5cm of a3] (b3);
    \vertex[right=2cm of a3, square dot, red] (a4) {};
    \vertex[right=1cm of a4] (a5); 
    \vertex[below=0.4cm of a3] (e3) {\(Q_{mn}\)};
    \vertex[below=0.4cm of a2] (f3) {\(Q_{ij}\)};
    \vertex[below=0.4cm of a4] (g3) {\(Q_{kl}\)};
                                \diagram* { 
                                (a1) -- [double,thick] (a5),
                                (a2) -- [boson, red,half left] (a3),
                                (a3) -- [scalar, blue,half left] (a4),
                                };
                                \end{feynman} 
                                \end{tikzpicture}}
\end{eqnarray}
The corresponding contribution to the effective action is given by: 
\begin{equation}
    S_{EFF \ 5PN}^{Q^3 2} = -i \int_{-\infty}^{\infty}\frac{dk_0}{2\pi}\int_{-\infty}^{\infty}\frac{dq_0}{2\pi}\mathcal{M}^{Q^3 2} \ , 
\end{equation}
where: 
\begin{equation}
    \mathcal{M}^{Q^3 2} = \biggl( \mathcal{M}^{Q^3 2}_{\sigma^2}+\mathcal{M}^{Q^3 2}_{A^2} + \mathcal{M}^{Q^3 2}_{\phi^2}+\mathcal{M}^{Q^3 2}_{\sigma A } + \mathcal{M}^{Q^3 2}_{\sigma \phi} + \mathcal{M}^{Q^3 2}_{A \phi} \biggr) \ .
\end{equation}

Once we substitute the Feynman rules, apply tensor contractions and tensor decomposition, it turns out that only $\mathcal{M}^{Q^3 2 }_{\sigma^2}$ is non-vanishing, all the other diagrams give a zero result, and we get: 
\begin{eqnarray}
\mathcal{M}^{Q^3 2 }=Q_{ij}(k_0)Q_{jk}(q_0)Q_{ki}(p_0)\tilde{\mathcal{M}}^{Q^3 2 }
\end{eqnarray}
\subsection{IBP decomposition}
All the scalar integrals appearing belong to the same topology that we encountered for the memory terms \eqref{eq:topology_memory}. The total amplitude can then be written in terms of 1 Master Integral:
\begin{eqnarray}
    \tilde{\mathcal{M}}^{Q^32}= C^{Q^3 2}\ \scalebox{0.9}{\begin{tikzpicture}[baseline=(current bounding box.center)]             \begin{feynman}
            \vertex (a) ;
            \vertex[right=0.2cm of a] (b);
            \vertex[right=1cm of b] (c); 
            \vertex[right=1cm of c] (d);
            \diagram* { 
            (b) -- [half left,very thick ,edge label'=\(1\)] (c),
            (b) -- [half right,very thick ] (c),
            (c) -- [half left,very thick ,edge label'=\(3\)] (d),
            (c) -- [half right,very thick ] (d),
            };
            \end{feynman} \end{tikzpicture}}\ ,
\label{eq:double_emission_decomposed}
\end{eqnarray}

\subsubsection{Coefficient}
The coefficient, writing $d=3+\epsilon$ is given by: 
\begin{equation}
    C^{Q^3 2} = \frac{ik_0^4q_0^4}{128\Lambda^2}
\end{equation}
\subsubsection{Master Integrals}
The master integral is the same appearing in \eqref{eq:MI_memory}
\subsection{Result}
We can obtain the final expression for the effective action by substituting the explicit expression for the MI and by taking the limit $\epsilon\to 0$: 
\begin{equation}
    S^{Q^3 2}_{eff \ 5PN}= -\frac{G_N^2}{2}\int \frac{dk_0}{(2\pi)}\frac{dq_0}{(2\pi)}k_0^4q_0^4Q_{ij}(k_0)Q_{jk}(q_0)Q_{kl}(p_0)
    \label{eq:INOUT_double}
\end{equation}
If we move to time-domain we get: 
\begin{equation}
    S^{Q^3 2}_{eff \ 5PN}= -\frac{G_N^2}{2}\int dt \ddddot{Q}_{ij}\ddddot{Q}_{jk}Q_{kl} \ .
\end{equation}
This result seems not to agree with Eq.(189) of \cite{Blumlein:2021txe}. Being more careful we can notice that substituting Eq.(201) of \cite{Blumlein:2021txe} in Eq.(197) we do not get an expression compatible with \eqref{eq:INOUT_double}. Hence, Eq.(201) of \cite{Blumlein:2021txe} is only an example of calculation, and not the final result for the double emission diagram in the causal formalism. 
\chapter{Hereditary Effects in the In-In formalism}
\label{chapter:inin}
In the previous chapter we used Feynman diagrams to evaluate Hereditary effects in post-Newtonian theory up to 5PN order. However, the dynamic of a binary coalescing system is non-conservative, since it involves the emission of radiation, and just dictated by initial conditions in time. The use of Feynman diagrams instead is based on symmetric in time boundary conditions, and it is not the correct approach to deal with dissipative problems. As emphasized in \cite{Galley:2009px}, we should study radiation problems like this using the so-called \textit{Schwinger-Keldysh} (or \textit{In-In} or \textit{CTP}) formalism, which was introduced by Schwinger in \cite{Schwinger:1960qe,Keldysh:1964ud}, to compute expectation values in quantum mechanics from a path integral formalism. \\ Since its introduction this formalism has been widely used in problems where an initial value formulation is important to describe the dynamical evolution of a system, like in inflationary cosmology \cite{Weinberg:2005vy}.
In Sec.\ref{Sec:back_scattering} we saw that the back-reaction diagram at 1-loop gives a total derivative once we move to position space, and does not yield a contribution to the equations of motion. \\ But being interested in studying dissipative effects, we expect to have non-vanishing contributions arising.
Let us notice that if we had distinguished between the two quadrupole moments appearing in Sec.\ref{Sec:back_scattering}, let us call them $Q_{+}$ and $Q_{-}$, then we would have obtained a non-zero result: 
\begin{equation}
    S_{eff \ 2.5 PN}^{Q_\pm}=-\frac{G_N}{5}\int dt Q_{+ij}(t)\frac{d^5 Q_{-ij}(t)}{dt^5} \ ,
\end{equation}
where we multiplied by $2$ to remove the symmetry factor inserted in the previous computation. \\ 
The In-Out approach that characterizes scattering processes is not well suited to deal with open systems where dissipative effects are acting. We can overcome this problem by doubling the field variables, using an approach known as In-In formalism, that will be described it in the next section. 

\subsubsection{State-of-the-Art}
The In-In formalism has been used to compute radiation reaction effects at 2.5PN in \cite{Galley:2009px}, at 3.5PN in \cite{Galley:2012qs}. It has also been used for tail  at 4PN in \cite{Foffa:2011np}\cite{Foffa:2013qca}\cite{Galley:2015kus}, at 5PN in \cite{Blumlein:2021txe}, for tail of tail effects up to 7PN order in \cite{Edison:2022cdu}, and for memory effects up to 5PN in \cite{Blumlein:2021txe}. For reviews of In-In formalism in classical systems one can consult \cite{Galley:2012hx,Galley:2014wla}. 
\subsubsection{Goals of the Chapter:}
The chapter will be organized as follows: 
\begin{itemize}
    \item it will be introduced the in-in formalism in classical mechanics, working on an example made by two harmonic oscillators, where we will show that it is necessary to double the field variables in order to obtain the correct equations of motions which respect causality.
    \item Then we will describe the In-In formalism in NRGR, and we will develop a smart procedure to obtain In-In amplitudes using Feynman amplitudes as building blocks, so that only the MIs appearing have to be computed case by case according to the boundary conditions appearing in the propagators.
    \item With a generalization of the computational algorithm used in the previous chapter, we will recompute the Hereditary effects in the CTP formalism, comparing the results with the ones appearing in literature.
\end{itemize}
\section{In-In Formalism}
We usually formulate physical theories using an action principle, where all the information about the system is contained at the level of the action: degrees of freedom, interactions, symmetries. Moreover, thanks to the EFT techniques, one can approximate a theory directly at the level of the action, obtaining simplified theories. \\ 
Unfortunately this formulation is not well suited to deal with system characterized by generic interactions and irreversible processes that arise from dissipation causal, radiation and so on, as discussed in \cite{Galley:2012hx}, \cite{Galley:2014wla}, and they fail to describe the causal dynamics of such problems.

\subsection{An example: two coupled harmonic oscillators} 
In order to understand why action principles in general cannot give us a causal description, let us consider a system made of two coupled harmonic oscillators $q(t)$ and $Q(t)$, of masses $m$ and $M$ and frequencies $w$ and $\Omega$, defined by the following conservative action: 
\begin{equation}
    S[q,Q]\ =\ \int_{t_i}^{t_f}dt\biggl\{\frac{m}{2}(\dot{q}^2-w^2q^2)+\lambda \ q \ Q + \frac{M}{2}(\dot{Q}^2-\Omega^2Q^2)\biggr\} \ . 
\end{equation}
The total system is conservative, since the corresponding Lagrangian is explicitly independent of time. However, if instead we consider only the dynamics of $q(t)$, then the system is open, and it may gain or lose energy since we do not have access to the dynamics of $Q(t)$. \\ 
Within an EFT approach, we can find an effective action which depends only on $q(t)$ by integrating out the $Q$ modes via equation of motions: 
\begin{eqnarray}
    Q(t)\  & = & \ Q^{(h)}(t)+\frac{\lambda}{M}\int_{t_i}^{t_f}\ dt' G_{ret}(t-t')q(t') \ , \\ 
    S_{eff}[q] & = & \int_{t_i}^{t_f}\ dt \biggl\{\frac{m}{2}(\dot{q}^2-w^2q^2)+\lambda q Q^{(h)}(t)+\frac{\lambda^2}{2M}\int_{t_i}^{t_f}dt'q(t)G_{ret}(t-t')q(t')\biggr\} \ . 
    \label{eq:Seff_ho}
\end{eqnarray}
In the new effective action \eqref{eq:Seff_ho}, the factor $q(t)q(t')$ is symmetric under $t\leftrightarrow t'$, so that only the \textit{time-symmetric} piece of the retarded Green's function contributes to the last term in the effective action, and using the identity $G_{ret}(t'-t)=G_{adv}(t-t')$ we can rewrite it as: 
\begin{equation}
    \frac{\lambda^2}{2M}\int_{t_i}^{t_f}\ dt dt'q(t)\biggl[\frac{G_{ret}(t-t')+G_{adv}(t-t')}{2}\biggr]q(t') \ . 
\end{equation}
We can then apply Hamilton's principle of stationary action to \eqref{eq:Seff_ho} to get the equations of motion for $q$: 
\begin{equation}
    m\ddot{q}+mw^2q=\lambda Q^{(h)}(t)+\frac{\lambda^2}{2M}\int_{t_i}^{t_f}dt'\biggl[G_{ret}(t-t')+G_{adv}(t-t')\biggr]q(t') \ .
\end{equation}
The problem is non-causal since the advanced Green's function appears explicitly in the equations of motion, and we cannot solve them by knowing only the initial conditions of the system. Moreover, the sum of the retarded and advanced Green's functions is symmetric in time, and implies that the integral accounts only for interactions which are conserving the energy between $q$ and $Q$. \\ 
From an action principle we can have only time-symmetric potentials and interactions for non-conservative systems. However, if we choose to work at the level of the equations of motion and integrate out $Q$ by substituting its solution into the equation of motion of $q$, we would obtain the causal, non-conservative dynamics for $q$. In order to understand why Hamilton's principle seems unable to describe the causal interactions and how we can recover causality, one can look at the last term in \eqref{eq:Seff_ho}:
\begin{itemize}
    \item advanced Green's function appears in the effective action and in the equation of motion for $q$ because the retarded Green's functions couples to the oscillator in a time-symmetric way via $q(t)q(t')$ \ , 
    \item if one breaks this symmetry by introducing two sets of variables $q\to (q_1,q_2)$, then $q_1(t)q_2(t')$ couples to the full retarded Green's function, and varying with respect to only $q_1$, say, gives the correct force if we set $q_2=q_1$ after the variation is performed.
\end{itemize}
This doubling of the field variables is known as In-In formalism or CTP (closed-time-path) formalism, and we will explain it in the next subsection. 

\subsection{Lagrangian formalism for non-conservative systems}
\begin{figure}[H]
    \centering
    \includegraphics[width=0.6\textwidth]{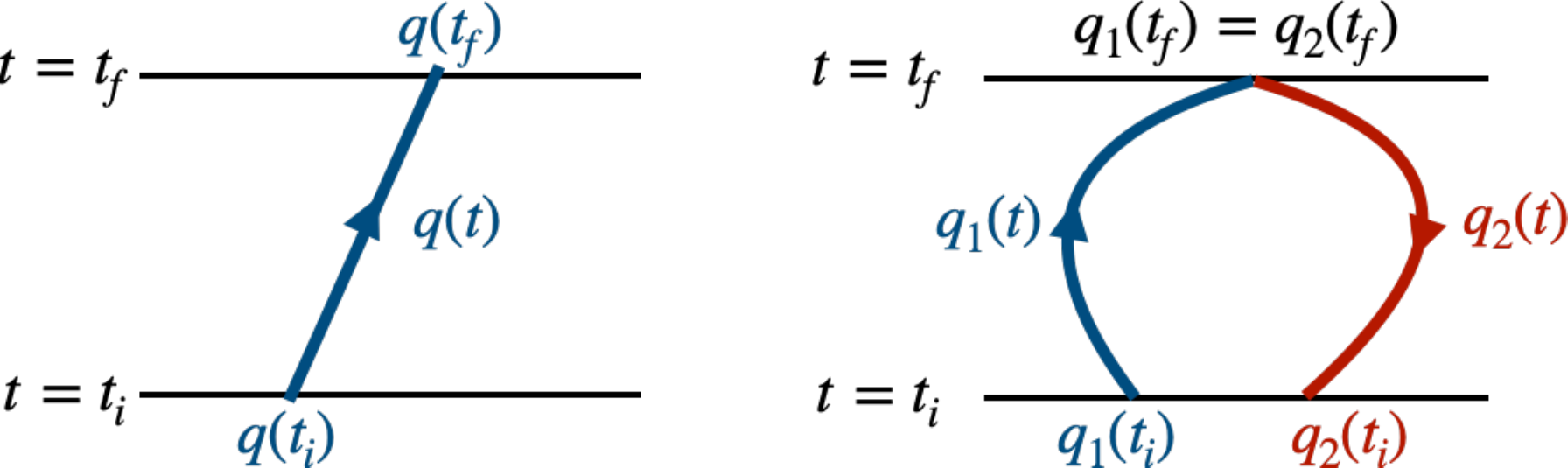}
    \caption{Time evolution of a coordinate $q(t)$ in the Lagrangian formalism and in the In-In formalism. On the left the coordinate $q(t)$ evolves from $t=t_i$ to $t=t_f$, whereas on the right we double the field variables into $q_1$ and $q_2$, where only the initial condition are fixed, and at the final time $q_1(t_f)=q_2(t_f)$}
    \label{fig:In_In}
\end{figure}
Let us consider a set of $N$ generalized coordinates and velocities of a general system:  $\mathbf{q}(t)=\{q^I(t)\}_{I=1}^N$ and $\dot{\mathbf{q}}(t)=\{\dot{q}^I(t)\}_{I=1}^N$. In conservative mechanics, one considers the evolution of the system from some initial time $t=t_i$ to a final time $t=t_f$, so that the degrees of freedom trace out a trajectory in coordinate space, as appearing on the left of Fig.\ref{fig:In_In}. One can write down an action $S[\mathbf{q}]$, which is the time integral of a Lagrangian $L[\mathbf{q},\dot{\mathbf{q}},t]$ along a trajectory $\mathbf{q}(t)$ that passes through $\mathbf{q}(t_i)=\mathbf{q}_i$, and  $\mathbf{q}(t_f)=\mathbf{q}_f$ at the initial and final times, respectively:  
\begin{equation}
    S[\mathbf{q}]=\int_{t_i}^{t_f}\ dt \ L(\mathbf{q},\dot{\mathbf{q}},t) \ . 
\end{equation}
One can then get the equations of motions as: 
\begin{equation}
    \frac{d}{dt}\frac{\partial L}{\partial \dot{q}^I}=\frac{\partial L }{\partial q^I}
\end{equation}
At this point, following the intuition of the previous subsection, let us double the degrees of freedom as: 
\begin{equation}
    q^I(t)\to \bigl(q_1^I(t),q^I_2(t)\bigr) \ , 
\end{equation}
and similarly for the velocities. 
The new variables $\mathbf{q}_1^I$ and $\mathbf{q}_2^I$ evolve from some initial value $\mathbf{q}_{1i},\mathbf{q}_{2i}$ to the final states $\mathbf{q}_{1f}=\mathbf{q}_{2f}$, where we are not fixing the actual value of the final states. 
The action for this theory is given by: 
\begin{equation}
    S[\mathbf{q}_1,\mathbf{q}_2]=\int_{t_i}^{t_f}\ dt\  L(\mathbf{q}_1,\dot{\mathbf{q}}_1,t)-\int_{t_i}^{t_f}\ dt\  L(\mathbf{q}_2,\dot{\mathbf{q}}_2,t) \ , 
\end{equation}
where $\mathbf{q}_1$ and $\mathbf{q}_2$ are decoupled one from each other. 
In a general picture, we can have an arbitrary function $K(\mathbf{q}_1,\mathbf{q}_2,\dot{\mathbf{q}}_1,\dot{\mathbf{q}}_2)$ that does couple the doubled variables, and so we can have an action of the form: 
\begin{equation}
    S[\mathbf{q}_1,\mathbf{q}_2]= \int_{t_i}^{t_f}\ dt\biggl[ L(\mathbf{q}_1,\dot{\mathbf{q}}_1,t)- L(\mathbf{q}_2,\dot{\mathbf{q}}_2,t)+K(\mathbf{q}_1,\mathbf{q}_2,\dot{\mathbf{q}}_1,\dot{\mathbf{q}}_2,t)\biggr] \ .
    \label{eq:In-In-Action}
\end{equation}
$K(\mathbf{q}_1,\mathbf{q}_2,\dot{\mathbf{q}}_1,\dot{\mathbf{q}}_2,t)$ represents a non-conservative potential, and without its presence there is no need to keep variables doubled because that means that the system is conservative. \\ 
 This action can be rewritten in terms of a Lagrangian $\Lambda(\mathbf{q}_1,\mathbf{q}_2,\dot{\mathbf{q}}_1,\dot{\mathbf{q}}_2,t)$ for the doubled degrees of freedom: 
\begin{equation}
    \Lambda = L(\mathbf{q}_1,\dot{\mathbf{q}}_1,t)-L(\mathbf{q}_2,\dot{\mathbf{q}}_2,t)+K(\mathbf{q}_1,\mathbf{q}_2,\dot{\mathbf{q}}_1,\dot{\mathbf{q}}_2,t) \ , 
\end{equation}
as: 
\begin{equation}
    S[\mathbf{q}_1,\mathbf{q}_2]=\int_{t_i}^{t_f}\Lambda(\mathbf{q}_1,\mathbf{q}_2,\dot{\mathbf{q}}_1,\dot{\mathbf{q}}_2,t) \ . 
\end{equation}
As described in \cite{Galley:2014wla}, one can define a variational principle under which the action $S$ is stationary, which is specified by four conditions: the coordinates of the two variables at the initial time: $\mathbf{q}_{1,2}(t_i)$, and the equality conditions: $\mathbf{q}_{1}(t_f)=\mathbf{q}_{2}(t_f)$,  $\pi_{1}(t_f)=ì\pi_{2}(t_f)$, where $\pi_{1,2}$ are the conjugated momenta, defined as: 
\begin{equation}
    \pi_{2I}(\mathbf{q}_{1,2},\dot{\mathbf{q}}_{1,2}) \ = \ -\frac{\partial \Lambda}{\partial \dot{q}_2^I(t)} \ = \ \frac{\partial L(\mathbf{q}_2,\dot{\mathbf{q}}_2)}{\partial \dot{q}_2^I(t)}-\frac{\partial K }{\partial \dot{q}_2^I(t)} \ . 
\end{equation}
Notice that with the equality conditions we are imposing that the two variables and the two conjugated momenta are equal at finite time, but are left unspecified. These conditions ensure that our variational principle is consistent with our ignorance of the final state of the accessible degrees of freedom. The equations of motion are given by: 
\begin{equation}
    \frac{d}{dt}\frac{\partial \Lambda}{\partial \dot{q}_\alpha^I}=\frac{\partial \Lambda}{\partial q_\alpha^I} \ , 
\end{equation}
where $a=1,2$. However, the resulting equations are not physical until we take the physical limit (PL), where the variables are identified: 
\begin{equation}
    q_1^I=q_2^I=q^I \ , \qquad \dot{q}_1^I=\dot{q}_2^I=\dot{q}^I \ , 
\end{equation}
after all variations and derivatives of the Lagrangian are taken. \\ 
The PL of both equations reduces to: 
\begin{equation}
    \frac{d}{dt}\frac{\partial L }{\partial \dot{q}^I}-\frac{\partial L }{\partial q^I}  = \biggl[\frac{\partial K}{\partial q_1^I}-\frac{d}{dt}\frac{\partial K}{\partial \dot{q}_1^I}\biggr]_{PL}=- \biggl[\frac{\partial K}{\partial q_2^I}-\frac{d}{dt}\frac{\partial K}{\partial \dot{q}_2^I}\biggr]_{PL}
\end{equation}
A more convenient parametrization of the coordinates that yields some important physical insight is given by the average and relative difference of the two histories: 
\begin{eqnarray}
    q_+^I  =  \frac{q_1^I+q_2^I}{2} \  , \qquad q_-^I= q_1^I-q_2^I \ , 
    \label{eq:doubling_q}
\end{eqnarray}
and the physical limit is then given by: 
\begin{equation}
    q_+^I \to q^I \ , \qquad q_-^I \to 0 
\end{equation}
In these coordinates, the non-conservative Lagrangian is: 
\begin{equation}
    \Lambda = \Lambda(\mathbf{q}_+,\mathbf{q}_-,\dot{\mathbf{q}}_+,\dot{\mathbf{q}}_-,t) \ .
\end{equation}
The equality condition becomes:
\begin{equation}
    \mathbf{q}_-(t_f)=0 \ , \qquad \pi_- (t_f) =  0 \  ,
\end{equation}
implying that the physical relevant average $(+)$ quantities are not specified at the final time in order to have a well-defined variational principle. 
The resulting equations of motion are: 
\begin{equation}
    \frac{d}{dt}\frac{\partial \Lambda}{\partial \dot{q}_\alpha^I}=\frac{\partial \Lambda}{\partial q_\alpha^I} \ , 
\end{equation}
where $\alpha = +,-$. Taking the Physical limit $\alpha=+$ is identically zero, while $\alpha=-$ survives: 
\begin{equation}
    \frac{d}{dt}\frac{\partial L }{\partial \dot{q}^I}-\frac{\partial L }{\partial q^I}= \biggl[\frac{\partial K}{\partial q_-^I}-\frac{d}{dt}\frac{\partial K}{\partial\dot{q}_-^I}\biggr]_{PL} \  .
\end{equation}
More generally, the equations of motion can be derived by computing: 
\begin{equation}
    0= \biggl[\frac{\delta S}{\delta q_-^I(t)}\biggr]_{PL} \  ,
    \label{eq:in_in_eom} 
\end{equation}
where $\delta/\delta q_-^I(t)$ is a functional derivative with respect to $q_-^I(t)$. 

\subsection{Coupled harmonic oscillators revisited}
Let us show that the formalism that we have just introduced reproduce the correct equations of motion for $q(t)$ after integrating out $Q(t)$ from the original action. \\ 
Let us impose the initial conditions for both oscillators: 
\begin{equation}
    q(t_i)=q_i \ , \ \ \dot{q}(t_i) = v_i \ , \ \ Q(t_i) = Q_i \ , \ \ \dot{Q}(t_i)=V_i \ . 
    \label{eq:initial_conditions_ho} 
\end{equation}
The two oscillators form a closed system, which conserves the total energy, and they can be described using an action. Since we are interested in integrating out $Q$ from the action, we first need to double the degrees of freedom to ensure that the proper causal conditions on the dynamics of $Q$ are respected and maintained, obtaining the following action in the $\pm$ basis: 
\begin{equation}
    S[q_{\pm},Q_{\pm}]=\int_{t_i}^{t_f} \ dt \biggl\{ m\dot{q}_-\dot{q}_+ -m w^2 q_-q_+ + \lambda\  q_- Q_+ + \lambda\  q_+ Q_- + M\ \dot{Q}_-\dot{Q}_+ -M\ \Omega^2 Q_-Q_+    \biggr\} \ . 
\end{equation}
The generic term $K$ is not present since the system is closed, but once we integrate out $Q$ the system becomes open, and we expect to get an effective $K$ in the action. The effective action for the open dynamics of $q$ is found by eliminating the $Q_\pm$ variables via EOM, which are given by: 
\begin{equation}
    M\ddot{Q}_\pm + M\Omega^2Q_\pm = \lambda q_\pm  \ . 
\end{equation}
We associate the initial conditions in \eqref{eq:initial_conditions_ho} with the initial conditions for $Q_+$, whereas from the equality condition at the final time we get the final conditions for $Q_-$, given by: 
\begin{equation}
    Q_-(t_f)=\dot{Q}_-(t_f) =  0
\end{equation}
The resulting solutions are: 
\begin{eqnarray}
    Q_+(t)& = & Q^{(h)}(t)+\frac{\lambda}{M}\int_{t_i}^{t_f}\ dt'\  G_{ret}(t-t')q_+(t') \\ 
    Q_-(t)&=& \frac{\lambda}{M}\int_{t_i}^{t_f}\ dt\  G_{adv}(t-t')q_-(t')  \ .
    \label{eq:solutions_ho_pm} 
\end{eqnarray}
Notice that because the solution to the (physical) $Q_+$ equation satisfies initial data while the solution to the (unphysical) $Q_-$ equation satisfies final data, then the former evolves forward in time while the latter evolves backward, hence the appearance of the advanced Green's function. Substituting \eqref{eq:solutions_ho_pm} back into the action we get the effective action for $q_\pm(t)$: 
\begin{equation}
    S_{eff}[q_{\pm}]= \int_{t_i}^{t_f}\ dt \biggl\{m\dot{q}_-\dot{q}_+-mw^2q_-q_++\lambda q_-Q^{(h)}+\frac{\lambda^2}{M}\int_{t_i}^{t_f}\ dt' q_-(t)G_{ret}(t-t')q_+(t')\biggr\}
\end{equation}
from which, by rewriting $q_{\pm}$ in terms of $q_{1,2}$, as defined in \eqref{eq:doubling_q}, and expecting a generic action of the form \eqref{eq:In-In-Action},  we read off that: 
\begin{eqnarray}
    L & = & \frac{1}{2}\dot{q}^2-\frac{1}{2}mw^2q^2 \\ 
    K & = & \lambda q_- Q^{(h)}(t)+\frac{\lambda^2}{M}\int_{t_i}^{t_f}\ dt' q_-(t)G_{ret}(t-t')q_+(t') \ ,
\end{eqnarray}
 where the last term above contains a factor $q_-(t)q_+(t')$ that is not symmetric in $t\leftrightarrow t'$ and so it couples to the full retarded Green's function and not just the time-symmetric piece, unlike the effective action constructed using the usual Hamilton's principle in \eqref{eq:Seff_ho}. \\ 
From \eqref{eq:in_in_eom}, one can get the equations of motion for $q(t)$: 
\begin{equation}
    m\ddot{q}+mw^2q = \lambda Q^{(h)}(t)+\frac{\lambda^2}{M}\int_{t_i}^{t_f}dt' G_{ret}(t-t')q(t') \ ,
\end{equation}
where only the retarded Green's function appears in the equation, and so causality is respected. Therefore, we have shown that the use of the In-In formalism gives the proper causal evolution for the open system given only initial data and does not require to fix the configuration of the degrees of freedom at final time.

\subsection{An example from classical field theory: two coupled scalar fields}
Let us now consider two relativistic scalar fields, $\phi(x)$ and $\chi(x)$ non-linearly coupled to each other and mutually evolving in a flat spacetime from initial data specified at a given instant of time. 
The action for the closed system is given by: 
\begin{equation}
    S[\phi,\chi]= \int d^4x \biggl\{\frac{1}{2}\partial_\alpha\phi\partial^\alpha\phi +\frac{1}{2}\partial_\alpha\chi\partial^\alpha\chi +\frac{g}{2}\phi^2\chi\biggr\} \ ,
\end{equation}
where $g$ is a coupling constant. We choose to integrate out the $\chi$ field at the level of the action, by working in the In-In formalism. 
We can double the degrees of freedom, and switch to the $\pm$ representation, and integrating out the field $\chi$ we get: 
\begin{equation}
    S[\phi_a,\chi_a]=\int \ d^4x \biggl\{\partial_\alpha\phi_-\partial^\alpha\phi_+ +g\phi_-\phi_+\chi^{(h)}\biggr\}+\frac{g^2}{2}\int d^4 x d^4 x'\phi_-(x)\phi_+(x)G_{ret}(x,x')\biggl[\phi_+^2(x')+\frac{1}{4}\phi_-^2(x')\biggr]
\end{equation}
from which, by rewriting $\phi_{\pm}$ in terms of $\phi_{1,2}$, as defined in \eqref{eq:doubling_q}, and expecting a generic action of the form \eqref{eq:In-In-Action}, we read off that:
\begin{eqnarray}
    \mathcal{L} & = & \int d^4 x \frac{1}{2} \  ,\partial_\alpha\phi\partial^\alpha\phi \\ 
    \mathcal{K} & = & \frac{g^2}{2}\int d^4x d^4x' \phi_-(x)\phi_+(x)G_{ret}(x,x')\biggl[\phi_+^2(x')+\frac{1}{4}\phi_-^2(x')\biggr] \ . 
\end{eqnarray}
Furthermore, we can obtain the equations of motion for $\phi$ by varying $S_{eff}$ with respect to $\phi_-$, and taking the PL, obtaining: 
\begin{equation}
    \Box\phi(x) = \chi^{(h)}(x)\phi(x)+\frac{g^2}{2}\phi(x)\int d^4x' G_{ret}(x,x')\phi(x') , 
\end{equation}
which respect causality since only the retarded propagator is appearing. 

\section{NRGR in the IN-IN Formalism}
Let us now consider the evolution of a compact binary system in the EFT approach in the far zone. The total system formed by the radiation graviton fields $\bar{h}_{\mu\nu}$ and 
the worldlines of the compact bodies $\mathbf{x}_{1,2}(t)$, is closed. However, once gravitational perturbations are integrated out, the dynamic of the worldlines is open, and one should use In-In formalism since the problem is formulated by specifying only initial conditions, as discussed in  \cite{Galley:2009px}\cite{Galley:2012qs}. We need to double the degrees of freedom as: 
\begin{equation}
    \bar{h}_{\mu\nu}\ \to \ (\bar{h}_{\mu\nu1},\bar{h}_{\mu\nu2}) \ , \qquad \mathbf{x}_k \ \to \ (\mathbf{x}_{k1},\mathbf{x}_{k2}) \ ,
\end{equation}
and construct the following action: 
\begin{equation}
    S[\mathbf{x}_{k1},\mathbf{x}_{k2},\bar{h}_{\mu\nu 1},\bar{h}_{\mu\nu 2}] \ = \ S_{rad}[\mathbf{x}_{k1},\bar{h}_{\mu\nu 1 }]-S_{rad}[\mathbf{x}_{k2},\bar{h}_{\mu\nu 2 }] \ , 
\end{equation}
where each action on the right side is of the form of \eqref{eq:radiative_action}.  \\ 
One can then integrate out the radiation gravitons using a diagrammatic approach at the desired PN order, obtaining an effective action for the open dynamics of the binary's inspiral: 
\begin{eqnarray}
    e^{iS_{eff}[\mathbf{x}_{k1},\mathbf{x}_{k2}]}=\int \ D\Bar{h}_{\mu\nu 1}\Bar{h}_{\mu\nu 2}e^{i S[\mathbf{x}_{k1},\mathbf{x}_{k2},\bar{h}_{\mu\nu 1},\bar{h}_{\mu\nu 2}]} \ . 
\end{eqnarray}
To get the equations of motions one has to take the field variations, and then to take the \textbf{physical limit} by identifying the doubled variables with the physical ones as: $\mathbf{x}_{k1}=\mathbf{x}_{k2}=\mathbf{x}_{k}$. \\
We will work in the $\pm$ variables, defined as: 
\begin{equation}
    \mathbf{x}_{k+} = \frac{\mathbf{x}_{k1}+\mathbf{x}_{k2}}{2}\, \qquad \mathbf{x}_{k-}=\mathbf{x}_{k1}-\mathbf{x}_{k2} \ ,
\end{equation}
and in these variables the physical limit is given by: $\mathbf{x}_{k+}=\mathbf{x}_{k}, \ \mathbf{x}_{k-}= 0$. \\ 
One can obtain worldline equations of motion, that properly incorporates radiation reaction effects, by computing the following variation:
\begin{equation}
    \frac{\delta S_{eff}[\mathbf{x}_{1\pm},\mathbf{x}_{2\pm}]}{\mathbf{x}_{K-}}\bigg|_{\mathbf{x}_{k-}=0, \mathbf{x}_{k+}=\mathbf{x}_{k}}=0
    \label{eq:eom_in_in}
\end{equation}
It is important to notice that \eqref{eq:eom_in_in} receives non-zero contributions from terms in the effective action that are perturbatively linear in $\mathbf{x}_{k-}$ or its time derivatives. Therefore, all terms of quadratic order or higher in any $"-"$ variables can be dropped out from the effective action for the purposes of deriving the worldline equations of motion. We will take advantage of this property in the course of the calculations. 
\subsection{Feynman rules in the In-In formalism}
The diagrammatic  structure of In-in perturbation theory is nearly identical to the in-out approach. However, when drawing Feynman diagrams in the in-in formalism one needs to:
\begin{itemize}
    \item include CTP labels $a,b,c,\cdots$ at each vertex to keep track of the forward and backward branches of time in the CTP path integral, where $a,b,c,\ldots=\pm$; 
    \item rise and lower label indices with the following matrix $C^{ab}$:
    \begin{equation}
        C^{ab} = C_{ab} = \begin{pmatrix}  
 0 & 1 \\
 1 & 0
        \end{pmatrix} \ ;
    \end{equation}
    \item substitute Feynman propagators appearing in the graviton two-point functions with a matrix $G^{ab}$ connecting particle-field vertices labeled by CTP indices $a$ and $b$, defined as: 
    \begin{equation}
        G^{ab} = \begin{pmatrix}  
 0 & -iG_{adv} \\
 -iG_{red} & 0
        \end{pmatrix} \ , 
    \end{equation} 
    \item sum over all CTP indices. 
\end{itemize}

\subsection{Computational strategy}
As noticed in the previous subsection, the only differences between In-Out and In-In approaches are the appearances of \textbf{CTP indices} and of the \textbf{propagator matrix} $G^{ab}$.  \\ 
It is possible to decouple the In-In algebra from the diagram computation, and one can use the \textbf{Feynman amplitudes} computed in the previous chapter as \textbf{building blocks} to get \textbf{In-In Amplitudes}. More precisely, since all the operations done in early stages of the computations are purely algebraic and not sensitive to the boundary conditions for the propagators, we can consider the Feynman amplitudes after IBP decomposition, where we distinguish identical sources by labeling them as  $Q^{(a)},\ Q^{(b)}, \ Q^{(c)}, \ \ldots$, and we do not specify the boundary conditions for the propagators: 
\begin{eqnarray}
    \mathcal{M}_{her} & = & Tr[Q^{(a)}\ldots Q^{(n)}]\tilde{\mathcal{M}}_{her} \label{eq:her_amplitudes_generic}\\ 
    \tilde{\mathcal{M}}_{her}&=& \sum_iC_iI_i^{MI}\label{eq:her_amplitudes_generic_IBP}
\end{eqnarray}
Then one can proceed in the computation of the hereditary effects in the In-In formalism as follows: 
\begin{itemize}
    \item rewrite each Feynman rule in $\pm$ formalism, first by doubling the field variables and then by rewriting them in $\pm$ variables, in a symbolic fashion.
    \item Write down each diagram written in terms of the previously expanded Feynman rules, then take Wick contractions among the graviton fields using the $G^{ab}$ matrices. As a result one obtains a linear combination of terms, each one characterized by:  a choice of $\pm$ sources, a choice of propagators, and a symmetry factor in front. 
    \item As discusses before, we keep only linear terms in $"-"$ in the multipole sources, since they are linear in $x_{K-}$, and so they are the only one surviving in the equations of motions. 
    \item Each term can be obtained from Eqns.\eqref{eq:her_amplitudes_generic}\eqref{eq:her_amplitudes_generic_IBP}, by substituting the $\pm$ sources, by computing the Master Integrals using the right prescriptions for the propagators, and by adding the appropriate symmetry factor in front. 
    \item The total result is given by the sum of these $\pm$ amplitudes. 
\end{itemize}

\subsubsection{Doubling the Feynman rules}
One can double the vertices appearing in the diagrams in a symbolic way, following \cite{Blumlein:2021txe}, as: 
\begin{eqnarray}
    V_{Qh} & = & 
    Q_1h_1-Q_2h_2 = Q_+h_-+Q_-h_+ \\ 
    V_{Qh^2} & = &
    Q_1 h_1^2 - Q_2 h_2^2 = 2Q_+h_+h_- +Q_- h_+^2+\frac{1}{4}Q_-h_-^2 \\ 
    V_{h^3} & = & 
    h_1^3-h_2^3 = \frac{1}{4}h_-^3+3h_-h_+^2
\end{eqnarray}

\subsubsection{Hereditary terms in the Schwinger-Keldysh variables}

Following \cite{Foffa:2021pkg}, one can take the Wick contractions of the symbolic expression of the amplitudes written in doubled vertices, and keep only terms linear in the multipole source, to know which diagrams are contributing. \\ 
For the \textbf{radiation reaction}, in terms of Schwinger-Keldysh variables the relevant couplings are: 
\begin{eqnarray}
    \scalebox{0.4}{\begin{tikzpicture} [baseline=(a1)] 
        \begin{feynman}
        \vertex (a1) ;
        \vertex[right=1cm of a1,square dot, red] (a2) {}; 
        \vertex[right=2cm of a2] (a3); 
        \vertex[above=1.5cm of a3] (b3);
        \vertex[right=2cm of a3, square dot, red] (a4) {};
        \vertex[right=1cm of a4] (a5); 
        \vertex[below=0.2cm of a3] (e3);
        \vertex[below=0.4cm of a2] (f3) {\(Q_{ij}\)};
        \vertex[below=0.4cm of a4] (g3) {\(Q_{kl}\)};
        \diagram* { 
        (a1) -- [double,thick] (a5),
        (a2) -- [boson,quarter left] (b3),
        (b3) -- [boson,quarter left] (a4),
        };
        \end{feynman} 
        \end{tikzpicture}} &= &  <V_{Qh}(t)V_{Qh}(t')> \nonumber \\ 
   & = &  <(Q_-h_++Q_+h_-)(t)(Q_-h_++Q_+h_-)(t')> \nonumber \\ 
    & = &  Q_-(t)<h^-(t)h^+(t')>Q_+(t') + Q_+(t)<h^+(t)h^-(t')>Q_-(t') \nonumber \\ 
    & = &  Q_-(t)\bigl(-iG_{ret}(t-t')\bigr)Q_+(t')+ Q_+(t)\bigl(-iG_{adv}(t-t')\bigr)Q_-(t') \ , 
    \label{eq:double_radiation_reaction}
\end{eqnarray}
where in the last line we kept only linear terms in $Q_-$. \\
For the \textbf{tail terms} we have a triple graviton vertex where one of the graviton is coupled to a conserved source, hence it has no energy component. Feynman, advanced or retarded propagators collapse to the same quantity, so there is no meaning in doubling that graviton leg, and we have for the triple graviton vertex: 
\begin{equation}
    V_{h^2\phi} = h_1^2\phi-h_2^2\phi=2h_+h_-\phi
\end{equation}
Writing the tail terms in Schwinger-Keldysh variables, following \cite{Foffa:2011np}  we get: 
\begin{eqnarray}
    \scalebox{0.4}{\begin{tikzpicture}[baseline=(a1)] 
        \begin{feynman}
        \vertex (a1) ;
        \vertex[right=1cm of a1, square dot, red] (a2) {}; 
        \vertex[right=2cm of a2, square dot, red] (a3) {}; 
        \vertex[above=1.5cm of a3] (b3);
        \vertex[right=2cm of a3, square dot, red] (a4) {};
        \vertex[right=1cm of a4] (a5); 
        \vertex[below=0.4cm of a3] (e3) {\(E\)};
        \vertex[below=0.4cm of a2] (f3) {\(Q_{ij}\)};
        \vertex[below=0.4cm of a4] (g3) {\(Q_{kl}\)};
        \diagram* { 
        (a1) -- [double,thick] (a5),
        (a2) -- [boson, black!60!green,quarter left] (b3),
        (b3) -- [boson, black!60!green,quarter left] (a4),
        (b3) -- [scalar, blue] (a3)
        };
        \end{feynman} 
        \end{tikzpicture}} & = & 
    <V_{Qh}(t)V_{Qh}(t')V_{E\phi}(t")V_{h^2\phi}(t_i)> \nonumber \\ 
   & = & 
    <(Q_-h_++Q_+h_-)(t)(Q_-h_++Q_+h_-)(t')(Q\phi)(t")(2h_+h_-\phi)(t_i)>E \nonumber \\ 
    & = & 
     \{Q_-(t)[2(-iG_{ret}(t-t_i))(-iG_{ret}(t'-t_i))]Q_+(t')\nonumber \\ 
     & & +Q_+(t)[2(-iG_{adv}(t-t_i))(-iG_{adv}(t'-t_i))]Q_-(t')\ \}(-iG_{ret}(t''-t_i))E
     \label{eq:double_tail}
\end{eqnarray}
Similarly for the memory term we have: 
\begin{eqnarray}
    \scalebox{0.4}{\begin{tikzpicture}[baseline=(a1)] 
        \begin{feynman}
        \vertex (a1) ;
        \vertex[right=1cm of a1,square dot, red] (a2) {}; 
        \vertex[right=2cm of a2,square dot, red] (a3) {}; 
        \vertex[above=1.5cm of a3] (b3);
        \vertex[right=2cm of a3,square dot, red] (a4) {};
        \vertex[right=1cm of a4] (a5); 
        \vertex[below=0.4cm of a3] (e3) {\(Q_{mn}\)};
        \vertex[below=0.4cm of a2] (f3) {\(Q_{ij}\)};
        \vertex[below=0.4cm of a4] (g3) {\(Q_{kl}\)};
        \diagram* { 
        (a1) -- [double,thick] (a5),
        (a2) -- [boson, black!60!green, quarter left] (b3),
        (b3) -- [boson, black!60!green, quarter left] (a4),
        (b3) -- [boson, black!60!green] (a3)
        };
        \end{feynman} 
        \end{tikzpicture}} & = & 
 <V_{Qh}(t)V_{Qh}(t')V_{Qh}(t")V_{h^3}(t_i)> \nonumber \\ 
 & = & 
 3<(Q_-h_+)(t)(Q_+h_-)(t')(Q_+h_-)(t")(h_+h+h_-)(t_i)> \nonumber \\ 
 & = & 
 3Q_-(t)Q_+(t')Q_+(t")(-iG_{adv})(t'-t_i)(-iG_{adv})(t"-t_i)(-iG_{ret})(t-t_i)
 \label{eq:double_memory}
\end{eqnarray}
Eventually, if one consider the double emission diagram we get: 
\begin{eqnarray}
    \scalebox{0.4}{\begin{tikzpicture}[baseline = (a1)] 
        \begin{feynman}
        \vertex (a1) ;
        \vertex[right=1cm of a1,square dot, red] (a2) {}; 
        \vertex[right=2cm of a2,square dot, red] (a3) {}; 
        \vertex[above=1.5cm of a3] (b3);
        \vertex[right=2cm of a3,square dot, red] (a4) {};
        \vertex[right=1cm of a4] (a5); 
        \vertex[below=0.2cm of a3] (e3) {\(Q_{mn}\)};
        \vertex[below=0.2cm of a2] (f3) {\(Q_{ij}\)};
        \vertex[below=0.2cm of a4] (g3) {\(Q_{kl}\)};
        \diagram* { 
        (a1) -- [double,thick] (a5),
        (a2) -- [boson, black!60!green,half left] (a3),
        (a3) -- [boson, black!60!green,half left] (a4),
        };
        \end{feynman} 
        \end{tikzpicture}} & = &  <V_{Qh}(t)V_{Qh^2}(t')V_{Qh}(t")>  \nonumber \\
 & = & < (Q_-h_++Q_+h_-)(t) ( 2Q_+h_+h_- +Q_- h_+^2+\frac{1}{4}Q_-h_-^2 )(t')(Q_-h_++Q_+h_-)(t")> \nonumber \\ 
 & = & 2Q_+(t)Q_+(t')Q_-(t") (-iG_{adv})(t-t')(-iG_{adv})(t"-t') \nonumber \\ 
 & & +\  2Q_-(t)Q_-(t')Q_+(t") (-iG_{ret})(t-t')(-iG_{ret})(t"-t')\nonumber \\ 
 & & +\  Q_+(t)Q_-(t')Q_+(t")(-iG_{ret})(t-t')(-iG_{adv})(t"-t')
 \label{eq:double_double_emission}
\end{eqnarray}
\subsubsection{Master Integrals evaluation}
As shown in the previous chapter, all master integrals appearing are factorizable in products of multi-loop 1-point functions. 
Hence, the analytic form of the solution of each tadpole is always the same:
\begin{equation}
        j_1(k_0)=\scalebox{0.8}{\begin{tikzpicture}[baseline=(current bounding box.center)]             \begin{feynman}
            \vertex (a) ;
            \vertex[right=0.2cm of a] (b);
            \vertex[right=1cm of b] (c); 
            \vertex[right=1cm of c] (d);
            \diagram* { 
            (b) -- [half left,very thick ] (c),
            (b) -- [half right,very thick ] (c)
            };
            \end{feynman} \end{tikzpicture}}= \int \frac{d^d\mathbf{k}}{(2\pi)^d}\frac{1}{\mathbf{k^2}-k_0^2+i\epsilon}=  \frac{\Gamma[1-d/2]}{(4\pi)^{d/2-1}}(-k_0^2)^{d/2-1} \ . 
\end{equation}
We just need to take care of the square roots appearing for which we can have:
\begin{equation}
    \sqrt{-(k_0\pm i\epsilon)^2}= \pm i k_0
\end{equation}
according to the fact that we have \textbf{advanced/retarded propagators}. \\ 
If at higher PN order non-factorizable integrals appear, one should solve them using differential equations, case by case depending on the boundary conditions for the propagators.

\section{Evaluation of Hereditary effects in the In-In formalism}
In this section we will evaluate again the Hereditary effects, this time in the in-in formalism, following the computational strategy underlined before.

\subsection{Back reaction in the In-In formalism}
In order to understand better what we are going to do let us compute in detail the simple 1-loop back-reaction diagram, given in the In-In formalism as: 
\begin{equation}
    \scalebox{0.6}{\begin{tikzpicture} [baseline=(a1)] 
        \begin{feynman}
        \vertex (a1) ;
        \vertex[right=1cm of a1,square dot, red] (a2) {}; 
        \vertex[right=2cm of a2] (a3); 
        \vertex[above=1.5cm of a3] (b3);
        \vertex[right=2cm of a3,square dot, red] (a4) {};
        \vertex[right=1cm of a4] (a5); 
        \vertex[below=0.2cm of a3] (e3);
        \vertex[below=0.4cm of a2] (f3) {\(Q_{-ij}\)};
        \vertex[below=0.4cm of a4] (g3) {\(Q_{+kl}\)};
        \diagram* { 
        (a1) -- [double,thick] (a5),
        (a2) -- [boson,quarter left] (b3),
        (b3) -- [boson,quarter left] (a4)
        };
        \end{feynman} 
        \end{tikzpicture}} = \langle V_{Qh}^{(-)}V_{Qh}^{(+)}\rangle= Q_-Q_+(-iG_{ret})
\end{equation}
using as building block the result found the previous chapter: 
\begin{equation}
    \scalebox{0.6}{\begin{tikzpicture} [baseline=(a1)] 
        \begin{feynman}
        \vertex (a1) ;
        \vertex[right=1cm of a1,square dot, red] (a2) {}; 
        \vertex[right=2cm of a2] (a3); 
        \vertex[above=1.5cm of a3] (b3);
        \vertex[above=0.2cm of b3] (l3)  {\(G_1\)};
        \vertex[right=2cm of a3,square dot, red] (a4) {};
        \vertex[right=1cm of a4] (a5); 
        \vertex[below=0.2cm of a3] (e3);
        \vertex[below=0.4cm of a2] (f3) {\(Q_{ij}^{(a)}\)};
        \vertex[below=0.4cm of a4] (g3) {\(Q_{-kl}^{(b)}\)};
        \diagram* { 
        (a1) -- [double,thick] (a5),
        (a2) -- [boson,quarter left] (b3),
        (b3) -- [boson,quarter left] (a4)
        };
        \end{feynman} 
        \end{tikzpicture}} = Tr[Q^{(a)}Q^{(b)}]\tilde{\mathcal{M}}^{Q^2}(G_1)
\end{equation}
where:
\begin{equation}
    \tilde{\mathcal{M}}^{Q^2}=\frac{(d-2)(d+1)}{16(d-1)(d+2)\Lambda^2}k_0^4 \scalebox{0.8}{\begin{tikzpicture}[baseline=(current bounding box.center)]             \begin{feynman}
        \vertex (a) ;
        \vertex[right=0.2cm of a] (b);
        \vertex[right=1cm of b] (c); 
        \vertex[right=0.2cm of b] (e) {\(G_1\)};
        \vertex[right=1cm of c] (d);
        \diagram* { 
        (b) -- [half left,very thick ] (c),
        (b) -- [half right,very thick ] (c)
        };
        \end{feynman} \end{tikzpicture}} \ .
\end{equation}
According to \eqref{eq:double_radiation_reaction}, one can compute the diagram with multipole sources $Q_{-},Q_{+}$ by substituting $Q^{(a)}Q^{(b)}\to Q_-Q_+$, and considering a retarded propagator $G_1\to G_{ret}$, obtaining: 
\begin{equation}
    \scalebox{0.6}{\begin{tikzpicture} [baseline=(a1)] 
        \begin{feynman}
        \vertex (a1) ;
        \vertex[right=1cm of a1,square dot, red] (a2) {}; 
        \vertex[right=2cm of a2] (a3); 
        \vertex[above=1.5cm of a3] (b3);
        \vertex[right=2cm of a3,square dot, red] (a4) {};
        \vertex[right=1cm of a4] (a5); 
        \vertex[below=0.2cm of a3] (e3);
        \vertex[below=0.4cm of a2] (f3) {\(Q_{-ij}\)};
        \vertex[below=0.4cm of a4] (g3) {\(Q_{+kl}\)};
        \diagram* { 
        (a1) -- [double,thick] (a5),
        (a2) -- [boson,quarter left] (b3),
        (b3) -- [boson,quarter left] (a4)
        };
        \end{feynman} 
        \end{tikzpicture}} = \langle V_{Qh}^{(-)}V_{Qh}^{(+)}\rangle= Tr[Q^{(-)}Q^{(+)}]\tilde{\mathcal{M}}^{Q^2}(G_{ret}) \ . 
\end{equation}
After substituting the corresponding value for the MIs, and moving to position space we get:
\begin{equation}
    S_{eff}^{Q_\pm}= \frac{G_N}{5}\int dt\  Q_{-ij}(t)\frac{d^5Q_{+ij}(t)}{dt^5} \ , 
\end{equation}

that is exactly the result given at the beginning of this chapter. This result agrees with Eq.(A21) of \cite{Galley:2012qs}

\subsection{Tail terms}

\begin{eqnarray}
     \scalebox{0.6}{\begin{tikzpicture}[baseline=(a1)] 
    \begin{feynman}
    \vertex (a1) ;
    \vertex[right=1cm of a1,square dot, red] (a2) {}; 
    \vertex[right=2cm of a2,square dot, red] (a3) {}; 
    \vertex[above=1.5cm of a3] (b3);
    \vertex[right=2cm of a3,square dot, red] (a4) {};
    \vertex[right=1cm of a4] (a5); 
    \vertex[below=0.4cm of a3] (e3) {\(E\)};
    \vertex[below=0.4cm of a2] (f3) {\(Q_{+ ij}\)};
    \vertex[below=0.4cm of a4] (g3) {\(Q_{- kl}\)};
    \diagram* { 
    (a1) -- [double,thick] (a5),
    (a2) -- [boson, black!60!green,quarter left] (b3),
    (b3) -- [boson, black!60!green,quarter left] (a4),
    (b3) -- [scalar, blue] (a3)
    };
    \end{feynman} 
    \end{tikzpicture}}&   =  & 
    <(V_{Qh})^{(+)}(t)V_{Qh}(t')^{(-)}V_{E\phi}(t")V_{h^2\phi}(t_i)>\nonumber \\ 
    &=& Q_+[2(-iG_{adv})(-iG_{adv})] Q_-(-iG_{ret}) E \label{eq:ININ_tail}
    \end{eqnarray}
According to \eqref{eq:ININ_tail}, one can recompute the tail effects diagram with multipole sources $Q_{-},Q_{+}E $, from \eqref{eq:tail_decomposed}: 
\begin{eqnarray}
    \scalebox{0.7}{\begin{tikzpicture}[baseline=(a1)] 
    \begin{feynman}
    \vertex (a1) ;
    \vertex[right=1cm of a1,square dot, red] (a2) {}; 
    \vertex[right=2cm of a2,square dot, red] (a3) {}; 
    \vertex[above=1.5cm of a3] (b3);
    \vertex[right=2cm of a3,square dot, red] (a4) {};
    \vertex[right=1cm of a4] (a5); 
    \vertex[below=0.4cm of a3] (e3) {\(E\)};
    \vertex[below=0.4cm of a2] (f3) {\(Q_{ij}^{(a)}\)};
    \vertex[below=0.4cm of a4] (g3) {\(Q_{kl}^{(b)}\)};
    \diagram* { 
    (a1) -- [double,thick] (a5),
    (a2) -- [boson, black!60!green,quarter left, edge label'=\(G_{1}\)] (b3),
    (b3) -- [boson, black!60!green,quarter left, edge label'=\(G_{2}\)] (a4),
    (b3) -- [scalar, blue] (a3)
    };
    \end{feynman} 
    \end{tikzpicture}}=  Tr[Q^{(a)}Q^{(b)}]E \tilde{\mathcal{M}}^{EQ^2}(G_1,G_2)
\end{eqnarray}
where: 
\begin{eqnarray}
    \tilde{\mathcal{M}}^{EQ^2}(G_1,G_2) & = & C^{EQ^2} \scalebox{0.8}{\begin{tikzpicture}[baseline=(current bounding box.center)]             \begin{feynman}
        \vertex (a) ;
        \vertex[right=0.2cm of a] (b);
        \vertex[right=1cm of b] (c); 
        \vertex[right=0.2cm of b] (e1) {\(G_1\)}; 
        \vertex[right=1cm of c] (d);
        \vertex[right=0.2cm of c] (e2) {\(G_2\)}; 
        \diagram* { 
        (b) -- [half left,very thick ] (c),
        (b) -- [half right,very thick ] (c),
        (c) -- [half left,very thick ] (d),
        (c) -- [half right,very thick ] (d),
        };
        \end{feynman} \end{tikzpicture}}
\end{eqnarray}
by multiplying by a factor $2$,by substituting $Q^{a}Q^{b}\to Q_-Q_+$, and by considering two advanced propagators $G_1,G_2\to G_{adv}$,$G_{adv}$, we get: 
\begin{eqnarray}
    \scalebox{0.7}{\begin{tikzpicture}[baseline=(a1)] 
    \begin{feynman}
    \vertex (a1) ;
    \vertex[right=1cm of a1,square dot, red] (a2) {}; 
    \vertex[right=2cm of a2,square dot, red] (a3) {}; 
    \vertex[above=1.5cm of a3] (b3);
    \vertex[right=2cm of a3,square dot, red] (a4) {};
    \vertex[right=1cm of a4] (a5); 
    \vertex[below=0.4cm of a3] (e3) {\(E\)};
    \vertex[below=0.4cm of a2] (f3) {\(Q_{-ij}\)};
    \vertex[below=0.4cm of a4] (g3) {\(Q_{+kl}\)};
    \diagram* { 
    (a1) -- [double,thick] (a5),
    (a2) -- [boson, black!60!green,quarter left, edge label'=\(G_{1}\)] (b3),
    (b3) -- [boson, black!60!green,quarter left, edge label'=\(G_{2}\)] (a4),
    (b3) -- [scalar, blue] (a3)
    };
    \end{feynman} 
    \end{tikzpicture}}=  Tr[Q_-Q_+]E \tilde{\mathcal{M}}^{EQ^2}(G_{adv},G_{adv}) \ . 
\end{eqnarray}
After substituting the corresponding expression for the MI we get:
\begin{equation}
    S^{EQ\pm}= -\frac{2 G_N^2 E}{5}\int_{-\infty}^{\infty}\frac{dk_0}{(2\pi)}k_0^6\left[\frac{1}{\epsilon}-\frac{41}{30}-i \pi + \log\left(\frac{k_0^2}{\bar{\mu}^2}\right)+\mathcal{O}(\epsilon)\right]Q_{+ ij}(k_0)Q^{ij}_-(-k_0)
\end{equation}
One can notice that the evaluation of the tail effects in the In-In formalism leads for the same equation of motions, that one could get using In-Out formalism. This is peculiar of these kinds of diagrams, and the evaluation is analogous for higher order tails with the same diagrammatic structure. This result coincide with Eq.(3.3) of \cite{Galley:2015kus} and with Eq.(12) of \cite{Foffa:2011np}.

\subsection{Memory  effects}
\begin{eqnarray}
     \scalebox{0.7}{\begin{tikzpicture}[baseline=(a1)] 
    \begin{feynman}
    \vertex (a1) ;
    \vertex[right=1cm of a1,square dot, red] (a2) {}; 
    \vertex[right=2cm of a2,square dot, red] (a3) {}; 
    \vertex[above=1.5cm of a3] (b3);
    \vertex[right=2cm of a3,square dot, red] (a4) {};
    \vertex[right=1cm of a4] (a5); 
    \vertex[below=0.4cm of a3] (e3) {\(Q_{- mn}\)};
    \vertex[below=0.4cm of a2] (f3) {\(Q_{+ ij}\)};
    \vertex[below=0.4cm of a4] (g3) {\(Q_{+ kl}\)};
    \diagram* { 
    (a1) -- [double,thick] (a5),
    (a2) -- [boson, black!60!green,quarter left] (b3),
    (b3) -- [boson, black!60!green,quarter left] (a4),
    (b3) -- [boson, black!60!green] (a3)
    };
    \end{feynman} 
    \end{tikzpicture}}&   =  & 
 <(V_{Qh})^{(+)}(V_{Qh})^{(-)}(V_{Qh})^{(+)}V_{h^3}> \nonumber \\ 
 & = & 3Q_-Q_+Q_+(-iG_{adv})(-iG_{adv})(-iG_{ret})\label{eq:ININ_memory}
    \end{eqnarray}
According to \eqref{eq:ININ_memory}, one can recompute the memory effect diagram from \eqref{eq:memory_decomposed}: 
\begin{eqnarray}
    \scalebox{0.7}{\begin{tikzpicture}[baseline=(a1)] 
    \begin{feynman}
    \vertex (a1) ;
    \vertex[right=1cm of a1,square dot, red] (a2) {}; 
    \vertex[right=2cm of a2,square dot, red] (a3) {}; 
    \vertex[above=1.5cm of a3] (b3);
    \vertex[right=2cm of a3,square dot, red] (a4) {};
    \vertex[right=1cm of a4] (a5); 
    \vertex[below=0.4cm of a3] (e3) {\(Q_{mn}^{(a)}\)};
    \vertex[below=0.4cm of a2] (f3) {\(Q_{ij}^{(b)}\)};
    \vertex[below=0.4cm of a4] (g3) {\(Q_{kl}^{(c)}\)};
    \diagram* { 
    (a1) -- [double,thick] (a5),
    (a2) -- [boson, black!60!green,quarter left, edge label'=\(G_{1}\)] (b3),
    (b3) -- [boson, black!60!green,quarter left, edge label'=\(G_{2}\)] (a4),
    (b3) -- [boson, black!60!green, edge label'=\(G_{3}\)] (a3)
    };
    \end{feynman} 
    \end{tikzpicture}}=  Tr[Q^{(a)}Q^{(b)}Q^{(c)}] \tilde{\mathcal{M}}^{Q^3}(G_1,G_2,G_3)
\end{eqnarray}
by multiplying by a factor $3$, and by considering one retarded and two advanced propagators, obtaining: 
\begin{eqnarray}
    S^{Q_+Q_+Q_-}_{eff\ 5PN}& =& 
    -\frac{G_N^2}{105} \int_{-\infty}^{\infty}\frac{dk_0}{(2\pi)}\frac{dq_0}{(2\pi)} \biggl(k_0^4 \left(-32 p_0^2 q_0^2+28 p_0^3 q_0+21 p_0^4+74 p_0 q_0^3-21 q_0^4\right)\nonumber \\
    & & 
    +32
      k_0^2 p_0^3 q_0^3+2 k_0^3 p_0 q_0^3 (2 p_0+23 q_0)+28 k_0 p_0^4 q_0^3+21 p_0^4
      q_0^4\biggr)Q_{+ij}(k_0)Q_{+jk}Q_{-ki} 
   \end{eqnarray}
And in position space we have: 
\begin{equation}
    S_{eff \ 5PN }^{Q_+Q_+Q_-}= \frac{1}{35}\int dt \  Tr\biggl[8\bigl(\dddot{Q}_+\bigr)^2\ddot{Q}_-+7\bigl(\ddddot{Q}_+\bigr)^2Q_- -12\dddot{Q}_+\ddot{Q}_+\dddot{Q}_--14\ddddot{Q}_+Q_+\ddddot{Q}_-\biggr]
    \label{eq:memory_ININ}
\end{equation}
Let us notice that for the memory term we have obtained a different result for the equation of motions from the one that we would have obtained working in In-Out formalism. This result agrees with $S_2 $ in Eq.(203) given in \cite{Blumlein:2021txe}. Notice that in \cite{Blumlein:2021txe} the authors are using a different parametrization of the $\pm$ variables with respect to ours. For this reason we should multiply Eq.(203) by an overall factor $\sqrt{2}/2$ in order to compare it to our result.

\subsection{Double emission}

Let us consider again the double emission diagram in order to evaluate it in the In-In formalism. According to \eqref{eq:double_double_emission}, there are 2 terms linear in $Q_-$: 
\begin{eqnarray}
 \scalebox{0.7}{\begin{tikzpicture}[baseline=(a1)] 
    \begin{feynman}
    \vertex (a1) ;
    \vertex[right=1cm of a1, square dot, red] (a2) {}; 
    \vertex[below=0.2cm of a2] (e2) {\(Q_+\)};
    \vertex[right=2cm of a2, square dot, red] (a3) {}; 
    \vertex[below=0.2cm of a3] (32) {\(Q_-\)};
    \vertex[above=1.5cm of a3] (b3);
    \vertex[right=2cm of a3,square dot, red] (a4) {};
    \vertex[below=0.2cm of a4] (e4) {\(Q_+\)};
    \vertex[right=1cm of a4] (a5);
    \vertex[below=0.2cm of a2] (f3) {};
    \vertex[below=0.2cm of a4] (g3) {};
    \diagram* { 
    (a1) -- [double,thick] (a5),
    (a2) -- [boson, black!60!green, ultra thick,half left] (a3),
    (a3) -- [boson, black!60!green, ultra thick,half left] (a4),
    };
    \end{feynman} 
    \end{tikzpicture}} & = & Q_+Q_-Q_+(-iG_{ret})(-iG_{adv})\\
 \scalebox{0.7}{\begin{tikzpicture}[baseline= (a1)] 
    \begin{feynman}
    \vertex (a1) ;
    \vertex[right=1cm of a1, square dot, red] (a2) {}; 
    \vertex[below=0.2cm of a2] (e2) {\(Q_+\)};
    \vertex[right=2cm of a2, square dot, red] (a3) {}; 
    \vertex[below=0.2cm of a3] (32) {\(Q_+\)};
    \vertex[above=1.5cm of a3] (b3);
    \vertex[right=2cm of a3,square dot, red] (a4) {};
    \vertex[below=0.2cm of a4] (e4) {\(Q_-\)};
    \vertex[right=1cm of a4] (a5);
    \vertex[below=0.2cm of a2] (f3) {};
    \vertex[below=0.2cm of a4] (g3) {};
    \diagram* { 
    (a1) -- [double,thick] (a5),
    (a2) -- [boson, black!60!green, ultra thick,half left] (a3),
    (a3) -- [boson, black!60!green, ultra thick,half left] (a4),
    };
    \end{feynman} 
    \end{tikzpicture}} & = & 2Q_+Q_+Q_- (-iG_{adv})(-iG_{adv})
\end{eqnarray}
We can compute the total contributions from \eqref{eq:double_emission_decomposed}:
\begin{equation}
 \scalebox{0.7}{\begin{tikzpicture}[baseline=(a1)] 
    \begin{feynman}
    \vertex (a1) ;
    \vertex[right=1cm of a1, square dot, red] (a2) {}; 
    \vertex[below=0.2cm of a2] (e2) {\(Q^{(a)}\)};
    \vertex[right=2cm of a2, square dot, red] (a3) {}; 
    \vertex[below=0.2cm of a3] (32) {\(Q^{(b)}\)};
    \vertex[above=1.5cm of a3] (b3);
    \vertex[right=2cm of a3,square dot, red] (a4) {};
    \vertex[below=0.2cm of a4] (e4) {\(Q^{(c)}\)};
    \vertex[right=1cm of a4] (a5);
    \vertex[below=0.2cm of a2] (f3) {};
    \vertex[below=0.2cm of a4] (g3) {};
    \diagram* { 
    (a1) -- [double,thick] (a5),
    (a2) -- [boson, black!60!green, ultra thick,half left,edge label'=\(G_{1}\)] (a3),
    (a3) -- [boson, black!60!green, ultra thick,half left,edge label'=\(G_{2}\)] (a4),
    };
    \end{feynman} 
    \end{tikzpicture}}  =  Tr[Q^{(a)}Q^{(b)}Q^{(c)}]\tilde{\mathcal{M}}^{Q^32} \ , \\
\end{equation} 
substituting appropriately symmetry factors, multipoles and propagators, and summing the two terms, obtaining:
\begin{eqnarray}
    S_{eff \ 5PN}^{Q_+Q_+Q_- 2 } & =&  
    G_N^2\int \frac{dk_0}{(2\pi)}\frac{dq_0}{(2\pi)}k_0^3q_0^3Q_{-ij}(k_0)Q_{+jk}(q_0)Q_{+kl}(-k_0-q_0) \nonumber \\ 
    & & 
    - \frac{G_N^2}{2}\int \frac{dk_0}{(2\pi)}\frac{dq_0}{(2\pi)}k_0^3q_0^3Q_{+ij}(k_0)Q_{+jk}(q_0)Q_{-kl}(-k_0-q_0)\ ,
\end{eqnarray}
and in position space we have: 
\begin{equation}
    S_{eff \ 5PN}^{Q_+Q_+Q_- 2 }\ = \ \int dt \ Tr\biggl[\ddddot{Q}_-\ddddot{Q}_+Q_+ -\frac{1}{2}\ddddot{Q}_+\ddddot{Q}_+Q_- \biggr]
    \label{eq:double_emission_ININ}
\end{equation}
This result does not agree with $S1$, given in Eq.(186) of \cite{Blumlein:2021txe}, in particular we need to divide our result by $3$ to get Eq.(186). This may be due to a missing symmetry factor, and this apparent mismatch have to be sorted out in a future work.

\section{Evaluation of the conservative Lagrangian for memory effects}
On this section we will use the analysis done so far to evaluate the conservative Lagrangian in the In-In formalism. As pointed out in Eq.\eqref{eq:In-In-Action}, once we integrate out degrees of freedom in a radiation system, and we perform the following change of variables:
\begin{equation}
    Q_+\ = \ \frac{Q_1+Q_2}{2} \qquad Q_- \ = \ Q_1 - Q_2 \ , 
    \label{eq:change_qpm}
\end{equation}
the generic action that we obtain is given by the sum of three pieces: 
\begin{eqnarray}
    S_{eff\ 5PN}^{Q^3} \ = \ \int \ dt \ \biggl(L[Q_1]-L[Q_2]+R[Q_1,Q_2] \biggr)\ , 
\end{eqnarray}
where $L[Q_i]$ can be identified as the conservative Lagrangian, whereas $R[Q_1,Q_2]$ is a generic non-conservative potential.
Hence, from the computation of the memory effect in Eq.\eqref{eq:memory_ININ}, we can perform the change of variables of Eq.\eqref{eq:change_qpm}, obtaining: 
\begin{eqnarray}
    S_{eff\ 5PN}^{Q_+Q_+Q_-} \ &=&  \ \int \ dt \ \biggl(L^{(1)}[Q_1]-L^{(1)}[Q_2]+R^{(1)}[Q_1,Q_2] \biggr) \ ,  \\ 
    L^{(1)}[Q_1]& = &  
    G_N^2 \ Tr \biggl[-\frac{1}{20}Q_1Q_1^{(4)}Q_1^{(4)}-\frac{1}{35}Q_1^{(2)}Q_1^{(3)}Q_1^{(3)}\biggr]\ ,  \\ 
    R^{(1)}[Q_1,Q_2] & = & 
    G_N^2 \ Tr\biggl[\frac{1}{10}Q_1Q_1^{(4)}Q_2^{(4)}-\frac{3}{20}Q_2Q_1^{(4)}Q_1^{(4)}+\frac{4}{35}Q_1^{(2)}Q_1^{(3)}Q_2^{(3)}-\frac{1}{7}Q_2^{(2)}Q_1^{(3)}Q_1^{(3)}\biggr]\nonumber \\ 
    & &  -(1\leftrightarrow 2) \ ,
\end{eqnarray}
where $L^{(1)}[Q]$ can be identified as the conservative Lagrangian.
This result is in agreement with Eq.206 and Eq.208 of \cite{Blumlein:2021txe}. Analogously, we can rewrite the result for the double-emission diagram from Eq.\eqref{eq:double_emission_ININ}, as: 
\begin{eqnarray}
    S_{eff\ 5PN}^{Q_+Q_+Q_-2 } \ &=&  \ \int \ dt \ \biggl(L^{(2)}[Q_1]-L^{(2)}[Q_2]+R^{(2)}[Q_1,Q_2] \biggr) \ ,  \\ 
    L^{(2)}[Q_1]& = &  \frac{G_N^2}{8} Tr\biggl[Q_1Q_1^{(4)}Q_1^{(4)}\biggr] \ ,  \\ 
    R^{(2)}[Q_1,Q_2] & = &  G_N^2 Tr\biggl[-\frac{1}{4}Q_1Q_1^{(4)}Q_2^{(4)}+\frac{3}{8}Q_2Q_1^{(4)}Q_1^{(4)}\biggr] -(1\leftrightarrow 2 ) \ ,
\end{eqnarray}
where $L^{(2)}[Q]$ is the corresponding conservative Lagrangian.
This result does not agree with Eq.205 and Eq.207 of \cite{Blumlein:2021txe}, as noticed in the previous section. However, we cannot say that these conservative Lagrangians $L^{(1)}[Q],L^{(2)}[Q]$ really encodes the conservative dynamics. In fact the In-In formalism has been designed to get the equations of motions of the system and not the conservative Lagrangians. The conservative part is not uniquely defined, and one can always add terms in the action that vanish in the equations of motion, that modify the conservative part.

\chapter{GR as an EFT and Post-Minkowskian scheme}
\label{Chapter:GR_PM}
In the previous chapters we described a systematic procedure to study the classical dynamics of a binary coalescing system using a Post-Newtonian expansion. On this chapter instead we want to develop an alternative, but complementary, framework, that allow us to obtain both classical and quantum General Relativistic corrections to the Newtonian potential between two massive objects. This is possible by working in the Post-Minkowskian perturbative scheme, which is a weak field expansion where at $n$-PM order we consider corrections that scale as $G_N^n$. By working in this relativistic framework, one obtains expressions valid at all order in velocity powers, and can recover post-Newtonian results by focusing on the classical contributions, and performing a suitable non-relativistic limit, as first noticed in 1971 by Iwasaki in \cite{Iwasaki:1971vb}. In order to develop this framework, we have to look again to the full theory of General Relativity, this time from a QFT perspective, considering it not as a fundamental theory but rather as a low-energy effective field theory (EFT)\cite{Donoghue:1993eb, Donoghue:1995cz} of an unknown ultraviolet theory (UV) of gravity (see \cite{Burgess:2003jk,Donoghue:1995cz} for general reviews). Within the EFT approach, it is possible to study a binary coalescing system as a scattering process, involving particles that can have different spins, which are gravitationally interacting. Moreover, from these scattering amplitudes it is possible to compute post-Minkowskian corrections to the Newtonian potential between two massive objects.\cite{Bjerrum-Bohr:2002gqz}\cite{Cristofoli:2019neg}\cite{Bjerrum-Bohr:2022blt}. In order to compute them, it is not necessary to evaluate the full loop amplitudes, but we can focus only on the long-range contributions which arise from non-analytical terms in the amplitude. In the treatment of the PM approach to GR we will adopt the "mostly minus" convention: $\eta_{\mu\nu}=diag(+,-,-,-)$.
\subsubsection{Goals of the Chapter:}
The chapter will be divided as follows: 
\begin{itemize}
    \item We will introduce GR from a QFT prospective, showing that the EFT approach is nowadays the right approach to study GR. We will then quantize the theory and derive the relevant Feynman rules from it.
    \item Then, we will show how one can obtain a non-relativistic potential from scattering amplitudes in the so-called Born series, obtained from the \textit{Lippman-Schwinger} equation, as an example we will then compute the Newtonian potential from the low-energy limit a tree-level calculation;
    \item Eventually, we will explain a smart method to compute 1-loop corrections to the Newtonian potential, for which is not necessary to compute the full 1-loop amplitude, but one can focus only on the massless double-cut of the amplitude.
\end{itemize}
\section{General Relativity from a QFT perspective}
Let us suppose to be interested in the study of General Relativity from a Quantum Field Theory prospective, instead of the usual Geometrical interpretation. 
We want to construct a theory for a massless spin-2 particle, the \textit{graviton} $h_{\mu\nu}(x)$, which is invariant under a gauge symmetry, which is the invariance under local coordinate transformations: 
\begin{equation}
    x^\mu \to x^\mu +\xi^\alpha(x) \ . 
\end{equation}
As showed on Chapter\ref{Ch:grav_waves} the Noether current associated to this symmetry is the energy momentum tensor $T_{\mu\nu}$, and the associated charges are energy and momentum. \\ 
Under such gauge transformation the graviton field transforms as:
\begin{equation}
  h_{\mu\nu}'(x)=h_{\mu\nu}+\partial_\mu\xi_\nu+\partial_\nu\xi_\mu \ . 
\end{equation}
If we try to build a Lagrangian for the field $h_{\mu\nu}$ which respects Poincarè and gauge invariance then we obtain the \textit{Fierz-Pauli} action, that we introduced in the previous chapter, defined as: 
\begin{equation}
      S_{FP}=\frac{-1}{64\pi G_N}\int d^4x(\partial_{\rho}h_{\alpha\beta}\partial^{\rho}h^{\alpha\beta}-\partial_{\rho}h\partial^{\rho}h+2\partial_{\rho}h^{\rho\alpha}\partial_{\alpha}h-2\partial_{\rho}h_{\alpha\beta}\partial^{\beta}h^{\rho\alpha})\ . 
\end{equation}
One can then start to add self-interactions between graviton fields, and it is possible to show \cite{Feynman:1996kb} that the minimal set of the interactions can be combined in the Einstein-Hilbert action: 
\begin{equation}
    S_{EH}=\frac{2}{\kappa^2}\int d^4 x \sqrt{-g}R \ , 
    \label{eq:eh_action}
\end{equation}
 where $R$ is the Ricci scalar, and $\kappa=\sqrt{32\pi G_N}$.
 As remarked on chapter.\ref{Ch:grav_waves}, this action is invariant under the whole group of diffeomorphisms: 
 \begin{equation}
     x^\mu\to x^{'\mu}(x)
 \end{equation}
 and the metric tensor transforms as: 
\begin{equation}
    g_{\mu\nu}(x)\to g_{\mu\nu}'(x)=\frac{\partial x^\rho}{\partial x^{' \mu}}\frac{\partial x^\sigma}{\partial x^{' \nu}}g_{\rho\sigma}(x) \ .
\end{equation}
If we perform a weak field expansion around the flat Minkowski metric:
\begin{equation}
    g_{\mu\nu}=\eta_{\mu\nu}+ \frac{h_{\mu\nu}}{\kappa} \ , 
\end{equation}
one recovers the Fierz-Pauli action and an infinite number of self-interactions among graviton fields. \\ 
The quantization of this theory has been studied for a long time \cite{Feynman:1963ax,DeWitt:1967yk,DeWitt:1967ub,DeWitt:1967uc}, and was De Witt in \cite{DeWitt:1967uc} the first one who derived the Feynman rules in a Lorentz covariant form.
From dimensional analysis one can notice that $[\kappa]=-1$, so this is a non-renormalizable theory. 
Hence, if one is interested in studying scattering processes involving gravitons, increasing the number of loops one encounters an increasing number of divergences, which can be removed by adding a set of counter-terms to the action. More precisely, if we decide to work in dimensional regularization, with $d$ the continuous dimensions of spacetime,  an appropriate suppression of all the divergences at one loop order can be effectively obtained by adding to $S_{EH}$ the following counterterm \cite{tHooft:1974toh}: 
\begin{equation}
    \Delta S =  \frac{1}{8\pi^2 \epsilon}\int d^4 x \sqrt{-g}\left(\frac{R^2}{120}+\frac{7}{20}R_{\mu\nu}R^{\mu\nu}\right) \ , 
\end{equation}
where $\epsilon=4-d$. \\ 
At two loops we should add accordingly other counterterms, with cubic curvature invariants like $R_{\mu\nu\rho\sigma}R^{\rho\sigma\alpha\beta}R_{\alpha\beta}^{\mu\nu}$ multiplied by appropriate divergent coefficients. \\ 
Given the form of these counterterms, it is impossible to reabsorb divergences into a renormalization of the fields, and of the coupling constant $\kappa$. The finite part of the coefficients that multiplies each counterterm must be fixed by comparison with experiments, limiting the predictive power of the theory. 
\subsection{GR as an Effective Field Theory}
Going up with the loop order the situation gets even worse, and our theory of gravity looses completely its predictive character. \\We need to look for an alternative approach to the problem. A hint on how to proceed comes from the chiral perturbation theory: just like the QFT of gravity discussed above, this theory is non-renormalizable and non-linear, being the low energy limit of QCD. Nonetheless, predictive and experimentally verified calculations have been accomplished within it, in the theoretical framework of the effective field theories (EFT), which basically involves a separation between the low-energy, well-behaved degrees of freedom and the diverging high energy sector. \\ 
As first noticed in  \cite{Donoghue:1993eb, Donoghue:1995cz},  the key to overcome this problem is to treat General Relativity as a low-energy effective field theory of an unknown ultraviolet (UV) theory of Quantum Gravity, with the Plank mass $M_p=\frac{1}{\kappa}\sim 1.2 \times 10^{19} \ Gev$ as the UV cutoff of this theory, that is the energy above which the EFT loses its validity. In order to consistently build an effective action describing this problem from a bottom-up approach we need to:
\begin{itemize}
    \item Identify the low-energy degrees of freedom of the theory,
    \item define the symmetries of the action,
    \item define appropriate power counting rules\ , 
\end{itemize}
and then construct the most general Lagrangian respecting the symmetries of the system. \\ 
If we consider a pure gravitational theory, the only low-energy degrees of freedom are the metric tensors, $g_{\mu\nu}$, and the action should be invariant under Poincarè symmetry and under generic diffeomorphisms.  \\
The most general action satisfying these symmetries contains the Einstein Hilbert term, plus an infinite tower of terms:
\begin{equation}
    S_{grav}=\int d^4x\sqrt{-g}\biggl(\Lambda+\frac{2}{\kappa^2}R+c_1R^2+c_2R_{\mu\nu}R^{\mu\nu}+\cdots\biggr)
    \label{eq:s_grav}
\end{equation}
where $\Lambda,c_1,c_2$ are constants and the ellipses denote higher order powers of $R,R_{\mu\nu}$ and $R_{\mu\nu\alpha\beta}$.\\
The constant $\Lambda$ is proportional to the cosmological constant, and it can be safely set to zero by experimental evidence. \\
The coefficients $c_1,c_2,\cdots$ represent the appearance of the effective couplings in the action, and they are called \textit{Wilson Coefficients}. \\
One can associate any divergence arising in the computation of loop diagrams with some component of the action \eqref{eq:s_grav}, and hence absorb it via a simple redefinition of the respective coupling constant. 
We need then to define an appropriate energy expansion that allow us to identify which terms are relevant at low-energies. \\
Since by definition the curvature $R$ is proportional to second order derivatives, and in momentum space derivatives turn into momentum terms $i\partial_\mu \sim p_\mu$, each curvature is said to be of order $p^2$: 
\begin{equation}
    \{R,R_{\mu\nu},R_{\mu\nu\rho\sigma}\}\sim \partial^2\sim p^2 \ .
\end{equation}
That means that terms in the action containing two powers of curvature are of order $p^4$. Since we are interested in the low-energy theory, the graviton energy can be arbitrary small, and such terms are much smaller with respect to those of order $p^2$. \\
Thus, for long distance or equivalently low energy computations such as the one we are interested in, where the energy scale is well below $M_p$, all the terms in \eqref{eq:s_grav} except for the Einstein-Hilbert one provide extremely little contributions. \\ 
In conclusion then, by following the EFT prescription, one can describe the pure gravitational sector by truncating of $S_{grav}$ to the simple Einstein-Hilbert action, without the concern for singularities, which affect only the short-distance, high-energy scale we neglect.  \\ 
Let us now consider the EFT of a scalar field coupled to gravity.
The gravitational effective field theory action of a scalar field coupled to gravity is made up of two pieces:
\begin{equation}
    S=S_{grav}+S_{matter}
\end{equation}
where $S_{grav}$ is the gravitational action described in the previous section whereas $S_{matter}$ is the action of a scalar field coupled to gravity and in general it is given by:
\begin{equation}
    S_{matter}= \int d^4x \sqrt{-g}\frac{1}{2}\biggl(g^{\mu\nu}\partial_{\mu}\phi\partial_{\nu}\phi-m^2\phi^2\biggr)+d_1R^{\mu\nu}\partial_\mu\phi\partial_\nu\phi + R\biggl(d_2\partial_\mu\phi\partial^\mu\phi+d_3m^2\phi^2\biggr) \ , 
\end{equation}
where the derivatives acting on a massive matter field $\phi$ are not small quantities, since they will generate powers of the interacting masses, which are usually orders of magnitude higher than the momentum terms. \\ Hence, the derivative expansion only counts derivatives which act on the light field, which in this case are only the gravitons. \\ The coefficients $d_1,d_2,d_3$ denote terms with higher order derivative couplings. \\
At lowest order only the minimal coupling term is relevant, and we will focus only on that in the following. \\
We can summarize the action as the sum of three pieces:
\begin{equation}
    S=S_{EH}+S_{m} +S_{EF} \ , 
\end{equation}
where $S_{EH}$ is the Einstein-Hilbert action, $S_m$ is the action of a scalar field minimally coupled to gravity and $S_{EF}$ contains an infinite series of higher order derivative operators associated with new gravitational couplings, and ensure that, order by order in the energy expansion, any UV divergence due to loop effects can be absorbed in the effective action. \\ 
The theory is UV consistent up to the cut-off determined by the validity of the energy expansion, that we assume to be of the order of the Plank mass: $M_{Plank}\approx 10^{19}GeV$. \\

\section{Quantization of General Relativity}
\label{sec:Feynman_Rules_PM}
Now that we have defined an effective field theory, we can extract the Feynman rules for the theory, following \cite{Holstein:2008sx}, by considering only the low energy effective action, and working in dimensional regularization:
\begin{equation}
    S=S_{EH}+S_{m} \ , 
\end{equation}
where:
\begin{equation}
    S_{EH}=\frac{2}{\kappa^2}\int d^dx\sqrt{-g}R \qquad S_{m}= -\frac{1}{2}\int d^dx\sqrt{-g}(\partial_\mu\phi\partial^\mu\phi-m^2\phi^2) \ ,
\end{equation}
where $d$ is the continuous dimension of spacetime, $\kappa=\sqrt{32\pi G_d}$, with $G_d=G_N\mu^{4-d}$ the $d$-dimensional coupling constant.
We need first to quantize our theory, and we can do that following the background field method, in which the metric tensor is expanded as:
\begin{equation}
    g_{\mu\nu}=\Bar{g}_{\mu\nu}+\kappa h_{\mu\nu} \ , 
\end{equation}
where $\Bar{g}_{\mu\nu}$ is a classical background metric and $h_{\mu\nu}$ is the quantum field. \\
Calculations are greatly simplified with this method because gravitons running in the loops must be derived from an expansion involving the quantum field $h_{\mu\nu}$ whereas gravitons that are not within a loop may be derived from expanding the background field:
\begin{equation}
    \Bar{g}_{\mu\nu}= \eta_{\mu\nu}+\kappa H_{\mu\nu} \ , 
\end{equation}
where $H_{\mu\nu}$ denotes an external graviton that is not inside a loop. \\
Being the theory redundant under a gauge symmetry, we need to impose gauge fixing condition only on the quantum field $h_{\mu\nu}$, leaving the general covariance of the background unaffected. We will be interested in one loop calculations, where at most two gravitons involved in a vertex are propagating within a loop, which allows us to use a gauge fixing condition linear in the quantum field $h_{\mu\nu}$ and greatly simplifies the derivation of both the triple graviton vertex and the vertex that couples one graviton to the ghost fields. We work in the harmonic gauge:
\begin{equation}
    \Bar{D}^{\nu}h_{\mu\nu}-\frac{1}{2}\Bar{D}_{\mu}h=0 \ , 
\end{equation}
where $\Bar{D}_{\mu}$ denotes the covariant derivative on the background metric. \\
This condition leads to a gauge fixing piece of the action:
\begin{equation}
    S_{GF}=\int d^dx\sqrt{-\Bar{g}}\biggl(\Bar{D}^{\nu}h_{\mu\nu}-\frac{1}{2}\Bar{D}_{\mu}h\biggr)\biggl(\Bar{D}_{\rho}h^{\mu\rho}-\frac{1}{2}\Bar{D}^{\mu}h\biggr) \ , 
\end{equation}
as well as the ghost action:
\begin{equation}
    S_{Ghost}=\int d^dx\sqrt{-\Bar{g}}\Bar{\eta}^{\mu}\bigl(\Bar{D}_{\mu}\Bar{D}_{\nu}-R_{\mu\nu}\bigr)\eta^{\nu}
\end{equation}
where with $\eta^{\mu}$ and $\Bar{\eta}^{\mu}$ we denote respectively the ghost and anti-ghost particle.\\
We can now derive the Feynman rules for the effective field theory of gravity. \\
The total gravitational action is made of three pieces:
\begin{equation}
    S_{g}=S_{EH}+S_{GF}+S_{Ghost} \ . 
\end{equation}
By expanding $S_{g}$ in terms of $h_{\mu\nu}$ and $H_{\mu\nu}$  it is possible to derive the Feynman rules for the propagator of the graviton field, all the pure gravitational vertices, the ghost propagator, and the ghosts-graviton interactions, whereas expanding $S_{matter}$ we can get the scalar propagator and scalar-graviton interaction vertices. 
The Feynman rule describing the interaction of 2-scalars and 1 graviton is given by:
\begin{equation}
\scalebox{0.8}{
\begin{tikzpicture}[baseline=(current bounding box.center)] \begin{feynman}
\vertex (a) ;
\vertex[below=1.5cm of a] (b);
\vertex[below=1.5cm of b] (c); 
\vertex[right=1.5cm of b] (d){\(\mu\nu\)};
\vertex[below=1cm of d] (e) ;
\vertex[above=1cm of d] (f);
\diagram* { 
(c) -- [fermion, edge label'=\(p_1\)] (b)-- [fermion, edge label'=\(p_2\)] (a),
(b) -- [gluon] (d) ,
};
\end{feynman} \end{tikzpicture}}= \tau_{1\mu\nu}(p_1,p_2,m) =  -\frac{i\kappa}{2}\biggl[p_{1\mu}p_{2\nu}+p_{1\nu}p_{2\mu}-\eta_{\mu\nu}(p_1\cdot p_2 -m^2)\biggr]
\end{equation}\\
The 2-scalar 2-graviton vertex reads:

\begin{equation*}
\centering
\scalebox{0.8}{
\begin{tikzpicture}[baseline=(current bounding box.center)] \begin{feynman}
\vertex (a) ;
\vertex[below=1.5cm of a] (b);
\vertex[below=1.5cm of b] (c); 
\vertex[right=1.5cm of b] (d);
\vertex[below=1cm of d] (e) {\(\mu\nu\)};
\vertex[above=1cm of d] (f){\(\rho\sigma\)};
\diagram* { 
(c) -- [fermion, edge label'=\(p_1\)] (b)-- [fermion, edge label'=\(p_2\)] (a),
(b) -- [gluon] (e) ,
(b) -- [gluon] (f) ,
};
\end{feynman} \end{tikzpicture}}
\end{equation*}
\begin{eqnarray}
\tau_{2\mu\nu\rho\sigma}(p_1,p_2) &=& \frac{i\kappa^2}{2}\biggl[
2I_{\mu\nu\alpha\gamma}I^{\gamma}_{\beta,\rho\sigma}(p_{1}^{\alpha}p_{2}^{\beta}+p_{1}^{\beta}p_{2}^{\alpha})-(\eta_{\mu\nu}I_{\alpha\beta,\rho\sigma}+\eta_{\rho\sigma}I_{\alpha\beta,\mu\nu})p_1^\alpha p_2^\beta\nonumber \\
& & - \mathcal{P}_{\mu\nu,\rho\sigma}(p_1\cdot p_2 -m^2)\biggr]
\end{eqnarray}
with:
\begin{eqnarray}
   I_{\alpha\beta,\gamma\delta}&=& \frac{1}{2}(\eta_{\alpha\gamma}\eta_{\beta\delta}+\eta_{\alpha\delta}\eta_{\beta\gamma}) \\
   P_{\alpha\beta,\gamma\delta}&=&\frac{1}{2}(\eta_{\alpha\gamma}\eta_{\beta\delta}+\eta_{\alpha\delta}\eta_{\beta\gamma}-\frac{2}{d-2}\eta_{\alpha\beta}\eta_{\gamma\delta})\\
   \mathcal{P}_{\alpha\beta,\gamma\delta}&=&\frac{1}{2}(\eta_{\alpha\gamma}\eta_{\beta\delta}+\eta_{\alpha\delta}\eta_{\beta\gamma}-\eta_{\alpha\beta}\eta_{\gamma\delta})
\end{eqnarray}
The scalar field propagator is given by:
\begin{equation}
\begin{tikzpicture}[baseline=(current bounding box.center)] \begin{feynman}
\vertex (a) {\(\mu\)};
\vertex[right =2 cm of a] (b) {\(\nu\)};
\diagram* { 
(a) -- [fermion , edge label'=\(q\)] (b),
};
\end{feynman} \end{tikzpicture}=\frac{i}{q^2+i\epsilon}
\end{equation}
The graviton propagator is given by:
\begin{equation}
    \begin{tikzpicture}[baseline=(current bounding box.center)] \begin{feynman}
\vertex (a) {\(\mu\nu\)};
\vertex[right =2 cm of a] (b) {\(\rho\sigma\)};
\diagram* { 
(a) -- [gluon, edge label'=\(q\)] (b),
};\end{feynman} \end{tikzpicture}= iD_{\mu\nu\rho\sigma}(q)=\frac{iP_{\mu\nu\rho\sigma}}{q^2} \qquad  
\end{equation}
The ghost propagator is:
\begin{equation}
    \begin{tikzpicture}[baseline=(current bounding box.center)] \begin{feynman}
\vertex (a) {\(\mu\)};
\vertex[right =2 cm of a] (b) {\(\nu\)};
\diagram* { 
(a) -- [charged scalar , edge label'=\(q\)] (b),
};\end{feynman} \end{tikzpicture}=iD_{\mu\nu}(q)=\frac{i\eta_{\mu\nu}}{q^2}
\end{equation}
The triple graviton vertex reads:
\begin{equation*}
\scalebox{0.8}{
    \begin{tikzpicture}[baseline=(current bounding box.center)] \begin{feynman}
\vertex (a) {\(\gamma\delta\)};
\vertex[below=1.5cm of a] (b) ;
\vertex[below=1.5cm of b] (c) {\(\alpha\beta\)}; 
\vertex[right=1.5cm of b] (d){\(\mu\nu\)};
\vertex[below=1cm of d] (e) ;
\vertex[above=1cm of d] (f);
\diagram* { 
(c) -- [gluon, momentum=\(k+q\)] (b)-- [gluon, momentum=\(k\)] (a),
(b) -- [gluon, momentum =\(q\)] (d) ,
};
\end{feynman} \end{tikzpicture}}
\end{equation*}
\begin{eqnarray}
    \tau^{\mu\nu}_{3\alpha\beta,\gamma\delta}&=&\frac{-i\kappa}{2}\biggl\{\mathcal{P}_{\alpha\beta,\gamma\delta}\bigl[k^\mu k^\nu+(k+q)^\mu(k+q)^\nu+q^\mu q^\nu -\frac{3}{2}\eta^{\mu\nu}q^2\bigr]\nonumber\\
    & & +2q_\lambda q_\sigma \bigl[I^{\lambda\sigma}_{\alpha\beta}I^{\mu\nu}_{\gamma\delta}+I^{\lambda\sigma}_{\gamma\delta}I^{\mu\nu}_{\alpha\beta}-I^{\lambda\mu}_{\alpha\beta}I^{\sigma\nu}_{\gamma\delta}-I^{\sigma\nu}_{\alpha\beta}I^{\lambda\mu}_{\gamma\delta}\bigr] \nonumber\\
    & &+\bigl[q_{\lambda}q^\mu(\eta_{\alpha\beta}I^{\lambda\nu}_{\gamma\delta}+\eta_{\gamma\delta}I^{\lambda\nu}_{\alpha\beta})+q_\lambda q^\nu(\eta_{\alpha\beta}I^{\lambda\mu}_{\gamma\delta}+\eta_{\gamma\delta}I^{\lambda\mu}_{\alpha\beta})\nonumber\\
    & &-q^2(\eta_{\alpha\beta}I^{\mu\nu}_{\gamma\delta}+\eta_{\gamma\delta}I^{\mu\nu}_{\alpha\beta})-\eta^{\mu\nu}q^\lambda q^\sigma (\eta_{\alpha\beta}I_{\gamma\delta,\lambda\sigma}+\eta_{\gamma\delta}I_{\alpha\beta,\lambda\sigma})\bigr] \nonumber\\
    & &+\bigl[2q^\lambda(I^{\sigma\nu}_{\gamma\delta}I_{\alpha\beta\lambda\sigma}k^\mu+I^{\sigma\mu}_{\gamma\delta}I_{\alpha\beta\lambda\sigma}k^\nu-I^{\sigma\nu}_{\alpha\beta}I_{\gamma\delta,\lambda\sigma}(k+q)^\mu-I^{\sigma\mu}_{\alpha\beta}I_{\gamma\delta,\lambda\sigma}(k+q)^\nu) \nonumber\\
    & &+q^2(I^{\sigma\mu}_{\alpha\beta}I_{\gamma\delta,\sigma}^\nu I^{\sigma\mu}_{\gamma\delta})+\eta^{\mu\nu}q^\lambda q_\sigma(I_{\alpha\beta,\lambda\rho}I^{\rho\sigma}_{\gamma\delta}+I_{\gamma\delta,\lambda\rho}I^{\rho\sigma}_{\alpha\beta})\bigr]\nonumber\\ 
    & &\bigl[(k^2+(k+q)^2)\bigl(I^{\sigma\mu}_{\alpha\beta}I_{\gamma\delta,\sigma}^{\nu}+I^{\sigma\nu}_{\alpha\beta}I_{\gamma\delta,\sigma}^\mu-\frac{1}{2}\eta^{\mu\nu}\mathcal{P}_{\alpha\beta,\gamma\delta}\bigr)\nonumber \\
    & & -((k+q)^2\eta_{\alpha\beta}I^{\mu\nu}_{\gamma\delta}+k^2\eta_{\gamma\delta}I^{\mu\nu}_{\alpha\beta})\bigr]\biggr\}
\end{eqnarray}
The Feynman rule describing the interaction between one graviton field and two ghost fields is given by:
\begin{equation*}
\centering
\scalebox{0.8}{
\begin{tikzpicture}[baseline=(current bounding box.center)] \begin{feynman}
\vertex (a) ;
\vertex[below=1.5cm of a] (b);
\vertex[below=1.5cm of b] (c); 
\vertex[right=1.5cm of b] (d){\(\mu\nu\)};
\vertex[below=1cm of d] (e) ;
\vertex[above=1cm of d] (f);
\diagram* { 
(c) -- [ghost, momentum=\(k+q\)] (b)-- [ghost, momentum =\(k\)] (a),
(b) -- [gluon, momentum =\(q\)] (d) ,
};
\end{feynman} \end{tikzpicture}}
\end{equation*}
\begin{eqnarray}
\tau^{\mu\nu}_{G\alpha,\beta}(k,q)&=&\frac{i\kappa}{2}\biggl[ (k^2+(k+q)^2+q^2)I_{\alpha\beta}^{\mu\nu}+2\eta_{\alpha\beta}k^\lambda(k+q)^\sigma P_{\lambda\sigma}^{\mu\nu}+2q_\alpha k^\lambda I_{\beta\lambda}^{\mu\nu}\nonumber \\
& & -2q_\beta(k+q)^\lambda I_{\alpha\lambda}^{\mu\nu}+q_\alpha q_\beta \eta^{\mu\nu}\biggr]
\end{eqnarray}
In the case of $\tau^{\mu\nu}_{3\alpha\beta,\gamma\delta}$ and $\tau^{\mu\nu}_{G\alpha,\beta} $ the graviton with Lorentz indices $\mu\nu$ represents a background graviton, and therefore it cannot carry loop momenta. \\
Now that we have obtained the Feynman rules, we want to understand how to use scattering amplitudes techniques to compute General Relativity corrections to physical observables. 
\section{Interaction potential from Scattering amplitudes}
\label{sec:scattering_amplitudes_intro}
In this section we will show how it is possible to obtain a two-body non-relativistic potential from scattering amplitudes \cite{Cristofoli:2019neg,Grignani:2020ahv}.\\ 
This method can be seen as a relativistic generalization of the connection between amplitudes and interaction potentials that is typically found in non-relativistic quantum mechanics books.
Let us consider the one-particle Hamiltonian for a two-body system of massive particles: 
\begin{equation}
    \hat{\mathcal{H}}=\hat{\mathcal{H}}_0+\hat{V} \ , 
\end{equation}
where
\begin{equation}
    \hat{\mathcal{H}}_0 = c\sqrt{\hat{p}^2+m_1^2c^2}+c\sqrt{\hat{p}^2+m_2^2c^2} \ , 
\end{equation}
is the free Hamiltonian, whereas $\hat{V}$ is a potential containing the gravitational interactions. \\ 
Then we can define the $\mathbb{C}-$valued Green's operator: 
\begin{equation}
    \hat{G}_0(z)=\bigl(z-\hat{\mathcal{H}}_0^{-1}\bigr) \ , \qquad \hat{G}(z)=\bigl(z-\hat{\mathcal{H}}^{-1}\bigr) \ . 
\end{equation}
These operators are analytic throughout the complex plane apart from the spectra of the respective Hamiltonians.
Therefore, by knowing $\hat{G}_0(z)$ and $\hat{G}(z)$ for all $z\in\mathbb{C}$ we have a complete solution to the eigenvalue problem of $\hat{\mathcal{H}}_0$ and $\hat{\mathcal{H}}$. We can observe that  $\hat{G}_0(z)$ and $\hat{G}(z)$ can be related by using the following identity:
\begin{equation}
    \hat{A}^{-1}=\hat{B}^{-1}+\hat{B}^{-1}(\hat{B}-\hat{A})\hat{A}^{-1} \ .
\end{equation}
For $\hat{A}^{-1}=\hat{G}$ and $\hat{B}^{-1}=G_0$ one obtains: 
\begin{equation}
    \hat{G}=\hat{G}_0 + \hat{G}_0\hat{V}\hat{G} \ . 
    \label{eq:G}
\end{equation}
Along with these Green's operators, we introduce also the off-shell scattering operator: 
\begin{equation}
    \hat{T}(z)= \hat{V}+\hat{V}\hat{G}(z)\hat{V} \ ,
    \label{eq:T}
\end{equation}
whose on-shell matrix elements correspond to the non-trivial components of the scattering S-matrix. \\
Moreover, by inserting \eqref{eq:G} in \eqref{eq:T} one obtains the well-known \textit{Lippmann-Schwinger equation}: 
\begin{eqnarray}
    \hat{T}(z)&= & \hat{V}+\hat{V}\bigl(\hat{G}_0 + \hat{G}_0\hat{V}\hat{G}\bigr)\hat{V} \nonumber \\ 
    & = & \hat{V}+\hat{V}\hat{G}_0 \bigl(\hat{V} + \hat{V}\hat{G}\hat{V} \bigr) \nonumber \\ 
    & = & \hat{V} + \hat{V}\hat{G}_0\hat{T} \ . 
\end{eqnarray}
If we consider a two-particle scattering process, which evolves from the state $\vert p_1, p_2 \rangle$ to the state $\vert p_3,p_4 \rangle$, the on-shell matrix element of $\hat{T}$ assumes the following integral form: 
\begin{eqnarray}
    \langle p_3,p_4 \vert \hat{T}(z)\vert p_1,p_2 \rangle & = &  \langle p_3,p_4 \vert \hat{V}\vert p_1, p_2 \rangle + \int \frac{d^3\mathbf{k}_1}{(2\pi)^3}\frac{d^3\mathbf{k}_2}{(2\pi)^3}\frac{\langle p_3,p_4 \vert \hat{V}\vert k_1, k_2 \rangle\langle k_1,k_2 \vert \hat{T}(z)\vert p_1, p_2 \rangle}{z-E_{\mathbf{k}_1}-E_{\mathbf{k}_2}} \ ,
    \label{eq:interaction_pot} 
\end{eqnarray}
where: 
\begin{equation}
    E_{\mathbf{k}_1}+E_{\mathbf{k}_2}=c\sqrt{\mathbf{k}_1^2+m_1^2c^2}+c\sqrt{\mathbf{k}_2^2+m_2^2c^2}
\end{equation}
is the total energy of the intermediate on-shell states $\vert k_1 k_2 \rangle$ over which the integral in \eqref{eq:interaction_pot} spans. \\ 
Eq. \eqref{eq:interaction_pot} is used to determine scattering amplitudes from a known interaction potential. Indeed, considering: 
\begin{equation}
    E_{\mathbf{p}_i}=c\sqrt{\mathbf{p}_i^2+m_i^2c^2} \ , \qquad (i=1,2) \ , 
\end{equation}
one can put $z=  E_{\mathbf{p}_1}+ E_{\mathbf{p}_2}+i\epsilon$, while taking the limit $\epsilon\to 0^+ $, and turn to the relation: 
\begin{equation}
    \lim_{\epsilon\to 0^+}\langle p_3,p_4 \vert \hat{T}( E_{\mathbf{p}_1}+ E_{\mathbf{p}_2}+i\epsilon)\vert p_1, p_2 \rangle= iM(p_1,p_2,p_3,p_4) \ , 
\end{equation}
where $M$ is the non-relativistic scattering amplitude. In this way eq.\eqref{eq:interaction_pot} becomes:  
\begin{eqnarray}
    iM(p_1,p_2,p_3,p_4)  =   \langle p_3,p_4 \vert \hat{V}\vert p_1, p_2 \rangle + \lim_{\epsilon\to 0^+}\int \frac{d^3\mathbf{k}_1}{(2\pi)^3}\frac{d^3\mathbf{k}_2}{(2\pi)^3}\frac{\langle p_3,p_4 \vert \hat{V}\vert k_1, k_2 \rangle  M(k_1,k_2,p_1,p_2)}{E_{\mathbf{p}_1}+ E_{\mathbf{p}_2}-E_{\mathbf{k}_1}-E_{\mathbf{k}_2}+i\epsilon}
    \label{eq:interaction_m}
\end{eqnarray}
However, we want to express the interaction potential in terms of the non-relativistic amplitude $M$. In order to do so, we can invert the recursive Eq. \eqref{eq:interaction_m}, and solve it iteratively for the matrix element of $\hat{V}$, obtaining the so-called \textit{Born series}: 
\begin{eqnarray}
    \langle p_3,p_4 \vert \hat{V}\vert p_1, p_2 \rangle  =  iM(p_1,p_2,p_3,p_4)-\lim_{\epsilon\to 0^+}\int  \frac{d^3\mathbf{k}_1}{(2\pi)^3}\frac{d^3\mathbf{k}_2}{(2\pi)^3} \frac{iM(p_1,p_2,k_1,k_2)iM(k_1,k_2,p_3,p_4)}{E_{\mathbf{p}_1}+ E_{\mathbf{p}_2}-E_{\mathbf{k}_1}-E_{\mathbf{k}_2}+i\epsilon}
    \label{eq:born_series}
\end{eqnarray}
At the lowest possible order, in the so-called \textit{Born approximation}, the potential in momentum space is directly given by the tree-level non-relativistic scattering amplitude. \\ 
The second term, known as the \textit{Born subtraction}, provides important corrections for higher order computations. Eventually, in the case of growing precision both in PN and PM framework, even other terms of the Born series would happen to offer non-negligible contributions. 
Defining the momentum transfer: 
\begin{equation}
    q^\mu=p_1^\mu +p_2^\mu=-p_3^\mu-p_4^\mu
\end{equation}
we can rewrite the potential as: 
\begin{equation}
    \mathcal{V}(p_1,p_2)= \langle p_3,p_4 \vert \hat{V}\vert p_1, p_2 \rangle \ , 
\end{equation}
and we can get its expression in position space with a Fourier transform of the type $\mathbf{q}\to\mathbf{r}$. \\ 
Now that we have found the connection between interaction potential and scattering amplitudes, we will proceed to outline the quantum field theory of gravity coupled to matter in an Effective Field Theory approach, from which the gravitational amplitudes can be evaluated. \\ 
The scattering amplitude obtained from the EFT of gravity is not $M$, but a relativistic amplitude $\mathcal{M}$, which is related to the former by a non-relativistic normalization: 
\begin{equation}
    M(p_1,p_2,p_3,p_4)=\frac{\mathcal{M}(p_1,p_2,p_3,p_4)}{4c^2\sqrt{p_1^0p_2^0p_3^0p_4^0}} \ . 
\end{equation}
When computing scattering amplitudes, one needs to eliminate by hand the antiparticle sector. Moreover, we will treat the classical objects that compose a binary system as point particles. Hence, we can neglect from the start any process in the $t$ and $u$ channels, where particle annihilation takes place. Let us apply now these results to the case of the Effective Field Theory of gravity.

\section{Corrections to the Newtonian potential from Scattering Amplitudes}
\label{sec:pm_potential}
Let us consider two massive scalars of masses $m_1$ and $m_2$ gravitationally interacting, the potential between them in Newtonian gravitational theory is given by the Newtonian potential:
\begin{equation}
    V(r)=-\frac{G_N m_1m_2}{r} \ .
\end{equation} 
This is modified in general relativity by higher order effects in $v^2/c^2$ and by nonlinear terms in the field equations of order $\frac{G_N\ m}{rc^2}$. General corrections to the potential would be of the form:
\begin{equation}
    V(r)=-\frac{G_N m_1m_2}{r}\biggl(1+a\frac{G_N (m_1+m_2)}{rc^2}+b\frac{G_N\hbar}{r^2c^3}+\ldots\biggr) \ , 
\end{equation}
and we would like to compute $a$ and $b$ within a scattering amplitude approach.
\begin{figure}[H]
    \centering 
    \begin{tikzpicture} \begin{feynman} 
        \vertex[blob](a1) {}; 
        \vertex[left=1.5cm of a1] (a2); 
        \vertex[right=1.5cm of a1] (a3);
        \vertex[above=1.5cm of a2] (b1);
        \vertex[below=1.5cm of a2] (b2);
        \vertex[above=1.5cm of a3] (b3);
        \vertex[below=1.5cm of a3] (b4);
        \diagram* { 
        (b1) -- [fermion, edge label'=\(p_1\)] (a1) ,
        (b2) -- [fermion, edge label'=\(p_2\)] (a1),
        (b3) -- [fermion, edge label'=\(p_3\)] (a1),
        (b4) -- [fermion, edge label'=\(p_4\)] (a1),
        };
        \end{feynman} \end{tikzpicture}
\caption{4-point scattering amplitude between two massive scalar fields $\psi_1$ and $\psi_2$\label{fig:4point_massive}}
\end{figure}
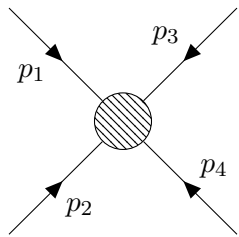
In order to see this problem in the amplitude framework, let us consider a 4-point process describing the interaction between two massive scalar fields $\psi_1,\psi_2$ of masses $m_1$ and $m_2$, as the one reported in Fig.\ref{fig:4point_massive}: 
\begin{equation}
    \psi_1(p_1)+\psi_1(p_2)\to \psi_2(p_3)+\psi_2(p_4)
\end{equation}
We choose all momenta incoming, and we work in the center of mass frame, more precisely:
\begin{equation}
    p_1=(E_1,\mathbf{p}) \quad p_2=(-E_1,\mathbf{p}') \quad p_3=(E_2,-\mathbf{p})\quad p_4=(-E_2,-\mathbf{p}')
\end{equation}
where $q=(p_1+p_2)=-p_3-p_4=(0,\mathbf{q})=(0,\mathbf{p}+\mathbf{p'})$ is the transferred momentum in the process, and $\vert\mathbf{p}\vert= \vert\mathbf{p}'\vert$.
Both particles are massive, and so we have the following conditions on the external momenta:
\begin{equation}
    p_1^2=m_1^2 \ , \quad p_2^2=m_1^2 \ ,  \quad p_3^2=m_2^2 \ , \quad p_4^2=m_2^2 \ , \qquad p_1+p_2+p_3+p_4 = 0 \ , 
\end{equation}
and the Mandelstam variables of the scattering process are: 
\begin{equation}
    s=q^2=(p_1+p_2)^2 \quad t=(p_1+p_3)^2 \quad u=(p_1+p_4)^2 \qquad s+t+u=2m_1^2+2m_2^2
\end{equation}
We are interested in computing the  amplitude $\mathcal{M}$ for the process, up to 1-loop order, within the effective field theory of General Relativity, defined as:
\begin{equation}
    \mathcal{M}=\frac{1}{\hbar}\mathcal{M}^{(0)}+\mathcal{M}^{(1)}
\end{equation}
where $\mathcal{M}^{0}$ denotes the tree level process whereas $\mathcal{M}^{(1)}$ represents the one-loop corrections.\\ We want then to use it to compute a non-relativistic potential. According to the analysis done in Sec.\ref{sec:scattering_amplitudes_intro}, we can define a non-relativistic potential in momentum space $\mathcal{V}(\mathbf{p},\mathbf{q})$ in terms of the non-relativistic scattering amplitude $M(\mathbf{p},\mathbf{p'})$, as:
\begin{equation}
    \mathcal{V}(\mathbf{p},\mathbf{q})=i M(\mathbf{p},\mathbf{p'}) +\ldots \ ,
\end{equation}
where the ellipses denote higher order terms, and in the low-energy limit the non-relativistic amplitude is defined as:
\begin{equation}
    M(\mathbf{p},\mathbf{p'})=\frac{\mathcal{M}(\mathbf{p},\mathbf{p'})}{4 m_1 m_2} \ . 
\end{equation}
The corresponding expression for the potential at tree level in position space is:
\begin{equation}
\mathcal{V}(r)=\frac{1}{4m_1m_2}\int \frac{d^3\mathbf{q}}{(2\pi)^3}e^{i\mathbf{q}\cdot \mathbf{r}}i \mathcal{M}(\mathbf{p},\mathbf{p'})+\ldots \ . 
\end{equation}
For its evaluation it will be crucial to use the scalar integral evaluated in Eq.\eqref{eq:scalar_integral_fourier}.

\subsection{Newtonian potential from a tree Level calculation}
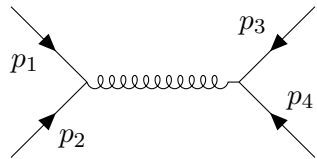
\begin{figure}[H]
    \centering
        \begin{tikzpicture} \begin{feynman}
    \vertex (a1) ; 
    \vertex[right=1cm of a1] (a2); 
    \vertex[right=1cm of a2] (a3); 
    \vertex[right=1cm of a3] (a4);
    \vertex[right=1cm of a4] (a5);
    \vertex[below= 1cm of a1] (b1);
     \vertex[below= 2cm of b1] (d1); 
    \vertex[below= 1cm of a5] (b5); 
    \vertex[below= 2cm of b5] (d5); 
    \vertex[below=2cm of a1] (c1); 
    \vertex[right=1cm of c1] (c2);
     \vertex[right = 2cm of c2] (c4);
     \vertex[right=1 of c4] (c5) ;
    \vertex[below=4cm of a1] (e1) ; 
    \vertex[right=4cm of e1] (e5) ;
    \diagram* { 
    (b1) -- [fermion, edge label'=\(p_1\)] (c2),
    (d1) -- [fermion, edge label'=\(p_2\)] (c2),
    (c2) -- [gluon,] (c4),
    (b5) -- [fermion, edge label'=\(p_3\)] (c4),
    (d5) -- [fermion, edge label'=\(p_4\)] (c4),
    };
    \end{feynman} \end{tikzpicture}
    \caption{Tree-level Feynman diagrams for the process. Thin lines indicate massive scalars whereas curly ones correspond to gravitons.\label{fig:tree_level_massive_1}}
    \end{figure}

Before proceeding further, let us show that this method works by computing the potential arising from a simple tree-level calculation. We work in dimensional regularization, with $\epsilon=\frac{4-d}{2}$, where $d$ denotes the continuous dimension of the spacetime. At tree level only the s-channel is contributing to the process, since the three point vertex between two scalars and one graviton can couple only identical particles, and the relevant diagram is reported in Fig.\ref{fig:tree_level_massive_1}.
The amplitude is given by: 
\begin{equation}
    \mathcal{M}^{(0)}= \tau_1^{\mu\nu}(p_1,-p_2)\frac{iP_{\mu\nu\rho\sigma}}{q^2+i\epsilon}\tau_1^{\rho\sigma}(p_3,-p_4)
\end{equation}
Making tensor contractions, simplifying the expression, and writing $d=4-2\epsilon$, we get:
\begin{equation}
     \mathcal{M}^{(0)}=\frac{-i \kappa^2}{4 q^2 (\epsilon -1)}\left[m_2^2\left(2 m_1^2 (\epsilon -2)-(\epsilon -1) (t+u)\right)+m_2^4
     (\epsilon -1)+(\epsilon -1) \left(m_1^2-t\right)
     \left(m_1^2-u\right)\right]
\end{equation}
Taking the limit $\epsilon\to 0 $ we get:
\begin{eqnarray}
    \mathcal{M}^{(0)}=\frac{-i \kappa^2}{4 q^2}\biggl(m_2^2 \left(4 m_1^2-t-u\right)+m_2^4+\left(m_1^2-t\right)
    \left(m_1^2-u\right)\biggr)+\mathcal{O}[\epsilon]
\end{eqnarray} 
In order to connect this amplitude to the 0PM potential studied in the first part of this project, we take the non-relativistic limit of this amplitude $\mathbf{q}^2 \ll m_1^2,m_2^2$, which is obtained with the following approximations:
\begin{equation}
    s\approx -\mathbf{q}^2\ ,  \qquad t\approx (m_1+m_2)^2  \ . 
\end{equation}
Then the amplitude becomes:
\begin{equation}
    \mathcal{M}^{(0)}(\mathbf{p},\mathbf{p}')= i \frac{16 \pi G_N m_1^2m_2^2}{\mathbf{q^2}} \ . 
    \label{eq:tree_amplitude_PM}
\end{equation}
We can obtain the corresponding potential by taking the Fourier transform:
\begin{eqnarray}
    \mathcal{V}^{tree}(r)& = &  \frac{1}{4m_1m_2}\int \frac{d^{3}\mathbf{q}}{(2\pi)^{3}}e^{i\mathbf{q}\cdot \mathbf{r}}\left( \  \scalebox{0.6}{\begin{tikzpicture}[baseline=(c2)] \begin{feynman}
        \vertex (a1) ; 
        \vertex[right=1cm of a1] (a2); 
        \vertex[right=1cm of a2] (a3); 
        \vertex[right=1cm of a3] (a4);
        \vertex[right=1cm of a4] (a5);
        \vertex[below= 1cm of a1] (b1);
         \vertex[below= 2cm of b1] (d1); 
        \vertex[below= 1cm of a5] (b5); 
        \vertex[below= 2cm of b5] (d5); 
        \vertex[below=2cm of a1] (c1); 
        \vertex[right=1cm of c1] (c2);
         \vertex[right = 2cm of c2] (c4);
         \vertex[right=1 of c4] (c5) ;
        \vertex[below=4cm of a1] (e1) ; 
        \vertex[right=4cm of e1] (e5) ;
        \diagram* { 
        (b1) -- [fermion, edge label'=\(p_1\)] (c2),
        (d1) -- [fermion, edge label'=\(p_2\)] (c2),
        (c2) -- [gluon,] (c4),
        (b5) -- [fermion, edge label'=\(p_3\)] (c4),
        (d5) -- [fermion, edge label'=\(p_4\)] (c4),
        };
        \end{feynman} \end{tikzpicture}} \ \right)\bigg|_{NR} \nonumber \\ 
    & = & 
    \frac{1}{4m_1m_2}\int \frac{d^{3}\mathbf{q}}{(2\pi)^{3}}e^{i\mathbf{q}\cdot \mathbf{r}}i \mathcal{M}^{(0)}(\mathbf{p},\mathbf{p}')\nonumber \\ 
    & = & 
    -4 \pi G_N m_1 m_2 \int \frac{d^3\mathbf{q}}{(2\pi)^3}e^{i \mathbf{q}\cdot \mathbf{r}}\frac{1}{\mathbf{q^2}}\nonumber \\ 
    & = & 
    -\frac{G_Nm_1m_2}{r} 
    \label{eq:newtonian_potential_tree}
\end{eqnarray}
where in the last line we used Eq.\eqref{eq:scalar_integral_fourier} to perform the Fourier transform.
Eq.\eqref{eq:newtonian_potential_tree} is the Newtonian potential between two massive particles, and match the 0PN calculation reported in Eq.\eqref{eq:0PN}. In the next section we will show how to generalize this computational method at 1-loop level in a smart way. 

\section{Focus on non-analytic contributions to the Scattering Amplitudes}
It seems natural now to continue our study of the gravitational potential arising from loop amplitudes proceeding as follows:
\begin{itemize}
    \item evaluate all the 1-loop diagrams from the lowest order effective Lagrangian $\mathcal{O}(p^2)$,
    \item combine the effects of the order $p^2$ and $p^4$ Lagrangian at tree level with one-loop corrections,
    \item absorb divergences of the one-loop diagrams into renormalized coefficients of the Lagrangian ($m,c_i,d_i$), using an appropriate renormalization scheme which does not violate general covariance,
    \item measure the unknown coefficients by comparison with experiments,
    \item having determined the parameters of the theory, we can make predictions for other experimental observables, valid to $\mathcal{O}(p^4) $ in the energy expansion. 
\end{itemize}
However, no observables are sensitive at $\mathcal{O}(p^4)$, and it seems impossible to have some physical predictions. We need to focus on the class of quantum corrections, uncovered in such procedure, which is independent of these unknown coefficients.\\
The general form of a 1-loop gravitational amplitude is the following:
\begin{equation}
    i \mathcal{M}^{(1)}=Aq^2\left(1+\alpha\kappa q^2 +\beta \kappa ^2 q^2 \log(-q^2) +\beta \kappa ^2 \frac{m}{\sqrt{-q^2}} \ . +\cdots\right)
\end{equation}
We should remember that the ultimate goal of this computation is to provide general relativistic corrections to the Newtonian potential. \\  
Among the various terms appearing, the contributions we are interested in are exclusively the long range ones. Specifically, with respect to the momentum transfer $q$, our interest is entirely focused on those terms which dominate the infrared regime $q^2\to 0$. \\  
 In order to identify them, is useful to distinguish between non-analytical and analytical contributions to the scattering amplitudes. \\ 
Analytical terms are polynomials in $q$, therefore they vanish in the infrared regime. On the other side, leading non-analytic contributions at low energy satisfy: 
\begin{equation}
    \bigg|\log(-q^2)\bigg|  \gg 1 \qquad \bigg| \frac{m}{\sqrt{-q^2}} \bigg| \gg 1  \ , 
    \label{eq:non_analytic_terms}
\end{equation}
\noindent thereby dominating over the analytic effects. \\
Moreover, this situation is obviously unchanged after a Fourier transform $\mathbf{q}\to\mathbf{r}$, since analytic terms give rise to unwanted ultra-local contributions, which are $r-$dependent $\delta$-functions or their derivatives, whereas non-analytical terms produce long-distance corrections, proportional to some positive powers of $1/r$. \\ More precisely when we take the Fourier transform of the potential, the term with the square-root yields the classical $Gm/r^2$ general relativistic correction to the potential whereas the term with the logarithm produces a long-distance $G\hbar/r^3$ quantum correction. The distinction between classical and quantum corrections from loop amplitudes is deeply discussed in \cite{Holstein:2004dn,Kosower:2018adc}. \\
\noindent It should be noticed that such non-analytical contributions come from the propagation of massless particle modes. \\ 
 The difference between massive and massless particle modes originates from the impossibility of expanding a massless propagator $\left( \sim 1/q^2\right)$, while for a massive one we can make a Taylor expansion for $q \ll m^2$:
 \begin{eqnarray}
     \frac{1}{q^2-m^2}\sim -\frac{1}{m^2}\left(1+\frac{q^2}{m^2}+\cdots \right).
 \end{eqnarray}
 Notice that no $\left(\sim 1/q^2\right)$ terms are generated in the above expansion of the massive propagator, thus such terms all arise from the propagation of massless modes.
The low energy propagation of massless particles, leads to long-distance classical and quantum corrections which are distinct from the effects of the local effective Lagrangian. \\
\noindent The leading non-analytic effects \cite{Donoghue:1994dn,Donoghue:1993eb} involve only massless degrees of freedom and the low energy couplings of the theory,  since higher order effects at the vertices introduce more powers of $q^2$, both of which are known independently on the ultimate high-energy theory. \\
Moreover, the separation between these two classes of terms is always clear in our theory and the cumbersome renormalization process we have mentioned in the previous section ends up affecting only negligible analytic sector.
So in distinction to the analytic contributions, which depends on the unknown parameters $c_1,c_2,...$, the leading quantum corrections are parameter free. \\
Let us notice that if other massless particles are present in our theory, such as photons in QED, they also can generate non-analytic behavior in loop amplitudes when they are coupled to gravity.  \\
Since we want to focus only on non-analytic terms, we need to improve our computation in order to catch only the relevant part of the scattering amplitudes.

\subsection{Relevant part of the 1-loop corrections}
It is not necessary to compute the full 1-loop amplitude to get the lowest order quantum corrections. Non-analytical terms in Eq.\eqref{eq:non_analytic_terms}  can develop an imaginary component once we analytically extend the amplitude in the complex plane. The imaginary part of the amplitude contains the branch-cut information, especially the $1/s^{\alpha}$ poles, with $s=-\mathbf{q}^2$, with $\mathbf{q}$ momentum transfer. \\
It is possible to compute the imaginary part of the amplitude, and hence obtain the information about non-analytic terms, by focusing on the massless double cut of the 1-loop amplitude. \\ 
Nowadays modern amplitudes techniques are used to compute the amplitudes in the massless double-cut. 
Simplifications come from the use of \textit{generalized unitarity}\cite{Bjerrum-Bohr:2013bxa}, a method originally developed in the context of Yang-Mills theories which outlines a systematic scheme for building loop amplitudes from simpler tree-level ones.  On the same note, the \textit{double copy construction} \cite{Bern:2010ue} is used to establish a systematic connection between gravity and gauge theory scattering amplitudes, giving the chance to work with the latter, which are more manageable, and recover with contained effort the corresponding results for gravity.  \\ 
For our calculations, we will not use these modern amplitude techniques, but we will perform the calculations using Feynman diagrams, by focusing only on the massless double cut. 
\section{Classical Gravity from Loop Amplitudes}
In this section we will show how classical terms can arise from loop Amplitudes.
The belief that a loop expansion is a $\hbar$ expansion, and so the fact that loop corrections are purely quantum corrections, is wrong in general.\footnote{For a complete discussion about the topic, with several examples, see \cite{Holstein:2004dn}\cite{Kosower:2018adc} }
Both classical and quantum terms are generated at every order in a loop expansion. \\ 
It is possible to count the $\hbar$ dependency of the various contributions to amplitudes in quantum field theory to separate classical (non-linear) general relativity corrections from pure quantum contributions, at one and two-loop order. \\ 
In a typical Feynman diagram, 
 a vertex arises from the expansion of:
\begin{equation}
    e^{\frac{i}{\hbar}\int d^4x \mathcal{L}_{int}(\phi_{int})}
\end{equation}
and so carries with it a factor of $\hbar^{-1}$.\\
On the other hand the field commutation relations: 
\begin{equation}
    [\phi(\Vec{x}),\Pi(\Vec{y})]=i\hbar \delta^3(\Vec{x}-\Vec{y})
\end{equation}
lead to a factor $\hbar$ in each propagator:
\begin{equation}
    \langle 0 \vert T(\phi(x)\phi(y))\vert 0\rangle =\int \frac{d^4k}{(2\pi)^4}\frac{i\hbar e^{ik(x-y)}}{k^2-\frac{m^2}{\hbar^2}+i\epsilon} \ . 
\end{equation}
In order to count factors of $\hbar$ then, one needs to compute the number of vertices and propagators in a given diagram. For a generic diagram the Euler's formula relate the number of loops $L$, of vertices $V$ and of propagators $P$ as: $L=P-V+1$. Associating a factor of $\hbar^{-1}$ for the $V$ vertices and $\hbar^{+1}$ for the $P$ propagators yields and overall factor: 
\begin{equation}
    \hbar^{P-V+1}=\hbar^{L} \ , 
\end{equation}
which is at the origin of the claim that a loop expansion coincides with a $\hbar$ expansion. \\
However, this assumption is not valid, and it can be shown explicitly that it fails in several calculations. \\
One loophole to this argument is visible in the propagator, which contains $\hbar$ in more than one location. \\
When the propagator is written in momentum space, the mass carries an inverse factor of $\hbar$. \\
This is because the Klein-Gordon equation reads:
\begin{equation}
    \left(\square +\frac{m^2}{\hbar^2}\right)\phi(x)=0 \ , 
\end{equation}
when $\hbar$ is made visible. \\
This means that the counting of $\hbar$ from the vertices and the propagators is incomplete, and one needs also to know how mass factors enter the result, because there are factors of $\hbar$ attached there in also. \\
Let us now focus on non-analytic terms that will be useful in the following since, as mentioned in the previous section, they provide long-range corrections to the Newtonian potential. \\
In particular, we will be interested in two kinds of structures:
\begin{equation}
    \sqrt{\frac{m^2}{-q^2}} \qquad \log\left(-\frac{q^2}{m^2}\right)
\end{equation}
Restoring the factors of $\hbar$ it is easy to see that the coefficient of the square root non-analytic behavior scales as $\hbar^{-1}$, while the logarithmic term is independent on $\hbar$. \\
Then we can see that one loop results carry different powers of $\hbar$ because they contain different powers of the factor $q^2/m^2$.\\
With the general expectation of one factor of $\hbar$ at one loop, there is a specific combination of the mass and momentum that eliminate $\hbar$ in order to produce a classical results. \\
In order to remove one power of $\hbar$ requires a factor of:
\begin{equation}
\sqrt{\frac{m^2}{-q^2}} \ .
\end{equation}
The emergence of the power of $\hbar^{-1}$ involves an interplay between the massive particle (whose mass carries a factor of $\hbar$) and the massless one (which generates the required non-analytic form).

\chapter{Bending of light in perturbative quantum Gravity}
\label{chapter:bending}

In the previous chapter we have shown that is possible to obtain the gravitational potential between two compact objects from the Fourier transformation of a 4-point scattering amplitude.
On this chapter we want to use the multi-loop techniques introduced in chapter\ref{Ch:multiloop} and already applied in chapters\ref{chapter:hereditary}\ref{chapter:inin}, to evaluate 1-loop amplitudes which are necessary to compute post-Minkowskian corrections to physical processes. In the first part of this chapter, motivated by analogous studies done in other gauge theories such as QED or QCD \cite{Bonciani:2021okt,Mandal:2022vju},  we will be interested evaluation of 1-loop classical and quantum corrections to the process of light bending under a heavy object. Working in the effective field theory of General Relativity introduced in the previous chapter, we will study the scattering process of two scalar particles, one massive and one massless, up to one loop order. We will develop a computational algorithm in \texttt{Mathematica} to compute the 1-loop amplitude, and we will use it to obtain a prediction for the bending angle. We will confront our results with those appearing in literature \cite{Bjerrum-Bohr:2014zsa,Bjerrum-Bohr:2016hpa,Bjerrum-Bohr:2017dxw,Bai:2016ivl}, finding extra pieces contributing to the quantum corrections to the bending angle, with respect to the known ones. Eventually, we will adapt this machinery to the evaluation of 1-loop amplitudes between two massive scalar particles, from which one can obtain the 1-loop corrections to the Newtonian potential. 
\subsubsection{State-of-the-art}
As already seen in Chapter\ref{Chapter:GR_PM} the QFT treatment of General Relativity is not something new. The evaluation of 1-loop corrections to the Newtonian potential has been done in \cite{Bjerrum-Bohr:2002gqz}, and studied for different spins in \cite{Holstein:2008sx}, as well as using new scattering amplitudes techniques in \cite{Bjerrum-Bohr:2013bxa}. An increasing interest has received also the evaluation of the bending angle under a massive object \cite{Bjerrum-Bohr:2014zsa,Bjerrum-Bohr:2016hpa,Bjerrum-Bohr:2016hpa,Bjerrum-Bohr:2018xdl,Bai:2016ivl,Chi:2019owc}, which has paved the way to new analytical studies of the amplitudes including also massless particles. GWs detection require precision predictions, and in the last few years a systematic framework for PM computations has been developed, which uses modern amplitudes techniques such as double-copy and generalized unitarity. The $1PM$ and $2PM$ corrections have been evaluated in \cite{Cheung:2018wkq,Cristofoli:2019neg}, the $3PM$ in \cite{Bern:2019nnu,Bern:2019crd,Cheung:2020gyp,Kalin:2020fhe,Bjerrum-Bohr:2021din} and the $4PM$ in \cite{Bern:2021dqo,Bern:2021yeh,Dlapa:2021npj}. 
\subsubsection{Structure of the chapter:}
The chapter will be divided as follows: 
\begin{itemize}
    \item after a short review of the bending angle in General Relativity, we will consider the scattering amplitude of a massive scalar and a massless scalar in the EFT of General Relativity, and we will evaluate it up to 1-loop order by developing a computational algorithm in \texttt{Mathematica} based on multi-loop techniques. 
    \item We will take the low-energy limit of the 1-loop amplitude, and we will confront our result with the ones appearing in literature, which have been obtained using a different approach, underlying the differences.
    \item Then we will use the 1-loop amplitude to get a prediction for the bending angle using two different, but equivalent, approaches: a semiclassical approach, and an Eikonal method. 
    \item In the last section we will adapt the analysis done so far to evaluate the 1-loop scattering amplitude between two massive scalars, and we will use it to obtain the low-energy 1-loop corrections to the Newtonian potential. 
\end{itemize}
\section{Bending of light in General Relativity}
One of the first tests of General Relativity has been the deflection of light by the Sun, which has been first measured in 1919 during a total solar eclipse, finding an agreement with the theoretical predictions. Since then a lot of measurements of such effect have been done, using radio telescopes to observe thousands of radio galaxies and quasars over the entire celestial physics.  Moreover, a related phenomenon known as gravitational lensing, is nowadays a standard astrophysical tool used to study the distribution of dark matter in the universe and for imaging the most distance galaxies. Nowadays, there is also the Event Horizon Telescope which uses the light bending phenomenon to study supermassive black holes \cite{EventHorizonTelescope:2019dse} \cite{EventHorizonTelescope:2019ggy}. \\ 
In this section we want to evaluate the light deflection formula in General Relativity at NLO in a Schwarzschild metric, following \cite{doi:10.1119/1.1570416}.
Let us consider a static, spherically symmetric metric of the form:
\begin{equation}
    ds^2=A(r)dt^2-B(r)dr^2-r^2d\Omega^2 \ , 
\end{equation}
In the case of Schwarzschild metric we have:
\begin{equation}
    A(r)=\frac{1}{B(r)}=1-\frac{2G_N m}{r} \ . 
\end{equation}
The deflection angle is given by: 
\begin{equation}
    \theta = 2\int_0^1 du \frac{\sqrt{B/C}}{\sqrt{\frac{C}{A}\sqrt{R^2}{b^2}-u^2}}
    \label{eq:deflection_angle_GR}
\end{equation}
where: $u=R/r$, with $R$ the radius of closest approach, $b=c|L/E|$ is the impact parameter defined as the ratio between the angular momentum $L$ and the energy $E$. \\ 
$R$ is defined as: 
\begin{equation}
    \frac{1}{b^2}= \frac{A(R)}{C(R)}\frac{1}{R^2}
    \label{eq:bending_impact}
\end{equation}
where $b$ is a coordinate independent quantity whereas $R$ depends on the coordinate system. 
If we specify Eq.\eqref{eq:deflection_angle_GR} to the case of a Schwarzschild metric we obtain:
\begin{equation}
    \theta = 2\int_0^1du\frac{1}{\sqrt{1-u^2-\frac{2G_N m}{R}(1-u^3)}}-\pi
\end{equation}
The solution can be found in a perturbative way by noticing that near the sun $2 G_N m/R\approx 10^{-3}\ll 1$, and so we have: 
\begin{eqnarray}
    \theta 
    & = & 
    2\int_0^1 \frac{1}{\sqrt{1-u^2}}+\frac{1}{2}\frac{2G_N m}{R}\frac{1+u+u^2}{(1+u)\sqrt{1-u^2}}+\frac{3}{8}\biggl(\frac{2G_N m }{R}\biggr)^2\frac{(1+u+u^2)^2}{(1+u)^2\sqrt{1-u^2}}+\ldots \nonumber \\ 
    & = & \frac{4G_Nm}{R}+\frac{4 G_N^2 m^2}{R^2}\biggl(\frac{15\pi }{16}-1\biggr)+\ldots \label{eq:bending_GR} \ . 
\end{eqnarray}
Eq.\eqref{eq:bending_GR} can be rewritten in terms of the impact parameter $b$, which from Eq.\eqref{eq:bending_impact} can be expanded as:
\begin{equation}
    b=\sqrt{B(R)R}=\frac{R}{\sqrt{1-\frac{2G m}{R}}}=R+G_N m + \ldots \ \ , 
\end{equation}
obtaining:
\begin{equation}
\theta= \frac{4 G_N m}{b}+\frac{15\pi G_N^2 m^2}{4 b^2}+\mathcal{O}\biggl(\frac{1}{b^3}\biggr)
\label{eq:bending_classical}
\end{equation}
This is the standard derivation and arises from considering light as particles traversing a classical trajectory. We want now to obtain a classical and quantum prediction for the bending angle following an alternative approach based on scattering amplitudes. 
\section{Scattering amplitude}
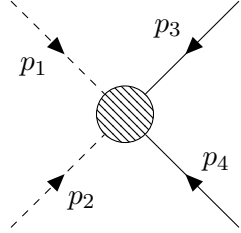
\begin{figure}[H]
    \centering 
    \begin{tikzpicture} \begin{feynman} 
        \vertex[blob](a1) {}; 
        \vertex[left=1.5cm of a1] (a2); 
        \vertex[right=1.5cm of a1] (a3);
        \vertex[above=1.5cm of a2] (b1);
        \vertex[below=1.5cm of a2] (b2);
        \vertex[above=1.5cm of a3] (b3);
        \vertex[below=1.5cm of a3] (b4);
        \diagram* { 
        (b1) -- [charged scalar, edge label'=\(p_1\)] (a1) ,
        (b2) -- [charged scalar, edge label'=\(p_2\)] (a1),
        (b3) -- [fermion, edge label'=\(p_3\)] (a1),
        (b4) -- [fermion, edge label'=\(p_4\)] (a1),
        };
        \end{feynman} \end{tikzpicture}
\caption{4-point scattering amplitude between a massless scalar $\phi$ and a massive one $\psi$}
    \end{figure}
Let us consider the scattering amplitude of the following process:
\begin{equation}
    \phi(p_1)+\phi(p_2)\to \psi(p_3)+\psi(p_4)
\end{equation}
where $\phi$ stands for a massless scalar particle whereas $\psi$ represents a massive scalar particle of mass $m$, in the effective field theory of general relativity. \\
Let us take all momenta as incoming, more precisely:
\begin{equation}
    p_1=(E_1,\mathbf{p}_1) \quad p_2=(-E_1,\mathbf{q}-\mathbf{p}_1) \quad p_3=(E_2,\mathbf{p}_3)\quad p_4=(-E_2,-\mathbf{q}-\mathbf{p}_3)
\end{equation}
with the conditions:
\begin{equation}
    p_1^2=0 \quad p_2^2=0 \quad p_3^2=m^2 \quad p_4^2=m^2 \qquad p_1+p_2+p_3+p_4 = 0
\end{equation}
where $q=(p_1+p_2)=-p_3-p_4=(0,\mathbf{q})$ is the transferred momentum in the process. \\ 
The Mandelstam invariants of the scattering process are: 
\begin{equation}
    s=(p_1+p_2)^2 \qquad t=(p_1+p_3)^2 \qquad u=(p_1+p_4)^2 \qquad 
\end{equation}
satisfying the condition: $s+t+u=2m^2$.\\
We denote the scattering amplitude $\mathcal{M}$ of the process, up to one-loop order, as:
\begin{equation}
    \mathcal{M}=\frac{1}{\hbar}\mathcal{M}^{(0)}+\mathcal{M}^{(1)}
\end{equation}
where $\mathcal{M}^{0}$ denotes the tree level process whereas $\mathcal{M}^{(1)}$ represents the one-loop corrections.\\
We are interested in the evaluation of the tree level amplitude and of that part of the one-loop corrections which give rise to non-analytical terms, within the effective field theory of general relativity introduced in the previous chapter, using Feynman diagrams.
The relevant Feynman rules needed for the evaluation of the diagrams are given Sec.\ref{sec:Feynman_Rules_PM}. Calculations will be done in dimensional regularization, with $\epsilon=\frac{4-d}{2}$, where $d$ denotes the continuous dimension of the spacetime. 

\subsection{Tree-level amplitude}

\begin{figure}[H]
    \centering
        \begin{tikzpicture} \begin{feynman}
    \vertex (a1) ; 
    \vertex[right=1cm of a1] (a2); 
    \vertex[right=1cm of a2] (a3); 
    \vertex[right=1cm of a3] (a4);
    \vertex[right=1cm of a4] (a5);
    \vertex[below= 1cm of a1] (b1);
     \vertex[below= 2cm of b1] (d1); 
    \vertex[below= 1cm of a5] (b5); 
    \vertex[below= 2cm of b5] (d5); 
    \vertex[below=2cm of a1] (c1); 
    \vertex[right=1cm of c1] (c2);
     \vertex[right = 2cm of c2] (c4);
     \vertex[right=1 of c4] (c5) ;
    \vertex[below=4cm of a1] (e1) ; 
    \vertex[right=4cm of e1] (e5) ;
    \diagram* { 
    (b1) -- [charged scalar, edge label'=\(p_1\)] (c2),
    (d1) -- [charged scalar, edge label'=\(p_2\)] (c2),
    (c2) -- [gluon,] (c4),
    (b5) -- [fermion, edge label'=\(p_3\)] (c4),
    (d5) -- [fermion, edge label'=\(p_4\)] (c4),
    };
    \end{feynman} \end{tikzpicture}
    \caption{Tree-level Feynman diagram for the process. Dashed lines indicate a massless scalar particle ($\phi$), whilst thin ones indicate a massive scalar ($\psi$). Curly and dotted lines correspond to gravitons and ghosts, respectively.\label{fig:tree_level}}
    \end{figure}
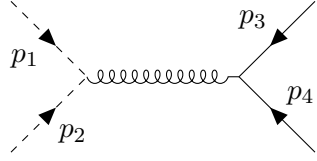

At tree level only the s-channel is contributing to the process, since the three point vertex between two scalars and one graviton can couple only identical particles, and the relevant diagram is present in Fig.\ref{fig:tree_level}.\\
The amplitude can be easily evaluated as: 
\begin{equation}
    \mathcal{M}^{(0)}= \tau_1^{\mu\nu}(p_1,-p_2)\frac{iP_{\mu\nu\rho\sigma}}{q^2+i\epsilon}\tau_1^{\rho\sigma}(p_3,-p_4)
\end{equation}
Substituting the expressions for the Feynman rules, making the tensor contractions and expressing the result in terms of Mandelstam variables we obtain in $d$-dimensions: 
\begin{equation}
     \mathcal{M}^{(0)}= i \kappa^2 \frac{(t-m^2)(u-m^2)}{4s}
\end{equation}
We can now get this expression in the rest frame of the massive scalar particle in the limit in which the energy $E$ of the massless particle satisfies: $E \ll m$. \\
We have to apply the following approximations:
\begin{equation}
    s=(p_1+p_2)^2 \approx -\mathbf{q^2} \qquad t=(p_1+p_3)^2\approx m^2+2E m  \ . 
\end{equation}
Moreover, we will be interested in the small angle scattering approximation then: $s \ll E^2$. \\
The amplitude in the low-energy limit becomes:
\begin{equation}
    \mathcal{M}^{(0)}= i \kappa^2\frac{E^2m^2}{\mathbf{q}^2}
\end{equation}
\section{The $s$-channel double-cut of one-loop amplitude}
We will now focus on the evaluation of one-loop corrections to this process. \\
We will not consider the full one-loop amplitude, but only the part arising from the s-channel massless double-cut. This is the part that gives rise to non-analytic terms, which generate long-range interactions once we take the Fourier transform.\\
We will denote this part of the 1-loop amplitude as:
\begin{equation}
    \mathcal{C}^{(1)}=Cut_{12}(\mathcal{M}^{(1)})
\end{equation}
We will study two different sets of diagrams: 
\begin{enumerate}
    \item the first set contains diagrams with at least two graviton propagators in the s-channel, namely on the graviton double cut,
    \item the second set contains diagrams with at least two massless scalar propagators in the s-channel, namely on the massless scalar double cut.
\end{enumerate}
We will perform separately the analysis of the two sets in order to compare them with results present in literature, and we will then obtain the total amplitude in the s-channel double-cut as:
\begin{eqnarray}
\mathcal{C}^{(1)}=\mathcal{C}^{(g)}+\mathcal{C}^{(\phi)} \ , 
\end{eqnarray}
where $\ \mathcal{C}^{(g)}\ $ denotes the 1-loop amplitude in the graviton double-cut whereas $\ \mathcal{C}^{(\phi)}\ $ denotes the 1-loop amplitude in the massless scalar double-cut.

\subsection{Graviton double-cut}

The relevant diagrams needed to evaluate $\mathcal{C}^{(g)}$ are the ones containing at least two massless graviton propagators in the s-channel, and are given in Fig.\ref{fig:1loop_grav},  : 
\begin{figure}[H]
\begin{subfigure}{0.99\textwidth}
    \scalebox{0.8}{
    \begin{tikzpicture} \begin{feynman}
    \vertex (a1) ; \vertex[right=1cm of a1] (a2); \vertex[right=1cm of a2] (a3); \vertex[right=1cm of a3] (a4);\vertex[right=1cm of a4] (a5);
    \vertex[below= 1cm of a1] (b1); \vertex[right= 2cm of b1] (b3); \vertex[below= 2cm of b1] (d1); \vertex[right= 2cm of d1] (d3);
    \vertex[below= 1cm of a5] (b5); \vertex[below= 2cm of b5] (d5); 
    \vertex[below=2cm of a1] (c1); \vertex[right=1cm of c1] (c2); \vertex[right = 2cm of c2] (c4);\vertex[right=1 of c4] (c5) ;
    \vertex[below=4cm of a1] (e1); \vertex[right=2cm of e1] (e3) ; \vertex[right=4cm of e1] (e5) ;
    \vertex[below=0.5 cm of d3] (f3) {\((a)\)}; ;
    \diagram* { 
    (b1) -- [charged scalar] (c2),
    (d1) -- [charged scalar] (c2),
    (c2) -- [gluon, quarter left] (b3) -- [gluon, quarter left] (c4),
    (c2) -- [gluon, quarter right] (d3) -- [gluon, quarter right] (c4),
    (b5) -- [fermion] (c4),
    (d5) -- [fermion] (c4),
    };
    \end{feynman} 
    \end{tikzpicture}}
    \quad
    \scalebox{0.8}{
        \begin{tikzpicture} \begin{feynman}
    \vertex (a1) ;
    \vertex[right=1cm of a1] (a2);
    \vertex[right=1cm of a2] (a3); 
    \vertex[right=1cm of a3] (a4);
    \vertex[right=1cm of a4] (a5);
    \vertex[below= 1cm of a1] (b1);
    \vertex[right= 1cm of b1] (b2);
    \vertex[right= 2cm of b1] (b3);
    \vertex[right= 2cm of b3] (b5);
    \vertex[below= 2cm of b1] (d1); 
    \vertex[right= 2cm of d1] (d3);
    \vertex[right= 1cm of d1] (d2);
    \vertex[below= 2cm of b5] (d5); 
    \vertex[below=2cm of a1] (c1); 
    \vertex[right=1cm of c1] (c2); 
    \vertex[right = 2cm of c2] (c4);
    \vertex[right=1 of c4] (c5) ;
    \vertex[below=4cm of a1] (e1); 
    \vertex[right=2cm of e1] (e3)  {\((b)\)}; 
    \vertex[right=4cm of e1] (e5) ;
    
    \diagram* { 
    (a1) -- [charged scalar] (b2),
    (b2) -- [scalar] (d2),
    (e1) -- [charged scalar] (d2),
    (b2) -- [gluon] (c4),
    (d2) -- [gluon] (c4),
    (b5) -- [fermion] (c4),
    (d5) -- [fermion] (c4),
    };
    \end{feynman} 
    \end{tikzpicture}}
    \quad 
    \scalebox{0.8}{
    \begin{tikzpicture} 
    \begin{feynman}
    \vertex (a1) ;
    \vertex[right=1cm of a1] (a2);
    \vertex[right=1cm of a2] (a3); 
    \vertex[right=1cm of a3] (a4);
    \vertex[right=1cm of a4] (a5);
    \vertex[below= 1cm of a1] (b1);
    \vertex[right= 1cm of b1] (b2);
    \vertex[right= 2cm of b1] (b3);
    \vertex[right= 1cm of b3] (b4);
    \vertex[right= 2cm of b3] (b5);
    \vertex[below= 2cm of b1] (d1); 
    \vertex[right= 2cm of d1] (d3);
    \vertex[right= 1cm of d1] (d2);
    \vertex[right= 1cm of d2] (d3);
    \vertex[right= 1cm of d3] (d4);
    \vertex[below= 2cm of b5] (d5); 
    \vertex[below=2cm of a1] (c1); 
    \vertex[right=1cm of c1] (c2); 
    \vertex[right = 2cm of c2] (c4);
    \vertex[right=1 of c4] (c5) ;
    \vertex[below=4cm of a1] (e1); 
    \vertex[right=2cm of e1] (e3)  {\((c)\)}; 
    \vertex[right=4cm of e1] (e5) ;
    
    \diagram* { 
    (b1) -- [charged scalar] (c2),
    (d1) -- [charged scalar] (c2),
    (b4) -- [gluon] (c2),
    (d4) -- [gluon] (c2),
    (a5) -- [fermion] (b4),
    (b4) -- [] (d4),
    (e5) -- [fermion] (d4),
    };
    \end{feynman} 
    \end{tikzpicture}}
    \quad 
    \scalebox{0.8}{\begin{tikzpicture} 
    \begin{feynman}
    \vertex (a1) ;
    \vertex[right=1cm of a1] (a2);
    \vertex[right=1cm of a2] (a3); 
    \vertex[right=1cm of a3] (a4);
    \vertex[right=1cm of a4] (a5);
    \vertex[below= 1cm of a1] (b1);
    \vertex[right= 1cm of b1] (b2);
    \vertex[right= 2cm of b1] (b3);
    \vertex[right= 1cm of b3] (b4);
    \vertex[right= 2cm of b3] (b5);
    \vertex[below= 2cm of b1] (d1); 
    \vertex[right= 2cm of d1] (d3);
    \vertex[right= 1cm of d1] (d2);
    \vertex[right= 1cm of d2] (d3);
    \vertex[right= 1cm of d3] (d4);
    \vertex[below= 2cm of b5] (d5); 
    \vertex[below=2cm of a1] (c1); 
    \vertex[right=1cm of c1] (c2); 
    \vertex[right = 2cm of c2] (c4);
    \vertex[right=1 of c4] (c5) ;
    \vertex[below=4cm of a1] (e1); 
    \vertex[right=2cm of e1] (e3)  {\((d)\)}; 
    \vertex[right=4cm of e1] (e5) ;
    
    \diagram* { 
    (a1) -- [charged scalar] (b2),
    (b2) -- [scalar] (d2),
    (e1) -- [charged scalar] (d2),
    (b2) -- [gluon] (b4),
    (d2) -- [gluon] (d4),
    (a5) -- [fermion] (b4),
    (b4) -- [] (d4),
    (e5) -- [fermion] (d4),
    };
    \end{feynman} 
    \end{tikzpicture}}
    \quad
    \scalebox{0.8}{\begin{tikzpicture} 
    \begin{feynman}
    \vertex (a1) ;
    \vertex[right=1cm of a1] (a2);
    \vertex[right=1cm of a2] (a3); 
    \vertex[right=1cm of a3] (a4);
    \vertex[right=1cm of a4] (a5);
    \vertex[below= 1cm of a1] (b1);
    \vertex[right= 1cm of b1] (b2);
    \vertex[right= 2cm of b1] (b3);
    \vertex[right= 1cm of b3] (b4);
    \vertex[right= 2cm of b3] (b5);
    \vertex[below= 2cm of b1] (d1); 
    \vertex[right= 2cm of d1] (d3);
    \vertex[right= 1cm of d1] (d2);
    \vertex[right= 1cm of d2] (d3);
    \vertex[right= 1cm of d3] (d4);
    \vertex[below= 2cm of b5] (d5); 
    \vertex[below=2cm of a1] (c1); 
    \vertex[right=1cm of c1] (c2); 
    \vertex[right = 2cm of c2] (c4);
    \vertex[right=1 of c4] (c5) ;
    \vertex[below=4cm of a1] (e1); 
    \vertex[right=2cm of e1] (e3)  {\((e)\)}; 
    \vertex[right=4cm of e1] (e5) ;
    
    \diagram* { 
    (a1) -- [charged scalar] (b2),
    (b2) -- [scalar] (d2),
    (e1) -- [charged scalar] (d2),
    (b2) -- [gluon] (d4),
    (d2) -- [gluon] (b4),
    (a5) -- [fermion] (b4),
    (b4) -- [] (d4),
    (e5) -- [fermion] (d4),
    };
    \end{feynman} 
    \end{tikzpicture}}
    \quad 
    \scalebox{0.8}{\begin{tikzpicture} 
    \begin{feynman}
    \vertex (a1) ;
    \vertex[right=1cm of a1] (a2);
    \vertex[right=1cm of a2] (a3); 
    \vertex[right=1cm of a3] (a4);
    \vertex[right=1cm of a4] (a5);
    \vertex[below= 1cm of a1] (b1);
    \vertex[right= 1cm of b1] (b2);
    \vertex[right= 2cm of b1] (b3);
    \vertex[right= 1cm of b3] (b4);
    \vertex[right= 2cm of b3] (b5);
    \vertex[below= 2cm of b1] (d1); 
    \vertex[right= 2cm of d1] (d3);
    \vertex[right= 1cm of d1] (d2);
    \vertex[right= 1cm of d2] (d3);
    \vertex[right= 1cm of d3] (d4);
    \vertex[below= 2cm of b5] (d5); 
    \vertex[below=2cm of a1] (c1); 
    \vertex[right=1cm of c1] (c2); 
    \vertex[right=1cm of c2] (c3); 
    \vertex[right = 2cm of c2] (c4);
    \vertex[right=1 of c4] (c5) ;
    \vertex[below=4cm of a1] (e1); 
    \vertex[right=2cm of e1] (e3)  {\((f)\)}; 
    \vertex[right=4cm of e1] (e5) ;
    
    \diagram* { 
    (a1) -- [charged scalar] (b2),
    (b2) -- [scalar] (d2),
    (e1) -- [charged scalar] (d2),
    (b2) -- [gluon] (c3),
    (d2) -- [gluon] (c3),
    (c3) -- [gluon] (c4),
    (b5) -- [fermion] (c4),
    (d5) -- [fermion] (c4),
    };
    \end{feynman} 
    \end{tikzpicture}}
    \quad 
    \scalebox{0.8}{\begin{tikzpicture} 
    \begin{feynman}
    \vertex (a1) ;
    \vertex[right=1cm of a1] (a2);
    \vertex[right=1cm of a2] (a3); 
    \vertex[right=1cm of a3] (a4);
    \vertex[right=1cm of a4] (a5);
    \vertex[below= 1cm of a1] (b1);
    \vertex[right= 1cm of b1] (b2);
    \vertex[right= 2cm of b1] (b3);
    \vertex[right= 1cm of b3] (b4);
    \vertex[right= 2cm of b3] (b5);
    \vertex[below= 2cm of b1] (d1); 
    \vertex[right= 2cm of d1] (d3);
    \vertex[right= 1cm of d1] (d2);
    \vertex[right= 1cm of d2] (d3);
    \vertex[right= 1cm of d3] (d4);
    \vertex[below= 2cm of b5] (d5); 
    \vertex[below=2cm of a1] (c1); 
    \vertex[right=1cm of c1] (c2); 
    \vertex[right=1cm of c2] (c3); 
    \vertex[right = 2cm of c2] (c4);
    \vertex[right=1 of c4] (c5) ;
    \vertex[below=4cm of a1] (e1); 
    \vertex[right=2cm of e1] (e3)  {\((g)\)}; 
    \vertex[right=4cm of e1] (e5) ;
    
    \diagram* { 
    (b1) -- [charged scalar] (c2),
    (d1) -- [charged scalar] (c2),
    (c2) -- [gluon] (c3),
    (c3) -- [gluon] (b4),
    (c3) -- [gluon] (d4),
    (a5) -- [fermion] (b4),
    (b4) -- [] (d4),
    (e5) -- [fermion] (d4),
    };
    \end{feynman} 
    \end{tikzpicture}}
    \quad 
    \scalebox{0.8}{\begin{tikzpicture} 
    \begin{feynman}
    \vertex (a1) ;
    \vertex[right=1cm of a1] (a2);
    \vertex[right=1cm of a2] (a3); 
    \vertex[right=1cm of a3] (a4);
    \vertex[right=1cm of a4] (a5);
    \vertex[below= 1cm of a1] (b1);
    \vertex[right= 1cm of b1] (b2);
    \vertex[right= 2cm of b1] (b3);
    \vertex[right= 1cm of b3] (b4);
    \vertex[right= 2cm of b3] (b5);
    \vertex[below= 2cm of b1] (d1); 
    \vertex[right= 2cm of d1] (d3);
    \vertex[right= 1cm of d1] (d2);
    \vertex[right= 1cm of d2] (d3);
    \vertex[right= 1cm of d3] (d4);
    \vertex[below= 2cm of b5] (d5); 
    \vertex[below=2cm of a1] (c1); 
    \vertex[right=1cm of c1] (c2); 
    \vertex[right = 2cm of c2] (c4);
    \vertex[right=1 of c4] (c5) ;
    \vertex[right=1cm of c2] (c3); 
    \vertex[right=0.5 of c2] (c6);
    \vertex[right=0.5 of c3] (c7);
    \vertex[below=4cm of a1] (e1); 
    \vertex[right=2cm of e1] (e3)  {\((h)\)}; 
    \vertex[right=4cm of e1] (e5) ;
    \diagram* { 
    (b1) -- [charged scalar] (c2),
    (d1) -- [charged scalar] (c2),
    (c2) -- [gluon, half left] (c3),
    (c2) -- [gluon, half right] (c3),
    (c3) -- [gluon] (c4),
    (b5) -- [fermion] (c4),
    (d5) -- [fermion] (c4),
    };
    \end{feynman} 
    \end{tikzpicture}}
    \quad
    \scalebox{0.9}{\begin{tikzpicture} 
    \begin{feynman}
    \vertex (a1) ;
    \vertex[right=1cm of a1] (a2);
    \vertex[right=1cm of a2] (a3); 
    \vertex[right=1cm of a3] (a4);
    \vertex[right=1cm of a4] (a5);
    \vertex[below= 1cm of a1] (b1);
    \vertex[right= 1cm of b1] (b2);
    \vertex[right= 2cm of b1] (b3);
    \vertex[right= 1cm of b3] (b4);
    \vertex[right= 2cm of b3] (b5);
    \vertex[below= 2cm of b1] (d1); 
    \vertex[right= 2cm of d1] (d3);
    \vertex[right= 1cm of d1] (d2);
    \vertex[right= 1cm of d2] (d3);
    \vertex[right= 1cm of d3] (d4);
    \vertex[below= 2cm of b5] (d5); 
    \vertex[below=2cm of a1] (c1); 
    \vertex[right=1cm of c1] (c2); 
    \vertex[right=1cm of c2] (c3); 
    \vertex[right = 2cm of c2] (c4);
    \vertex[right=1 of c4] (c5);
    \vertex[right=0.5 of c2] (c6);
    \vertex[right=0.5 of c3] (c7);
    \vertex[below=4cm of a1] (e1); 
    \vertex[right=2cm of e1] (e3)  {\((i)\)}; 
    \vertex[right=4cm of e1] (e5) ;
    \diagram* { 
    (b1) -- [charged scalar] (c2),
    (d1) -- [charged scalar] (c2),
    (c3) -- [gluon, half left] (c4),
    (c3) -- [gluon, half right] (c4),
    (c2) -- [gluon] (c3),
    (b5) -- [fermion] (c4),
    (d5) -- [fermion] (c4),
    };
    \end{feynman} 
    \end{tikzpicture}}
    \qquad
    \scalebox{0.9}{\begin{tikzpicture} 
    \begin{feynman}
    \vertex (a1) ;
    \vertex[right=1cm of a1] (a2);
    \vertex[right=1cm of a2] (a3); 
    \vertex[right=1cm of a3] (a4);
    \vertex[right=1cm of a4] (a5);
    \vertex[below= 1cm of a1] (b1);
    \vertex[right= 1cm of b1] (b2);
    \vertex[right= 2cm of b1] (b3);
    \vertex[right= 1cm of b3] (b4);
    \vertex[right= 2cm of b3] (b5);
    \vertex[below= 2cm of b1] (d1); 
    \vertex[right= 2cm of d1] (d3);
    \vertex[right= 1cm of d1] (d2);
    \vertex[right= 1cm of d2] (d3);
    \vertex[right= 1cm of d3] (d4);
    \vertex[below= 2cm of b5] (d5); 
    \vertex[below=2cm of a1] (c1); 
    \vertex[right=1cm of c1] (c2); 
    \vertex[right = 2cm of c2] (c4);
    \vertex[right=1 of c4] (c5) ;
    \vertex[right=1cm of c2] (c3); 
    \vertex[right=0.5 of c2] (c6);
    \vertex[right=0.5 of c3] (c7);
    \vertex[below=4cm of a1] (e1); 
    \vertex[right=2cm of e1] (e3)  {\((j)\)}; 
    \vertex[right=4cm of e1] (e5) ;
    \diagram* { 
    (b1) -- [charged scalar] (c2),
    (d1) -- [charged scalar] (c2),
    (c2) -- [gluon] (c6),
    (c6) -- [gluon, half left] (c7),
    (c6) -- [gluon, half right] (c7),
    (c7) -- [gluon] (c4),
    (b5) -- [fermion] (c4),
    (d5) -- [fermion] (c4),
    };
    \end{feynman} 
    \end{tikzpicture}}
    \qquad 
    \scalebox{0.9}{\begin{tikzpicture} 
    \begin{feynman}
    \vertex (a1) ;
    \vertex[right=1cm of a1] (a2);
    \vertex[right=1cm of a2] (a3); 
    \vertex[right=1cm of a3] (a4);
    \vertex[right=1cm of a4] (a5);
    \vertex[below= 1cm of a1] (b1);
    \vertex[right= 1cm of b1] (b2);
    \vertex[right= 2cm of b1] (b3);
    \vertex[right= 1cm of b3] (b4);
    \vertex[right= 2cm of b3] (b5);
    \vertex[below= 2cm of b1] (d1); 
    \vertex[right= 2cm of d1] (d3);
    \vertex[right= 1cm of d1] (d2);
    \vertex[right= 1cm of d2] (d3);
    \vertex[right= 1cm of d3] (d4);
    \vertex[below= 2cm of b5] (d5); 
    \vertex[below=2cm of a1] (c1); 
    \vertex[right=1cm of c1] (c2); 
    \vertex[right = 2cm of c2] (c4);
    \vertex[right=1 of c4] (c5) ;
    \vertex[right=1cm of c2] (c3); 
    \vertex[right=0.5 of c2] (c6);
    \vertex[right=0.5 of c3] (c7);
    \vertex[below=4cm of a1] (e1); 
    \vertex[right=2cm of e1] (e3)  {\((k)\)}; 
    \vertex[right=4cm of e1] (e5) ;
    \diagram* { 
    (b1) -- [charged scalar] (c2),
    (d1) -- [charged scalar] (c2),
    (c2) -- [gluon] (c6),
    (c6) -- [ghost, half left] (c7),
    (c6) -- [ghost, half right] (c7),
    (c7) -- [gluon] (c4),
    (b5) -- [fermion] (c4),
    (d5) -- [fermion] (c4),
    };
    \end{feynman} 
    \end{tikzpicture}}
    \end{subfigure}
    \caption{One-loop Feynman diagrams contributing to the process in the s-channel graviton double-cut. (set 1) \label{fig:1loop_grav}}
    \end{figure}
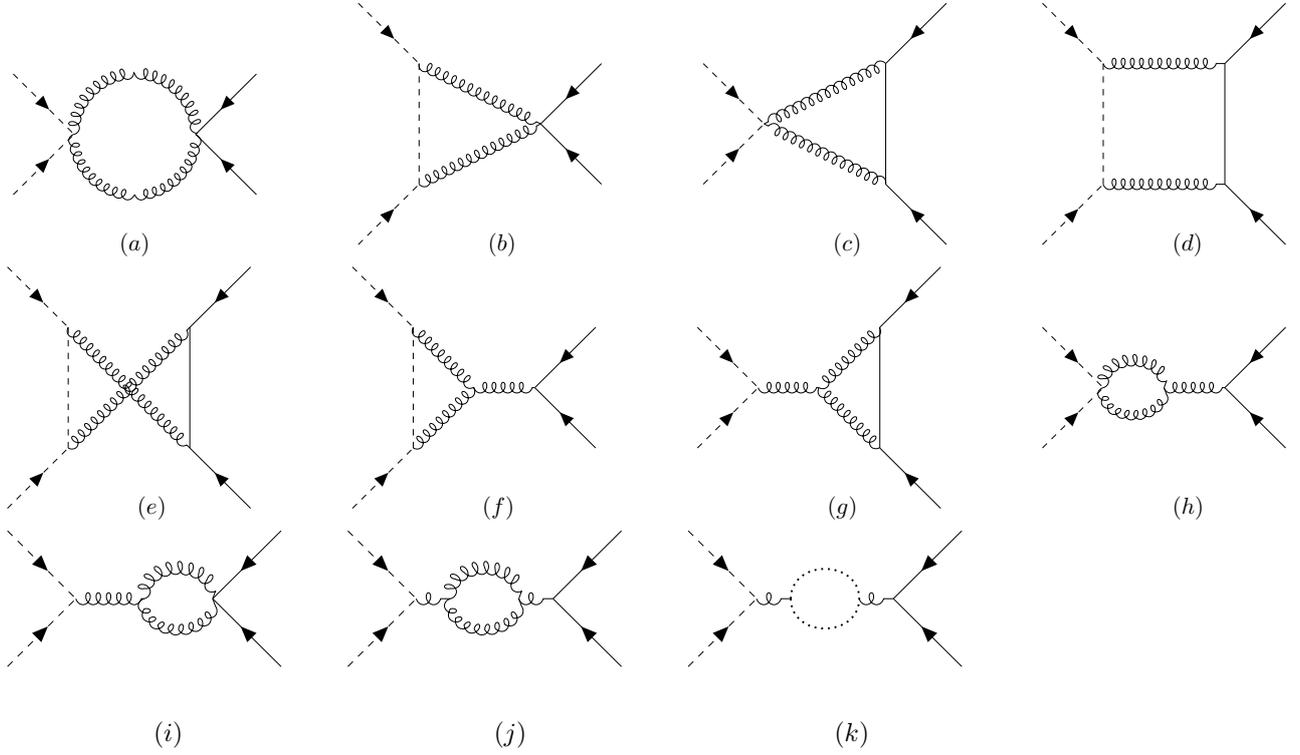
The Feynman amplitudes for each diagram, taking $k$ as loop momenta are:
\begin{eqnarray}
   \mathcal{M}_a &=& \frac{1}{2!}\int \frac{d^dk}{(2\pi)^d}\frac{\tau_2^{\mu\nu\rho\sigma}(p_1,-p_2)P_{\mu\nu\alpha\beta}P_{\rho\sigma\gamma\delta}\tau_2^{\alpha\beta\gamma\delta}(p_3,-p_4)}{(k^2+i\epsilon)[(k+q)^2+i\epsilon]}\nonumber \\
   \mathcal{M}_b &=&  \int \frac{d^dk}{(2\pi)^d}\frac{\tau_1^{\mu\nu}(p_1,k+p_1)\tau_1^{\rho\sigma}(p_2,-k-p_1)\tau_2^{\alpha\beta\gamma\delta}(p_3,-p_4)P_{\mu\nu\alpha\beta}P_{\rho\sigma\gamma\delta}}{(k^2+i\epsilon)[(k+q)^2+i\epsilon][(k+p_1)^2+i\epsilon]}\nonumber \\
   \mathcal{M}_c &=&  \int \frac{d^dk}{(2\pi)^d}\frac{\tau_1^{\mu\nu}(p_3,-k+p_3)\tau_1^{\rho\sigma}(p_4,k-p_3)\tau_2^{\alpha\beta\gamma\delta}(p1,-p2)P_{\mu\nu\alpha\beta}P_{\rho\sigma\gamma\delta}}{(k^2+i\epsilon)[(k+q)^2+i\epsilon][(k-p_3)^2-m^2+i\epsilon]}\nonumber \\
   \mathcal{M}_d & = & \int\frac{d^dk}{(2\pi)^d}\frac{\tau_1^{\mu\nu}(p_1,k+p_1)\tau_1^{\rho\sigma}(p_2,-k-p_1)\tau_1^{\alpha\beta}(p_3,-k+p_3)\tau_1^{\gamma\delta}(p_4,k-p_3)P_{\mu\nu\alpha\beta}P_{\rho\sigma\gamma\delta}}{(k^2+i\epsilon)[(k+q)^2+i\epsilon][(k+p_1)^2+i\epsilon][(k-p_3)^2-m^2+i\epsilon]}\nonumber \\
   \mathcal{M}_e &=& \int \frac{d^dk}{(2\pi)^d}\frac{\tau_1^{\mu\nu}(p_1,k+p_1)\tau_1^{\rho\sigma}(p_2,-k-p_1)\tau_1^{\alpha\beta}(p_3,k-p_4)\tau_1^{\gamma\delta}(p_4,-k+p_4)P_{\mu\nu\gamma\delta}P_{\rho\sigma\alpha\beta}}{(k^2+i\epsilon)[(k+q^2)+i\epsilon][(k+p_1)^2+i\epsilon][(k-p_4)^2-m^2+i\epsilon]}\nonumber \\
   \mathcal{M}_f & =& \int \frac{d^dk}{(2\pi)^d}\frac{\tau_1^{\mu\nu}(p_1,k+p_1)\tau_1^{\rho\sigma}(p_2,-k-p_1)\tau_{3\alpha\beta\gamma\delta}^{\tau\eta}(k,q)\tau_1^{\xi\phi}(p_3,-p_4)P_{\mu\nu}^{\alpha\beta}P_{\rho\sigma}^{\gamma\delta}P_{\tau\eta\xi\phi}}{(k^2+i\epsilon)[(k+q)^2+i\epsilon][(k+p_1)^2+i\epsilon](q^2+i\epsilon)}\nonumber \\
   \mathcal{M}_g &= & \int\frac{d^dk}{(2\pi)^d}\frac{\tau_1^{\mu\nu}(p_3,-k+p_3)\tau_1^{\rho\sigma}(p_4,k-p_3)\tau_{3\alpha\beta\gamma\delta}^{\mu\nu}(-k,-q)\tau_1^{\xi\phi}(p_1,p_2)P_{\mu\nu}^{\alpha\beta}P_{\rho\sigma}^{\gamma\delta}P_{\tau\eta\xi\phi}}{(k^2+i\epsilon)[(k+q)^2-m^2+i\epsilon][(k-p_3)^2-m^2+i\epsilon](q^2+i\epsilon)}\nonumber \\
   \mathcal{M}_h &= &\int\frac{d^dk}{(2\pi)^d}\frac{d^dk}{(2\pi)^d}\frac{\tau_2^{\mu\nu}(p_3,-p_4)\tau_{3\gamma\delta\rho\sigma}^{\alpha\beta}(k,q)\tau_2^{\tau\xi\eta\chi}(p_1,-p_2)P_{\mu\nu\alpha\beta}P^{\gamma\delta}_{\tau\xi}P^{\rho\sigma}_{\eta\chi}}{(k^2+i\epsilon)[(k+q)^2+i\epsilon](q^2+i\epsilon)}\nonumber\\
   \mathcal{M}_i & = &\frac{1}{2!}\int\frac{d^dk}{(2\pi)^d}\frac{\tau_2^{\mu\nu}(p_1,-p_2)\tau_{3\gamma\delta\rho\sigma}^{\alpha\beta}(k,q)\tau_2^{\tau\xi\eta\chi}(p_3,-p_4)P_{\mu\nu\alpha\beta}P^{\gamma\delta}_{\tau\xi}P^{\rho\sigma}_{\eta\chi}}{(k^2+i\epsilon)[(k+q)^2+i\epsilon](q^2+i\epsilon)}\nonumber \\
   \mathcal{M}_j &=& \frac{1}{2!} \int\frac{d^dk}{(2\pi)^d}\frac{\tau_1^{\mu\nu}(p_1,-p_2)P_{\mu\nu\alpha\beta}\tau_{3\gamma\delta\rho\sigma}^{\alpha\beta}P^{\gamma\delta\tau\lambda}P^{\rho\sigma\xi\phi}\tau_{3\tau\lambda\xi\phi}^{\psi\epsilon}P_{\psi\epsilon\zeta\chi}\tau_2^{\zeta\chi}(p_3,-p_4)}{(k^2+i\epsilon)[(k+q)^2+i\epsilon](q^2+i\epsilon)^2}\nonumber \\
   \mathcal{M}_k &=&  -\int\frac{d^dk}{(2\pi)^d}\frac{\tau_1^{\mu\nu}(p_1,-p_2)\tau_G^{\alpha\beta\gamma\delta}\tau_{G\alpha\beta}^{\rho\sigma}\tau_1^{\epsilon\tau}(p_3,-p_4)P_{\mu\nu\gamma\delta}P_{\mu\nu\epsilon\tau}}{(k^2+i\epsilon)[(k+q)^2+i\epsilon](q^2+i\epsilon)^2}
\end{eqnarray}
Notice that for each internal graviton loop we have a $1/2!$ symmetry factor. The evaluation of these diagrams is very tricky due to the presence of triple graviton vertices and so it cannot be done simply by hand. In order to efficiently compute the diagrams we developed a computational algorithm that will be described in the next paragraph.
\subsection{Evaluation of the diagrams}
\begin{figure}[!ht]
\centering
\tikzstyle{decision} = [diamond, draw, fill=red!20,
   , text badly centered, inner sep=0pt]
\tikzstyle{block} = [rectangle, draw, fill=red!20,
    text centered, rounded corners, minimum height=4em]
\tikzstyle{inoroutput} = [trapezium, draw, fill=red!20,
    text centered, minimum height=4em,trapezium left angle=60, trapezium right angle=120]
\tikzstyle{inoroutput} = [trapezium, draw, fill=red!20,
   text centered, minimum height=4em,trapezium left angle=60, trapezium right angle=120]
\tikzstyle{line} = [draw, -latex']
\tikzstyle{cloud} = [draw, ellipse,fill=red!20, node distance=3cm,
    minimum height=2em]
\begin{tikzpicture}[auto]
\node at (0,0) [block, text width=20em] (start) {Integrand generation (\texttt{QGRAF})};
\node [below=0.8cm of start,block, text width=20em] (tcontr) {Tensor Contractions via \texttt{xAct}\begin{equation*}
\mathcal{M}_i=\int \frac{\mathcal{N}_i}{\prod_j D_j}
\end{equation*}};
\node [below=0.8cm of tcontr, block, text width=20em] (intred) {Integrand reduction
\begin{equation*}
\mathcal{M}_i=\int \sum_{k=i}^{\text{$\sharp$ of den}} \sum_{\{j_1,\cdots,j_k\}} \frac{\Delta_{j_1,\cdots,j_k}}{D_{j_1}\cdots D_{j_k}}
\end{equation*}
};
\node [below=0.8cm of intred, block, text width=20em](ibpdec){IBP decomposition (\texttt{Litered}) \begin{equation*}
\mathcal{M}_i=\sum_j c_{ij} I_{j}^{M.I.}
\end{equation*}};
\node [below=0.8cm of ibpdec, block, text width=20em](mi){Master Integrals (\texttt{PackageX})};
\node [below= 0.8cm of mi, block,text width=20em](results){Results
\begin{equation*}
\mathcal{M}= \sum_i \mathcal{M}_i = \sum_j c_j I^{M.I.}_j
\end{equation*}
};
\path [line] (start) -- (tcontr);
\path [line] (tcontr) -- (intred);
\path [line] (intred) -- (ibpdec);
\path [line] (ibpdec) -- (mi);
\path [line] (mi) -- (results);
\end{tikzpicture}
\captionof{figure}{Flow chart of the algorithm for the evaluation of scattering amplitudes}\label{fig:flowalgoritmo}
\end{figure}
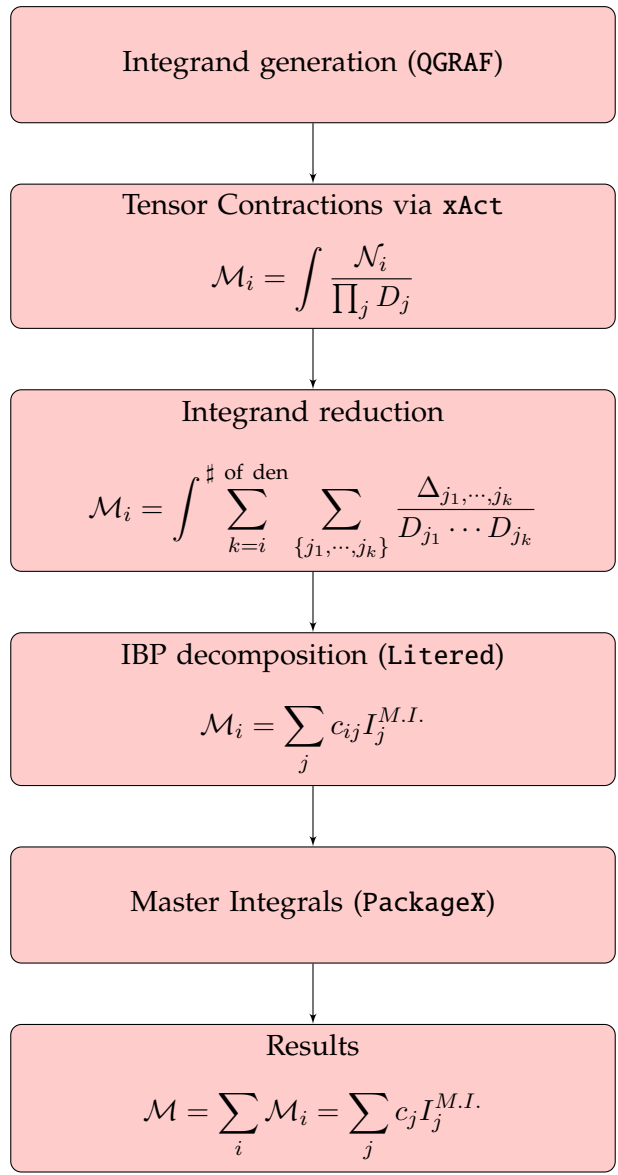

The complete evaluation of the one-loop amplitude has been performed by developing an in-house code in \texttt{Mathematica}, following a systematic procedure that can be divided in different steps which are summarized in figure Fig.\ref{fig:flowalgoritmo}.
\subsubsection{Integrand generation}
The diagrams relevant for the process were generated using \texttt{QGRAF}\cite{Nogueira:1991ex}, a Fortran-based program which automatically generates diagrams at a given loop order, once the Feynman rules have been defined. It gave as output a set of diagrams written in terms of propagators and vertices, with corresponding symmetry factors.
From that expressions we recover the diagrams written in terms of Feynman rules with appropriate substitution rules written using the \texttt{xAct} package. 
\subsubsection{Tensor Contractions} 
Since it is very difficult to deal with three graviton vertices due to the high number of free indices appearing, it was necessary to develop a smart procedure to optimize the tensor contractions.
We divided each diagram in subdiagrams, made by one vertex and one propagator, and for each one of them we perform tensor contractions using \texttt{xTensor} package. 
To use the symmetries of the appearing tensors, contractions has been done twice for each subdiagram, first by keeping the factors $P_{\mu\nu\rho\sigma}$, that appear in  graviton propagators, intact, and then by substituting the explicit expression for it. \\ 
Afterwards, we multiplied the different sub-diagrams one with each other to reconstruct the full expression of the integrand, and we performed a final tensor contraction.
\subsubsection{Integrand Reduction and transcription to Litered}
In order to get the result in a format readable by the package \texttt{Litered}, that has been used for IBP decomposition, we had to perform several operations.\\
First we needed to rewrite contracted tensors appearing in the numerator of the integrand in $xAct$ format in terms of ordinary scalar products, and then we expressed these scalar products in terms of the denominators appearing in our integrand.\\
At this point we expanded the integrands, and we performed an integrand reduction, finding an expression of the integrand in terms of scalar integrals, which could then be easily rewritten in a Litered form. \\

\subsubsection{Master Integrals}
As already mentioned, the IBP decomposition has been done using the package \texttt{LiteRed}. \\
In order to do that we had to find a basis of Master Integrals in which our integrals could be decomposed into.
It is easy to see that all the diagrams in figure \eqref{fig:1loop_grav}, except the crossed-box diagram, that is the one indicated with letter $(e)$, belong to the same topology of integrals, that we will refer as \textit{s-topology}. The crossed-box diagram instead belongs to the \textit{u-topology}. \\
That means that we can use only two basis of Master Integrals to decompose all the diagrams.\\
We made the following choice of denominators for the diagrams in the s-topology:
\begin{equation}
    D_1= k^2 \qquad D_2 = (k+p_1+p_2)^2 \qquad D_3= (k+p_1)^2 \qquad D_4= (k-p_3)^2-m^2
\end{equation}
Instead the denominators for the u-topology are:
\begin{equation}
    D_1= k^2 \qquad D_2 = (k+p_1+p_2)^2 \qquad D_3= (k+p_1)^2 \qquad D_{4u}= (k+p_1+p_2+p_3)^2-m^2
\end{equation}
Then a generic integral in the s-topology can be written as: 
\begin{equation} 
    j^s_{n_1,n_2,n_3,n_4}= \begin{tikzpicture}[baseline=(current bounding box.center)] 
        \begin{feynman}
            \vertex(a1) ;
            \vertex[right= 1.4 cm of a1] (a2);
            \vertex[below =0.2 cm of a1] (b1);
            \vertex[right = 0.2 cm of b1] (b2);
            \vertex[right = 1cm of b2] (b3);
            \vertex[below = 1 cm of b1] (c1);
            \vertex[below = 0.2 cm of c1] (d1);
            \vertex[right = 1.4cm of d1] (d2);
            \vertex[right = 0.2 cm of c1] (c2);
            \vertex[right = 1cm of c2] (c3);
            \vertex[below = 0.5 cm of b1] (e1);
             \vertex[right = 1 cm of e1] (e2);
              \vertex[right = 0.2 cm of e2] (e3);
            \diagram* { 
            (a1) --  (b2),
            (b2) --  (c2),
            (d1)--(c2),
            (a2) -- [very thick](b3),
            (d2) -- [very thick](c3),
            (b2)-- (b3),
            (c2) -- (c3),
            (b3) -- [very thick](c3),
            };
            \end{feynman} \end{tikzpicture} = \int \frac{d^dk}{(2\pi)^d}\frac{1}{D_1^{n_1}D_2^{n_2}D_3^{n_3}D_4^{n_4}} \ , 
\end{equation} and of the u-topology as:
\begin{equation}
    j^u_{n_1,n_2,n_3,n_4}= \begin{tikzpicture}[baseline=(current bounding box.center)] 
        \begin{feynman}
            \vertex(a1) ;
            \vertex[right= 1.4 cm of a1] (a2);
            \vertex[below =0.2 cm of a1] (b1);
            \vertex[right = 0.2 cm of b1] (b2);
            \vertex[right = 1cm of b2] (b3);
            \vertex[below = 1 cm of b1] (c1);
            \vertex[below = 0.2 cm of c1] (d1);
            \vertex[right = 1.4cm of d1] (d2);
            \vertex[right = 0.2 cm of c1] (c2);
            \vertex[right = 1cm of c2] (c3);
            \vertex[below = 0.5 cm of b1] (e1);
             \vertex[right = 1 cm of e1] (e2);
              \vertex[right = 0.2 cm of e2] (e3);
            \diagram* { 
            (a1) --  (b2),
            (b2) --  (c2),
            (d1)--(c2),
            (a2) -- [very thick](b3),
            (d2) -- [very thick](c3),
            (b2)-- (c3),
            (c2) -- (b3),
            (b3) -- [very thick](c3),
            };
            \end{feynman} \end{tikzpicture} = \int \frac{d^dk}{(2\pi)^d}\frac{1}{D_1^{n_1}D_2^{n_2}D_3^{n_3}D_{4u}^{n_4}} \ . 
\end{equation}
Using Integration-by-parts identities we could find the Master Integrals for each family. Moreover, by putting a cut in the first two denominators, which corresponds to the graviton propagators, we could select only the diagrams giving non-analytical contributions to the amplitude, and we could find also relations between the two basis.
By doing so we found four different Master Integrals.
\begin{equation}
\begin{split}
&  I_{1}= j^s_{1,1,0,0}=\int \frac{d^dk}{(2\pi)^d}\frac{1}{D_1D_2}=      \begin{tikzpicture}[baseline=(current bounding box.center)]             \begin{feynman}
    \vertex (a) ;
    \vertex[right=0.2cm of a] (b);
    \vertex[right=1cm of b] (c); 
    \vertex[right=0.2cm of c] (d);
    \diagram* { 
    (a) --  (b),
    (b) -- [half left ] (c),
    (b) -- [half right ] (c),
    (c) -- (d),
    };
    \end{feynman} \end{tikzpicture}\\
& I_{2}= j^s_{1,1,0,1}=\int \frac{d^dk}{(2\pi)^d}\frac{1}{D_1D_2D_4}=
    \begin{tikzpicture}[baseline=(current bounding box.center)] 
\begin{feynman}
   \vertex(a1) ;
    \vertex[right= 1.4 cm of a1] (a2);
    \vertex[below =0.2 cm of a1] (b1);
    \vertex[below =0.5 cm of b1] (f1);
    \vertex[right =0.2 cm of f1] (f2);
    \vertex[right = 0.2 cm of b1] (b2);
     \vertex[right = 1cm of b2] (b3);
    \vertex[below = 1 cm of b1] (c1);
    \vertex[below = 0.2 cm of c1] (d1);
    \vertex[right = 0.2 cm of c1] (c2);
     \vertex[right = 1 cm of c2] (c3);
    \vertex[below = 0.5 cm of b1] (e1);
     \vertex[right = 1 cm of e1] (e2);
      \vertex[right = 0.2 cm of e2] (e3);
      \vertex[right = 1.4cm of d1] (d2);
    \diagram* { 
    (f1)-- (f2),
    (f2)-- (b3),
    (f2) -- (c3),
    (b3) -- [very thick](c3),
    (a2) -- [very thick](b3),
    (d2) -- [very thick](c3),
    };
    \end{feynman} \end{tikzpicture}   \\
& I_{3}= j^s_{1,1,1,1}=\int \frac{d^dk}{(2\pi)^d}\frac{1}{D_1D_2D_3D_4}=   \begin{tikzpicture}[baseline=(current bounding box.center)] 
\begin{feynman}
    \vertex(a1) ;
    \vertex[right= 1.4 cm of a1] (a2);
    \vertex[below =0.2 cm of a1] (b1);
    \vertex[right = 0.2 cm of b1] (b2);
    \vertex[right = 1cm of b2] (b3);
    \vertex[below = 1 cm of b1] (c1);
    \vertex[below = 0.2 cm of c1] (d1);
    \vertex[right = 1.4cm of d1] (d2);
    \vertex[right = 0.2 cm of c1] (c2);
    \vertex[right = 1cm of c2] (c3);
    \vertex[below = 0.5 cm of b1] (e1);
     \vertex[right = 1 cm of e1] (e2);
      \vertex[right = 0.2 cm of e2] (e3);
    \diagram* { 
    (a1) --  (b2),
    (b2) --  (c2),
    (d1)--(c2),
    (a2) -- [very thick](b3),
    (d2) -- [very thick](c3),
    (b2)-- (b3),
    (c2) -- (c3),
    (b3) -- [very thick](c3),
    };
    \end{feynman} \end{tikzpicture}\\
& I_{4}= j^u_{1,1,1,1}= \int \frac{d^dk}{(2\pi)^d}\frac{1}{D_1D_2D_3D_{4u}}=  
\begin{tikzpicture}[baseline=(current bounding box.center)] 
\begin{feynman}
    \vertex(a1) ;
    \vertex[right= 1.4 cm of a1] (a2);
    \vertex[below =0.2 cm of a1] (b1);
    \vertex[right = 0.2 cm of b1] (b2);
    \vertex[right = 1cm of b2] (b3);
    \vertex[below = 1 cm of b1] (c1);
    \vertex[below = 0.2 cm of c1] (d1);
    \vertex[right = 1.4cm of d1] (d2);
    \vertex[right = 0.2 cm of c1] (c2);
    \vertex[right = 1cm of c2] (c3);
    \vertex[below = 0.5 cm of b1] (e1);
     \vertex[right = 1 cm of e1] (e2);
      \vertex[right = 0.2 cm of e2] (e3);
    \diagram* { 
    (a1) --  (b2),
    (b2) --  (c2),
    (d1)--(c2),
    (a2) -- [very thick](b3),
    (d2) -- [very thick](c3),
    (b2)-- (c3),
    (c2) -- (b3),
    (b3) -- [very thick](c3),
    };
    \end{feynman} \end{tikzpicture}\\
\end{split}
\end{equation}
\subsubsection{IBP decomposition}
Using \texttt{Litered} we can then apply IBP decomposition to each diagram to decompose it in terms of Master Integrals as:
\begin{eqnarray}
    \mathcal{C}^{(g)}_j
     = \sum_{i=1}^4 c_{j i}^{(g)} \, I_i \
     =
        c_{j 1}^{(g)} \scalebox{0.5}{\begin{tikzpicture}[baseline=(current bounding box.center)]\begin{feynman}
    \vertex (a) ;
    \vertex[right=0.2cm of a] (b);
    \vertex[right=1cm of b] (c); 
    \vertex[right=0.2cm of c] (d);
    \diagram* { 
    (a) --  (b),
    (b) -- [half left ] (c),
    (b) -- [half right ] (c),
    (c) -- (d),
    };
    \end{feynman} \end{tikzpicture}}
      + c_{j2}^{(g)} \scalebox{0.5}{ \begin{tikzpicture}[baseline=(current bounding box.center)] 
\begin{feynman}
   \vertex(a1) ;
    \vertex[right= 1.4 cm of a1] (a2);
    \vertex[below =0.2 cm of a1] (b1);
    \vertex[below =0.5 cm of b1] (f1);
    \vertex[right =0.2 cm of f1] (f2);
    \vertex[right = 0.2 cm of b1] (b2);
     \vertex[right = 1cm of b2] (b3);
    \vertex[below = 1 cm of b1] (c1);
    \vertex[below = 0.2 cm of c1] (d1);
    \vertex[right = 0.2 cm of c1] (c2);
     \vertex[right = 1 cm of c2] (c3);
    \vertex[below = 0.5 cm of b1] (e1);
     \vertex[right = 1 cm of e1] (e2);
      \vertex[right = 0.2 cm of e2] (e3);
      \vertex[right = 1.4cm of d1] (d2);
    \diagram* { 
    (f1)-- (f2),
    (f2)-- (b3),
    (f2) -- (c3),
    (b3) -- [very thick](c3),
    (a2) -- [very thick](b3),
    (d2) -- [very thick](c3),
    };
    \end{feynman} \end{tikzpicture} }
      + c_{j3}^{(g)} \scalebox{0.5}{\begin{tikzpicture}[baseline=(current bounding box.center)]
\begin{feynman}
    \vertex(a1) ;
    \vertex[right= 1.4 cm of a1] (a2);
    \vertex[below =0.2 cm of a1] (b1);
    \vertex[right = 0.2 cm of b1] (b2);
    \vertex[right = 1cm of b2] (b3);
    \vertex[below = 1 cm of b1] (c1);
    \vertex[below = 0.2 cm of c1] (d1);
    \vertex[right = 1.4cm of d1] (d2);
    \vertex[right = 0.2 cm of c1] (c2);
    \vertex[right = 1cm of c2] (c3);
    \vertex[below = 0.5 cm of b1] (e1);
     \vertex[right = 1 cm of e1] (e2);
      \vertex[right = 0.2 cm of e2] (e3);
    \diagram* { 
    (a1) --  (b2),
    (b2) --  (c2),
    (d1)--(c2),
    (a2) -- [very thick](b3),
    (d2) -- [very thick](c3),
    (b2)-- (b3),
    (c2) -- (c3),
    (b3) -- [very thick](c3),
    };
    \end{feynman} \end{tikzpicture}} 
      + c_{j4}^{(g)} \scalebox{0.5}{\begin{tikzpicture}[baseline=(current bounding box.center)] 
\begin{feynman}
    \vertex(a1) ;
    \vertex[right= 1.4 cm of a1] (a2);
    \vertex[below =0.2 cm of a1] (b1);
    \vertex[right = 0.2 cm of b1] (b2);
    \vertex[right = 1cm of b2] (b3);
    \vertex[below = 1 cm of b1] (c1);
    \vertex[below = 0.2 cm of c1] (d1);
    \vertex[right = 1.4cm of d1] (d2);
    \vertex[right = 0.2 cm of c1] (c2);
    \vertex[right = 1cm of c2] (c3);
    \vertex[below = 0.5 cm of b1] (e1);
     \vertex[right = 1 cm of e1] (e2);
      \vertex[right = 0.2 cm of e2] (e3);
    \diagram* { 
    (a1) --  (b2),
    (b2) --  (c2),
    (d1)--(c2),
    (a2) -- [very thick](b3),
    (d2) -- [very thick](c3),
    (b2)-- (c3),
    (c2) -- (b3),
    (b3) -- [very thick](c3),
    };
    \end{feynman} \end{tikzpicture}} \ . 
      \end{eqnarray} 
\subsubsection{Results}
We can get the final expression for the amplitude at 1-loop by summing all the different diagrams, and we can express it as a linear combination of the Master Integrals that we found in the previous section:
\begin{eqnarray}
    \mathcal{C}^{(g)}
     = \sum_{i=1}^4 c_{i}^{(g)} \, I_i \
     =
        c_{1}^{(g)} \scalebox{0.5}{\begin{tikzpicture}[baseline=(current bounding box.center)]\begin{feynman}
    \vertex (a) ;
    \vertex[right=0.2cm of a] (b);
    \vertex[right=1cm of b] (c); 
    \vertex[right=0.2cm of c] (d);
    \diagram* { 
    (a) --  (b),
    (b) -- [half left ] (c),
    (b) -- [half right ] (c),
    (c) -- (d),
    };
    \end{feynman} \end{tikzpicture}}
      + c_{2}^{(g)} \scalebox{0.5}{ \begin{tikzpicture}[baseline=(current bounding box.center)] 
\begin{feynman}
   \vertex(a1) ;
    \vertex[right= 1.4 cm of a1] (a2);
    \vertex[below =0.2 cm of a1] (b1);
    \vertex[below =0.5 cm of b1] (f1);
    \vertex[right =0.2 cm of f1] (f2);
    \vertex[right = 0.2 cm of b1] (b2);
     \vertex[right = 1cm of b2] (b3);
    \vertex[below = 1 cm of b1] (c1);
    \vertex[below = 0.2 cm of c1] (d1);
    \vertex[right = 0.2 cm of c1] (c2);
     \vertex[right = 1 cm of c2] (c3);
    \vertex[below = 0.5 cm of b1] (e1);
     \vertex[right = 1 cm of e1] (e2);
      \vertex[right = 0.2 cm of e2] (e3);
      \vertex[right = 1.4cm of d1] (d2);
    \diagram* { 
    (f1)-- (f2),
    (f2)-- (b3),
    (f2) -- (c3),
    (b3) -- [ultra thick](c3),
    (a2) -- [ultra thick](b3),
    (d2) -- [ultra thick](c3),
    };
    \end{feynman} \end{tikzpicture} }
      + c_{3}^{(g)} \scalebox{0.5}{\begin{tikzpicture}[baseline=(current bounding box.center)]
\begin{feynman}
    \vertex(a1) ;
    \vertex[right= 1.4 cm of a1] (a2);
    \vertex[below =0.2 cm of a1] (b1);
    \vertex[right = 0.2 cm of b1] (b2);
    \vertex[right = 1cm of b2] (b3);
    \vertex[below = 1 cm of b1] (c1);
    \vertex[below = 0.2 cm of c1] (d1);
    \vertex[right = 1.4cm of d1] (d2);
    \vertex[right = 0.2 cm of c1] (c2);
    \vertex[right = 1cm of c2] (c3);
    \vertex[below = 0.5 cm of b1] (e1);
     \vertex[right = 1 cm of e1] (e2);
      \vertex[right = 0.2 cm of e2] (e3);
    \diagram* { 
    (a1) --  (b2),
    (b2) --  (c2),
    (d1)--(c2),
    (a2) -- [ultra thick](b3),
    (d2) -- [ultra thick](c3),
    (b2)-- (b3),
    (c2) -- (c3),
    (b3) -- [ultra thick](c3),
    };
    \end{feynman} \end{tikzpicture}} 
      + c_{4}^{(g)} \scalebox{0.5}{\begin{tikzpicture}[baseline=(current bounding box.center)] 
\begin{feynman}
    \vertex(a1) ;
    \vertex[right= 1.4 cm of a1] (a2);
    \vertex[below =0.2 cm of a1] (b1);
    \vertex[right = 0.2 cm of b1] (b2);
    \vertex[right = 1cm of b2] (b3);
    \vertex[below = 1 cm of b1] (c1);
    \vertex[below = 0.2 cm of c1] (d1);
    \vertex[right = 1.4cm of d1] (d2);
    \vertex[right = 0.2 cm of c1] (c2);
    \vertex[right = 1cm of c2] (c3);
    \vertex[below = 0.5 cm of b1] (e1);
     \vertex[right = 1 cm of e1] (e2);
      \vertex[right = 0.2 cm of e2] (e3);
    \diagram* { 
    (a1) --  (b2),
    (b2) --  (c2),
    (d1)--(c2),
    (a2) -- [ultra thick](b3),
    (d2) -- [ultra thick](c3),
    (b2)-- (c3),
    (c2) -- (b3),
    (b3) -- [ultra thick](c3),
    };
    \end{feynman} \end{tikzpicture}}
    \label{eq:ampdecomis} \ .
      \end{eqnarray}
\subsection{An example: The Triangle diagram}
\begin{figure}[!ht]
    \centering
    \includegraphics[width=0.99\textwidth]{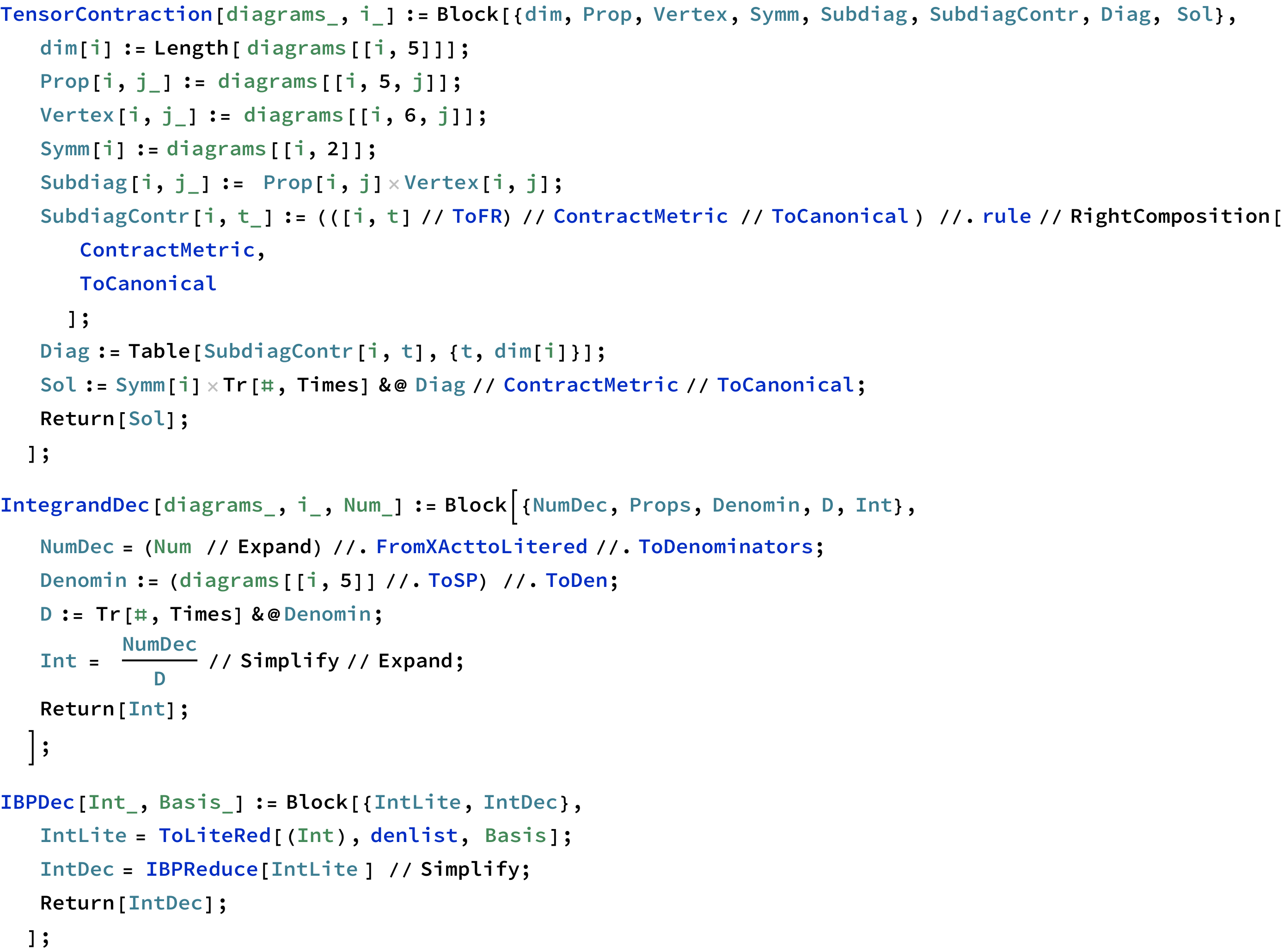}
    \caption{Functions used to evaluate diagrams in \texttt{Mathematica}}
    \label{fig:triangle_diagram_QGRAF}
\end{figure}

As a pedagogical example, in this subsection we present the step-by-step explicit evaluation of the massless triangle diagram indicated with letter $(c)$ in \ref{fig:1loop_grav}. The expression for the amplitude is:
\begin{equation}
    \mathcal{M}_c= \scalebox{0.5}{
            \begin{tikzpicture}[baseline=(current bounding box.center)]  
            \begin{feynman}
            \vertex (a1) ;
            \vertex[right=1cm of a1] (a2);
            \vertex[right=1cm of a2] (a3); 
            \vertex[right=1cm of a3] (a4);
            \vertex[right=1cm of a4] (a5);
            \vertex[below= 1cm of a1] (b1);
            \vertex[right= 1cm of b1] (b2);
            \vertex[right= 2cm of b1] (b3);
            \vertex[right= 1cm of b3] (b4);
            \vertex[right= 2cm of b3] (b5);
            \vertex[below= 2cm of b1] (d1); 
            \vertex[right= 2cm of d1] (d3);
            \vertex[right= 1cm of d1] (d2);
            \vertex[right= 1cm of d2] (d3);
            \vertex[right= 1cm of d3] (d4);
            \vertex[below= 2cm of b5] (d5); 
            \vertex[below=2cm of a1] (c1); 
            \vertex[right=1cm of c1] (c2); 
            \vertex[right = 2cm of c2] (c4);
            \vertex[right=1 of c4] (c5) ;
            \vertex[below=4cm of a1] (e1); 
            \vertex[right=2cm of e1] (e3); 
            \vertex[right=4cm of e1] (e5) ;
            
            \diagram* { 
            (b1) -- [charged scalar] (c2),
            (d1) -- [charged scalar] (c2),
            (b4) -- [gluon] (c2),
            (d4) -- [gluon] (c2),
            (a5) -- [fermion] (b4),
            (b4) -- [] (d4),
            (e5) -- [fermion] (d4),
            };
            \end{feynman} 
            \end{tikzpicture}}=\int \frac{d^dk}{(2\pi)^d}\frac{\tau_1^{\mu\nu}(p_3,-k+p_3)\tau_1^{\rho\sigma}(p_4,k-p_3)\tau_2^{\alpha\beta\gamma\delta}(p1,-p2)P_{\mu\nu\alpha\beta}P_{\rho\sigma\gamma\delta}}{(k^2+i\epsilon)[(k+q)^2+i\epsilon][(k-p_3)^2-m^2+i\epsilon]} \ , 
\end{equation}
which has been generated using \texttt{QGRAF}, as: 
\begin{lstlisting}[language=Mathematica,caption={Example of QGRAF output for a diagram}]
    Diagram[Dia["photonphoton", "scalscal", 1, 5], 1, 
    {F["photon", -1, 3, p1], F["photon", -3, 3, p2]}, {F["scal", -2, 1, q1], F["scal", -4, 2, q2]}, 
    {P["scal", 2, 1, 2, 1, -k1], P["grav", 4, 3, 3, 1, k1 + q1], P["grav", 6, 5, 3, 2, -k1 + q2]}, 
    {V[1, "scalscalgrav", -2, -q1, 1, -k1, 3, k1 + q1], V[2, "scalscalgrav", -4, -q2, 2, k1, 5, -k1 + q2], 
    V[3, "photonphotongravgrav", -1, p1, -3, p2, 4, -k1 - q1, 6,k1 - q2]}]
\end{lstlisting}
where in the first line there is the symmetry factor, in the second the external states, in the third the propagators, in the fourth and in the fifth the vertices. 
In order to decompose it in terms of Master Integrals we use the procedure outlined in the previous subsection, and we present the code used in fig.\ref{fig:triangle_diagram_QGRAF}, which is made of 3 functions:
\begin{enumerate}
    \item The function \texttt{TensorContraction} subdivide the numerator of the diagram in different subdiagrams, each one made by one vertex and one propagator. For each one of them we substitute the Feynman rules (\texttt{ToFR}) and perform the contractions (\texttt{ContractMetric, ToCanonical}). Then the subdiagrams are multiplied together and contracted into a single expression, which is then multiplied by the symmetry factor.
\item  The function \texttt{IntegrandDec} rewrite the integrand in a format readable by $Litered$. First, we rewrite the contracted vectors in terms of scalar products (\texttt{FromXActoToLitered}), and then we rewrite scalar products in terms of denominators \texttt{ToDenominators}. Then it adds the denominator in order to form the total integrand, which is then expanded in terms of scalar integrals.
\item The function \texttt{IBPDec} write the integrals in a \texttt{Litered} format as defined in the previous subsection (\texttt{ToLiteRed}) and then perform BP Reduction to rewrite the integrals in terms of Master Integrals (\texttt{IBPReduce}). 
\end{enumerate}
This same procedure has been implemented to evaluate all the diagrams in figure \ref{fig:1loop_grav}.
\subsection{Results in d-dimensions}
In this section we present the results obtained with this procedure, and we compare them with the ones appearing in \cite{Bjerrum-Bohr:2014zsa}\cite{Bjerrum-Bohr:2016hpa}. 
In the previous section we explained how to get the expression for the 1-loop amplitude using Feynman diagrams, obtaining an expression for the total amplitude in terms of four Master Integrals. \\
The final result in d-dimensions can be expressed in terms of four Master Integrals and has the form:

\begin{eqnarray}
    \kappa^{-4}\mathcal{C}^{(g)}
     = \sum_{i=1}^4 c_{i}^{(g)} \, I_i \
     =
        c_{1}^{(g)} \scalebox{0.5}{\begin{tikzpicture}[baseline=(current bounding box.center)]\begin{feynman}
    \vertex (a) ;
    \vertex[right=0.2cm of a] (b);
    \vertex[right=1cm of b] (c); 
    \vertex[right=0.2cm of c] (d);
    \diagram* { 
    (a) --  (b),
    (b) -- [half left ] (c),
    (b) -- [half right ] (c),
    (c) -- (d),
    };
    \end{feynman} \end{tikzpicture}}
      + c_{2}^{(g)} \scalebox{0.5}{ \begin{tikzpicture}[baseline=(current bounding box.center)] 
\begin{feynman}
   \vertex(a1) ;
    \vertex[right= 1.4 cm of a1] (a2);
    \vertex[below =0.2 cm of a1] (b1);
    \vertex[below =0.5 cm of b1] (f1);
    \vertex[right =0.2 cm of f1] (f2);
    \vertex[right = 0.2 cm of b1] (b2);
     \vertex[right = 1cm of b2] (b3);
    \vertex[below = 1 cm of b1] (c1);
    \vertex[below = 0.2 cm of c1] (d1);
    \vertex[right = 0.2 cm of c1] (c2);
     \vertex[right = 1 cm of c2] (c3);
    \vertex[below = 0.5 cm of b1] (e1);
     \vertex[right = 1 cm of e1] (e2);
      \vertex[right = 0.2 cm of e2] (e3);
      \vertex[right = 1.4cm of d1] (d2);
    \diagram* { 
    (f1)-- (f2),
    (f2)-- (b3),
    (f2) -- (c3),
    (b3) -- [very thick](c3),
    (a2) -- [very thick](b3),
    (d2) -- [very thick](c3),
    };
    \end{feynman} \end{tikzpicture} }
      + c_{3}^{(g)} \scalebox{0.5}{\begin{tikzpicture}[baseline=(current bounding box.center)]
\begin{feynman}
    \vertex(a1) ;
    \vertex[right= 1.4 cm of a1] (a2);
    \vertex[below =0.2 cm of a1] (b1);
    \vertex[right = 0.2 cm of b1] (b2);
    \vertex[right = 1cm of b2] (b3);
    \vertex[below = 1 cm of b1] (c1);
    \vertex[below = 0.2 cm of c1] (d1);
    \vertex[right = 1.4cm of d1] (d2);
    \vertex[right = 0.2 cm of c1] (c2);
    \vertex[right = 1cm of c2] (c3);
    \vertex[below = 0.5 cm of b1] (e1);
     \vertex[right = 1 cm of e1] (e2);
      \vertex[right = 0.2 cm of e2] (e3);
    \diagram* { 
    (a1) --  (b2),
    (b2) --  (c2),
    (d1)--(c2),
    (a2) -- [very thick](b3),
    (d2) -- [very thick](c3),
    (b2)-- (b3),
    (c2) -- (c3),
    (b3) -- [very thick](c3),
    };
    \end{feynman} \end{tikzpicture}} 
      + c_{4}^{(g)} \scalebox{0.5}{\begin{tikzpicture}[baseline=(current bounding box.center)] 
\begin{feynman}
    \vertex(a1) ;
    \vertex[right= 1.4 cm of a1] (a2);
    \vertex[below =0.2 cm of a1] (b1);
    \vertex[right = 0.2 cm of b1] (b2);
    \vertex[right = 1cm of b2] (b3);
    \vertex[below = 1 cm of b1] (c1);
    \vertex[below = 0.2 cm of c1] (d1);
    \vertex[right = 1.4cm of d1] (d2);
    \vertex[right = 0.2 cm of c1] (c2);
    \vertex[right = 1cm of c2] (c3);
    \vertex[below = 0.5 cm of b1] (e1);
     \vertex[right = 1 cm of e1] (e2);
      \vertex[right = 0.2 cm of e2] (e3);
    \diagram* { 
    (a1) --  (b2),
    (b2) --  (c2),
    (d1)--(c2),
    (a2) -- [very thick](b3),
    (d2) -- [very thick](c3),
    (b2)-- (c3),
    (c2) -- (b3),
    (b3) -- [very thick](c3),
    };
    \end{feynman} \end{tikzpicture}}
    \label{eq:1_loop_grav_dec}
      \end{eqnarray}
The coefficients $c_{i}^{(g)}$, substituting $d=4-2\epsilon$, are given by:
\begin{eqnarray}
    c_{1}^{(g)}& =& \frac{1}{128\epsilon
 \left(\epsilon -1\right) \left(2 \epsilon -5\right) \left(2 \epsilon -3\right) \left(s-4 m^2\right)^2}2 m^4 \biggl(s^2 \bigl(288 \epsilon ^{12}-1728 \epsilon ^{11}+4024 \epsilon ^{10}\nonumber\\
 & & -7232 \epsilon ^9+17200 \epsilon ^8-31800 \epsilon ^7 +31817 \epsilon
 ^6-14893 \epsilon ^5+1632 \epsilon ^4+5651 \epsilon ^3-12737 \epsilon ^2\nonumber\\ 
 & &
 +10334 \epsilon 
 -2580\bigr)  -8 s t \bigl(132 \epsilon ^6-640 \epsilon ^5+1378 \epsilon
 ^4-2787 \epsilon ^3  +4275 \epsilon ^2-3015 \epsilon \nonumber \\
 & & +720\bigr)
 -8 t^2 \biggl(70 \epsilon ^6-330 \epsilon ^5 +650 \epsilon ^4-1237 \epsilon ^3+1969 \epsilon ^2-1455
 \epsilon +360\biggr)\biggr) \nonumber \\
 & & 
 -2 m^2 s \biggl(s^2 \bigl(144 \epsilon ^{12}-972 \epsilon ^{11}+2660 \epsilon ^{10}-5161 \epsilon ^9+11492 \epsilon ^8-22407 \epsilon
 ^7\nonumber \\
 & &
 +26492 \epsilon ^6-15530 \epsilon ^5+1473 \epsilon ^4+4100 \epsilon ^3-4281 \epsilon ^2+2650 -660\bigr)\epsilon\nonumber\\
 & & -2 s t \left(155 \epsilon ^6-791 \epsilon ^5+1916
 \epsilon ^4-4071 \epsilon ^3+5917 \epsilon ^2-3930 \epsilon +900\right)\nonumber \\
 & & -8 t^2 \biggl(31 \epsilon ^6-155 \epsilon ^5+364 \epsilon ^4-775 \epsilon ^3+1153 \epsilon
 ^2-780 \epsilon +180\biggr)\biggr)\nonumber \\
 & & 
 -16 m^6 \biggl(s \bigl(72 \epsilon ^{11}-432 \epsilon ^{10}+1030 \epsilon ^9-1928 \epsilon ^8+4338 \epsilon ^7-7079 \epsilon ^6+5503
 \epsilon ^5\nonumber \\
 & &
 -680 \epsilon ^4 -381 \epsilon ^3-2099 \epsilon ^2+2178 \epsilon -540\bigr)\nonumber \\
 & & -2 t \bigl(70 \epsilon ^6-330 \epsilon ^5+650 \epsilon ^4-1237 \epsilon ^3 +1969
 \epsilon ^2-1455 \epsilon +360\bigr)\biggr)-16 m^8 \biggl(70 \epsilon ^6\nonumber \\
 & & -330 \epsilon ^5+650 \epsilon ^4-1237 \epsilon ^3+1969 \epsilon ^2-1455 \epsilon
 +360\biggr)+s^2 \biggl(s^2 \bigl(36 \epsilon ^{12}-252 \epsilon ^{11}\nonumber \\
 & & +719 \epsilon ^{10}-1419 \epsilon ^9+3114 \epsilon ^8-6144 \epsilon ^7+7503 \epsilon ^6-4545
 \epsilon ^5+396 \epsilon ^4+1176 \epsilon ^3\nonumber\\
 & & -948 \epsilon ^2+484 \epsilon -120\bigr)-2 s t \biggl(31 \epsilon ^6-171 \epsilon ^5+460 \epsilon ^4-971 \epsilon ^3+1305
 \epsilon ^2\nonumber\\
 & & -810 \epsilon +180\biggr)-2 t^2 \bigl(31 \epsilon ^6-171 \epsilon ^5+460 \epsilon ^4-971 \epsilon ^3+1305 \epsilon ^2-810 \epsilon +180\bigr)\biggr) \ , \\
   c_{2}^{(g)} &=& \frac{1}{16 \left(\epsilon -1\right) \left(s-4 m^2\right)^2}\biggl(2 m^6 \biggl(s^2 \left(12 \epsilon ^5-12 \epsilon ^4-29 \epsilon ^3+42 \epsilon ^2-84 \epsilon +70\right)\nonumber \\
   & & +2 s t \left(\epsilon ^3-25 \epsilon ^2-41 \epsilon
 +60\right)-2 t^2 \left(15 \epsilon ^2+\epsilon -15\right)\biggr)+m^4 s \bigl(s^2 \biggl(-3 \epsilon ^5\nonumber \\
 & & +3 \epsilon ^4+13 \epsilon ^3-22 \epsilon ^2+72 \epsilon
 -63\biggr)-2 s t \left(\epsilon ^3-10 \epsilon ^2-72 \epsilon +75\right)\nonumber \\
 & & -2 t^2 \left(\epsilon ^3-10 \epsilon ^2-40 \epsilon +45\right)\bigr)+m^2 s^2
 \biggl(-\left(s^2 \left(\epsilon ^3-2 \epsilon ^2+14 \epsilon -13\right)\right)\nonumber \\
 & & +s t \left(36-38 \epsilon \right)+2 t^2 \left(15-16 \epsilon \right)\biggr)-2 m^8
 \biggl(s \left(24 \epsilon ^5-24 \epsilon ^4-43 \epsilon ^3+46 \epsilon ^2-80 \epsilon +73\right)\nonumber \\
 & & -4 t \left(15 \epsilon ^2+\epsilon -15\right)\biggr)-4 m^{10}
 \left(15 \epsilon ^2+\epsilon -15\right)+s^3 \left(\epsilon -1\right) \left(s^2+3 s t+3 t^2\right)\biggr) \ , \\ 
  c_{3}^{(g)}&=& \frac{1}{16}(m^2-t)^4 \ , \\
  c_{4}^{(g)}&=& \frac{1}{16}(m^2-u)^4 \ .  
\end{eqnarray}
The explicit expressions for the Master Integrals used for this work have been obtained using \texttt{Package-X}, and are given by:
\begin{eqnarray}
      I_{1}&=& N_G\biggl(\frac{-\mu^2}{s}\biggr)^{\epsilon}\frac{1}{\epsilon(1-2\epsilon)}, \\
     I_{2}&=& N_G\frac{1}{s\beta}\biggl[\frac{4\pi^2}{6}+2Li_2\biggl(\frac{\beta-1}{\beta+1}\biggr)+\frac{1}{2}\log^2\biggl(\frac{\beta-1}{\beta+1}\biggr)+\mathcal{O}[\epsilon]\biggr]\\
    I_{3} &=& -N_G\frac{1}{s(m^2-t)}\biggl(\frac{\mu^2}{m^2}\biggr)^{\epsilon}\biggl[\frac{2}{\epsilon^2}-\frac{1}{\epsilon}\biggl(2\log\biggl(\frac{m^2-t}{m^2}\biggr)+\log\biggl(\frac{-s}{m^2}\biggr)\biggr)\nonumber\\
    & & +2\log\biggl(\frac{m^2-t}{m^2}\biggr)\log\biggl(\frac{-s}{m^2}\biggr)-\frac{\pi^2}{2}+\mathcal{O}[\epsilon]\biggr] \\
     I_{4} &=&-N_G\frac{1}{s(m^2-u)}\biggl(\frac{\mu^2}{m^2}\biggr)^{\epsilon}\biggl[\frac{2}{\epsilon^2}-\frac{1}{\epsilon}\biggl(2\log\biggl(\frac{m^2-u}{m^2}\biggr)+\log\biggl(\frac{-s}{m^2}\biggr)\biggr)\nonumber\\
     & & +2\log\biggl(\frac{m^2-u}{m^2}\biggr)\log\biggl(\frac{-s}{m^2}\biggr)-\frac{\pi^2}{2}+\mathcal{O}[\epsilon]\biggr]
\end{eqnarray}
where 
\begin{equation}
 N_G=\frac{i}{(4\pi)^{d/2}}r_G\qquad r_G= \frac{(\Gamma[1-\epsilon])^2\Gamma[1+\epsilon]}{\Gamma[1-2\epsilon]} \qquad \beta^2=1-\frac{4m^2}{s}
\end{equation}
It is worth noticing that in order to get the correct result, compatible with our normalization of the loop integrals, it was necessary to add a normalization factor  $N_G $ in front of each expression given by \texttt{PackageX}. We can express the result, substituting the explicit expression for the master integrals and expanding in powers of $\epsilon$ up to order $\mathcal{O}[\epsilon^0]$, as:
\begin{eqnarray}
 \kappa^{-4}\mathcal{C}^{(g)}=\sum_{i=-2}^{0}\epsilon^i\mathcal{C}^{(g)}_i
\end{eqnarray}
where:
\begin{eqnarray}
 \mathcal{C}_{-2}^{(g)} &=& \ -\frac{i \left(-3 m^2 (s+2 t)+3 m^4+s^2+3 s t+3 t^2\right)}{256 \pi ^2}, \\ 
 \mathcal{C}_{-1}^{(g)} &=& \ -\frac{i}{7680 \pi ^2} \biggl(\frac{30}{s} \biggl(-s \left(-3 m^2 (s+2 t)+3 m^4+s^2+3 s t+3 t^2\right) \log \left(-\frac{\bar{\mu}^2}{s}\right)\nonumber\\
 & & +s
   \left(-3 m^2 (s+2 t)+3 m^4+s^2+3 s t+3 t^2\right) \log \left(-\frac{m^2}{s}\right)\nonumber \\
   & & +2 \left(m^2-u\right)^3 \log \left(\frac{\bar{\mu}^2}{m^2-u}\right)+2 \left(m^2-t\right)^3 \log \left(\frac{\bar{\mu}^2}{m^2-t}\right)\biggr)\nonumber \\
   & & +\frac{1}{\left(s-4
   m^2\right)^2} \bigl(m^4 \left(-79 s^2+348 s
   t-t^2\right)+m^2 s \left(17 s^2-270 s t-192 t^2\right) \nonumber\\ 
   & & +72 m^6 (s+t)-36 m^8+s^2 \left(-s^2+39 s t+39 t^2\right)\bigr)\biggr), \\ 
 \mathcal{C}_{0}^{(g)} &=& 
 \ \frac{i}{230400 \pi ^2 s} \biggl(\frac{1}{\sqrt{1-\frac{4 m^2}{s}} \left(s-4 m^2\right)^2}\Biggl(150 \biggl(-20 m^6 \left(7 s^2+12 s t+3 t^2\right)\nonumber \\
 & & 
 +3 m^4 s \left(21 s^2+50 s t+30 t^2\right)
 -m^2 s^2 \left(13 s^2+36 s
   t+30 t^2\right)+2 m^8 (73 s+60 t)\nonumber \\
   & & 
   -60 m^{10}+s^3 \left(s^2+3 s t+3 t^2\right)\biggr) \biggl(12 \text{Li}_2\left(\frac{\left(\sqrt{1-\frac{4
   m^2}{s}}-1\right) s}{2 m^2}+1\right)\nonumber\\
   & & +3 \log ^2\left(\frac{s \left(\sqrt{1-\frac{4 m^2}{s}}-1\right)}{2 m^2}+1\right)+4 \pi
   ^2\biggr)\Biggr)+s \biggl(-3 m^2 (s+2 t)+3 m^4+s^2\nonumber\\
   & & 
   +3 s t+3 t^2\biggr) \biggl(450 \log
   ^2\left(-\frac{\bar{\mu}^2}{s}\right)+4560 \log \left(-\frac{\bar{\mu}^2}{s}\right)-75 \pi ^2+14524\biggr)\nonumber \\
   & &
   -\frac{1}{\left(s-4 m^2\right)^2}\biggl(2 s \biggl(m^4
   \left(6457 s^2+14940 s t+7260 t^2\right)-5 m^2 s \left(331 s^2+966 s t+768 t^2\right)\nonumber \\
   & & 
   -24 m^6 (453 s+605 t)+7260 m^8+s^2 \bigl(151 s^2+495 s
   t +495 t^2\bigr)\biggr) \biggl(15 \log \left(-\frac{\bar{\mu}^2}{s}\right)\nonumber \\
   & & 
   +76\biggr)\biggr)+\frac{15 s}{\left(s-4 m^2\right)^2} \biggl(m^4
   \left(23747 s^2+56768 s t+26736 t^2\right)-m^2 s \biggl(6109 s^2\nonumber \\
   & & 
   +19014 s t+15016 t^2\biggr)-13368 m^6 (3 s+4 t)+26736 m^8+s^2 \biggl(560 s^2+1999
   s t\nonumber\\
   & & 
   +1999 t^2\biggr)\biggr)+300 \left(m^2-t\right)^3 \biggl(-3 \log \left(-\frac{m^2}{s}\right) \left(\log
   \left(\frac{\bar{\mu}^2}{m^2-t}\right)+\log \left(\frac{m^2}{m^2-t}\right)\right) \nonumber \\
   & & 
   -3 \log \left(\frac{\bar{\mu}^2}{m^2}\right)
   \log \left(\frac{m^2}{m^2-t}\right)-3 \log \left(\frac{\bar{\mu}^2}{m^2}\right) \log \left(\frac{\bar{\mu}^2}{m^2-t}\right)+2 \pi
   ^2\biggr) \nonumber \\
   & & +300 \left(m^2-u\right)^3 \biggl(-3 \log \left(\frac{\bar{\mu}^2}{m^2}\right) \log \left(\frac{m^2}{m^2-u}\right)-3 \log
   \left(\frac{\bar{\mu}^2}{m^2}\right) \log \biggl(\frac{\bar{\mu}^2}{m^2-u}\biggr)\nonumber \\
   & &-3 \log \left(-\frac{m^2}{s}\right) \left(\log
   \left(\frac{\bar{\mu}^2}{m^2-u}\right)+\log \left(\frac{m^2}{m^2-u}\right)\right)+2 \pi ^2\biggr)\biggr). 
\end{eqnarray}
We defined ${\bar{\mu}}^2=4\pi e^{-\gamma_E} \mu^2$, with $\gamma_E$ being the Euler-Mascheroni constant.
\subsection{Massless scalar double-cut}
\begin{figure}[H]
\centering
\scalebox{0.45}{\begin{tikzpicture} 
\begin{feynman}
\vertex (a1) ;
\vertex[right=1cm of a1] (a2);
\vertex[right=1cm of a2] (a3); 
\vertex[right=1cm of a3] (a4);
\vertex[right=1cm of a4] (a5);
\vertex[below= 1cm of a1] (b1);
\vertex[right= 1cm of b1] (b2);
\vertex[right= 2cm of b1] (b3);
\vertex[right= 1cm of b3] (b4);
\vertex[right= 2cm of b3] (b5);
\vertex[below= 2cm of b1] (d1); 
\vertex[right= 2cm of d1] (d3);
\vertex[right= 1cm of d1] (d2);
\vertex[right= 1cm of d2] (d3);
\vertex[right= 1cm of d3] (d4);
\vertex[below= 2cm of b5] (d5); 
\vertex[below=2cm of a1] (c1); 
\vertex[right=1cm of c1] (c2); 
\vertex[right=1cm of c2] (c3); 
\vertex[right = 2cm of c2] (c4);
\vertex[right=1 of c4] (c5) ;
\vertex[below=4cm of a1] (e1); 
\vertex[right=2cm of e1] (e3)  {\((l)\)}; 
\vertex[right=4cm of e1] (e5) ;

\diagram* { 
(a1) -- [charged scalar] (b2),
(b2) -- [gluon] (d2),
(e1) -- [charged scalar] (d2),
(b2) -- [scalar] (c3),
(d2) -- [scalar] (c3),
(c3) -- [gluon] (c4),
(b5) -- [fermion] (c4),
(d5) -- [fermion] (c4),
};
\end{feynman} 
\end{tikzpicture}} \qquad 
\scalebox{0.45}{\begin{tikzpicture} 
\begin{feynman}
\vertex (a1) ;
\vertex[right=1cm of a1] (a2);
\vertex[right=1cm of a2] (a3); 
\vertex[right=1cm of a3] (a4);
\vertex[right=1cm of a4] (a5);
\vertex[below= 1cm of a1] (b1);
\vertex[right= 1cm of b1] (b2);
\vertex[right= 2cm of b1] (b3);
\vertex[right= 1cm of b3] (b4);
\vertex[right= 2cm of b3] (b5);
\vertex[below= 2cm of b1] (d1); 
\vertex[right= 2cm of d1] (d3);
\vertex[right= 1cm of d1] (d2);
\vertex[right= 1cm of d2] (d3);
\vertex[right= 1cm of d3] (d4);
\vertex[below= 2cm of b5] (d5); 
\vertex[below=2cm of a1] (c1); 
\vertex[right=1cm of c1] (c2); 
\vertex[right = 2cm of c2] (c4);
\vertex[right=1 of c4] (c5) ;
\vertex[below=4cm of a1] (e1); 
\vertex[right=2cm of e1] (e3)  {\((m)\)}; 
\vertex[right=4cm of e1] (e5) ;
\diagram* { 
(b1) -- [charged scalar] (c2),
(d1) -- [charged scalar] (c2),
(c2) -- [gluon] (c6),
(c6) -- [scalar, half left] (c7),
(c6) -- [scalar, half right] (c7),
(c7) -- [gluon] (c4),
(b5) -- [fermion] (c4),
(d5) -- [fermion] (c4),
};
\end{feynman} 
\end{tikzpicture}}
    \caption{One-loop Feynman diagrams contributing to the process in the s-channel massless scalar double-cut. (set 2)}
    \label{fig3}
\end{figure}
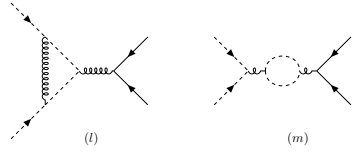

The diagrams needed to evaluate $\mathcal{C}^{(\phi)}$ are given in Fig.\ref{fig3} \\
\noindent As done before we can evaluate the diagrams and express the amplitude $\mathcal{C}^{(\phi)}$ as a linear combination of Master Integrals:
\begin{equation}
    \kappa^{-4}\mathcal{C}^{(\phi)}= c_{1}^{(\phi)} \ \scalebox{0.5}{\begin{tikzpicture}[baseline=(current bounding box.center)]\begin{feynman}
    \vertex (a) ;
    \vertex[right=0.2cm of a] (b);
    \vertex[right=1cm of b] (c); 
    \vertex[right=0.2cm of c] (d);
    \diagram* { 
    (a) --  (b),
    (b) -- [half left ] (c),
    (b) -- [half right ] (c),
    (c) -- (d),
    };
    \end{feynman}
    \end{tikzpicture}}
\end{equation}
where the coefficients are given by:
\begin{equation}
\begin{split}
c_{1}^{(\phi)}=&-\frac{3}{128\epsilon \left(\epsilon -1\right) \left(2 \epsilon -5\right) \left(2 \epsilon -3\right)}\biggl(m^2 \biggl(s \left(6 \epsilon ^3-36 \epsilon ^2+70 \epsilon -40\right)\\
&-4 t \left(5 \epsilon ^2-21 \epsilon +20\right)\biggr)+2
   m^4 \left(5 \epsilon ^2-21 \epsilon +20\right)+\epsilon ^2 \left(-27 s^2+10 s t+10 t^2\right)\\
   &+14 \epsilon  \left(s^2-3 s t-3 t^2\right)-3 s^2
   \epsilon ^4+16 s^2 \epsilon ^3+40 t (s+t)\biggr)
\end{split}
\end{equation}
As done in the previous case, we can express the result, substituting the explicit expression for the master integrals and expanding in powers of $\epsilon$ up to order $\mathcal{O}[\epsilon^0]$, as:
\begin{eqnarray}
 \kappa^{-4}\mathcal{C}^{(\phi)}=\sum_{i=-2}^{0}\epsilon^i\mathcal{C}^{(\phi)}_i
\end{eqnarray}
where:
\begin{eqnarray}
 \mathcal{C}_{-2}^{(\phi)} &=& \ \frac{i \left(m^2-t\right) \left(u-m^2\right)}{256 \pi ^2}, \\ 
 \mathcal{C}_{-1}^{(\phi)} &=& \ \frac{i}{15360 \pi ^2} \biggl(60 \left(m^2-t\right) \left(u-m^2\right) \log \left(-\frac{\bar{\mu}^2}{s}\right)-m^2 (139 s+362 t)+181 m^4+21 s^2\nonumber\\ 
 & &+181 s
   t+181 t^2\biggr),\\
 \mathcal{C}_{0}^{(\phi)} &=& \ -\frac{i}{460800 \pi ^2} \biggl(30 \log \left(-\frac{\bar{\mu}^2}{s}\right) \biggl(-30 \left(m^2-t\right) \left(u-m^2\right) \log \left(-\frac{\bar{\mu}^2}{s}\right)\nonumber\\ 
   & &+m^2 (139 s+362 t) -181 m^4-21 s^2-181 s t-181 t^2\biggr)\nonumber \\
   & & -2 \left(75 \pi ^2-6346\right) t \left(2 m^2-s\right)
   -150 \pi ^2 m^2
   s+8738 m^2 s+150 \pi ^2 m^4\nonumber \\
   & & -12692 m^4-1347 s^2+2 \left(75 \pi ^2-6346\right) t^2\biggr).
\end{eqnarray}

\subsection{Total contribution}
The total 1-loop amplitude in the massless double-cut can be obtained by summing the two contributions, namely:
\begin{eqnarray}
 \mathcal{C}^{(1)}=\mathcal{C}^{(g)}+\mathcal{C}^{(\phi)}
\end{eqnarray}
We can express the result, substituting the explicit expression for the master integrals and expanding in powers of $\epsilon$ up to order $\mathcal{O}[\epsilon^0]$, as:
\begin{eqnarray}
 \kappa^{-4}\mathcal{C}^{(1)}=\sum_{i=-2}^{0}\epsilon^i\mathcal{C}^{(1)}_i
\end{eqnarray}
where:
\begin{eqnarray}
 \mathcal{C}_{-2}^{(1)} &=& \ -\frac{i \left(-2 m^2 s-4 m^2 t+2 m^4+s^2+2 s t+2 t^2\right)}{256 \pi ^2}, \\ 
 \mathcal{C}_{-1}^{(1)} &=& \ \frac{i}{15360 \pi ^2} \biggl(\frac{1}{\left(s-4 m^2\right)^2}\biggl(m^4 \left(1787 s^2+5096 s t+2968 t^2\right)-424 m^6 (9 s+14 t)\nonumber\\
 & & 
 -m^2 s (11 s+14 t) (31 s+76 t)+2968 m^8+s^2 \left(23
   s^2+103 s t+103 t^2\right)\biggr)\nonumber \\
   & & -60 \left(-3 m^2 (s+2 t)+3 m^4+s^2+3 s t+3 t^2\right) \log
   \left(-\frac{m^2}{s}\right) \nonumber \\
   & & 
   +\frac{60}{s} \biggl(s \left(-2 m^2+s+2 t\right)^2 \log \left(-\frac{\bar{\mu}^2}{s}\right)+2
   \left(u-m^2\right)^3 \log \left(\frac{\bar{\mu}^2}{m^2-u}\right)\nonumber\\
   & & 
   -2 \left(m^2-t\right)^3 \log \left(\frac{\bar{\mu}^2}{m^2-t}\right)\biggr)\biggr), \\ 
 \mathcal{C}_{0}^{(1)} &=& \ \frac{i}{460800 \pi ^2 s} \biggl(\frac{1}{\sqrt{1-\frac{4 m^2}{s}} \left(s-4 m^2\right)^2}\biggl(300 \biggl(-20 m^6 \left(7 s^2+12 s t+3 t^2\right)\nonumber \\
 & & +3 m^4 s \left(21
   s^2+50 s t+30 t^2\right)-m^2 s^2 \left(13 s^2+36 s t+30 t^2\right)+2 m^8 (73 s+60
   t)\nonumber \\
   & & 
   -60 m^{10}+s^3 \left(s^2+3 s t+3 t^2\right) \Biggl(12
   \text{Li}_2\left(\frac{\left(\sqrt{1-\frac{4 m^2}{s}}-1\right) s}{2 m^2}+1\right)\nonumber \\
   & & 
   +3\log ^2\left(\frac{s \left(\sqrt{1-\frac{4 m^2}{s}}-1\right)}{2 m^2}+1\right)+4 \pi
   ^2\Biggr)\biggr)\nonumber\\
   & & -\frac{1}{\left(s-4 m^2\right)^2}\biggl(2 s \biggl(m^4
   \left(14021 s^2+33816 s t+16488 t^2\right)-m^2 s \biggl(3307 s^2\nonumber \\
   & & +10890 s t+8664
   t^2\biggr)-24 m^6 (1057 s+1374 t)+16488 m^8+s^2 \biggl(281 s^2+1113 s t\nonumber \\
   & & +1113
   t^2\biggr)\biggr) \left(15 \log \left(-\frac{\bar{\mu}^2}{s}\right)+76\right)\biggr) +\frac{1}{\left(s-4 m^2\right)^2}\biggl(15 s \biggl(2 m^4
   \biggl(24113 s^2\nonumber \\
   & & 
   +59264 s t+27984 t^2\biggr)-4 m^2 s \left(2888 s^2+9897 s t+7820
   t^2\right)-48 m^6 (1803 s\nonumber \\
   & & +2332 t)+55968 m^8+s^2 \left(997 s^2+4154 s t+4154
   t^2\right)\biggr)\biggr)\nonumber \\
   & & +600 \left(m^2-t\right)^3 \biggl(2 \pi ^2-3
   \biggl(\log \left(\frac{\bar{\mu}^2}{m^2}\right) +\log
   \left(-\frac{m^2}{s}\right)\biggr) \biggl(\log \left(\frac{\bar{\mu}^2}{m^2-t}\right)\nonumber \\
   & & 
   +\log \left(\frac{m^2}{m^2-t}\right)\biggr)\biggr)+2 s \left(-2
   m^2+s+2 t\right)^2 \biggl(30 \log \left(-\frac{\bar{\mu}^2}{s}\right) \biggl(15
   \log \left(-\frac{\bar{\mu}^2}{s}\right)\nonumber \\
   & & +152\biggr) -75 \pi ^2+14524\biggr)+600
   \left(m^2-u\right)^3 \biggl(2 \pi ^2-3 \biggl(\log \left(\frac{\bar{\mu}^2}{m^2}\right)\nonumber \\
   & & +\log \left(-\frac{m^2}{s}\right)\biggr) \biggl(\log
   \left(\frac{\bar{\mu}^2}{m^2-u}\right)+\log
   \left(\frac{m^2}{m^2-u}\right)\biggr)\biggr)\biggr).  
\end{eqnarray}

\section{Low-energy limit}

To obtain an estimate of the bending angle we have to study the low-energy non-relativistic limit of this amplitude, where the energy $E$ of the massless particle is much smaller than the mass of the massive scalar $E\ll m$ and the momentum transfer $s\approx -\mathbf{q}^2$ is tiny as well. \\ 
The Mandelstam parameters in this specific kinematic limit become:
\begin{equation}
    s=(p_1+p_2)^2 \approx -\mathbf{q^2} \quad t=(p_1+p_3)^2\approx m^2+2E m 
\end{equation}
Moreover we will be interested in the small angle scattering approximation: $\mathbf{q^2} \ll E^2$. \\
Our problem is characterized by three different scales, and we must be very careful in taking the series expansion:
\begin{equation}
    \mathbf{q}^2\ll E^2 \ll m^2
\end{equation}  
In order to obtain the low energy limit, we introduce two a-dimensional variables $\alpha,\tau$, defined as:
\begin{equation}
    \alpha= \frac{E}{m} \qquad \tau= \frac{\mathbf{q^2}}{m^2}
 \end{equation}
Then we perform multiple series expansions in the following order: 
\begin{enumerate}
    \item expansion around $d=4$ dimensions, namely around $\epsilon=0$, up to the finite term, ${\cal O}(\epsilon^0)$;
    \item expansion around $\tau = 0$ up to ${\cal O}(\tau^0)$, for isolating the leading non-analytic terms;
    \item expansion around $ \alpha=0$, up to
    ${\cal O}(\alpha^3)$. 
\end{enumerate}
    \subsection{Graviton double-cut}
    We can express the 1-loop amplitude in the graviton double-cut $\mathcal{C}^{(g)}$ in the low-energy limit in powers of $\epsilon$ as:
    \begin{eqnarray}
    \kappa^{-4}\mathcal{C}^{(g)}=\sum_{i=-2}^{0}\epsilon^i\mathcal{C}^{(g)}_{i}
    \end{eqnarray}
    where 
    \begin{eqnarray}
    \mathcal{C}^{(g)}_{-2} & = & \ -\frac{3 i E^2 m^2}{ 64 \pi ^2}, \\
    \mathcal{C}^{(g)}_{-1} & = & -\frac{3 i E^2 m^2 }{64 \pi ^2}\log \left(\frac{\bar{\mu}^2}{4
       E^2}\right)-\frac{E^3 m^3}{16 \pi  s}+\frac{83 i E^2 m^2}{2560 \pi ^2} \ , \\
    \mathcal{C}^{(g)}_{0}& =& \mathcal{C}^{(g)}_{0 \ a}
    +\mathcal{C}^{(g)}_{0  \ n.a.} 
    \end{eqnarray}
    We wrote $\mathcal{C}^{(g)}_{0}$ as the sum of a part made by analytic terms $\mathcal{C}^{(g)}_{0 \ a}$ and a part containing non-analytic contributions 
    $\mathcal{C}^{(g)}_{0  \ n.a.}$, given by:
    \begin{eqnarray}
    \mathcal{C}^{(g)}_{0 \ a} & = & \  -\frac{3 i E^2
       m^2 }{64 \pi ^2}\log \left(\frac{\bar{\mu}^2}{4 E^2}\right) \log \left(\frac{\bar{\mu}^2}{m^2}\right)+\frac{i E^2 m^2
       }{32 \pi ^2}\log \left(\frac{\bar{\mu}^2}{m^2}\right)\nonumber\\
       & &-\frac{E^3 m^3 }{16 \pi  s}\log \left(\frac{\bar{\mu}^2}{m^2}\right) + \frac{7}{256} i E^2 m^2-\frac{103 i E^2 m^2}{1600 \pi ^2} \\
        \mathcal{C}^{(g)}_{0 \ n.a.} &=& 
       \frac{15 i E^2 m^3}{512
       \sqrt{-s}} \nonumber \\
        & & 
        \ +\frac{3 i E^2 m^2}{128 \pi
       ^2} \log ^2\left(-\frac{\bar{\mu}^2}{s}\right)+\frac{3 i E^2 m^2 }{2560 \pi ^2}\log \left(-\frac{\bar{\mu}^2}{s}\right)\nonumber\\
       & & 
       +\log \left(-\frac{s}{m^2}\right) \left(\frac{3 i E^2 m^2 }{64 \pi ^2}\log \left(\frac{\bar{\mu}^2}{4
       E^2}\right)+\frac{E^3 m^3}{16 \pi  s}-\frac{i E^2 m^2}{512 \pi ^2}\right) \ .
    \label{eq:M10_non_analytic}
    \end{eqnarray}
    We can try to give an interpretation to the non-analytic components appearing in \eqref{eq:M10_non_analytic}: 
    \begin{itemize}
        \item the first term represent a classical term, which is the first post-Newtonian correction;
        \item The first logarithmic contribution, as well as the other are quantum gravity correction,
        \item The second Logarithmic contribution arises from the one-loop ultraviolet divergence of the amplitude 
        \item The first term in the last line is new, the second one contributes to the phase of the amplitude, and will not be considered in the following. The last term is a quantum correction of the metric.
    \end{itemize}
    This non-analytic  contribution $\mathcal{C}^{(g)}_{0\ n.a.}$ was earlier considered in
    \cite{Bjerrum-Bohr:2014zsa,Bjerrum-Bohr:2016hpa}. We observe that the coefficient of  ${\rm log}(-s/m^2)$, appearing in the last line of Eq.(\ref{eq:M10_non_analytic}),
    is different from Eq.(11) of  \cite{Bjerrum-Bohr:2014zsa}, and Eq.(4.9) of \cite{Bjerrum-Bohr:2016hpa}:
    in particular, the first term 
    proportional to ${\rm log}({\bar \mu}^2/(4E^2))$
    is new; and the numerical factors of the last two terms, respectively proportional to $E^3 m^2/s$ and $E^2m^2$, are not the same. 
    More details on the source of the difference will be provided in Sec.  \ref{sec:gravitoncutcheck}. \\ 
    In the following, we will be interested only in the non-analytic part $\mathcal{C}^{(g)}_{0\ n.a.}$ as it is the one that gives rise to long-range contributions once we take the Fourier transform of the amplitude.

    \subsection{Massless scalar double-cut}
    
    As done for the graviton double-cut case, we can expand the amplitude $\mathcal{C}^{(\phi)}$ in the low-energy non-relativistic limit and express it in powers of $\epsilon$ as:
    \begin{eqnarray}
    \kappa^{-4}\mathcal{C}^{(\phi)}=\sum_{i=-2}^{0}\epsilon^i\mathcal{C}^{(\phi)}_{i}
    \end{eqnarray}
    where 
    \begin{eqnarray}
    \mathcal{C}^{(\phi)}_{-2} & = & \ \frac{i E^2 m^2}{64 \pi ^2}, \\
    \mathcal{C}^{(\phi)}_{-1} & = & \ \frac{i E^2 m^2 }{64 \pi ^2}\log \left(-\frac{\bar{\mu}^2}{s}\right)+\frac{181 i E^2 m^2}{3840 \pi ^2}, \label{eq:massless_scalar_Eps}\\
    \mathcal{C}^{(\phi)}_{0} & = & \ -\frac{i E^2 m^2}{57600 \pi ^2} \left(-450 \log ^2\left(-\frac{\bar{\mu}^2}{s}\right)-2715 \log \left(-\frac{\bar{\mu}^2}{s}\right)+75
       \pi ^2-6346\right), \\
    \nonumber
    \end{eqnarray}
    We can notice that there is a problem arising in the massless scalar double cut: in Eq.\eqref{eq:massless_scalar_Eps} one can notice that there is a non-analytic term proportional to $1/\epsilon$, which will give a divergent contribution that do not vanish once we take the Fourier transform. This term was not noticed in \cite{Bai:2016ivl}, but it is mandatory to deal with it because it is affecting a physical quantity. Since this result needs a renormalization procedure and a more accurate analysis, we will not use it to predict physical observables, but this problem need to be sorted out in future works.
    
    \subsection{Total contribution in the Low-Energy limit}
    
    For further analysis, we report here the total amplitude $\mathcal{C}^{(1)}$ in the non-relativistic limit given as sum of the contributions coming from the two sets and expressed in powers of $\epsilon$ as:
    \begin{eqnarray}
     \kappa^{-4}\mathcal{C}^{(1)}= \sum_{i=-2}^{0}\epsilon^i\mathcal{C}^{(1)}_{i}
    \end{eqnarray}
    where:
    \begin{eqnarray}
     \mathcal{C}^{(1)}_{-2} & = & \ -\frac{i E^2 m^2}{32 \pi ^2}, \\
    \mathcal{C}^{(1)}_{-1} & = & \ -\frac{3 i E^2 m^2 }{64 \pi ^2}\log \left(\frac{\bar{\mu}^2}{4 E^2}\right)+\frac{i E^2 m^2 }{64 \pi ^2}\log
       \left(-\frac{\bar{\mu}^2}{s}\right)  \nonumber\\
       & & -\frac{E^3 m^3}{16 \pi  s}+\frac{611 i E^2 m^2}{7680 \pi
       ^2} , \\
    \mathcal{C}^{(1)}_{0}&=& \mathcal{C}^{(1)}_{0 \ a}+\mathcal{C}^{(1)}_{0  \ n.a.}    \\
    \end{eqnarray}
    where we wrote $\mathcal{C}^{(1)}_{0}$ as the sum of an analytic part $\mathcal{C}^{(1)}_{0 \ a }$ and of a non-analytic part $\mathcal{C}^{(1)}_{0 \ n.a. }$ which are given by:
    \begin{eqnarray}
    \mathcal{C}^{(1)}_{0 \ a} & = & \ -\frac{3 i E^2 m^2 }{64 \pi ^2}\log \left(\frac{\bar{\mu}^2}{4 E^2}\right) \log \left(\frac{\bar{\mu}^2}{m^2}\right)+\frac{i E^2 m^2 }{32 \pi ^2}\log \left(\frac{\bar{\mu}^2}{m^2}\right)\nonumber\\
    & &  -\frac{E^3 m^3 }{16 \pi  s}\log \left(\frac{\bar{\mu}^2}{m^2}\right)+\frac{5}{192} i E^2 m^2+\frac{1319 i E^2 m^2}{28800 \pi ^2}, \\
        \mathcal{C}^{(1)}_{0 \
    n.a.}& = &  +\frac{i
       E^2 m^2 }{32 \pi ^2}\log ^2\left(-\frac{\bar{\mu}^2}{s}\right) +\frac{371 i E^2 m^2 }{7680 \pi ^2}\log \left(-\frac{\bar{\mu}^2}{s}\right) \nonumber\\
       & &+\log
       \left(-\frac{s}{m^2}\right) \left(\frac{3 i E^2 m^2 }{64 \pi
       ^2}\log \left(\frac{\bar{\mu}^2}{4 E^2}\right)+\frac{E^3 m^3}{16 \pi  s}-\frac{i E^2 m^2}{512 \pi ^2}\right)\nonumber\\
       & &  +\frac{15 i E^2 m^3}{512
       \sqrt{-s}} 
       \label{eq:C10notan}
    \end{eqnarray}

    \section{Discussion and Checks}

    \subsection{Graviton double-cut in the Low-Energy Limit}
    \label{sec:gravitoncutcheck}
    
    The $\mathcal{C}^{(g)}_{0\ n.a.}$ reported in eq.\eqref{eq:M10_non_analytic} was earlier computed in Eq.(11) of \cite{Bjerrum-Bohr:2014zsa} and in Eq.(4.9) of \cite{Bjerrum-Bohr:2016hpa}. In order to understand why our result is different with respect to those, let us perform a symbolic analysis of the 1-loop amplitude, in order to understand which terms are contributing.
    Within our calculation, we can make series expansions of coefficients
    and MIs in the three parameters: $\epsilon,s,E$:
    \begin{eqnarray}
    c^{(g)}_{i} &=& \sum_{j,k,\ell} c^{(g)}_{i,j,k,l} \, \epsilon^j s^k E^\ell \ , \\
    I_i &=& \sum_{j,k,\ell} \mathcal{I}_{i,j,k,l} \, \epsilon^j s^k E^\ell \ , 
    \end{eqnarray}
    the amplitude reported in eq.(\ref{eq:ampdecomis}) admits the following decomposition:
    \begin{eqnarray}
    \mathcal{C}^{(g)} &=& \sum_{j,k,\ell} \, \mathcal{C}^{(g)}_{j,k,l} \, \, \epsilon^j s^k E^\ell \ .
    \end{eqnarray}
   One can observe that the contribution to the term proportional to $E^2 \log(-s/m^2)$, comes from the combination of various terms that scale as $\epsilon^0s^0E^2$, and is given by (dropping the vanishing terms):
    \begin{eqnarray} 
    \mathcal{C}^{(g)}_{0,0,2}= c^{(g)}_{2,0,0,2} \, \mathcal{I}_{2,0,0,0} + c^{(g)}_{4,0,0,4} \, \mathcal{I}_{4,0,0,-2} +c^{(g)}_{4,0,1,3} \, \mathcal{I}_{4,0,-1,-1} 
    \end{eqnarray}
    where
    \begin{eqnarray} 
    c^{(g)}_{2,0,0,2} &=& \ -\frac{15}{16}m^2, \\
    c^{(g)}_{4,0,0,4} &=& \ m^4, \\
    c^{(g)}_{4,0,1,3} &=& \ 2m^3, \\
    \mathcal{I}_{2,0,0,0} &=& \ \frac{i}{16 \pi ^2 m^2}-\frac{i }{32 \pi ^2 m^2}\log \left(-\frac{s}{m^2}\right), \\
    \mathcal{I}_{4,0,0,-2} &=&\  +\log
       \left(-\frac{s}{m^2}\right) \left(-\frac{i }{64 \pi ^2 m^2}\log \left(\frac{\bar{\mu}^2}{4 E^2}\right)-\frac{i}{32 \pi ^2
       m^2}\right)\nonumber\\ 
    & &  +\frac{i }{32 \pi ^2 m^2}\log \left(\frac{\bar{\mu}^2}{m^2}\right)-\frac{i}{96 m^2}+ \frac{i }{64 \pi ^2 m^2}\log \left(\frac{\bar{\mu}^2}{4 E^2}\right) \log \left(\frac{\bar{\mu}^2}{m^2}\right), \\
    \mathcal{I}_{4,0,-1,-1} &=&\  +\frac{i}{32 \pi ^2 m} \log
       \left(\frac{\bar{\mu}^2}{4 E^2}\right) \log
       \left(-\frac{s}{m^2}\right)\nonumber\\
       & & -\frac{i}{32 \pi ^2 m} \log \left(\frac{\bar{\mu}^2}{4
       E^2}\right) \log \left(\frac{\bar{\mu}^2}{m^2}\right)+\frac{i}{48 m}\ .
    \end{eqnarray}

   \noindent Our novel result for $\mathcal{C}^{(g)}_{0 \ n.a.}$ completes the one available in the literature \cite{Bjerrum-Bohr:2014zsa,Bjerrum-Bohr:2016hpa}, whose coefficient $\mathcal{C}^{(g)}_{0,0,2}$ was determined by considering only the contribution of the term 
    $c^{(g)}_{2,0,0,2} \mathcal{I}_{2,0,0,0} $, coming from the 3-point function, and dropping the other two contributions $c^{(g)}_{4,0,0,4} \mathcal{I}_{4,0,0,-2} $, and $c^{(g)}_{4,0,1,3} \mathcal{I}_{4,0,-1,-1} $, which are instead coming from the crossed 4-point function. 
    In particular, our result for the term
    proportional to $\log\left(\frac{-s}{m^2}\right)$
    in $\mathcal{C}_{0,0,2}^{(g)}$, reads:
    \begin{eqnarray}
         \mathcal{C}_{0,0,2}^{(g)}
         \bigg|_{\log\left(\frac{-s}{m^2}\right)}&=&\frac{3 i E^2 m^2 }{64 \pi ^2}\log \left(\frac{\bar{\mu}^2}{4
       E^2}\right)+\frac{E^3 m^3}{16 \pi  s}-\frac{i E^2 m^2}{512 \pi ^2}
    \end{eqnarray}
    whereas in \cite{Bjerrum-Bohr:2014zsa,Bjerrum-Bohr:2016hpa} it reads:
    \begin{eqnarray}
         \mathcal{C}_{0,0,2}^{(g)} \bigg|_{\log\left(\frac{-s}{m^2}\right)}
         &=&+\frac{E^3 m^3}{8 \pi  s}+\frac{i 15  E^2 m^2}{512 \pi ^2} \ .
    \end{eqnarray}

    \subsection{Total double-cut in the Low-Energy Limit }
    The total double-cut 1-loop amplitude $\mathcal{C}^{(1)}_{0 \ n.a.}$ was earlier computed in \cite{Bai:2016ivl}, making use of the results of \cite{Bjerrum-Bohr:2014zsa,Bjerrum-Bohr:2016hpa}. 
    Therefore, also in this case, our result differs from the one presented in \cite{Bai:2016ivl}, because of the missing contributions coming from $c^{(g)}_{4,0,0,4} I_{4,0,0,-2}$ and $c^{(g)}_{4,0,1,3} \mathcal{I}_{4,0,-1,-1} $.
    After adding the latter to the expression of 
    $\mathcal{C}^{(1)}_{0 \ n.a.}$ in \cite{Bai:2016ivl}, the resulting combination agrees with our \eqref{eq:C10notan}, therefore implying that the contribution of the massless-scalar double-cut 
    computed in \cite{Bai:2016ivl} agrees with our result. However, as already mentioned above, in \cite{Bai:2016ivl} was not included the divergent non-analytic term which scales as $1/\epsilon$, and it needs to be treated appropriately with a renormalization procedure. \\ 
    


\section{Bending angle from a Scattering Amplitude Approach}

The amplitude in the non-relativistic low-energy limit can be used to estimate the bending angle of a light scalar field under the influence of a heavy scalar. In order to do that let us consider only the non-analytic part of the 1-loop amplitude in the graviton double-cut, where we subtract the complex contribution since it is not contributing to the scattering angle as pointed out in  \cite{Weinberg:1965nx,Donoghue:1996mt},  which in the small momentum transfer limit $s\approx -\mathbf{q}^2$ has the form: 
\begin{eqnarray}
    i\mathcal{M}^{(g)}(\mathbf{q})
    &=& 
    i\mathcal{M}^{(0)}+i\mathcal{C}^{(g)}_{0\ n.a.} +\frac{i \kappa^4E^3m^3}{16\pi}\frac{1}{\mathbf{q}^2}\log\biggl(\frac{\mathbf{q}^2}{m^2}\biggr) \ , \nonumber \\ 
    & = & 
    E^2 m^2\biggl[-\frac{\kappa^2}{\mathbf{q}^2}-\kappa^4\frac{15}{512\pi^2}\frac{m}{\vert\mathbf{q}\vert}\nonumber \\ 
    & &-\kappa^4\frac{3}{128\pi^2}\log^2\biggl(\frac{\mathbf{q}^2}{\Bar{\mu}^2}\biggr)+\kappa^4\frac{b_u}{(8\pi)^2}\log\biggl(\frac{\mathbf{q}^2}{\Bar{\mu}^2}\biggr)\nonumber \\ \
    & & 
    +\kappa^4\log\biggl(\frac{\mathbf{q}^2}{m^2}\biggr)\biggl(\frac{1}{512\pi^2}-\frac{3}{64\pi^2}\log\biggl(\frac{\Bar{\mu}^2}{4E^2}\biggr)\biggr)\biggr] \ ,
    \label{eq:non_analytic_grav}
\end{eqnarray}
where we defined the coefficient $b_u=3/40$ to compare our analysis with the ones appearing in literature. 
The bending angle can be obtained in two different equivalent ways, with a semi-classical approximation or following an Eikonal approach, and in the following we will compute it using both methods.

\subsection{Bending angle via semiclassical approximation}

 From  Eq.\eqref{eq:non_analytic_grav}, we can define a semiclassical potential for a massless scalar interacting with a massive scalar object by use of the Born approximation: 
\begin{eqnarray}
    V^{(g)} (r)
    & = & 
    \frac{1}{4 E m}\int\frac{d^3\mathbf{q}}{(2\pi)^3}e^{i\mathbf{q}\cdot \mathbf{r}} i\mathcal{M}^{(g)}(\mathbf{q}) \nonumber \\ 
    & = & 
    -\frac{2 G_N E m}{r}-\frac{15}{4}\frac{G_N^2 E m^2}{r^2} \nonumber \\ 
    & & 
    -2 b_u \frac{G_N^2 E m}{\pi r^3}-\frac{1}{4}\frac{G_N^2 E m}{\pi r^3}-12\frac{G_N^2 E m }{\pi r^3}\log\biggl(\frac{r}{r_0}\biggr)+6\frac{G_N^2 E m}{\pi r^3}\log\biggl(\frac{\Bar{\mu}^2}{4E^2}\biggr) \ , 
\end{eqnarray}
where $r_0$ is an infrared scale. To perform the Fourier transforms we used Eq.\eqref{eq:scalar_integral_fourier} and the following formulas, that can be found in \cite{Bai:2016ivl,Chi:2019owc}: 
\begin{eqnarray}
    \int\frac{d^3\mathbf{q}}{(2\pi)^3}e^{i\mathbf{q}\cdot \mathbf{r}}\log (\mathbf{q}^2)=-\frac{1}{2\pi r^3}\ ,  \qquad \int\frac{d^3\mathbf{q}}{(2\pi)^3}e^{i\mathbf{q}\cdot \mathbf{r}}\log^2 \biggl(\frac{\mathbf{q}^2}{\mu^2}\biggr)=\frac{2}{\pi r^3}\log\biggl(\frac{r}{r_0}\biggr) \ .
    \label{eq:fourier_logs} 
\end{eqnarray}
Using naively the semiclassical formula for angular deflection given in Eq. 12 of \cite{Bjerrum-Bohr:2016hpa}, we find the bending angle of a massless scalar: 
\begin{eqnarray}
    \theta^{(g)}
    &\approx&
     -\frac{b}{E}\int_{-\infty}^{+\infty}du \frac{V'(b\sqrt{1+u^2})}{\sqrt{1+u^2}} \nonumber \\ 
    &\approx & 
    \frac{4G_N m }{b} +\frac{15 G_N^2 m^2 \pi }{4 b^2}+\frac{G_N^2 m \hbar}{\pi b^3}\biggl(8b_u + 9 +\Delta + 48\log\biggl(\frac{b}{2b_0}\biggr) -24\log\biggl(\frac{\Bar{\mu}^2}{4E^2}\biggr)\biggr) \ , 
    \label{eq:bending_angle_semi}
\end{eqnarray}
where $\Delta=16$, $b$ is a gauge-invariant impact parameter, and $V'(r)=dV(r)/dr$. The first two terms give the correct classical values reported in Eq.\eqref{eq:bending_classical} whereas, the last term is a quantum gravity effect of order $G_N^2\hbar m/b^3= \ell_p^2 r_s/(2b^3)$. Our result is different from the one given in Eq.(12) of \cite{Bjerrum-Bohr:2014zsa}, and in Eq.(5.45) of \cite{Bjerrum-Bohr:2016hpa}, as expected from the analysis done in Sec.\ref{sec:gravitoncutcheck}, in particular: 
\begin{itemize}
    \item the $\Delta$ term, proportional to $G^2\hbar m/b^3$, is not present in \cite{Bjerrum-Bohr:2014zsa,Bjerrum-Bohr:2016hpa},
    \item There is a new term proportional to $\log\left(\frac{\Bar{\mu}^2}{4E^2}\right)$, which do not appear in previous computations.
    \item The term proportional to $\log(b/2b_0)$ presents a different sign with respect to the same one in \cite{Bjerrum-Bohr:2014zsa,Bjerrum-Bohr:2016hpa}, this point needs to be further clarified in future works.
\end{itemize}

\subsection{Bending angle via Eikonal approximation}
Another way to compute the bending angle is in the so-called \textit{Eikonal approximation}, a method that allow us to get classical observables directly from scattering amplitudes, without passing through intermediate, unphysical quantities. \\
As already mentioned before, in this approximation the momentum transfer $\vert \mathbf{q} \vert$ is taken to be much smaller than both the mass $m$ of the heavy scalar and the energy $E$ of the massless particle, more precisely: $\vert\mathbf{q}\vert\ll E \ll m$. Following \cite{Bjerrum-Bohr:2016hpa,Akhoury:2013yua}, in the small-angle scattering limit the dominant four momentum transfer is in the transverse spatial direction, and one can assume that the amplitude obtained in this limit $\mathcal{M(\mathbf{q})}$ can be transformed to impact parameter space via a two-dimensional Fourier transform with respect to the momentum transfer $\mathbf{q}$ defined as:
\begin{equation}
    i\tilde{\mathcal{M}}(\mathbf{b})=\frac{1}{4m E}\int \frac{d^{d-2}q}{(2\pi)^{d-2}}e^{i\mathbf{q}\cdot \mathbf{b}}i\mathcal{M}(\mathbf{q}) \ , 
\end{equation}
where $\mathbf{b}$ is the impact parameter.
As shown in \cite{Bjerrum-Bohr:2016hpa,Akhoury:2013yua}, in this space the 1 loop amplitude is expected to exponentiate into an eikonal phase, as:
\begin{eqnarray}
i\tilde{\mathcal{M}}(\mathbf{b})\approx 2(s-m^2)\biggl[e^{i(\chi_1+\chi_2)}-1\biggr]\  , 
\end{eqnarray}
where the two phases $\chi_1$ and $\chi_2$ are defined as:
\begin{eqnarray}
    \chi_1(\mathbf{b})& =& 
    \frac{1}{4 m E}\int \frac{d^2\mathbf{q}}{(2\pi)^2}e^{-i\mathbf{q}\cdot\mathbf{b}}i\mathcal{M}^{(0)}(\mathbf{q}) \ , \\
    \chi_2(\mathbf{b}) & = & \frac{1}{4 m E}\int \frac{d^2\mathbf{q}}{(2\pi)^2}e^{-i\mathbf{q}\cdot \mathbf{b}}i\mathcal{C}^{(g)}(\mathbf{q}) \ .
\end{eqnarray}
By substituting the explicit expressions for $\mathcal{M}^{(g)}(\mathbf{q})$ one gets: 
\begin{eqnarray}
    \chi_1 & = & G_N m E \biggl(\frac{1}{\epsilon}-\log(b)-\gamma_E-\frac{1}{2}\log(\pi)\biggr) \ , \\ 
    \chi_2 & = &  \frac{15 G_N^2m^2E \pi }{4 b}+\frac{G_N^2 m E}{2\pi b^2}\biggl(1+8b_u +48\log\biggl(\frac{2b_0}{b}\biggr)+24\log\biggl(\frac{\Bar{\mu}^2}{4E^2}\biggr)\biggr)\ . 
\end{eqnarray}
We want to point out that in \cite{Bjerrum-Bohr:2016hpa} there is a typo in Eq.(5.28): the second term in the sum should be proportional to $m$ and not to $m^2$. 
To perform the computation we made use of Eq.\eqref{eq:scalar_integral_fourier}, and of the following 2D Fourier integrals which can be found in \cite{Bjerrum-Bohr:2016hpa}: 
\begin{eqnarray}
    \int \frac{d^2q}{(2\pi)^2}e^{-i\mathbf{q}\cdot\mathbf{b}}\log \mathbf{q}^2  =  -\frac{1}{\pi b^2} \ ,\qquad 
    \int \frac{d^2q}{(2\pi)^2}e^{-i\mathbf{q}\cdot\mathbf{b}}\log^2 \mathbf{q}^2  = \frac{4}{\pi b^2}\log\frac{2}{b} \ . 
\end{eqnarray} 
One can get determine the deflection angle by evaluating the stationary phase of the exponent, which dominates the momentum space integration, as: 
\begin{eqnarray}
    \frac{\partial}{\partial b }(-\mathbf{q}\cdot\mathbf{b}+\chi_1(b)+\chi_2(b)+\ldots)=\frac{\partial}{\partial b }(qb+\chi_1(b)+\chi_2(b)+\ldots)=0 \ . 
\end{eqnarray}
Substituting $q=2Esin(\theta^{(g)}/2)$ we get:
\begin{eqnarray}
    2sin\frac{\theta^{(g)}}{2} \approx \theta^{(g)} = -\frac{1}{E}\frac{\partial}{\partial b}(\chi_1(b)+\chi_2(b)) \ . 
    \label{eq:defl_angle_1}
\end{eqnarray}
From \eqref{eq:defl_angle_1} we can substitute the explicit expressions for $\chi_1$ and $\chi_2$ to get the bending angle: 
\begin{eqnarray}
    \theta^{(g)} = \frac{G_Nm}{b}+\frac{15}{4}\frac{G_N^2m^2\pi}{b^2}+\biggl(8b_u+25-48\log\biggl(\frac{b}{2b_0}\biggr)-24\log\biggl(\frac{\Bar{\mu}^2}{4E^2}\biggr)\biggr)\frac{G_N^2 m}{\pi b^3} \ , 
    \label{eq:bending_angle_eikonal}
\end{eqnarray}
which does not agree with the result obtained in the previous subsection in Eq.\eqref{eq:bending_angle_semi}, in particular the term proportional to $\log(b/2b_0)$ presents a different sign with respect to the previous calculation. 
Since the two estimations of the bending angle $\theta^{(g)}$ (Eq.\eqref{eq:bending_angle_semi}\eqref{eq:bending_angle_eikonal}) apparently give different results, we cannot say that at quantum level the two methods are equivalent. These point needs to be further analyzed in future works.

\section{1-loop quantum corrections to the Newtonian potential}

We can easily adapt the analysis done so far to study the 1-loop quantum corrections to the Newtonian potential between two particles of masses $m_1$ and $m_2$. \\ 
Let us consider a 4-point process of the type:
\begin{figure}[H]
    \centering 
    \begin{tikzpicture} \begin{feynman} 
        \vertex[blob](a1) {}; 
        \vertex[left=1.5cm of a1] (a2); 
        \vertex[right=1.5cm of a1] (a3);
        \vertex[above=1.5cm of a2] (b1);
        \vertex[below=1.5cm of a2] (b2);
        \vertex[above=1.5cm of a3] (b3);
        \vertex[below=1.5cm of a3] (b4);
        \diagram* { 
        (b1) -- [fermion, edge label'=\(p_1\)] (a1) ,
        (b2) -- [fermion, edge label'=\(p_2\)] (a1),
        (b3) -- [fermion, edge label'=\(p_3\)] (a1),
        (b4) -- [fermion, edge label'=\(p_4\)] (a1),
        };
        \end{feynman} \end{tikzpicture}
\caption{4-point scattering amplitude between two massive scalar fields $\psi_1$ and $\psi_2$}
    \end{figure}
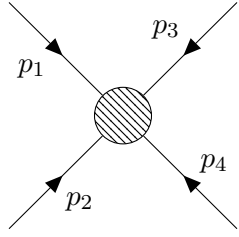
\noindent Let us choose the following configurations of momenta, where we work in the center of mass frame:
\begin{equation}
    p_1=(E_1,\mathbf{p}) \ ,   \quad p_2=(-E_1,\mathbf{p}') \ ,  \quad p_3=(E_2,-\mathbf{p}) \,  \quad p_4=(-E_2,-\mathbf{p}')  \ , 
\end{equation}
where this time both the particles are massive, and so we have the following conditions on the external momenta:
\begin{equation}
    p_1^2=m_1^2 \ , \quad p_2^2=m_1^2 \ , \quad p_3^2=m_2^2 \ , \quad p_4^2=m_2^2 \ , \qquad p_1+p_2+p_3+p_4 = 0 \ ,
\end{equation}
where $q=(p_1+p_2)=-p_3-p_4=(0,\mathbf{q})=(0,\mathbf{p}+\mathbf{p}')$ is the transferred momentum in the process, where we work in the center of mass frame.
The Mandelstam variables for the scattering process are: 
\begin{equation}
    s=q^2=(p_1+p_2)^2 \quad t=(p_1+p_3)^2 \quad u=(p_1+p_4)^2 \qquad s+t+u=2m_1^2+2m_2^2 \ . 
\end{equation}
We denote the scattering amplitude $\mathcal{M}$ of the process, up to one-loop order, as:
\begin{equation}
    \mathcal{M}=\frac{1}{\hbar}\mathcal{M}^{(0)}+\mathcal{M}^{(1)}
\end{equation}
where $\mathcal{M}^{(0)}$ denotes the tree level process whereas $\mathcal{M}^{(1)}$ represents the one-loop corrections. As explained in sec.\ref{sec:pm_potential}, we can obtain General Relativity corrections to the Newtonian potential using a scattering amplitude approach. 
In the previous chapter in Eq.\eqref{eq:newtonian_potential_tree} we have shown that by taking the Fourier transform of the 4-point scattering amplitude $\mathcal{M}^{(0)}(\mathbf{p},\mathbf{p}')$: 
\begin{eqnarray}
    \mathcal{V}^{tree}(r)= \frac{1}{4 m_1m_2}\int \frac{d^{3}\mathbf{q}}{(2\pi)^{3}}e^{i\mathbf{q}\cdot \mathbf{r}}i \mathcal{M}^{(0)}(\mathbf{p},\mathbf{p}') \ , 
    \label{eq:potential_amplitudes}
\end{eqnarray} 
we can obtain the Newtonian potential from a tree level calculation.
This can be generalized at 1-loop order: by taking into account the Born subtraction term defined in Eq.\eqref{eq:born_series}, we can define the 1-loop potential in momentum space as: 
\begin{equation}
    \mathcal{V}^{1loop}(\mathbf{p},\mathbf{p'})=i M^{(1)}(\mathbf{p},\mathbf{p}') + \mathcal{B} \ , 
\end{equation}
where $M^{(1)}(\mathbf{p},\mathbf{p}')$ is the non-relativistic generalization of the amplitude, which has the form: 
\begin{equation}
    M^{(1)}(\mathbf{p},\mathbf{p}')=\frac{1}{4 m_1 m_2}\mathcal{M}^{(1)}(\mathbf{p},\mathbf{p}') \ , 
\end{equation}
and $\mathcal{B}$ is the Born subtraction term, defined as: 
\begin{equation}
   \mathcal{B} \ = \  -\lim_{\epsilon\to 0 }\int \frac{d^{d-1}\mathbf{k}}{(2\pi)^{d-1}}\frac{i M^{(0)}(\mathbf{p},\mathbf{k})i M^{(0)}(\mathbf{k},\mathbf{p}')}{E_p-E_k+i\epsilon}
   \label{eq:Born_subtraction}
\end{equation} 
where $E_p=E_{p1}+E_{p2}$, with $E_{p1,2}=\sqrt{m_{1,2}^2+\mathbf{p}^2}$, and similarly for $E_k$. \\ 
The corresponding expression for the potential in position space can then be obtained via the Fourier transform: 
\begin{equation}
    \mathcal{V}(r)= \int    \frac{d^{3}\mathbf{q}}{(2\pi)^3}e^{i \mathbf{q}\cdot \mathbf{r}}\mathcal{V}(\mathbf{p},\mathbf{p'}) \ . 
\end{equation}
Let us now move forward and compute the 1-loop corrections to the amplitude. 

\subsection{1-loop corrections to the Newtonian potential}
We can obtain the 1-loop corrections to the Newtonian potential by adapting the calculations done in the previous section. 
Again, we will not consider the full one-loop amplitude, but only the part arising from the s-channel massless double-cut. This is the part that gives rise to non-analytic terms, which generate long-range interactions once we take the Fourier transform.\\
We will denote this part of the 1-loop amplitude as:
\begin{equation}
    \mathcal{C}^{(g)}=Cut_{12}(\mathcal{M}^{(1)})
\end{equation}
This time only the set of diagrams with at least two graviton propagators in the s-channel, namely on the graviton double cut, is contributing.
The relevant diagrams are exactly the one reported in figure \eqref{fig:1loop_grav}, where we substitute the dashed lines denoting the massless scalar particle with thin lines denoting the massive scalar of mass $m_1$. \\
In this case we have to redefine the basis of Master Integrals since both scalars are massive. \\
For the s-topology we made the following choice of denominators:
\begin{equation}
     D_1= k^2 \qquad D_2 = (k+p_1+p_2)^2 \qquad D_3= (k+p_1)^2-m_1^2 \qquad D_4= (k-p_3)^2-m_2^2 \ , 
\end{equation}
and so a generic integral can be written, as in the previous case as:
\begin{equation}
     j^{ms}_{n_1,n_2,n_3,n_4}=\begin{tikzpicture}[baseline=(current bounding box.center)] 
        \begin{feynman}
            \vertex(a1) ;
            \vertex[right= 1.4 cm of a1] (a2);
            \vertex[below =0.2 cm of a1] (b1);
            \vertex[right = 0.2 cm of b1] (b2);
            \vertex[right = 1cm of b2] (b3);
            \vertex[below = 1 cm of b1] (c1);
            \vertex[below = 0.2 cm of c1] (d1);
            \vertex[right = 1.4cm of d1] (d2);
            \vertex[right = 0.2 cm of c1] (c2);
            \vertex[right = 1cm of c2] (c3);
            \vertex[below = 0.5 cm of b1] (e1);
             \vertex[right = 1 cm of e1] (e2);
              \vertex[right = 0.2 cm of e2] (e3);
            \diagram* { 
            (a1) --  [very thick](b2),
            (b2) --  [very thick](c2),
            (d1)--[very thick](c2),
            (a2) -- [very thick](b3),
            (d2) -- [very thick](c3),
            (b2)-- (b3),
            (c2) -- (c3),
            (b3) -- [very thick](c3),
            };
            \end{feynman} \end{tikzpicture}= \int \frac{d^dk}{(2\pi)^d}\frac{1}{D_1^{n_1}D_2^{n_2}D_3^{n_3}D_4^{n_4}} \ . 
\end{equation}
For the u-topology instead the denominators are:
\begin{equation}
     D_1= k^2 \qquad D_2 = (k+p_1+p_2)^2 \qquad D_3= (k+p_1)^2-m_1^2 \qquad D_{4u}= (k+p_1+p_2+p_3)^2-m_2^2
\end{equation}
and so a generic integral can be written as:
\begin{equation}
     j^{mu}_{n_1,n_2,n_3,n_4}=\begin{tikzpicture}[baseline=(current bounding box.center)] 
        \begin{feynman}
            \vertex(a1) ;
            \vertex[right= 1.4 cm of a1] (a2);
            \vertex[below =0.2 cm of a1] (b1);
            \vertex[right = 0.2 cm of b1] (b2);
            \vertex[right = 1cm of b2] (b3);
            \vertex[below = 1 cm of b1] (c1);
            \vertex[below = 0.2 cm of c1] (d1);
            \vertex[right = 1.4cm of d1] (d2);
            \vertex[right = 0.2 cm of c1] (c2);
            \vertex[right = 1cm of c2] (c3);
            \vertex[below = 0.5 cm of b1] (e1);
             \vertex[right = 1 cm of e1] (e2);
              \vertex[right = 0.2 cm of e2] (e3);
            \diagram* { 
            (a1) --  [very thick](b2),
            (b2) --  [very thick](c2),
            (d1)--[very thick](c2),
            (a2) -- [very thick](b3),
            (d2) -- [very thick](c3),
            (b2)-- (c3),
            (c2) -- (b3),
            (b3) -- [very thick](c3),
            };
            \end{feynman} \end{tikzpicture}= \int \frac{d^dk}{(2\pi)^d}\frac{1}{D_1^{n_1}D_2^{n_2}D_3^{n_3}D_{4u}^{n_4}} \ . 
\end{equation}
By performing an IBP decomposition, with cuts in the first two denominators, and looking for relations between the two basis we found five different Master Integrals:
\begin{equation}
\begin{split}
&  I_{1}= j^{ms}_{1,1,0,0}=\int \frac{d^dk}{(2\pi)^d}\frac{1}{D_1D_2}=      \begin{tikzpicture}[baseline=(current bounding box.center)]             \begin{feynman}
    \vertex (a) ;
    \vertex[right=0.2cm of a] (b);
    \vertex[right=1cm of b] (c); 
    \vertex[right=0.2cm of c] (d);
    \diagram* { 
    (a) --  (b),
    (b) -- [half left ] (c),
    (b) -- [half right ] (c),
    (c) -- (d),
    };
    \end{feynman} \end{tikzpicture}\\
& I_{2}= j^{ms}_{1,1,1,0}=\int \frac{d^dk}{(2\pi)^d}\frac{1}{D_1D_2D_3}=
\begin{tikzpicture}[baseline=(current bounding box.center)] 
\begin{feynman}
    \vertex(a1) ;
    \vertex[right= 1.4 cm of a1] (a2);
    \vertex[below =0.2 cm of a1] (b1);
    \vertex[right = 0.2 cm of b1] (b2);
    \vertex[right = 1cm of b2] (b3);
    \vertex[below = 1 cm of b1] (c1);
    \vertex[below = 0.2 cm of c1] (d1);
    \vertex[right = 1.4cm of d1] (d2);
    \vertex[right = 0.2 cm of c1] (c2);
    \vertex[right = 1cm of c2] (c3);
    \vertex[below = 0.5 cm of b1] (e1);
     \vertex[right = 1 cm of e1] (e2);
      \vertex[right = 0.2 cm of e2] (e3);
    \diagram* { 
    (a1) --  [very thick](b2),
    (b2) --  [very thick](c2),
    (b2) -- (e2),
    (c2) -- (e2),
    (e2) -- (e3),
    (d1)--[very thick](c2),
    };
    \end{feynman} \end{tikzpicture}        \\
& I_{3}= j^{ms}_{1,1,0,1}=\int \frac{d^dk}{(2\pi)^d}\frac{1}{D_1D_2D_4}=
\begin{tikzpicture}[baseline=(current bounding box.center)] 
\begin{feynman}
    \vertex(a1) ;
    \vertex[right= 1.4 cm of a1] (a2);
    \vertex[below =0.2 cm of a1] (b1);
    \vertex[below =0.5 cm of b1] (f1);
    \vertex[right =0.2 cm of f1] (f2);
    \vertex[right = 0.2 cm of b1] (b2);
     \vertex[right = 1cm of b2] (b3);
    \vertex[below = 1 cm of b1] (c1);
    \vertex[below = 0.2 cm of c1] (d1);
    \vertex[right = 0.2 cm of c1] (c2);
     \vertex[right = 1 cm of c2] (c3);
    \vertex[below = 0.5 cm of b1] (e1);
     \vertex[right = 1 cm of e1] (e2);
      \vertex[right = 0.2 cm of e2] (e3);
      \vertex[right = 1.4cm of d1] (d2);
    \diagram* { 
    (f1)-- (f2),
    (f2)-- (b3),
    (f2) -- (c3),
    (b3) -- [very thick](c3),
    (a2) -- [very thick](b3),
    (d2) -- [very thick](c3),
    };
    \end{feynman} \end{tikzpicture} \\
& I_{4}= j^{ms}_{1,1,1,1}=\int \frac{d^dk}{(2\pi)^d}\frac{1}{D_1D_2D_3D_4}=   \begin{tikzpicture}[baseline=(current bounding box.center)] 
\begin{feynman}
    \vertex(a1) ;
    \vertex[right= 1.4 cm of a1] (a2);
    \vertex[below =0.2 cm of a1] (b1);
    \vertex[right = 0.2 cm of b1] (b2);
    \vertex[right = 1cm of b2] (b3);
    \vertex[below = 1 cm of b1] (c1);
    \vertex[below = 0.2 cm of c1] (d1);
    \vertex[right = 1.4cm of d1] (d2);
    \vertex[right = 0.2 cm of c1] (c2);
    \vertex[right = 1cm of c2] (c3);
    \vertex[below = 0.5 cm of b1] (e1);
     \vertex[right = 1 cm of e1] (e2);
      \vertex[right = 0.2 cm of e2] (e3);
    \diagram* { 
    (a1) --  [very thick](b2),
    (b2) --  [very thick](c2),
    (d1)--[very thick](c2),
    (a2) -- [very thick](b3),
    (d2) -- [very thick](c3),
    (b2)-- (b3),
    (c2) -- (c3),
    (b3) -- [very thick](c3),
    };
    \end{feynman} \end{tikzpicture}\\
& I_{5}= j^{mu}_{1,1,1,1}= \int \frac{d^dk}{(2\pi)^d}\frac{1}{D_1D_2D_3D_{4u}}=  
\begin{tikzpicture}[baseline=(current bounding box.center)] 
\begin{feynman}
    \vertex(a1) ;
    \vertex[right= 1.4 cm of a1] (a2);
    \vertex[below =0.2 cm of a1] (b1);
    \vertex[right = 0.2 cm of b1] (b2);
    \vertex[right = 1cm of b2] (b3);
    \vertex[below = 1 cm of b1] (c1);
    \vertex[below = 0.2 cm of c1] (d1);
    \vertex[right = 1.4cm of d1] (d2);
    \vertex[right = 0.2 cm of c1] (c2);
    \vertex[right = 1cm of c2] (c3);
    \vertex[below = 0.5 cm of b1] (e1);
     \vertex[right = 1 cm of e1] (e2);
      \vertex[right = 0.2 cm of e2] (e3);
    \diagram* { 
    (a1) --  [very thick](b2),
    (b2) --  [very thick](c2),
    (d1)--[very thick](c2),
    (a2) -- [very thick](b3),
    (d2) -- [very thick](c3),
    (b2)-- (c3),
    (c2) -- (b3),
    (b3) -- [very thick](c3),
    };
    \end{feynman} \end{tikzpicture}\\
\end{split}
\end{equation}
As done in the previous case, we can sum all the diagrams and obtain the amplitudes, expressed in terms of the Master Integrals, as:
\begin{eqnarray}
    \kappa^{-4}\mathcal{C}^{(g)}
     = \sum_{i=1}^4 c_{i}^{(g)} \, I_i \
     =
        c_{1}^{(g)} \scalebox{0.5}{\begin{tikzpicture}[baseline=(current bounding box.center)]\begin{feynman}
    \vertex (a) ;
    \vertex[right=0.2cm of a] (b);
    \vertex[right=1cm of b] (c); 
    \vertex[right=0.2cm of c] (d);
    \diagram* { 
    (a) --  (b),
    (b) -- [half left ] (c),
    (b) -- [half right ] (c),
    (c) -- (d),
    };
    \end{feynman} \end{tikzpicture}}
    +c_{2}^{(g)} \scalebox{0.5}{\begin{tikzpicture}[baseline=(current bounding box.center)] 
        \begin{feynman}
            \vertex(a1) ;
            \vertex[right= 1.4 cm of a1] (a2);
            \vertex[below =0.2 cm of a1] (b1);
            \vertex[right = 0.2 cm of b1] (b2);
            \vertex[right = 1cm of b2] (b3);
            \vertex[below = 1 cm of b1] (c1);
            \vertex[below = 0.2 cm of c1] (d1);
            \vertex[right = 1.4cm of d1] (d2);
            \vertex[right = 0.2 cm of c1] (c2);
            \vertex[right = 1cm of c2] (c3);
            \vertex[below = 0.5 cm of b1] (e1);
             \vertex[right = 1 cm of e1] (e2);
              \vertex[right = 0.2 cm of e2] (e3);
            \diagram* { 
            (a1) --  [very thick](b2),
            (b2) --  [very thick](c2),
            (b2) -- (e2),
            (c2) -- (e2),
            (e2) -- (e3),
            (d1)--[very thick](c2),
            };
            \end{feynman} \end{tikzpicture}}
            + c_{3}^{(g)} \scalebox{0.5}{ \begin{tikzpicture}[baseline=(current bounding box.center)] 
\begin{feynman}
   \vertex(a1) ;
    \vertex[right= 1.4 cm of a1] (a2);
    \vertex[below =0.2 cm of a1] (b1);
    \vertex[below =0.5 cm of b1] (f1);
    \vertex[right =0.2 cm of f1] (f2);
    \vertex[right = 0.2 cm of b1] (b2);
     \vertex[right = 1cm of b2] (b3);
    \vertex[below = 1 cm of b1] (c1);
    \vertex[below = 0.2 cm of c1] (d1);
    \vertex[right = 0.2 cm of c1] (c2);
     \vertex[right = 1 cm of c2] (c3);
    \vertex[below = 0.5 cm of b1] (e1);
     \vertex[right = 1 cm of e1] (e2);
      \vertex[right = 0.2 cm of e2] (e3);
      \vertex[right = 1.4cm of d1] (d2);
    \diagram* { 
    (f1)-- (f2),
    (f2)-- (b3),
    (f2) -- (c3),
    (b3) -- [very thick](c3),
    (a2) -- [very thick](b3),
    (d2) -- [very thick](c3),
    };
    \end{feynman} \end{tikzpicture} }
      + c_{4}^{(g)} \scalebox{0.5}{\begin{tikzpicture}[baseline=(current bounding box.center)]
\begin{feynman}
    \vertex(a1) ;
    \vertex[right= 1.4 cm of a1] (a2);
    \vertex[below =0.2 cm of a1] (b1);
    \vertex[right = 0.2 cm of b1] (b2);
    \vertex[right = 1cm of b2] (b3);
    \vertex[below = 1 cm of b1] (c1);
    \vertex[below = 0.2 cm of c1] (d1);
    \vertex[right = 1.4cm of d1] (d2);
    \vertex[right = 0.2 cm of c1] (c2);
    \vertex[right = 1cm of c2] (c3);
    \vertex[below = 0.5 cm of b1] (e1);
     \vertex[right = 1 cm of e1] (e2);
      \vertex[right = 0.2 cm of e2] (e3);
    \diagram* { 
    (a1) --  [very thick] (b2),
    (b2) -- [very thick] (c2),
    (d1)-- [very thick] (c2),
    (a2) -- [very thick](b3),
    (d2) -- [very thick](c3),
    (b2)-- (b3),
    (c2) -- (c3),
    (b3) -- [very thick](c3),
    };
    \end{feynman} \end{tikzpicture}} 
      + c_{5}^{(g)} \scalebox{0.5}{\begin{tikzpicture}[baseline=(current bounding box.center)] 
\begin{feynman}
    \vertex(a1) ;
    \vertex[right= 1.4 cm of a1] (a2);
    \vertex[below =0.2 cm of a1] (b1);
    \vertex[right = 0.2 cm of b1] (b2);
    \vertex[right = 1cm of b2] (b3);
    \vertex[below = 1 cm of b1] (c1);
    \vertex[below = 0.2 cm of c1] (d1);
    \vertex[right = 1.4cm of d1] (d2);
    \vertex[right = 0.2 cm of c1] (c2);
    \vertex[right = 1cm of c2] (c3);
    \vertex[below = 0.5 cm of b1] (e1);
     \vertex[right = 1 cm of e1] (e2);
      \vertex[right = 0.2 cm of e2] (e3);
    \diagram* { 
    (a1) -- [very thick] (b2),
    (b2) -- [very thick] (c2),
    (d1)-- [very thick] (c2),
    (a2) -- [very thick](b3),
    (d2) -- [very thick](c3),
    (b2)-- (c3),
    (c2) -- (b3),
    (b3) -- [very thick](c3),
    };
    \end{feynman} \end{tikzpicture}}
    \label{eq:1_loop_grav_dec_mass}
      \end{eqnarray}
where the coefficients are, writing $d=4-2\epsilon$:
\begin{eqnarray}
    c_1^{(g)} &=& 
    \frac{\left(2 \epsilon -1\right)}{128 \left(\epsilon -1\right)^2 \left(2 \epsilon
       -5\right) \left(2 \epsilon -3\right) \left(s-4 m_2^2\right)^2 \left(s-4 m_1^2\right)^2} \biggl(-16 \biggl(32 \biggl(8 \epsilon ^4-61 \epsilon ^3+168 \epsilon ^2\nonumber \\ 
       & & 
       -199 \epsilon +87\biggr) m_1^4-8 \left(16 \epsilon ^4-98
       \epsilon ^3+212 \epsilon ^2-196 \epsilon +69\right) s m_1^2+\bigl(16 \epsilon ^4-82 \epsilon ^3+132 \epsilon ^2\nonumber \\ 
       & & 
       -72 \epsilon +9\bigr) s^2\biggr) m_2^8-16 \biggl(64
       \left(8 \epsilon ^4-78 \epsilon ^3+228 \epsilon ^2-232 \epsilon +59\right) m_1^6-8 \bigl(\bigl(60 \epsilon ^4-542 \epsilon ^3\nonumber \\ 
       & & 
       +1576 \epsilon ^2-1732 \epsilon
       +587\biggr) s+8 \left(8 \epsilon ^4-61 \epsilon ^3+168 \epsilon ^2-199 \epsilon +87\right) t\bigr) m_1^4+2 s \biggl(\bigl(72 \epsilon ^4 \nonumber \\ 
       & & 
       -528 \epsilon ^3+1324
       \epsilon ^2-1310 \epsilon +415\bigr) s+8 \left(16 \epsilon ^4-98 \epsilon ^3+212 \epsilon ^2-196 \epsilon +69\right) t\biggr) m_1^2\nonumber \\ 
       & & 
       -s^2 \left(\left(14 \epsilon
       ^4-79 \epsilon ^3+146 \epsilon ^2-99 \epsilon +18\right) s+2 \left(16 \epsilon ^4-82 \epsilon ^3+132 \epsilon ^2-72 \epsilon +9\right) t\right)\bigr) m_2^6\nonumber \\ 
       & & 
       -2 \biggl(256
       \left(8 \epsilon ^4-61 \epsilon ^3+168 \epsilon ^2-199 \epsilon +87\right) m_1^8-64 \bigl(\bigl(60 \epsilon ^4-542 \epsilon ^3+1576 \epsilon ^2-1732 \epsilon
       \nonumber \\ 
       & & 
       +587\bigr) s+8 \left(8 \epsilon ^4-61 \epsilon ^3+168 \epsilon ^2-199 \epsilon +87\right) t\bigr) m_1^6-16 \bigl(\bigl(8 \epsilon ^5-184 \epsilon ^4+1354
       \epsilon ^3\nonumber \\ 
       & & 
       -3804 \epsilon ^2+4240 \epsilon -1521\bigr) s^2-16 \left(24 \epsilon ^4-159 \epsilon ^3+380 \epsilon ^2-395 \epsilon +156\right) t s-16 \bigl(8 \epsilon
       ^4\nonumber \\ 
       & & 
       -61 \epsilon ^3+168 \epsilon ^2-199 \epsilon +87\bigr) t^2\bigr) m_1^4+8 s \bigl(\bigl(8 \epsilon ^5-110 \epsilon ^4+603 \epsilon ^3-1458 \epsilon ^2+1503
       \epsilon \nonumber \\ 
       & & 
       -522\bigr) s^2-2 \left(144 \epsilon ^4-770 \epsilon ^3+1332 \epsilon ^2-832 \epsilon +141\right) t s-8 \bigl(16 \epsilon ^4-98 \epsilon ^3+212 \epsilon
       ^2\nonumber \\ 
       & & 
       -196 \epsilon +69\bigr) t^2\bigr) m_1^2+s^2 \bigl(\left(-8 \epsilon ^5+88 \epsilon ^4-357 \epsilon ^3+652 \epsilon ^2-533 \epsilon +158\right) s^2\nonumber \\ 
       & & 
       +8
       \left(32 \epsilon ^4-140 \epsilon ^3+140 \epsilon ^2+58 \epsilon -87\right) t s+8 \left(16 \epsilon ^4-82 \epsilon ^3+132 \epsilon ^2-72 \epsilon +9\bigr)
       t^2\right)\biggr) m_2^4\nonumber \\
       & & 
       +2 s \bigl(64 \left(16 \epsilon ^4-98 \epsilon ^3+212 \epsilon ^2-196 \epsilon +69\right) m_1^8-16 \bigl(\bigl(72 \epsilon ^4-528 \epsilon
       ^3+1324 \epsilon ^2\nonumber \\ 
       & & 
       -1310 \epsilon +415\bigr) s+8 \left(16 \epsilon ^4-98 \epsilon ^3+212 \epsilon ^2-196 \epsilon +69\right) t\bigr) m_1^6-8 \bigl(\bigl(8
       \epsilon ^5-110 \epsilon ^4+603 \epsilon ^3\nonumber \\ 
       & & -1458 \epsilon ^2+1503 \epsilon -522\bigr) s^2-2 \left(144 \epsilon ^4-770 \epsilon ^3+1332 \epsilon ^2-832 \epsilon
       +141\right) t s-8 \bigl(16 \epsilon ^4\nonumber \\ 
       & & -98 \epsilon ^3+212 \epsilon ^2-196 \epsilon +69\bigr) t^2\bigr) m_1^4+4 s \bigl(\left(8 \epsilon ^5-72 \epsilon ^4+287
       \epsilon ^3-580 \epsilon ^2+549 \epsilon -186\right) s^2\nonumber \\ 
       & & 
       -8 \left(24 \epsilon ^4-103 \epsilon ^3+92 \epsilon ^2+71 \epsilon -84\right) t s-16 \left(8 \epsilon ^4-37
       \epsilon ^3+44 \epsilon ^2+3 \epsilon -18\right) t^2\bigr) m_1^2\nonumber \\ 
       & & 
       -\left(\epsilon -1\right) s^2 \bigl(\left(4 \epsilon ^4-27 \epsilon ^3+70 \epsilon ^2-81
       \epsilon +34\right) s^2+2 \left(-40 \epsilon ^3+97 \epsilon ^2+117 \epsilon -270\right) t s\nonumber \\ 
       & &
       -8 \left(8 \epsilon ^3-21 \epsilon ^2-17 \epsilon +48\right)
       t^2\bigr)\bigr) m_2^2+s^2 \bigl(-16 \left(16 \epsilon ^4-82 \epsilon ^3+132 \epsilon ^2-72 \epsilon +9\right) m_1^8\nonumber \\ 
       & & 
       +16 \left(\left(14 \epsilon ^4-79 \epsilon
       ^3+146 \epsilon ^2-99 \epsilon +18\right) s+2 \left(16 \epsilon ^4-82 \epsilon ^3+132 \epsilon ^2-72 \epsilon +9\right) t\right) m_1^6\nonumber \\ 
       & & 
       +2 \bigl(\left(8 \epsilon
       ^5-88 \epsilon ^4+357 \epsilon ^3-652 \epsilon ^2+533 \epsilon -158\right) s^2-8 \bigl(32 \epsilon ^4-140 \epsilon ^3+140 \epsilon ^2+58 \epsilon \nonumber \\ 
       & & 
       -87\bigr) t s-8
       \left(16 \epsilon ^4-82 \epsilon ^3+132 \epsilon ^2-72 \epsilon +9\right) t^2\bigr) m_1^4-2 \left(\epsilon -1\right) s \bigl(\bigl(4 \epsilon ^4-27 \epsilon
       ^3+70 \epsilon ^2\nonumber \\ 
       & & 
       -81 \epsilon +34\bigr) s^2+2 \left(-40 \epsilon ^3+97 \epsilon ^2+117 \epsilon -270\right) t s-8 \left(8 \epsilon ^3-21 \epsilon ^2-17 \epsilon
       +48\right) t^2\bigr) m_1^2 \nonumber \\ 
       & & 
       +\left(\epsilon -1\right) s^2 \bigl(\left(\epsilon ^2-3 \epsilon +2\right)^2 s^2-2 \left(8 \epsilon ^3-13 \epsilon ^2-49 \epsilon
       +78\right) t s\nonumber \\ 
       & & +2 \left(-8 \epsilon ^3+13 \epsilon ^2+49 \epsilon -78\right) t^2\bigr)\bigr)\biggr) \ , \\ 
    c_2^{(g)} &=&  
    -\frac{1}{16 \left(\epsilon -1\right)^2 \left(s-4
       m_1^2\right)^2} \bigl(m_2^2 \bigl(-2 m_1^2 s^2 \left(\epsilon -1\right) \left(19 s \epsilon +32 t \epsilon -17 s-30 t\right)\nonumber \\ 
       & & 
       -4 m_1^6 \left(s \left(66 \epsilon
       ^2-119 \epsilon +45\right)+2 t \left(16 \epsilon ^2-30 \epsilon +15\right)\right)+2 m_1^4 s \bigl(s \left(82 \epsilon ^2-149 \epsilon +65\right)\nonumber \\ 
       & & 
       +10 t \left(10
       \epsilon ^2-19 \epsilon +9\right)\bigr)+8 m_1^8 \left(16 \epsilon ^2-34 \epsilon +9\right)+3 s^3 \left(\epsilon -1\right)^2 (s+2 t)\bigr)\nonumber \\ 
       & & 
       +m_2^4 \bigl(2
       m_1^2 s^2 \left(16 \epsilon ^2-31 \epsilon +15\right)-10 m_1^4 s \left(10 \epsilon ^2-19 \epsilon +9\right)+4 m_1^6 \left(16 \epsilon ^2-30 \epsilon
       +15\right)\nonumber \\ 
       & & 
       -3 s^3 \left(\epsilon -1\right)^2\bigr)+4 m_1^6 \left(35 s^2 \left(\epsilon -1\right)^2+s t \left(66 \epsilon ^2-125 \epsilon +60\right)+t^2 \left(16
       \epsilon ^2-30 \epsilon +15\right)\right)\nonumber \\ 
       & & 
       -m_1^4 s \left(\epsilon -1\right) \left(63 s^2 \left(\epsilon -1\right)+2 s t \left(82 \epsilon -75\right)+10 t^2
       \left(10 \epsilon -9\right)\right)\nonumber \\ 
       & & 
       +m_1^2 s^2 \left(\epsilon -1\right) \left(13 s^2 \left(\epsilon -1\right)+2 s t \left(19 \epsilon -18\right)+2 t^2 \left(16
       \epsilon -15\right)\right)-2 m_1^8 \bigl(s \bigl(74 \epsilon ^2\nonumber \\ 
       & & 
       -147 \epsilon +73\bigr)+4 t \left(16 \epsilon ^2-30 \epsilon +15\right)\bigr)+4 m_1^{10}
       \left(16 \epsilon ^2-30 \epsilon +15\right)\nonumber \\ 
       & & -s^3 \left(\epsilon -1\right)^2 \left(s^2+3 s t+3 t^2\right)\bigr)\ , \\ 
    c_3^{(g)} &=&  
    -\frac{1}{16 \left(\epsilon -1\right)^2 \left(s-4 m_2^2\right)^2} \bigl(4 m_2^6 \bigl(-m_1^2 \left(s \left(66 \epsilon ^2-119 \epsilon +45\right)+2 t \left(16 \epsilon ^2-30 \epsilon +15\right)\right)\nonumber \\ 
    & & 
    +m_1^4
       \left(16 \epsilon ^2-30 \epsilon +15\right)+35 s^2 \left(\epsilon -1\right)^2+s t \left(66 \epsilon ^2-125 \epsilon +60\right)+t^2 \left(16 \epsilon ^2-30 \epsilon
       +15\right)\bigr)\nonumber \\ 
       & & 
       -m_2^4 s \bigl(-2 m_1^2 \left(s \left(82 \epsilon ^2-149 \epsilon +65\right)+10 t \left(10 \epsilon ^2-19 \epsilon +9\right)\right)+10
       m_1^4 \left(10 \epsilon ^2-19 \epsilon +9\right)\nonumber \\ 
       & & 
       +\left(\epsilon -1\right) \left(63 s^2 \left(\epsilon -1\right)+2 s t \left(82 \epsilon -75\right)+10 t^2
       \left(10 \epsilon -9\right)\right)\bigr)+m_2^2 s^2 \bigl(-2 m_1^2 \left(\epsilon -1\right) \bigl(19 s \epsilon \nonumber \\
       & & 
       +32 t \epsilon -17 s-30 t\bigr)+m_1^4
       \left(32 \epsilon ^2-62 \epsilon +30\right)+\left(\epsilon -1\right) \bigl(13 s^2 \left(\epsilon -1\right)+2 s t \left(19 \epsilon -18\right)\nonumber \\ 
       & & 
       +2 t^2 \left(16 \epsilon
       -15\right)\bigr)\bigr)+2 m_2^8 \bigl(4 m_1^2 \left(16 \epsilon ^2-34 \epsilon +9\right)+s \left(-74 \epsilon ^2+147 \epsilon -73\right)-4 t \bigl(16 \epsilon
       ^2\nonumber \\ 
       & & 
       -30 \epsilon +15\bigr)\bigr)+4 m_2^{10} \left(16 \epsilon ^2-30 \epsilon +15\right)-s^3 \left(\epsilon -1\right)^2 \bigl(-3 m_1^2 (s+2 t)+3 m_1^4\nonumber \\ 
       & & +s^2+3
       s t+3 t^2\bigr)\bigr)\ , \\ 
    c_4^{(g)} &=&  \frac{1}{16 \left(\epsilon -1\right)^2} \bigl(2 m_2^2 \left(\epsilon  \left(m_1^2-t\right)+t\right)+m_2^4 \left(\epsilon -1\right)+\left(\epsilon -1\right)
       \left(m_1^2-t\right)^2\bigr)^2\ , \\ 
    c_5^{(g)} &=&  \frac{1}{16 \left(\epsilon -1\right)^2} \left(2 m_2^2 \left(\epsilon  \left(m_1^2-s-t\right)+s+t\right)+m_2^4 \left(\epsilon -1\right)+\left(\epsilon -1\right)
       \left(-m_1^2+s+t\right)^2\right)^2\ . 
    \end{eqnarray}
The Master Integrals have been obtained using \texttt{PackageX} and their expressions are:
\begin{eqnarray}
    I_1 & = & N_G\biggl(\frac{-\mu^2}{s}\biggr)^{\epsilon}\frac{1}{\epsilon(1-2\epsilon)}\ , \\ 
    I_2 & = &  N_G\frac{1}{s\beta_1}\biggl[\frac{4\pi^2}{6}+2Li_2\biggl(\frac{\beta_1-1}{\beta_1+1}\biggr)+\frac{1}{2}\log^2\biggl(\frac{\beta_1-1}{\beta_1+1}\biggr)+\mathcal{O}[\epsilon]\biggr] \ , \\ 
    I_3 & = & N_G\frac{1}{s\beta_2}\biggl[\frac{4\pi^2}{6}+2Li_2\biggl(\frac{\beta_2-1}{\beta_2+1}\biggr)+\frac{1}{2}\log^2\biggl(\frac{\beta_2-1}{\beta_2+1}\biggr)+\mathcal{O}[\epsilon]\biggr]\ , \\ 
    I_4 & = & \frac{2 N_G}{s \sqrt{-2 m_2^2 \left(m_1^2+t\right)+m_2^4+\left(m_1^2-t\right)^2}}\biggl[\nonumber \\ 
    & & 
    \frac{1}{\epsilon} \log \left(\frac{\sqrt{-2 m_2^2 \left(m_1^2+t\right)+m_2^4+\left(m_1^2-t\right)^2}+m_2^2+m_1^2-t}{2 m_1
   m_2}\right)\nonumber \\ 
   & & 
   +\log \left(-\frac{\Bar{\mu}^2e^{\gamma_E}}{4 \pi  s}\right) \log \left(\frac{\sqrt{-2 m_2^2
   \left(m_1^2+t\right)+m_2^4+\left(m_1^2-t\right)^2}+m_1^2+m_2^2-t}{2 m_1 m_2}\right)\nonumber \\ 
   & & 
   +O\left(\epsilon ^1\right)\biggr]\ , \\ 
    I_5 & = & 
    \frac{2 N_G}{s \sqrt{-2 m_2^2 \left(m_1^2+u\right)+m_2^4+\left(m_1^2-u\right)^2}}\biggl[\nonumber \\
    & & 
    \frac{1}{\epsilon} \log \left(\frac{\sqrt{-2 m_2^2 \left(m_1^2+u\right)+m_2^4+\left(m_1^2-u\right)^2}+m_2^2+m_1^2-u}{2 m_1
   m_2}\right)\nonumber \\ 
   & & 
   +\log \left(-\frac{\Bar{\mu}^2e^{\gamma_E}}{4 \pi  s}\right) \log \left(\frac{\sqrt{-2 m_2^2
   \left(m_1^2+u\right)+m_2^4+\left(m_1^2-u\right)^2}+m_1^2+m_2^2-u}{2 m_1 m_2}\right)\nonumber \\ 
   & & 
   +O\left(\epsilon\right)\biggr]\ ,
\end{eqnarray}
where $\beta_1^2=1-\frac{4m_1^2}{s}$ and  $\beta_2^2=1-\frac{m_2^2}{s}$. 
We can find the result, by substituting the explicit expressions for the master integrals, and expanding in powers of $\epsilon$ up to $\mathcal{O}[\epsilon^0]$.

\subsection{Non-relativistic limit}
We want now to consider the non-relativistic limit of the 1-loop amplitude, which can be obtained by considering the transferred momentum much smaller than the masses of the scalar particles: $-s\approx\mathbf{q^2}\ll m_1^2,m_2^2$, and the center of mass momentum vanishing as-well $\mathbf{p}\ll 1$. The Mandelstam parameters in this specific kinematic limit become: 
\begin{eqnarray}
    s=(p_1+p_2)^2&  \approx &  -\mathbf{q}^2 \ , \\ 
    t=(p1+p_3)^2 &=  & \ m_1^2 + m_2^2 +2\sqrt{m_1^2+\mathbf{p}^2}\sqrt{m_2^2+\mathbf{p}^2} +2\mathbf{p}^2 \ ,  \\ 
\end{eqnarray}
where $\mathbf{p}^2\gg\mathbf{q}^2$. 
In order to obtain the low-energy limit of the amplitude, we introduce one a-dimensional variables $\tau$, as before, defined as: 
\begin{equation}
    \tau= -\frac{s}{m_1 m_2}  \ , 
\end{equation}
and we obtain the non-relativistic limit with the following series expansions: 
\begin{itemize}
    \item expansion around $d=4$ dimensions, namely around $\epsilon=0$, up to the finite term, $\mathcal{O}(\epsilon^0)$ ;
    \item expansion around $\tau=0$ up to $\mathcal{O}(\tau^0)$;
    \item expansion around $p=0$, up to $\mathcal{O}(p^0)$
\end{itemize}
We can express the 1-loop amplitude in the graviton double-cut $\mathcal{C}^{(g)}$ in the low-energy limit in powers of $\epsilon$ as:
    \begin{eqnarray}
    \mathcal{C}^{(g)}=\sum_{i=-1}^{0}\epsilon^i\mathcal{C}^{(g)}_{i}
    \end{eqnarray}
    where 
    \begin{eqnarray}
        \mathcal{C}^{(g)}_{-1} & = &  \frac{101 \kappa ^4 m_1^2 m_2^2}{1280 \pi ^2}-\frac{\kappa ^4 m_1^4 m_2^4}{64 \pi  \vert\mathbf{p}\vert s (m_1+m_2)}\ , \\
        \mathcal{C}^{(g)}_{0}& =& \mathcal{C}^{(g)}_{0 \ a}
        +\mathcal{C}^{(g)}_{0  \ n.a.} 
        \end{eqnarray}
        We wrote $\mathcal{C}^{(g)}_{0}$ as the sum of a part made by analytic terms $\mathcal{C}^{(g)}_{0 \ a}$ and a part containing non-analytic contributions 
        $\mathcal{C}^{(g)}_{0  \ n.a.}$, given by:
        \begin{eqnarray}
        \mathcal{C}^{(g)}_{0 \ a} & = & -\frac{17149 i \kappa ^4 m_1^2 m_2^2}{115200 \pi ^2}+\frac{\kappa ^4 m_1^4 m_2^4 }{32 \pi  \vert \mathbf{p} \vert  s (m_1+m_2)}\ , \\
            \mathcal{C}^{(g)}_{0 \ n.a.} &=& \frac{3 i \kappa ^4 m_1^2 m_2^2 (m_1+m_2)}{128 \sqrt{-s}} -\log\left(-\frac{m_1 m_2}{s}\right)\frac{3 i \kappa ^4 m_1^2 m_2^2}{64 \pi ^2}\nonumber \\ 
            & & \log\left(-\frac{\Bar{\mu}^2}{s}\right)\biggl[\frac{101 i \kappa ^4 m_1^2 m_2^2}{1280 \pi ^2}-\frac{\kappa ^4 m_1^4 m_2^4}{64 \pi  \vert\mathbf{p}\vert \ s( m_1+m_2)}\biggr] \ . 
        \label{eq:M10_non_analytic_massive}
        \end{eqnarray}
The non-analytic part, written in terms of the transferred momentum $\mathbf{q}$, takes the form: 
\begin{eqnarray}
            i \mathcal{C}^{(g)}_{0 \ n.a.} &=& 
            -\frac{3  \kappa ^4 m_2^2 m_1^2 (m_2+m_1)}{128 \vert\mathbf{q}\vert} +\frac{3  \kappa ^4 m_2^2 m_1^2}{64 \pi ^2}\log\left(\frac{m_1 m_2}{\mathbf{q}^2}\right) \nonumber \\ 
            & & - \log\left(\frac{\Bar{\mu}^2}{\mathbf{q}^2}\right)\frac{101  \kappa ^4 m_2^2 m_1^2}{1280 \pi ^2} + i \frac{\kappa^4 m_1^4 m_2^4}{64 \pi \vert\mathbf{p}\vert \  \mathbf{q}^2 (m_1+m_2)}\log\biggl(\frac{\bar{\mu}^2}{\mathbf{q}^2}\biggr)\ .
            \label{eq:1loop_non_analytic_massive}
\end{eqnarray}  
We can notice that the term that scale as $1/p$, with $p$ the center of mass momentum, constitute complex contributions that will not contribute to the potential, as pointed out in \cite{Weinberg:1965nx}, and in \cite{Donoghue:1996mt}. 
One can remove them by subtracting the Born subtraction term.
\subsection{Born-subtraction term}
Let us compute the Born-subtraction term defined in Eq.\eqref{eq:Born_subtraction}.
We already computed the tree-level amplitude $\mathcal{M}^{(0)}(\mathbf{p},\mathbf{p'})$ in Eq.\eqref{eq:tree_amplitude_PM}, and its non-relativisic version has the form: 
\begin{equation} 
    M^{(0)}(\mathbf{p},\mathbf{p}')\ =  \  i \frac{4 \pi G_N m_1 m_2 }{\mathbf{q}^2} \ . 
\end{equation}
Hence, the two terms appearing in the integrand of \eqref{eq:Born_subtraction} have the form: 
\begin{equation}
    M^{(0)}(\mathbf{p},\mathbf{k})\ =  \  
    i \frac{4 \pi G_N m_1 m_2 }{(\mathbf{p}+\mathbf{k})^2} \ , \qquad 
    M^{(0)}(\mathbf{k},\mathbf{p'})\  =  \ 
    i \frac{4 \pi G_N m_1 m_2 }{(\mathbf{p}'+\mathbf{k})^2} \ .
\end{equation}
Moreover, by Laurent expanding the loop momentum $\mathbf{k}$, we have: 
\begin{equation}
    \frac{1}{E_p-E_k+i\epsilon} \ \approx \ 2 \frac{m_1 m_2}{m_1 + m_2} \frac{1}{(\mathbf{p}^2-\mathbf{k}^2)} \ , 
\end{equation}
and the Born term becomes: 
\begin{equation}
    \mathcal{B} \ = \ -\frac{32 \pi^2 G_N^2 m_1^3 m_2^3}{(m_1 + m_2)}\  I_{\mathcal{B}} \ ,
\end{equation}
where the integral appearing has been evaluated in \cite{Holstein:2008sx}, and it takes the form (keeping only the non-analytic term): 
\begin{equation}
    I_{\mathcal{B}}\ =\ \lim_{d\to 3 } \int \frac{d^{d}\mathbf{k}}{(2\pi)^{d}}\frac{1}{(\mathbf{p}+\mathbf{k})^2(\mathbf{p}^{'}+\mathbf{k})^2(\mathbf{p}^2-\mathbf{k}^2)} \ = \ \frac{i }{8 \pi \vert\mathbf{p}\vert \ \mathbf{q}^2}\log(\mathbf{q}^2) +\ldots \ . 
\end{equation}
By substituting the value of the integral we get: 
\begin{equation}
    \mathcal{B} \ = \  - i  \frac{4 \pi G_N^2 m_1^3 m_2^3}{(m_1+m_2)}\frac{\log(\mathbf{q}^2)}{\vert\mathbf{p}\vert\  \mathbf{q}^2}
    \label{eq:born_subtraction_final} \ . 
\end{equation}
One can evaluate the corresponding potential in momentum space  using Eq.\eqref{eq:potential_amplitudes}, by substituting Eq.\eqref{eq:1loop_non_analytic_massive} and by subtracting the Born term (Eq.\eqref{eq:born_subtraction_final}),  which cancels out the non-analytic term proportional to $1/|\mathbf{p}|$ in Eq.\eqref{eq:M10_non_analytic_massive},  obtaining:
\begin{eqnarray}
    \mathcal{V}^{(g)}(\mathbf{p},\mathbf{q}) &=&  \frac{i\mathcal{C}_{0\ n.a.}^{(g)}}{4m_1m_2} + \mathcal{B} \nonumber \\ 
    & = & 
    \frac{6 \pi^2 G_N^2 m_1 m_2 (m_1+m_2)}{\vert\mathbf{q}\vert} -12 G_N^2 m_1 m_2 \log\left(\frac{m_1 m_2}{\mathbf{q}}\right) \nonumber \\ 
    & & +\frac{101  G_N^2 m_1 m_2}{5} \log\left(\frac{\Bar{\mu}^2}{\mathbf{q}^2}\right) \ .
\end{eqnarray} 
\subsection{1-loop corrections to the Newtonian potential}
The corresponding potential in position space is obtained by taking the Fourier transform:
\begin{equation}
    \mathcal{V}^{(g)}(r)=\int\frac{d^3\mathbf{q}}{(2\pi)^3}e^{i\mathbf{q}\cdot r}\mathcal{V}^{(g)}(\mathbf{p},\mathbf{q})=-G_N^2m_1m_2\biggl(\frac{3(m_1+m_2)}{r^2}+\frac{41}{10\pi r^3}\biggr) \ , 
\end{equation}
where we used Eq.\eqref{eq:scalar_integral_fourier}\eqref{eq:fourier_logs}. This result matches the ones obtained in \cite{Bjerrum-Bohr:2002gqz,Bjerrum-Bohr:2004qcf,Bjerrum-Bohr:2013bxa,Holstein:2008sx}. \\ 
This analysis can be extended in future works, including higher order terms in $p$, in order to get the $2PM$ potential obtained in \cite{Cheung:2018wkq,Cristofoli:2019neg}.
\chapter*{Conclusions}
\addcontentsline{toc}{chapter}{Conclusions}

The discovery of gravitational waves emitted by a binary coalescing system has opened new possibilities to probe our Universe and to test General Relativity. 
The increase measurements of gravitational waves signals, and the higher experimental precision expected from the next generation of gravitational waves detectors, require a more accurate knowledge of the dynamics of the two-body problem in General Relativity, which has been studied within this thesis following two distinctive, but complementary, perturbative schemes, which are the PN (post-Newtonian) and the PM (post-Minkowskian) expansions. Calculations within the two approaches can be done in an innovative way, by taking advantage of Effective Field Theories and of Scattering Amplitudes multi-loop techniques, that for many years have been developed in Particle Physics, and that only very recently have been applied to this new sector.
The conservative dynamics of a slow inspiral binary can be described by taking deviations from the Newton's two-body potential in terms of Post-Newtonian contributions, where at $n$-PN order one considers corrections that scale as $G_N^{n-l}v^{2l}$ where $0\leq \ell \leq n-1$. PN physics can be studied using an Effective Field theory approach, where the binary dynamics can be described as a system of point particles which are gravitationally interacting in a flat spacetime. In this approach, PN corrections can be organized in terms of Feynman diagrams, which have a clear power counting in terms of $v/c$. Such diagrams, can be topologically mapped into multi-loop two-points and one-point functions, and we took advantage of this property to perform high precision calculations using modern multi-loop techniques. Following the method of regions, the computations can be divided into near and far zone. Focusing on the far-zone dynamics, the binary system can be described as a single source, expanded in terms of multipole moments, and emitting radiation gravitons. Within this framework we considered the so-called Hereditary Effects, which are the effects of the past evolution of a material system on its present gravitational dynamics. They can be described as diagrams containing radiation gravitons emitted by the system and then scattered back into the same GWs source, and they can be divided into back-scattering, tail and memory terms. Moreover, in dealing with radiation problems like this, one should use the so-called \textit{Schwinger-Keldysh} formalism (or In-In or CTP formalism), instead of Feynman formalism. The first original contribution of this thesis has been the development of an automatic computational algorithm in \textit{Mathematica} for the evaluation of Hereditary Effects in PN Physics, simultaneously in Feynman formalism and CTP Formalism. Starting from the effective Lagrangian for the theory, the code evaluates the Feynman rules, generates the diagrams and evaluates them using powerful identities, such as Integration-by-parts identities, to decompose difficult Feynman amplitudes appearing in terms of a basis of few integrals, known as \textit{Master Integrals}, leaving us to the computation of the MIs only. Moreover, we implemented a new way to deal with In-In diagrams, which uses Feynman amplitudes as building blocks for In-In ones, where only the Master Integrals have to be evaluated case-by-case according to the boundary conditions appearing in the propagators of the diagrams. We used this algorithm to evaluate the Hereditary Effects up to 5PN order in both formalism, which involved the evaluation of several 2-loop integrals, and we were able to check our results with those appearing in literature. This powerful algorithm can be easily generalized to compute higher order Hereditary Effects. In literature appear different derivations of the 5PN dynamics, obtained with different techniques, which provide incompatible predictions for physical observables. These differences have to be sorted out, and future works starting from this project can try to solve this puzzle by better understanding the origin of each term appearing, and by comparing them with the ones entering in other computations, paving the way to the derivation of higher order PN effects. 
During the second part of this project, we studied a framework to deal with relativistic gravitational scattering processes, which is based on the Post-Minkowskian approach to General Relativity, which is a weak-field expansion only, where at $n$-PM order we consider contributions that scale as $G_N^n$. By treating General Relativity as the low-energy EFT of a quantum theory of gravity, this formalism gave us the possibility to obtain both classical and quantum contributions to physical observables using Feynman diagrams. With a generalization of the \textit{Lippman-Schwinger} equation, one can study the interaction potential between two bodies from the Fourier transform of the low-energy limit of PM amplitudes. Within this formalism we studied the classical and quantum corrections to the bending angle of a light-like scalar under the influence of a massive object at 1-loop order, which in terms of scattering amplitudes could be described as a gravitational scattering between a massive and a massless scalar object. The second original contribution of this thesis has been the development of a novel computational algorithm which automatically generates the diagrams, and then computes them in dimensional regularization using multi-loop techniques, and we used it to evaluate the amplitude for this process at 1-loop. Since the bending angle evaluation involves a Fourier transform, it was not necessary to compute the full 1-loop amplitude, but we could focus only on the massless double-cut, since it captures the non-analytic pieces, which are the only ones surviving after the Fourier transform.  From the low-energy limit of this amplitude, we obtained the classical and quantum corrections to the bending angle, which was computed using two different techniques, and we found new quantum terms arising which were not covered in previous analyses. As discussed on Chapter\ref{chapter:bending}, the presence of such terms opens the need to review the analyses done so far in \cite{Bjerrum-Bohr:2014zsa,Bai:2016ivl,Chi:2019owc}. In particular, it is necessary to better understand the role of the new logarithmic term arising from the computation, how it is affecting the physical observables, and if the factorization theorems proved in \cite{Akhoury:2013yua,Bjerrum-Bohr:2016hpa} are valid also including such new terms. Considering also the massless particle double-cut \cite{Bai:2016ivl} the situation gets even worse, since there is a divergence appearing in the non-analytic part of the amplitude, which makes any physical prediction meaningless.  Moreover, the appearance of such contributions can be used to put constrains on Wilson coefficients, considering higher order terms in the EFT of gravity. This analysis can be extended also considering different types of external fields, as already done in \cite{Bjerrum-Bohr:2014zsa,Bjerrum-Bohr:2016hpa}, and using modern scattering amplitudes techniques, such as Generalized Unitarity. More in general, one needs to understand which is the physical meaning of the quantum corrections appearing, and such analysis can start only from a full control of the 1-loop quantum amplitude, which can start from this work. In the last part of this master thesis we made use of this 1-loop computation to get another relevant physical result, which is the computation of the 1-loop classical and quantum corrections to the Newtonian potential between two massive scalar fields. Just by making the other field massive, we were able to use the same computational algorithm to get the 1-loop amplitude and the corrections to the potential, finding agreement with previous results appearing in literature. This was an important calculation to check the validity of our computation also for the massless case. Moreover, the computational algorithm can be generalized to higher order in $G_N$, and can be used as an additional test for higher order PM calculations.
Recently, a new interesting method has been proposed \cite{Kosower:2018adc}, which allows obtaining predictions for classical gravitational waves observables directly from scattering amplitudes, without going through intermediate un-physical steps, such as the evaluation of effective actions and potentials. By analyzing in detail the appearance of classical contributions from quantum mechanical calculations, this new way of thinking about gravity could be leading the sector in the next years. \\  The field of Gravitational Waves is very exciting nowadays, and there is an urgent need from the theoretical side to better understand how to compute physical observables with higher precision. The use of multi-loop techniques, and of advanced scattering amplitudes methods, provide a very efficient way to achieve such goals. The gain in using these methods is twofold: on one side one can advance in the phenomenological study of gravitational waves, and of the astrophysical sources that generate them, thanks to the precision predictions available, whereas on the other side one can have a better understanding of Quantum Field Theory and of Scattering amplitudes, and on how to obtain classical physics from quantum mechanics. New properties can be discovered, which can be used to connect the hidden sector of gravity to gauge theories, exploring similarities and differences. The scattering amplitude community is putting a lot of effort in this new promising field of Gravitational Waves, and we expect that new interesting results will arise that will allow us to better understand the fundamental laws of gravity.

\chapter*{Acknowledgements}
First, I want to thank my supervisor Prof. Pierpaolo Mastrolia, for all the time and the patience invested in me during these six months and throughout all my master degree, with the only intent of forming me as a new young physicist, giving me all the best skills required nowadays in this sector. He was able to transmit me the passion for research, making be an active member of his group, with a day-by-day supervision and long discussions where fantastic ideas emerged. He encouraged me to come to the University every day, to participate to group meetings, to have lunches together, and in that breeding environment I was able to find my path in Physics and to start many scientific collaborations. I think that I could not find a better person to guide me through this experience, he was really capable to teach me how to work as a group, and how to deal with new physical problems, and I am really happy to continue working with him as my PhD supervisor. Then I want to thank my co-supervisor Dr. Manoj K. Mandal, for being always present during this project, and for investing a lot of time in my training. He was able to teach me not only how to reason on new physical problems, but also how to efficiently implement my computations, devoting a lot of hours tutoring me in the code development. He helped me also with important choices that I had to make during this period, and I hope to continue collaborating with him in the future. I want also to thank Raj Patil, for collaborating with me during these six months, and for the projects that we are carrying on together with the rest of the group, for introducing me to new topics in many occasions with clear presentations, and for all the insightful discussion done together to understand things. Then I wish to thank all the persons that helped me during my master work, in particular: Prof. Thibault Damour, Prof. Donato Bini, Dr. Andrea Geralico, Prof. Riccardo Sturani, Prof. Stefano Foffa, Gabriel Luz Almeida, Prof. Emil Bjerrum-Bohr, Prof. Pierre Vanhove, Prof. Tiziano Peraro, and Dr. Claudia Lazzaro. I want to thank also the rest of the Padova Amplitude group, for welcoming me inside the team, as a collaborator and as a friend, and for helping me in many occasions.  \\ 
Voglio poi ringraziare tutti quelli che sono stati accanto a me durante questi cinque anni, che mi hanno sostenuto e sopportato anche nei periodi peggiori. Voglio ringraziare Bea, la mia ancora e la mia metà, per la felicità che mi regala ogni giorno, per sostenermi in tutte le mie scelte, per avermi dato la forza per raggiungere i miei obiettivi con la sua presenza, gentilezza e amore, e con cui spero di costruire un futuro insieme. Voglio ringraziare i miei genitori, Meris e Massimo, per aver sempre creduto in me facendomi sentire sempre speciale, per non avermi mai fatto mancare nulla, e senza i quali non sarei arrivato a questo traguardo, e mio fratello Marco, che mi ricorda che ogni tanto bisogna saper vivere con leggerezza. Voglio ringraziare la Zia Nadia, per avermi motivato e aiutato in questi anni, e per essermi stata accanto con tutta la sua bontà e gentilezza. Voglio ringraziare anche i nonni Lidia e Mario, che mi hanno cresciuto dandomi tutto l'amore del mondo, che mi hanno visto iniziare questo percorso, e sono sicuro che continuano a sostenermi sempre. Voglio ringraziare anche Zia Valeria, Zio Franco, Zio Tiziano e i nonni Lina e Franco per la loro presenza durante questo percorso. Voglio infine ringraziare gli amici che mi sono stati vicini, e con i quali ho passato molte belle serate, tra cui: Christian, Filippo, Alberto, Matteo, Brook, Tommaso, Stefano, Lorenzo, Sara e Andrea. 

\appendix
\chapter{Automatic evaluation of Feynman rules}
\label{chapter:feynman_rules_pn}
This chapter will be dedicated to the evaluation of the Feynman rules in momentum space useful to perform the computations in the Near and the Far zone Effective Field Theories. The rules have been obtained by developing an automatic code in \texttt{Mathematica} which perform the evaluation in two stages: 
\begin{itemize}
   \item selects the piece of the action needed for a Feynman rule and 
   \item apply functional derivatives necessary for the derivation of the corresponding rule. 
\end{itemize}
The results have been compared with the ones reported in Appendix.A of \cite{Blumlein:2021txe}, where we found some typos such as in Eq.(114) and in Eq.(118). For simplicity, we will take all momenta as incoming.
\section{Bulk gravitational interactions} 
In order to extract the bulk graviton interactions we need to consider the pure gravitational action:
\begin{eqnarray}
 S_{bulk}= S_{EH}+S_{GF} = 2\Lambda^2\int d^d x dt \sqrt{-g}\biggl(R(g)-\frac{1}{2}\Gamma_\mu\Gamma^\mu\biggr)\ , 
\end{eqnarray}
which has been expanded in $d$-dimensions in terms of $\phi,A,\sigma$,  containing all the relevant interactions up to the relevant PN order as: 
\begin{eqnarray}
   S_{bulk} 
   & \supset &
   \int d^{d+1}x\sqrt{-\gamma}\Biggl\{ \frac{1}{4}\biggl[\bigl(\Vec{\nabla}\sigma\bigr)^2-2\bigl(\Vec{\nabla}\sigma_{ij}\bigr)^2-\bigl(\dot{\sigma}^2-2(\dot{\sigma}_{ij})^2\bigr)^{-\frac{c_d\phi}{\Lambda}}\biggr]\nonumber \\ 
   & & 
   -c_d\biggl[\bigl(\Vec{\nabla}\phi\bigr)^2-\dot{\phi}^2-\dot{\phi}^2e^{-\frac{c_d\phi}{\Lambda}}\biggr] +\biggl[\frac{F_{ij}^2}{2}+\bigl(\Vec{\nabla}\cdot\Vec{A}\bigr)^2-\dot{\Vec{A}}^2e^{-\frac{c_d\phi}{\Lambda}}\biggr]e^{\frac{c_d\phi}{\Lambda}}\nonumber \\ 
   & & 
   +\frac{2}{\Lambda}\biggl[\bigl(F_{ij}A^{i}\dot{A}^j+\Vec{A}\cdot\dot{\Vec{A}}(\Vec{\nabla}\cdot\Vec{A})\bigr)-c_d\dot{\phi}\Vec{A}\cdot\Vec{\nabla}\phi\biggr]+2c_d\biggl(\dot{\phi}\Vec{\nabla}\cdot\Vec{A}-\dot{\Vec{A}}\cdot\Vec{\nabla}\phi\biggr)\nonumber \\ 
   & & 
   +\frac{\dot{\sigma}_{ij}}{\Lambda}\biggl(-\delta^{ij}A_l\hat{\Gamma}^l_{k k}+2A_k\hat{\Gamma}^k_{ij}-2A^i\hat{\Gamma}^j_{k k }\biggr)\nonumber \\ 
   & & 
   -\frac{1}{\Lambda}\biggl(\frac{\sigma}{2}\delta^{ij}-\sigma^{ij}\biggr)\biggl(\sigma_{ik}^{,l}\sigma_{jl}^{,k}-\sigma_{ik}^{,k}\sigma_{jl}^{,l}+\sigma_{, i }\sigma_{jk}^{,k}-\sigma_{ik,j}\sigma^{,k}\biggr)\Biggr\}
\end{eqnarray}
where: $\Lambda=\sqrt{32\pi G_d}^-1$, $G_d=G_N^{d-3} $ is the $d$-dimensional coupling constant, $F_{ij}=A_{j,i}-A_{i,j}$ and $\hat{\Gamma}^i_{jk}$ is the connection of the purely spatial $d-$dimensional metric $\gamma_{ij}=\delta_{ij}+\sigma_{ij}/\Lambda$, which is also used above to raise and contract spatial indices. All spatial derivatives are understood as simple (not covariant) derivatives and when ambiguities might rise gradients are always meant to act on controvariant fields, e.g. $\Vec{\nabla}\cdot\Vec{A}=\gamma^{ij}A_{i,j}$, $F_{ij}^2=\gamma^{ik}\gamma^{jl}F_{ij}F_{kl}$, and $\sigma=\sigma^i_i=\gamma^{ij}\sigma_{ij}$. 
\subsection{Propagators}
From quadratic terms in $S_{bulk}$, we can extract the propagators:
\begin{eqnarray}
 \scalebox{0.9}{
\begin{tikzpicture} 
\begin{feynman}
\vertex (a1) ;
\vertex[right=3cm of a1] (a2); 
\diagram* { 
(a1) -- [scalar,blue] (a2)
};
\end{feynman} 
\end{tikzpicture}} & = & 
-\frac{i}{2c_d}D(p)\ ,\\
 \scalebox{0.9}{\begin{tikzpicture} 
\begin{feynman}
\vertex (a1) {\(i \)} ;
\vertex[right=3cm of a1] (a2) {\(j\)}; 
\diagram* { 
(a1) -- [boson,red] (a2)
};
\end{feynman} 
\end{tikzpicture}} & = & \frac{i\delta_{ij}}{2}D(p)\ ,\\
\scalebox{0.9}{\begin{tikzpicture} 
\begin{feynman}
\vertex (a1) {\( ab \)} ;
\vertex[right=3cm of a1] (a2) {\(cd \)}; 
\diagram* { 
(a1) -- [ double_boson, black!60!green] (a2)
};
\end{feynman} 
\end{tikzpicture}} & = & -\frac{i}{2}D(p)[\delta_{ac}\delta_{bd}+\delta_{ad}\delta_{bc}+(2-c_d)\delta_{ab}\delta_{cd}]\ ,
\end{eqnarray}
where $D(p)=\frac{1}{\mathbf{k}^2-k_0^2}$. \\ 
When performing calculations in the near-zone, since we are dealing with potential gravitons we should take a non-relativistic expansion of the propagator given by: 
\begin{equation}
   D(p)=\frac{1}{\mathbf{k}^2-k_0^2}\approx \frac{1}{\mathbf{k^2}}\biggl(1+\frac{k_0^2}{\mathbf{k}^2}+\ldots\biggr)
\end{equation}
On the other side, when performing far zone calculations only radiation gravitons appear, and we have to keep the full propagator. 
\subsection{Triple graviton vertices}
From the triple terms in $S_{bulk}$ one can extract the "bulk" triple vertices: 
\begin{eqnarray}
 \scalebox{0.9}{\begin{tikzpicture}[baseline=(a2)]
\begin{feynman}
\vertex (a1) {};
\vertex[below=0.75cm of a1] (a2); 
\vertex[below=0.75cm of a2] (a3);
\vertex[left=0.5cm of a3] (a4) {};
\vertex[right=0.5cm of a3] (a5) {};
\vertex[above= 0.1cm of a1] (a6) {\(p_1\)};
\vertex[below= 0.1cm of a4] (a7) {\(p_2\)};
\vertex[below= 0.1cm of a5] (a8) {\(p_3\)};
\diagram* { 
(a1) -- [scalar,blue] (a2),
(a2) -- [scalar,blue] (a4),
(a2) -- [scalar,blue] (a5)
};
\end{feynman} 
\end{tikzpicture}} & = & \frac{2ic_d^2}{\Lambda}(p_{10}p_{20}+p_{10}p_{30}+p_{20}p_{30})\ , \\
 \scalebox{0.9}{\begin{tikzpicture}[baseline=(a2)]
\begin{feynman}
\vertex (a1) ;
\vertex[below=0.75cm of a1] (a2); 
\vertex[below=0.75cm of a2] (a3);
\vertex[left=0.5cm of a3] (a4);
\vertex[right=0.5cm of a3] (a5) ;
\vertex[above= 0.1cm of a1] (a6) {\(p_1\)};
\vertex[below= 0.1cm of a4] (a7) {\( p_2\)};
\vertex[below= 0.1cm of a5] (a8) {\(a , p_3\)};
\diagram* { 
(a1) -- [scalar,blue] (a2),
(a2) -- [scalar,blue] (a4),
(a2) -- [boson,red] (a5)
};
\end{feynman} 
\end{tikzpicture}} & = &  -\frac{2ic_d}{\Lambda}(p_{10}p_{2a}+p_{1a}p_{20})\ , \\
 \scalebox{0.9}{\begin{tikzpicture}[baseline=(a2)]
\begin{feynman}
\vertex (a1);
\vertex[below=0.75cm of a1] (a2); 
\vertex[below=0.75cm of a2] (a3);
\vertex[left=0.5cm of a3] (a4);
\vertex[right=0.5cm of a3] (a5);
\vertex[above= 0.1cm of a1] (a6) {\(p_1\)};
\vertex[below= 0.1cm of a4] (a7) {\( p_2\)};
\vertex[below= 0.1cm of a5] (a8) {\(ab , \ p_3\)};
\diagram* { 
(a1) -- [scalar,blue] (a2),
(a2) -- [scalar,blue] (a4),
(a2) -- [ double_boson, black!60!green] (a5)
};
\end{feynman} 
\end{tikzpicture}} & = & \frac{ic_d}{\Lambda}(-p_{1a}p_{2b}-p_{1b}p_{2a}+\delta_{ab}p_1\cdot p_2) \ , \\
 \scalebox{0.9}{\begin{tikzpicture}[baseline=(a2)]
\begin{feynman}
\vertex (a1) ;
\vertex[below=0.75cm of a1] (a2); 
\vertex[below=0.75cm of a2] (a3);
\vertex[left=0.5cm of a3] (a4);
\vertex[right=0.5cm of a3] (a5);
\vertex[above= 0.1cm of a1] (a6) {\(p_1\)};
\vertex[below= 0.1cm of a4] (a7) {\(a, p_2\)};
\vertex[below= 0.1cm of a5] (a8) {\(b , p_3\)};
\diagram* { 
(a1) -- [scalar,blue] (a2),
(a2) -- [boson, red] (a4),
(a2) -- [boson,red] (a5)
};
\end{feynman} 
\end{tikzpicture}} & = & 
\frac{2 i c_d}{\Lambda}(-p_{2a}p_{3b}+p_{2b}p_{3a}-\delta_{ab}p_2\cdot p_3)\ , \\
 \scalebox{0.9}{\begin{tikzpicture}[baseline=(a2)]
\begin{feynman}
\vertex (a1);
\vertex[below=0.75cm of a1] (a2); 
\vertex[below=0.75cm of a2] (a3);
\vertex[left=0.5cm of a3] (a4);
\vertex[right=0.5cm of a3] (a5);
\vertex[above= 0.1cm of a1] (a6) {\(p_1\)};
\vertex[below= 0.1cm of a4] (a7) {\(a, p_2\)};
\vertex[below= 0.1cm of a5] (a8) {\(bc , p_3\)};
\diagram* { 
(a1) -- [scalar,blue] (a2),
(a2) -- [boson, red] (a4),
(a2) -- [ double_boson, black!60!green] (a5)
};
\end{feynman} 
\end{tikzpicture}} & = & \frac{ic_d}{\Lambda}(-\delta_{ab}p_{10}p_{2c}+\delta_{ab}p_{1c}p_{20}+\delta_{bc}p_{10}p_{2a} -\delta_{bc}p_{1a}p_{20}-\delta_{ac}p_{10}p_{2b}+\delta_{ac}p_{1b}p_{20})\ , \\
 \scalebox{0.9}{\begin{tikzpicture}[baseline=(a2)]
\begin{feynman}
\vertex (a1) ;
\vertex[below=0.75cm of a1] (a2); 
\vertex[below=0.75cm of a2] (a3);
\vertex[left=0.5cm of a3] (a4);
\vertex[right=0.5cm of a3] (a5);
\vertex[above= 0.1cm of a1] (a6) {\(p_1\)};
\vertex[below= 0.1cm of a4] (a7) {\(ab, p_2\)};
\vertex[below= 0.1cm of a5] (a8) {\(cd , p_3\)};
\diagram* { 
(a1) -- [scalar,blue] (a2),
(a2) -- [ double_boson, black!60!green] (a4),
(a2) -- [ double_boson, black!60!green] (a5)
};
\end{feynman} 
\end{tikzpicture}} & = &  \frac{ic_d}{2\Lambda}p_{20}p_{30}(\delta_{ac}\delta_{bd}-\delta_{ab}\delta_{cd}+\delta_{ad}\delta_{bc}) \label{eq:FR_triple_pss}\ , \\
 \scalebox{0.9}{\begin{tikzpicture}[baseline=(a2)]
\begin{feynman}
\vertex (a1) ;
\vertex[below=0.75cm of a1] (a2); 
\vertex[below=0.75cm of a2] (a3);
\vertex[left=0.5cm of a3] (a4);
\vertex[right=0.5cm of a3] (a5);
\vertex[above= 0.1cm of a1] (a6) {\(a, p_1\)};
\vertex[below= 0.1cm of a4] (a7) {\(b, p_2\)};
\vertex[below= 0.1cm of a5] (a8) {\(c , p_3\)};
\diagram* { 
(a1) -- [boson,red] (a2),
(a2) -- [boson,red] (a4),
(a2) -- [boson,red] (a5)
};
\end{feynman} 
\end{tikzpicture}} & = & \frac{2 i}{\Lambda } \biggl(-p_{20} p_{1b} \delta_{ac}+p_{20} p_{1a} \delta_{bc}+p_{1c} (p_{20}-p_{30}) \delta_{ab}+p_{30} p_{1b}
\delta_{ac}+p_{30} p_{1a} \delta_{bc}\nonumber \\
& & +p_{10} p_{2b} \delta_{ac}-p_{10} p_{2a} \delta_{bc}+p_{2c} (p_{10}-p_{30})
\delta_{ab}+p_{10} p_{3c} \delta_{ab}+p_{10} p_{3b} \delta_{ac}\nonumber \\
& &-p_{10} p_{3a} \delta_{bc} +p_{30} p_{2b} \delta_{ac}+p_{30} p_{2a} \delta_{bc}+p_{20} p_{3c} \delta_{ab}-p_{20} p_{3b} \delta_{ac}+p_{20} p_{3a} \delta_{bc}\biggr)\ , \\
 \scalebox{0.9}{\begin{tikzpicture}[baseline=(a2)]
\begin{feynman}
\vertex (a1) ;
\vertex[below=0.75cm of a1] (a2); 
\vertex[below=0.75cm of a2] (a3);
\vertex[left=0.5cm of a3] (a4);
\vertex[right=0.5cm of a3] (a5);
\vertex[above= 0.1cm of a1] (a6) {\(a, p_1\)};
\vertex[below= 0.1cm of a4] (a7) {\(b, p_2\)};
\vertex[below= 0.1cm of a5] (a8) {\(cd , p_3\)};
\diagram* { 
(a1) -- [boson,red] (a2),
(a2) -- [boson,red] (a4),
(a2) -- [ double_boson, black!60!green] (a5)
};
\end{feynman} 
\end{tikzpicture}} & = & \frac{i}{\Lambda } \biggl(p_{1g} p_2^{g} \delta_{ac} \delta_{bd}+p_{1g} p_2^{g} \delta_{ad} \delta_{bc}-p_{1g} p_2^{g} \delta_{ab} \delta_{cd}-p_{1b} p_{2d} \delta_{ac}-p_{1b} p_{2c} \delta_{ad}\nonumber \\
& & +p_{1a} p_{2d} \delta_{bc}+p_{1d} (p_{2c} \delta_{ab}+p_{2b} \delta_{ac}-p_{2a} \delta_{bc})+p_{1a}
   p_{2c} \delta_{bd}\nonumber \\ 
   & & +p_{1c} (p_{2d} \delta_{ab}+p_{2b}
   \delta_{ad}-p_{2a} \delta_{bd})+p_{2a} p_{1b} \delta_{cd}-p_{1a} p_{2b} \delta_{cd}-p_{10} p_{20} \delta_{ac}
   \delta_{bd}\nonumber \\
   & & -p_{10} p_{20} \delta_{ad} \delta_{bc}+p_{10} p_{20}
   \delta_{ab} \delta_{cd}\biggr)\ , \\
 \scalebox{0.9}{\begin{tikzpicture}[baseline=(a2)]
\begin{feynman}
\vertex (a1) ;
\vertex[below=0.75cm of a1] (a2); 
\vertex[below=0.75cm of a2] (a3);
\vertex[left=0.5cm of a3] (a4);
\vertex[right=0.5cm of a3] (a5);
\vertex[above= 0.1cm of a1] (a6) {\(a, \ p_1\)};
\vertex[below= 0.1cm of a4] (a7) {\(bc, p_2\)};
\vertex[below= 0.1cm of a5] (a8) {\(de , p_3\)};
\diagram* { 
(a1) -- [boson, red] (a2),
(a2) -- [ double_boson, black!60!green] (a4),
(a2) -- [ double_boson, black!60!green] (a5)
};
\end{feynman} 
\end{tikzpicture}} & = &\frac{i}{2 \Lambda} \biggl(p_{30} p_{2e} \delta_{ad} \delta_{bc}+p_{30} p_{2a}
   \delta_{bc} \delta_{ed}-p_{30} p_{2c} \delta_{ad} \delta_{be}+p_{30} p_{2e} \delta_{ac} \delta_{bd}-p_{30} p_{2c}
   \delta_{ae} \delta_{bd}\nonumber \\
   & & -p_{30} p_{2b} \delta_{ad} \delta_{ce}-p_{30} p_{2a} \delta_{bd} \delta_{ce}+p_{30} p_{2d}
   (\delta_{ae} \delta_{bc}+\delta_{ac} \delta_{be}+\delta_{ab} \delta_{ce})+p_{30} p_{2e} \delta_{ab} \delta_{cd}  \nonumber \\
   & & -p_{30} p_{2b}
   \delta_{ae} \delta_{cd}-p_{30} p_{2a} \delta_{be} \delta_{cd}-p_{30} p_{2c} \delta_{ab} \delta_{ed}-p_{30} p_{2b}
   \delta_{ac} \delta_{ed}-p_{20} p_{3e} \delta_{ad} \delta_{bc}\nonumber \\
   & & +p_{20} p_{3a} \delta_{bc} \delta_{ed}  +p_{20} p_{3c}
   \delta_{ad} \delta_{be}-p_{20} p_{3e} \delta_{ac} \delta_{bd}+p_{20} p_{3c} \delta_{ae} \delta_{bd}+p_{20} p_{3b}
   \delta_{ad} \delta_\nonumber \\
   & & -p_{20} p_{3a} \delta_{bd} \delta_{ce}-p_{20} p_{3d} (\delta_{ae} \delta_{bc}+\delta_{ac} \delta_{be}+\delta_{ab} \delta_{ce})-p_{20} p_{3e} \delta_{ab} \delta_{cd}\nonumber \\ 
   & & +p_{20} p_{3b} \delta_{ae}\delta_{cd}-p_{20} p_{3a}
   \delta_{be} \delta_{cd}+p_{20} p_{3c} \delta_{ab} \delta_{ed} +p_{20} p_{3b} \delta_{ac} \delta_{ed}\biggr) \ , \\
      \scalebox{0.9}{\begin{tikzpicture}[baseline=(a2)]
      \begin{feynman}
      \vertex (a1) ;
      \vertex[below=0.75cm of a1] (a2); 
      \vertex[below=0.75cm of a2] (a3);
      \vertex[left=0.5cm of a3] (a4);
      \vertex[right=0.5cm of a3] (a5);
      \vertex[above= 0.1cm of a1] (a6) {\(ab, \ p_1\)};
\vertex[below= 0.1cm of a4] (a7) {\(cd, \  p_2\)};
\vertex[below= 0.1cm of a5] (a8) {\(ef , \  p_3\)};
      \diagram* { 
      (a1) -- [ double_boson, black!60!green] (a2),
      (a2) -- [ double_boson, black!60!green] (a4),
      (a2) -- [ double_boson, black!60!green] (a5)
      };
      \end{feynman} 
      \end{tikzpicture}} & = & \frac{i}{32\Lambda}\bigl(\tilde{V}^{\sigma\sigma\sigma}_{ab,cd,ef}+\tilde{V}^{\sigma\sigma\sigma}_{ba,cd,ef}\bigr)
\end{eqnarray}
where: 
\begin{eqnarray}
      \tilde{V}^{\sigma\sigma\sigma}_{ab,cd,ef}&=& 
      V^{\sigma\sigma\sigma}_{ab,cd,ef}+V^{\sigma\sigma\sigma}_{ab,dc,ef}+V^{\sigma\sigma\sigma}_{ab,cd,fe}+V^{\sigma\sigma\sigma}_{ab,dc,fe} \\ 
      V^{\sigma\sigma\sigma}_{ab,cd,ef} & = & 
      (p_1^2+p_1\cdot p_1+p_2^2)\biggl(-\delta_{cd}\bigl(2\delta_{ae}\delta_{bf}-\delta_{ab}\delta_{ef}\bigr) \nonumber \\ 
      & & 
      +2\bigl[\delta_{ac}(4\delta_{be}\delta_{df}-\delta_{bd}\delta_{ef}-\delta_{ab}\delta_{ce}\delta_{df}\bigr]\biggr)\nonumber \\ 
      & & 
      +2\biggl\{ 4\bigl(p_{1f}p_{2b}-p_{1b}p_{2f}\bigr)\delta_{ac}\delta_{de} \nonumber \\ 
      & & 
      +2\bigl[(p_{1a}+p_{2a})p_{2b}\delta_{ce}\delta_{df}-p_{1e}p_{2f}\delta_{ac}\delta_{bd}\bigr]\nonumber \\ 
      & & 
      +\delta_{cd}\bigl[p_{1e}p_{2f}\delta_{ab}+2(p_{1f}p_{2b}-p_{1b}p_{2f})\delta_{ae} \nonumber \\ 
      & & 
      -(p_{1a}+p_{2a})p_{2b}\delta_{ef}\bigr] \nonumber \\ 
      & & 
      p_{2d}\biggl(4p_{1b}\delta_{ae}\delta_{cf}+p_{1c}\bigl(2\delta_{ae}\delta_{bf}-\delta_{ab}\delta_{ef}\bigr)\nonumber \\ 
      & & 
      +2\bigl[\delta_{ac}(p_{1b}\delta_{ef}-2p_{1f}\delta_{be})-p_{1f}\delta_{ab}\delta_{ce}\bigr]\biggr) \nonumber \\ 
      & & 
      + p_{1d}\biggl( p_{1c}\bigl(2\delta_{ae}\delta_{bf}-\delta_{ab}\delta_{ef}\bigr) -4p_{2b}\delta_{ae}\delta_{cf} \nonumber \\ 
      & & 
      +2\biggl[p_{2f}\delta_{ab}\delta_{ce}+\delta_{ac}\bigl(2p_{2f}\delta_{be}-p_{2b}\delta_{ef}\bigr)\biggr]\biggr)\biggr\}
\ .
\end{eqnarray}

\section{Coupling with multipole source}
The coupling with the multipole source can be extracted from $S_{mult}$ (Eq.\eqref{eq:multipole_action}) introduced in the previous chapter and given by: 
\begin{eqnarray}
S_{mult}[\Bar{h},\{Q_a\}]=\int dt \biggl[\frac{1}{2}E \bar{h}_{00}-\frac{1}{2}\epsilon_{ijk}L^i\bar{h}_{0j,k}-\frac{1}{2}Q^{ij}\mathcal{E}_{ij}-\frac{1}{6}O^{ijk}\mathcal{E}_{ij,k}-\frac{2}{3}J^{ij}B_{ij}+\ldots\biggr]
\label{eq:multipole_action_FR}
\end{eqnarray}
\subsection{Linear Emission}
We can get the Linear Coupling with the source, by expanding the electric and magnetic component of the Riemann Tensor at linear order and then substituting the expressions in terms of $\phi,A,\sigma$ fields, obtaining: 
\begin{eqnarray}
   \mathcal{E}_{ij}&=& 
   -\frac{1}{2\Lambda}\biggl(\ddot{\sigma}_{ij}-\dot{A}_{i,j}-\dot{A}_{j,i}-2\phi_{,ij}-\frac{2\delta_{ij}}{d-2}\ddot{\phi}+\mathcal{O}(\Bar{h}^2)\biggr) \\
   B_{ij}&=&
   \frac{1}{4\Lambda}\epsilon_{ikl}\biggl[\frac{1}{2}(\dot{\sigma}_{jk,l}-\dot{\sigma}_{jl,k})+(A_{l,kj}-A_{k,lj})+\frac{2}{d-2}\bigl(\dot{\phi}_{,k}\delta_{jl}-\dot{\phi}_{,l}\delta_{jk}\bigr)+\mathcal{O}(\Bar{h}^2)\biggr] \ .
\end{eqnarray}
 
The Feynman rules are given by:
\begin{eqnarray}
\scalebox{0.9}{
\begin{tikzpicture} 
\begin{feynman}
\vertex (a1) ; 
\vertex[right=1cm of a1,square dot, red ] (a2) {}; 
\vertex[below= 0.1cm of a2] (c1);
\vertex[right=1cm of a2] (a3); 
\vertex[above=1cm of a2] (b2) {\( p_1\)};
\diagram* { 
(a1)-- [double,thick] (a3),
(b2) -- [scalar,blue] (a2)
};
\end{feynman} 
\end{tikzpicture}} &=& \frac{1}{6 \Lambda }\biggl(p_{1a}p_{1b}p_{1c}
  O^{abc}+3 i p_{1a}p_{1b}
  Q^{ab}-6 i E \biggr)\ , \\
\scalebox{0.9}{\begin{tikzpicture} \begin{feynman}
\vertex (a1) ; 
\vertex[right=1cm of a1,square dot, red ] (a2) {}; 
\vertex[below= 0.1cm of a2] (c1) ;
\vertex[right=1cm of a2] (a3); 
\vertex[above=1cm of a2] (b2) {\(a, \ p_1\)};
\diagram* { 
(a1)-- [double,thick] (a3),
(b2) -- [boson,red] (a2)
};
\end{feynman} \end{tikzpicture}} & = & -\frac{1}{6 \Lambda }\biggl(-3 L^b p_{1c}
   \epsilon_{a}^{bc}+p_{1b} (2 i
   p_{1c} J^b_d
   \epsilon_{a}^{cd}+p_{10} p_{1c}
   O_a^{bc}+3 i p_{10} Q_a^b)\biggr) \ , \\
\scalebox{0.9}{\begin{tikzpicture} 
\begin{feynman}
\vertex (a1) ; 
\vertex[right=1cm of a1,square dot, red ] (a2) {}; 
\vertex[below= 0.1cm of a2] (c1);
\vertex[right=1cm of a2 ] (a3) ; 
\vertex[above=1cm of a2] (b2) {\(ab, \ p_1\)};
\diagram* { 
(a1)-- [double,thick] (a3),
(b2) -- [ double_boson, black!60!green] (a2)
};
\end{feynman} \end{tikzpicture}} & = & -\frac{p_{10}}{12 \Lambda } \biggl(2 i p_{1c} J_{ad}
   \epsilon_b^{cd}+2 i p_{1c}J_{bd}
  \epsilon_a^{cd}+p_{10} p_{1c}
   O_{ab}^{c}+3 i p_{10} Q_{ab}\biggr) \ .
\end{eqnarray}

\subsection{Double Emission}

In the computation of the hereditary effects we will need the Feynman rules with two gravitons emitted by an electric quadrupole source. This can be done by expanding the electric component of the Riemann tensor at second order, as: 
\begin{eqnarray}
   \mathcal{E}_{ij}&=& 
  - \frac{1}{\Lambda^2}\biggl(-\frac{1}{4} \partial_aA_j \partial^aA_i - 
    \frac{1}{2} \partial_a\sigma_{ij} \partial_a\phi + (\delta_{ij}\partial_a\phi \partial^a\phi)/(-2 + d)
     + \frac{1}{4} \partial^aA_j \partial_iA_a  
  \nonumber \\ 
     & &   + \frac{1}{2} \partial^a\phi \partial_i\sigma_{ja} 
     - \phi \partial_i
     \partial_tA_j - \frac{1}{2} A^a \partial_i\partial_t\sigma_{ja}
     + 
    \frac{1}{2} A_a \partial_i\partial_t\sigma_{j}^a - 
    A_j \partial_i\partial_t\phi \nonumber \\ 
     & & + \frac{1}{4} \partial^aA_i \partial_jA_a - 
    \frac{1}{2} \partial_iA^a \partial_jA_a
     + \frac{1}{4} \partial_iA_a \partial_jA^a + 
    \frac{1}{2} \partial^a\phi \partial_j\sigma_{ia} - 
    \partial_i\phi \partial_j\phi\nonumber \\ 
     & &   - \frac{2 \partial_i\phi \partial_j\phi}{-2 + d}
    + \frac{1}{2} A^a \partial_j\partial_iA_a - \frac{1}{2} A_a \partial_j\partial_iA^a - 
    2 \phi \partial_j\partial_i\phi - \phi \partial_j
     \partial_tA_i - A_i \partial_j\partial_t\phi \nonumber \\ 
     & & + 
    \frac{1}{2} \partial_i\sigma_{j}^a\partial_tA_a + 
    \frac{1}{2} \partial_j\sigma_i^a\partial_tA_a - 
    \frac{1}{2} \partial_a\sigma_{ij}\partial_t
      A^a + \frac{\delta_{ij}\partial_a\phi\partial_tA^a}{-2 + 
     d}
     - \partial_j\phi\partial_tA_i \nonumber \\ 
     & & - \frac{
    \partial_j\phi\partial_tA_i}{-2 + d} - 
    \partial_i\phi\partial_tA_j - \frac{
    \partial_i\phi\partial_tA_j}{-2 + d}  - 
   \partial_tA_i\partial_tA_j 
    + 
    \frac{1}{4} \partial^aA_j\partial_t\sigma_{ia} - 
    \frac{1}{4} \partial_jA^a\partial_t\sigma_{ia}\nonumber \\ 
   & &  + 
    \frac{1}{4} \partial^aA_i\partial_t\sigma_{ja} - 
    \frac{1}{4} \partial_iA^a\partial_t\sigma_{ja} 
    - \frac{1}{4}\partial_t\sigma_i^a\partial_t\sigma_{ja} - 
    \frac{1}{2} \partial_iA_j\partial_t\phi - 
    \frac{1}{2} \partial_jA_i\partial_t\phi\nonumber \\ 
    & &  - 
    \frac{1}{2}\partial_t\sigma_{ij}\partial_t\phi 
    - \frac{
    3\partial_t\sigma_{ij}\partial_t\phi}{2 (-2 + d)} + \frac{
   \partial_t\sigma_{ji}\partial_t\phi}{
    2 (-2 + d)} + \frac{\delta_{ij}\partial_t\phi^2}{(-2 + 
      d)^2} + \frac{\delta_{ij}\partial_t\phi^2}{-2 + d}\nonumber \\
      & & - 
    \frac{1}{2} A_j \partial_t^2A_i - 
    \frac{1}{2} A_i \partial_t^2
      A_j - \frac{\phi \partial_t^2\sigma_{ij}}{-2 + 
     d} - \frac{\sigma_{ij} \partial_t^2\phi}{-2 + d} + \frac{
    2 \delta_{ij}\phi \partial_t^2\phi}{-2 + d}^2\biggr)
   \end{eqnarray}
   Then, by substituting this expression in \eqref{eq:multipole_action_FR}, we can derive the corresponding Feynman rules for double emission from an electric quadrupole source: 
   \begin{eqnarray}
      \scalebox{0.9}{\begin{tikzpicture}[baseline=(a1)] 
         \begin{feynman}
         \vertex (a1) ; 
         \vertex[right=1cm of a1,square dot, red ] (a2) {}; 
     \vertex[below= 0.2cm of a2] (c1) {\(Q_{ij}\)};
         \vertex[right=1cm of a2 ] (a3); 
         \vertex[above=1cm of a2] (b2);
         \vertex[left=0.5cm of b2] (b1);
         \vertex[right=0.5cm of b2] (b3);
         \vertex[left=0.4cm of b2] (b4) {\(p_1\)};
         \vertex[right=0.1cm of b3] (b5) {\(p_2\)};
         \diagram* { 
         (a1)-- [double,thick] (a3),
         (b1) -- [ scalar,blue] (a2),
         (b3) -- [ scalar,blue] (a2)
         };
         \end{feynman} \end{tikzpicture}} & = & 
         \frac{i}{(d-2) \Lambda ^2} Q_{aa1} \biggl(p_1^a ((d-2) p_{1}^b+d p_{2}^b)+(d-2) p_2^a p_{2}^b\biggr) \\ 
          \scalebox{0.9}{\begin{tikzpicture}[baseline=(a1)]  
         \begin{feynman}
         \vertex (a1) ; 
         \vertex[right=1cm of a1,square dot, red ] (a2) {}; 
         \vertex[below= 0.2cm of a2] (c1) {\(Q_{ij}\)};
         \vertex[right=1cm of a2 ] (a3); 
         \vertex[above=1cm of a2] (b2);
         \vertex[left=0.5cm of b2] (b1) ;
         \vertex[right=0.5cm of b2] (b3);
         \vertex[left=0.6cm of b2] (b4) {\(a, p_1\)};
         \vertex[right=0.1cm of b3] (b5) {\(b, p_2\)};
         
         \diagram* { 
         (a1)-- [double,thick] (a3),
         (b1) -- [boson ,red, ] (a2),
         (b3) -- [boson ,red] (a2)
         };
         \end{feynman} \end{tikzpicture}} & = & 
         \frac{i}{4\Lambda^2} \biggl(p_1^c (p_{2}^d \delta^{ab}Q_{cd}-p_2^a Q_{d}^b)+Q^{ab} \bigl(p_1^c
   p_{2c}+2 (p_{10}+p_{20})^2\bigr)\nonumber \\ 
   & & -p_2^cp_1^b Q^a_c\biggr)\\ 
    \scalebox{0.9}{\begin{tikzpicture}[baseline=(a1)]  
         \begin{feynman}
         \vertex (a1) ; 
         \vertex[right=1cm of a1,square dot, red ] (a2) {}; 
         \vertex[below= 0.2cm of a2] (c1) {\(Q_{ij}\)};
         \vertex[right=1cm of a2 ] (a3); 
         \vertex[above=1cm of a2] (b2);
         \vertex[left=0.5cm of b2] (b1);
         \vertex[right=0.5cm of b2] (b3);
         \vertex[left=0.4cm of b2] (b4) {\(ab, p_1\)};
         \vertex[right=0.1cm of b3] (b5) {\(cd, p_2\)};
         \diagram* { 
         (a1)-- [double,thick] (a3),
         (b1) -- [double_boson ,black!60!green] (a2),
         (b3) -- [double_boson ,black!60!green] (a2)
         };
         \end{feynman} \end{tikzpicture}} & = & 
         \frac{i p_{10} p_{20}}{16 \Lambda ^2} \biggl(\delta^{ac} Q^{bd}+\delta^{ad} Q^{bc}+Q^{ad} \delta^{bc}+Q^{ac} \delta^{bd}\biggr)\\
         \scalebox{0.9}{\begin{tikzpicture}[baseline=(a1)]  
         \begin{feynman}
         \vertex (a1) ; 
         \vertex[right=1cm of a1,square dot, red ] (a2) {}; 
         \vertex[below= 0.2cm of a2] (c1) {\(Q_{ij}\)};
         \vertex[right=1cm of a2 ] (a3); 
         \vertex[above=1cm of a2] (b2);
         \vertex[left=0.5cm of b2] (b1);
         \vertex[right=0.5cm of b2] (b3);
         \vertex[left=0.4cm of b2] (b4) {\(ab, p_2\)};
         \vertex[right=0.1cm of b3] (b5) {\(p_1\)};
         \diagram* { 
         (a1)-- [double,thick] (a3),
         (b1) -- [double_boson ,black!60!green] (a2),
         (b3) -- [scalar ,blue] (a2)
         };
         \end{feynman} \end{tikzpicture}} & = & 
       -\frac{i}{4 (d-2) \Lambda ^2} \biggl(-Q^{ab} \left((d-2) p_1^c p_{2c}+d p_{10} p_{20}+2 p_{10}^2+2 p_{20}^2\right)\nonumber \\ 
       & & +(d-2)p_{2}^c
  p_{1}^b Q^a_c+(d-2) p_1^a p_{2}^c Q^b_c\biggr) \\ 
    \scalebox{0.9}{\begin{tikzpicture}[baseline=(a1)]  
         \begin{feynman}
         \vertex (a1) ; 
         \vertex[right=1cm of a1,square dot, red ] (a2) {}; 
         \vertex[below= 0.2cm of a2] (c1) {\(Q_{ij}\)};
         \vertex[right=1cm of a2 ] (a3); 
         \vertex[above=1cm of a2] (b2);
         \vertex[left=0.5cm of b2] (b1);
         \vertex[right=0.5cm of b2] (b3);
         \vertex[left=0.4cm of b2] (b4) {\(bc, p_2\)};
         \vertex[right=0.1cm of b3] (b5) {\(a, p_1\)};
         \diagram* { 
         (a1)-- [double,thick] (a3),
         (b1) -- [double_boson ,black!60!green] (a2),
         (b3) -- [photon ,red] (a2)
         };
         \end{feynman} \end{tikzpicture}} & = & 
       \frac{i}{8 \Lambda ^2} (-p_{20} p_1^d \delta^{ab} Q^c_d-p_{20} p_1^d \delta^{ac} Q^b_c+2 p_{10} p_2^d
   \delta^{ab} Q^c_d+2 p_{10} p_2^d \delta^{ac} Q^b_d\nonumber \\ 
   & & 
   +p_{20} p_1^c Q^{ab}+p_{20}p_{1}^b Q^{ac}-2
   p_{10} p_2^a Q^{bc}) \\ 
   \scalebox{0.9}{\begin{tikzpicture}[baseline=(a1)]  
         \begin{feynman}
         \vertex (a1) ; 
         \vertex[right=1cm of a1,square dot, red ] (a2) {}; 
         \vertex[below= 0.2cm of a2] (c1) {\(Q_{ij}\)};
         \vertex[right=1cm of a2 ] (a3); 
         \vertex[above=1cm of a2] (b2);
         \vertex[left=0.5cm of b2] (b1);
         \vertex[right=0.5cm of b2] (b3);
         \vertex[left=0.4cm of b2] (b4) {\(a, p_2\)};
         \vertex[right=0.1cm of b3] (b5) {\( p_1\)};
         \diagram* { 
         (a1)-- [double,thick] (a3),
         (b1) -- [boson ,red] (a2),
         (b3) -- [scalar ,blue] (a2)
         };
         \end{feynman} \end{tikzpicture}} & = & 
       -\frac{i}{2 (d-2) \Lambda ^2} Q^a_b \biggl(2 (d-1) p_{20} p_1^b+(d-2) p_{10} p_2^b\biggr) 
\end{eqnarray}
In Eq.(111) of \cite{Blumlein:2021txe} the ellipses denote higher order multipoles coupled to 2 sigma fields.

\section{Couplings with worldlines}
 For the evaluation of the Near zone contributions to the conservative dynamics one needs the coupling between graviton fields and worldlines, which can be extracted from the point-particle action \eqref{eq:pp_action_kk}, by expanding it in terms of $\phi$, $A$,$\sigma$:
 \begin{equation}
   S_{pp}=-m_a\int dt e^{\frac{\phi}{\Lambda}}\sqrt{1-\frac{2v_iA^i}{\Lambda}-\gamma_{ij}v^{i}v^{j}e^{-c_d\frac{\phi}{\Lambda}}+\frac{(A_iv^i)^2}{\Lambda^2}} \ .
 \end{equation}
The single emission rules are given by: 
\begin{eqnarray}
      \begin{tikzpicture}[baseline=(b)]
      \begin{feynman}
      \vertex (a);
      \vertex[right=1cm of a] (b);
      \vertex[right=1cm of b] (c);
      \vertex[above=1cm of b] (d) ;
      \diagram* {
      (a) -- [] (b) -- [] (c),
      (b) -- [blue, scalar] (d),};
      \end{feynman}
      \end{tikzpicture}
      & = & 
      -\frac{im_a}{\Lambda}\frac{1}{\sqrt{1-v^2}}\biggl(1+\frac{c_d-2}{2}\vec{v}^2\biggr) \approx -\frac{im_a}{\Lambda}\biggl(1+\frac{c_d-1}{2}\vec{v}^2+\mathcal{O}[v^4]\biggr)\ ,  \\  
      \begin{tikzpicture}[baseline=(b)]
         \begin{feynman}
         \vertex (a);
         \vertex[right=1cm of a] (b);
         \vertex[right=1cm of b] (c);
         \vertex[above=1cm of b] (d) {\(i \)};
         \diagram* {
         (a) -- (b) --  (c),
         (b) -- [red , photon] (d),};
         \end{feynman}
         \end{tikzpicture}
        & = & 
        \frac{im_av_i}{\Lambda}\frac{1}{\sqrt{1-\vec{v}^2}}\approx \frac{im_av_i}{\Lambda}\biggl(1+\frac{1}{2}\vec{v}^2
         +\frac{3}{8}\vec{v}^4+\mathcal{O}[v^6]\biggr) \ ,  \\ 
         \begin{tikzpicture}[baseline=(b)]
            \begin{feynman}
            \vertex (a);
            \vertex[right=1cm of a] (b);
            \vertex[right=1cm of b] (c);
            \vertex[above=1cm of b] (d) {\( ij \)};
            \diagram* {
            (a) --  (b) --  (c),
            (b) -- [double_boson, black!60!green] (d),};
            \end{feynman}
            \end{tikzpicture}
           & =& \frac{im_av_iv_j}{2\Lambda}\frac{1}{\sqrt{1-\vec{v}^2}} \approx \frac{im_av_iv_j}{2\Lambda}\biggl(1+\frac{1}{2}\vec{v}^2
            +\frac{3}{8}\vec{v}^4+\mathcal{O}[v^6]\biggr) \ ,
\end{eqnarray}
which can then be expanded in powers of velocities according to the desired PN order.
Analogously, one can compute the Feynman rules for multiple graviton emissions, such as: 
\begin{equation}
      \begin{tikzpicture}[baseline=(b)]
      \begin{feynman}
      \vertex (a);
      \vertex[right=1cm of a] (b);
      \vertex[right=1cm of b] (c);
      \vertex[above=1cm of b] (d);
      \vertex[left=0.5cm of d] (e);
      \vertex[right=0.5cm of d] (f);
      \diagram* {
      (a) --  (b) --  (c),
      (b) -- [blue, scalar] (d),
      (b) -- [blue, scalar] (e),
      (b) -- [blue, scalar] (f),};
      \end{feynman}
      \end{tikzpicture}
      =-\frac{im_a}{\Lambda^nn!}+\mathcal{O}[v^2]
\end{equation}
\section{An example: the triple graviton interaction $\phi\sigma^2$}
Let us give an example of the evaluation of the Feynman rules in order to understand how the code works. 
Let us suppose to be interested in the evaluation of the triple graviton vertex $\phi,\sigma^2$, given in Eq.\eqref{eq:FR_triple_pss}. \\
In order to do so we have to identify the corresponding piece of the action, with an appropriate function given by:
\begin{figure}[H]
   \centering
   \includegraphics[width=0.8\textwidth]{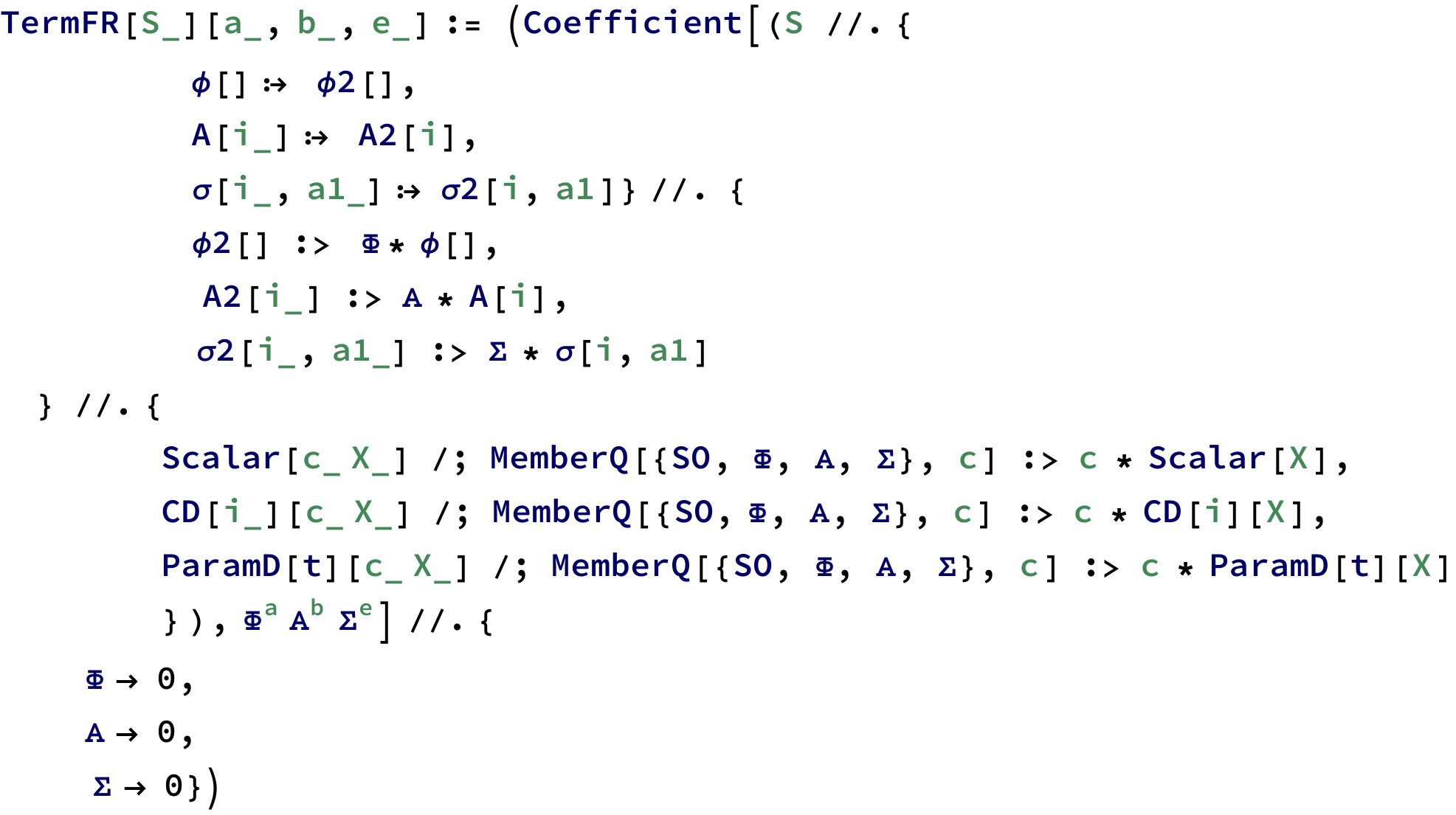}
\end{figure} 
So the term corresponding to the $\phi\sigma^2$ interaction will be given by \textit{TermFR[1,0,2]}
Then we have to take functional derivatives with appropriate definitions, in momentum space, and
the Feynman rule is given by: 
\begin{figure}[H]
   \centering
   \includegraphics[width=0.8\textwidth]{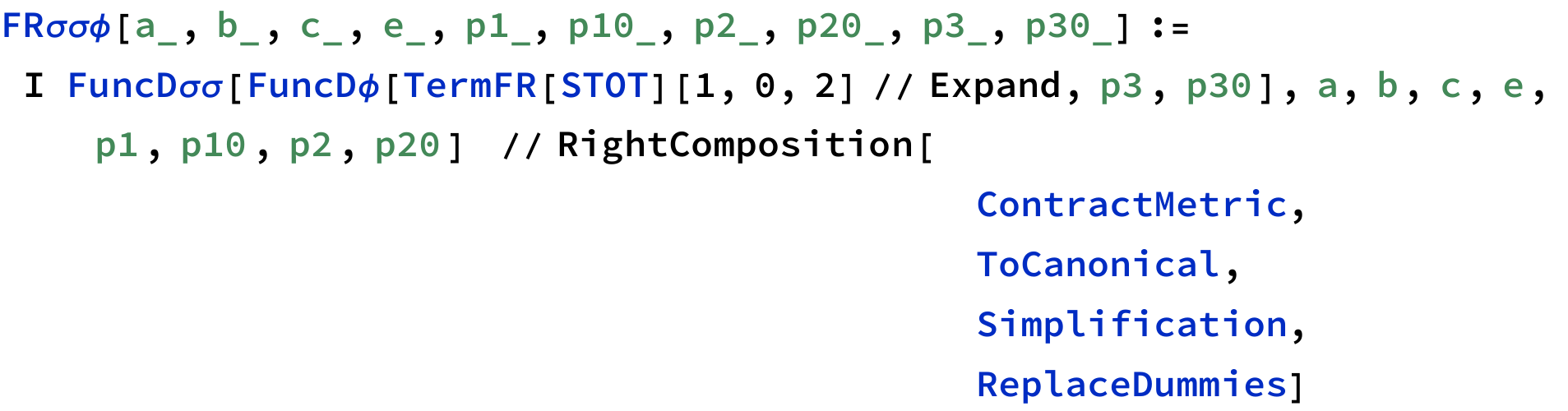}
\end{figure}
\bibliographystyle{unsrt}
\bibliography{bibliography}

\begin{thebibliography}{100}

\bibitem{LIGOScientific:2014pky}
J.~Aasi et~al.
\newblock {Advanced LIGO}.
\newblock {\em Class. Quant. Grav.}, 32:074001, 2015.

\bibitem{VIRGO:2014yos}
F.~Acernese et~al.
\newblock {Advanced Virgo: a second-generation interferometric gravitational
  wave detector}.
\newblock {\em Class. Quant. Grav.}, 32(2):024001, 2015.

\bibitem{LIGOScientific:2016aoc}
B.~P. Abbott et~al.
\newblock {Observation of Gravitational Waves from a Binary Black Hole Merger}.
\newblock {\em Phys. Rev. Lett.}, 116(6):061102, 2016.

\bibitem{KAGRA:2020agh}
T.~Akutsu et~al.
\newblock {Overview of KAGRA: Calibration, detector characterization, physical
  environmental monitors, and the geophysics interferometer}.
\newblock {\em PTEP}, 2021(5):05A102, 2021.

\bibitem{LIGOScientific:2018mvr}
B.~P. Abbott et~al.
\newblock {GWTC-1: A Gravitational-Wave Transient Catalog of Compact Binary
  Mergers Observed by LIGO and Virgo during the First and Second Observing
  Runs}.
\newblock {\em Phys. Rev. X}, 9(3):031040, 2019.

\bibitem{LIGOScientific:2020ibl}
R.~Abbott et~al.
\newblock {GWTC-2: Compact Binary Coalescences Observed by LIGO and Virgo
  During the First Half of the Third Observing Run}.
\newblock {\em Phys. Rev. X}, 11:021053, 2021.

\bibitem{LIGOScientific:2021qlt}
R.~Abbott et~al.
\newblock {Observation of Gravitational Waves from Two Neutron
  Star\textendash{}Black Hole Coalescences}.
\newblock {\em Astrophys. J. Lett.}, 915(1):L5, 2021.

\bibitem{LIGOScientific:2021djp}
R.~Abbott et~al.
\newblock {GWTC-3: Compact Binary Coalescences Observed by LIGO and Virgo
  During the Second Part of the Third Observing Run}.
\newblock 11 2021.

\bibitem{Usman:2015kfa}
Samantha~A. Usman et~al.
\newblock {The PyCBC search for gravitational waves from compact binary
  coalescence}.
\newblock {\em Class. Quant. Grav.}, 33(21):215004, 2016.

\bibitem{LISA:2017pwj}
Pau Amaro-Seoane et~al.
\newblock {Laser Interferometer Space Antenna}.
\newblock 2 2017.

\bibitem{Saleem:2021iwi}
M.~Saleem et~al.
\newblock {The science case for LIGO-India}.
\newblock {\em Class. Quant. Grav.}, 39(2):025004, 2022.

\bibitem{Reitze:2019iox}
David Reitze et~al.
\newblock {Cosmic Explorer: The U.S. Contribution to Gravitational-Wave
  Astronomy beyond LIGO}.
\newblock {\em Bull. Am. Astron. Soc.}, 51(7):035, 2019.

\bibitem{Punturo:2010zz}
M.~Punturo et~al.
\newblock {The Einstein Telescope: A third-generation gravitational wave
  observatory}.
\newblock {\em Class. Quant. Grav.}, 27:194002, 2010.

\bibitem{Blanchet:2013haa}
Luc Blanchet.
\newblock {Gravitational Radiation from Post-Newtonian Sources and Inspiralling
  Compact Binaries}.
\newblock {\em Living Rev. Rel.}, 17:2, 2014.

\bibitem{Porto:2016pyg}
Rafael~A. Porto.
\newblock {The effective field theorist\textquoteright{}s approach to
  gravitational dynamics}.
\newblock {\em Phys. Rept.}, 633:1--104, 2016.

\bibitem{Schafer:2018kuf}
Gerhard Sch\"afer and Piotr Jaranowski.
\newblock {Hamiltonian formulation of general relativity and post-Newtonian
  dynamics of compact binaries}.
\newblock {\em Living Rev. Rel.}, 21(1):7, 2018.

\bibitem{Barack:2018yvs}
Leor Barack and Adam Pound.
\newblock {Self-force and radiation reaction in general relativity}.
\newblock {\em Rept. Prog. Phys.}, 82(1):016904, 2019.

\bibitem{Buonanno:1998gg}
A.~Buonanno and T.~Damour.
\newblock {Effective one-body approach to general relativistic two-body
  dynamics}.
\newblock {\em Phys. Rev. D}, 59:084006, 1999.

\bibitem{Einstein:1938yz}
Albert Einstein, L.~Infeld, and B.~Hoffmann.
\newblock {The Gravitational equations and the problem of motion}.
\newblock {\em Annals Math.}, 39:65--100, 1938.

\bibitem{Ohta:1973je}
T.~Ohta, H.~Okamura, T.~Kimura, and K.~Hiida.
\newblock {Physically acceptable solution of einstein's equation for many-body
  system}.
\newblock {\em Prog. Theor. Phys.}, 50:492--514, 1973.

\bibitem{Jaranowski:1997ky}
Piotr Jaranowski and Gerhard Schaefer.
\newblock {Third postNewtonian higher order ADM Hamilton dynamics for two-body
  point mass systems}.
\newblock {\em Phys. Rev. D}, 57:7274--7291, 1998.
\newblock [Erratum: Phys.Rev.D 63, 029902 (2001)].

\bibitem{Damour:1999cr}
Thibault Damour, Piotr Jaranowski, and Gerhard Schaefer.
\newblock {Dynamical invariants for general relativistic two-body systems at
  the third postNewtonian approximation}.
\newblock {\em Phys. Rev. D}, 62:044024, 2000.

\bibitem{Blanchet:2000nv}
Luc Blanchet and Guillaume Faye.
\newblock {Equations of motion of point particle binaries at the third
  postNewtonian order}.
\newblock {\em Phys. Lett. A}, 271:58, 2000.

\bibitem{Damour:2001bu}
Thibault Damour, Piotr Jaranowski, and Gerhard Schaefer.
\newblock {Dimensional regularization of the gravitational interaction of point
  masses}.
\newblock {\em Phys. Lett. B}, 513:147--155, 2001.

\bibitem{Damour:2014jta}
Thibault Damour, Piotr Jaranowski, and Gerhard Sch\"afer.
\newblock {Nonlocal-in-time action for the fourth post-Newtonian conservative
  dynamics of two-body systems}.
\newblock {\em Phys. Rev. D}, 89(6):064058, 2014.

\bibitem{Jaranowski:2015lha}
Piotr Jaranowski and Gerhard Sch\"afer.
\newblock {Derivation of local-in-time fourth post-Newtonian ADM Hamiltonian
  for spinless compact binaries}.
\newblock {\em Phys. Rev. D}, 92(12):124043, 2015.

\bibitem{Damour:2016abl}
Thibault Damour, Piotr Jaranowski, and Gerhard Sch\"afer.
\newblock {Conservative dynamics of two-body systems at the fourth
  post-Newtonian approximation of general relativity}.
\newblock {\em Phys. Rev. D}, 93(8):084014, 2016.

\bibitem{Bini:2019nra}
Donato Bini, Thibault Damour, and Andrea Geralico.
\newblock {Novel approach to binary dynamics: application to the fifth
  post-Newtonian level}.
\newblock {\em Phys. Rev. Lett.}, 123(23):231104, 2019.

\bibitem{Bini:2020rzn}
Donato Bini, Thibault Damour, Andrea Geralico, Stefano Laporta, and Pierpaolo
  Mastrolia.
\newblock {Gravitational scattering at the seventh order in $G$: nonlocal
  contribution at the sixth post-Newtonian accuracy}.
\newblock {\em Phys. Rev. D}, 103(4):044038, 2021.

\bibitem{Bini:2020uiq}
Donato Bini, Thibault Damour, Andrea Geralico, Stefano Laporta, and Pierpaolo
  Mastrolia.
\newblock {Gravitational dynamics at $O(G^6)$: perturbative gravitational
  scattering meets experimental mathematics}.
\newblock 8 2020.

\bibitem{Georgi:1990um}
Howard Georgi.
\newblock {An Effective Field Theory for Heavy Quarks at Low-energies}.
\newblock {\em Phys. Lett. B}, 240:447--450, 1990.

\bibitem{Kaplan:2005es}
David~B. Kaplan.
\newblock {Five lectures on effective field theory}.
\newblock 10 2005.

\bibitem{Goldberger:2007hy}
Walter~D. Goldberger.
\newblock {Les Houches lectures on effective field theories and gravitational
  radiation}.
\newblock In {\em {Les Houches Summer School - Session 86: Particle Physics and
  Cosmology: The Fabric of Spacetime}}, 1 2007.

\bibitem{Manohar:2018aog}
Aneesh~V. Manohar.
\newblock {Introduction to Effective Field Theories}.
\newblock 4 2018.

\bibitem{Damour:1995kt}
Thibault Damour and Gilles Esposito-Farese.
\newblock {Testing gravity to second postNewtonian order: A Field theory
  approach}.
\newblock {\em Phys. Rev. D}, 53:5541--5578, 1996.

\bibitem{Goldberger:2004jt}
Walter~D. Goldberger and Ira~Z. Rothstein.
\newblock {An Effective field theory of gravity for extended objects}.
\newblock {\em Phys. Rev. D}, 73:104029, 2006.

\bibitem{Porto:2005ac}
Rafael~A. Porto.
\newblock {Post-Newtonian corrections to the motion of spinning bodies in
  NRGR}.
\newblock {\em Phys. Rev. D}, 73:104031, 2006.

\bibitem{Foffa:2013qca}
Stefano Foffa and Riccardo Sturani.
\newblock {Effective field theory methods to model compact binaries}.
\newblock {\em Class. Quant. Grav.}, 31(4):043001, 2014.

\bibitem{Rothstein:2014sra}
Ira~Z. Rothstein.
\newblock {Progress in effective field theory approach to the binary inspiral
  problem}.
\newblock {\em Gen. Rel. Grav.}, 46:1726, 2014.

\bibitem{Levi:2018nxp}
Mich\`ele Levi.
\newblock {Effective Field Theories of Post-Newtonian Gravity: A comprehensive
  review}.
\newblock {\em Rept. Prog. Phys.}, 83(7):075901, 2020.

\bibitem{Goldberger:2022ebt}
Walter~D. Goldberger.
\newblock {Effective field theories of gravity and compact binary dynamics: A
  Snowmass 2021 whitepaper}.
\newblock In {\em {2022 Snowmass Summer Study}}, 6 2022.

\bibitem{Blumlein:2022qjy}
J.~Bl\"umlein, A.~Maier, P.~Marquard, and G.~Sch\"afer.
\newblock {Gravity in binary systems at the fifth and sixth post-Newtonian
  order}.
\newblock In {\em {16th DESY Workshop on Elementary Particle Physics: Loops and
  Legs in Quantum Field Theory 2022}}, 8 2022.

\bibitem{Kol:2007bc}
Barak Kol and Michael Smolkin.
\newblock {Non-Relativistic Gravitation: From Newton to Einstein and Back}.
\newblock {\em Class. Quant. Grav.}, 25:145011, 2008.

\bibitem{Kol:2007rx}
Barak Kol and Michael Smolkin.
\newblock {Classical Effective Field Theory and Caged Black Holes}.
\newblock {\em Phys. Rev. D}, 77:064033, 2008.

\bibitem{Kol:2010ze}
Barak Kol, Michele Levi, and Michael Smolkin.
\newblock {Comparing space+time decompositions in the post-Newtonian limit}.
\newblock {\em Class. Quant. Grav.}, 28:145021, 2011.

\bibitem{Kol:2010si}
Barak Kol and Michael Smolkin.
\newblock {Einstein's action and the harmonic gauge in terms of Newtonian
  fields}.
\newblock {\em Phys. Rev. D}, 85:044029, 2012.

\bibitem{Beneke_1998}
M.~Beneke and V.A. Smirnov.
\newblock Asymptotic expansion of feynman integrals near threshold.
\newblock {\em Nuclear Physics B}, 522(1-2):321--344, jun 1998.

\bibitem{Smirnov:2002pj}
Vladimir~A. Smirnov.
\newblock {Applied asymptotic expansions in momenta and masses}.
\newblock {\em Springer Tracts Mod. Phys.}, 177:1--262, 2002.

\bibitem{Manohar:2006nz}
Aneesh~V. Manohar and Iain~W. Stewart.
\newblock {The Zero-Bin and Mode Factorization in Quantum Field Theory}.
\newblock {\em Phys. Rev. D}, 76:074002, 2007.

\bibitem{Jantzen:2011nz}
Bernd Jantzen.
\newblock {Foundation and generalization of the expansion by regions}.
\newblock {\em JHEP}, 12:076, 2011.

\bibitem{RevModPhys.52.299}
Kip~S. Thorne.
\newblock Multipole expansions of gravitational radiation.
\newblock {\em Rev. Mod. Phys.}, 52:299--339, Apr 1980.

\bibitem{Goldberger:2009qd}
Walter~D. Goldberger and Andreas Ross.
\newblock {Gravitational radiative corrections from effective field theory}.
\newblock {\em Phys. Rev. D}, 81:124015, 2010.

\bibitem{Ross:2012fc}
Andreas Ross.
\newblock {Multipole expansion at the level of the action}.
\newblock {\em Phys. Rev. D}, 85:125033, 2012.

\bibitem{Foffa:2016rgu}
Stefano Foffa, Pierpaolo Mastrolia, Riccardo Sturani, and Christian Sturm.
\newblock {Effective field theory approach to the gravitational two-body
  dynamics, at fourth post-Newtonian order and quintic in the Newton constant}.
\newblock {\em Phys. Rev. D}, 95(10):104009, 2017.

\bibitem{Chetyrkin:1981qh}
K.~G. Chetyrkin and F.~V. Tkachov.
\newblock {Integration by Parts: The Algorithm to Calculate beta Functions in 4
  Loops}.
\newblock {\em Nucl. Phys. B}, 192:159--204, 1981.

\bibitem{Tkachov:1981wb}
F.~V. Tkachov.
\newblock {A Theorem on Analytical Calculability of Four Loop Renormalization
  Group Functions}.
\newblock {\em Phys. Lett. B}, 100:65--68, 1981.

\bibitem{Laporta:2000dsw}
S.~Laporta.
\newblock {High precision calculation of multiloop Feynman integrals by
  difference equations}.
\newblock {\em Int. J. Mod. Phys. A}, 15:5087--5159, 2000.

\bibitem{Laporta:2000dc}
S.~Laporta.
\newblock {Calculation of master integrals by difference equations}.
\newblock {\em Phys. Lett. B}, 504:188--194, 2001.

\bibitem{Kotikov:1990kg}
A.~V. Kotikov.
\newblock {Differential equations method: New technique for massive Feynman
  diagrams calculation}.
\newblock {\em Phys. Lett. B}, 254:158--164, 1991.

\bibitem{Henn:2013pwa}
Johannes~M. Henn.
\newblock {Multiloop integrals in dimensional regularization made simple}.
\newblock {\em Phys. Rev. Lett.}, 110:251601, 2013.

\bibitem{Argeri:2007up}
Mario Argeri and Pierpaolo Mastrolia.
\newblock {Feynman Diagrams and Differential Equations}.
\newblock {\em Int. J. Mod. Phys. A}, 22:4375--4436, 2007.

\bibitem{Argeri:2014qva}
Mario Argeri, Stefano Di~Vita, Pierpaolo Mastrolia, Edoardo Mirabella, Johannes
  Schlenk, Ulrich Schubert, and Lorenzo Tancredi.
\newblock {Magnus and Dyson Series for Master Integrals}.
\newblock {\em JHEP}, 03:082, 2014.

\bibitem{Gilmore:2008gq}
James~B. Gilmore and Andreas Ross.
\newblock {Effective field theory calculation of second post-Newtonian binary
  dynamics}.
\newblock {\em Phys. Rev. D}, 78:124021, 2008.

\bibitem{Foffa:2011ub}
Stefano Foffa and Riccardo Sturani.
\newblock {Effective field theory calculation of conservative binary dynamics
  at third post-Newtonian order}.
\newblock {\em Phys. Rev. D}, 84:044031, 2011.

\bibitem{Foffa:2012rn}
Stefano Foffa and Riccardo Sturani.
\newblock {Dynamics of the gravitational two-body problem at fourth
  post-Newtonian order and at quadratic order in the Newton constant}.
\newblock {\em Phys. Rev. D}, 87(6):064011, 2013.

\bibitem{Foffa:2019hrb}
Stefano Foffa, Pierpaolo Mastrolia, Riccardo Sturani, Christian Sturm, and
  William~J. Torres~Bobadilla.
\newblock {Static two-body potential at fifth post-Newtonian order}.
\newblock {\em Phys. Rev. Lett.}, 122(24):241605, 2019.

\bibitem{Foffa:2019ahl}
Stefano Foffa, Pierpaolo Mastrolia, Riccardo Sturani, Christian Sturm, and
  William~J. Torres~Bobadilla.
\newblock {Calculating the static gravitational two-body potential to fifth
  post-Newtonian order with Feynman diagrams}.
\newblock {\em PoS}, RADCOR2019:027, 2019.

\bibitem{Foffa:2019rdf}
Stefano Foffa and Riccardo Sturani.
\newblock {Conservative dynamics of binary systems to fourth Post-Newtonian
  order in the EFT approach I: Regularized Lagrangian}.
\newblock {\em Phys. Rev. D}, 100(2):024047, 2019.

\bibitem{Mandal:2022nty}
Manoj~K. Mandal, Pierpaolo Mastrolia, Raj Patil, and Jan Steinhoff.
\newblock {Gravitational Spin-Orbit Hamiltonian at NNNLO in the post-Newtonian
  framework}.
\newblock 9 2022.

\bibitem{Levi:2022SO}
Jung-Wook Kim, Michèle Levi, and Zhewei Yin.
\newblock N$^3$lo spin-orbit interaction via the eft of spinning gravitating
  objects, 2022.

\bibitem{Blumlein:2019zku}
J.~Bl\"umlein, A.~Maier, and P.~Marquard.
\newblock {Five-Loop Static Contribution to the Gravitational Interaction
  Potential of Two Point Masses}.
\newblock {\em Phys. Lett. B}, 800:135100, 2020.

\bibitem{Blumlein:2020pog}
J.~Bl\"umlein, A.~Maier, P.~Marquard, and G.~Sch\"afer.
\newblock {Fourth post-Newtonian Hamiltonian dynamics of two-body systems from
  an effective field theory approach}.
\newblock {\em Nucl. Phys. B}, 955:115041, 2020.

\bibitem{Blumlein:2020pyo}
J.~Bl\"umlein, A.~Maier, P.~Marquard, and G.~Sch\"afer.
\newblock {The fifth-order post-Newtonian Hamiltonian dynamics of two-body
  systems from an effective field theory approach: potential contributions}.
\newblock {\em Nucl. Phys. B}, 965:115352, 2021.

\bibitem{Blumlein:2020znm}
J.~Bl\"umlein, A.~Maier, P.~Marquard, and G.~Sch\"afer.
\newblock {Testing binary dynamics in gravity at the sixth post-Newtonian
  level}.
\newblock {\em Phys. Lett. B}, 807:135496, 2020.

\bibitem{Blumlein:2021txe}
J.~Bl\"umlein, A.~Maier, P.~Marquard, and G.~Sch\"afer.
\newblock {The fifth-order post-Newtonian Hamiltonian dynamics of two-body
  systems from an effective field theory approach}.
\newblock {\em Nucl. Phys. B}, 983:115900, 2022.

\bibitem{Blumlein:2021txj}
J.~Bl\"umlein, A.~Maier, P.~Marquard, and G.~Sch\"afer.
\newblock {The 6th post-Newtonian potential terms at $O(G_N^4)$}.
\newblock {\em Phys. Lett. B}, 816:136260, 2021.

\bibitem{Blanchet:1987wq}
Luc Blanchet and Thibault Damour.
\newblock {Tail Transported Temporal Correlations in the Dynamics of a
  Gravitating System}.
\newblock {\em Phys. Rev. D}, 37:1410, 1988.

\bibitem{Blanchet:1992br}
Luc Blanchet and Thibault Damour.
\newblock {Hereditary effects in gravitational radiation}.
\newblock {\em Phys. Rev. D}, 46:4304--4319, 1992.

\bibitem{Blanchet:1993ec}
L.~Blanchet and Gerhard Schaefer.
\newblock {Gravitational wave tails and binary star systems}.
\newblock {\em Class. Quant. Grav.}, 10:2699--2721, 1993.

\bibitem{Galley:2009px}
Chad~R. Galley and Manuel Tiglio.
\newblock {Radiation reaction and gravitational waves in the effective field
  theory approach}.
\newblock {\em Phys. Rev. D}, 79:124027, 2009.

\bibitem{Galley:2012qs}
Chad~R. Galley and Adam~K. Leibovich.
\newblock {Radiation reaction at 3.5 post-Newtonian order in effective field
  theory}.
\newblock {\em Phys. Rev. D}, 86:044029, 2012.

\bibitem{Foffa:2011np}
S.~Foffa and Riccardo Sturani.
\newblock {Tail terms in gravitational radiation reaction via effective field
  theory}.
\newblock {\em Phys. Rev. D}, 87(4):044056, 2013.

\bibitem{Galley:2015kus}
Chad~R. Galley, Adam~K. Leibovich, Rafael~A. Porto, and Andreas Ross.
\newblock {Tail effect in gravitational radiation reaction: Time nonlocality
  and renormalization group evolution}.
\newblock {\em Phys. Rev. D}, 93:124010, 2016.

\bibitem{Foffa:2019eeb}
Stefano Foffa and Riccardo Sturani.
\newblock {Hereditary terms at next-to-leading order in two-body gravitational
  dynamics}.
\newblock {\em Phys. Rev. D}, 101(6):064033, 2020.
\newblock [Erratum: Phys.Rev.D 103, 089901 (2021)].

\bibitem{Henry:2021cek}
Quentin Henry, Guillaume Faye, and Luc Blanchet.
\newblock {The current-type quadrupole moment and gravitational-wave mode
  (\ensuremath{\ell}, m) = (2, 1) of compact binary systems at the third
  post-Newtonian order}.
\newblock {\em Class. Quant. Grav.}, 38(18):185004, 2021.

\bibitem{Almeida:2021xwn}
Gabriel~Luz Almeida, Stefano Foffa, and Riccardo Sturani.
\newblock {Tail contributions to gravitational conservative dynamics}.
\newblock {\em Phys. Rev. D}, 104(12):124075, 2021.

\bibitem{Foffa:2021pkg}
Stefano Foffa and Riccardo Sturani.
\newblock {Near and far zones in two-body dynamics: An effective field theory
  perspective}.
\newblock {\em Phys. Rev. D}, 104(2):024069, 2021.

\bibitem{Edison:2022cdu}
Alex Edison and Mich\`ele Levi.
\newblock {A tale of tails through generalized unitarity}.
\newblock 2 2022.

\bibitem{Porto:2017dgs}
Rafael~A. Porto and Ira~Z. Rothstein.
\newblock {Apparent ambiguities in the post-Newtonian expansion for binary
  systems}.
\newblock {\em Phys. Rev. D}, 96(2):024062, 2017.

\bibitem{Porto:2017shd}
Rafael~A. Porto.
\newblock {Lamb shift and the gravitational binding energy for binary black
  holes}.
\newblock {\em Phys. Rev. D}, 96(2):024063, 2017.

\bibitem{Foffa:2019yfl}
Stefano Foffa, Rafael~A. Porto, Ira Rothstein, and Riccardo Sturani.
\newblock {Conservative dynamics of binary systems to fourth Post-Newtonian
  order in the EFT approach II: Renormalized Lagrangian}.
\newblock {\em Phys. Rev. D}, 100(2):024048, 2019.

\bibitem{Blanchet:2019rjs}
Luc Blanchet, Stefano Foffa, Fran\c{c}ois Larrouturou, and Riccardo Sturani.
\newblock {Logarithmic tail contributions to the energy function of circular
  compact binaries}.
\newblock {\em Phys. Rev. D}, 101(8):084045, 2020.

\bibitem{Bini:2021gat}
Donato Bini, Thibault Damour, and Andrea Geralico.
\newblock {Radiative contributions to gravitational scattering}.
\newblock {\em Phys. Rev. D}, 104(8):084031, 2021.

\bibitem{Bern:2021yeh}
Zvi Bern, Julio Parra-Martinez, Radu Roiban, Michael~S. Ruf, Chia-Hsien Shen,
  Mikhail~P. Solon, and Mao Zeng.
\newblock {Scattering Amplitudes, the Tail Effect, and Conservative Binary
  Dynamics at O(G4)}.
\newblock {\em Phys. Rev. Lett.}, 128(16):161103, 2022.

\bibitem{Keldysh:1964ud}
L.~V. Keldysh.
\newblock {Diagram technique for nonequilibrium processes}.
\newblock {\em Zh. Eksp. Teor. Fiz.}, 47:1515--1527, 1964.

\bibitem{Galley:2012hx}
Chad~R. Galley.
\newblock {Classical Mechanics of Nonconservative Systems}.
\newblock {\em Phys. Rev. Lett.}, 110(17):174301, 2013.

\bibitem{Galley:2014wla}
Chad~R. Galley, David Tsang, and Leo~C. Stein.
\newblock {The principle of stationary nonconservative action for classical
  mechanics and field theories}.
\newblock 12 2014.

\bibitem{Schwinger:1960qe}
Julian~S. Schwinger.
\newblock {Brownian motion of a quantum oscillator}.
\newblock {\em J. Math. Phys.}, 2:407--432, 1961.

\bibitem{Iwasaki:1971vb}
Y.~Iwasaki.
\newblock {Quantum theory of gravitation vs. classical theory. - fourth-order
  potential}.
\newblock {\em Prog. Theor. Phys.}, 46:1587--1609, 1971.

\bibitem{Donoghue:1993eb}
John~F. Donoghue.
\newblock {Leading quantum correction to the Newtonian potential}.
\newblock {\em Phys. Rev. Lett.}, 72:2996--2999, 1994.

\bibitem{Donoghue:1994dn}
John~F. Donoghue.
\newblock {General relativity as an effective field theory: The leading quantum
  corrections}.
\newblock {\em Phys. Rev. D}, 50:3874--3888, 1994.

\bibitem{Donoghue:1995cz}
John~F. Donoghue.
\newblock {Introduction to the effective field theory description of gravity}.
\newblock In {\em {Advanced School on Effective Theories}}, 6 1995.

\bibitem{Burgess:2003jk}
C.~P. Burgess.
\newblock {Quantum gravity in everyday life: General relativity as an effective
  field theory}.
\newblock {\em Living Rev. Rel.}, 7:5--56, 2004.

\bibitem{Holstein:2004dn}
Barry~R. Holstein and John~F. Donoghue.
\newblock {Classical physics and quantum loops}.
\newblock {\em Phys. Rev. Lett.}, 93:201602, 2004.

\bibitem{Bjerrum-Bohr:2002gqz}
N.~E.~J Bjerrum-Bohr, John~F. Donoghue, and Barry~R. Holstein.
\newblock {Quantum gravitational corrections to the nonrelativistic scattering
  potential of two masses}.
\newblock {\em Phys. Rev. D}, 67:084033, 2003.
\newblock [Erratum: Phys.Rev.D 71, 069903 (2005)].

\bibitem{Bjerrum-Bohr:2002fji}
Niels Emil~Jannik Bjerrum-Bohr, John~F. Donoghue, and Barry~R. Holstein.
\newblock {Quantum corrections to the Schwarzschild and Kerr metrics}.
\newblock {\em Phys. Rev. D}, 68:084005, 2003.
\newblock [Erratum: Phys.Rev.D 71, 069904 (2005)].

\bibitem{Holstein:2008sx}
Barry~R. Holstein and Andreas Ross.
\newblock {Spin Effects in Long Range Gravitational Scattering}.
\newblock 2 2008.

\bibitem{Holstein:2008sw}
Barry~R. Holstein and Andreas Ross.
\newblock {Spin Effects in Long Range Electromagnetic Scattering}.
\newblock 2 2008.

\bibitem{Khriplovich:2004cx}
I.~B. Khriplovich and G.~G. Kirilin.
\newblock {Quantum long range interactions in general relativity}.
\newblock {\em J. Exp. Theor. Phys.}, 98:1063--1072, 2004.

\bibitem{Holstein:2006pq}
Barry~R. Holstein.
\newblock {Factorization in graviton scattering and the 'natural' value of the
  g-factor}.
\newblock 7 2006.

\bibitem{Bjerrum-Bohr:2013bxa}
N.~E.~J. Bjerrum-Bohr, John~F. Donoghue, and Pierre Vanhove.
\newblock {On-shell Techniques and Universal Results in Quantum Gravity}.
\newblock {\em JHEP}, 02:111, 2014.

\bibitem{Bern:1994zx}
Zvi Bern, Lance~J. Dixon, David~C. Dunbar, and David~A. Kosower.
\newblock {One loop n point gauge theory amplitudes, unitarity and collinear
  limits}.
\newblock {\em Nucl. Phys. B}, 425:217--260, 1994.

\bibitem{Bertotti:1956pxu}
B.~Bertotti.
\newblock {On gravitational motion}.
\newblock {\em Nuovo Cim.}, 4(4):898--906, 1956.

\bibitem{Bel:1981be}
LLuis Bel, T.~Damour, N.~Deruelle, J.~Ibanez, and J.~Martin.
\newblock {Poincar\'e-invariant gravitational field and equations of motion of
  two pointlike objects: The postlinear approximation of general relativity}.
\newblock {\em Gen. Rel. Grav.}, 13:963--1004, 1981.

\bibitem{Damour:2016gwp}
Thibault Damour.
\newblock {Gravitational scattering, post-Minkowskian approximation and
  Effective One-Body theory}.
\newblock {\em Phys. Rev. D}, 94(10):104015, 2016.

\bibitem{Cheung:2018wkq}
Clifford Cheung, Ira~Z. Rothstein, and Mikhail~P. Solon.
\newblock {From Scattering Amplitudes to Classical Potentials in the
  Post-Minkowskian Expansion}.
\newblock {\em Phys. Rev. Lett.}, 121(25):251101, 2018.

\bibitem{Cristofoli:2019neg}
Andrea Cristofoli, N.~E.~J. Bjerrum-Bohr, Poul~H. Damgaard, and Pierre Vanhove.
\newblock {Post-Minkowskian Hamiltonians in general relativity}.
\newblock {\em Phys. Rev. D}, 100(8):084040, 2019.

\bibitem{Bern:2019nnu}
Zvi Bern, Clifford Cheung, Radu Roiban, Chia-Hsien Shen, Mikhail~P. Solon, and
  Mao Zeng.
\newblock {Scattering Amplitudes and the Conservative Hamiltonian for Binary
  Systems at Third Post-Minkowskian Order}.
\newblock {\em Phys. Rev. Lett.}, 122(20):201603, 2019.

\bibitem{Bern:2019crd}
Zvi Bern, Clifford Cheung, Radu Roiban, Chia-Hsien Shen, Mikhail~P. Solon, and
  Mao Zeng.
\newblock {Black Hole Binary Dynamics from the Double Copy and Effective
  Theory}.
\newblock {\em JHEP}, 10:206, 2019.

\bibitem{Cheung:2020gyp}
Clifford Cheung and Mikhail~P. Solon.
\newblock {Classical gravitational scattering at $ \mathcal{O} $(G$^{3}$) from
  Feynman diagrams}.
\newblock {\em JHEP}, 06:144, 2020.

\bibitem{Kalin:2020fhe}
Gregor K\"alin, Zhengwen Liu, and Rafael~A. Porto.
\newblock {Conservative Dynamics of Binary Systems to Third Post-Minkowskian
  Order from the Effective Field Theory Approach}.
\newblock {\em Phys. Rev. Lett.}, 125(26):261103, 2020.

\bibitem{Bjerrum-Bohr:2021din}
N.~Emil~J. Bjerrum-Bohr, Poul~H. Damgaard, Ludovic Plant\'e, and Pierre
  Vanhove.
\newblock {The amplitude for classical gravitational scattering at third
  Post-Minkowskian order}.
\newblock {\em JHEP}, 08:172, 2021.

\bibitem{Bern:2021dqo}
Zvi Bern, Julio Parra-Martinez, Radu Roiban, Michael~S. Ruf, Chia-Hsien Shen,
  Mikhail~P. Solon, and Mao Zeng.
\newblock {Scattering Amplitudes and Conservative Binary Dynamics at ${\cal
  O}(G^4)$}.
\newblock {\em Phys. Rev. Lett.}, 126(17):171601, 2021.

\bibitem{Dlapa:2021npj}
Christoph Dlapa, Gregor K\"alin, Zhengwen Liu, and Rafael~A. Porto.
\newblock {Dynamics of binary systems to fourth Post-Minkowskian order from the
  effective field theory approach}.
\newblock {\em Phys. Lett. B}, 831:137203, 2022.

\bibitem{Bjerrum-Bohr:2014zsa}
N.~E.~J. Bjerrum-Bohr, John~F. Donoghue, Barry~R. Holstein, Ludovic Plant\'e,
  and Pierre Vanhove.
\newblock {Bending of Light in Quantum Gravity}.
\newblock {\em Phys. Rev. Lett.}, 114(6):061301, 2015.

\bibitem{Bjerrum-Bohr:2016hpa}
N.~E.~J. Bjerrum-Bohr, John~F. Donoghue, Barry~R. Holstein, Ludovic Plante, and
  Pierre Vanhove.
\newblock {Light-like Scattering in Quantum Gravity}.
\newblock {\em JHEP}, 11:117, 2016.

\bibitem{Bjerrum-Bohr:2017dxw}
N.~Emil~J. Bjerrum-Bohr, Barry~R. Holstein, John~F. Donoghue, Ludovic Plant\'e,
  and Pierre Vanhove.
\newblock {Illuminating Light Bending}.
\newblock {\em PoS}, CORFU2016:077, 2017.

\bibitem{Bai:2016ivl}
Dong Bai and Yue Huang.
\newblock {More on the Bending of Light in Quantum Gravity}.
\newblock {\em Phys. Rev. D}, 95(6):064045, 2017.

\bibitem{Chi:2019owc}
Huan-Hang Chi.
\newblock {Graviton Bending in Quantum Gravity from One-Loop Amplitudes}.
\newblock {\em Phys. Rev. D}, 99(12):126008, 2019.

\bibitem{AccettulliHuber:2020oou}
Manuel Accettulli~Huber, Andreas Brandhuber, Stefano De~Angelis, and Gabriele
  Travaglini.
\newblock {Eikonal phase matrix, deflection angle and time delay in effective
  field theories of gravity}.
\newblock {\em Phys. Rev. D}, 102(4):046014, 2020.

\bibitem{Carroll:2004st}
Sean~M. Carroll.
\newblock {\em {Spacetime and Geometry}}.
\newblock Cambridge University Press, 7 2019.

\bibitem{Maggiore:2007ulw}
Michele Maggiore.
\newblock {\em {Gravitational Waves. Vol. 1: Theory and Experiments}}.
\newblock Oxford Master Series in Physics. Oxford University Press, 2007.

\bibitem{Taracchini:2013rva}
Andrea Taracchini et~al.
\newblock {Effective-one-body model for black-hole binaries with generic mass
  ratios and spins}.
\newblock {\em Phys. Rev. D}, 89(6):061502, 2014.

\bibitem{Bohe:2016gbl}
Alejandro Boh\'e et~al.
\newblock {Improved effective-one-body model of spinning, nonprecessing binary
  black holes for the era of gravitational-wave astrophysics with advanced
  detectors}.
\newblock {\em Phys. Rev. D}, 95(4):044028, 2017.

\bibitem{Becher:2014oda}
Thomas Becher, Alessandro Broggio, and Andrea Ferroglia.
\newblock {\em {Introduction to Soft-Collinear Effective Theory}}, volume 896.
\newblock Springer, 2015.

\bibitem{Cristofoli:2018bex}
Andrea Cristofoli.
\newblock {An Effective Field Theory Approach to the Two-Body Problem in
  General Relativity}.
\newblock Master's thesis, Padua U., 3 2018.

\bibitem{Zeldovich:1974gvh}
Y.~B. Zel'dovich and A.~G. Polnarev.
\newblock {Radiation of gravitational waves by a cluster of superdense stars}.
\newblock {\em Sov. Astron.}, 18:17, 1974.

\bibitem{Christodoulou:1991cr}
D.~Christodoulou.
\newblock {Nonlinear nature of gravitation and gravitational wave experiments}.
\newblock {\em Phys. Rev. Lett.}, 67:1486--1489, 1991.

\bibitem{Bini:2020hmy}
Donato Bini, Thibault Damour, and Andrea Geralico.
\newblock {Sixth post-Newtonian nonlocal-in-time dynamics of binary systems}.
\newblock {\em Phys. Rev. D}, 102(8):084047, 2020.

\bibitem{Kalin:2022hph}
Gregor K\"alin, Jakob Neef, and Rafael~A. Porto.
\newblock {Radiation-Reaction in the Effective Field Theory Approach to
  Post-Minkowskian Dynamics}.
\newblock 7 2022.

\bibitem{Jakobsen:2022psy}
Gustav~Uhre Jakobsen, Gustav Mogull, Jan Plefka, and Benjamin Sauer.
\newblock {All Things Retarded: Radiation-Reaction in Worldline Quantum Field
  Theory}.
\newblock 7 2022.

\bibitem{xactref}
J.~M.~M. Garc\'ia.
\newblock {xAct: Efficient tensor computer algebra for Mathematica.}

\bibitem{Lee:2013mka}
Roman~N. Lee.
\newblock {LiteRed 1.4: a powerful tool for reduction of multiloop integrals}.
\newblock {\em J. Phys. Conf. Ser.}, 523:012059, 2014.

\bibitem{Weinberg:2005vy}
Steven Weinberg.
\newblock {Quantum contributions to cosmological correlations}.
\newblock {\em Phys. Rev. D}, 72:043514, 2005.

\bibitem{Bjerrum-Bohr:2022blt}
N.~E.~J. Bjerrum-Bohr, P.~H. Damgaard, L.~Plante, and P.~Vanhove.
\newblock {The SAGEX Review on Scattering Amplitudes, Chapter 13:
  Post-Minkowskian expansion from Scattering Amplitudes}.
\newblock 3 2022.

\bibitem{Feynman:1996kb}
R.~P. Feynman.
\newblock {\em {Feynman lectures on gravitation}}.
\newblock 1996.

\bibitem{Feynman:1963ax}
R.~P. Feynman.
\newblock {Quantum theory of gravitation}.
\newblock {\em Acta Phys. Polon.}, 24:697--722, 1963.

\bibitem{DeWitt:1967yk}
Bryce~S. DeWitt.
\newblock {Quantum Theory of Gravity. 1. The Canonical Theory}.
\newblock {\em Phys. Rev.}, 160:1113--1148, 1967.

\bibitem{DeWitt:1967ub}
Bryce~S. DeWitt.
\newblock {Quantum Theory of Gravity. 2. The Manifestly Covariant Theory}.
\newblock {\em Phys. Rev.}, 162:1195--1239, 1967.

\bibitem{DeWitt:1967uc}
Bryce~S. DeWitt.
\newblock {Quantum Theory of Gravity. 3. Applications of the Covariant Theory}.
\newblock {\em Phys. Rev.}, 162:1239--1256, 1967.

\bibitem{tHooft:1974toh}
Gerard 't~Hooft and M.~J.~G. Veltman.
\newblock {One loop divergencies in the theory of gravitation}.
\newblock {\em Ann. Inst. H. Poincare Phys. Theor. A}, 20:69--94, 1974.

\bibitem{Grignani:2020ahv}
Gianluca Grignani, Troels Harmark, Marta Orselli, and Andrea Placidi.
\newblock {Fixing the non-relativistic expansion of the 1PM potential}.
\newblock {\em JHEP}, 12:142, 2020.

\bibitem{Kosower:2018adc}
David~A. Kosower, Ben Maybee, and Donal O'Connell.
\newblock {Amplitudes, Observables, and Classical Scattering}.
\newblock {\em JHEP}, 02:137, 2019.

\bibitem{Bern:2010ue}
Zvi Bern, John Joseph~M. Carrasco, and Henrik Johansson.
\newblock {Perturbative Quantum Gravity as a Double Copy of Gauge Theory}.
\newblock {\em Phys. Rev. Lett.}, 105:061602, 2010.

\bibitem{Bonciani:2021okt}
R.~Bonciani et~al.
\newblock {Two-Loop Four-Fermion Scattering Amplitude in QED}.
\newblock {\em Phys. Rev. Lett.}, 128(2):022002, 2022.

\bibitem{Mandal:2022vju}
Manoj~K. Mandal, Pierpaolo Mastrolia, Jonathan Ronca, and William~J.
  Bobadilla~Torres.
\newblock {Two-loop scattering amplitude for heavy-quark pair production
  through light-quark annihilation in QCD}.
\newblock 4 2022.

\bibitem{Bjerrum-Bohr:2018xdl}
N.~E.~J. Bjerrum-Bohr, Poul~H. Damgaard, Guido Festuccia, Ludovic Plant\'e, and
  Pierre Vanhove.
\newblock {General Relativity from Scattering Amplitudes}.
\newblock {\em Phys. Rev. Lett.}, 121(17):171601, 2018.

\bibitem{EventHorizonTelescope:2019dse}
Kazunori Akiyama et~al.
\newblock {First M87 Event Horizon Telescope Results. I. The Shadow of the
  Supermassive Black Hole}.
\newblock {\em Astrophys. J. Lett.}, 875:L1, 2019.

\bibitem{EventHorizonTelescope:2019ggy}
Kazunori Akiyama et~al.
\newblock {First M87 Event Horizon Telescope Results. VI. The Shadow and Mass
  of the Central Black Hole}.
\newblock {\em Astrophys. J. Lett.}, 875(1):L6, 2019.

\bibitem{doi:10.1119/1.1570416}
Jeremiah Bodenner and Clifford~M. Will.
\newblock Deflection of light to second order: A tool for illustrating
  principles of general relativity.
\newblock {\em American Journal of Physics}, 71(8):770--773, 2003.

\bibitem{Nogueira:1991ex}
Paulo Nogueira.
\newblock {Automatic Feynman graph generation}.
\newblock {\em J. Comput. Phys.}, 105:279--289, 1993.

\bibitem{Weinberg:1965nx}
Steven Weinberg.
\newblock {Infrared photons and gravitons}.
\newblock {\em Phys. Rev.}, 140:B516--B524, 1965.

\bibitem{Donoghue:1996mt}
John~F. Donoghue and Tibor Torma.
\newblock {On the power counting of loop diagrams in general relativity}.
\newblock {\em Phys. Rev. D}, 54:4963--4972, 1996.

\bibitem{Akhoury:2013yua}
Ratindranath Akhoury, Ryo Saotome, and George Sterman.
\newblock {High Energy Scattering in Perturbative Quantum Gravity at Next to
  Leading Power}.
\newblock {\em Phys. Rev. D}, 103(6):064036, 2021.

\bibitem{Bjerrum-Bohr:2004qcf}
Niels~EmilJannik Bjerrum-Bohr.
\newblock {\em {Quantum gravity, effective fields and string theory}}.
\newblock PhD thesis, Bohr Inst., 2004.

\end{thebibliography}
\end{document}